\documentclass[traditabstract]{aa}  

\usepackage{graphicx}
\usepackage{txfonts}
\usepackage[toc,title,page]{appendix}

\usepackage{caption}
\usepackage{subcaption}
\usepackage{epstopdf}
\usepackage{amsmath}
\usepackage[english]{babel}
\usepackage{amssymb}
\usepackage{wrapfig}
\usepackage{pdflscape}
\usepackage{lscape}
\usepackage{longtable}
\usepackage{appendix}
\usepackage{textcomp}
\usepackage{multirow}
\bibpunct{(}{)}{;}{a}{}{,}
\usepackage{array}
\usepackage{ragged2e}
\usepackage{adjustbox}
\usepackage{gensymb}

\begin{document}

   \title{Unravelling the origin of extended radio emission in narrow-line Seyfert 1 galaxies with the JVLA}


   \author{E. Järvelä\inst{1,2}\fnmsep\thanks{ejarvela@sciops.esa.int}
          \and
          R. Dahale\inst{3}
          \and 
          L. Crepaldi\inst{4}
          \and 
          M. Berton\inst{5,6}
          \and
          E. Congiu\inst{7}
          \and
          R. Antonucci\inst{2}
          }

   \institute{European Space Agency (ESA), European Space Astronomy Centre (ESAC), Camino Bajo del Castillo s/n, 28692 Villanueva de la Ca\~nada, Madrid, Spain;
         \and
           Department of Physics, University of California, Santa Barbara, CA 93106, USA
         \and
             Indian Institute of Science Education and Research Kolkata, Mohanpur, Nadia, West Bengal - 741246, India;
        \and
            Dipartimento di Fisica e Astronomia ``G. Galilei'', Universit\`a di Padova, Vicolo dell'Osservatorio 3, 35122, Padova, Italy;
        \and
            Finnish Centre for Astronomy with ESO - University of Turku, Vesilinnantie 5, FIN-20500, Turku, Finland;
        \and 
            Aalto University Metsähovi Radio Observatory, Metsähovintie 114, FIN-02450, Kylmälä, Finland;
        \and
            Departamento de Astronom\'{i}a, Universidad de Chile, Camino del Observatorio 1515, Las Condes, Santiago, Chile
             }

   \date{Received; accepted}

  \abstract
   {Narrow-line Seyfert 1 (NLS1) galaxies are believed to be active galactic nuclei (AGN) in the early stages of their evolution. Some dozens of them have been found to host relativistic jets, whilst the majority of NLS1s has not even been detected in radio, emphasising the heterogeneity of the class in this band. In this paper, our aim is to determine the predominant source of radio emission in a sample of 44 NLS1s, selected based on their extended kpc-scale radio morphologies at 5.2~GHz. We accomplish this by analysing their spatially resolved radio spectral index maps, centred at 5.2~GHz, as the spectral index carries information about the production mechanisms of the emission. In addition, we utilise several diagnostics based on mid-infrared emission to estimate the star formation activity of their host galaxies. These data are complemented by archival data to draw a more complete picture of each source. We find an extraordinary diversity among our sample. Approximately equal fractions ($\sim$10-12 sources) of our sources can be identified as AGN-dominated, composite, and host-dominated. Among the AGN-dominated sources are a few NLS1s with very extended jets, reaching distances of tens of kpc from the nucleus. One of these, J0814+5609, hosts the most extended jets found in an NLS1 so far. We also identify five NLS1s that could be classified as compact steep-spectrum sources. In addition, one source shows a possible kpc-scale relic reaching well outside the host galaxy and restarted nuclear activity, and one could belong to the NLS1 subclass hosting relativistic jets that seems to be absorbed at lower radio frequencies (< 10~GHz). We further conclude that due to the variety seen in NLS1s simple proxies, such as the star formation diagnostics also employed in this paper, and the radio loudness parameter, are not ideal tools for characterising NLS1s. We emphasise the necessity of examining NLS1s as individuals, instead of making assumptions based on their classification. When these issues are properly taken into account, NLS1s offer an exceptional environment to study the interplay of the host galaxy and several AGN-related phenomena, such as jets and outflows.}


   \maketitle
   
\newcommand{\kms}{km s$^{-1}$}
\section{Introduction}
\defcitealias{2018berton1}{B18}

Among active galactic nuclei (AGN), the class of narrow-line Seyfert 1 (NLS1) galaxies has been in the spotlight for the last decade. In the optical spectra of NLS1s the emission lines originating from the high-density broad-line region (BLR) and the low-density narrow-line region (NLR) are of comparable width, unlike in broad-line AGN. By definition, the full-width at half maximum of the broad H$\beta$, FWHM(H$\beta$) $<$ 2000~\kms \citep{1985osterbrock1}, whereas the emission lines arising from the NLR usually have widths of a few hundred \kms. However, we remark that this limit is used mostly for historical reasons, since no real threshold is present in the FWHM(H$\beta$) distribution at least up to 4000 \kms\ \citep[e.g., see][]{2018marziani1}. To ensure that they are Type 1 AGN, and we have a direct view of the BLR an additional criterion for an NLS1 classification requires the flux ratio of [O~III] and total (broad $+$ narrow) H$\beta$ to be $<$ 3. This threshold has been found to be determinative in separating Type 1 and 2 AGN \citep[e.g., ][]{1981shuder1}. NLS1s also often exhibit strong Fe~II multiplets \citep{1989goodrich1}, confirming the unobstructed view of the central engine.

In the quasar main sequence \citep[MS, Fig. 2 in][]{2018marziani1}, originally derived by means of principal component analysis \citep[e.g.,][]{1992boroson1}, and believed to be mainly driven by the Eddington ratio and orientation, NLS1s are population A sources (FWHM(H$\beta$) $<$ 4000~\kms). The horizontal branch in the MS is driven by the Eddington ratio, and R4570\footnote{R4570 is the flux ratio of Fe~II and H$\beta$.}, that also anticorrelates with the [O~III] line strength, is used as its proxy. The spread on the vertical branch (FWHM(H$\beta$)), is though to be mainly dominated by orientation if R4570 is kept fixed (\citealp{2014shen1}, but see, \citealp{2018panda1}). NLS1s show very diverse values of R4570, and since it correlates with the Eddington ratio, this implies that a considerable fraction of NLS1s are accreting close to or even above the Eddington limit \citep{1992boroson1}. Naturally, due to their classification, NLS1s do not show considerable spread in the FWHM(H$\beta$) values, but interestingly they, and other population A sources, do show different emission line profiles (Lorentzian) compared to the population B sources (FWHM(H$\beta$) $>$ 4000~\kms and Gaussian) \citep{2000sulentic2, 2003marziani1, 2020berton1}. Lorentzian profiles are believed to be dominated by turbulent motion in the BLR \citep{2011kollatschny1,2012goad1}, rather than rotation, and thus orientation might not play a crucial role in most of the population A sources. If this is the case, the narrowness \footnote{If the BLR is virialised, the velocity dispersion of the gas clouds, i.e. the FWHM of the permitted lines, depends quadratically on the mass of the central object. Narrow permitted lines, then, may originate because of an undermassive black hole.} of the broad emission lines in NLS1s would be caused by a relatively low-mass black hole, typically $<10^8$ M$_\odot$ \citep{2011peterson1}, compared to, for example, broad-line Seyfert 1 galaxies, that usually have M$_\textrm{BH}$ $>10^8$ M$_\odot$. This seems to be confirmed by reverberation mapping campaigns focusing on NLS1s \citep{2016wang1,2018du1,2019du1}. Interestingly, it has been found that the H$\beta$ lags compared to changes in the continuum in high Eddington ratio sources are shorter than expected, suggesting that this quality affects the distance of the BLR clouds from the continuum \citep{2018du1, 2020dallabonta1}.

Assuming that the black hole masses in NLS1s are $<10^8$ M$_\odot$, and since the mass can only increase with time, NLS1s may constitute an early stage of AGN life cycle and they will eventually grow into fully developed BLS1s \citep{2000mathur1, 2000sulentic1, 2017fraixburnet1}. However, an alternative hypothesis is that the narrowness of permitted lines originates due to the BLR geometry. If the BLR is flattened, when observed pole-on we only observe the part of the velocity vector directed towards us, and since the rotation happens on a plane misaligned with our line of sight we do not observe considerable Doppler broadening of the lines, and the emission lines appear narrow. In this scenario, the black hole mass of NLS1s can be significantly higher than $10^8$ M$_\odot$, and NLS1s would be no different from BLS1s and other broad-line AGN \citep{2008decarli1}. However, the earlier mentioned reverberation mapping studies, and other observational properties of NLS1s, such as their host galaxy morphologies (\citealp[e.g.,][but see]{2001krongold1, 2006deo1, 2008anton1, 2016kotilainen1, 2018jarvela1, 2019berton1, 2020olguiniglesias1, 2021hamilton1} \citealp{2017dammando1, 2018dammando1}), seem to indicate that the black hole mass is genuinely low when compared to BLS1s and other broad-line AGN.

The central engine of an AGN can launch a host of different effluxes, spanning from highly collimated powerful relativistic jet, through lower power non-relativistic jets, to wide-angle outflows generated by nuclear winds. Traditionally, the most powerful relativistic jets have been associated with the most massive supermassive black holes residing in old elliptical galaxies \citep{2000laor1}, whereas low-power jets and outflows can be seen in a wider variety of AGN. Interestingly, some NLS1s ($\sim$7\%, \citealp{2006komossa1}) show prominent radio emission and several blazar-like properties, such as high brightness temperature, prominent variability, and a double-humped spectral energy distribution \citep{2008yuan1}. The discovery of gamma-ray emission from a handful of them ($\sim$20, \citealp{2018romano1, 2020jarvela1, 2021rakshit1}) proved that, just like blazars and radio galaxies, NLS1s can harbour powerful relativistic jets. As the NLS1s that do not host relativistic jets may be the progenitors of BLS1s, it has been suggested that NLS1s with relativistic jets are an early stage of the life cycle of flat-spectrum radio quasars (FSRQs) \citep{2015foschini1, 2017foschini1, 2020foschini1}. Furthermore, several authors hypothesise that when seen at a larger angle, relativistic jetted NLS1s may appear as kinematically young radio galaxies, such as compact steep-spectrum sources (CSS, \citealp{2001oshlack1, 2006gallo1, 2006komossa1, 2014caccianiga1, 2016berton1, 2017berton1, 2017foschini1, 2017caccianiga1, 2021yao1}). At radio frequencies, when their relativistic jet axis is close to the line of sight, these NLS1s typically show a flat spectrum and a compact morphology on kpc-scale \citepalias[\citealp{2018berton1}, from now on ][]{2018berton1}. When instead the relativistic jet has a larger inclination, they tend to become fainter due to the decreasing impact of boosting effects, more extended, and to show, at least in a few cases, radio lobes \citepalias[][Vietri et al. in prep.]{2018berton1}.

However, the vast majority of NLS1s is radio-quiet or radio-silent\footnote{Radio-quiet AGN by definition have a ratio $S_{\rm{radio}}/S_{\rm{optical}} < 10$, where $S$ are the flux densities at 5~GHz and in B-band, respectively \citep{1989kellermann1}. Radio-silent sources have no known detection in radio.}. The origin of the radio emission in these NLS1s is still debated. The presence of jets, even relativistic ones, in them cannot be ruled out. Because of the non-linear scaling between jet power and black hole mass \citep{2003heinz1}, and the still widely unexplored impact of the magnetic flux density \citep{2021chamani1}, jets, even relativistic ones, in NLS1s can be weak, and barely dominate the total radio emission produced by the galaxy. In some extreme cases, they may even be completely invisible at low radio frequencies due to absorption by a screen of ionised gas, making the AGN appear as radio-quiet or -silent \citep{2018lahteenmaki1,2020berton2,2021jarvela1}. Alternative sources of radio emission on scales smaller than $\sim$0.1~kpc can be the accretion disk corona or weakly collimated disk winds. Also wide-angle outflows due to nuclear winds \citep{2007proga1} are common in these sources. In radio frequencies these outflows are characterised by non-collimated morphologies and steep spectral indices \citep{2012fauchergiguere1,2019panessa1} that can show steepening toward high frequencies \citep{2010jiang1}. At the host galaxy scale, instead, the radio emission from star formation activity typically dominates, in the form of synchrotron emission from supernova remnants and free-free emission from H~II regions \citep{2018lister1, 2019panessa1}. All of these components are often present at the same time, and to distinguish them is not an easy task. \citetalias{2018berton1} found that, on average, radio-quiet objects often show an extended morphology with a spectral index steeper than what is found in NLS1s with known relativistic jets, but they could not draw many conclusions on the sample as a whole. When the radio morphology is extended, a detailed analysis of each source is the preferred way to understand exactly the mechanisms at play in each individual source. 

The goal of this paper is to accurately analyse the sources in \citetalias{2018berton1} with an extended morphology to understand the nature and the origins of their radio emission. A possible way to do this, and to tell apart the different sources of radio emission, is by studying spatially resolved spectral index maps. Such technique can provide some indications on the nature of the extended emission in sources without jets, while in jetted NLS1s it can reveal the presence of interaction between the relativistic jets and the interstellar medium. Furthermore, we carried out a detailed research in the literature for each source, to obtain as much useful information as possible to paint a complete picture of all of our targets. The paper is organised as follows: in Sect.~\ref{sec:sample} we briefly overview the sample, in Sect.~\ref{sec:data-analysis} we describe the data reduction and the production of the radio and spectral index maps, in Sect.~\ref{sec:radiofromsf} we overview star formation mechanisms that produce radio emission. We also introduce some diagnostic tools we can use to estimate its contribution in our source, and also discuss the possible issues with these tools. Then, in Sect.~\ref{sec:results} we summarise the star formation results for the whole sample, and discuss each source individually. Finally, in Sect.~\ref{sec:discussion} we discuss our results and their implications, and we conclude with a brief summary in Sect.~\ref{sec:summary}. Throughout the paper, we use the standard $\Lambda$CDM cosmology, with $H_0$ = 70 km s$^{-1}$ Mpc$^{-1}$, and $\Omega_{\Lambda}$ = 0.73 \citep{2011komatsu1}. For spectral indices we adopt the convention of $S_{\nu} \propto \nu ^{\alpha}$ at frequency $\nu$.

\section{Sample}
\label{sec:sample}

This paper is continuation to \citetalias{2018berton1}, where the original sample selection is explained. Our sources were selected from their sample based on the radio morphology. \citetalias{2018berton1} presents Karl G. Jansky Very Large Array (JVLA) A-configuration observations of 74 NLS1s obtained in project 15A-283 (P.I. J. Richards). The observations are centred at 5.2~GHz, with a bandwidth of 2~GHz. From the original sample we selected for further analysis all sources whose radio morphology was classified either as extended (E) or intermediate (I) (see Table A.1. in \citetalias{2018berton1}). In extended sources the ratio between the peak flux density and the integrated flux density is < 0.75, and in intermediate sources the ratio is between 0.75 and 0.95. Sources with a ratio > 0.95 were classified as compact (C). We decided to leave out the compact sources in this study since most of them show a flat radio spectrum, and they lack the extended radio emission we are especially interested in. This selection criterion resulted in a sample of 20 extended and 26 intermediate sources. Two extended sources were left out due to bad data quality, so our final sample size is 44 sources. The compact sources will be a subject of a future study. 

\begin{figure}
    \centering
    \includegraphics[width=9cm]{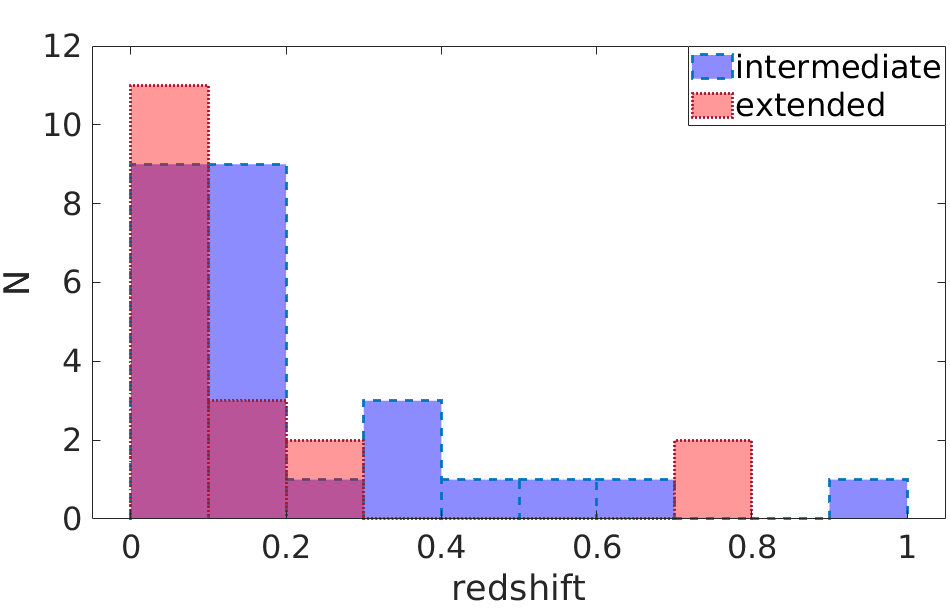}
    \caption{Redshift distribution of our sample, divided based on the radio morphology classification. Sources with intermediate radio morphology marked in blue with a dashed edge line, sources with extended radio morphology marked in red with a dotted edge line. }
    \label{fig:zdist}
\end{figure}

\begin{figure}
    \centering
    \includegraphics[width=9cm]{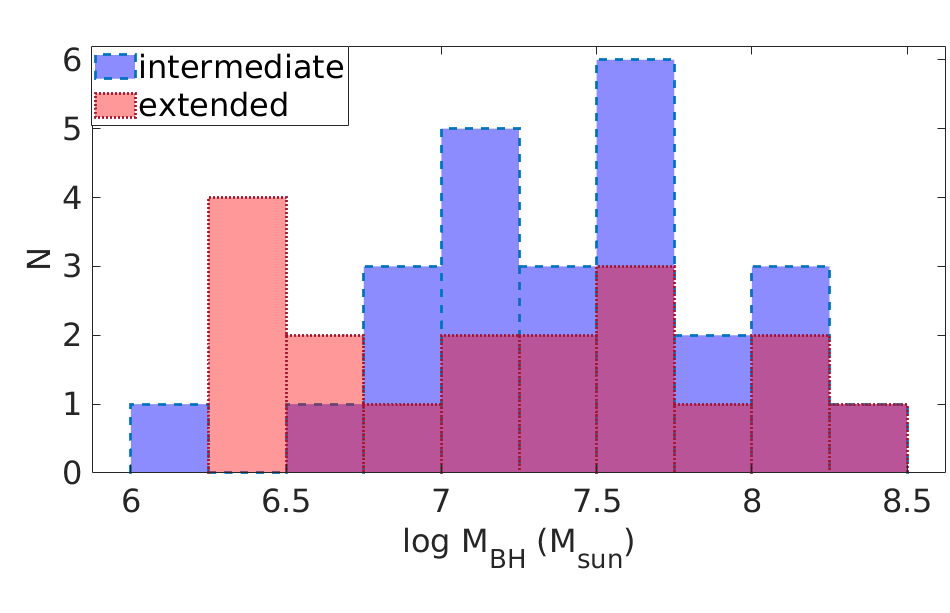}
    \caption{Logarithmic black hole mass distribution of our sample. Division to subsamples, and colours and edge styles as in Fig.~\ref{fig:zdist}.}
    \label{fig:bhdist}
\end{figure}

\begin{figure}
    \centering
    \includegraphics[width=9cm]{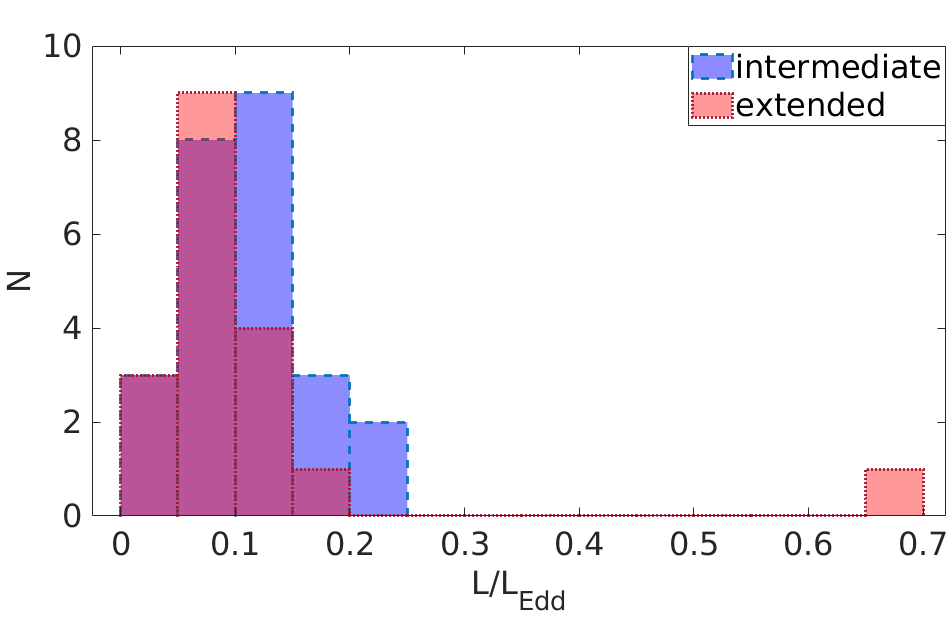}
    \caption{Eddington ratio distribution of our sample. Division to subsamples, and colours and edge styles as in Fig.~\ref{fig:zdist}.}
    \label{fig:ledddist}
\end{figure}

Basic information of the sample, including name, coordinates, redshift, scale, and radio morphological type, is listed in Table~\ref{tab:basicdata}. The redshift, black hole mass, Eddington ratio, and radio luminosity distributions of our sample, taken from \citetalias{2018berton1} and \citet{2020berton2}, divided to intermediate and extended source sample, are shown in Figs.~\ref{fig:zdist}-\ref{fig:ledddist}.  

\begin{figure}
    \centering
    \includegraphics[width=9cm]{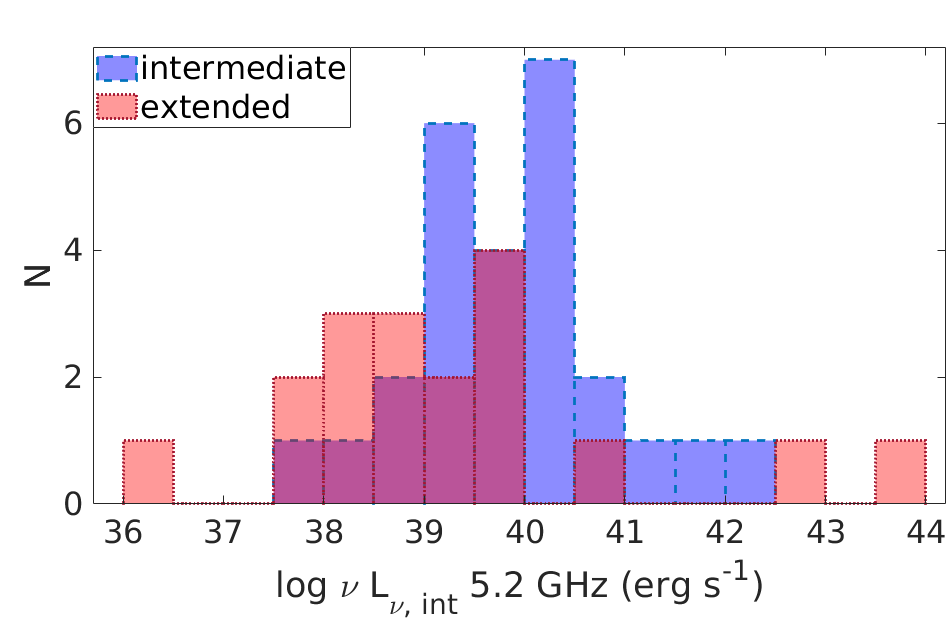}
    \caption{Distribution of the 5.2~GHz integrated radio luminosities of our sources. Division to subsamples, and colours and edge styles as in Fig.~\ref{fig:zdist}.}
    \label{fig:lradiodist}
\end{figure}


\section{Data analysis}
\label{sec:data-analysis}

\subsection{Radio data calibration}

We reduced the radio data again, and used the standard Extended VLA (EVLA) pipeline 5.0.0 for the calibration. Each data set had its own flux calibrator. After running the pipeline we split the individual measurement sets of each source from the main data sets, averaging the data over the 64 channels of each of the 16 spectral windows and over ten seconds of exposure time. For further data reduction and analysis after splitting we used CASA version 5.6.2-3 due to its enhanced data reduction capabilities, for example, in producing spectral index maps.

\subsection{Spectral index maps}

We constructed the spectral index maps and their respective error maps following the procedure described in \citet{2015wiegert1}, that we also summarise here. The procedure includes the actual cleaning of the data as well as several post-imaging corrections.

\subsubsection{Cleaning}

The clean algorithm in CASA (\texttt{tclean}) enables multi-term (multi-scale) multi-frequency synthesis, \texttt{mt-mfs} \citep{2011rau1}. The multi-term feature allows a simultaneous fitting of a spectral index over the whole band-width and as a function of the position within the image using a simple power-law. The multi-scale algorithm enables an enhanced modelling of flux components at different scales by using both delta-functions and circular Gaussians in the deconvolution, instead of just delta-functions as in the other clean algorithms. We experimented with the multi-scale feature using some of the most extended sources in our sample, but concluded that using different scales does not significantly alter the resulting spectral index or error maps in comparison to `traditional clean' with scales = 0, and thus we decided to use only the multi-term option.  We used a natural weighting scheme for the majority of our sources, with a few exceptions: for J0347+0105, J1209+3217, and J1317+6010 we used Briggs weighting with robust = 0.0, and for J1047+4725 we used uniform weighting. These four sources suffered from strong sidelobes whose effects we suppressed by using an alternative weighting scheme. 

The \texttt{mt-mfs} algorithm performs the fitting by modelling the spectrum of each flux component (pixel) by a Taylor series expansion about the reference frequency, $\nu_0$. A specific intensity at frequency $\nu$, $I_{\nu}$, can be fit with:

\begin{equation}
    I_{\nu} = I_{\nu_0} \left( \frac{\nu}{\nu_0} \right) ^{\alpha + \beta (\nu / \nu_0)}
\end{equation} \label{eq:taylor}

where $\alpha$ is the spectral index and $\beta$ is the curvature of the power-law. Expanding Eq.~\ref{eq:taylor} in a Taylor series about $\nu_0$ results in Taylor terms, or maps, equal to the number of terms in the Taylor polynomial used. The first map, TT0, corresponds to the specific intensities at $\nu_0$ and is thus equal to the normal radio map. The second map, TT1, is defined so that $\alpha$ = TT1 / TT0. The third term, TT2, describes the spectral curvature and is defined so that $ \rm \beta = TT2 / TT0 -\alpha(\alpha-1)/2$.

The number of Taylor terms is not limited, but in practise using more than two or three terms is usually not needed. Furthermore, the data quality does not usually allow it, since the more terms are used, the higher the required S/N of the data is. Most of our sources are quite faint, and the extended parts, where some curvature could be seen are usually regions of low S/N ($\sim$3-6), and thus not of adequate quality to fit the $\beta$ term. Therefore, we chose to use two Taylor terms to maximise the quality of the $\alpha$ maps. The end products of the cleaning procedure are a TT0 map, centred in the middle of the observing bandwidth at $\sim$5.2~GHz, a corresponding TT1 map, and $\alpha$, and $\Delta\alpha$ maps. The $\Delta\alpha$ map is an error map describing the empirical error estimate based on the errors of the TT0 and TT1 residual images.

In the \texttt{mt-mfs} method the fit is performed on the flux component, resulting in a uniform resolution and S/N over the whole band. This is clearly superior to the traditional way of forming a spectral index map by splitting the observed bandwidth to two and then estimating the spectral index based on these two maps that intrinsically have different resolutions and S/N. However, \texttt{mt-mfs} is not perfect either, and $\alpha$ and $\Delta\alpha$ maps require some post-processing steps, which are described in the next Sections.

\subsubsection{Wide-band primary beam correction and 5$\sigma$ cut-off}
\label{sec:wbpbcorr}

The primary beam varies with frequency and thus imposes its own spectral index onto the Taylor-coefficient images, TT0 and TT1, and the $\alpha$ map. This can be corrected with the CASA task \texttt{widebandpbcor}, which computes a set of primary beams at given frequencies, calculates the Taylor-coefficient images representing the primary beam spectrum, performs the primary beam correction of the Taylor-coefficient images, and finally computes the primary beam corrected $\alpha$ map using the corrected Taylor-coefficient images.

However, this correction cannot account for any variations during a specific observation, including, for example, the slightly changing shape of each telescope, and the rotation of the primary beam on the sky when tracking a source. The primary beam errors also increase with distance from the pointing centre, and it is necessary to understand the impact of these effects on the final $\alpha$ maps to determine their accuracy. \citet{2013bhatnagar1} showed that the effects are insignificant up to the half-power beam width (HPBW), after which they cause considerable errors in total intensity as well as spectral index maps. The HPBW of the JVLA at 5.2~GHz is $\sim$8~arcmin, giving a radius of $\sim$4~arcmin. All of our sources are clearly less extended than this -- the most extended ones being $<$ 6~arcsec -- and the total intensity, as well as the spectral index maps, should be accurate.

Unfortunately we cannot perform any further tests on the correctness of the $\alpha$ maps since our data is limited; we do not have any observations with multiple pointing, or observations carried out in more than one observing sessions. What we can do, however, is to compare the $\alpha$ maps to the spectral index values obtained in \citetalias{2018berton1} where the traditional way of estimating the spectral indices was used. This comparison is made in Section~\ref{sec:spindcomparison}. 

In addition to wide-band primary beam correction, we masked pixels with a S/N < 5$\sigma$, where $\sigma$ is the rms of the corresponding TT0 image of the source. We chose 5$\sigma$ instead of a more traditional 3$\sigma$ threshold since the $\alpha$ maps consistently show extreme or clearly erroneous values when the S/N is low, and thus these peripheral regions in general do not hold valuable information.

\subsubsection{$\Delta\alpha$ cut-off}
\label{sec:deltaalphacutoff}

The primary beam corrected $\alpha$ maps can still show considerable variations, especially near the edges. Also the $\Delta\alpha$ map correspondingly shows very high errors in these regions, implying that the data quality might not have been adequate to accurately estimate the spectral index. We removed the most extreme values by creating an additional mask based on the values of the $\Delta\alpha$ map, and then applying it onto the $\alpha$ map. Following \citet{2015wiegert1} we decided to cut off all the data that has $\Delta\alpha >$ 1. We also experimented with lower values, but the cut-off of the data would have been too drastic, so we decided to use the threshold of $\Delta\alpha >$ 1, even if some less accurate values might remain.

\subsubsection{Smoothing of the $\alpha$ and $\Delta\alpha$ maps}
\label{sec:alphasmoothing}

The Taylor coefficient maps are convolved with the clean beam, but as the $\alpha$ and $\Delta\alpha$ maps are derived from them using mathematical operations their final resolution is not the same as that of the Taylor coefficient maps. We thus smoothed the $\alpha$ and $\Delta\alpha$ maps using the parameters of the clean beam of each source. This considerably decreases the small-scale variance of the $\alpha$ maps and the errors of the $\Delta\alpha$ maps.

\subsubsection{Remaining issues and additional tests}

In summary, the post-imaging correction steps performed to achieve the final $\alpha$ and $\Delta\alpha$ maps were:
\begin{enumerate}
    \item wide-band primary beam correction;
    \item 5$\sigma$ cut-off;
    \item $\Delta\alpha >$ 1 cut-off;
    \item convolution with the clean beam.
\end{enumerate}

The final $\alpha$ and $\Delta\alpha$ maps, overlaid with the normal radio contours, are shown in Appendix~\ref{app:maps}, from Figure~\ref{fig:J0347} to \ref{fig:J2314}, in panels a) and b). Individual sources are discussed in Sec.~\ref{sec:results}. The rms of the maps and the peak and integrated flux densities are listed in Table~\ref{tab:measurements}. We obtained the peak flux density by fitting a 2D Gaussian to the data, and the integrated flux density by summing up all the emission within the 3$\sigma$ contour. The peak flux density error is the one given by CASA when fitting a 2D Gaussian to the data, the errors of the integrated flux densities were estimated with rms per beam $\times$ the square root of the area of emission expressed in beams. In addition, for each source we calculated an average spectral index over the whole region that the $\alpha$ map covers weighted with the surface brightness, and in the core in a region with a radius of 2~px. For some interesting sources we also calculated a spectral index in a region of interest outside the core. Also these regions had a radius of 2~px. The total, core, and region of interest spectral indices, as well as the coordinates of the regions of interest are listed in Table~\ref{tab:spinds}.

Even after these passages some issues remain with the maps. The peripheral regions of some $\alpha$ maps exhibit extreme values, that are usually accompanied by higher-than-average errors in the $\Delta\alpha$ maps. Thus any seemingly drastic changes in the spectral index in the edge regions of the maps should be taken with a grain of salt. In case of some of the faintest and smallest sources these effects seem to dominate the whole $\alpha$ map.

Even though the resolution over the whole band is uniform when using the \texttt{mt-mfs} method, it is in principle possible that a source might have some structures that are resolved-out at higher frequencies, thus artificially steepening the spectral index. In practise these structures would have to be on a scale of several kpc, and generally such extended emission is diffuse and faint, so we considered their contribution negligible. Furthermore, it is also worth noting that in NLS1s extended emission at scales much larger than a few kpc is very rarely observed \citep{2020chen1}.

Finally, the frequency-dependent changes in the uv-range might have an impact on the $\alpha$ maps, especially in very extended sources. We examined its effects by selecting a few of the most extended sources and doing a comparison $\alpha$ map where the uv-range was selected to span only the range where all spectral windows had data. Examples of an $\alpha$ and a $\Delta\alpha$ map produced with a common uv-range for J1302+1624 are shown in Figures~\ref{fig:J1302spindcommon} and \ref{fig:J1302spinderrcommon}. The $\alpha$ and $\Delta\alpha$ maps of the same source without limiting the uv-range are shows in Figures~\ref{fig:J1302spindnoncommon} and \ref{fig:J1302spinderrnoncommon}. It is clear that the differences between the $\alpha$ maps are marginal, and it thus seems safe to assume that the effect of the uv-range to the $\alpha$ maps is insignificant.

\subsection{Comparison to classical spectral indices}
\label{sec:spindcomparison}

Since our sources are a subsample of the NLS1 sample studied in \citetalias{2018berton1}, where also their spectral indices were estimated, we can compare our $\alpha$ maps and weighted average spectral indices with their results. They used the traditional way of estimating the spectral index by splitting the 2~GHz band to two 1~GHz wide windows, cleaning them using a common uvrange, and then measuring their emission properties separately. Effectively this gives an in-band two-point spectral index with the central frequencies of 4.7 and 5.7~GHz. In most cases the spectral indices in \citetalias{2018berton1} agree with our results, but there are a few curious cases in \citetalias{2018berton1} where the in-band spectral index is very close to zero, but these sources look steep with $\alpha$ closer to -1 in our new maps. The in-band spectral indices of these sources also do not correlate with the 1.4-5~GHz spectral index derived in \citetalias{2018berton1} (recalculated and shown also here in Table~\ref{tab:spinds}), whereas the majority of their sources show a correlation of these two spectral indices. This raised our suspicion and we decided to estimate again the traditional spectral indices of some of these rogue sources using a newer CASA version (5.6.2-3), since an older version (4.7.2) was used in \citetalias{2018berton1}. We did the cleaning and the estimation of the spectral indices exactly as in \citetalias{2018berton1} and it turns out that the new results we got are in agreement with the $\alpha$ maps, i.e. these sources do show steep indices also when deriving it the classical way. We thus believe that our $\alpha$ maps are reliable, and the spectral index estimation of the few peculiar sources in \citetalias{2018berton1} is influenced by some undetermined issue.


\subsection{Tapered maps}

In addition to normal maps we produced tapered maps of our sources to enhance the sensitivity to extended structures. In this procedure, in addition to the selected weighting scheme, a Gaussian taper, that decreases the weights of the outermost baselines in the uv-plane, is applied on the data. We produced the tapered maps using two different tapers, 60k$\lambda$ and 90k$\lambda$, to examine the extended structures with a bit more detail (90k$\lambda$), and to try to bring out even the faintest extended emission (60k$\lambda$). Tapered maps are shown in Appendix~\ref{app:maps}, from Figure~\ref{fig:J0347} to \ref{fig:J2314}, in panels c) and d). The rms and the integrated flux densities of the tapered maps are listed in Table~\ref{tab:measurements}. The errors of the integrated flux densities were estimated with rms per beam $\times$ the square root of the area of emission expressed in beams.

\section{Radio emission from star formation}
\label{sec:radiofromsf}

In addition to the processes connected to the nuclear activity, also star formation activity, via free-free emission and synchrotron emission from supernova remnants, can significantly contribute to the radio emission of galaxies \citep{1992condon1, 2019panessa1}. In some cases, the radio emission produced this way is even strong enough to make a source appear radio-loud\footnote{Radio loudness is defined as the ratio between the 5~GHz flux density and the $B$ band flux density \citep{1989kellermann1}. Sources with radio loudness $>$ 10 are classified as radio-loud, and sources with radio loudness $<$ 10 as radio-quiet.} \citep{2019ganci1}, which is usually considered a sign of strong nuclear activity, or even of the presence of jets. Especially NLS1s are known to often show enhanced star formation \citep{2010sani1}. Indeed, \citet{2015caccianiga1} claim that the mid-infrared colours of some sources in a sample of flat-spectrum NLS1s suggest that star formation is actually the predominant source of their radio emission. The combination of the almost flat spectral index of the free-free emission and the steep spectral index of the supernova remnant synchrotron emission results in a total spectral index of $\sim$-0.7, close to what is observed in several of our sources. We thus wanted to estimate the star formation related radio emission in our sources, to better understand their nature and to aid the interpretation of our results. The best way to do this is to use mid- or far-infrared observations, since star formation manifests strongly in these bands. Only a few NLS1s have far-infrared data, therefore we are limited to the mid-infrared data from the Wide-Field Infrared Survey Explorer \citep[WISE, ][]{2010wright1} AllWISE data release. WISE performed observations in four bands: W1, W2, W3, and W4, with respective wavelengths of 3.4, 4.6, 12, and 22~$\mu$m. In particular, the longer wavelength bands, W3 and W4, are relevant to star formation studies in AGN. To achieve a comprehensive picture we decided to use several different methods, described below.

1) Investigating a sample of flat-spectrum NLS1s, \citet{2015caccianiga1} concluded that WISE colours can be used as a proxy of their star formation activity, because they are sensitive to the comparable strengths of different mid-infrared emitting components in active galaxies. Especially the W3-W4 colour is sensitive to the strength of star formation because increasing star formation gets stronger toward longer wavelengths thus it especially affects the W4 band emission, making the colour redder. They determine that colours redder than W3-W4 $>$ 2.5 cannot be explained by AGN spectral energy distribution templates, but require a strong star formation component (see Figs. 2 and 3 in their paper). We calculated the W3-W4 colour of our sources, and it is shown in Table~\ref{tab:sf}, as well as the W3 and W4 magnitudes, and the W3 flux density. 

2) Another parameter \citet{2015caccianiga1} use is a variant of the widely used $q24$ parameter, that reflects the strength of the 24$\mu$m emission relative to the 1.4~GHz emission. \citet{2015caccianiga1} use a $q22$ parameter instead, defined as

\begin{equation}
q22 = \textrm{log } (S_{22 \mu \textrm{m}} / S_{1.4 \textrm{GHz}})
\end{equation}

where $S_{22 \mu \textrm{m}}$ is the W4 band flux density and $S_{1.4 \textrm{GHz}}$ is the 1.4~GHz flux density. We only have 5.2~GHz flux densities from our observations, so we extrapolate the 1.4~GHz flux densities from the 5.2~GHz values using the weighted total spectral index from the $\alpha$ maps of the sources. \citet{2015caccianiga1} define that a major star formation contribution to the radio emission can be expected in sources with $q22 >$ 1, especially when combined with a red W3-W4 colour.

3) Another method we used is the estimation of the star formation related radio emission using mid-infrared observations. In \citet{2007boyle1} the authors derive relations between the 20~cm (1.5~GHz) and the 24~$\mu$m flux densities using two different data sets. The WISE W4 band observations are at 22~$\mu$m but the 2~$\mu$m difference is most probably negligible in this case since the variation in the restframe wavelengths due to redshifts is considerably larger, and the observed bands largely overlap. Moreover, \citet{2007boyle1} did not take redshift into account either, and their redshift range is comparable to ours, so their data is influenced by the same redshift bias. Thus we decided not to correct the 22~$\mu$m flux densities for redshifts. The relations used are

\begin{equation} 
S_{20\textrm{cm}} = (0.041 \pm 0.002) S_{24 \mu \textrm{m}} + (1.35 \pm 0.8)~\mu \textrm{Jy}  
\end{equation} \label{eq:CDFS}

and

\begin{equation}
S_{20\textrm{cm}} = (0.039 \pm 0.004) S_{24\mu \textrm{m}} + (7.1 \pm 3.3)~\mu \textrm{Jy}
\end{equation} \label{eq:ELAIS}

where $S_{20\textrm{cm}}$ is the estimated flux density at 20~cm, or 1.5~GHz, and $S_{24\mu \textrm{m}}$ is the observed 24~$\mu$m flux density. 

The star formation related radio emission estimates calculated using the above equations are shown in Table~\ref{tab:sf}. Equation~3 was derived using the observations of the $Chandra$ Deep Field South (CDFS) region, and we thus call it the $S_{20\textrm{cm}}$ CDFS estimate, whereas the observations of the European Large Area $ISO$ Survey (ELAIS) field was used to derive Equation~\ref{eq:ELAIS}, and we correspondingly call it the $S_{20\textrm{cm}}$ ELAIS estimate. It should be noted that the aforementioned Equations give the flux density estimate at 1.5~GHz, whereas our observations are centred at 5.2~GHz. If we assume for star formation related emission the characteristic spectral index of -0.7, the flux density at 5.2~GHz, $S_{5.2\textrm{GHz}}$, is 0.42 $\times$ $S_{20\textrm{cm}}$. 

4) The fourth method we implemented is the direct comparison of the radio and mid-infrared flux densities. \citet{2021kozielwierzbowska1} studied the radio - mid-infrared connection in a large sample of radio AGN and pure star forming galaxies, and concluded that a simple separation based on $S_{W3} = S_{1.4GHz}$ correctly classifies 98-99\% of galaxies, with the star forming galaxies lying above the line, and the radio AGN below. Choosing W3 instead of W4 is based on the fact that the polycyclic aromatic hydrocarbon (PAH) features, which manifest the strongest around $\sim$11~$\mu$m, for sources with $z <$ 0.6 fall in the W3 band \citep{2011jarrett1}, which covers 7-17~$\mu$m. The PAH emission is often associated with the presence of massive young stars, and it can thus be used as a proxy for the star formation in the galaxy \citep{2008dacunha1}. In our case we do not expect the division to be so straightforward since it is expected that in many of our sources the radio emission is produced by both, the AGN and star formation. Furthermore, according to some studies \citep[e.g.,][]{2012lamassa1} the PAH features are considerably suppressed in AGN-dominated systems, which might affect the usefulness of this proxy in some of our sources. On the other hand, \citet{2014esquej1} did not find proof of PAH suppression in the vicinity of low luminosity AGN, which most of the sources in our sample are. The feasibility of this method is further supported by the study in \citet{2010sani1} where they successfully use the PAH features to investigate star formation in NLS1s. In \citet{2021kozielwierzbowska1} the radio flux density at 1.4~GHz is used so we extrapolated it for our sources from our 5.2~GHz observations the weighted total spectral index measured from their $\alpha$ maps.

\subsection{Remarks on mid-infrared emission}

However, some caveats should be kept in mind when using the aforementioned methods. They have been formulated using large samples of AGN, and reflect their general properties, but are not necessarily as accurate for smaller samples with peculiar properties, such as NLS1s. Naturally, the most important question to ask is: how certain are we of the origin of the mid-infrared emission in AGN, and especially in NLS1s? The above-mentioned diagnostics assume that enhanced mid-infrared emission and certain colours are associated with star formation, which necessarily is not always the case. Indeed, in some cases features similar to those caused by star formation can be caused by the AGN itself heating the surrounding dust.

\subsubsection{Polar dust} 

One significant phenomenon causing possible deviation from these relations is the presence of polar dust in AGN \citep[e.g.,][]{2009burtscher1,2016asmus1,2019leftley1}. Polar dust comprises of warm ($\sim$a few hundred K) dust that is situated along the polar direction of an AGN, i.e. perpendicular to the conventional dusty torus, and usually spans some tens of pc. It is believed to be maintained by a nuclear wind driven by radiation pressure \citep{2019leftley1}. Its spectral energy distribution ($\nu F_{\nu}$ vs. $\lambda$) peaks in mid-infrared, around 10-30$\mu$m \citep[][]{2018lyu1}, depending on the temperature, and in some cases its contribution to mid- to far-infrared emission is significanty stronger than the contribution of the AGN, or star formation, reaching even a dominance of 90\% \citep{2013honig1, 2019asmus1}. Its spectral shape is somewhat similar to the shape caused by strong star formation at mid-infrared wavelengths \citep[see Fig. 9 in][]{2018lyu1}. This is relevant in our case since polar dust emission can significantly enhance the emission in both W3 and W4 bands, and due to its properties it relatively enhances W4 band emission more than W3, mimicking mid-infrared properties conventionally associated with star formation activity. An example is NGC 3782, a Seyfert 1 galaxy that does not show significant star formation, but has a W3 flux density 600~mJy higher than its 1.4~GHz flux density, and its W3-W4 $>$ 2.5. Indeed, its mid-infrared emission was found to be dominated by dust in the polar region \citep{2013honig1}. Such sources are plenty \citep[see e.g.][]{2013zhang1,2018lyu1}, and unfortunately we have no way of estimating the polar dust contribution without careful spectral energy distribution (SED) modelling or direct high-resolution mid-infrared observations of the emitting region, since such data is not available. Moreover, there are indications that high Eddington ratio sources, such as NLS1s, have a tendency to show more polar dust \citep{2019leftley1}. 

However, \citet{2013zhang1} found that the ratio of the [O~III] wing and the bolometric luminosity is correlated with the mid-infrared covering factor. The wing is believed to rise from turbulent polar outflows in the inner narrow-line region \citep{2014peng1}, and thus they propose that this indicates that a considerable fraction of the warm dust producing mid-infrared emission in AGN is likely embedded in polar outflows. The [O~III] lines of all of our objects were analysed in \citet{2021berton1}, which reports their core and wing properties, including velocity and width of both components. The [O~III]$\lambda$5007 was fitted with two Gaussians, one for the core component, and one to reproduce the blue wing. The same Gaussians, rescaled, were used to fit the [O~III]$\lambda$4959 simultaneously. The errors were calculated via a Monte Carlo method, by randomly varying the noise on the line and fitting them 1000 times. Unfortunately we do not have the information about the luminosity of the wing, but the presence of a wind alone already indicates the presence of nuclear outflows, and increases the probability those sources might host a prominent polar dust component.

\subsubsection{Dusty torus}

Another emission source whose contribution to the infrared emission is not totally clear is the more conventional AGN structure, the dusty torus. It is known that the hot (1000-1500~K) dust, in the inner parts of the torus manifests as a bump in near-infrared bands, around 2-4~$\mu$m (in $\nu F_{\nu}$ vs. $\lambda$), and in general seems to dominate the infrared spectrum between 1 and 10~$\mu$m \citep[e.g.][]{1986edelson1, 2006rodriguezardila1}. The overall torus emission instead peaks at mid- to far-infrared wavelengths \citep{2006fritz1,2018zhuang1}, and as demonstrated by \citet{2018zhuang1} contribution from different AGN- and host-related elements significantly changes when modelling the same SED with different models \citep[see also][]{2019gonzalezmartin1}, leaving a lot of space for interpretation. The implications of this are two-fold: the torus emission can result in enhanced mid-infrared emission, without any contribution from star formation, and, if the mid-infrared emission is dominated by the AGN, it means that W3 and W4 might not reflect the properties of the host galaxy, but of the AGN itself, which would render some of the diagnostic tools we are using unreliable.

In the literature, the results regarding the AGN and host galaxy contributions differ considerably, from 90\% AGN domination to only $\sim$15\% AGN contribution \citep{2009dicken1,2013rosario1,2018zhuang1,2019gonzalezmartin1}. However, a trend of decreasing AGN contribution from brighter quasars to more moderate Seyfert galaxies can be observed. \citet{2013rosario1} studied a sample of 13000 Type 2 AGN using the W3 and W4 bands, and 1.4~GHz radio luminosities. They found that Seyfert galaxies almost exclusively lie in the mid-infrared-bright region, and conclude that their mid-infrared emission is mainly related to star formation activity, and that only $\sim$15\% of the W4 band emission has an origin in the AGN-heated dust. Interestingly they find that the W3 band emission in Seyfert galaxies is suppressed compared to pure star-forming galaxies, possibly due to the PAH destruction due to the AGN emission \citep{2012lamassa1}.

\subsubsection{Summary of mid-infrared issues}

All these components -- the torus, polar dust, and star formation -- can, and often do, co-exist \citep[see Fig. 13 in][]{2018lyu1} complicating the situation even more. Before moving forward, we summarise how these different mid-infrared emission production scenarios can affect our diagnostics, and what we can do to mitigate their effects. 

\begin{enumerate}
 \item W3-W4 mid-infrared colour:
 \begin{itemize}
     \item Polar dust emission can show properties similar to star formation \textrightarrow the presence of [O~III] wing can be used to estimate whether polar winds are present and might contribute to the mid-infrared emission and colours.
     \item PAH features can be suppressed, which will redden the W3-W4 colour \textrightarrow no effect to the diagnostic value.
 \end{itemize}

 \item $q22$ parameter:
  \begin{itemize}
      \item Cannot account for additional mechanisms producing 22$\mu$m emission, for example, polar dust \textrightarrow other diagnostics need to be used to estimate the probability of contamination.
      \item Cannot account for strong AGN radio emission \textrightarrow $q22$ values below 1 should not be used for diagnostic purposes. 
  \end{itemize}

 \item $S_{\textrm{W3}}$ - $S_{1.4\textrm{GHz}}$ -relation:
       \begin{itemize}
      \item The W3 emission can be enhanced by AGN-heated dust emission that can make the source look star-forming \textrightarrow the presence of [O~III] wing can be used to estimate if polar dust is present.
      \item AGN can suppress the 11.3~$\mu$m PAH feature \textrightarrow other diagnostics must be used.
      \end{itemize}

 \item $S_{\textrm{CDFS}}$ - $S_{\textrm{int}}$ -relation:
       \begin{itemize}
      \item Has been calibrated using star forming galaxies, does not take into account the contribution of the AGN in W4 band \textrightarrow the AGN contribution in Seyferts seems to be small.
      \end{itemize}
\end{enumerate}

Since the contribution of the AGN to the mid-infrared emission of Seyferts, and thus also NLS1s, seems to be small \citep{2013rosario1}, and NLS1s are known to be strongly star-forming galaxies, we can assume that in most cases the diagnostics we are using are pointing in the right direction. However, all the sources will be analysed individually, and all diagnostic tools, supplemented by any other information we have of these sources, will be used to draw a more complete picture of the situation.

\section{Results}
\label{sec:results}

\subsection{Significance of star formation}

We implemented several diagnostics to study the contribution of the radio emission produced by star formation related processes in our sources, as described in Sect.~\ref{sec:radiofromsf}. For the radio data we used the total intensity measured from the 60k$\lambda$ map to include also possible faint extended emission, not visible in the normal or the 90k$\lambda$ map. Keeping in mind the remarks in the previous section, we summarise the results here.

1) 25 out of 44 sources in our sample have W3-W4 $>$ 2.5, indicating mid-infrared colours so red that they might be hard to explain only by AGN activity. If the red colour is indeed caused by star formation, it is probable that in these sources its contribution to the radio emission is significant as well. The result that more than half of our sources seem to have considerable star formation is well in agreement with the previous study by \citet{2010sani1}. This result on its own does not mean that there is no strong AGN activity in these sources. 

2) The second method was to use the $q22$ parameter to estimate whether the star formation can significantly contribute to the observed radio emission. We found that 26 out of 44 sources have $q22 >$ 1. Most of the sources with W3-W4 $>$ 2.5 also have $q22 >$ 1, but there are some curious sources where this is not the case (see Table~\ref{tab:sf}). These sources on average have high radio flux densities, and whereas the W3-W4 colour indicates star formation in these sources, the strong radio emission distorts the $q22$ value which remains low, or even negative. Indeed, the more widely used $q24$ is expected to be very low for radio-loud AGN \citep{2008ibar1}, and it cannot account for the effects of both AGN and star formation activity in a source. Thus, especially for NLS1s, where often a star forming host galaxy and jets can co-exist, no conclusions should be drawn from $q22$ values $<$ 1.

\begin{figure}
    \centering
    \includegraphics[width=9cm]{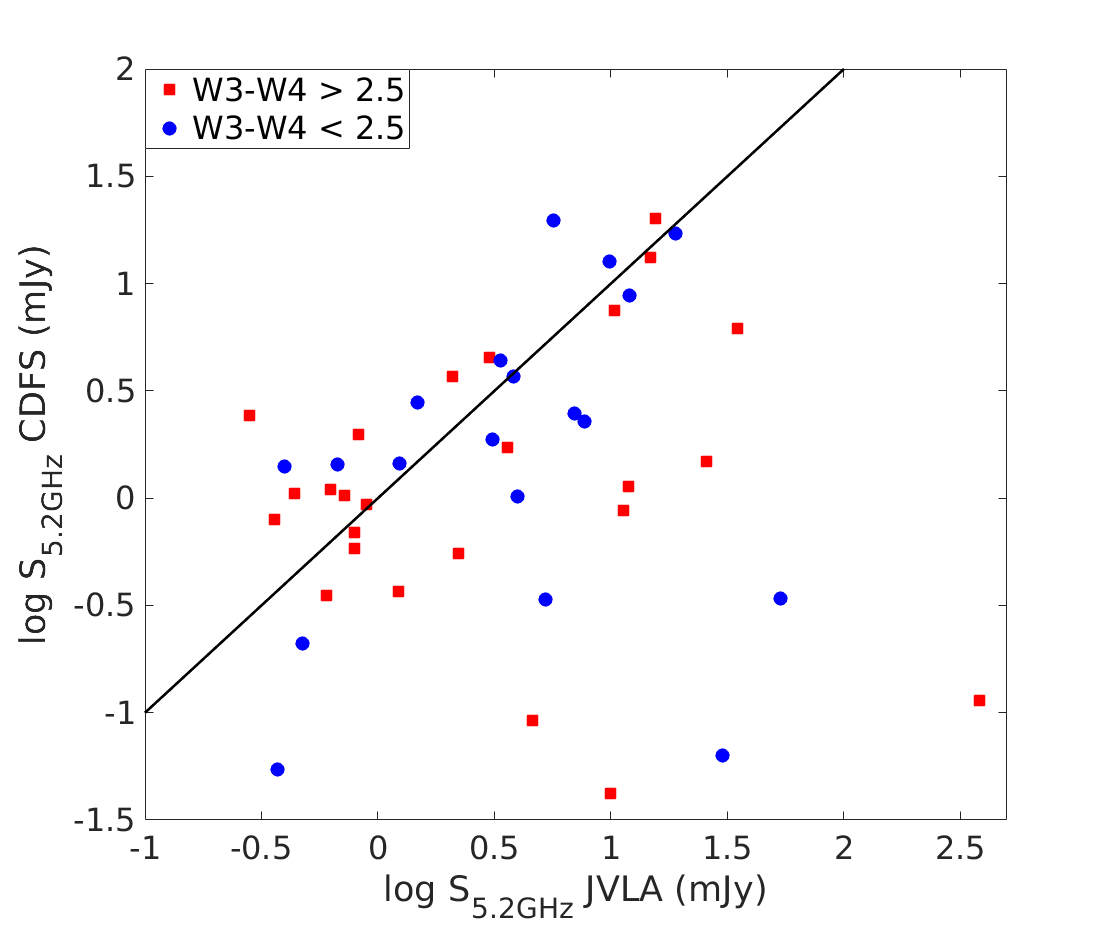}
    \caption{5.2~GHz flux densities estimated from the 22$\mu$m flux densities, and extrapolated from 1.5~GHz using $\alpha$=-0.7 vs. measured JVLA 5.2~GHz flux densities. The errorbars of the flux densities are smaller than the size of the markers and not shown. Sources with the WISE W3-W4 colour $>$ 2.5, that is sources that are unusually red and most likely to have enhanced star formation, are shown with filled red squares, and sources with the WISE W3-W4 colour $<$ 2.5, that is, sources that are not unusually red, as filled blue circles. The black line denotes equal flux densities. $S_{\textrm{5.2GHz}}$ CDFS and $S_{\textrm{5.2GHz}}$ JVLA values close to each other means that the bulk of the radio emission can be explained by star formation processes, whereas the sources lying below the black line have excess radio emission.}
    \label{fig:jvlacdfs}
\end{figure}

3) The third method we used was to estimate the radio emission produced by star formation using mid-infrared data \citep{2007boyle1}. We used the WISE W3 flux density to estimate the radio emission from star formation at 1.5~GHz. We used a spectral index characteristic for star formation ($\alpha$=-0.7) to extrapolate the 1.5~GHz estimate to 5.2~GHz. The comparison of the measured 5.2~GHz flux densities and the estimated 5.2~GHz flux densities is shown in Fig.~\ref{fig:jvlacdfs}. We decided to use only the CDFS estimate, since both estimates, CDFS and ELAIS, are very close to each other. It can be seen that a majority of our sources are clustered around the  $S_{\textrm{5.2GHz, CDFS}}$ = $S_{\textrm{5.2GHz, JVLA}}$ line. This suggests that, assuming that the predominant source of the W3 band emission is star formation, in most of our sources these processes are enough to explain the bulk of the observed radio emission. Only in the sources clearly below the black line the CDFS estimate underestimates the radio emission, and these sources should have a major contribution from the AGN to explain the observed flux density. Interestingly, the mid-infrared colour W3-W4 does not seem to correlate with the position of a source in Fig.~\ref{fig:jvlacdfs}, indicating that there are some discrepancies between the diagnostics.

\begin{figure}
    \centering
    \includegraphics[width=9cm]{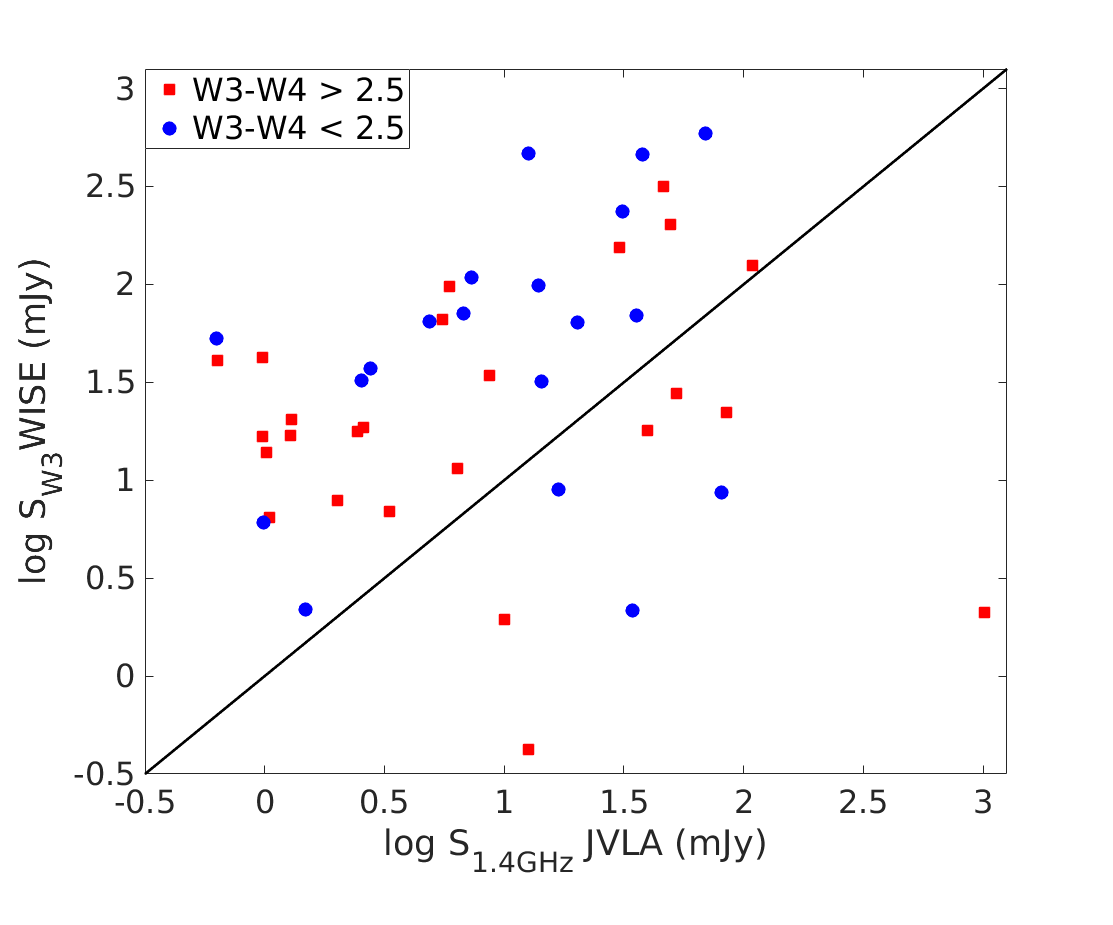}
    \caption{WISE W3 flux densities vs. JVLA 1.4~GHz flux densities extrapolated from the 5.2~GHz flux densities using the total spectral indices from the $\alpha$ maps. The errorbars of the flux densities are smaller than the size of the markers and not shown. Sources with the WISE W3-W4 colour $>$ 2.5, that is sources that are unusually red and most likely to have enhanced star formation, are shown with filled red squares, and sources with the WISE W3-W4 colour $<$ 2.5, that is, sources that are not unusually red, as filled blue circles. The black line denotes equal flux densities. Sources above the black line should have more contribution from star formation, whereas the sources below the black line should be dominated by AGN activity.}
    \label{fig:jvlaw3}
\end{figure}

4) The last diagnostic is based on the direct comparison of the W3 and 1.4~GHz flux densities \citep{2021kozielwierzbowska1}. Since our observations are at 5.2~GHz we extrapolated them to 1.4~GHz using the weighted total spectral index from the $\alpha$ map of each source. The comparison of the 1.4~GHz and the W3 flux densities is shown in Figure~\ref{fig:jvlaw3}. The sources above the black line ought to have strong star formation contribution, whereas only the sources below the line are expected to be dominated by the AGN radio emission. Also in this case the mid-infrared colour does not seem to play a determinative role. This result is also in agreement with the other diagnostics implying that a majority of NLS1s show enhanced star formation, even at levels where it can be the predominant source of the radio emission in the galaxy. However, it must be taken into account that the W3 flux density can be enhanced also due to excess mid-infrared emission from the torus or the polar dust.

Furthermore, methods 3) and 4) are in agreement with each other since all sources that are below the threshold shown by the black line in Figure~\ref{fig:jvlaw3} are below the threshold also in Figure~\ref{fig:jvlacdfs}. So, whereas it might be hard to tell the exact origin of the mid-infrared and the radio emission in sources with high W3 and W4 flux densities, the two diagnostics agree on the sources that are AGN-dominated. The properties of individual sources will be discussed in the next sections.

\subsection{Notes on individual sources}

The results for all sources in this study will be individually discussed in this section. Their general radio emission and radio morphological properties were extensively discussed in \citetalias{2018berton1}, thus in this paper we will concentrate on their spectral properties, and investigating the origin of their radio emission. We note that throughout the rest of the paper we use the term ``jet'' to refer to collimated outflows in general, since in most cases we have no way of estimating whether the jets are relativistic or not. We use ``outflow'' to refer to non-collimated efflux. The luminosities of the sources are taken from \citetalias{2018berton1} (their Table A.5). In addition we obtained the PanSTARRS-1 $i$ band optical images \citep{2020flewelling1} of the host galaxies of our sources. We overlaid them with the 90k$\lambda$ tapered radio map, and in some cases also the normal radio map, to give a better sense of the extent of their radio morphologies, and to investigate if there are some optical structures that match with the radio structures. For the sources with the highest redshifts this does not significantly add new information because the scales we are probing at high redshift are several kpc, and the host galaxies are usually not resolved either, but for consistency we decided to include them.

We use the radio loudness parameter to make our results comparable with the literature. However, in general we do not encourage its usage since it is an arbitrarily chosen threshold, and ambiguous for most of the sources \citep{2017padovani1,2017jarvela1}. Whereas very radio-loud sources are more likely to host relativistic jet, there are examples of sources with no previous radio detection at low frequencies, but that were found to have relativistic jets \citep{2018lahteenmaki1,2020berton2,2021jarvela1}, and also of sources whose star formation activity is so enhanced they appear radio-loud \citep{2015caccianiga1}. Thus, as convenient as it might be, we cannot use the radio loudness to deduce the properties of sources, but have to treat and study all of them individually.

Before going into the detailed discussion a few remarks should be made. First, it turned out that the star formation diagnostics are not very reliable when it comes to NLS1s, unless the source is clearly dominated by the AGN. Thus we do not use these tools alone to draw any conclusions. This is further discussed in Sect.~\ref{sec:discussion}. In general, star forming galaxies, even starbursts, do not exceed a radio luminosity of log $\nu L_{\nu, \textrm{int}}$ = 40.0 erg s$^{-1}$ \citep{2009sargsyan1}, which can thus be used as an additional proxy for the AGN dominance. Second, we use log $\nu L_{\nu, \textrm{int}}$ = 40.60 erg s$^{-1}$ at 5.2~GHz as a threshold for `traditional' CSS sources \citep{1998odea1}. Low-luminosity CSS sources do not have a defined lower luminosity limit but we remark that, for example, the low-luminosity CSS sources in \citet{2002kunert1} were chosen to have a 5~GHz flux density $>$ 150~mJy. Only one of our sources exceeds this, which makes the comparison between the low-luminosity CSS sources and NLS1s in this paper uncertain.

We list all the flux density and luminosity variables we use in Table~\ref{tab:acronyms}. We give a summary of the radio morphology properties and the mid-infrared diagnostics in Table~\ref{tab:summary}. We remark that the data in this table is based only on this paper, for example, if a jet has been detected in previous observations, we do not list it in the table unless we detect it also at 5.2~GHz. We tentatively classify sources based on $R_\textrm{CDFS}$ = ($S_{\textrm{5.2GHz, CDFS}}$ - $S_{\textrm{int}}$) / $S_{\textrm{5.2GHz, CDFS}}$), and $R_\textrm{W3}$ = ($S_{\textrm{W3}}$ - $S_{\textrm{1.4GHz, JVLA}}$) / $S_{\textrm{W3}}$. Large negative values of $R_\textrm{CDFS}$ indicate that there is more radio emission in the source than what can be explained by star formation, while positive values suggest that the radio emission could be explained by star formation only. Large positive values of $R_\textrm{W3}$ possibly indicate star formation as the predominant source of radio emission, while large negative values suggest that the AGN is the main radio emission source. The thresholds are given in Table~\ref{tab:rclass}.

\begin{table*}[!h]
\caption[]{Variables used in this paper.}
\centering
\begin{tabular}{l l p{12cm}}
\hline\hline
Variable                                & Units           & Description \\ \hline
$S_{\textrm{peak}}$                     & mJy beam$^{-1}$ & Peak flux density of the normal radio map\\
$S_{\textrm{int}}$                      & mJy             & Integrated flux density of the normal radio map within the 3$\sigma$ contour  \\
$S_{\textrm{90k}\lambda, \textrm{int}}$ & mJy             & Integrated flux density of the 90k$\lambda$ radio map within the 3$\sigma$ contour \\
$S_{\textrm{60k}\lambda, \textrm{int}}$ & mJy             & Integrated flux density of the 60k$\lambda$ radio map within the 3$\sigma$ contour \\
$S_{\textrm{1.4GHz, JVLA}}$             & mJy             & Flux density at 1.4~GHz, extrapolated from the 5.2~GHz using the total spectral index \\
$S_{\textrm{5.2GHz, CDFS}}$             & mJy             & Flux density at 5.2~GHz, extrapolated (with $\alpha$ = -0.7) from the 1.5~GHz flux density estimated from the 22~$\mu$m flux density \\
$S_{\textrm{W3}}$                       & mJy             & WISE W3 flux density \\
log $\nu L_{\nu, \textrm{int}}$         & erg s$^{-1}$    & Logarithm of the luminosity at 5.2~GHz \\ \hline
\end{tabular}
\tablefoot{(1) Variable name, (2) variable units, (3) description of the variable.}
\label{tab:acronyms}
\end{table*}

\begin{table}
\caption[]{Threshold for classifying the predominant source of the radio emission.}
\centering
\begin{tabular}{l l}
\hline\hline
Criterion                         & Classification \\ \hline
0.0 $ < R_\textrm{CDFS}$          & star formation dominated     \\
-0.5 $ < R_\textrm{CDFS} <$ 0.0   & inconclusive    \\
-1.0 $ < R_\textrm{CDFS} <$ -0.5  & possibly AGN dominated      \\
$R_\textrm{CDFS} <$ -1.0          & AGN dominated      \\
1.0 $ < R_\textrm{W3}$            & star formation dominated      \\
0.5 $ < R_\textrm{W3} <$ 1.0    & possibly star formation dominated  \\
-0.5 $ < R_\textrm{W3} <$ 0.5   & inconclusive    \\
-1.0 $ < R_\textrm{W3} <$ -0.5  & possibly AGN dominated \\
$R_\textrm{W3} <$ -1.0          &  AGN dominated   \\ \hline
\end{tabular} \label{tab:rclass}
\end{table}

\subsubsection{J0347+0105} 
\label{sec:J0347}

J0347+0105 is a radio-quiet NLS1 at $z$ = 0.031, previously studied with the JVLA in A configuration at 8.4~GHz by \citet{2000thean1}. They found only a marginally resolved core with a flux density of 6.8~mJy. It was not detected at 22~GHz with VLBI \citep{2016doi1}, but the sensitivity of those observations was poor with a detection limit around 7~mJy. J0347+0105 has also been found to exhibit water maser emission \citep{2011tarchi1}, though its origin remains unclear. 

Our radio map in Figure~\ref{fig:J0347spind} does not show any significant structure. The integrated flux density is 11.65~mJy. Using the 8.4~GHz flux density by \citet{2000thean1}, we derive a spectral index of -1.12. Our $\alpha$ map shows a consistently steep, but not quite that extreme, total spectral index of -0.73. The observations are almost 20 years apart and the beam sizes are different, which probably causes the difference in the spectral indices. The tapered maps (Figures~\ref{fig:J0347-90k} and \ref{fig:J0347-60k}) show a slightly elongated north-east/south-west structure, but the integrated flux density is only $\sim$0.5~mJy higher than in the normal map, so any extended emission is very faint. The host galaxy does not show any clear features either, and the radio emission is confined inside the host (Fig.~\ref{fig:J0347-host}).

The W3 flux density ($S_{\textrm{W3}}$) of this source is 203.5~mJy higher than the extrapolated 1.4~GHz flux density ($S_{\textrm{1.4GHz, JVLA}}$ = 31.3~mJy) while the extrapolated CDFS 5.2~GHz flux density ($S_{\textrm{5.2GHz, CDFS}}$ = 8.8~mJy) is slightly lower than the JVLA 5.2~GHz flux density ($S_{\textrm{int}}$). The $q22$ parameter is 1.22, and the W3-W4 is 2.29, not indicating extreme colours, although W3-W4 might be underestimated since the W3 emission is unusually high. The $\alpha$ map, with a spectral index characteristic of star formation, the high W3 flux density, and $q22 >$ 1 imply that the contribution of star formation is very strong in this source. Since the star formation rate in this source has not been studied, an alternative explanation for the high mid-infrared flux densities can be equatorial or polar dust, since this source shows a prominent and turbulent [O~III] wing with a velocity of -371~\kms\ and FWHM of 1074\kms\. A CSS-like nature cannot be ruled out either without higher resolution observations, but considering its luminosity, log $\nu L_{\nu, \textrm{int}}$ = 39.17 erg s$^{-1}$, which is clearly below the usual CSS threshold, star formation seems the more plausible explanation.


\subsubsection{J0629-0545} 
\label{sec:J0629}

J0629-0545 (IRAS 06269-0543) is a radio-quiet NLS1 at $z$ = 0.117 that has been identified as an ultra-luminous infrared galaxy (ULIRG, \citealp{2002zheng1}). Its integrated flux density is 14.58~mJy, with the flux density of the tapered maps approximately similar. It exhibits starburst level star formation at a rate of 183 $M_{\odot}$ yr$^{-1}$ derived from the CO(1-0) luminosity \citep{2019tan1}. This is evident also in our star formation diagnostics, as the W3-W4 colour is 2.88, $q22$ = 1.23, $S_{\textrm{W3}}$ is 153.7~mJy higher than $S_{\textrm{1.4GHz, JVLA}}$ (45.3~mJy). $S_{\textrm{5.2GHz, CDFS}}$ and $S_{\textrm{int}}$ are quite similar, 13.2 and 14.6~mJy, respectively, indicating that the radio emission might be dominated by star formation activity. It also shows a strong [O~III] wing, with a velocity of -464~\kms\ and a FWHM of 1623~\kms, so a contribution of AGN-heated dust to the mid-infrared brightness cannot be ruled out either. However, its radio luminosity, log $\nu L_{\nu, \textrm{int}}$ = 40.43 erg s$^{-1}$, would be high even for a starburst galaxy, and thus it is plausible that radio emission is a combination of nuclear and star formation related processes. \citet{2019tan1} in ALMA observations found a rotating gas disk with ordered velocity gradient, indicating that the host (Figs.~\ref{fig:J0629-host-zoom} and \ref{fig:J0629-host}) is a disk galaxy which has not undergone any recent merger. They estimate the inclination of the disk to be 51$\pm$10~\degree.

The radio map of J0629-0545 (Fig.~\ref{fig:J0629spind}) clearly shows two distinct components with a separation of 2.7~kpc. We assume that the brighter component is the core, since the north-eastern one is close to the edge of the host galaxy, as seen in Fig.~\ref{fig:J0629-host}. These same features were also detected at 8.4~GHz in \citet{2000moran1}, but interestingly the north-eastern component is brighter in their observations. Unfortunately they do not give the flux densities of the single components so we cannot do any comparison. J0629-0545 was also detected in NVSS with a flux density of 32.7~mJy, which gives a 1.4-5.2~GHz spectral index of -0.62. The $\alpha$ map in Fig.~\ref{fig:J0629spind} indicates steeper spectral indices: both the total and the core spectral indices are -0.92. We cannot say anything definitive about the spectral index of the north-eastern component, because the S/N of the data was not high enough to model it reliably. \citet{2000moran1} reported an extremely steep spectral index of -2.2 between 1.4 and 8.4~GHz. They also found the source to exhibit variability, which might partly account for the discrepancy between their and our results. The nature of the north-eastern component is unclear but a strong blueshift of -550~km s$^{-1}$ of the [O~III] lines found in \citet{2002zheng1} suggests that the source has a jet, and, indeed, the asymmetry of the radio morphology might indicate an AGN origin for the north-east region, rather than star formation related origin. 


\subsubsection{J0632+6340} 
\label{sec:0632}

J0632+6340 (UGC 3478) is a radio-quiet, nearby ($z$ = 0.013) NLS1 with a modest integrated flux density of 2.83~mJy, and a low luminosity of log $\nu L_{\nu, \textrm{int}}$ = 37.76 erg s$^{-1}$. The peak flux density is only 1.69~mJy beam$^{-1}$, so a considerable fraction of the radio emission comes from the regions of extended emission. The radio map (Fig~\ref{fig:J0632spind}) shows a two-sided extended morphology, approximately at a PA of 40~\degree. The extent of the whole emitting region is $\sim$1~kpc end-to-end, it is well confined within the bulge of its spiral host galaxy (Figs.~\ref{fig:J0632-host} and \ref{fig:J0632-host-zoom}), and seems to be aligned with its major axis. The tapered maps do not reveal any additional structures. The total spectral index of the source is -0.51 and the core spectral index is -0.63 (Fig.~\ref{fig:J0632spind}).

J0632+6340 has a few archival detections, 22~mJy at 325~MHz in Westerbork Northern Sky Survey (WENSS), 12.8~mJy at 1.4~GHz in NVSS, and 1.4~mJy at 8.4~GHz in VLA observations \citep{2000kinney1}, based on which they also report seeing only an unresolved core, with an extent $<$ 25~pc \citep{2001schmitt3}. The spectral index seems to be steepening toward higher frequencies: based on these archival observations the spectral indices are -0.37 between 325~MHz and 1.4~GHz, -1.15 between 1.4 and 5.2~GHz, and -1.47 between 5.2 and 8.4~GHz. The 8.4~GHz observations were performed in the same configuration as ours, but their rms was higher (10 vs. 32~$\mu$Jy beam$^{-1}$), so they might have lost some of the extended emission seen in our map, which might steepen the spectral index. In addition, the beam difference might affect the measurements, making the higher frequency flux densities appear lower. None of these observations are simultaneous either, so variability might affect the indices.

\citet{2003schmitt1} studied the extent of [O~III] emission in this source, and found a morphology resembling ionisation cones (their Fig.~8). The PA of the [O~III] emission is 55 degrees, roughly in agreement with the extended radio emission in our map. No blueshift is detected in the [O~III] line profiles. The star formation rate of the host galaxy was found to be $\sim$1.9 $\pm$ 1~$M_{\odot}$ yr$^{-1}$ \citep{2020jackson1}. Our diagnostics support that there is ongoing star formation in the galaxy: the W3-W4 colour is 2.52, $q22$ = 1.65, $S_{\textrm{W3}}$ is $\sim$91.5~mJy higher than $S_{\textrm{1.4GHz, JVLA}}$, and $S_{\textrm{5.2GHz, CDFS}}$ is 1.49~mJy higher than $S_{\textrm{int}}$. J0632+6340 exhibits an [O~III] wing with a shift of -290~\kms\ and a FWHM of 1000~\kms, and thus polar, or otherwise AGN-heated, dust can be responsible for a fraction of the observed mid-infrared emission. Due to the low redshift of the source, we are able to examine only the centremost region in our observations. Because of the array configuration used for our observations, any radio emission originating further out in the galaxy may be resolved out, and thus the flux density might be underestimated. Moreover, the peculiar radio morphology is not compatible with the circumnuclear star formation usually seen in NLS1s. Since it does not seem probable that this source hosts relativistic jets, the radio morphology can be explained either by non-relativistic jets, or outflows produced by nuclear winds. The non-collimated morphology, and the spectral index steepening toward higher frequencies would point to the presence of an outflow in J0632+6340. Also the presence of an [O~III] wing supports the existence of nuclear outflows in this source.

\subsubsection{J0706+3901} 

J0706+3901 (FBQS J0706+3901) is a radio-quiet NLS1 at $z$ = 0.086 with a modest $S_{\textrm{int}}$ of 2.14~mJy. The flux densities of the tapered maps are approximately the same, and the luminosity is moderate with log $\nu L_{\nu, \textrm{int}}$ = 39.33~erg s$^{-1}$. It does not show any resolved structure in any of the maps (Figures~\ref{fig:J0706spind},\ref{fig:J0706-90k}, and \ref{fig:J0706-60k}). Its total and core spectral indices are around -0.8, consistent with either optically thin synchrotron emission or star formation. The W3-W4 colour is 2.55, suggesting star formation activity. $S_{\textrm{W3}}$ is $\sim$5.1~mJy higher than $S_{\textrm{1.4GHz, JVLA}}$, which is a considerable difference taking into account the modest flux density values of the source, but it does not necessarily imply very strong star formation. Also $q22$ is only 0.70. $S_{\textrm{5.2GHz, CDFS}}$ is only 0.55~mJy and is thus somewhat lower than $S_{\textrm{int}}$. The [O~III] of J0706+3901 does not have a wing component at all, so polar dust might not be present, but equatorial dust still can be. The radio emission is confined within a seemingly featureless host galaxy, as seen in Fig.~\ref{fig:J0706-host}, consistent with the star formation explanation. However, with these data it is impossible to distinguish the AGN and star formation -related components from each other; higher resolution observations will be needed to achieve that.

\subsubsection{J0713+3820}  

J0713+3820 (FBQS J0713+3820) is a radio-quiet NLS1 at $z$ = 0.123 with a peak flux density of 2.32~mJy, integrated flux density of 3.39~mJy, and 60k$\lambda$ tapered map flux density of 3.80~mJy, thus a considerable amount of the emission originates from extended regions (Fig.~\ref{fig:J0713}). The radio luminosity is moderate with log $\nu L_{\nu, \textrm{int}}$ = 39.82 erg s$^{-1}$. The $\alpha$ map (Fig.~\ref{fig:J0713spind}) shows a steep spectral index around $\sim$-1.0 throughout the emitting region. This source has been detected in WENSS (25~mJy) and FIRST (10.78~mJy), yielding spectral indices of -0.57 between 325~MHz and 1.4~GHz, and -0.88 between 1.4 and 5.2~GHz. Therefore, based on these spectral indexes, it seems like the spectrum steepens toward higher frequencies, but the effect of different beam sizes is unknown. J0713+3820 was not detected at 22~GHz in VLBI observations \citep{2016doi1}.

Mid-infrared data indicates that there is star formation going on in the host as the W3 flux density is 85.2~mJy higher than the extrapolated 1.4~GHz flux density, and the $q22$ parameter is 1.19. $S_{\textrm{5.2GHz, CDFS}}$ and $S_{\textrm{int}}$ are about the same (3.7 vs. 3.8~mJy), so based on this diagnostic there is no need for other sources of radio emission beyond star formation. The W3-W4 colour though is only 2.28. The extended, somewhat patchy, morphology of the radio emission, covering the whole host galaxy (Fig.~\ref{fig:J0713-host}), supports the presence of star formation in the host. The source shows an [O~III] wing component shifted by -742~\kms\ with respect to the line core, thus the contribution of polar dust to the mid-infrared flux densities is a possibility. It is worth noting that the FWHM of the wing component is 1495~\kms, indicating strong turbulence in the gas where the line originates. 

Interestingly this source exhibits a considerable \textit{redshift} of 300~\kms\ in its [O~III] lines \citep{2021berton1}. Like mentioned before, the [O~III] wing is blueshifted relative to the [O~III] line core. It is known that [O~III] wings often arise due to nuclear winds, but they are usually accompanied by blueshifted [O~III] lines, which is not the case with J0713+3820. It might be that the redshifted [O~III] core and the blueshifted [O~III] wing arise from kinematically separate regions. Unfortunately, our data are not enough to pinpoint and resolve the [O~III] emitting regions, and further observations, especially by means of integral field spectroscopy, will be necessary to unravel the nature of this source.

\subsubsection{J0804+3853} 

J0804+3853 (FBQS J0804+3853) is a radio-quiet NLS1 at $z$ = 0.212 with a low peak flux density of 0.37~mJy beam$^{-1}$, and an integrated flux density of 0.81~mJy, so more than half of the radio emission comes from extended regions. This is evident also in the radio map in Fig.~\ref{fig:J0804spind}, where an asymmetric emitting region with an extent of 4.1~kpc toward east - south-east is seen. The host galaxy does not have any distinct features, and the radio emission is confined within it (Fig.~\ref{fig:J0804-host}). J0804+3853 was detected in FIRST with a flux density of 2.68~mJy, giving a 1.4-5.2~GHz spectral index of -0.91, which is consistent with the $\alpha$ map of the source in Fig.~\ref{fig:J0804spind}: the total spectral index of the source is -0.98. J0804+3853 has a moderate luminosity of log $\nu L_{\nu, \textrm{int}}$ = 39.69 erg s$^{-1}$.

\citet{2015caccianiga1} estimate the star formation rate in this galaxy to be as high as 89 $M_{\odot}$ yr$^{-1}$, and our star formation proxies are in good agreement with this. The W3-W4 colour is 2.70, $q22$ is 1.36, $S_{\textrm{W3}}$ is 16.0~mJy higher than $S_{\textrm{1.4GHz, JVLA}}$, and $S_{\textrm{5.2GHz, CDFS}}$ and $S_{\textrm{int}}$ are roughly the same. It seems like the star formation activity can account for most of the radio emission in this source, but the asymmetric radio morphology seems unusual for a star forming galaxy, and resembles a morphology seen in sources with nuclear winds instead. Indeed this source shows an [O~III] wing with a velocity of -388~\kms\ and a FWHM of 760~\kms, which supports the presence of nuclear winds. If the mid-infrared emission is enhanced by the AGN-heated dust, the star formation might not be that strong after all. More observations will be needed to determine the components that contribute to the radio emission in this source.


\subsubsection{J0806+7248} 

J0806+7248 (RGB J0806+728) is a moderately radio-loud NLS1 at $z$ = 0.098 with a rather steep total spectral index of -0.96, as seen in its $\alpha$ map in Figure~\ref{fig:J0806spind}. Its integrated flux density of 11.46~mJy is similar to the flux density in the tapered maps (Figures~\ref{fig:J0806-90k} and \ref{fig:J0806-60k}). J0806+7248 does not show any resolved structure in any of the maps. Its host galaxy does not show any resolved features, and the radio emission is constrained within the host (Fig.~\ref{fig:J0806-host}).

This source has been extensively studied in the past. \citet{1991gregory1} performed observations with the 91~m Green Bank (GB) radio telescope (GBT) at 4.85~GHz and reported that J0806+7248 had a flux density of 31~mJy. The observations were done in 1987. A few years later \citet{1991becker1} reported a flux density of 29~mJy at 4.85~GHz with the GBT. In the early and mid-1990's J0806+7248 was observed in the NVSS at 1.4~GHz with a flux density of 50~mJy \citep{1998condon1}, and at 5~GHz with the VLA showing a flux density of 20~mJy \citep{1997laurentmuehleisen1}. We can estimate that the spectral index of J0806+7248 in the early 1990's was $\sim$-0.7.

\citet{2007doi1} observed J0806+7248 with the Japanese VLBI Network (JVN) at 8.4~GHz. They obtained a flux density of 6.9~mJy, and a brightness temperature, $T_{\textrm{B}}$, of $> 10^7$~K, indicating a non-thermal origin of the radio emission. The source shows a slightly extended structure at the mas-scale of the observations, but no clear morphology is seen. J0806+7248 was later observed with the VLBA at 1.7~GHz \citep{2011doi1}, obtaining a flux density of 23~mJy, and $T_{\textrm{B}}$ of $> 10^{8.7}$, unquestionably confirming the non-thermal origin of the radio emission. It exhibits a remarkable elongated structure (100~pc) with regions having $T_{\textrm{B}} > 10^{8}$~K, identifying this structure as a jet. Also some signs of a possible counter-jet are seen. What is interesting is that \citet{2011doi1} estimate its spectral index using JVN and VLBA observations to be -0.94$\pm$0.42, which is well in agreement with our core spectral index of -0.99, and with what is seen in its $\alpha$ map. Furthermore, J0806+7248 was found to be a blue outlier by \citet{2021berton1} with the [O~III] lines blueshifted -300~\kms, which is consistent with the presence of a relativistic jet. Also a wing is present with a velocity of -426~\kms, and a FWHM of 733~\kms.

It is striking that it looks like the flux density of J0806+7248 has been decreasing since the first observations, whereas its spectral index seems to have consistently remained very steep. Combined with the non-thermal origin of its radio emission and a rather high luminosity (log $\nu L_{\nu, \textrm{int}}$ = 40.17 erg s$^{-1}$) it seems safe to assume that it really is a CSS-like source. Also the star formation indicators support this conclusion. The W3-W4 colour is at the threshold with 2.57, but $q22$ is very low (0.11) and $S_{\textrm{int}}$ is some tens of mJy higher than $S_{\textrm{W3}}$ or $S_{\textrm{5.2GHz, CDFS}}$, confirming that the radio emission cannot be explained only by star formation.


\subsubsection{J0814+5609} 
\label{sec:J0814}

J0814+5609 (SDSS J081432.11+560956.6) is a radio-loud NLS1 galaxy ($z$ = 0.510) that shows an impressive core-jet morphology as seen in Fig.~\ref{fig:J0814spind}. The peak flux density is 25.85~mJy, the integrated flux density is 29.27~mJy, and the flux densities of the tapered maps (Figs.~\ref{fig:J0814-90k} and \ref{fig:J0814-60k}) are less than a mJy more. It is one of the most luminous sources in our sample with log $\nu L_{\nu, \textrm{int}}$ = 42.19~erg s$^{-1}$. The core shows a flat spectral index, very close to zero (Fig.~\ref{fig:J0814spind}), and the jet shows throughout spectral indices in agreement with optically thin synchrotron emission, or slightly steeper. The spectral index of the brightening in the middle of the jet (RA 08:14:32.48, Dec 56:09:55.07) is -0.45. This region seems to be slighly flatter than the jet in general, possibly due to interaction or shocks that are also responsible for the increased radio emission. Interestingly there seems to be a hotspot with an inverted spectral index near the unresolved core, but taking into account the edge effects and that it is within only a 6$\sigma$ contour its presence needs to be verified with further observations. Most of the extended emission is clearly outside the seemingly featureless host galaxy, as seen in Figs.~\ref{fig:J0814-host-zoom} and \ref{fig:J0814-host}. The SDSS $r$ band 25.0~mag arcsec$^{-2}$ isophotal semimajor axis of the host is 8.6~kpc, whereas the maximum extent of the jet toward south-east is $\sim$34~kpc. The jet exhibits a considerably bent structure outside the host, but without a proper characterisation of the morphology of both the jet and the counter-jet, it is impossible to say whether the bend is due to jet precession, or dissipation of the jet when it fails to be fed enough to maintain its direction. The [O~III] lines of J0814+5609 show a significant blueshift of -590~\kms, but no wing component is present \citep{2021berton1}.

In the tapered maps in Figs~\ref{fig:J0814-90k} and \ref{fig:J0814-60k} also a counter-jet is seen, but due to the decreased resolution caused by tapering nothing can be said about its morphology. However, since both jets are detected, we can estimate the inclination of the system by using their flux density ratio \citep{1979scheuer1}:

\begin{equation}
\frac{ S_{j} }{ S_{cj} } = \left( \frac{1 + \beta cos \theta}{1 - \beta cos \theta} \right) ^{2-\alpha_{jet}}
\end{equation} \label{eq:jetcounterjet}

where $S_{\textrm{j}}$ and $S_{\textrm{cj}}$ are the flux density of the jet and the counter-jet, respectively, $\beta$ = $v/c$, i.e. the speed of the jet relative to the speed of light, $\theta$ is the viewing angle, and $\alpha_{\textrm{jet}}$ is the spectral index of the jet, which we assume to be -1. We fitted the core of the 60k$\lambda$ tapered map with a Gaussian component using the \texttt{imfit} task in CASA, and measured the flux densities of the jet and the counter-jet from the residual map. The flux density ratio we obtain is $\sim$5.34. Even if a jet would be relativistic when launched, it usually decelerates to non-relativistic velocities at kpc scales. Unfortunately we do not have a way to estimate its velocity, but based on its morphology that does not stay properly collimated for long, we can assume that the jet is not relativistic at kpc scales. Estimating the PA using reasonable velocities can still give us information about the inclination of the source and the range of the possible deprojected extents of the jets. Assuming a very moderate $\beta$ = 0.3~c the viewing angle is 24.9\degree, and assuming $\beta$ = 0.5~c gives $\theta$ = 57.0\degree. Without more exact measurements of the jet speed, for example, with higher resolution monitoring closer to the nucleus, we cannot determine the inclination accurately. However, these approximate results show that J0814+5609 cannot be a very low inclination source, whose permitted emission lines would be narrow due to orientation effects, but is a real NLS1. The projected length of the south-east jet in the 60k$\lambda$ tapered map is 41~kpc, and of the western counter-jet 61~kpc. It is interesting that the counter-jet seems more extended than the approaching jet. This might happen due to the bend that shortens the maximum extent of the approaching jet, whereas it might be that the counter-jet is straighter. Assuming a viewing angle of 24.9\degree, the deprojected sizes of the jet and the counter-jet would be 97.4 and 144.9~kpc, respectively, and with $\theta$ = 57.0\degree the deprojected sizes would be 48.9 and 72.7~kpc. With a somewhat unrealistically high $\beta$ of 0.8~c $\theta$ would be 70.1\degree, and the deprojected sizes still 43.6 and 64.9~kpc, making J0814+5609 the most extended NLS1s known to date \citep{2018rakshit1}. 

As expected from a jetted source the mid-infrared emission is swamped by the radio emission in this source: the W3-W4 is 1.98, $q22$ is -0.99 and $S_{\textrm{int}}$ and $S_{\textrm{1.4GHz, JVLA}}$ are both considerably higher than the respective $S_{\textrm{5.2GHz, CDFS}}$ and $S_{\textrm{W3}}$.

In previous studies J0814+5609 has been found to show a flat spectral index all the way from 325~MHz to at least 8.4~GHz. \citet{2015gu1} observed J0814+5609 with the VLBA at 5~GHz, and also analysed archival data at 2.3, 5.0, 8.4~GHz. The source exhibits at all frequencies an elongated structure toward east, consistent with our map, but in their observations any larger-scale extended emission was probably resolved out. Furthermore, they detect considerable variability at 5~GHz, that, assuming a flat spectral index above 8.4~GHz, seems to be present also at higher frequencies since \citet{2016doi1} detected the source with the JVN at 22~GHz with a flux density of 117~mJy.


\subsubsection{J0913+3658} 

J0913+3658 (RX J0913.2+3658) is a radio-quiet NLS1 at $z$ = 0.107 with sub-mJy radio emission: $S_{\textrm{int}}$ is only 0.30~mJy, and $S_{\textrm{90k}\lambda, \textrm{int}}$ is 0.41~mJy. The total spectral index of the source is -0.77 and of the core -0.99 (Fig.~\ref{fig:J0913spind}). The source has been earlier observed at 1.4~GHz in FIRST with a flux density of 1.08~mJy \citep{1995becker1}. This gives a spectral index of -0.78 between 1.4 and 5.2~GHz, in agreement with our $\alpha$ map. The tapered maps in Figs.~\ref{fig:J0913-90k} and \ref{fig:J0913-60k} show diffuse emission, more extended than what is seen in the normal map (Fig.~\ref{fig:J0913spind}). The source has a low luminosity (log $\nu L_{\nu, \textrm{int}}$ = 38.66 erg s$^{-1}$), a reddish W3-W4 colour (2.55), and quite high $q22$ (1.67). Its $S_{\textrm{5.2GHz, CDFS}}$ is about double of its $S_{\textrm{int}}$, whereas its W3 flux density is 15.7~mJy higher than its $S_{\textrm{1.4GHz, JVLA}}$. This is indicative of strong star formation, which is probably the predominant source of radio emission in this source. Nonetheless, an [O~III] wing is present with a velocity of -307~\kms\ and a FWHM of 800~\kms, possibly indicating also the presence of polar dust in this source. Its host galaxy, seen in Fig.~\ref{fig:J0913-host}, appears to be a barred spiral galaxy. The radio emission is concentrated around the centre, possibly the bulge, and partly along the bar to the south from the bulge. The overlapping optical and radio features are in agreement with the star formation scenario. The host galaxy might be accompanied by a small companion on its south side, but this should be confirmed by means of spectroscopic observations.


\subsubsection{J0925+5217} 

J0925+5217 (Mrk 110) is a radio-quiet NLS1 at $z$ = 0.035 with quite a low luminosity of log $\nu L_{\nu, \textrm{int}}$ = 38.72~erg s$^{-1}$. At 5.2~GHz we detected a peak flux density of 1.70~mJy beam$^{-1}$, an integrated flux density of 2.25~mJy for the central emitting region, and of 0.17~mJy for the northern emitting region (Fig.~\ref{fig:J0925spind}). The flux density of the tapered maps is 3.10~mJy. J0925+5217 shows considerable extended emission in the radio maps. In the normal map (Fig.~\ref{fig:J0925spind}) a separate emitting region is seen toward north, and in the tapered maps (Figs.~\ref{fig:J0925-90k} and \ref{fig:J0925-60k}) the radio emission is especially extended toward south. The projected separation of the northern region is 1.7~kpc, and it corresponds to the structures seen already in \citet{1993miller1} at 4.86~GHz with the VLA. They detect a total flux density of 3.8~mJy, slightly higher than ours, and propose that the radio morphology is consistent with circumnuclear star formation. Also \citet{1994kellermann1} detect the same structure at 5~GHz with the VLA and discuss its nature as ``a highly curved jet or ringlike structure''. The extended northern emission is also seen at 1.4~GHz with the VLA, but only the central source is seen at 8.4~GHz \citep{1998kukula1}, indicating it might have a steep spectral index. At 1.7~GHz with VLBA only an unresolved core with a flux density of 1.2~mJy was detected \citep{2013doi1}. They derived the brightness temperature to be at least 10$^{7.8}$~K, indicating an origin of non-thermal nature.

In the 60k$\lambda$ tapered map the extent of the radio emission toward north is 2.8~kpc, and toward south 4.7~kpc. Also possible emission toward west at a separation of 3.3~kpc is seen. It is hard to say whether it is real, but it is also seen in the normal map, at a 5$\sigma$ level. The SDSS $r$ band 25~mag arcsec$^{-1}$ isophotal semimajor and semiminor axes of the host galaxy are 8.9~kpc and 4.9~kpc, and the major axis of the galaxy is almost perpendicular to the radio emission. Toward south the radio emission is reaching the very outskirts of the host, even though it looks like the emission is more extended in radio than in optical. This is because the outskirts of the galaxy are faint, and higher contrast image would be needed to see it clearly. The host itself does not seem to exhibit any clear features (Fig.~\ref{fig:J0925-host}), although in some higher contrast images a disturbed morphology with apparent tidal features can be seen, possibly indicating a recent merger. The other bright source seen in the optical image is a foreground star \citep{1988hutchings1}.

In our $\alpha$ map, the northern region has an inverted spectral index, which may not be real but possibly caused by a low S/N ratio in such a small emitting region. The core is flat with a spectral index of -0.33, consistent with the non-thermal origin of the radio emission. J0925+5217 was detected in FIRST with a flux density of 8.19~mJy, yielding a very steep spectral index of -0.93 between 1.4 and 5.2~GHz. Thus, whereas the core is flat, the overall spectral index of the source seems to be rather steep. Although, if the radio emission is dominated by the central AGN, variability can explain the spectral index since the observations are not simultaneous.

\citet{1993miller1} identified the extended radio emission toward north as possible circumnuclear star formation and there is indeed an excess of mid-infrared emission compared to radio emission, W3 flux density is 59.8~mJy higher than $S_{\textrm{1.4GHz, JVLA}}$, $q22$ is 1.35,  and $S_{\textrm{5.2GHz, CDFS}}$ and $S_{\textrm{int}}$ are roughly the same. However, the colour W3-W4 is not very red with a value of 2.01. The radio emission probably has contribution from both, the star formation and the AGN, since also a non-thermal core was detected and the spectral index of the central region is flat, uncharacteristically for a star-forming region. However, based on these data it is impossible to determine the origin of the large-scale radio emission seen in the tapered maps. Its radio morphology is not common for star-forming regions, unless it is somehow related to the possible recent merger of the host galaxy. Furthermore, an accretion disk wind causing an outflow of the BLR has been detected \citep{2003kollatschny1}, though [O~III] lines are not shifted nor show a wing, suggesting that the wind has not affected the NLR, at least yet, or it is not energetic enough to cause bulk motion of the NLR. Further observations of J0925+5217 are needed to determine the origins of the radio emission at different scales.

\subsubsection{J0926+1244} 

J0926+1244 (Mrk 705) is a radio-quiet NLS1 ($z$ = 0.029) with a low luminosity of log $\nu L_{\nu, \textrm{int}}$ = 38.52 erg s$^{-1}$. The peak flux density is 2.81~mJy beam$^{-1}$, the integrated flux density 3.36~mJy, and the flux densities of the tapered maps minimally higher. The radio morphology seen in Fig.~\ref{fig:J0926spind} looks peculiar: the central component seems to be surrounded by patches of radio emission on all sides. The spectral index of the core is -0.75, and the total spectral index is -0.59. The FIRST flux density for J0926+1244 is 8.52~mJy, giving a 1.4-5.2~GHz spectral index of -0.71, well in agreement with our result. The spectral indices of the surrounding patchy emitting regions seem flat, but most likely are not reliable due to edge effects and low S/N ratio. The extended emission seen in the map is approximately within a kpc from the centre, except the south-west emission, better seen in the tapered maps in Figs.~\ref{fig:J0926-90k} and \ref{fig:J0926-60k}, that is more extended.

VLBA observations at 1.7~GHz show a core-jet structure, with the jet extending 26~pc east from the core \citep{2013doi1}. The total flux density detected was 3.3~mJy, so a significant fraction of the flux was resolved-out or is seen only at larger scales. They derived a brightness temperature of at least $10^{7.9}$~K, confirming the non-thermal nature of the emission. In archival Multi-Element Radio Linked Interferometer Network (MERLIN) observations at 1.7~GHz at 150~mas resolution only single component with a flux density of 5.7~mJy beam$^{-1}$ was seen. Also 8.46~GHz VLA observations yield only an unresolved core with a flux density of 2~mJy \citep{2001schmitt1}. This gives a 5.2-8.46~GHz spectral index of -1.07, steeper than what we observe. Although it should be noted that the observations are more than a decade apart, and since the source hosts a jet, variability is likely to affect the flux densities. Interestingly, the source does not show shift of [O~III] lines \citep{2021berton1}, possibly due to the small scale or low power of the jet.

The host galaxy of J0926+1244 (Fig.~\ref{fig:J0926-host}) was indeed classified as (R)SA(r)0+ in the CVRHS classification scheme in \citet{2017buta1}. This classification indicates that it is a SA type spiral galaxy that has an outer as well as an inner ring. The 5.2~GHz radio emission is concentrated within the innermost part of the host, probably a bulge. The star formation rate of the host galaxy based on 11.3~$\mu$m PAH features was estimated to be as low as 0.6 $\pm$ 0.2~$M_{\odot}$ yr$^{-1}$ \citep{2019martinezparedes1}. On the other hand, based on the relation between the star formation rate and the luminosity of the [C~II] line \citep{2012sargsyan1,2014sargsyan1} it can be estimated to be much higher, 6.6~$M_{\odot}$ yr$^{-1}$. Our star formation diagnostics support the presence of star formation as the W3 flux density is $\sim$101.6~mJy higher than $S_{\textrm{1.4GHz, JVLA}}$, $q22$ is 1.54, and $S_{\textrm{5.2GHz, CDFS}}$ and $S_{\textrm{int}}$ are almost the same. However, the mid-infrared colour with W3-W4 = 2.36 does not necessarily indicate strong star formation. The [O~III] shows a low-velocity, -79~\kms, but turbulent, FWHM = 846~\kms, wing component, and thus the contribution of polar, or equatorial, dust cannot be ruled out either.

In conclusion, despite the presence of a small-scale jet, the star formation seems to be the predominant source of radio emission in this source. However, further observations are needed to disentangle the contribution of different components.


\subsubsection{J0937+3615} 

J0937+3615 (SDSS J093703.03+361537.2) is a radio-quiet NLS1 at $z$ = 0.180 with a moderate integrated flux density of 1.03~mJy, and the flux densities of the tapered maps (Figs.~\ref{fig:J0937-90k} and \ref{fig:J0937-60k}) slightly lower. It is a moderate luminosity source with log $\nu L_{\nu, \textrm{int}}$ = 39.71 erg s$^{-1}$ and does not show any distinct morphology in any of the maps. The radio emission is constrained within its host galaxy, that does not show distinguishable features either. J0937+3615 was detected in FIRST with a flux density of 3.21~mJy, yielding a 1.4-5.2~GHz spectral index of -0.87. This is in agreement with our $\alpha$ map in Fig.~\ref{fig:J0937spind}: the total spectral index is -0.77, and the core spectral index is -0.95. \citet{2015caccianiga1} estimated that the star formation rate in J0937+3615 is extremely high (83~$M_{\odot}$ yr$^{-1}$), and possibly the predominant source of its radio emission. Other star formation diagnostics support this conclusion: its W3-W4 colour is 2.66, $q22$ is 1.34, $S_{\textrm{5.2GHz, CDFS}}$ and $S_{\textrm{int}}$ are roughly the same, but $S_{\textrm{W3}}$ is 15.2~mJy higher than $S_{\textrm{1.4GHz, JVLA}}$. However, it also might exhibit nuclear winds and polar dust, since it has an [O~III] wing with a velocity of -250~\kms, and with a FWHM of 1000~\kms.

\subsubsection{J0952-0136} 

J0952-0136 (Mrk 1239) is a radio-quiet NLS1 at $z$ = 0.020 that has been widely studied in the past. It has one of the highest flux densities in our sample, the peak flux density is 17.49~mJy beam$^{-1}$, the integrated flux density is 18.54~mJy, and the flux densities of the tapered maps about 0.5~mJy higher. Due to its low redshift this translates to a moderate luminosity of log $\nu L_{\nu, \textrm{int}}$ = 39.04 erg s$^{-1}$. Our radio map (Fig.~\ref{fig:J0952spind}) shows a compact component and almost symmetrical extended emission toward north-east and south-west at a position angle of 32\degree. The extended emission is particularly pronounced in the tapered maps in Figs.~\ref{fig:J0952-90k} and \ref{fig:J0952-60k}. The whole extent of the emission in the 60k$\lambda$ tapered map is 4.2~kpc. The spectral index of the source is very steep, the total spectral index is -0.98, and the core spectral index is -1.07.

J0952-0136 was detected at 2.25~GHz with the Parkes-Tidbinbilla Interferometer (PTI) with a flux density of 32~mJy \citep{1990norris1}, and with the VLA in A configuration at frequencies of 1.5, 5, and 8.4~GHz with flux densities of 56.5, 19.5, and 7.9~mJy, respectively \citep{1995ulvestad1}. The source remained unresolved in all these observations. These observations point to a spectral index steepening toward higher frequencies, a feature that was explained to be possibly due to the combination of different resolution at each frequency, and the dominance of diffuse components in this source \citep{2013doi1}. Our 5.2~GHz flux density is in agreement with the 5~GHz observation in \citet{1995ulvestad1}, and the FIRST detection of the source (59.84~mJy) at 1.4~GHz is roughly in agreement with their 1.5~GHz result.

\citet{2010orienti1} reduced archival VLA data of J0952-0136 at 8.4 and 14.9~GHz, and found an unresolved morphology at both frequencies. Further, they produced a VLBA map at 1.6~GHz and were able to see two separate components with a separation of $\sim$30~pc at a position angle of 40 deg. The total flux density they obtained is only 10.2~mJy, meaning that most of the flux is missing at the VLBA scale.

Later, also \citet{2013doi1} observed the source with the VLBA at 1.7~GHz obtaining a core flux density of 4.8mJy, and a total flux density of 20~mJy. Their map shows an elongated 40~pc long structure consisting of several components at a position angle of 47\degree. They estimated the brightness temperature to be 10$^{7.9}$~K, proving the non-thermal origin of the emission. They also claimed to have found a corresponding structure at 8.5~GHz using archival VLA observations. They further studied the source in \citet{2015doi1} where they reported that it shows a two-sided jet-like morphology, extended $\sim$34~pc to both directions. They also re-analysed archival 1.6 and 8.5~GHz VLA data and found an elongation on one side of 1.4~kpc and 85~pc, respectively, with a position angle matching the pc-scale jet. Since most of the radiated power is concentrated within the innermost 100~pc, they conclude it is alike to Fanaroff-Riley I radio galaxies, i.e. edge-darkened{\footnote{Assuming a spectral index of -1 the diffuse emission luminosity threshold between Fanaroff-Riley I and II radio galaxies at 5~GHz is $\sim$log $\nu L_{\nu}$ = 39.5~erg s$^{-1}$.} }. They suggest that the edge-darkened morphology may be caused by the subsonic speed of advancing radio lobes, and the origin of the diffuse radio emission might be a dissipated jet, not powerful enough to penetrate the dense circumnuclear matter. J0952-0136 was not detected at 22~GHz with the JVN \citep{2016doi1}, which could have been expected taking into account its steep spectral index.

It seems plausible that we are seeing the same structure that was detected in previous observations, in which case the main origin of the radio emission would be a low-power two-sided jet. However, also the presence of star formation in this galaxy has been studied quite extensively. The host galaxy (Fig.~\ref{fig:J0952-host}) was originally classified as compact \citep{1992whittle1}, and later as S0/E \citep{1994heisler1}. Its SDSS $r$ band 25~mag arcsec$^{-2}$ isophotal semimajor axis is 2.9~kpc, and its almost perpendicular to the radio axis. \citet{2003rodriguezardila1} studied the near-infrared spectrum of J0952-0136 and detected a 3.43~$\mu$m feature whose origin remained unclear. They re-observed the 0.8-4.5~$\mu$m near-infrared spectrum and did not detect the 3.43~$\mu$m line, or any PAH features \citep{2006rodriguezardila1}. Based on the 11.3~$\mu$m PAH feature \citet{2017ruscheldutra1} estimated the star formation rate to be less than 7.5~$M_{\odot}$ yr$^{-1}$. Consistent with this its star formation rate was estimated to be 3.47~$M_{\odot}$ yr$^{-1}$ based on SED modelling \citep{2016gruppioni1}. \citet{2020buhariwalla1} modelled the X-ray spectrum of J0952-0136 with alternative models and found that the $<$ 3~keV emission cannot be modelled without a starburst component, whose extent they estimate to be less than 400~pc from the nucleus. This is interesting since the other star formation proxies do not predict starburst level star formation. However, the PAH emission can be suppressed in presence of an AGN \citep{2012lamassa1}. The presence of star formation seems to be supported by the W3 flux density that is 520.2~mJy higher than the $S_{\textrm{1.4GHz, JVLA}}$, and $q22$ of 1.16, while the $S_{\textrm{5.2GHz, CDFS}}$ and $S_{\textrm{int}}$ are roughly the same. On the other hand, the W3-W4 colour is not very red (2.02), and the presence of an [O~III] wing with a velocity of -686~\kms\ and a FWHM of 1586~\kms\ indicates that AGN-heated dust could be partly responsible for the very high mid-infrared flux densities. In \citet{1996mulchaey1} the [O~III] emission was found to be clearly elongated in the northeast-southwest direction, matching the position angle of the radio emission.

\subsubsection{J1034+3938} 

J1034+3938 (KUG 1031+398) is a radio-quiet NLS1 at $z$ = 0.042 with a peak flux density of 7.05~mJy beam$^{-1}$, integrated flux density of 7.37~mJy, and the tapered map flux densities only slightly higher. It has a moderate luminosity of log $\nu L_{\nu, \textrm{int}}$ = 39.24 erg s$^{-1}$. Its FIRST flux density is 25.94~mJy, yielding quite a steep 1.4-5.2~GHz spectral index of -0.96. However, it is consistent with the $\alpha$ map (Fig.~\ref{fig:J1034spind}) of the source: the total spectral index of the source is -1.17, and the core spectral index is -1.13. The VLBA the flux density at 1.4~GHz was found to be 8.34~mJy \citep{2014deller1}, indicating that most of the emission was resolved-out and thus comes from larger scales. The normal map (Fig.~\ref{fig:J1034spind}) shows a morphology only slightly extended toward south-east. In the tapered maps (Figs.~\ref{fig:J1034-90k} and \ref{fig:J1034-60k}) an asymmetric east-west structure is seen. Especially the eastern extended emission seems to trace the morphology of the spiral host galaxy, seen in Fig.~\ref{fig:J1034-host}, as it seems to overlap with a spiral arm. The most obvious explanation for the extended emission seems to be star formation, even though using simple stellar population synthesis the star formation rate of J1034+3938 was found to be as low as 0.23~$M_{\odot}$ yr$^{-1}$ \citep{2010bian1}. The W3 flux density is indeed 33.7~mJy higher than $S_{\textrm{1.4GHz, JVLA}}$, but could be also explained by the dust heated by the AGN. \citet{2019smethurst1} detected an [O~III] outflow in this source, and estimate the outflow rate to be $>$ 0.07~$M_{\odot}$ yr$^{-1}$. \citet{2021berton1} did not detect a blueshift of the [O~III] lines but an [O~III] wing is present with a velocity of -253~\kms, and a FWHM of 901~\kms. Other star formation proxies do not indicate strong activity, $W3-W4$ is 2.14, $q22$ is 0.57, and $S_{\textrm{5.2GHz, CDFS}}$ is 5.4~mJy lower than $S_{\textrm{int}}$. Thus the radio emission seems to be dominated by the AGN, with star formation -- based on the morphology -- having a small contribution.

J1034+3938 is one of the NLS1s showing the most prominent soft excess in X-rays \citep{1995pounds1}, and interestingly, it is one of the rare AGN where a quasi-periodic oscillation (QPO) has been reported in X-rays, and, in fact, the only source showing the QPO with high significance and repeatability \citep[][ and references therein]{2021jin1}.

\subsubsection{J1038+4227} 

J1038+4227 (SDSS J103859.58+422742.3) is a radio-loud source at $z$ = 0.220 with very pronounced diffuse extended emission (Fig.~\ref{fig:J1038}). Its integrated flux density is 3.75~mJy, and $S_{\textrm{90k}\lambda, \textrm{int}}$ is 5.36~mJy, showing that a large fraction of the emission is very low surface brightness. Its central region exhibits a steep spectral index with $\alpha$ = -0.77, and the total spectral index is -0.89. Unfortunately, the extended emission does not have a high enough S/N to estimate the spectral index outside the centre. J1038+4227 was detected at 1.4~GHz in the FIRST and in the NVSS with flux densities of 2.44~mJy and 7~mJy, respectively. The FIRST measurement underestimates the flux density because it might not have had high enough sensitivity to detect the faint extended emission.

The star formation diagnostics do not indicate strong activity: the W3-W4 colour is 2.29, $q22$ is 0.57, and both $S_{\textrm{W3}}$ (9.0~mJy) and $S_{\textrm{5.2GHz, CDFS}}$ (0.4~mJy) are lower than the respective JVLA values. However, this might be partly due to the sensitivity limit of WISE, whose 5$\sigma$ point source sensitivity is estimated to be 0.86 and 5.4~mJy in W3 and W4 bands, respectively \citep{2010wright1}. Considering the diffuse nature of the emission seen in J1038+4227, a significant fraction of the mid-infrated emission could have gone undetected, leading to underestimated mid-infrared flux densities. 

The spatial extent of the emission is huge, its most extended diameter in the normal map (Fig.~\ref{fig:J1038spind}) is 9.4~arcsec, which corresponds to 33.4~kpc, and in the 60k$\lambda$ tapered map 12.5~arcsec, corresponding to 44.4~kpc. An SDSS $r$ band 25~mag arcsec$^{-2}$ isophotal major axis estimate for its host galaxy is only $\sim$4~arcsec, or 14.2~kpc, clearly smaller than the extent of the radio emission. In the image of the host galaxy, overlaid with the 90k$\lambda$ tapered radio map (Fig.~\ref{fig:J1038-host}) it can be seen that the radio emission seems to reach well outside the host, forming a halo-like structure.

The source is bright, with log $\nu L_{\nu, \textrm{int}}$ = 40.67 erg s$^{-1}$. In the normal map most of the emission comes from the core, and thus its luminosity is at the CSS threshold. Also the steep spectral index of the core would agree with this classification. The origin of the extended emission remains unknown, and this source certainly needs to be studied further to reveal its nature.



\subsubsection{J1047+4725} 

J1047+4725 (SDSS J104732.68+472532.0) was cleaned using the uniform weighting to suppress the very strong sidelobes, but the normal radio map (Fig.~\ref{fig:J1047spind}) still suffers from some artefacts around its peripheries. J1047+4725 at $z$ = 0.799 is very radio-loud and the most luminous source in our sample with log $\nu L_{\nu, \textrm{int}}$ = 43.77~erg s$^{-1}$. Its peak flux density is 190.90~mJy beam$^{-1}$, and the integrated flux density is 381.75~mJy, with the integrated flux densities of the tapered maps being approximately the same, thus about half of its radio emission comes from the unresolved core. Its W3-W4 colour is red (2.69) but clearly the AGN is the main origin of the radio emission since the $q22$ is -2.18, and the $S_{\textrm{int}}$ and $S_{\textrm{1.4GHz, JVLA}}$ are both several hundred mJy higher than the corresponding $S_{\textrm{5.2GHz, CDFS}}$ and $S_{\textrm{W3}}$ values. Its host galaxy does not exhibit any clear structures, and the radio emission seems to be confined within it (Fig.~\ref{fig:J1047-host}).

Several radio measurements of J1047+4725 exist in the literature. \citet{2015gu1} observed it at 5~GHz with the VLBA, but were not able to resolve it. However, they also analysed archival VLBA data where a clear core-jet structure can be seen (see their Fig. 6). Furthermore, at 8.4~GHz observed with the JVLA the source exhibits a two-sided jet structure, implying a large viewing angle. However, the [O~III] lines are not shifted, so the jets have not been powerful enough to cause bulk motion in the NLR. At 22.4~GHz only an unresolved core is seen. \citet{2015gu1} also note that based on archival data, which extends all the way down to 74~MHz \citep{2007cohen1}, it looks like the break frequency of the radio spectrum lies in the MHz regime, as often seen in CSS sources \citep[see Fig. 17 in][]{2015gu1}. Above 1.4~GHz the spectrum steepens to -0.5/-0.6, which is consistent with our $\alpha$ map in Fig.~\ref{fig:J1047spind} which has a core spectral index of -0.63 and a total spectral index of -0.75, and also with the CSS characteristics \citep{1998odea1}. It thus seems evident that J1047+4725 is a jetted NLS1 seen at quite a large angle, and could also be classified as a CSS.


\subsubsection{J1048+2222} 

J1048+2222 (SDSS J104816.57+222238.9) is a radio-quiet NLS1 at $z$ = 0.330 with a very low integrated flux density of 0.24~mJy. The flux density of the 60k$\lambda$ tapered map is 0.47~mJy, almost double compared to the normal map. Due to its redshift the low flux density corresponds to a moderate luminosity of log $\nu L_{\nu, \textrm{int}}$ = 39.75~erg s$^{-1}$. It was detected in FIRST with a flux density of 1.49~mJy. This gives a very steep spectral index of -1.39 between 1.4 and 5.2~GHz. Our $\alpha$ map (Fig.~\ref{fig:J1048spind}) does not suggest this steep an index, as the total spectral index is -0.56 and the core spectral index is -0.71. The map shows flattening toward north, but in such a faint source it can be artificial. The normal map does not show considerable extended structure, but the tapered maps (Figs.~\ref{fig:J1048-90k} and \ref{fig:J1048-60k}) show an extensive, though very faint, region of emission toward south-west. The maximum extent of the emission is 42.8~kpc, which places it clearly outside the host galaxy (see Fig.~\ref{fig:J1048-host}) whose SDSS $r$ band 25~mag arcsec$^{-2}$ isophotal semimajor axis is 7.5~kpc. The bulk of the radio emission is confined within the host. What is the origin of the extended emission remains unclear based on the currently available data.

The star formation diagnostics of this source are not conclusive either. The $q22$ parameter is 1.09, the W3 flux density is 5.1~mJy higher than $S_{\textrm{1.4GHz, JVLA}}$, and $S_{\textrm{5.2GHz, CDFS}}$ and $S_{\textrm{int}}$ are about the same, but the W3-W4 colour is only 2.2. J1048+2222 is a blue outlier with the [O~III] lines shifted by -209~\kms, indicating that it might host a jet or a powerful nuclear outflow. The existence of an [O~III] wing with a velocity of -531~\kms, and a FWHM of 1239~\kms\ indicates that a nuclear wind is present in this source. The blueshifted [O~III] lines, inconclusive star formation diagnostics, and the morphology of the radio emission indicate that the AGN is responsible for the most of the radio emission seen in this source.

\subsubsection{J1102+2239} 

J1102+2239 (SDSS J110223.38+223920.7) is a radio-quiet NLS1 with a low integrated flux density of 0.70~mJy, and a slightly higher 60k$\lambda$ tapered map flux density of 1.22~mJy. Owing to its redshift of 0.453 its luminosity is quite high, with log $\nu L_{\nu, \textrm{int}}$ = 40.47~erg s$^{-1}$. It was considered as a possible gamma-ray emitting NLS1 in \citet{2011foschini1}, but the detection has not been confirmed. Its only archival radio detection is 1.80~mJy in FIRST. This gives a 1.4-5.2~GHz spectral index of -0.72, in agreement with the $\alpha$ map of the source (Fig.~\ref{fig:J1102spind}) that has a total spectral index of -0.76. The $\alpha$ map is actually indicating even steeper spectral indices, flattening toward north, but these might be edge effects combined with a low S/N ratio. However, it does exhibit blueshifted [O~III] lines with a velocity of -565~\kms, and a FWHM of 880~\kms, as well as an [O~III] wing with a velocity and FWHM of -750 and 1320~\kms, respectively. So the NLR experiences bulk motions, probably induced by a jet or a wind. 

The normal map in Fig.~\ref{fig:J1102spind} is mostly featureless. The tapered maps (Figs.~\ref{fig:J1102-90k} and \ref{fig:J1102-60k}) show extended, but faint, radio emission toward west - south-west, but the emission is so weak that we cannot be certain it is real. The host is a late-type galaxy, significantly perturbed by merging with another disk galaxy \citep{2020olguiniglesias1}. \citet{2020olguiniglesias1} also detect an H~II region that based on its spectrum belongs to the interacting system. Indeed, \citet{2015caccianiga1} found the star formation rate of this source to be at a starburst level with 302~$M_{\odot}$ yr$^{-1}$. This is somewhat supported by the star formation diagnostics: the W3-W4 colour is 2.66, $S_{\textrm{W3}}$ is 3.6~mJy higher than $S_{\textrm{1.4GHz, JVLA}}$, and $S_{\textrm{5.2GHz, CDFS}}$ and $S_{\textrm{int}}$ are comparable, but $q22$ is only 0.81.

Based on these data, with the exception of the high luminosity and the outflow, it seems like star formation has a major contribution to the radio emission of J1102+2239. In principle, the [O~III] outflow could have been triggered by enhanced star formation that causes stellar winds resulting in outflows, or be a feature resulting from the interaction. However, usually the velocities and FWHMs of outflows induced by these processes are more moderate -- from a few hundred to $\sim$500~ km s$^{-1}$ -- than what is seen in the AGN-induced outflows \citep{2012harrison1}. It thus seems likely that the outflow is of AGN origin. This does not mean that a jet needs to be present, but taking into account the luminosity that is higher than usually seen in pure starburst galaxies, it seems that also the AGN is required to explain the observed radio emission.

\subsubsection{J1110+3653} 

J1110+3653 (SDSS J111005.03+365336.3) is a radio-loud NLS1 at $z$ = 0.629 with a peak flux density of 8.12~mJy beam$^{-1}$, integrated flux density of 9.34~mJy, and the flux densities of the tapered maps about 0.5~mJy higher. It is one of the most luminous sources in our sample with log $\nu L_{\nu, \textrm{int}}$ = 41.89~erg s$^{-1}$. Its radio map exhibits considerably extended emission toward south - south-east (Fig.~\ref{fig:J1110spind}), and the tapered maps (Figs.~\ref{fig:J1110-90k} and \ref{fig:J1102-60k}) show that the source is also extended toward north, but the emission does not seem to be exactly axisymmetric, but lopsided to east. The extent of the southern emission in the normal map is 21.2~kpc, and the north-south extent of the whole emitting region in the 60k$\lambda$ tapered map is 61.5~kpc. The SDSS $r$ band 25.0~mag arcsec$^{-2}$ isophotal semimajor axis of the host is 7.0~kpc, so the radio emission is reaching way outside the host (Fig.~\ref{fig:J1110-host-zoom}).

J1110+3653 has been previously observed in the Westerbork Northern Sky Survey (WENNS) at 325~MHz and in FIRST at 1.4~GHz with flux densities of 45 and 20.83~mJy, respectively. This gives a spectral index of -0.53 between 325~MHz and 1.4~GHz, and -0.61 between 1.4 and 5.2~GHz, so the source seems to have an overall steep radio spectral index. In our $\alpha$ map (Fig.~\ref{fig:J1110spind}) the core of the source is flat, with a spectral index of -0.10. The extended emission is steeper, with spectral indices around -0.5, or steeper.

In 5~GHz observations with the VLBA the extended emission is resolved-out and the source shows only an unresolved core with a flux density of 8.8~mJy \citep{2015gu1}. The brightness temperature was found to be at least 10$^{10}$~K, suggesting non-thermal processes being responsible for the emission, but no extreme Doppler boosting. The source was not detected at 22~GHz with the JVN with a detection limit of 7~mJy \citep{2016doi1}. 

J1110+3653 was not detected in W3 or W4 bands so we can not say anything about the star formation activity in the host galaxy, as it has not been studied in the past either. However, all things considered, it seems rather evident that the AGN is responsible for the bulk of the observed radio emission. The extended emission resembles a jet that has propagated through the host and it possibly dissipating when reaching the edge of the host. However, the possibility that the morphology could instead be caused by an outflow cannot be ruled out with these data. Interestingly, J1110+3653 does not show a blueshift in its [O~III] lines, nor an [O~III] wing component.

\subsubsection{J1121+5351} 

J1121+5351 (SBS 1118+541) is a radio-quiet NLS1 at $z$ = 0.103 with a moderate luminosity of log $\nu L_{\nu, \textrm{int}}$ = 39.25~erg s$^{-1}$. Its peak flux density is 1.02~mJy beam$^{-1}$, and the integrated flux density is 1.22~mJy, with the values for the tapered maps being only marginally higher. It was detected in the International Low-Frequency Array (LOFAR) survey at 144~MHz with a flux density of 11.49~mJy \citep{2019gurkan1}, and in FIRST with a flux density of 2.14~mJy. These yield a spectral index of -0.74 between 144~MHz and 1.4~GHz, and of -0.43 between 1.4 and 5.2~GHz. The $\alpha$ map of the source in Fig.~\ref{fig:J1121spind} shows a generally steep spectral index: the total spectral index is -0.61, and the core spectral index is -0.76. The normal radio map shows a slightly extended structure, but due to its triangular shape it is hard to say whether it could be artificial and caused by sidelobes. There is no hint of it in the tapered maps (Figs.~\ref{fig:J1121-90k} and \ref{fig:J1121-60k}), which instead show a faint extended structure toward east - south-east. The bulk of the radio emission is confined within the featureless host galaxy (Fig.~\ref{fig:J1121-host}).

The mid-infrared emission of J1121+5351 is enchanced, the W3 band flux density is 34.6~mJy higher than $S_{\textrm{1.4GHz, JVLA}}$, and $q22$ is 1.43. Also $S_{\textrm{5.2GHz, CDFS}}$ and $S_{\textrm{int}}$ are comparable, but the W3-W4 colour is only 2.33. According to the conventional Baldwin-Phillips-Terlevich (BPT) diagnostic diagram \citep{1981baldwin1} this source would be classified as a star forming galaxy \citep{2013stern1}, but in a revised classification scheme by \citet{2012shirazi1} it is classified as an AGN-dominated galaxy. However, even in the latter scheme it is very close to the threshold of composite galaxies, and taking into account the earlier classification it seems safe to assume that both the AGN and star formation contribute to the ionising continuum. The mid-infared emission can be further enhanced by the AGN-heated dust emission, since an [O~III] wing, with a velocity of -208~\kms\ and a FWHM of 735~\kms, can be seen in its optical spectrum. It seems like star formation could be also the predominant source of the radio emission, but without further studies the possible AGN contribution cannot be ruled out.

\subsubsection{J1138+3653} 

J1138+3653 (SDSS J113824.54+365327.1) is a radio-loud NLS1 at $z$ = 0.356 with a peak flux density of 4.37~mJy  beam$^{-1}$, an integrated flux density of 4.64~mJy, with similar integrated flux densities of the tapered maps. Its luminosity is quite high with log $\nu L_{\nu, \textrm{int}}$ = 41.02 erg s$^{-1}$. The radio map in Fig.~\ref{fig:J1138spind} does not show any resolved structures, and the radio emission is contained within the host galaxy (Fig.~\ref{fig:J1138-host}). The total spectral index is -0.59 and the core spectral index is -0.68. J1138+3653 was detected in FIRST with a flux density of 12.64~mJy, which gives a spectral index of -0.76 between 1.4 and 5.2~GHz, consistent with what we find.

J1138+3653 shows considerable variability, since its flux density at 5~GHz in VLBA observations, a few years prior to our observations, was found to be 9.5~mJy \citep{2015gu1}. The source was not resolved in these observations. The brightness temperature was found to be at least 10$^{9.5}$~K, indicative of non-thermal processes being responsible for the radio emission, but with only mild beaming, possibly due to slow jet speed or large viewing angle. Were the latter true, it means that the spatial extent of J1138+3653 must have been very small at the time of the VLBA observations, since the resolution of their observations was $\sim$20~pc. Thus the former explanation seems more plausible. J1138+3653 does not show a shift in its [O~III] lines either, indicating that powerful, large-scale jet action is not present.

Interestingly, J1138+3653 was found to exhibit a significant amount of star formation (14~$M_{\odot}$ yr$^{-1}$,  \citealp{2015caccianiga1}), though the star formation diagnostics employed in this paper do not reflect this. The W3-W4 colour is red (2.52), but the $q22$ is -0.28, and $S_{\textrm{int}}$ is $\sim$4.5~mJy higher than $S_{\textrm{5.2GHz, CDFS}}$, and $S_{\textrm{1.4GHz, JVLA}}$ $\sim$8.0~mJy higher than $S_{\textrm{W3}}$, indicating that the radio emission is dominated by the AGN. According to its properties this source seems to be CSS-like.

\subsubsection{J1159+2838} 

J1159+2838 (SDSS J115917.32+283814.5) is a radio-quiet NLS1 at $z$ = 0.210 with a peak flux density of 0.79~mJy  beam$^{-1}$, and an integrated flux density of 0.84~mJy, with the tapered map showing approximately the same flux densities. Its radio maps (Fig.~\ref{fig:J1159}) show a slightly elongated structure toward south-west, but only at 3$\sigma$ significance level. No structures in the featureless host galaxy overlap with these radio contours (Fig.~\ref{fig:J1159-host}). According to the $\alpha$ map (Fig.~\ref{fig:J1159spind}) the spectral index is steep: the total spectral index is -0.71, and the core spectral index is -0.93. Calculating the spectral index between 1.4 and 5.2~GHz using the archival FIRST detection (2.01~mJy) gives -0.66, roughly in agreement with our $\alpha$ map.

\citet{2015caccianiga1} found that the star formation rate in J1159+2838 is 128~$M_{\odot}$ yr$^{-1}$, suggestive of starburst activity. The very red W3-W4 colour (3.03) supports this, as well as the $q22$ of 1.22, and the W3 flux density which is $\sim$5.8~mJy higher than the $S_{\textrm{1.4GHz, JVLA}}$. $S_{\textrm{int}}$ and $S_{\textrm{5.2GHz, CDFS}}$ are roughly the same, indicating that significant contribution from the AGN is not needed to explain the radio emission. The luminosity of J1159+2838 is moderate with log $\nu L_{\nu, \textrm{int}}$ = 39.76 erg s$^{-1}$. Starburst-level star formation activity seems to be the predominant source of radio emission is this galaxy.

\subsubsection{J1203+4431} 

J1203+4431 (NGC 4051) is one of the most studied sources in our sample due to its proximity ($z$ = 0.002), and also one of the six original Seyfert galaxies studied in \citet{1943seyfert1}. It is a radio-quiet NLS1 hosted by a spiral galaxy (Fig.~\ref{fig:J1203-host}). In our observations, its peak flux density is 0.78~mJy beam$^{-1}$, the integrated flux density is 5.47~mJy, and the flux densities of the tapered maps slightly higher. Due to its low redshift this translates to a very low luminosity of log $\nu L_{\nu, \textrm{int}}$ = 36.46 erg s$^{-1}$. Most of the emission originates from the extended regions, seen in Fig.~\ref{fig:J1203spind}. The emission is elongated in the north-east - south-west direction, and exhibits a curved S-shape. Also the core seems to be elongated with the same position angle. The morphologies seen in the tapered maps in Figs.~\ref{fig:J1203-90k} and \ref{fig:J1203-60k} are similar to the normal map. The maximum extent of the emission in the normal map is 0.42~kpc, and in the 60k$\lambda$ tapered map 0.44~kpc, thus the radio emission we see is confined within the very centre of the galaxy, as can be seen in Figs.~\ref{fig:J1203-host} and \ref{fig:J1203-host-zoom}. In the $\alpha$ map (Fig.~\ref{fig:J1203spind}) the spectral index of the core is -0.85, steepening to $\sim$-1 toward south-west. Interestingly it looks like the spectral index of the south-west extended emission is on average flat, with spectral indices around -0.5.

J1203+4431 was observed with VLA at 1.5 and 5~GHz already in \citet{1984ulvestad1}. At 5~GHz they observed an east-west -oriented structure, that resembles the central region seen in our radio map. At 1.4~GHz they see also the south-west extended region, with a morphology similar to what is seen in our map. \citet{1993baum1} observed J1203+4431 with the Westerbork Synthesis Radio Telescope (WSRT) at 5~GHz and with the VLA in D configuration at 1.5 and 5~GHz. In the WSRT observations they see a north-east - south-west structure, that is about twice in size compared to what is seen in our maps. In their 5~GHz VLA map they clearly see the radio emission originating in the spiral arms of the galaxy, and in the 1.5~GHz map they also see the radio emission of the disk.

At higher resolution, at least from $\sim$1.5 to 15~GHz, the core component splits into three separate components with sub-mJy flux densities, and that correspond to the structures seen at lower resolution \citep{2009giroletti1,2011king1,2018baldi1,2018saikia1}. Very high resolution observations with the VLBA at 1.7~GHz \citep{2013doi1} and with the JVN at 22~GHz \citep{2015doi1} do not detect the source. Several studies have suggested that the origin of the inner structure is the AGN activity \citep[e.g., ][]{2006gallimore1,2011maitra1,2017jones1}. Based on broadband SED modelling \citet{2011maitra1} conclude that the source hosts a mildly relativistic jet/outflow. Studying a sample of Seyfert galaxies with kpc-scale radio structures, including J1203+4431, \citet{2006gallimore1} argue that in general these structures originate when jet plasma interacts with the interstellar medium and gets decelerated. They argue that in most cases the jet loses most of its power within the innermost kpc. \citet{2017jones1} studied the radio variability of J1203+4431 but found no significant signs of it. They hypothesise that the jet-like structure results from an earlier period of higher activity, and that at the moment the source is at a low-activity state. This could be supported by a study by \citet{2020esparzaarredondo1} where they classify J1203+4431 as a possible early fading candidate, meaning that it looks like its nuclear activity is currently decreasing. These properties match well with what we observe: very close to the nucleus the morphology toward south-west appears somewhat collimated, but dissipates at a radius of $<$ 0.1~kpc. We assume that the north-east part is receding, and a similar nuclear jet cannot be seen due to Doppler deboosting. Interestingly the source was also detected by the \textit{Planck} satellite at 353, 545, and 857~GHz with flux densities of 0.74, 3.04, 10.50~Jy, respectively \citep{2013planck1}. This emission most probably originates from the dust in the host galaxy \citep{2011popescu1}.

A water maser was detected in this source by \citet{2018hagiwara1}. Also ultrafast outflows have been detected in J1203+4431 in X-rays in several studies \citep{2010tombesi1,2013pounds1,2020igo1} with velocities up to 0.12~c. \citet{2013pounds1} suggest that the outflow gets terminated at a rather small radius, losing much of its energy when interacting with the interstellar medium. However, the outflow is momentum-conserving, and thus enables momentum-driven AGN feedback in the galaxy. Outflows have been detected also in the optical spectrum. \citet{1997christopoulou} detected an [O~III] outflow with two blueshifted components, originating from opposite sides of the outflow cone. They model the outflow and determine its velocity to be 245~km s$^{-1}$. The outflow was later confirmed by \citet{2013fischer1}, who found a maximum velocity of 550~km s$^{-1}$. These models were further improved in \citet{2021meena1}, where they claim that the outflow is biconical, launched within 0.5~pc from the nucleus, and is driven by radiation. At a distance of 350~pc the outflow has a velocity of 680~km s$^{-1}$, and they argue that based on the gas kinematics and the host galaxy properties it can travel up to a distance of $\sim$1~kpc from the nucleus. The [O~III] outflow somewhat traces the north-east morphology seen in the radio maps, which is natural since it can be expected that the jet and the outflow are launched into the same direction. However, in the outflow model the north-east part is approaching and the south-west receding, which does not seem to be the case based on the radio morphology.

It is clear that the radio emission is of nuclear origin, and in any case the star formation diagnostics for J1203+44231 are unreliable since we see only a fraction of the whole radio emission of the source. However, the star formation rate of the host has been estimated to be a few $M_{\odot}$ yr$^{-1}$ \citep{2019lianou1,2020lamperti1}, and in the 5~GHz VLA radio map of \citet{1993baum1} the spiral arms are clearly seen, proving that across the host some of the radio emission does have an origin related to star formation.


\subsubsection{J1209+3217} 

J1209+3217 (RX J1209.7+3217) is another radio-quiet NLS1 galaxy ($z$ = 0.144) whose radio emission seems to be dominated by star formation. The radio map does not show significant structure, and the radio emission is confined inside the host galaxy (Fig.~\ref{fig:J1209-host}). The 5.2~GHz peak flux density is 0.60~mJy beam$^{-1}$, the integrated flux density is 0.72~mJy. The total spectral index is -0.55 and the core spectral index is -0.62 (Fig.~\ref{fig:J1209spind}). This is consistent with the 1.4-5.2~GHz spectral index of -0.76, estimated using archival 1.4~GHz FIRST data. The source has a moderate luminosity of log $\nu L_{\nu, \textrm{int}}$ = 39.31 erg s$^{-1}$. The W3-W4 colour is 2.88, the $q22$ value is 1.75, and the W3 flux density is significantly, $\sim$15.6~mJy, higher than $S_{\textrm{1.4GHz, JVLA}}$, indicative of strong star formation. $S_{\textrm{5.2GHz, CDFS}}$ is $\sim$0.5~mJy higher than $S_{\textrm{int}}$, suggesting that the radio emission can be explained by star formation. The [O~III] lines are not shifted, but a turbulent wing component with a velocity of -222~\kms\ and a FWHM of 1405~\kms\ is present.

\subsubsection{J1215+5442} 

J1215+5442 (SBS 1213+549A) is a radio-quiet NLS1 at $z$ = 0.150 with a peak flux density of 0.56~mJy beam$^{-1}$, an integrated flux density of 0.64~mJy, and a moderate luminosity of log $\nu L_{\nu, \textrm{int}}$ = 39.30 erg s$^{-1}$. Its radio maps (Fig.~\ref{fig:J1215}) do not exhibit any remarkable features, and according to the $\alpha$ map (Fig.~\ref{fig:J1215spind}) its total spectral index is -1.0. J1215+5442 was detected in FIRST with a flux density of 2.35~mJy, giving a 1.4-5.2~GHz spectral index of -0.99, confirming the steep spectral index of our map. 

Based on the star formation proxies, the radio emission of J1215+5442 can be explained by star formation activity and does not require a significant contribution from the AGN. Its mid-infrared colour, W3-W4, is 2.47, $q22$ is 1.52, and the W3 flux density is $\sim$29.6~mJy higher than the extrapolated JVLA 1.4~GHz flux density, suggesting strong star formation activity. Also $S_{\textrm{5.2GHz, CDFS}}$ is $\sim$0.8~mJy higher than $S_{\textrm{int}}$, supporting star formation as the main producer of radio emission in this galaxy. However, the contribution of AGN-heated dust to the mid-infrared emission cannot be ruled out, especially since an [O~III] wing is present, with a velocity of -500~\kms\ and a FWHM of 661~\kms.

\subsubsection{J1218+2948} 

J1218+2948 (Mrk 766) is a well-studied nearby ($z$ = 0.013) radio-loud NLS1 galaxy. Its peak flux density is 11.17~mJy beam$^{-1}$, the integrated flux density 15.46~mJy, and the flux densities of the tapered maps almost the same. Its luminosity is quite low with log $\nu L_{\nu, \textrm{int}}$ = 38.47~erg s$^{-1}$. The radio morphology (Fig.~\ref{fig:J1215spind}) is mostly featureless, but exhibits a slight elongation in south-east - north-west direction. The tapered maps (Figs.~\ref{fig:J1218-90k} and \ref{fig:J1218-60k}) reveal additional extended emission toward east. The radio emission is confined within the central parts of a barred spiral host (Fig.~\ref{fig:J1218-host} and \ref{fig:J1218-host-zoom}). The $\alpha$ map shows a steep core, with a spectral index of -0.84. The total spectral index is -0.83, so the source is thoroughly steep.

In archival radio observations obtained with the VLA the flux densities were found to be 35.9~mJy at 1.46~GHz, and 15.4~mJy at 4.89~GHz, in OVRO observations at 20~GHz the flux density was 5.3~mJy \citep{1987edelson1}, and observations with Arecibo yielded 29~mJy at 2.38~GHz \citep{1978dressel1}. \citet{1995bicay1} obtained 17 $\pm$ 4~mJy at 4.755~GHz with the GBT, and \citet{2010parra1} 14.5~mJy at 4.8~GHz with the VLA. All the archival values are consistent with what we obtain, and indicate that the emission is not variable, at least at these time-scales. Using the archival and our data the spectral index between 1.46 and 5.2~GHz is -0.66, and between 5.2 and 20~GHz -0.79, consistent with our $\alpha$ map. J1218+2948 has been unresolved in most radio imaging observations, but at 8.4~GHz with the VLA in A configuration a slight extension toward south-east was seen \citep{1995kukula1}, and 1.5~GHz VLA observations also detected extended emission on the north-west side of the nucleus, in addition to the south-east emission, was detected \citep{1999nagar1}. These structures correspond to the slightly elongated morphology we see in J1218+2948. The source was not detected at 22~GHz with the JVN \citep{2016doi1}, but was detected at 95~GHz with the Combined Array for Research in Millimeter-wave Astronomy (CARMA) with a flux density of 1.98~mJy \citep{2015behar1}. \citet{2015behar1} note that the high frequency emission follows the $L_{\textrm{R}}$ - $L_{\textrm{X}}$ relation found in stellar coronae, and argue that the high frequency radio emission is similarly produced in radio-quiet and moderately radio-loud AGN in the accretion disk corona closer to the black hole and within a smaller region than the large-scale radio emission. \citet{2020esparzaarredondo1} classify this source as a fading nucleus.

Structures resembling the radio morphology have also been observed in optical and IR emission lines. The [O~III] emission is found to be rather circular \citep{1996mulchaey1,2003schmitt1} and limited to the innermost $\sim$0.5~kpc of the galaxy, but an excitation map shows a biconical south-east -- north-west structure, with the south-east emission being brighter. \citet{1996gonzalesdelgado1} find that the H$\alpha$ emission is extended 1.5~kpc toward north-east, and conclude that the spectrum of this region can be fitted with H~II region models. They also find both a blue-shifted and a red-shifted component of [O~III] on opposite sides of the nucleus, and argue that the kinematics of this gas is a result of a nuclear outflow. \citet{2014schonell1} mapped the innermost 450~pc of the galaxy with the Gemini Near Infrared Integral Field Spectrograph in $J$ and $Ks$ bands, and found that also the [Fe~II] emission is extended toward south-east. Based on line ratios they deduce that they have been ionised by a mixture of nuclear and starburst emission. Toward south-east they detect signs of shocked gas and also an increase in the [Fe~II] velocity distribution, as well as red- and blue-shifted components they think are caused by a jet/ourflow. Also a water maser has been detected in J1218+2948, but whether it is associated with the nucleus is unclear \citep{2011tarchi1}.

Star formation diagnostics for this source can be unreliable due to the JVLA observation not mapping the whole galaxy, so the flux density can be somewhat underestimated. Furthermore, \citet{2005rodriguesardila1} found in the near-infrared spectrum an excess they attribute to hot dust with a blackbody temperature of 1200~K they think resides very close to the nucleus, at the dust sublimation radius. If also the toroidal/poloidal cooler dust is bright it can contribute to the observed mid-infrared flux densities. Keeping these in mind, the star formation diagnostics point to this source being star-forming. The W3-W4 is 2.87, $q22$ is 1.40, the W3 flux density is 269.8~mJy higher than $S_{\textrm{1.4GHz, JVLA}}$, and $S_{\textrm{5.2GHz, CDFS}}$ is 4.6~mJy higher than $S_{\textrm{int}}$. The star formation of the host is not extraordinarily high: based on far-infrared observations it is 5.42~$M_{\odot}$ yr$^{-1}$ \citep{2017terrazas1}, and based on the infrared luminosity it is 2.24~$M_{\odot}$ yr$^{-1}$ \citep{2020lamperti1}. However, \citet{2003rodriguezardila1} detect a 3.3~$\mu$m PAH feature within 150~pc from the nucleus at a level seen in circumnuclear starbursts in other sources. Also ultraviolet (UV) observations reveal several bright star-forming regions and star clusters, and the most pronounced star forming region seems to be associated with the east part of the galaxy bar \citep{2007munozmarin1}. This could explain the east-ward extended emission seen in the tapered maps.

Interestingly two QPOs with a frequency ratio of 3:2 have been reported in J1218+2948 in X-rays \citep{2001boller1,2017zhang1}. The periods of the QPOs are 4200 and 6450~s, and seem to be transient. Also ultrafast outflows (UFO) have been detected in this source, with velocities around $\sim$0.1~c \citep[][and references therein]{2020igo1}.

All in all, it seems like the radio emission of J1218+2948 is a mixture of star formation and AGN related emission. In the radio maps there are signs of possible jets or outflows in the south-east - north-west direction, but due to the lack of variability in the radio flux densities or radio morphology the jets/outflows appear lethargic. Furthermore, according to several studies the host galaxy is forming stars, but the level of this activity, and thus its contribution to the radio emission, is uncertain. Further observations would be needed to clarify the role of each component in this galaxy.

\subsubsection{J1227+3214} 

J1227+3214 (SDSS J122749.14+321458.9) is a radio-loud NLS1 ($z$ = 0.136) with a peak flux density of 2.96~mJy beam$^{-1}$, the integrated flux density of 3.34~mJy, and the flux densities of the tapered maps only slightly higher. This results in a moderate luminosity of log $\nu L_{\nu, \textrm{int}}$ = 39.94~erg s$^{-1}$. The source was detected in FIRST with a flux density of 6.42~mJy, yielding a spectral index of -0.50 between 1.4 and 5.2~GHz. Our $\alpha$ map in Fig.~\ref{fig:J1227spind} shows a steeper spectrum: the total spectral index is -0.67, and the core spectral index is -0.74. The difference might be due to variability since the mean epoch of the FIRST observations is in 1994, however, due to the lack of further radio observations we cannot investigate the possible variability. The radio morphology seems somewhat resolved, but without any clear structures. In the normal map the morphology seems slightly extended toward north-east and west, whereas the tapered maps reveal faint extended emission toward north and south-east. Whereas the bulk of the emission is confined within the host galaxy, this extended emission seems to be reaching outside it (Fig.~\ref{fig:J1227-host}).  

The star formation diagnostics indicate some activity in the host: the W3-W4 colour is 2.62, $q22$ is 1.06, and the W3 flux density is $\sim$25.4~mJy higher than the $S_{\textrm{1.4GHz, JVLA}}$. However, $S_{\textrm{int}}$ is 1.9~mJy higher than $S_{\textrm{5.2GHz, CDFS}}$. \citet{2015caccianiga1} find the star formation of this source to be high, 54~$M_{\odot}$ yr$^{-1}$, so star formation contribution to the radio emission can be expected. Interestingly, this source was labelled as a red quasar in \citet{2016lamassa1}. They argue that it is in a dusty, possibly post-merger state, when powerful nuclear winds start unveiling a heavily obscured nucleus. However, no shift was detected in [O~III] lines so a large-scale outflow does not seem to be present, and only a moderate wing component with a velocity of -209~\kms\ and a FWHM of 592~\kms\ was seen in \citet{2021berton1}. Also \citet{2017zhang2} studied its dust properties, and conclude that only the innermost parts of the AGN -- the accretion disk, the broad-line region, and the inner edge of the torus -- seem to be obscured, and not the narrow-line region. Based on Cloudy analysis they claim that the obscuring medium can actually be the torus itself. The host galaxy of J1227+3214 has been observed a few times, but despite its moderate redshift, the attemps to model the morphology of the host galaxy have been inconclusive, thus nothing can be said about the state of the host \citep{2018jarvela1,2020olguiniglesias1}. 

J1227+3214 would be very luminous for a galaxy dominated purely by star formation, so it is probable that also the AGN contributes to the overall radio emission. Based on its luminosity and the steep spectral index, it could be a low-luminosity CSS-like source. With these data it is impossible to say whether the extended emission that reaches outside the host is of nuclear or star formation origin, and further observations will be needed to reveal the nature of this source.

\subsubsection{J1242+3317} 

J1242+3317 (WAS 61) is a radio-quiet source at $z$ = 0.044 with a peak flux density of 1.32~mJy beam$^{-1}$, integrated flux density of 2.23~mJy, and the flux densities of the tapered maps similar. Its luminosity is quite low with log $\nu L_{\nu, \textrm{int}}$ = 38.74~erg s$^{-1}$. Its FIRST flux density is 6.32~mJy, giving a spectral index of -0.79 between 1.4 and 5.2~GHz. This is consistent with what we see in the $\alpha$ map in Fig.~\ref{fig:J1242spind}: the total spectral index is -0.74, and the core spectral index is -0.88. \citet{1995brinkmann1} report a value of 45~mJy at 4.85~GHz observed with the GBT in 1989, but this record is not found anywhere else, for example, in the NASA/IPAC Extragalactic Database (NED), the High Energy Astrophysics Science Archive Research Center's (HEASARC) Xamin, or SIMBAD's VizieR, so the value is highly dubious. The morphology of the source is clearly elongated toward east, and it seems like the spectral index flattens toward east as well, but this might be an edge-effect, especially since also the errors increase toward this direction. The tapered maps (Figs.~\ref{fig:J1242-90k} and \ref{fig:J1242-60k}) show the same extended emission toward east, and are additionally also slightly extended toward south-west. The maximum extent of the region of continuous emission is 1.3~kpc in the normal map and 2.9~kpc in the 90k$\lambda$ tapered map. The extended emission can be caused by either a low-power jet or a nuclear outflow. Higher-resolution observations would be needed to distinguish between these two options.

The host galaxy of J1242+3317 was classified as a possible S0 galaxy with a strong bar (SB) \citep{2007ohta1}, but modelling the host with a bulge and a disk yielded a S\'ersic index of 3.45 for the bulge, suggesting it is a classical bulge \citep{2012mathur1}, in which case the galaxy would not be a pristine late-type galaxy. However, the presence of a bar was not taken into account in the modelling, even if it is clearly visible in the residuals, and might affect the final results.

J1242+3317 is also a bright mid-infrared emitter, even though it is quite clear that the morphology is caused by an AGN. Its W3-W4 colour is 2.72, $q22$ is 1.59, the W3 flux density is 60.8~mJy higher than $S_{\textrm{1.4GHz, JVLA}}$, and $S_{\textrm{5.2GHz, CDFS}}$ is 1.6~mJy higher than $S_{\textrm{int}}$. \citet{2020yang1} estimated its star formation rate based on the 5.2~GHz radio luminosity ad concluded that the radio emission cannot be of star formation origin. However, they do not consider the possibility that it can be a mixture of the two. Using BPT diagrams \citet{2013stern1} classified J1242+3317 as an AGN, a star-forming galaxy, and a Seyfert galaxy in diagrams based on [N~II], [S~II], and [O~I], respectively. This result suggests that indeed probably also star formation is present in this source. Also an [O~III] wing is detectable in the spectrum, with a velocity of -213~\kms\ and a FWHM of 888~\kms, so additional contribution of the AGN-heated dust to the mid-infrared emission cannot be ruled out either.

\subsubsection{J1246+0222} 

J1246+0222 (PG 1244+026) is a radio-quiet, faint NLS1, at $z$ = 0.048, with a peak flux density of 0.43~mJy beam$^{-1}$, an integrated flux density of 0.71~mJy and the flux densities of the tapered maps of 0.85 and 0.82~mJy for the 90k$\lambda$ and the 60k$\lambda$ taper, respectively. It is also a low-luminosity source with log $\nu L_{\nu, \textrm{int}}$ = 38.31~erg s$^{-1}$. In \citet{1994kellermann1} its 5~GHz flux density was found to be 0.83~mJy, so it seems like it has not undergone major changes during the last decades. Its FIRST flux density is 2.23~mJy, giving a spectral index of -0.87 between 1.4 and 5.2~GHz. Upgraded Giant Metrewave Radio Telescope (uGMRT) observations at 685~MHz yielded a flux density of 3.1 $\pm$ 0.3~mJy, and they found a steep spectral index of -0.7 \citep{2020silpa1}. Furthermore, they found the source to be extended toward west - south-west on a kpc-scale. In the radio map in Fig.~\ref{fig:J1246spind} the source seems to be extended northward and west - southwestward. The spectral index varies over the emitting region, and the extreme values close to the edges are clearly a product of edge-effects and low S/N ratio data. The core spectral index is flat with $\alpha$ = 0.05, consistent with what was found in \citetalias{2018berton1}, although new observations are certainly required to confirm our result. The tapered maps show extended emission toward south-east, with an extent of 5.2~kpc in the 90k$\lambda$ tapered map. The host galaxy of J1246+0222 is featureless (Fig.~\ref{fig:J1246-host}), and the bulk of the radio emission is confined within it. Its SDSS $r$ band 25~mag arcsec$^{-2}$ isophotal semimajor axis is 3.42~kpc, implying that the extended emission reaches outside the host.

\citet{2013teng1} studied the H~I absorption properties of J1246+0222 and concluded that the blueshifted and broad H~I absorption features match an outflow caused by the nuclear activity, probably driven by a jet. Moreover, they found that the minima of the absorption troughs exceed the nuclear continuum flux densities, meaning that there is also off-nuclear radio emission contributing to the continuum in J1246+0222. Based on the resolution of their observations they argue that the emitting region needs to be at least 3~kpc from the core. The south-eastward extended emission visible in our tapered maps would fit the criterion of the required off-nuclear emission. Also fitting of the timing data of J1246+0222 in X-rays supports the presence of a jet or at least a jet base in this source. \citet{2017chainakun1} found that the data is best fitted with two X-ray sources, with the source further away from the black hole producing only small amounts of reflection X-rays. They interpret this component as a jet whose emission is beamed away from the accretion disk and thus does not get reflected. In any case, the radio morphology would be highly unusual for a star-forming galaxy, and thus it is probable that at least the extended emission is produced by nuclear activity, which implies that a jet or a nuclear outflow is or has been present. Interestingly, no [O~III] shift is seen in the optical spectrum, but a wing is present with a velocity of -503~\kms\ and a FWHM of 1245~\kms. Also a possible ultrafast outflow with a velocity of 0.08~c has been detected \citep{2020igo1}.

According to the star formation diagnostics the bulk of the radio emission could be from star formation as the W3-W4 colour is 2.56, $q22$ is 2.26, the W3 flux density is 40.3~mJy higher than the $S_{\textrm{1.4GHz, JVLA}}$, and $S_{\textrm{5.2GHz, CDFS}}$ is about the same as $S_{\textrm{int}}$. The star formation rates derived using different methods are roughly in agreement, ranging from 1 to 11~$M_{\odot}$ yr$^{-1}$ \citep{2014young1,2012sargsyan1,2020shangguan1}. However, \citet{2020yang1} claim that producing the amount of radio emission seen in J1246+0222 would require very improbable amounts of star formation. \citet{2007shi1} claim to have detected PAH features at 7.7 and 11.3$\mu$m in the infrared spectrum obtained with the Infrared Spectrograph (IRS) onboard Spitzer Space Telescope, whereas \citet{2018xie1} see only a flat, featureless spectrum in their data obtained with the same instrument. They model the spectrum and conclude that it can be reproduced by three continuum components with temperatures of about 394, 165, and 77~K. If in absence of strong star formation the hot, as well as partly the warm, dust emission could originate from the dust heated by the AGN, which could explain the excess mid-infrared emission.


\subsubsection{J1302+1624} 
\label{sec:J1302}

J1302+1624 (Mrk 783) is a radio-loud NLS1 at $z$ = 0.067 with a peak flux density of 3.28~mJy beam$^{-1}$, an integrated flux density of 11.47~mJy and the flux densities of the tapered maps about half a mJy higher. Its luminosity is log $\nu L_{\nu, \textrm{int}}$ = 39.91~erg s$^{-1}$. Its radio map in Fig.~\ref{fig:J1302spindnoncommon} reveals a core and considerable extended emission toward east and south-east, in a curved morphology. The tapered maps do not show any additional emission (Figs.~\ref{fig:J1302-90k} and \ref{fig:J1302-60k}). The extent of the emission from the core in the normal map toward south-east is 7~kpc, toward north 3.9~kpc, and toward west 2~kpc. In the $\alpha$ map (Fig.~\ref{fig:J1302spindnoncommon}) the spectral index of the core is $\sim$-0.83, and the total spectral index is -1.50. The extended emission is mostly very steep, with spectral indices $<$ -1.5. An exception is a region very south, at the tip of the extended emission: the spectral index at the centre of the highest contour (RA 13:02:58.04,Dec +16:24:24.49) is -0.67 , and thus considerably flatter than elsewhere. It is probable that these flatter spectral indices are real, since they are not only seen at edges, and this region is clearly brighter than the rest of the jet, resembling a hotspot, and indicating interaction with the interstellar medium.

J1302+1624 was detected with FIRST with a flux density of 28.72~mJy, but this value includes both the core and the extended emission, and thus deriving a spectral index using it does not tell us much. \citet{2013doi1} observed the source with the VLBA at 1.7~GHz finding it unresolved with a flux density of 1.3~mJy. It was not detected at 22~GHz using the JVN \citep{2016doi1}. J1302+1624 was extensively studied in \citet{2017congiu1}, \citet{2017congiu2}, and \citet{2020congiu1}. \citet{2017congiu2} found an extended narrow-line region in this source. This region has the same position angle as the radio emission extended south-east, but is much further away from the nucleus, reaching as far as 35~kpc. They confirmed with a BPT diagram that this region has been ionised by the AGN, and not star formation. Hints of gas ionised by star formation are seen closer to the nucleus, and between these there is a region where no line emission was detected at all. If the gas 35~kpc away from the nucleus was ionised by the AGN, we can use a simple light travel time argument to estimate a lower limit for the age of the source: it has taken the nuclear emission at least 35~kpc / $c$ = 10$^{5.05}$~yr to reach this region. 

\citet{2017congiu1} thoroughly analysed the same JVLA data we are using. Using tapering they estimated the extent of the emission to be 14~kpc south-eastward and 12~kpc north-westward. They measured the spectral indices using the same method as in \citetalias{2018berton1} and estimate the core spectral index to be -0.67 $\pm$ 0.13, and the spectral index of the extended emission to be -2.02 $\pm$ 0.74. We confirm these results with the $\alpha$ map. It should be noted that their extended spectral index includes both the very steep and the flatter region seen in our $\alpha$ map, thus being an average between them. Due to the very steep spectral index and the lack of clear collimated structures they hypothesise that the extended emission might be old, and is not being replenished anymore. They also propose that the S-shaped extended emission in the tapered map might suggest that the jets in J1302+1624 are precessing. They tested this scenario further in \citet{2020congiu1} observing the source with the VLBA and the enhanced Multi Element Remotely Linked Interferometer Network (e-MERLIN). The e-MERLIN 1.6~GHz observations confirmed the diffuse, non-collimated nature of the extended emission, as well as its steep spectral index. Interestingly, the source is partially resolved in VLBA observations at 5~GHz, but the pc-scale jet seems to be pointing north-east. Fitting a precession model to the JVLA 5.2~GHz data they confirm that it is a plausible scenario for the origin of the detected emission. Another possibility they bring up is intermittent activity, as J1302+1624 seems to be undergoing a merger, and mergers are known to cause this kind of behaviour. Furthermore, a merger can also be responsible for the change in the direction of the jet. Current data are not enough to distinguish between these scenarios. Interestingly, the flattened emission region we see in the $\alpha$ map coincides with the southern tidal structure seen in Fig. 6 of \citet{2020congiu1}, possibly indicating that the jet is interacting with the interstellar medium in the tidal tail. J1302+1624 is a blue outlier as its [O~III] lines are shifted by -191~\kms, but it is unclear whether the shift is due to the jet or possibly the merger.

The jet is clearly the main source of radio emission in this source, but as indicated by the BPT diagram, there probably is also some star formation going on in the host. The star formation diagnostics are dominated by the jet emission, and only the W3-W4 colour of 2.62 reveals possible star formation. $q22$ is -0.11, the W3 flux density is 63.0~mJy lower than the $S_{\textrm{1.4GHz, JVLA}}$, and $S_{\textrm{int}}$ is 10.8~mJy higher than $S_{\textrm{5.2GHz, CDFS}}$.

\subsubsection{J1305+5116} 

J1305+5116 (SDSS J130522.74+511640.2) is a radio-loud, bright NLS1 at $z$ = 0.788 with a peak flux density of 32.10~mJy beam$^{-1}$, an integrated flux density of 53.41~mJy and the flux densities of the tapered maps approximately the same. It is one of the most luminous sources in our sample with log $\nu L_{\nu, \textrm{int}}$ = 42.91~erg s$^{-1}$. J1305+5116 has also been detected in gamma-rays, which - when taking into account the quite high redshift of this source - strongly suggests it hosts a relativistic, beamed jet \citep{2015liao1}. Its radio map (Fig.~\ref{fig:J1305spind}) shows two components with a separation of 8.1~kpc, and at a position angle of 168\degree. The northern component is responsible for about 62\% of the emission (33.1~mJy). The $\alpha$ map reveals that this is the core, since it shows a flat spectral index ($\alpha$ = -0.07), whereas the spectral index of the southern component is -0.73 (RA 13:05:22.76, Dec 51:16:39.10), and consistent with optically thin synchrotron emission. The northern component is centred at the centre of the featureless host galaxy (Fig.~\ref{fig:J1305-host-zoom}). The SDSS $r$ band 25.0~mag arcsec$^{-2}$ isophotal semimajor axis of the host is 15.4~kpc, so also the jet is still confined within the host galaxy.

Based on archival data J1305+5116 was previously classified as a steep spectrum source: its flux densities at 150~MHz (6C/7C), 325~MHz (WENNS), and 1.4~GHz (FIRST) are 320, 211, and 86.9~mJy, respectively \citep{2008yuan1}. It was later observed with the Allen Telescope Array (ATA) at 1.4~GHz with a flux density of 92.7~mJy \citep{2010croft1}, indicating some degree of variability. Higher-resolution observations revealed a north-south -oriented core-jet structure \citep{2013petrov1}. In VLBA observations at 5~GHz the core flux density was found to be 15~mJy, and the jet flux density 8.9~mJy. These values are less than 50\% of what we observe at 5.2~GHz indicating that a substantial amount of the emission is resolved-out at the VLBA scales. The source was not detected at 22~GHz with the JVN with a detection limit of 9~mJy \citep{2016doi1}.

\citet{2018komossa1} investigated the optical spectrum of J1305+5116 and found extreme emission line shifts, with the largest velocities derived from them exceeding the escape velocity of the galaxy. They also found ionisation stratification -- the shifts of higher ionisation potential lines are larger -- and argue that this implies that the shifts are not caused by local interaction, but concern the whole narrow-line region. They conclude that the properties of the optical spectrum are most probably indicative of an early stage of jet evolution, where a young jet is still advancing through the dense interstellar medium and interacting with it. This scenario is in agreement with the radio morphology of the source, since the jet still seems to reside within the host. This source would be ideal for future observations about the jet - interstellar medium interaction.

The radio emission in J1305+5116 seems to be produced by a powerful nucleus and a relativistic, beamed jet. Nevertheless, for consistency the star formation diagnostics of this source are: the W3-W4 colour is 2.34, $q22$ is -0.61, $S_{\textrm{1.4GHz, JVLA}}$ is 72.3~mJy higher than $S_{\textrm{W3}}$, $S_{\textrm{int}}$ is 5.03~mJy higher than $S_{\textrm{5.2GHz, CDFS}}$.


\subsubsection{J1317+6010} 

J1317+6010 (SBS 1315+604) is a radio-quiet NLS1 ($z$ =0.137) with a peak flux density of 0.51~mJy beam$^{-1}$, an integrated flux density of 0.64~mJy, and the tapered flux densities slightly higher, 0.80 and 0.79~mJy for the 90 and 60k$\lambda$ taper, respectively. Due to its redshift the low flux density translates to a moderate luminosity of log $\nu L_{\nu, \textrm{int}}$ = 39.42~erg s$^{-1}$. The only archival radio detection is 1.82~mJy from FIRST, yielding a 1.4-5.2~GHz spectral index of -0.80. Interestingly, according to the $\alpha$ map in Fig.~\ref{fig:J1317spind} the core spectral index is quite flat with $\alpha$ = -0.34. This is consistent with what was found in \citetalias{2018berton1}, but since they used the same data set, this result should be checked. It can be an artefact due to the faintness of the source. The morphology seems to be somewhat extended, and the central region seems to be surrounded by patchy, diffuse emission. In the tapered maps (Figs.~\ref{fig:J1317-90k} and \ref{fig:J1317-60k}) these blobs form one, quite featureless, emission region. The radio emission is confined within the host galaxy, that does not show any distinguishable morphology either (Fig.~\ref{fig:J1317-host}).

The star formation proxies indicate star formation in the host, W3-W4 is 2.6, $q22$ is 1.60, $S_{\textrm{W3}}$ is 12.8~mJy higher than $S_{\textrm{1.4GHz, JVLA}}$, and $S_{\textrm{int}}$ and $S_{\textrm{5.2GHz, CDFS}}$ are about the same. However, the BPT diagrams using different emission line ratios all indicate that the gas has been ionised by the AGN \citep{2013stern1}. The [O~III] lines are not shifted, but the source does show an [O~III] wing with a velocity of -377~\kms\ and a FWHM of 1436~\kms, so the dust heated by the AGN could explain the mid-infrared emission.

\subsubsection{J1337+2423} 

J1337+2423 (IRAS 13349+2438) is a radio-quiet source at $z$ = 0.108, originally classified as an NLS1, but later found to be a broad-line Seyfert 1 (BLS1) with the FWHM(H$\beta$) $\sim$ 2500~\kms\ \citep{2013lee1}. However, its optical spectrum exhibits very strong Fe~II emission, unlike BLS1s in general, so it might be an ``overgrown'' NLS1. We decided to include it in the sample because it still shares a lot of properties with NLS1s, and might be interesting from a evolutionary point of view. It is defined as radio-quiet, mostly due to its strong optical/infrared emission. Its peak flux density is 9.64~mJy beam$^{-1}$, the integrated flux density is 9.96~mJy, and the tapered flux densities are about the same. It has a high luminosity of log $\nu L_{\nu, \textrm{int}}$ = 40.26~erg s$^{-1}$. Its FIRST flux density is 19.9~mJy, giving a 1.4-5.2~GHz spectral index of -0.53. The $\alpha$ map in Fig.~\ref{fig:J1337spind} shows generally steep spectral indices: the total spectral index is -1.03, and the core spectral index is -1.01. The source is slightly elongated toward east. The tapered maps do not reveal significant new emission. The radio emission is confined within the host galaxy, which shows hints of a bar and possible spiral arms or disturbed morphology (Fig.\ref{fig:J1337-host}). 

The star formation diagnostics, except for the W3-W4 colour, indicate strong star formation. The W3-W4 colour is 1.95, $q22$ is 1.29, the W3 flux density is 424.1~mJy higher than the $S_{\textrm{1.4GHz, JVLA}}$, and $S_{\textrm{5.2GHz, CDFS}}$ is 2.8~mJy higher than $S_{\textrm{int}}$. However, this source has been studied in infrared in detail, and it has been classified as a hot dust-obscured galaxy, where almost all the mid-infrared emission originates from the AGN-heated dust \citep{2018lyu1}. The AGN-fraction of the mid-infrared emission was studied also in \citep{2016alonsoherrero1} and \citet{2016gruppioni1} and was found to be 0.97 and 0.98, respectively, so it can be concluded that the very bright mid-infrared emission is of AGN origin. On the other hand, this source also shows considerable star formation, with the estimates ranging between $\sim$15 to 45~$M_{\odot}$ yr$^{-1}$ \citep{2012sargsyan1,2019mahajan1}, so the contribution of the star formation to the radio emission cannot be totally ruled out.

J1337+2423 has also been found to host strong multiphase ultrafast outflows with velocities of $\sim$0.14~c and $\sim$0.27~c \citep{2020parker1}, as well as two warm absorbers with velocities around -600~km s$^{-1}$ \citep{2018parker1}. The optical spectrum does not show [O~III] line shifts, but reveals a strong [O~III] wing component with a velocity of -598~\kms\ and a FWHM of 2241~\kms. Taking into account the proof of strong outflows in this source, it is possible that also the eastward extended emission is rather caused by an outflow than a jet. Also the steep spectral indices support this hypothesis.

\subsubsection{J1348+2622} 

J1348+2622 (SDSS J134834.28+262205.9) is a radio-quiet NLS1 at $z$ = 0.917 with a very low integrated flux density of 0.38~mJy. However, this source has the highest redshift among our sample, so the low $S_{\textrm{int}}$ translates to quite a high luminosity of log $\nu L_{\nu, \textrm{int}}$ = 40.84 erg s$^{-1}$. The source does not exhibit any distinct features in the radio maps (Fig.~\ref{fig:J1348}), and unsurprisingly also its host galaxy is unresolved due to the redshift (Fig.~\ref{fig:J1348-host}). The $\alpha$ map (Fig.~\ref{fig:J1348spind}) shows that the spectral index of the source is very steep, the total spectral index is -1.06, and the core spectral index is -1.30.

The star formation diagnostics do not imply strong star formation in this source, W3-W4 is 1.83, $q22$ is 0.32, and both $S_{\textrm{W3}}$ and $S_{\textrm{1.4GHz, JVLA}}$, and $S_{\textrm{int}}$ and $S_{\textrm{5.2GHz, CDFS}}$ are roughly the same. However, due to the high redshift of this source, the PAH features around 11.3~$\mu$m are not within the W3 band anymore, but actually fall into the W4 band, so the star formation proxies used are not necessarily reliable for this source. Since the PAH features should be within the W4 band we decided to use it instead of W3 in the $S_{\textrm{W3}}$ - $S_{\textrm{1.4GHz, JVLA}}$ relation. According to this the W4 flux density is $\sim$1.6~mJy higher than $S_{\textrm{1.4GHz, JVLA}}$, still not a very robust result in favour of star formation.

Based on these data it is not possible to determine the predominant source of the radio emission in J1348+2622. However, star forming, or even starburst galaxies rarely exceed the radio luminosity of log $\nu L_{\nu}$ = 40~erg s$^{-1}$ \citep[e.g.,][]{2009sargsyan1}, which might suggest that the radio emission is dominated by the AGN, and that J1348+2622 might be a CSS-like NLS1 galaxy.

\subsubsection{J1358+2658} 

J1358+2658 (SDSS J135845.38+265808.4) is a radio-quiet source at $z$ = 0.331 with a peak flux density of 0.52~mJy beam$^{-1}$, an integrated flux density of 0.61~mJy, with the flux densities of the tapered maps similar. Due to the redshift of this source the low flux density results in a high luminosity of log $\nu L_{\nu, \textrm{int}}$ = 40.05 erg s$^{-1}$. The radio map in Fig.~\ref{fig:J1358spind} shows a core, and faint extended emission north-eastward. A similar structure is seen also in the tapered maps (Figs.~\ref{fig:J1358-90k} and \ref{fig:J1358-60k}). However, with these data we cannot be certain the emission is real. The $\alpha$ map probably suffers from the faintness of the source, and shows variable, possibly not reliable, values. The general trend seems to be on the verge between flat and steep, but based on these data nothing conclusive can be said. The FIRST flux density is 1.17~mJy, and thus the 1.4-5.2~GHz spectral index -0.50, somewhat in agreement with the $\alpha$ map, that has a total spectral index of -0.42. The bulk of the radio emission comes from within the host galaxy, of whose morphology nothing can be said, but the north-eastward extension seems to be reaching outside it. The SDSS $r$ band 25~mag arcsec$^{-2}$ isophotal semimajor axis of the galaxy is 7.7~kpc, whereas the maximum extent of the radio emission is 17~kpc.

\citet{2015caccianiga1} estimate the star formation activity in the host to be at a starburst level with 129~$M_{\odot}$ yr$^{-1}$. Strong star formation is supported by the W3-W4 colour of 2.69, $q22$ of 1.30, the W3 flux density that is 5.4~mJy higher than $S_{\textrm{1.4GHz, JVLA}}$, and $S_{\textrm{int}}$ and $S_{\textrm{5.2GHz, CDFS}}$ that are about equal. Its optical spectrum also shows an [O~III] wing with a velocity of -253~\kms\ and a FWHM of 1366~\kms, so the contribution of the AGN-heated dust cannot be ruled out either.

The extended emission - if real - cannot easily be explained by star formation, and it is most likely of nuclear origin, but its nature cannot be determined with these data. Furthermore, the broad-band spectral index close to the flat/step border is not in agreement with a pure star formation scenario, which should result in a steeper spectral index. This indicates that whereas a fraction on the radio emission originates from the starburst, there needs to be also a flatter component present to account for the observed spectral index. Moreover, the luminosity of this source would be very high for a pure starburst galaxy, and thus the radio emission within the galaxy seems to be a combination of starburst and nuclear activity.

\subsubsection{J1402+2159} 

J1402+2159 (RX J1402.5+2159) is a radio-quiet NLS1 at $z$ = 0.066 with very fait radio emission, the peak flux density is 0.28~mJy beam$^{-1}$, the integrated flux density is 0.33~mJy, and the 90k$\lambda$ tapered flux density is 0.46~mJy. J1402+2159 is a quite nearby source, so also the resulting luminosity is low with log $\nu L_{\nu, \textrm{int}}$ = 38.34~erg s$^{-1}$. The source was detected in FIRST with a flux density of 0.87~mJy, giving a 1.4-5.2~GHz spectral index of -0.74. In addition to these there are not other radio observations of this source. The spectral index seen in the $\alpha$ map in Fig.~\ref{fig:J1402spind} is even steeper, the total spectral index is -0.95, and the core spectral index is -1.54. However, the spectral index values might be affected by the faintness and the low S/N of the data. The source morphology shows an extension toward north-east. This feature is more pronounced in the tapered maps (Figs.~\ref{fig:J1402-90k} and \ref{fig:J1402-60k}). Overlaid with the host galaxy (Fig.~\ref{fig:J1402-host}), it looks like the radio emission reaches slightly outside of it, or at least to the edges, even when considering that the elongated beam somewhat exaggerates the extent of the emission.

The host galaxy is probably of E/S0 type \citep{2007ohta1}, and all the star formation diagnostics indicate activity in this source, the W3-W4 colour is 2.75, $q22$ is 2.16, $S_{\textrm{W3}}$ is 41.2~mJy higher than $S_{\textrm{1.4GHz, JVLA}}$, and $S_{\textrm{5.2GHz, CDFS}}$ is 2.1~mJy higher than $S_{\textrm{int}}$. These diagnostics are supported by BPT diagrams based on various line ratios where this source is consistently classified as a star-forming galaxy, indicating that the main source of ionising emission in this galaxy are stars. The data are consistent with the star formation activity being the main source of radio emission in this source. However, the nature of the extended radio emission cannot be determined with these data, more, preferably higher S/N observations would be needed for that.

\subsubsection{J1536+5433} 

J1536+5433 (Mrk 486) is a radio-quiet NLS1 at $z$ = 0.039 with very faint radio emission: the peak flux density is 0.44~mJy beam$^{-1}$ and the integrated flux density is = 0.50~mJy. It also has a low luminosity of log $\nu L_{\nu, \textrm{int}}$ = 37.98~erg s$^{-1}$. It was detected at 5~GHz in the 1980's with a flux density of 0.47~mJy \citep{1989kellermann1}, indicating that it does not show significant variability. The FIRST detection at 1.4~GHz is 1.23~mJy, giving a 1.4-5.2~GHz spectral index of -0.69. Our $\alpha$ map in Fig.~\ref{fig:J1536spind} shows slightly flatter spectral indices, the total spectral index is -0.35, and the core spectral index is -0.50. Due to the faintness of this source we cannot be sure how reliable the spectral index range is but it seems like the general spectral index is rather flat, which is in agreement with the spectral index estimated in \citetalias{2018berton1}. The radio emission in confined within the inner regions of the host galaxy (Fig.~\ref{fig:J1536-host}), which appears to be late-type.

The PAH features of J1536+5433 at 11.3~$\mu$m were studied by \citet{2019martinezparedes1}, and whereas they were able to detect them they found no evidence of enhanced star formation; in fact the estimated star formation rate was only 0.24 $\pm$ 0.08~$M_{\odot}$ yr$^{-1}$. Although it should be kept in mind that an AGN can suppress the PAH features. Nevertheless, the mid-infrared colour agrees with this (W3-W4 = 1.92), but the other diagnostics do not. The $q22$ is 2.12, the W3 flux density is substantially higher, about 52.1~mJy, than $S_{\textrm{1.4GHz, JVLA}}$, and also $S_{\textrm{5.2GHz, CDFS}}$ is $\sim$1.0~mJy higher than $S_{\textrm{int}}$. Whereas $S_{\textrm{int}}$ and $S_{\textrm{5.2GHz, CDFS}}$ are comparable and indicate that the radio emission could be explained by modest star formation, the W3 flux density is surprisingly high, and hard to explain by the low star formation seen in this source. The [O~III] lines of J1536+5433 are not shifted, and do not show a wing either. This, of course, does not rule out the possibility that the excess mid-infrared emission might originate from the equatorial dust heated by the AGN. This would be in line with the flat spectral index -- uncommon for star formation emission -- of the source.

\subsubsection{J1555+1911} 

J1555+1911 (Mrk 291) is a faint, radio-quiet NLS1 ($z$ = 0.035) with a peak flux density of 0.12~mJy beam$^{-1}$ and an integrated flux density of 0.36~mJy. The flux densities of the tapered maps are 0.42 and 0.44~mJy, for the 90 and the 60k$\lambda$ taper, respectively. The normal map (Fig.~\ref{fig:J1555spind}) exhibits diffuse emission, and it is hard to say whether a proper core is even present. The FIRST detection of this source is 1.85~mJy, giving a 1.4-5.2~GHz spectral index of -1.25. Also our $\alpha$ map shows a very steep spectral index: the core spectral index is -1.13, and the total spectral index is -0.82, albeit the south-western part of the emission exhibits a flat spectral index, which might be due to the edge effects and thus not reliable. The tapered maps (Fig.~\ref{fig:J1555-90k} and \ref{fig:J1555-60k}) show a morphology slightly elongated toward north-west. J1555+1911 is hosted by a spiral galaxy (Fig.~\ref{fig:J1555-host}), and most of the radio emission is confined within the bulge of the galaxy.

Based on its mid-infrared properties, the diffuse radio morphology, and the low luminosity (log $\nu L_{\nu, \textrm{int}}$ = 37.85~erg s$^{-1}$) it seems rather evident that the predominant source of the radio emission is star formation. W3-W4 colour is 2.62, $q22$ is 1.67, W3 flux density is $\sim$19.2~mJy higher than $S_{\textrm{1.4GHz, JVLA}}$, and $S_{\textrm{int}}$ and $S_{\textrm{5.2GHz, CDFS}}$ are roughly in agreement.

\subsubsection{J1559+3501} 

J1559+3501 (Mrk 493) is a radio-quiet NLS1 at $z$ = 0.031 with faint radio emission, the peak flux density is 0.46~mJy beam$^{-1}$, the integrated flux density is 1.41~mJy, and the flux densities of the tapered maps are 1.47 and 1.48~mJy for the 90 and 60k$\lambda$, respectively. It is a nearby source so the flux density results in a low luminosity of log $\nu L_{\nu, \textrm{int}}$ = 38.19~erg s$^{-1}$. The radio map in Fig.~\ref{fig:J1559spind} shows a core surrounded by diffuse emission, slightly more elongated toward north-west and south-east. The tapered maps in Figs.~\ref{fig:J1559-90k} and \ref{fig:J1559-60k} do not show any additional structures. The $\alpha$ map suffers from low S/N, as can be seen especially close to the edges, but in general the map shows very steep spectral indices, and the total spectral index is as steep as -1.16. The FIRST detection of this source is 3.39~mJy, giving a 1.4-5.2~GHz spectral index of -0.67. \citet{2020yang1} report a JVLA A-configuration 8.84~GHz flux density of 0.56~mJy, resulting in a very steep 5.2-8.84~GHz spectral index of -1.74. They claim that the 5.2-8.84~GHz spectral index is flat, but their reported value for the integrated 5.2~GHz flux density is only 0.62~mJy, whereas their and ours peak flux densities are about the same. They obtained the value using the same data set we are using, but their rms is higher (11 vs. 15.64~$\mu$Jy beam$^{-1}$). However, the 1.4-5.2~GHz spectral index is steep, and it does not seem probable that it would flatten above 5.2~GHz.

It is worth noting that the detected radio emission is confined within the bulge of the host (Fig.~\ref{fig:J1559-host}) and probably does not represent the whole radio emission of the source. The host galaxy is a barred spiral that also exhibits a nuclear ring and a nuclear dust spiral \citep{2003crenshaw1,2006deo1,2007ohta1}. \citet{2009popovic1} studied the nuclear region in detail and concluded that the nuclear ring can indeed be a place of enhanced star formation. They estimated an unusually high star formation rate of 2~$M_{\odot}$ yr$^{-1}$ for the innermost $<$ 1~kpc. In the BPT diagrams J1559+3501 is classified as a star-forming galaxy, indicating that the narrow-line region lines as excited by star formation rather than the AGN. \citet{2010sani1} detected strong 6.2~$\mu$m PAH features in this source, confirming the presence of star formation. Our star formation diagnostics are somewhat in agreement with this, but it should be noted that the WISE flux densities cover the whole galaxy, whereas we likely are missing some faint, extended radio emission from host. The W3-W4 colour is 2.34, $q22$ is 1.38, W3 flux density is $\sim$64.3~mJy higher than $S_{\textrm{1.4GHz, JVLA}}$, and $S_{\textrm{5.2GHz, CDFS}}$ is 1.3~mJy higher than $S_{\textrm{int}}$. The [O~III] lines of J1559+3501 show a prominent wing with a velocity of -349~\kms\ and a FWHM of 807~\kms, so also nuclear winds are present.

The evidence seems to indicate that the star formation is a significant, if not the major, contributor to the radio emission. The nuclear, possibly star-forming ring found in previous studies falls within the radio emitting region in our map, and also the radio morphology seems to be more consistent with star formation than nuclear activity. This does not rule out the possibility that the AGN contributes too, but higher resolution observations will be needed to disentangle these two mechanisms.

\subsubsection{J1633+4718} 
\label{sec:J1633}

J1633+4718 (SDSS J163323.58+471858.9) is a radio-loud NLS1 ($z$ = 0.116) whose radio map in Fig.~\ref{fig:J1633spind} shows two distinct components. The integrated flux density of the core component is 24.48~mJy, and the integrated flux density of the north component is 0.79~mJy. The flux densities of the tapered maps are very similar to these. J1633+4718 is a high-luminosity source with log $\nu L_{\nu, \textrm{int}}$ = 40.67~erg s$^{-1}$. J1633+4718 resides in an interacting system \citep{2020olguiniglesias1}: the core component is associated with the disk-like host galaxy of J1633+4718, whereas the north component is associated with another disk galaxy (Fig.~\ref{fig:J1633-host}). The nuclei of the two galaxies are $\sim$8~kpc apart. The core spectral index of the component that hosts the NLS1 is -0.54, and thus on the verge between flat and steep. The $\alpha$ map of the companion galaxy suffers from low S/N and edge effects, but its ``core'' spectral index is -0.69.

Past radio observations of J1633+4718 indicate variability in the core. \citet{1994neumann1} found an inverted radio spectrum above 5~GHz with simultaneous observations, whereas \citet{2011doi1} found the 1.7 to 8.5~GHz spectral index to be steep with quasi-simultaneous observations. Also archival data at $\sim$5~GHz suggest considerable variability: \citet{1991gregory1} reported a flux density of 34~mJy at 4.85~GHz obtained with the GBT, \citet{1997laurentmuehleisen1} obtained 30~mJy with the VLA at 5~GHz, and \citet{2010gu1} reported 55.2~mJy with the VLBA at 5~GHz. This is indicative of jet activity, which has been confirmed by high brightness temperature of at least 10$^{11.3}$~K \citep{2007doi1}, and a pc-scale core-jet radio morphology in high resolution observations at 1.7 and 5~GHz \citep{2010gu1,2011doi1}. At 8.4~GHz with the JVN the source remains unresolved with a flux density of 21.2~mJy.

Archival FIRST 1.4~GHz flux density for J1633+4718 is 65.02~mJy, whereas \citet{2011doi1} were able to detect 55.5~mJy at 1.7~GHz with the VLBA. The discrepancy might be due to resolved-out emission in the VLBA observations, but also because of the variability. Several previous studies suggest that this might be a CSS-like source due to its small size and steep spectral incides, but taking into account that the source also shows flat or even inverted radio spectra, it seems more probable that the occasional steep indices have been due to the temporal variability in the source. Moreover, the 22~GHz flux density was found to be as high as 163~mJy observed with the JVN \citep{2016doi1}, also not consistent with the CSS scenario. 

The W3-W4 colour of 2.67 indicates that there might also be star formation present in the galaxy, and \citet{2015caccianiga1} estimated the star formation rate to be as high as 68~$M_{\odot}$ yr$^{-1}$. The other indicators do not favour star formation, since the radio emission is clearly dominated by the nucleus: $q22$ is 0.22, the W3 flux density is $\sim$24.4~mJy lower than $S_{\textrm{1.4GHz, JVLA}}$, and $S_{\textrm{5.2GHz, CDFS}}$ is 24.1~mJy lower than $S_{\textrm{int}}$. However, also a strong nuclear wind seen in the [O~III] wing is present with a velocity of -279~\kms\ and a FWHM of 1186~\kms\ so AGN-heated dust can contribute to the observed mid-infrared emission. The main source of the radio emission in J1633+4718 is the AGN, but the weak radio emission of the companion galaxy might be produced by star formation.

J1633+4718 is also an interesting X-ray source, since an ultrasoft X-ray component with a low blackbody temperature of $\sim$30~eV was detected in it \citep{2010yuan1,2016mallick1}. This was intepreted as a direct detection of the accretion disk, and so far similar components have not been detected in other NLS1s. \citet{2016mallick1} also detect excess UV emission, that requires the presence of a jet to be explained.

\subsubsection{J1703+4540} 

J1703+4540 (SDSS J170330.38+454047.1) is a radio-loud NLS1 at $z$ = 0.060 that was previously classified as a CSS with a turnover frequency below 150~MHz \citep{2004snellen1}. Our $\alpha$ map, shown in Fig.~\ref{fig:J1703spind}, agrees with this classification since the spectral index we find is steep: the total spectral index is -0.87, and the core spectral index is -0.91. The peak flux density is 32.26~mJy beam$^{-1}$, the integrated flux density of the normal map (Fig.~\ref{fig:J1703spind}) is 34.68~mJy, with log $\nu L_{\nu, \textrm{int}}$ = 40.20 erg s$^{-1}$, and it exhibits very mild elongation toward south-east. The integrated flux densities of the tapered maps (Figs.~\ref{fig:J1703-90k} and \ref{fig:J1703-60k}) are about a mJy higher, and both maps reveal extended, but faint radio emission toward south-east. The host of J1703+4540 is clearly a spiral galaxy, as seen in Fig.~\ref{fig:J1703-host}, and the radio emission is concentrated within the bulge of the galaxy, which is of a pseudo-bulge type \citep{2020olguiniglesias1}. The extended radio emission seems to somewhat trace the southern bar- or arm-like structure, possibly indicating the presence of star formation in the galaxy. The star formation diagnostics partly support this, as the mid-infrared colour W3-W4 is 2.59, and the W3 flux density 15.9~mJy higher than the $S_{\textrm{1.4GHz, JVLA}}$. $S_{\textrm{int}}$, on the other hand, is 28.8~mJy higher than $S_{\textrm{5.2GHz, CDFS}}$, and $q22$ is only 0.52, suggesting that star formation is not sufficient to explain all of the radio emission.

Indeed, analysing archival VLBA observations of J1703+4540 at 5~GHz \citet{2010gu1} found a one-sided core-jet structure directed toward south-west. They also reported considerable flux density variability, which might also affect our star formation diagnostics, since we do not know in which state the source was during our observations. However, our integrated flux density is $\sim$20~mJy lower than the 5~GHz VLBA flux density (56.8~mJy), suggesting that the source was in a rather low state during our observations. The presence of a jet was confirmed in VLBA observations at 1.7~GHz by \citet{2011doi1}, who found a similar one-sided structure, consistent with the jet of \citet{2010gu1}, and with a projected length of 35~pc. The jet emission is quite diffuse and does not appear Doppler boosted, and they conclude that the jet is either sub-relativistic or seen at a large angle. They also detect hints of a counter-jet but cannot confirm whether it is real or not. \citet{2017giroletti1} claim to have confirmed the presence of emission on the counter-side of the core at 5~GHz, and they classify J1703+4540 as an asymmetric double, and a low-power compact \citep[LPC,][]{2010kunertbajraszewska1} source, since it does not show considerable radio morphology evolution over time, as would be expected from a true CSS. 

A powerful molecular outflow with a bulk velocity of -660~km s$^{-1}$ \citep{2018longinotti1}, and a multi-component ultra-fast X-ray outflow showing velocities up to $\sim$0.1~c \citep{2018sanfrutos1}, consistent with each other, have also been found in this source. \citet{2018longinotti1} state that the outflow seen in J1703+4540 is consistent with an energy-conserving outflow propagating through the host galaxy, an instance of probably the most important and effective way of AGN feedback. In agreement with the outflows, also an [O~III] wing component is present, with a velocity of -336~\kms, and a FWHM of 646~\kms. Interestingly, the [O~III] emission line shows a \textit{redshift} of 210~\kms \citep{2021berton1}, the origin of which remains unclear. Since the wing velocity is defined relative to the [O~III] core, its real velocity, compared to the restframe, is smaller than the previously reported value. Spatially resolved spectra would be needed to study the morphology and kinematics of [O~III] in more detail.

\subsubsection{J1713+3523} 

J1713+3523 (FBQS J1713+3523, $z$ = 0.083) is an NLS1 that would be classified as radio-quiet based on the FIRST detection ($S_{\textrm{int, 1.4GHz}}$ = 11.24~mJy) or our data ($S_{\textrm{int}}$ = 3.81~mJy), but it was detected at 22~GHz with the JVN with a whopping flux density of 138~mJy \citep{2016doi1}, implying extreme variability and jet activity. Interestingly the 1.4-5.2~GHz spectral index is steep with $\alpha$ = -0.66, in agreement with our $\alpha$ map in Fig.~\ref{fig:J1713spind} whose total spectral index is -0.97. Assuming the less extreme spectral index of -0.66, the extrapolated 22~GHz flux density would be 1.47~mJy, requiring almost a hundred-fold increase in the flux density to achieve the observed 138~mJy. The spectral index between 5.2 and 22~GHz is 2.49, very close to the theoretical synchrotron self-absorption (SSA) limit. It should be noted that the observations are not simultaneous, the 22~GHz observations were performed in April 2014, and the JVLA observations in September 2015. However, radio flares usually propagate from higher to lower frequencies within time-scales from several months to a few years, so if anything the 22~GHz flare should have increased the 5.2~GHz flux density at the time of our observation, if it would have propagated to such low frequencies, so the estimated spectral index should be indicative of the real spectral index.

The 22~GHz observation was done with the JVN so it probes mas scale emission in the source. J1713+3523 lies at $z$ = 0.083, giving a scale of 1.562~pc mas$^{-1}$, thus the observed flux density must have come from a region of some tens of pcs across. It is interesting that we do not see any signs of the flattening of the spectral index in our $\alpha$ map, which would have been expected to happen if the 22~GHz flare would have propagated to lower frequencies. An alternative to a conventional AGN radio flare is that the 22~GHz flare was a sign of recently started jet activity in J1713+3523, as seen in high-frequency peakers (HFP). In this case there could exist a quasi-stationary feature at higher radio frequencies, with the inverted radio spectrum produced by SSA \citep{2021odea1} or free-free absorption by the ionised gas produced in the bow shock of the jet \citep{1997bicknell1}. The [O~III] lines of J1713+3523 are extremely blueshifted with $v$ = -674~\kms, providing further support of the influence of a jet still confined within its host galaxy and in interaction with the interstellar medium. Interestingly, no [O~III] wing is present in this source.

The whole radio emission seen in the JVLA observation of J1713+3523 could be explained by star formation, as the W3 flux density is $\sim$17.8~mJy higher than $S_{\textrm{1.4GHz, JVLA}}$, and $S_{\textrm{int}}$ and $S_{\textrm{5.2GHz, CDFS}}$ are roughly the same. On the other hand, the near-infrared colour, W3-W4 = 2.11, nor the $q22$ of 0.62, do not necessarily indicate strong star formation. However, the source is not extraordinarily bright, with log $\nu L_{\nu, \textrm{int}}$ = 39.54 erg s$^{-1}$, lower than for jetted sources in \citetalias{2018berton1} in general. The higher frequency observations confirm the presence of a jet, but it does not seem to dominate the 5.2~GHz radio emission.

\subsubsection{J2242+2943} 

J2242+2043 (Ark 564, $z$ = 0.025) is a radio-quiet NLS1 with a peak flux density of 5.61~mJy beam$^{-1}$, and an integrated flux density of 10.33~mJy that gives quite a low luminosity of log $\nu L_{\nu, \textrm{int}}$ = 38.88 erg s$^{-1}$. It was detected in NVSS with a flux density of 29.1~mJy, yielding a spectral index of -0.79 between 1.4 and 5.2~GHz, and at 95~GHz with a flux density of 1.14~mJy \citep{2015behar1}, giving a spectral index of -0.76 between 5.2 and 95~GHz. These results are consistent with our $\alpha$ map (Fig.~\ref{fig:J2242spind}), which has a total spectral index of -0.81, and a core spectral index of -0.85. The radio morphology of the source is clearly elongated toward north, as also found in previous studies \citep[e.g.][]{2000moran1}, and the spectral index seems to slightly flatten north-ward. The tapered maps (Figs.~\ref{fig:J2242-90k} and \ref{fig:J2242-60k}) do not show any additional structures, and the radio emission is confined within the bulge of a galaxy with a barred spiral morphology (Fig.~\ref{fig:J2242-host}). This source exhibits complex X-ray properties as both relativistic and sub-relativistic outflows have been detected \citep[e.g.,][]{2013gupta1, 2013gupta2, 2016khanna1}.

Whereas the asymmetric radio morphology indicates that J2242+2943 possesses a jet or a wind that produces radio emission, the star formation proxies suggest that also star formation is present. The W3-W4 colour is 2.57, $q22$ is 1.16, and the W3 flux density is as much as $\sim$123.8~mJy higher than $S_{\textrm{1.4GHz, JVLA}}$, and $S_{\textrm{int}}$ and $S_{\textrm{5.2GHz, CDFS}}$ are comparable. The low radio luminosity of the source also suggests that a possible jet cannot be very powerful, supported by the fact that no shift is seen in the [O~III] lines. However, an [O~III] wing is detected, with the velocity and the FWHM of -564 and 1090~\kms, respectively, so the mid-infrared emission might be enchanced by AGN-heated dust emission. It seems plausible that the radio emission is a combination of AGN and star formation related processes, but nothing definite can be said before more detailed observations.

\subsubsection{J2314+2243} 
\label{sec:J2314}

J2314+2243 (RX J2314.9+2243, $z$ = 0.169) is a radio-quiet source with a peak flux density of 6.17~mJy beam$^{-1}$, and an integrated flux density of 7.02~mJy. It was also detected in NVSS with a flux density of 18.7~mJy, giving a spectral index of -0.75 between 1.4 and 5.2~GHz. This value is in agreement with our $\alpha$ map in Fig.~\ref{fig:J2314spind}: the total spectral index is -0.81, and the core spectral index is -0.85. The radio maps (Fig.~\ref{fig:J2314}) do not show any distinctive features, nor does the host galaxy in Fig.~\ref{fig:J2314-host}, within which the radio emission is confined.

The W3-W4 colour (2.33) and $q22$ (0.85) do not indicate strong star formation, and also $S_{\textrm{int}}$ is $\sim$4.5~mJy higher than  $S_{\textrm{5.2GHz, CDFS}}$. However, $S_{\textrm{W3}}$ is 43.8~mJy higher than $S_{\textrm{1.4GHz, JVLA}}$. The origin of this excess emission can be either star formation in the host galaxy, or AGN-heated dust. The existence of a strong [O~III] wing with a velocity of -1031~\kms\ and a FWHM of 1579~\kms\ is in favour of the AGN-heated, possibly polar, dust being a major contributor to the mid-infrared emission. J2314+2243 would be unusually bright for a purely star forming galaxy, with log $\nu L_{\nu, \textrm{int}}$ = 40.48 erg s$^{-1}$, so it seems possible that the AGN is the main source of the radio emission. J2314+2243 could be another candidate for a CSS-like source.

\section{Extraordinary diversity of NLS1s}
\label{sec:discussion}

The most notable result is the remarkable diversity of NLS1s: they show a host of different components contributing to the radio, as well as the mid-infrared emission, and no two sources are alike. Spatially resolved spectral index map utilised in this paper offer a clear advantage in disentangling these components when compared to conventional ways of determining the spectral indices, that usually give one value for the whole emitting region. The $\alpha$ maps allow us to examine the spectral index and its changes over the whole emitting region, and determine the most plausible origin of the emission in different parts of the source. In addition, the associated errors of the $\alpha$ maps derived using the mt-mfs method are considerably smaller than the errors for conventional estimates. Using this method, complemented by data at other wavelengths, some NLS1 groups with similar properties can be identified, and are discussed below.

\subsection{AGN-dominated sources}

In about one third of the sources in our sample the radio emission is clearly dominated by the AGN. These sources are characterised by moderate to high radio luminosities (all but one have log $\nu L_{\nu, \textrm{int}} >$ 39.00 erg s$^{-1}$), asymmetric radio morphologies, and/or variable radio emission, and flat spectral indices in some cases. We can see considerable diversity even within this group, probably due to different ages, inclination angles, and environments of the sources.

\subsubsection{Large-scale jets} 

A few of our sources exhibit radio jets with extents of the order of tens of kpcs, with the jets reaching well outside the host galaxy. In some sources we are also able to detect counter-jet emission, indicating that these sources are seen at rather large inclination angles, and projection effects cannot be responsible for the narrowness of their permitted lines.

Based on simple travel-time arguments these sources cannot be kinematically very young, and must have maintained the jets for quite a long period of time for them to propagate through the host interstellar medium. \citet{2009czerny1} propose that intermittent activity due to radiation pressure instabilities of the accretion disk could be common in young radio sources accreting at high rates; the active, jetted phases would last for 10$^3$-10$^4$ years, separated by inactive, non-jetted periods of 10$^4$-10$^6$ years. It is unclear whether such short activity periods can be responsible for the kpc-scale emission seen in these impressive sources since the final extent of the radio emission after the jets have turned off depends on several factors, for example, the jet power, the length of the activity period, and the properties of the medium it is expanding into. 

Interestingly, one source in our sample, J1302+1624, shows possible relic emission with a very steep spectral index. This seems to be a consequence of its jets changing direction \citep{2020congiu1}, and with the current data we cannot say whether the jets turned off at some point or not. On the other hand, the most extended NLS1 discovered so far, J0814+5609, does not show signs of intermittent activity, and based on its flat core and observations at other radio frequencies the jets are currently active. However, this source might represent a transition object between NLS1s and broad-line radio galaxies, since its black hole mass is large for an NLS1 (10$^{8.44}$~M$_\odot$, \citealp{2016berton2}), and its FWHM(H$\beta$) on the threshold between NLS1s and BLS1s (2164~\kms, \citealp{2015foschini1}). The other sources with large-scale jet emission in our sample do not show signs of intermittency either, and their black hole masses are well below 10$^8$~M$_\odot$, implying that large-scale jets are not exclusively a property of high black hole mass sources. Morphologically, these sources resemble small broad-line radio galaxies, and are further proof that eventually some NLS1s might evolve into fully developed radio galaxies \citep{2016berton1}.


\subsubsection{Compact sources and small-scale jets}

Another group whose radio emission is dominated by the AGN are the sources with compact morphologies, showing no or moderate extended radio emission, or small-scale jets. In these sources the high AGN activity - usually in form of jets - was confirmed by, for example, flat spectrum, high luminosity, variability, or previous high resolution radio observations. Interestingly, many of these sources are nominally radio-quiet, casting yet another shadow over the use of radio-loudness parameter as a proxy of AGN activity. 

The compactness of these sources indicates that they are either less extended, and thus possibly younger sources, or seen at smaller inclination angles. The latter might be true especially in case of the most luminous sources, where the emission is probably enhanced by considerable beaming. The lower luminosity sources are probably either seen at large enough angles for the beaming not to have considerable effects and/or host slower, sub-relativistic jets.

An interesting individual is J1633+4718 since it resides in a late-type galaxy, hosts most likely relativistic jets, and is interacting with another late-type galaxy \citep{2020olguiniglesias1}. It has been hypothesised that interaction might play a role in triggering the nuclear activity \citep{1999taniguchi1, 2008barth1, 2008urrutia1, 2011ellison1}, and seems like jetted NLS1s might be a great example of this, since previous studies have shown that interaction and mergers are more common among jetted than non-jetted NLS1s \citep[e.g.,][]{2017olguiniglesias1, 2018jarvela1, 2019berton1}. However, it should be remembered that in general the jetted sources reside at higher redshifts than the non-jetted ones that have been studied, and the impact of this difference has not been properly addressed yet. 

\subsubsection{CSS-like sources}

We identify five new NLS1s that exhibit properties consistent with the CSS classification: J0806+7238, J1047+4725, J1138+3653, J1348+2622, and J2314+2243. These sources show compact morphology, steep spectral index, and luminosity that is too high to be explained by star formation or even a starburst. Three out of five sources exceed the traditional luminosity threshold for CSS sources at 5~GHz (log $\nu L_{\nu, \textrm{int}}$ = 40.60 erg s$^{-1}$), and the luminosities of the other two are log $\nu L_{\textrm{int}} >$ 40.00 erg s$^{-1}$. 

A link between NLS1s and CSS sources has already been suggested several times, and some NLS1s have been identified as CSS sources in the literature \citep[e.g., ][]{2001oshlack1, 2006gallo1, 2014caccianiga1, 2016berton1}. These results lead to a tentative model where some CSS sources, especially the low-luminosity compact objects (LLCs) could be a part of the parent population of jetted NLS1s \citep{2016berton1}. According to \citet{2010kunertbajraszewska2} some LLCs can be classified as high-excitation sources, similarly to NLS1s, and their intrinsic properties, such as black hole mass and Eddington ratio distributions, show significant overlap with beamed NLS1s. Also their radio luminosity functions are consistent with each other, when beaming is taken into account, and they show similar, simple radio morphologies, without considerable diffuse emission.

What is not yet known is if their host galaxies, and larger, group- and cluster-scale environments, are similar, and if so, to what extent \citep{2021odea1}. The environments of high-excitation LCCs and most CSS-like NLS1s have not been studied systematically, and also the knowledge of the environments of beamed NLS1s is far from comprehensive. An additional issue that must be taken into account in future studies is the selection criteria of LLCs: in \citet{2010kunertbajraszewska1} the sources have 5~GHz flux densities $>$ 150~mJy, and in \citet{2010kunertbajraszewska1} they have 1.4~GHz flux densities $>$ 70~mJy. Most CSS-like NLS1s are fainter than this, thus in order to have comparable luminosities they must be at higher redshifts, which might induce some selection biases in the samples. However, the unification scenario of sub-populations of NLS1s and CSS sources should definitely be studied further. 

\subsubsection{Large-scale outflows}

A third group of AGN-dominated sources are the ones with large-scale outflows. These sources are rather scarce in our sample, but it might partially be because without higher-resolution observations they are hard to distinguish from sources with slow, poorly collimated jets. Also their luminosities are quite low, so unless they can be identified based on the morphology, or actual observations of the outflow, they can easily go undetected. Outflows and jets can also co-exist (for example, J1703+4540, \citealp{2010gu1,2018longinotti1}), and since the radio emission of the jet is generally brighter than the outflow radio emission, the jet dominates the radio band.

J0804+7248 is an example of a source than could host an outflow instead of a jet: it is only moderately luminous, its radio morphology it clearly asymmetric, but does not look collimated, and its [O~III] lines are blueshifted, implying that the NLR is experiencing bulk motion. Another similar source, J0632+6340, shows a two-sided non-collimated radio morphology at the same PA as [O~III] ionisation cones, and has a very low luminosity implying that the low-speed jets or outflows have very low power. 

\subsection{Host-dominated sources}

About a dozen of our sources seem to be dominated by radio emission that originates from the host galaxy, that is, from star formation related processes. They do not show any clear signs of AGN activity in radio, and the star formation activity is sufficient to explain their radio emission. Their radio morphologies are generally featureless, and the emission is confined within the host, or even the host bulge, hinting at circumnuclear star formation. In some cases the radio morphology is scattered in patches around the core, and in some cases it seems to trace the host galaxy features. The sources for which BPT diagram data exists are consistently classified as star-forming galaxies, suggesting that the stars are the main source of the ionising continuum. Interestingly, more than half of these sources show a prominent [O~III] wing, indicating that nuclear winds reaching the inner NLR are present.

\subsection{Composite sources}

Approximately one third of the sources can be classified as composite: in these sources both the AGN and the host galaxy have a significant contribution to the total radio emission. These sources show a mixed variety of properties characteristic for AGN-dominated as well as for star formation dominated sources. They can show, for example, a flat spectrum core, radio morphology resembling a jet, or very high luminosity, and on the other hand, for example, patchy radio emission reminiscent of circumnuclear star formation, PAH features, or other positive star formation proxies, indicating star formation even at a starburst level. The different BPT diagrams also often give mixed results for these sources, indicating that both the AGN and the starlight continuum are significant. In addition, different kinds of outflows have been detected in these sources, including UFOs, BLR and NLR outflows, and molecular outflows. As a cherry on top, one of these sources was identified as a water maser source.

The ensemble of properties seen in these sources beautifully demonstrates why NLS1s should not be treated as a class consisting of similar sources, but should be treated as individuals. Caution should be exercised also when using simple diagnostic tools, since they cannot properly take into account the versatility seen in these sources, and might give misleading results when used on their own. This matter is further discussed in Sect.\ref{sec:issues}.

\subsubsection{Peculiar sources}

There are some unique sources that do not fit any of the categories described earlier. An obvious one is J1337+2423 that in closer inspection turned out not to be an NLS1, but a BLS1 galaxy \citep{2013lee1}. This source highlights a possibly bigger issue in the identification on different types of Seyferts: most of them have been classified based on quite low signal-to-noise (S/N) ratio spectra, and often using automated algorithms. Subtle, but crucial, details of the emission lines, such as the wings, cannot be properly modelled using low S/N data, but can easily lead to misclassification of NLS1s and BLS1s, and confusion with intermediate Seyferts \citep{2020jarvela1, 2021peruzzi1}. Mixed samples will add undesired noise to statistical population studies, and hinder our efforts to understand the nature of the different classes of AGN.

J1038+4227 is another extremely interesting object, since its radio morphology does not resemble anything we see in other NLS1s. However, its optical spectrum clearly identifies it as one: its FWHM(H$\beta$) = 1979~\kms, and it shows very strong Fe~II multiplets. Were it a more evolved source, for example, an FSRQ with a flattened BLR seen face-on, we would not expect to see iron in the spectrum \citep{2018marziani2,2020berton1}. J1038+4227 is a luminous source, and since the core is responsible for most of its radio emission, and shows a steep spectral index it would be consistent with a CSS classification. The symmetrical, almost circular, faint radio emission reaching well outside the host galaxy is what sets J1038+4227 apart from other NLS1s. Due to its huge size its origin is probably the AGN, and it seems plausible that it is a radio lobe. It resembles the kpc-scale morphologies seen in some blazars \citep{2010kharb1}, but is considerably fainter. Unfortunately, this structure is too faint to estimate its spectral index, and derive any information about its age or whether its a relic or still replenished. However, taking into account its circular shape it seems like the jet must have pointed close to our line of sight, which does not seem to be the case right now, otherwise we would expect to see a flatter spectral index in the core. The nature of this source remains unclear for now, but it definitely deserves to be studied in detail in the future.

The third source that stands out is J1713+3523. It has a steep spectral index at least up to 5.2~GHz, and does not show strong signs of AGN activity at that frequency. In fact, its 5.2~GHz radio emission could be explained with star formation alone. However, a 22~GHz detection of 138~mJy is undeniable proof that this source hosts at least moderately powerful jets, and it furthermore yields a 5.2-22~GHz spectral index close to the theoretical SSA maximum. It is interesting that no signs of the jets are seen at lower frequencies. This source somewhat resembles an extraordinary group of previously radio-quiet or totally radio-silent NLS1s, first discovered at 37~GHz due to their bright (hundreds of mJy), and variable radio emission, characteristic for relativistic jets \citep{2018lahteenmaki1}. These sources were later observed with the JVLA, and all of them showed a steep 1.6-9.0~GHz radio spectrum, with no signs of possible jets \citep{2020berton2}. The nature of these sources is still somewhat unclear, but absorption, either SSA or free-free has been suggested as the culprit behind their unusual radio spectra. More observations of J1713+2523 are certainly needed to explain the observed properties of this source. 

\subsection{Implications and issues}
\label{sec:issues}

The results of this paper strengthen the view of NLS1s as a wonderfully diverse class of sources. However, this versatility might also cause additional trouble, especially when studying larger samples of sources, rather than individuals, and thus caution is required when dealing with large samples and/or statistical studies. Below we discuss the biggest issues resulting from this diversity. 

\subsubsection{NLS1s are \emph{not} a homogeneous class}

It cannot be stressed enough that the variety of sources under the NLS1 umbrella is vast. After all, the classification is based solely on the narrowness of the permitted lines, that indicates an undermassive black hole, and the oxygen-to-hydrogen ratio, that ensures they are Type 1 sources. This does not tell us anything about the activity state of the AGN, and it ranges from radio-silent sources to rare individuals that are able to maintain fully developed relativistic jets. It also does not reflect the properties of the host galaxy, and in NLS1s the AGN rarely dominates the whole electromagnetic spectrum over the host, unlike in quasars or blazars. Even if this might be true also for some NLS1s, the majority does not fit one mould, as demonstrated in this paper. It should also be kept in mind that the radio view of NLS1s is very different than the optical view, and whereas their optical spectra are similar, the radio properties absolutely are not.

\subsubsection{Advice against using simple proxies}

The amasing diversity of NLS1s, and the varying contributions of the AGN and the host along the electromagnetic spectrum can have unexpected effects when using relations and proxies possibly calibrated with more homogeneous or simple samples of sources. This has been continuously demonstrated regarding the use of the radio-loudness parameter, that might have originally served as an acceptable first order estimate of the activity of the AGN, but fails to perform when more complicated sources are included in the scenario \citep{2017jarvela1,2017padovani1}. 

In this paper we used several star formation diagnostics based on mid-infrared data, or relations between radio and mid-infrared. It soon became evident that these proxies were not ideal tools to estimate the strength of star formation in our sources. They often show results contradicting each other, as well as results in the literature obtained with more reliable methods, for example, PAH features. It is possibly the complexity of NLS1s that fool these proxies, and especially in mid-infrared the presence of an emphasised dust component that can mimic the properties of emission produced by star formation. Nevertheless, we decided to include these diagnostics in our results to highlight their behaviour in the case of NLS1s, and as a general caution for future, especially statistical, studies using similar diagnostics.

\section{Summary and conclusions}
\label{sec:summary}

In this paper we analysed 44 NLS1 galaxies by means of spatially resolved radio spectral index maps. In addition, we employed mid-infrared diagnostics to estimate the star formation activity in their hosts. With these tools, complemented by archival data, we estimated the predominant source or sources of radio emission in them. Based on these analyses our main conclusions can be summarised as:

\begin{enumerate}
    \item NLS1s are an exceptionally heterogeneous class of AGN, and in radio band they range from individuals with blazar- and radio galaxy -like characteristics to sources whose radio emission is dominated by the star formation activity in the host galaxy. Furthermore, some of them could actually be classified as CSS sources in addition to the NLS1 classification, thus providing further confirmation of the scenario in which NLS1s are AGN in an early evolutionary stage.
    \item Due to this diversity and the varying contributions of different sources of radio as well as mid-infrared emission NLS1s are not ideal targets for simple proxies, such as the radio loudness parameter, or star formation diagnostics.
    \item As the NLS1 classification is based solely on their optical spectrum, and includes all Type 1 sources with narrow permitted lines, it does not reflect their other physical properties, for example, in the radio band. As a result the NLS1 classification does not tell us much as these sources do not necessarily share similar physical properties. Also the threshold of FWHM(H$\beta$) = 2000~\kms is somewhat arbitrary, and should not be regarded as a real physical limit. These remarks should be kept in mind especially when studying them as a class, or using the general NLS1 population in wider AGN studies.
\end{enumerate}
  
NLS1s are complicated sources, and any conclusions regarding the whole class cannot and should not be drawn lightly. Instead, these sources should be first and foremost studied as individuals, at least until we understand their nature and distinctive subclasses and their characteristics better. As demonstrated by this study, the majority of them require very high-quality data and detailed data analysis to disentangle the contribution of the different elements. And even still, many of the sources in our sample will require further observations, for example, in radio at higher resolution, or integral field spectroscopy, to unquestionably determine their nature, and their physical properties. Future facilities, such as the Cherenkov Telescope Array (CTA), the Advanced Telescope for High Energy Astrophysics (ATHENA), and Euclid, will also be essential to improve our understanding of this peculiar class of objects. Despite the complexity, the NLS1s definitely are a class worth investigating. The jetted NLS1s are an ideal laboratory for studying the launching of the jets, and the first stages of the evolution of more powerful FSRQs and high-excitation radio galaxies. In general in NLS1s, the interplay of the host and several AGN-related phenomena, for example, jets and outflows, offers a great opportunity to examine the AGN feedback in action.

\begin{acknowledgements}
E.J. is a current ESA research fellow. E.C. acknowledges support from ANID project Basal AFB-170002. The National Radio Astronomy Observatory is a facility of the National Science Foundation operated under cooperative agreement by Associated Universities, Inc. This publication makes use of data products from the Wide-field Infrared Survey Explorer, which is a joint project of the University of California, Los Angeles, and the Jet Propulsion Laboratory/California Institute of Technology, funded by the National Aeronautics and Space Administration. The Pan-STARRS1 Surveys (PS1) and the PS1 public science archive have been made possible through contributions by the Institute for Astronomy, the University of Hawaii, the Pan-STARRS Project Office, the Max-Planck Society and its participating institutes, the Max Planck Institute for Astronomy, Heidelberg and the Max Planck Institute for Extraterrestrial Physics, Garching, The Johns Hopkins University, Durham University, the University of Edinburgh, the Queen's University Belfast, the Harvard-Smithsonian Center for Astrophysics, the Las Cumbres Observatory Global Telescope Network Incorporated, the National Central University of Taiwan, the Space Telescope Science Institute, the National Aeronautics and Space Administration under Grant No. NNX08AR22G issued through the Planetary Science Division of the NASA Science Mission Directorate, the National Science Foundation Grant No. AST-1238877, the University of Maryland, Eotvos Lorand University (ELTE), the Los Alamos National Laboratory, and the Gordon and Betty Moore Foundation. Funding for the Sloan Digital Sky Survey (SDSS) has been provided by the Alfred P. Sloan Foundation, the Participating Institutions, the National Aeronautics and Space Administration, the National Science Foundation, the U.S. Department of Energy, the Japanese Monbukagakusho, and the Max Planck Society. The SDSS Web site is http://www.sdss.org/. The SDSS is managed by the Astrophysical Research Consortium (ARC) for the Participating Institutions. The Participating Institutions are The University of Chicago, Fermilab, the Institute for Advanced Study, the Japan Participation Group, The Johns Hopkins University, the Korean Scientist Group, Los Alamos National Laboratory, the Max-Planck-Institute for Astronomy (MPIA), the Max-Planck-Institute for Astrophysics (MPA), New Mexico State University, University of Pittsburgh, University of Portsmouth, Princeton University, the United States Naval Observatory, and the University of Washington. Thank you coffee and sparkling wine. This research has made use of the NASA/IPAC Extragalactic Database (NED), which is operated by the Jet Propulsion Laboratory, California Institute of Technology, under contract with the National Aeronautics and Space Administration. This research has made use of the SIMBAD database, operated at CDS, Strasbourg, France.

\end{acknowledgements}

\bibliographystyle{aa}
\bibliography{artikkeli.bib}

\begin{thebibliography}{240}
\expandafter\ifx\csname natexlab\endcsname\relax\def\natexlab#1{#1}\fi

\bibitem[{{Alonso-Herrero} {et~al.}(2016){Alonso-Herrero}, {Poulton}, {Roche},
  {Hern{\'a}n-Caballero}, {Aretxaga}, {Mart{\'\i}nez-Paredes}, {Ramos Almeida},
  {Pereira-Santaella}, {D{\'\i}az-Santos}, {Levenson}, {Packham}, {Colina},
  {Esquej}, {Gonz{\'a}lez-Mart{\'\i}n}, {Ichikawa}, {Imanishi}, {Rodr{\'\i}guez
  Espinosa}, \& {Telesco}}]{2016alonsoherrero1}
{Alonso-Herrero}, A., {Poulton}, R., {Roche}, P.~F., {et~al.} 2016, \mnras,
  463, 2405

\bibitem[{{Ant{\'o}n} {et~al.}(2008){Ant{\'o}n}, {Browne}, \&
  {March{\~a}}}]{2008anton1}
{Ant{\'o}n}, S., {Browne}, I.~W.~A., \& {March{\~a}}, M.~J. 2008, \aap, 490,
  583

\bibitem[{{Asmus}(2019)}]{2019asmus1}
{Asmus}, D. 2019, \mnras, 489, 2177

\bibitem[{{Asmus} {et~al.}(2016){Asmus}, {H{\"o}nig}, \& {Gandhi}}]{2016asmus1}
{Asmus}, D., {H{\"o}nig}, S.~F., \& {Gandhi}, P. 2016, \apj, 822, 109

\bibitem[{{Baldi} {et~al.}(2018){Baldi}, {Williams}, {McHardy}, {Beswick},
  {Argo}, {Dullo}, {Knapen}, {Brinks}, {Muxlow}, {Aalto}, {Alberdi}, {Bendo},
  {Corbel}, {Evans}, {Fenech}, {Green}, {Kl{\"o}ckner}, {K{\"o}rding}, {Kharb},
  {Maccarone}, {Mart{\'\i}-Vidal}, {Mundell}, {Panessa}, {Peck},
  {P{\'e}rez-Torres}, {Saikia}, {Saikia}, {Shankar}, {Spencer}, {Stevens},
  {Uttley}, \& {Westcott}}]{2018baldi1}
{Baldi}, R.~D., {Williams}, D.~R.~A., {McHardy}, I.~M., {et~al.} 2018, \mnras,
  476, 3478

\bibitem[{{Baldwin} {et~al.}(1981){Baldwin}, {Phillips}, \&
  {Terlevich}}]{1981baldwin1}
{Baldwin}, J.~A., {Phillips}, M.~M., \& {Terlevich}, R. 1981, \pasp, 93, 5

\bibitem[{{Barth} {et~al.}(2008){Barth}, {Bentz}, {Greene}, \&
  {Ho}}]{2008barth1}
{Barth}, A.~J., {Bentz}, M.~C., {Greene}, J.~E., \& {Ho}, L.~C. 2008, \apjl,
  683, L119

\bibitem[{{Baum} {et~al.}(1993){Baum}, {O'Dea}, {Dallacassa}, {de Bruyn}, \&
  {Pedlar}}]{1993baum1}
{Baum}, S.~A., {O'Dea}, C.~P., {Dallacassa}, D., {de Bruyn}, A.~G., \&
  {Pedlar}, A. 1993, \apj, 419, 553

\bibitem[{{Becker} {et~al.}(1991){Becker}, {White}, \& {Edwards}}]{1991becker1}
{Becker}, R.~H., {White}, R.~L., \& {Edwards}, A.~L. 1991, \apjs, 75, 1

\bibitem[{{Becker} {et~al.}(1995){Becker}, {White}, \& {Helfand}}]{1995becker1}
{Becker}, R.~H., {White}, R.~L., \& {Helfand}, D.~J. 1995, \apj, 450, 559

\bibitem[{{Behar} {et~al.}(2015){Behar}, {Baldi}, {Laor}, {Horesh}, {Stevens},
  \& {Tzioumis}}]{2015behar1}
{Behar}, E., {Baldi}, R.~D., {Laor}, A., {et~al.} 2015, \mnras, 451, 517

\bibitem[{{Berton} {et~al.}(2020{\natexlab{a}}){Berton}, {Bj{\"o}rklund},
  {L{\"a}hteenm{\"a}ki}, {Congiu}, {J{\"a}rvel{\"a}}, {Terreran}, \& {La
  Mura}}]{2020berton1}
{Berton}, M., {Bj{\"o}rklund}, I., {L{\"a}hteenm{\"a}ki}, A., {et~al.}
  2020{\natexlab{a}}, Contributions of the Astronomical Observatory Skalnate
  Pleso, 50, 270

\bibitem[{{Berton} {et~al.}(2016{\natexlab{a}}){Berton}, {Caccianiga},
  {Foschini}, {Peterson}, {Mathur}, {Terreran}, {Ciroi}, {Congiu}, {Cracco},
  {Frezzato}, {La Mura}, \& {Rafanelli}}]{2016berton1}
{Berton}, M., {Caccianiga}, A., {Foschini}, L., {et~al.} 2016{\natexlab{a}},
  \aap, 591, A98

\bibitem[{{Berton} {et~al.}(2019){Berton}, {Congiu}, {Ciroi}, {Komossa},
  {Frezzato}, {Di Mille}, {Ant{\'o}n}, {Antonucci}, {Caccianiga}, {Coppi},
  {J{\"a}rvel{\"a}}, {Kotilainen}, {L{\"a}hteenm{\"a}ki}, {Mathur}, {Chen},
  {Cracco}, {La Mura}, \& {Rafanelli}}]{2019berton1}
{Berton}, M., {Congiu}, E., {Ciroi}, S., {et~al.} 2019, \aj, 157, 48

\bibitem[{{Berton} {et~al.}(2018){Berton}, {Congiu}, {J{\"a}rvel{\"a}},
  {Antonucci}, {Kharb}, {Lister}, {Tarchi}, {Caccianiga}, {Chen}, {Foschini},
  {L{\"a}hteenm{\"a}ki}, {Richards}, {Ciroi}, {Cracco}, {Frezzato}, {La Mura},
  \& {Rafanelli}}]{2018berton1}
{Berton}, M., {Congiu}, E., {J{\"a}rvel{\"a}}, E., {et~al.} 2018, \aap, 614,
  A87

\bibitem[{{Berton} {et~al.}(2017){Berton}, {Foschini}, {Caccianiga}, {Ciroi},
  {Congiu}, {Cracco}, {Frezzato}, {La Mura}, \& {Rafanelli}}]{2017berton1}
{Berton}, M., {Foschini}, L., {Caccianiga}, A., {et~al.} 2017, Frontiers in
  Astronomy and Space Sciences, 4, 8

\bibitem[{{Berton} {et~al.}(2016{\natexlab{b}}){Berton}, {Foschini}, {Ciroi},
  {Cracco}, {La Mura}, {Di Mille}, \& {Rafanelli}}]{2016berton2}
{Berton}, M., {Foschini}, L., {Ciroi}, S., {et~al.} 2016{\natexlab{b}}, \aap,
  591, A88

\bibitem[{{Berton} \& {J{\"a}rvel{\"a}}(2021)}]{2021berton1}
{Berton}, M. \& {J{\"a}rvel{\"a}}, E. 2021, Universe, 7, 188

\bibitem[{{Berton} {et~al.}(2020{\natexlab{b}}){Berton}, {J{\"a}rvel{\"a}},
  {Crepaldi}, {L{\"a}hteenm{\"a}ki}, {Tornikoski}, {Congiu}, {Kharb},
  {Terreran}, \& {Vietri}}]{2020berton2}
{Berton}, M., {J{\"a}rvel{\"a}}, E., {Crepaldi}, L., {et~al.}
  2020{\natexlab{b}}, \aap, 636, A64

\bibitem[{{Bhatnagar} {et~al.}(2013){Bhatnagar}, {Rau}, \&
  {Golap}}]{2013bhatnagar1}
{Bhatnagar}, S., {Rau}, U., \& {Golap}, K. 2013, \apj, 770, 91

\bibitem[{{Bian} \& {Huang}(2010)}]{2010bian1}
{Bian}, W.-H. \& {Huang}, K. 2010, \mnras, 401, 507

\bibitem[{{Bicay} {et~al.}(1995){Bicay}, {Kojoian}, {Seal}, {Dickinson}, \&
  {Malkan}}]{1995bicay1}
{Bicay}, M.~D., {Kojoian}, G., {Seal}, J., {Dickinson}, D.~F., \& {Malkan},
  M.~A. 1995, \apjs, 98, 369

\bibitem[{{Bicknell} {et~al.}(1997){Bicknell}, {Dopita}, \&
  {O'Dea}}]{1997bicknell1}
{Bicknell}, G.~V., {Dopita}, M.~A., \& {O'Dea}, C. P.~O. 1997, \apj, 485, 112

\bibitem[{{Boller} {et~al.}(2001){Boller}, {Keil}, {Tr{\"u}mper}, {O'Brien},
  {Reeves}, \& {Page}}]{2001boller1}
{Boller}, T., {Keil}, R., {Tr{\"u}mper}, J., {et~al.} 2001, \aap, 365, L146

\bibitem[{{Boroson} \& {Green}(1992)}]{1992boroson1}
{Boroson}, T.~A. \& {Green}, R.~F. 1992, \apjs, 80, 109

\bibitem[{{Boyle} {et~al.}(2007){Boyle}, {Cornwell}, {Middelberg}, {Norris},
  {Appleton}, \& {Smail}}]{2007boyle1}
{Boyle}, B.~J., {Cornwell}, T.~J., {Middelberg}, E., {et~al.} 2007, \mnras,
  376, 1182

\bibitem[{{Brinkmann} {et~al.}(1995){Brinkmann}, {Siebert}, {Reich}, {Fuerst},
  {Reich}, {Voges}, {Truemper}, \& {Wielebinski}}]{1995brinkmann1}
{Brinkmann}, W., {Siebert}, J., {Reich}, W., {et~al.} 1995, \aaps, 109, 147

\bibitem[{{Buhariwalla} {et~al.}(2020){Buhariwalla}, {Waddell}, {Gallo},
  {Grupe}, \& {Komossa}}]{2020buhariwalla1}
{Buhariwalla}, M.~Z., {Waddell}, S. G.~H., {Gallo}, L.~C., {Grupe}, D., \&
  {Komossa}, S. 2020, \apj, 901, 118

\bibitem[{{Burtscher} {et~al.}(2009){Burtscher}, {Jaffe}, {Raban},
  {Meisenheimer}, {Tristram}, \& {R{\"o}ttgering}}]{2009burtscher1}
{Burtscher}, L., {Jaffe}, W., {Raban}, D., {et~al.} 2009, \apjl, 705, L53

\bibitem[{{Buta}(2017)}]{2017buta1}
{Buta}, R.~J. 2017, \mnras, 471, 4027

\bibitem[{{Caccianiga} {et~al.}(2014){Caccianiga}, {Ant{\'o}n}, {Ballo},
  {Dallacasa}, {Della Ceca}, {Fanali}, {Foschini}, {Hamilton}, {Kraus},
  {Maccacaro}, {Mack}, {March{\~a}}, {Paulino-Afonso}, {Sani}, \&
  {Severgnini}}]{2014caccianiga1}
{Caccianiga}, A., {Ant{\'o}n}, S., {Ballo}, L., {et~al.} 2014, \mnras, 441, 172

\bibitem[{{Caccianiga} {et~al.}(2015){Caccianiga}, {Ant{\'o}n}, {Ballo},
  {Foschini}, {Maccacaro}, {Della Ceca}, {Severgnini}, {March{\~a}}, {Mateos},
  \& {Sani}}]{2015caccianiga1}
{Caccianiga}, A., {Ant{\'o}n}, S., {Ballo}, L., {et~al.} 2015, \mnras, 451,
  1795

\bibitem[{{Caccianiga} {et~al.}(2017){Caccianiga}, {Dallacasa}, {Ant{\'o}n},
  {Ballo}, {Berton}, {Mack}, \& {Paulino-Afonso}}]{2017caccianiga1}
{Caccianiga}, A., {Dallacasa}, D., {Ant{\'o}n}, S., {et~al.} 2017, \mnras, 464,
  1474

\bibitem[{{Chainakun} \& {Young}(2017)}]{2017chainakun1}
{Chainakun}, P. \& {Young}, A.~J. 2017, \mnras, 465, 3965

\bibitem[{{Chamani} {et~al.}(2021){Chamani}, {Savolainen}, {Hada}, \&
  {Xu}}]{2021chamani1}
{Chamani}, W., {Savolainen}, T., {Hada}, K., \& {Xu}, M.~H. 2021, \aap, 652,
  A14

\bibitem[{{Chen} {et~al.}(2020){Chen}, {J{\"a}rvel{\"a}}, {Crepaldi}, {Zhou},
  {Ciroi}, {Berton}, {Kharb}, {Foschini}, {Gu}, {La Mura}, \&
  {Vietri}}]{2020chen1}
{Chen}, S., {J{\"a}rvel{\"a}}, E., {Crepaldi}, L., {et~al.} 2020, \mnras, 498,
  1278

\bibitem[{{Christopoulou} {et~al.}(1997){Christopoulou}, {Holloway}, {Steffen},
  {Mundell}, {Thean}, {Goudis}, {Meaburn}, \& {Pedlar}}]{1997christopoulou}
{Christopoulou}, P.~E., {Holloway}, A.~J., {Steffen}, W., {et~al.} 1997,
  \mnras, 284, 385

\bibitem[{{Cohen} {et~al.}(2007){Cohen}, {Lane}, {Cotton}, {Kassim}, {Lazio},
  {Perley}, {Condon}, \& {Erickson}}]{2007cohen1}
{Cohen}, A.~S., {Lane}, W.~M., {Cotton}, W.~D., {et~al.} 2007, \aj, 134, 1245

\bibitem[{{Condon}(1992)}]{1992condon1}
{Condon}, J.~J. 1992, \araa, 30, 575

\bibitem[{{Condon} {et~al.}(1998){Condon}, {Cotton}, {Greisen}, {Yin},
  {Perley}, {Taylor}, \& {Broderick}}]{1998condon1}
{Condon}, J.~J., {Cotton}, W.~D., {Greisen}, E.~W., {et~al.} 1998, \aj, 115,
  1693

\bibitem[{{Congiu} {et~al.}(2017{\natexlab{a}}){Congiu}, {Berton}, {Giroletti},
  {Antonucci}, {Caccianiga}, {Kharb}, {Lister}, {Foschini}, {Ciroi}, {Cracco},
  {Frezzato}, {J{\"a}rvel{\"a}}, {La Mura}, {Richards}, \&
  {Rafanelli}}]{2017congiu1}
{Congiu}, E., {Berton}, M., {Giroletti}, M., {et~al.} 2017{\natexlab{a}}, \aap,
  603, A32

\bibitem[{{Congiu} {et~al.}(2017{\natexlab{b}}){Congiu}, {Contini}, {Ciroi},
  {Cracco}, {Di Mille}, {Berton}, {Frezzato}, {La Mura}, \&
  {Rafanelli}}]{2017congiu2}
{Congiu}, E., {Contini}, M., {Ciroi}, S., {et~al.} 2017{\natexlab{b}},
  Frontiers in Astronomy and Space Sciences, 4, 27

\bibitem[{{Congiu} {et~al.}(2020){Congiu}, {Kharb}, {Tarchi}, {Berton},
  {Caccianiga}, {Chen}, {Crepaldi}, {Di Mille}, {J{\"a}rvel{\"a}}, {Jarvis},
  {La Mura}, \& {Vietri}}]{2020congiu1}
{Congiu}, E., {Kharb}, P., {Tarchi}, A., {et~al.} 2020, \mnras, 499, 3149

\bibitem[{{Crenshaw} {et~al.}(2003){Crenshaw}, {Kraemer}, \&
  {Gabel}}]{2003crenshaw1}
{Crenshaw}, D.~M., {Kraemer}, S.~B., \& {Gabel}, J.~R. 2003, \aj, 126, 1690

\bibitem[{{Croft} {et~al.}(2010){Croft}, {Bower}, {Ackermann}, {Atkinson},
  {Backer}, {Backus}, {Barott}, {Bauermeister}, {Blitz}, {Bock}, {Bradford},
  {Cheng}, {Cork}, {Davis}, {DeBoer}, {Dexter}, {Dreher}, {Engargiola},
  {Fields}, {Fleming}, {Forster}, {Gutierrez-Kraybill}, {Harp}, {Helfer},
  {Hull}, {Jordan}, {Jorgensen}, {Keating}, {Kilsdonk}, {Law}, {van Leeuwen},
  {Lugten}, {MacMahon}, {McMahon}, {Milgrome}, {Pierson}, {Randall}, {Ross},
  {Shostak}, {Siemion}, {Smolek}, {Tarter}, {Thornton}, {Urry}, {Vitouchkine},
  {Wadefalk}, {Welch}, {Werthimer}, {Whysong}, {Williams}, \&
  {Wright}}]{2010croft1}
{Croft}, S., {Bower}, G.~C., {Ackermann}, R., {et~al.} 2010, \apj, 719, 45

\bibitem[{{Czerny} {et~al.}(2009){Czerny}, {Siemiginowska}, {Janiuk},
  {Nikiel-Wroczy{\'n}ski}, \& {Stawarz}}]{2009czerny1}
{Czerny}, B., {Siemiginowska}, A., {Janiuk}, A., {Nikiel-Wroczy{\'n}ski}, B.,
  \& {Stawarz}, {\L}. 2009, \apj, 698, 840

\bibitem[{{da Cunha} {et~al.}(2008){da Cunha}, {Charlot}, \&
  {Elbaz}}]{2008dacunha1}
{da Cunha}, E., {Charlot}, S., \& {Elbaz}, D. 2008, \mnras, 388, 1595

\bibitem[{{Dalla Bont{\`a}} {et~al.}(2020){Dalla Bont{\`a}}, {Peterson},
  {Bentz}, {Brandt}, {Ciroi}, {De Rosa}, {Fonseca Alvarez}, {Grier}, {Hall},
  {Hern{\'a}ndez Santisteban}, {Ho}, {Homayouni}, {Horne}, {Kochanek}, {Li},
  {Morelli}, {Pizzella}, {Pogge}, {Schneider}, {Shen}, {Trump}, \&
  {Vestergaard}}]{2020dallabonta1}
{Dalla Bont{\`a}}, E., {Peterson}, B.~M., {Bentz}, M.~C., {et~al.} 2020, \apj,
  903, 112

\bibitem[{{D'Ammando} {et~al.}(2018){D'Ammando}, {Acosta-Pulido}, {Capetti},
  {Baldi}, {Orienti}, {Raiteri}, \& {Ramos Almeida}}]{2018dammando1}
{D'Ammando}, F., {Acosta-Pulido}, J.~A., {Capetti}, A., {et~al.} 2018, \mnras,
  478, L66

\bibitem[{{D'Ammando} {et~al.}(2017){D'Ammando}, {Acosta-Pulido}, {Capetti},
  {Raiteri}, {Baldi}, {Orienti}, \& {Ramos Almeida}}]{2017dammando1}
{D'Ammando}, F., {Acosta-Pulido}, J.~A., {Capetti}, A., {et~al.} 2017, \mnras,
  469, L11

\bibitem[{{Decarli} {et~al.}(2008){Decarli}, {Dotti}, {Fontana}, \&
  {Haardt}}]{2008decarli1}
{Decarli}, R., {Dotti}, M., {Fontana}, M., \& {Haardt}, F. 2008, \mnras, 386,
  L15

\bibitem[{{Deller} \& {Middelberg}(2014)}]{2014deller1}
{Deller}, A.~T. \& {Middelberg}, E. 2014, \aj, 147, 14

\bibitem[{{Deo} {et~al.}(2006){Deo}, {Crenshaw}, \& {Kraemer}}]{2006deo1}
{Deo}, R.~P., {Crenshaw}, D.~M., \& {Kraemer}, S.~B. 2006, \aj, 132, 321

\bibitem[{{Dicken} {et~al.}(2009){Dicken}, {Tadhunter}, {Axon}, {Morganti},
  {Inskip}, {Holt}, {Gonz{\'a}lez Delgado}, \& {Groves}}]{2009dicken1}
{Dicken}, D., {Tadhunter}, C., {Axon}, D., {et~al.} 2009, \apj, 694, 268

\bibitem[{{Doi} {et~al.}(2013){Doi}, {Asada}, {Fujisawa}, {Nagai}, {Hagiwara},
  {Wajima}, \& {Inoue}}]{2013doi1}
{Doi}, A., {Asada}, K., {Fujisawa}, K., {et~al.} 2013, \apj, 765, 69

\bibitem[{{Doi} {et~al.}(2011){Doi}, {Asada}, \& {Nagai}}]{2011doi1}
{Doi}, A., {Asada}, K., \& {Nagai}, H. 2011, \apj, 738, 126

\bibitem[{{Doi} {et~al.}(2007){Doi}, {Fujisawa}, {Inoue}, {Wajima}, {Nagai},
  {Harada}, {Suematsu}, {Habe}, {Honma}, {Kawaguchi}, {Kawai}, {Kobayashi},
  {Koyama}, {Kuboki}, {Murata}, {Omodaka}, {Sorai}, {Sudou}, {Takaba},
  {Takashima}, {Takeda}, {Tamura}, \& {Wakamatsu}}]{2007doi1}
{Doi}, A., {Fujisawa}, K., {Inoue}, M., {et~al.} 2007, \pasj, 59, 703

\bibitem[{{Doi} {et~al.}(2016){Doi}, {Oyama}, {Kono}, {Yamauchi}, {Suzuki},
  {Matsumoto}, \& {Tazaki}}]{2016doi1}
{Doi}, A., {Oyama}, T., {Kono}, Y., {et~al.} 2016, \pasj, 68, 73

\bibitem[{{Doi} {et~al.}(2015){Doi}, {Wajima}, {Hagiwara}, \&
  {Inoue}}]{2015doi1}
{Doi}, A., {Wajima}, K., {Hagiwara}, Y., \& {Inoue}, M. 2015, \apjl, 798, L30

\bibitem[{{Dressel} \& {Condon}(1978)}]{1978dressel1}
{Dressel}, L.~L. \& {Condon}, J.~J. 1978, \apjs, 36, 53

\bibitem[{{Du} \& {Wang}(2019)}]{2019du1}
{Du}, P. \& {Wang}, J.-M. 2019, \apj, 886, 42

\bibitem[{{Du} {et~al.}(2018){Du}, {Zhang}, {Wang}, {Huang}, {Zhang}, {Lu},
  {Hu}, {Li}, {Bai}, {Bian}, {Yuan}, {Ho}, {Wang}, \& {SEAMBH
  Collaboration}}]{2018du1}
{Du}, P., {Zhang}, Z.-X., {Wang}, K., {et~al.} 2018, \apj, 856, 6

\bibitem[{{Edelson}(1987)}]{1987edelson1}
{Edelson}, R.~A. 1987, \apj, 313, 651

\bibitem[{{Edelson} \& {Malkan}(1986)}]{1986edelson1}
{Edelson}, R.~A. \& {Malkan}, M.~A. 1986, \apj, 308, 59

\bibitem[{{Ellison} {et~al.}(2011){Ellison}, {Patton}, {Mendel}, \&
  {Scudder}}]{2011ellison1}
{Ellison}, S.~L., {Patton}, D.~R., {Mendel}, J.~T., \& {Scudder}, J.~M. 2011,
  \mnras, 418, 2043

\bibitem[{{Esparza-Arredondo} {et~al.}(2020){Esparza-Arredondo},
  {Osorio-Clavijo}, {Gonz{\'a}lez-Mart{\'\i}n}, {Victoria-Ceballos},
  {Haro-Corzo}, {Reyes-Amador}, {L{\'o}pez-S{\'a}nchez}, \&
  {Pasetto}}]{2020esparzaarredondo1}
{Esparza-Arredondo}, D., {Osorio-Clavijo}, N., {Gonz{\'a}lez-Mart{\'\i}n}, O.,
  {et~al.} 2020, \apj, 905, 29

\bibitem[{{Esquej} {et~al.}(2014){Esquej}, {Alonso-Herrero},
  {Gonz{\'a}lez-Mart{\'\i}n}, {H{\"o}nig}, {Hern{\'a}n-Caballero}, {Roche},
  {Ramos Almeida}, {Mason}, {D{\'\i}az-Santos}, {Levenson}, {Aretxaga},
  {Rodr{\'\i}guez Espinosa}, \& {Packham}}]{2014esquej1}
{Esquej}, P., {Alonso-Herrero}, A., {Gonz{\'a}lez-Mart{\'\i}n}, O., {et~al.}
  2014, \apj, 780, 86

\bibitem[{{Faucher-Gigu{\`e}re} \& {Quataert}(2012)}]{2012fauchergiguere1}
{Faucher-Gigu{\`e}re}, C.-A. \& {Quataert}, E. 2012, \mnras, 425, 605

\bibitem[{{Fischer} {et~al.}(2013){Fischer}, {Crenshaw}, {Kraemer}, \&
  {Schmitt}}]{2013fischer1}
{Fischer}, T.~C., {Crenshaw}, D.~M., {Kraemer}, S.~B., \& {Schmitt}, H.~R.
  2013, \apjs, 209, 1

\bibitem[{{Flewelling} {et~al.}(2020){Flewelling}, {Magnier}, {Chambers},
  {Heasley}, {Holmberg}, {Huber}, {Sweeney}, {Waters}, {Calamida}, {Casertano},
  {Chen}, {Farrow}, {Hasinger}, {Henderson}, {Long}, {Metcalfe}, {Narayan},
  {Nieto-Santisteban}, {Norberg}, {Rest}, {Saglia}, {Szalay}, {Thakar},
  {Tonry}, {Valenti}, {Werner}, {White}, {Denneau}, {Draper}, {Hodapp},
  {Jedicke}, {Kaiser}, {Kudritzki}, {Price}, {Wainscoat}, {Chastel}, {McLean},
  {Postman}, \& {Shiao}}]{2020flewelling1}
{Flewelling}, H.~A., {Magnier}, E.~A., {Chambers}, K.~C., {et~al.} 2020, \apjs,
  251, 7

\bibitem[{{Foschini}(2011)}]{2011foschini1}
{Foschini}, L. 2011, in Narrow-Line Seyfert 1 Galaxies and their Place in the
  Universe, \url{https://pos.sissa.it/126/024/pdf}

\bibitem[{{Foschini}(2017)}]{2017foschini1}
{Foschini}, L. 2017, Frontiers in Astronomy and Space Sciences, 4, 6

\bibitem[{{Foschini}(2020)}]{2020foschini1}
{Foschini}, L. 2020, Universe, 6, 136

\bibitem[{{Foschini} {et~al.}(2015){Foschini}, {Berton}, {Caccianiga}, {Ciroi},
  {Cracco}, {Peterson}, {Angelakis}, {Braito}, {Fuhrmann}, {Gallo}, {Grupe},
  {J{\"a}rvel{\"a}}, {Kaufmann}, {Komossa}, {Kovalev}, {L{\"a}hteenm{\"a}ki},
  {Lisakov}, {Lister}, {Mathur}, {Richards}, {Romano}, {Sievers},
  {Tagliaferri}, {Tammi}, {Tibolla}, {Tornikoski}, {Vercellone}, {La Mura},
  {Maraschi}, \& {Rafanelli}}]{2015foschini1}
{Foschini}, L., {Berton}, M., {Caccianiga}, A., {et~al.} 2015, \aap, 575, A13

\bibitem[{{Fraix-Burnet} {et~al.}(2017){Fraix-Burnet}, {Marziani}, {D'Onofrio},
  \& {Dultzin}}]{2017fraixburnet1}
{Fraix-Burnet}, D., {Marziani}, P., {D'Onofrio}, M., \& {Dultzin}, D. 2017,
  Frontiers in Astronomy and Space Sciences, 4, 1

\bibitem[{{Fritz} {et~al.}(2006){Fritz}, {Franceschini}, \&
  {Hatziminaoglou}}]{2006fritz1}
{Fritz}, J., {Franceschini}, A., \& {Hatziminaoglou}, E. 2006, \mnras, 366, 767

\bibitem[{{Gallimore} {et~al.}(2006){Gallimore}, {Axon}, {O'Dea}, {Baum}, \&
  {Pedlar}}]{2006gallimore1}
{Gallimore}, J.~F., {Axon}, D.~J., {O'Dea}, C.~P., {Baum}, S.~A., \& {Pedlar},
  A. 2006, \aj, 132, 546

\bibitem[{{Gallo} {et~al.}(2006){Gallo}, {Edwards}, {Ferrero}, {Kataoka},
  {Lewis}, {Ellingsen}, {Misanovic}, {Welsh}, {Whiting}, {Boller}, {Brinkmann},
  {Greenhill}, \& {Oshlack}}]{2006gallo1}
{Gallo}, L.~C., {Edwards}, P.~G., {Ferrero}, E., {et~al.} 2006, \mnras, 370,
  245

\bibitem[{{Ganci} {et~al.}(2019){Ganci}, {Marziani}, {D'Onofrio}, {del Olmo},
  {Bon}, {Bon}, \& {Negrete}}]{2019ganci1}
{Ganci}, V., {Marziani}, P., {D'Onofrio}, M., {et~al.} 2019, \aap, 630, A110

\bibitem[{{Giroletti} \& {Panessa}(2009)}]{2009giroletti1}
{Giroletti}, M. \& {Panessa}, F. 2009, \apjl, 706, L260

\bibitem[{{Giroletti} {et~al.}(2017){Giroletti}, {Panessa}, {Longinotti},
  {Krongold}, {Guainazzi}, {Costantini}, \& {Santos-Lleo}}]{2017giroletti1}
{Giroletti}, M., {Panessa}, F., {Longinotti}, A.~L., {et~al.} 2017, \aap, 600,
  A87

\bibitem[{{Goad} {et~al.}(2012){Goad}, {Korista}, \& {Ruff}}]{2012goad1}
{Goad}, M.~R., {Korista}, K.~T., \& {Ruff}, A.~J. 2012, \mnras, 426, 3086

\bibitem[{{Gonzalez Delgado} \& {Perez}(1996)}]{1996gonzalesdelgado1}
{Gonzalez Delgado}, R.~M. \& {Perez}, E. 1996, \mnras, 278, 737

\bibitem[{{Gonz{\'a}lez-Mart{\'\i}n} {et~al.}(2019){Gonz{\'a}lez-Mart{\'\i}n},
  {Masegosa}, {Garc{\'\i}a-Bernete}, {Ramos Almeida},
  {Rodr{\'\i}guez-Espinosa}, {M{\'a}rquez}, {Esparza-Arredondo},
  {Osorio-Clavijo}, {Mart{\'\i}nez-Paredes}, {Victoria-Ceballos}, {Pasetto}, \&
  {Dultzin}}]{2019gonzalezmartin1}
{Gonz{\'a}lez-Mart{\'\i}n}, O., {Masegosa}, J., {Garc{\'\i}a-Bernete}, I.,
  {et~al.} 2019, \apj, 884, 11

\bibitem[{{Goodrich}(1989)}]{1989goodrich1}
{Goodrich}, R.~W. 1989, \apj, 342, 224

\bibitem[{{Gregory} \& {Condon}(1991)}]{1991gregory1}
{Gregory}, P.~C. \& {Condon}, J.~J. 1991, \apjs, 75, 1011

\bibitem[{{Gruppioni} {et~al.}(2016){Gruppioni}, {Berta}, {Spinoglio},
  {Pereira-Santaella}, {Pozzi}, {Andreani}, {Bonato}, {De Zotti}, {Malkan},
  {Negrello}, {Vallini}, \& {Vignali}}]{2016gruppioni1}
{Gruppioni}, C., {Berta}, S., {Spinoglio}, L., {et~al.} 2016, \mnras, 458, 4297

\bibitem[{{Gu} \& {Chen}(2010)}]{2010gu1}
{Gu}, M. \& {Chen}, Y. 2010, \aj, 139, 2612

\bibitem[{{Gu} {et~al.}(2015){Gu}, {Chen}, {Komossa}, {Yuan}, {Shen}, {Wajima},
  {Zhou}, \& {Zensus}}]{2015gu1}
{Gu}, M., {Chen}, Y., {Komossa}, S., {et~al.} 2015, \apjs, 221, 3

\bibitem[{{Gupta} {et~al.}(2013{\natexlab{a}}){Gupta}, {Mathur}, {Krongold}, \&
  {Nicastro}}]{2013gupta2}
{Gupta}, A., {Mathur}, S., {Krongold}, Y., \& {Nicastro}, F.
  2013{\natexlab{a}}, \apj, 768, 141

\bibitem[{{Gupta} {et~al.}(2013{\natexlab{b}}){Gupta}, {Mathur}, {Krongold}, \&
  {Nicastro}}]{2013gupta1}
{Gupta}, A., {Mathur}, S., {Krongold}, Y., \& {Nicastro}, F.
  2013{\natexlab{b}}, \apj, 772, 66

\bibitem[{{G{\"u}rkan} {et~al.}(2019){G{\"u}rkan}, {Hardcastle}, {Best},
  {Morabito}, {Prandoni}, {Jarvis}, {Duncan}, {Calistro Rivera}, {Callingham},
  {Cochrane}, {Croston}, {Heald}, {Mingo}, {Mooney}, {Sabater},
  {R{\"o}ttgering}, {Shimwell}, {Smith}, {Tasse}, \& {Williams}}]{2019gurkan1}
{G{\"u}rkan}, G., {Hardcastle}, M.~J., {Best}, P.~N., {et~al.} 2019, \aap, 622,
  A11

\bibitem[{{Hagiwara} {et~al.}(2018){Hagiwara}, {Doi}, {Hachisuka}, \&
  {Horiuchi}}]{2018hagiwara1}
{Hagiwara}, Y., {Doi}, A., {Hachisuka}, K., \& {Horiuchi}, S. 2018, \pasj, 70,
  54

\bibitem[{{Hamilton} {et~al.}(2021){Hamilton}, {Berton}, {Ant{\'o}n}, {Busoni},
  {Caccianiga}, {Ciroi}, {G{\"a}ssler}, {Georgiev}, {J{\"a}rvel{\"a}},
  {Komossa}, {Mathur}, \& {Rabien}}]{2021hamilton1}
{Hamilton}, T.~S., {Berton}, M., {Ant{\'o}n}, S., {et~al.} 2021, \mnras, 504,
  5188

\bibitem[{{Harrison} {et~al.}(2012){Harrison}, {Alexander}, {Swinbank},
  {Smail}, {Alaghband-Zadeh}, {Bauer}, {Chapman}, {Del Moro}, {Hickox},
  {Ivison}, {Men{\'e}ndez-Delmestre}, {Mullaney}, \&
  {Nesvadba}}]{2012harrison1}
{Harrison}, C.~M., {Alexander}, D.~M., {Swinbank}, A.~M., {et~al.} 2012,
  \mnras, 426, 1073

\bibitem[{{Heinz} \& {Sunyaev}(2003)}]{2003heinz1}
{Heinz}, S. \& {Sunyaev}, R.~A. 2003, \mnras, 343, L59

\bibitem[{{Heisler} \& {Vader}(1994)}]{1994heisler1}
{Heisler}, C.~A. \& {Vader}, J.~P. 1994, \aj, 107, 35

\bibitem[{{H{\"o}nig} {et~al.}(2013){H{\"o}nig}, {Kishimoto}, {Tristram},
  {Prieto}, {Gandhi}, {Asmus}, {Antonucci}, {Burtscher}, {Duschl}, \&
  {Weigelt}}]{2013honig1}
{H{\"o}nig}, S.~F., {Kishimoto}, M., {Tristram}, K.~R.~W., {et~al.} 2013, \apj,
  771, 87

\bibitem[{{Hutchings} \& {Craven}(1988)}]{1988hutchings1}
{Hutchings}, J.~B. \& {Craven}, S.~E. 1988, \aj, 95, 677

\bibitem[{{Ibar} {et~al.}(2008){Ibar}, {Cirasuolo}, {Ivison}, {Best}, {Smail},
  {Biggs}, {Simpson}, {Dunlop}, {Almaini}, {McLure}, {Foucaud}, \&
  {Rawlings}}]{2008ibar1}
{Ibar}, E., {Cirasuolo}, M., {Ivison}, R., {et~al.} 2008, \mnras, 386, 953

\bibitem[{{Igo} {et~al.}(2020){Igo}, {Parker}, {Matzeu}, {Alston}, {Alvarez
  Crespo}, {F{\"u}rst}, {Buisson}, {Lobban}, {Joyce}, {Mallick}, {Schartel}, \&
  {Santos-Lle{\'o}}}]{2020igo1}
{Igo}, Z., {Parker}, M.~L., {Matzeu}, G.~A., {et~al.} 2020, \mnras, 493, 1088

\bibitem[{{Jackson} {et~al.}(2020){Jackson}, {Rosario}, {Alexander}, {Scholtz},
  {McAlpine}, \& {Bower}}]{2020jackson1}
{Jackson}, T.~M., {Rosario}, D.~J., {Alexander}, D.~M., {et~al.} 2020, \mnras,
  498, 2323

\bibitem[{{Jarrett} {et~al.}(2011){Jarrett}, {Cohen}, {Masci}, {Wright},
  {Stern}, {Benford}, {Blain}, {Carey}, {Cutri}, {Eisenhardt}, {Lonsdale},
  {Mainzer}, {Marsh}, {Padgett}, {Petty}, {Ressler}, {Skrutskie}, {Stanford},
  {Surace}, {Tsai}, {Wheelock}, \& {Yan}}]{2011jarrett1}
{Jarrett}, T.~H., {Cohen}, M., {Masci}, F., {et~al.} 2011, \apj, 735, 112

\bibitem[{{J{\"a}rvel{\"a}} {et~al.}(2020){J{\"a}rvel{\"a}}, {Berton}, {Ciroi},
  {Congiu}, {L{\"a}hteenm{\"a}ki}, \& {Di Mille}}]{2020jarvela1}
{J{\"a}rvel{\"a}}, E., {Berton}, M., {Ciroi}, S., {et~al.} 2020, \aap, 636, L12

\bibitem[{{J{\"a}rvel{\"a}} {et~al.}(2021){J{\"a}rvel{\"a}}, {Berton}, \&
  {Crepaldi}}]{2021jarvela1}
{J{\"a}rvel{\"a}}, E., {Berton}, M., \& {Crepaldi}, L. 2021, arXiv e-prints,
  arXiv:2108.08521

\bibitem[{{J{\"a}rvel{\"a}} {et~al.}(2018){J{\"a}rvel{\"a}},
  {L{\"a}hteenm{\"a}ki}, \& {Berton}}]{2018jarvela1}
{J{\"a}rvel{\"a}}, E., {L{\"a}hteenm{\"a}ki}, A., \& {Berton}, M. 2018, \aap,
  619, A69

\bibitem[{{J{\"a}rvel{\"a}} {et~al.}(2017){J{\"a}rvel{\"a}},
  {L{\"a}hteenm{\"a}ki}, {Lietzen}, {Poudel}, {Hein{\"a}m{\"a}ki}, \&
  {Einasto}}]{2017jarvela1}
{J{\"a}rvel{\"a}}, E., {L{\"a}hteenm{\"a}ki}, A., {Lietzen}, H., {et~al.} 2017,
  \aap, 606, A9

\bibitem[{{Jiang} {et~al.}(2010){Jiang}, {Ciotti}, {Ostriker}, \&
  {Spitkovsky}}]{2010jiang1}
{Jiang}, Y.-F., {Ciotti}, L., {Ostriker}, J.~P., \& {Spitkovsky}, A. 2010,
  \apj, 711, 125

\bibitem[{{Jin} {et~al.}(2021){Jin}, {Done}, \& {Ward}}]{2021jin1}
{Jin}, C., {Done}, C., \& {Ward}, M. 2021, \mnras, 500, 2475

\bibitem[{{Jones} {et~al.}(2017){Jones}, {McHardy}, \&
  {Maccarone}}]{2017jones1}
{Jones}, S., {McHardy}, I., \& {Maccarone}, T.~J. 2017, \mnras, 465, 1336

\bibitem[{{Kellermann} {et~al.}(1989){Kellermann}, {Sramek}, {Schmidt},
  {Shaffer}, \& {Green}}]{1989kellermann1}
{Kellermann}, K.~I., {Sramek}, R., {Schmidt}, M., {Shaffer}, D.~B., \& {Green},
  R. 1989, \aj, 98, 1195

\bibitem[{{Kellermann} {et~al.}(1994){Kellermann}, {Sramek}, {Schmidt},
  {Green}, \& {Shaffer}}]{1994kellermann1}
{Kellermann}, K.~I., {Sramek}, R.~A., {Schmidt}, M., {Green}, R.~F., \&
  {Shaffer}, D.~B. 1994, \aj, 108, 1163

\bibitem[{{Khanna} {et~al.}(2016){Khanna}, {Kaastra}, \&
  {Mehdipour}}]{2016khanna1}
{Khanna}, S., {Kaastra}, J.~S., \& {Mehdipour}, M. 2016, \aap, 586, A2

\bibitem[{{Kharb} {et~al.}(2010){Kharb}, {Lister}, \& {Cooper}}]{2010kharb1}
{Kharb}, P., {Lister}, M.~L., \& {Cooper}, N.~J. 2010, \apj, 710, 764

\bibitem[{{King} {et~al.}(2011){King}, {Miller}, {Cackett}, {Fabian},
  {Markoff}, {Nowak}, {Rupen}, {G{\"u}ltekin}, \& {Reynolds}}]{2011king1}
{King}, A.~L., {Miller}, J.~M., {Cackett}, E.~M., {et~al.} 2011, \apj, 729, 19

\bibitem[{{Kinney} {et~al.}(2000){Kinney}, {Schmitt}, {Clarke}, {Pringle},
  {Ulvestad}, \& {Antonucci}}]{2000kinney1}
{Kinney}, A.~L., {Schmitt}, H.~R., {Clarke}, C.~J., {et~al.} 2000, \apj, 537,
  152

\bibitem[{{Kollatschny}(2003)}]{2003kollatschny1}
{Kollatschny}, W. 2003, \aap, 407, 461

\bibitem[{{Kollatschny} \& {Zetzl}(2011)}]{2011kollatschny1}
{Kollatschny}, W. \& {Zetzl}, M. 2011, \nat, 470, 366

\bibitem[{{Komatsu} {et~al.}(2011){Komatsu}, {Smith}, {Dunkley}, {Bennett},
  {Gold}, {Hinshaw}, {Jarosik}, {Larson}, {Nolta}, {Page}, {Spergel},
  {Halpern}, {Hill}, {Kogut}, {Limon}, {Meyer}, {Odegard}, {Tucker}, {Weiland},
  {Wollack}, \& {Wright}}]{2011komatsu1}
{Komatsu}, E., {Smith}, K.~M., {Dunkley}, J., {et~al.} 2011, \apjs, 192, 18

\bibitem[{{Komossa} {et~al.}(2006){Komossa}, {Voges}, {Xu}, {Mathur}, {Adorf},
  {Lemson}, {Duschl}, \& {Grupe}}]{2006komossa1}
{Komossa}, S., {Voges}, W., {Xu}, D., {et~al.} 2006, \aj, 132, 531

\bibitem[{{Komossa} {et~al.}(2018){Komossa}, {Xu}, \& {Wagner}}]{2018komossa1}
{Komossa}, S., {Xu}, D.~W., \& {Wagner}, A.~Y. 2018, \mnras, 477, 5115

\bibitem[{{Kotilainen} {et~al.}(2016){Kotilainen}, {Le{\'o}n-Tavares},
  {Olgu{\'{\i}}n-Iglesias}, {Baes}, {An{\'o}rve}, {Chavushyan}, \&
  {Carrasco}}]{2016kotilainen1}
{Kotilainen}, J.~K., {Le{\'o}n-Tavares}, J., {Olgu{\'{\i}}n-Iglesias}, A.,
  {et~al.} 2016, \apj, 832, 157

\bibitem[{{Kozie{\l}-Wierzbowska} {et~al.}(2021){Kozie{\l}-Wierzbowska}, {Vale
  Asari}, {Stasi{\'n}ska}, {Herpich}, {Sikora}, {{\.Z}ywucka}, \&
  {Goyal}}]{2021kozielwierzbowska1}
{Kozie{\l}-Wierzbowska}, D., {Vale Asari}, N., {Stasi{\'n}ska}, G., {et~al.}
  2021, \apj, 910, 64

\bibitem[{{Krongold} {et~al.}(2001){Krongold}, {Dultzin-Hacyan}, \&
  {Marziani}}]{2001krongold1}
{Krongold}, Y., {Dultzin-Hacyan}, D., \& {Marziani}, P. 2001, \aj, 121, 702

\bibitem[{{Kukula} {et~al.}(1998){Kukula}, {Dunlop}, {Hughes}, \&
  {Rawlings}}]{1998kukula1}
{Kukula}, M.~J., {Dunlop}, J.~S., {Hughes}, D.~H., \& {Rawlings}, S. 1998,
  \mnras, 297, 366

\bibitem[{{Kukula} {et~al.}(1995){Kukula}, {Pedlar}, {Baum}, \&
  {O'Dea}}]{1995kukula1}
{Kukula}, M.~J., {Pedlar}, A., {Baum}, S.~A., \& {O'Dea}, C.~P. 1995, \mnras,
  276, 1262

\bibitem[{{Kunert} {et~al.}(2002){Kunert}, {Marecki}, {Spencer}, {Kus}, \&
  {Niezgoda}}]{2002kunert1}
{Kunert}, M., {Marecki}, A., {Spencer}, R.~E., {Kus}, A.~J., \& {Niezgoda}, J.
  2002, \aap, 391, 47

\bibitem[{{Kunert-Bajraszewska} {et~al.}(2010){Kunert-Bajraszewska},
  {Gawro{\'n}ski}, {Labiano}, \& {Siemiginowska}}]{2010kunertbajraszewska1}
{Kunert-Bajraszewska}, M., {Gawro{\'n}ski}, M.~P., {Labiano}, A., \&
  {Siemiginowska}, A. 2010, \mnras, 408, 2261

\bibitem[{{Kunert-Bajraszewska} \& {Labiano}(2010)}]{2010kunertbajraszewska2}
{Kunert-Bajraszewska}, M. \& {Labiano}, A. 2010, \mnras, 408, 2279

\bibitem[{{L{\"a}hteenm{\"a}ki} {et~al.}(2018){L{\"a}hteenm{\"a}ki},
  {J{\"a}rvel{\"a}}, {Ramakrishnan}, {Tornikoski}, {Tammi}, {Vera}, \&
  {Chamani}}]{2018lahteenmaki1}
{L{\"a}hteenm{\"a}ki}, A., {J{\"a}rvel{\"a}}, E., {Ramakrishnan}, V., {et~al.}
  2018, \aap, 614, L1

\bibitem[{{LaMassa} {et~al.}(2012){LaMassa}, {Heckman}, {Ptak}, {Schiminovich},
  {O'Dowd}, \& {Bertincourt}}]{2012lamassa1}
{LaMassa}, S.~M., {Heckman}, T.~M., {Ptak}, A., {et~al.} 2012, \apj, 758, 1

\bibitem[{{LaMassa} {et~al.}(2016){LaMassa}, {Ricarte}, {Glikman}, {Urry},
  {Stern}, {Yaqoob}, {Lansbury}, {Civano}, {Boggs}, {Brandt}, {Chen},
  {Christensen}, {Craig}, {Hailey}, {Harrison}, {Hickox}, {Koss}, {Ricci},
  {Treister}, \& {Zhang}}]{2016lamassa1}
{LaMassa}, S.~M., {Ricarte}, A., {Glikman}, E., {et~al.} 2016, \apj, 820, 70

\bibitem[{{Lamperti} {et~al.}(2020){Lamperti}, {Saintonge}, {Koss}, {Viti},
  {Wilson}, {He}, {Shimizu}, {Greve}, {Mushotzky}, {Treister}, {Kramer},
  {Sanders}, {Schawinski}, \& {Tacconi}}]{2020lamperti1}
{Lamperti}, I., {Saintonge}, A., {Koss}, M., {et~al.} 2020, \apj, 889, 103

\bibitem[{{Laor}(2000)}]{2000laor1}
{Laor}, A. 2000, \apjl, 543, L111

\bibitem[{{Laurent-Muehleisen} {et~al.}(1997){Laurent-Muehleisen}, {Kollgaard},
  {Ryan}, {Feigelson}, {Brinkmann}, \& {Siebert}}]{1997laurentmuehleisen1}
{Laurent-Muehleisen}, S.~A., {Kollgaard}, R.~I., {Ryan}, P.~J., {et~al.} 1997,
  \aaps, 122, 235

\bibitem[{{Lee} {et~al.}(2013){Lee}, {Kriss}, {Chakravorty}, {Rahoui}, {Young},
  {Brandt}, {Hines}, {Ogle}, \& {Reynolds}}]{2013lee1}
{Lee}, J.~C., {Kriss}, G.~A., {Chakravorty}, S., {et~al.} 2013, \mnras, 430,
  2650

\bibitem[{{Leftley} {et~al.}(2019){Leftley}, {H{\"o}nig}, {Asmus}, {Tristram},
  {Gandhi}, {Kishimoto}, {Venanzi}, \& {Williamson}}]{2019leftley1}
{Leftley}, J.~H., {H{\"o}nig}, S.~F., {Asmus}, D., {et~al.} 2019, \apj, 886, 55

\bibitem[{{Lianou} {et~al.}(2019){Lianou}, {Barmby}, {Mosenkov}, {Lehnert}, \&
  {Karczewski}}]{2019lianou1}
{Lianou}, S., {Barmby}, P., {Mosenkov}, A.~A., {Lehnert}, M., \& {Karczewski},
  O. 2019, \aap, 631, A38

\bibitem[{Liao {et~al.}(2017)Liao, Liang, Weng, Berton, Gu, \& Fan}]{2015liao1}
Liao, N.-H., Liang, Y.-F., Weng, S.-S., {et~al.} 2017, Discovery of
  $\gamma$-ray emission from steep radio spectrum NLS1s

\bibitem[{{Lister}(2018)}]{2018lister1}
{Lister}, M. 2018, in Revisiting Narrow-Line Seyfert 1 Galaxies and their Place
  in the Universe, 22

\bibitem[{{Longinotti} {et~al.}(2018){Longinotti}, {Vega}, {Krongold},
  {Aretxaga}, {Yun}, {Chavushyan}, {Feruglio}, {G{\'o}mez-Ruiz}, {Monta{\~n}a},
  {Le{\'o}n-Tavares}, {Olgu{\'\i}n-Iglesias}, {Giroletti}, {Guainazzi},
  {Kotilainen}, {Panessa}, {Zapata}, {Cruz-Gonzalez}, {Pati{\~n}o-{\'A}lvarez},
  {Rosa-Gonzalez}, {Carrami{\~n}ana}, {Carrasco}, {Costantini}, {Dultzin},
  {Guichard}, {Puerari}, \& {Santos-Lleo}}]{2018longinotti1}
{Longinotti}, A.~L., {Vega}, O., {Krongold}, Y., {et~al.} 2018, \apjl, 867, L11

\bibitem[{{Lyu} \& {Rieke}(2018)}]{2018lyu1}
{Lyu}, J. \& {Rieke}, G.~H. 2018, \apj, 866, 92

\bibitem[{{Mahajan} {et~al.}(2019){Mahajan}, {Ashby}, {Willner}, {Barmby},
  {Fazio}, {Maragkoudakis}, {Raychaudhury}, \& {Zezas}}]{2019mahajan1}
{Mahajan}, S., {Ashby}, M.~L.~N., {Willner}, S.~P., {et~al.} 2019, \mnras, 482,
  560

\bibitem[{{Maitra} {et~al.}(2011){Maitra}, {Miller}, {Markoff}, \&
  {King}}]{2011maitra1}
{Maitra}, D., {Miller}, J.~M., {Markoff}, S., \& {King}, A. 2011, \apj, 735,
  107

\bibitem[{{Mallick} {et~al.}(2016){Mallick}, {Dewangan}, {Gandhi}, {Misra}, \&
  {Kembhavi}}]{2016mallick1}
{Mallick}, L., {Dewangan}, G.~C., {Gandhi}, P., {Misra}, R., \& {Kembhavi},
  A.~K. 2016, \mnras, 460, 1705

\bibitem[{{Mart{\'\i}nez-Paredes} {et~al.}(2019){Mart{\'\i}nez-Paredes},
  {Aretxaga}, {Gonz{\'a}lez-Mart{\'\i}n}, {Alonso-Herrero}, {Levenson}, {Ramos
  Almeida}, \& {L{\'o}pez-Rodr{\'\i}guez}}]{2019martinezparedes1}
{Mart{\'\i}nez-Paredes}, M., {Aretxaga}, I., {Gonz{\'a}lez-Mart{\'\i}n}, O.,
  {et~al.} 2019, \apj, 871, 190

\bibitem[{{Marziani} {et~al.}(2018{\natexlab{a}}){Marziani}, {del Olmo},
  {D'Onofrio}, {Dultzin}, {Negrete}, {Mart{\'{\i}}nez-Aldama}, {Bon}, {Bon}, \&
  {Stirpe}}]{2018marziani1}
{Marziani}, P., {del Olmo}, A., {D'Onofrio}, M., {et~al.} 2018{\natexlab{a}},
  in Proceedings of Science, vol. Revisiting narrow-line Seyfert 1 galaxies and
  their place in the Universe, 2

\bibitem[{{Marziani} {et~al.}(2018{\natexlab{b}}){Marziani}, {Dultzin},
  {Sulentic}, {Del Olmo}, {Negrete}, {Mart{\'\i}nez-Aldama}, {D'Onofrio},
  {Bon}, {Bon}, \& {Stirpe}}]{2018marziani2}
{Marziani}, P., {Dultzin}, D., {Sulentic}, J.~W., {et~al.} 2018{\natexlab{b}},
  Frontiers in Astronomy and Space Sciences, 5, 6

\bibitem[{{Marziani} {et~al.}(2003){Marziani}, {Zamanov}, {Sulentic}, \&
  {Calvani}}]{2003marziani1}
{Marziani}, P., {Zamanov}, R.~K., {Sulentic}, J.~W., \& {Calvani}, M. 2003,
  \mnras, 345, 1133

\bibitem[{{Mathur}(2000)}]{2000mathur1}
{Mathur}, S. 2000, \mnras, 314, L17

\bibitem[{{Mathur} {et~al.}(2012){Mathur}, {Fields}, {Peterson}, \&
  {Grupe}}]{2012mathur1}
{Mathur}, S., {Fields}, D., {Peterson}, B.~M., \& {Grupe}, D. 2012, \apj, 754,
  146

\bibitem[{{Meena} {et~al.}(2021){Meena}, {Crenshaw}, {Schmitt}, {Revalski},
  {Fischer}, {Polack}, {Kraemer}, \& {Dashtamirova}}]{2021meena1}
{Meena}, B., {Crenshaw}, D.~M., {Schmitt}, H.~R., {et~al.} 2021, \apj, 916, 31

\bibitem[{{Miller} {et~al.}(1993){Miller}, {Rawlings}, \&
  {Saunders}}]{1993miller1}
{Miller}, P., {Rawlings}, S., \& {Saunders}, R. 1993, \mnras, 263, 425

\bibitem[{{Moran}(2000)}]{2000moran1}
{Moran}, E.~C. 2000, NewAR, 44, 527

\bibitem[{{Mu{\~n}oz Mar{\'\i}n} {et~al.}(2007){Mu{\~n}oz Mar{\'\i}n},
  {Gonz{\'a}lez Delgado}, {Schmitt}, {Cid Fernandes}, {P{\'e}rez},
  {Storchi-Bergmann}, {Heckman}, \& {Leitherer}}]{2007munozmarin1}
{Mu{\~n}oz Mar{\'\i}n}, V.~M., {Gonz{\'a}lez Delgado}, R.~M., {Schmitt}, H.~R.,
  {et~al.} 2007, \aj, 134, 648

\bibitem[{{Mulchaey} {et~al.}(1996){Mulchaey}, {Wilson}, \&
  {Tsvetanov}}]{1996mulchaey1}
{Mulchaey}, J.~S., {Wilson}, A.~S., \& {Tsvetanov}, Z. 1996, \apjs, 102, 309

\bibitem[{{Nagar} {et~al.}(1999){Nagar}, {Wilson}, {Mulchaey}, \&
  {Gallimore}}]{1999nagar1}
{Nagar}, N.~M., {Wilson}, A.~S., {Mulchaey}, J.~S., \& {Gallimore}, J.~F. 1999,
  \apjs, 120, 209

\bibitem[{{Neumann} {et~al.}(1994){Neumann}, {Reich}, {Fuerst}, {Brinkmann},
  {Reich}, {Siebert}, {Wielebinski}, \& {Truemper}}]{1994neumann1}
{Neumann}, M., {Reich}, W., {Fuerst}, E., {et~al.} 1994, \aaps, 106, 303

\bibitem[{{Norris} {et~al.}(1990){Norris}, {Allen}, {Sramek}, {Kesteven}, \&
  {Troup}}]{1990norris1}
{Norris}, R.~P., {Allen}, D.~A., {Sramek}, R.~A., {Kesteven}, M.~J., \&
  {Troup}, E.~R. 1990, \apj, 359, 291

\bibitem[{{O'Dea}(1998)}]{1998odea1}
{O'Dea}, C.~P. 1998, \pasp, 110, 493

\bibitem[{{O'Dea} \& {Saikia}(2021)}]{2021odea1}
{O'Dea}, C.~P. \& {Saikia}, D.~J. 2021, \aapr, 29, 3

\bibitem[{{Ohta} {et~al.}(2007){Ohta}, {Aoki}, {Kawaguchi}, \&
  {Kiuchi}}]{2007ohta1}
{Ohta}, K., {Aoki}, K., {Kawaguchi}, T., \& {Kiuchi}, G. 2007, \apjs, 169, 1

\bibitem[{{Olgu{\'\i}n-Iglesias} {et~al.}(2020){Olgu{\'\i}n-Iglesias},
  {Kotilainen}, \& {Chavushyan}}]{2020olguiniglesias1}
{Olgu{\'\i}n-Iglesias}, A., {Kotilainen}, J., \& {Chavushyan}, V. 2020, \mnras,
  492, 1450

\bibitem[{{Olgu{\'{\i}}n-Iglesias} {et~al.}(2017){Olgu{\'{\i}}n-Iglesias},
  {Kotilainen}, {Le{\'o}n Tavares}, {Chavushyan}, \&
  {A{\~n}orve}}]{2017olguiniglesias1}
{Olgu{\'{\i}}n-Iglesias}, A., {Kotilainen}, J.~K., {Le{\'o}n Tavares}, J.,
  {Chavushyan}, V., \& {A{\~n}orve}, C. 2017, \mnras, 467, 3712

\bibitem[{{Orienti} \& {Prieto}(2010)}]{2010orienti1}
{Orienti}, M. \& {Prieto}, M.~A. 2010, \mnras, 401, 2599

\bibitem[{{Oshlack} {et~al.}(2001){Oshlack}, {Webster}, \&
  {Whiting}}]{2001oshlack1}
{Oshlack}, A.~Y.~K.~N., {Webster}, R.~L., \& {Whiting}, M.~T. 2001, \apj, 558,
  578

\bibitem[{{Osterbrock} \& {Pogge}(1985)}]{1985osterbrock1}
{Osterbrock}, D.~E. \& {Pogge}, R.~W. 1985, \apj, 297, 166

\bibitem[{{Padovani}(2017)}]{2017padovani1}
{Padovani}, P. 2017, Nature Astronomy, 1, 0194

\bibitem[{{Panda} {et~al.}(2018){Panda}, {Czerny}, {Adhikari}, {Hryniewicz},
  {Wildy}, {Kuraszkiewicz}, \& {{\'S}niegowska}}]{2018panda1}
{Panda}, S., {Czerny}, B., {Adhikari}, T.~P., {et~al.} 2018, \apj, 866, 115

\bibitem[{{Panessa} {et~al.}(2019){Panessa}, {Baldi}, {Laor}, {Padovani},
  {Behar}, \& {McHardy}}]{2019panessa1}
{Panessa}, F., {Baldi}, R.~D., {Laor}, A., {et~al.} 2019, Nature Astronomy, 3,
  387

\bibitem[{{Parker} {et~al.}(2020){Parker}, {Matzeu}, {Alston}, {Fabian},
  {Lobban}, {Miniutti}, {Pinto}, {Santos-Lle{\'o}}, \&
  {Schartel}}]{2020parker1}
{Parker}, M.~L., {Matzeu}, G.~A., {Alston}, W.~N., {et~al.} 2020, \mnras, 498,
  L140

\bibitem[{{Parker} {et~al.}(2018){Parker}, {Matzeu}, {Guainazzi},
  {Kalfountzou}, {Miniutti}, {Santos-Lle{\'o}}, \& {Schartel}}]{2018parker1}
{Parker}, M.~L., {Matzeu}, G.~A., {Guainazzi}, M., {et~al.} 2018, \mnras, 480,
  2365

\bibitem[{{Parra} {et~al.}(2010){Parra}, {Conway}, {Aalto}, {Appleton},
  {Norris}, {Pihlstr{\"o}m}, \& {Kewley}}]{2010parra1}
{Parra}, R., {Conway}, J.~E., {Aalto}, S., {et~al.} 2010, \apj, 720, 555

\bibitem[{{Peng} {et~al.}(2014){Peng}, {Chen}, {Gu}, \& {Zhang}}]{2014peng1}
{Peng}, Z.-X., {Chen}, Y.-M., {Gu}, Q.-S., \& {Zhang}, K. 2014, Research in
  Astronomy and Astrophysics, 14, 913

\bibitem[{{Peruzzi} {et~al.}(2021){Peruzzi}, {Pasquato}, {Ciroi}, {Berton},
  {Marziani}, \& {Nardini}}]{2021peruzzi1}
{Peruzzi}, T., {Pasquato}, M., {Ciroi}, S., {et~al.} 2021, \aap, 652, A19

\bibitem[{{Peterson}(2011)}]{2011peterson1}
{Peterson}, B.~M. 2011, ArXiv e-prints [\eprint[arXiv]{1109.4181}]

\bibitem[{{Petrov}(2013)}]{2013petrov1}
{Petrov}, L. 2013, \aj, 146, 5

\bibitem[{{Planck Collaboration} {et~al.}(2013){Planck Collaboration}, {Ade},
  {Aghanim}, {Arg{\"u}eso}, {Arnaud}, {Ashdown}, {Atrio-Barandela}, {Aumont},
  {Baccigalupi}, {Balbi}, {Banday}, {Barreiro}, {Battaner}, {Benabed},
  {Beno{\^\i}t}, {Bernard}, {Bersanelli}, {Bethermin}, {Bhatia}, {Bonaldi},
  {Bond}, {Borrill}, {Bouchet}, {Burigana}, {Cabella}, {Cardoso}, {Catalano},
  {Cay{\'o}n}, {Chamballu}, {Chary}, {Chen}, {Chiang}, {Christensen},
  {Clements}, {Colafrancesco}, {Colombi}, {Colombo}, {Coulais}, {Crill},
  {Cuttaia}, {Danese}, {Davis}, {de Bernardis}, {de Gasperis}, {de Zotti},
  {Delabrouille}, {Dickinson}, {Diego}, {Dole}, {Donzelli}, {Dor{\'e}},
  {D{\"o}rl}, {Douspis}, {Dupac}, {Efstathiou}, {En{\ss}lin}, {Eriksen},
  {Finelli}, {Forni}, {Fosalba}, {Frailis}, {Franceschi}, {Galeotta}, {Ganga},
  {Giard}, {Giardino}, {Giraud-H{\'e}raud}, {Gonz{\'a}lez-Nuevo}, {G{\'o}rski},
  {Gregorio}, {Gruppuso}, {Hansen}, {Harrison}, {Henrot-Versill{\'e}},
  {Hern{\'a}ndez-Monteagudo}, {Herranz}, {Hildebrandt}, {Hivon}, {Hobson},
  {Holmes}, {Jaffe}, {Jaffe}, {Jagemann}, {Jones}, {Juvela}, {Keih{\"a}nen},
  {Kisner}, {Kneissl}, {Knoche}, {Knox}, {Kunz}, {Kurinsky}, {Kurki-Suonio},
  {Lagache}, {L{\"a}hteenm{\"a}ki}, {Lamarre}, {Lasenby}, {Lawrence},
  {Leonardi}, {Lilje}, {L{\'o}pez-Caniego}, {Mac{\'\i}as-P{\'e}rez}, {Maino},
  {Mandolesi}, {Maris}, {Marshall}, {Mart{\'\i}nez-Gonz{\'a}lez}, {Masi},
  {Massardi}, {Matarrese}, {Mazzotta}, {Melchiorri}, {Mendes}, {Mennella},
  {Mitra}, {Miville-Desch{\`e}nes}, {Moneti}, {Montier}, {Morgante},
  {Mortlock}, {Munshi}, {Murphy}, {Naselsky}, {Nati}, {Natoli},
  {N{\o}rgaard-Nielsen}, {Noviello}, {Novikov}, {Novikov}, {Osborne}, {Pajot},
  {Paladini}, {Paoletti}, {Partridge}, {Pasian}, {Patanchon}, {Perdereau},
  {Perotto}, {Perrotta}, {Piacentini}, {Piat}, {Pierpaoli}, {Plaszczynski},
  {Pointecouteau}, {Polenta}, {Ponthieu}, {Popa}, {Poutanen}, {Pratt},
  {Prunet}, {Puget}, {Rachen}, {Reach}, {Rebolo}, {Reinecke}, {Renault},
  {Ricciardi}, {Riller}, {Ristorcelli}, {Rocha}, {Rosset}, {Rowan-Robinson},
  {Rubi{\~n}o-Mart{\'\i}n}, {Rusholme}, {Sajina}, {Sandri}, {Savini}, {Scott},
  {Smoot}, {Starck}, {Sudiwala}, {Suur-Uski}, {Sygnet}, {Tauber}, {Terenzi},
  {Toffolatti}, {Tomasi}, {Tristram}, {Tucci}, {T{\"u}rler}, {Valenziano}, {Van
  Tent}, {Vielva}, {Villa}, {Vittorio}, {Wade}, {Wandelt}, {White}, {Yvon},
  {Zacchei}, \& {Zonca}}]{2013planck1}
{Planck Collaboration}, {Ade}, P.~A.~R., {Aghanim}, N., {et~al.} 2013, \aap,
  550, A133

\bibitem[{{Popescu} {et~al.}(2011){Popescu}, {Tuffs}, {Dopita}, {Fischera},
  {Kylafis}, \& {Madore}}]{2011popescu1}
{Popescu}, C.~C., {Tuffs}, R.~J., {Dopita}, M.~A., {et~al.} 2011, \aap, 527,
  A109

\bibitem[{{Popovi{\'c}} {et~al.}(2009){Popovi{\'c}}, {Smirnova},
  {Kova{\v{c}}evi{\'c}}, {Moiseev}, \& {Afanasiev}}]{2009popovic1}
{Popovi{\'c}}, L.~{\v{C}}., {Smirnova}, A.~A., {Kova{\v{c}}evi{\'c}}, J.,
  {Moiseev}, A.~V., \& {Afanasiev}, V.~L. 2009, \aj, 137, 3548

\bibitem[{{Pounds} {et~al.}(1995){Pounds}, {Done}, \& {Osborne}}]{1995pounds1}
{Pounds}, K.~A., {Done}, C., \& {Osborne}, J.~P. 1995, \mnras, 277, L5

\bibitem[{{Pounds} \& {King}(2013)}]{2013pounds1}
{Pounds}, K.~A. \& {King}, A.~R. 2013, \mnras, 433, 1369

\bibitem[{{Proga}(2007)}]{2007proga1}
{Proga}, D. 2007, in Astronomical Society of the Pacific Conference Series,
  Vol. 373, The Central Engine of Active Galactic Nuclei, ed. L.~C. {Ho} \&
  J.~W. {Wang}, 267

\bibitem[{{Rakshit} {et~al.}(2021){Rakshit}, {Schramm}, {Stalin}, {Tanaka},
  {Paliya}, {Pal}, {Kotilainen}, \& {Shin}}]{2021rakshit1}
{Rakshit}, S., {Schramm}, M., {Stalin}, C.~S., {et~al.} 2021, \mnras
  [\eprint[arXiv]{2103.16521}]

\bibitem[{{Rakshit} {et~al.}(2018){Rakshit}, {Stalin}, {Hota}, \&
  {Konar}}]{2018rakshit1}
{Rakshit}, S., {Stalin}, C.~S., {Hota}, A., \& {Konar}, C. 2018, \apj, 869, 173

\bibitem[{{Rau} \& {Cornwell}(2011)}]{2011rau1}
{Rau}, U. \& {Cornwell}, T.~J. 2011, \aap, 532, A71

\bibitem[{{Rodr{\'\i}guez-Ardila} {et~al.}(2005){Rodr{\'\i}guez-Ardila},
  {Contini}, \& {Viegas}}]{2005rodriguesardila1}
{Rodr{\'\i}guez-Ardila}, A., {Contini}, M., \& {Viegas}, S.~M. 2005, \mnras,
  357, 220

\bibitem[{{Rodr{\'\i}guez-Ardila} \& {Mazzalay}(2006)}]{2006rodriguezardila1}
{Rodr{\'\i}guez-Ardila}, A. \& {Mazzalay}, X. 2006, \mnras, 367, L57

\bibitem[{{Rodr{\'\i}guez-Ardila} \& {Viegas}(2003)}]{2003rodriguezardila1}
{Rodr{\'\i}guez-Ardila}, A. \& {Viegas}, S.~M. 2003, \mnras, 340, L33

\bibitem[{{Romano} {et~al.}(2018){Romano}, {Vercellone}, {Foschini},
  {Tavecchio}, {Landoni}, \& {Kn{\"o}dlseder}}]{2018romano1}
{Romano}, P., {Vercellone}, S., {Foschini}, L., {et~al.} 2018, \mnras, 481,
  5046

\bibitem[{{Rosario} {et~al.}(2013){Rosario}, {Burtscher}, {Davies}, {Genzel},
  {Lutz}, \& {Tacconi}}]{2013rosario1}
{Rosario}, D.~J., {Burtscher}, L., {Davies}, R., {et~al.} 2013, \apj, 778, 94

\bibitem[{{Ruschel-Dutra} {et~al.}(2017){Ruschel-Dutra}, {Rodr{\'\i}guez
  Espinosa}, {Gonz{\'a}lez Mart{\'\i}n}, {Pastoriza}, \&
  {Riffel}}]{2017ruscheldutra1}
{Ruschel-Dutra}, D., {Rodr{\'\i}guez Espinosa}, J.~M., {Gonz{\'a}lez
  Mart{\'\i}n}, O., {Pastoriza}, M., \& {Riffel}, R. 2017, \mnras, 466, 3353

\bibitem[{{Saikia} {et~al.}(2018){Saikia}, {K{\"o}rding}, {Coppejans},
  {Falcke}, {Williams}, {Baldi}, {Mchardy}, \& {Beswick}}]{2018saikia1}
{Saikia}, P., {K{\"o}rding}, E., {Coppejans}, D.~L., {et~al.} 2018, \aap, 616,
  A152

\bibitem[{{Sanfrutos} {et~al.}(2018){Sanfrutos}, {Longinotti}, {Krongold},
  {Guainazzi}, \& {Panessa}}]{2018sanfrutos1}
{Sanfrutos}, M., {Longinotti}, A.~L., {Krongold}, Y., {Guainazzi}, M., \&
  {Panessa}, F. 2018, \apj, 868, 111

\bibitem[{{Sani} {et~al.}(2010){Sani}, {Lutz}, {Risaliti}, {Netzer}, {Gallo},
  {Trakhtenbrot}, {Sturm}, \& {Boller}}]{2010sani1}
{Sani}, E., {Lutz}, D., {Risaliti}, G., {et~al.} 2010, \mnras, 403, 1246

\bibitem[{{Sargsyan} {et~al.}(2012){Sargsyan}, {Lebouteiller}, {Weedman},
  {Spoon}, {Bernard-Salas}, {Engels}, {Stacey}, {Houck}, {Barry}, {Miles}, \&
  {Samsonyan}}]{2012sargsyan1}
{Sargsyan}, L., {Lebouteiller}, V., {Weedman}, D., {et~al.} 2012, \apj, 755,
  171

\bibitem[{{Sargsyan} {et~al.}(2014){Sargsyan}, {Samsonyan}, {Lebouteiller},
  {Weedman}, {Barry}, {Bernard-Salas}, {Houck}, \& {Spoon}}]{2014sargsyan1}
{Sargsyan}, L., {Samsonyan}, A., {Lebouteiller}, V., {et~al.} 2014, \apj, 790,
  15

\bibitem[{{Sargsyan} \& {Weedman}(2009)}]{2009sargsyan1}
{Sargsyan}, L.~A. \& {Weedman}, D.~W. 2009, \apj, 701, 1398

\bibitem[{{Scheuer} \& {Readhead}(1979)}]{1979scheuer1}
{Scheuer}, P.~A.~G. \& {Readhead}, A.~C.~S. 1979, \nat, 277, 182

\bibitem[{{Schmitt} {et~al.}(2001{\natexlab{a}}){Schmitt}, {Antonucci},
  {Ulvestad}, {Kinney}, {Clarke}, \& {Pringle}}]{2001schmitt1}
{Schmitt}, H.~R., {Antonucci}, R.~R.~J., {Ulvestad}, J.~S., {et~al.}
  2001{\natexlab{a}}, \apj, 555, 663

\bibitem[{{Schmitt} {et~al.}(2003){Schmitt}, {Donley}, {Antonucci},
  {Hutchings}, \& {Kinney}}]{2003schmitt1}
{Schmitt}, H.~R., {Donley}, J.~L., {Antonucci}, R.~R.~J., {Hutchings}, J.~B.,
  \& {Kinney}, A.~L. 2003, \apjs, 148, 327

\bibitem[{{Schmitt} {et~al.}(2001{\natexlab{b}}){Schmitt}, {Ulvestad},
  {Antonucci}, \& {Kinney}}]{2001schmitt3}
{Schmitt}, H.~R., {Ulvestad}, J.~S., {Antonucci}, R.~R.~J., \& {Kinney}, A.~L.
  2001{\natexlab{b}}, \apjs, 132, 199

\bibitem[{{Sch{\"o}nell} {et~al.}(2014){Sch{\"o}nell}, {Riffel},
  {Storchi-Bergmann}, \& {Winge}}]{2014schonell1}
{Sch{\"o}nell}, A.~J., {Riffel}, R.~A., {Storchi-Bergmann}, T., \& {Winge}, C.
  2014, \mnras, 445, 414

\bibitem[{{Seyfert}(1943)}]{1943seyfert1}
{Seyfert}, C.~K. 1943, \apj, 97, 28

\bibitem[{{Shangguan} {et~al.}(2020){Shangguan}, {Ho}, {Bauer}, {Wang}, \&
  {Treister}}]{2020shangguan1}
{Shangguan}, J., {Ho}, L.~C., {Bauer}, F.~E., {Wang}, R., \& {Treister}, E.
  2020, \apj, 899, 112

\bibitem[{{Shen} \& {Ho}(2014)}]{2014shen1}
{Shen}, Y. \& {Ho}, L.~C. 2014, \nat, 513, 210

\bibitem[{{Shi} {et~al.}(2007){Shi}, {Ogle}, {Rieke}, {Antonucci}, {Hines},
  {Smith}, {Low}, {Bouwman}, \& {Willmer}}]{2007shi1}
{Shi}, Y., {Ogle}, P., {Rieke}, G.~H., {et~al.} 2007, \apj, 669, 841

\bibitem[{{Shirazi} \& {Brinchmann}(2012)}]{2012shirazi1}
{Shirazi}, M. \& {Brinchmann}, J. 2012, \mnras, 421, 1043

\bibitem[{{Shuder} \& {Osterbrock}(1981)}]{1981shuder1}
{Shuder}, J.~M. \& {Osterbrock}, D.~E. 1981, \apj, 250, 55

\bibitem[{{Silpa} {et~al.}(2020){Silpa}, {Kharb}, {Ho}, {Ishwara-Chandra},
  {Jarvis}, \& {Harrison}}]{2020silpa1}
{Silpa}, S., {Kharb}, P., {Ho}, L.~C., {et~al.} 2020, \mnras, 499, 5826

\bibitem[{{Smethurst} {et~al.}(2019){Smethurst}, {Simmons}, {Lintott}, \&
  {Shanahan}}]{2019smethurst1}
{Smethurst}, R.~J., {Simmons}, B.~D., {Lintott}, C.~J., \& {Shanahan}, J. 2019,
  \mnras, 489, 4016

\bibitem[{{Snellen} {et~al.}(2004){Snellen}, {Mack}, {Schilizzi}, \&
  {Tschager}}]{2004snellen1}
{Snellen}, I.~A.~G., {Mack}, K.~H., {Schilizzi}, R.~T., \& {Tschager}, W. 2004,
  \mnras, 348, 227

\bibitem[{{Stern} \& {Laor}(2013)}]{2013stern1}
{Stern}, J. \& {Laor}, A. 2013, \mnras, 431, 836

\bibitem[{{Sulentic} {et~al.}(2000{\natexlab{a}}){Sulentic}, {Marziani}, \&
  {Dultzin-Hacyan}}]{2000sulentic2}
{Sulentic}, J.~W., {Marziani}, P., \& {Dultzin-Hacyan}, D. 2000{\natexlab{a}},
  \araa, 38, 521

\bibitem[{{Sulentic} {et~al.}(2000{\natexlab{b}}){Sulentic}, {Zwitter},
  {Marziani}, \& {Dultzin-Hacyan}}]{2000sulentic1}
{Sulentic}, J.~W., {Zwitter}, T., {Marziani}, P., \& {Dultzin-Hacyan}, D.
  2000{\natexlab{b}}, \apjl, 536, L5

\bibitem[{{Tan} {et~al.}(2019){Tan}, {Gao}, {Kohno}, {Xia}, {Omont}, {Hao},
  {Mao}, {Daddi}, {Shi}, {Zhao}, \& {Cox}}]{2019tan1}
{Tan}, Q.-H., {Gao}, Y., {Kohno}, K., {et~al.} 2019, \apj, 887, 24

\bibitem[{{Taniguchi}(1999)}]{1999taniguchi1}
{Taniguchi}, Y. 1999, \apj, 524, 65

\bibitem[{{Tarchi} {et~al.}(2011){Tarchi}, {Castangia}, {Columbano}, {Panessa},
  \& {Braatz}}]{2011tarchi1}
{Tarchi}, A., {Castangia}, P., {Columbano}, A., {Panessa}, F., \& {Braatz},
  J.~A. 2011, \aap, 532, A125

\bibitem[{{Teng} {et~al.}(2013){Teng}, {Veilleux}, \& {Baker}}]{2013teng1}
{Teng}, S.~H., {Veilleux}, S., \& {Baker}, A.~J. 2013, \apj, 765, 95

\bibitem[{{Terrazas} {et~al.}(2017){Terrazas}, {Bell}, {Woo}, \&
  {Henriques}}]{2017terrazas1}
{Terrazas}, B.~A., {Bell}, E.~F., {Woo}, J., \& {Henriques}, B. M.~B. 2017,
  \apj, 844, 170

\bibitem[{{Thean} {et~al.}(2000){Thean}, {Pedlar}, {Kukula}, {Baum}, \&
  {O'Dea}}]{2000thean1}
{Thean}, A., {Pedlar}, A., {Kukula}, M.~J., {Baum}, S.~A., \& {O'Dea}, C.~P.
  2000, \mnras, 314, 573

\bibitem[{{Tombesi} {et~al.}(2010){Tombesi}, {Cappi}, {Reeves}, {Palumbo},
  {Yaqoob}, {Braito}, \& {Dadina}}]{2010tombesi1}
{Tombesi}, F., {Cappi}, M., {Reeves}, J.~N., {et~al.} 2010, \aap, 521, A57

\bibitem[{{Ulvestad} {et~al.}(1995){Ulvestad}, {Antonucci}, \&
  {Goodrich}}]{1995ulvestad1}
{Ulvestad}, J.~S., {Antonucci}, R. R.~J., \& {Goodrich}, R.~W. 1995, \aj, 109,
  81

\bibitem[{{Ulvestad} \& {Wilson}(1984)}]{1984ulvestad1}
{Ulvestad}, J.~S. \& {Wilson}, A.~S. 1984, \apj, 285, 439

\bibitem[{{Urrutia} {et~al.}(2008){Urrutia}, {Lacy}, \&
  {Becker}}]{2008urrutia1}
{Urrutia}, T., {Lacy}, M., \& {Becker}, R.~H. 2008, \apj, 674, 80

\bibitem[{{Wang} {et~al.}(2016){Wang}, {Du}, {Hu}, {Bai}, {Wang}, {Yi}, {Wang},
  {Zhang}, {Xin}, {Lun}, {Chang}, \& {Fan}}]{2016wang1}
{Wang}, F., {Du}, P., {Hu}, C., {et~al.} 2016, \apj, 824, 149

\bibitem[{{Whittle}(1992)}]{1992whittle1}
{Whittle}, M. 1992, \apjs, 79, 49

\bibitem[{{Wiegert} {et~al.}(2015){Wiegert}, {Irwin}, {Miskolczi}, {Schmidt},
  {Mora}, {Damas-Segovia}, {Stein}, {English}, {Rand}, {Santistevan},
  {Walterbos}, {Krause}, {Beck}, {Dettmar}, {Kepley}, {Wezgowiec}, {Wang},
  {Heald}, {Li}, {MacGregor}, {Johnson}, {Strong}, {DeSouza}, \&
  {Porter}}]{2015wiegert1}
{Wiegert}, T., {Irwin}, J., {Miskolczi}, A., {et~al.} 2015, \aj, 150, 81

\bibitem[{{Wright} {et~al.}(2010){Wright}, {Eisenhardt}, {Mainzer}, {Ressler},
  {Cutri}, {Jarrett}, {Kirkpatrick}, {Padgett}, {McMillan}, {Skrutskie},
  {Stanford}, {Cohen}, {Walker}, {Mather}, {Leisawitz}, {Gautier}, {McLean},
  {Benford}, {Lonsdale}, {Blain}, {Mendez}, {Irace}, {Duval}, {Liu}, {Royer},
  {Heinrichsen}, {Howard}, {Shannon}, {Kendall}, {Walsh}, {Larsen}, {Cardon},
  {Schick}, {Schwalm}, {Abid}, {Fabinsky}, {Naes}, \& {Tsai}}]{2010wright1}
{Wright}, E.~L., {Eisenhardt}, P.~R.~M., {Mainzer}, A.~K., {et~al.} 2010, \aj,
  140, 1868

\bibitem[{{Xie} {et~al.}(2018){Xie}, {Ho}, {Li}, \& {Shangguan}}]{2018xie1}
{Xie}, Y., {Ho}, L.~C., {Li}, A., \& {Shangguan}, J. 2018, \apj, 860, 154

\bibitem[{{Yang} {et~al.}(2020){Yang}, {Yao}, {Yang}, {Ho}, {An}, {Wang},
  {Baan}, {Gu}, {Liu}, {Yang}, \& {Joshi}}]{2020yang1}
{Yang}, X., {Yao}, S., {Yang}, J., {et~al.} 2020, \apj, 904, 200

\bibitem[{{Yao} \& {Komossa}(2021)}]{2021yao1}
{Yao}, S. \& {Komossa}, S. 2021, \mnras, 501, 1384

\bibitem[{{Young} {et~al.}(2014){Young}, {Eracleous}, {Shemmer}, {Netzer},
  {Gronwall}, {Lutz}, {Ciardullo}, \& {Sturm}}]{2014young1}
{Young}, J.~E., {Eracleous}, M., {Shemmer}, O., {et~al.} 2014, \mnras, 438, 217

\bibitem[{{Yuan} {et~al.}(2010){Yuan}, {Liu}, {Zhou}, \& {Wang}}]{2010yuan1}
{Yuan}, W., {Liu}, B.~F., {Zhou}, H., \& {Wang}, T.~G. 2010, \apj, 723, 508

\bibitem[{{Yuan} {et~al.}(2008){Yuan}, {Zhou}, {Komossa}, {Dong}, {Wang}, {Lu},
  \& {Bai}}]{2008yuan1}
{Yuan}, W., {Zhou}, H.~Y., {Komossa}, S., {et~al.} 2008, \apj, 685, 801

\bibitem[{{Zhang} {et~al.}(2013){Zhang}, {Wang}, {Yan}, \& {Dong}}]{2013zhang1}
{Zhang}, K., {Wang}, T.-G., {Yan}, L., \& {Dong}, X.-B. 2013, \apj, 768, 22

\bibitem[{{Zhang} {et~al.}(2017{\natexlab{a}}){Zhang}, {Zhang}, {Yan}, {Fan},
  \& {Liu}}]{2017zhang1}
{Zhang}, P., {Zhang}, P.-f., {Yan}, J.-z., {Fan}, Y.-z., \& {Liu}, Q.-z.
  2017{\natexlab{a}}, \apj, 849, 9

\bibitem[{{Zhang} {et~al.}(2017{\natexlab{b}}){Zhang}, {Zhou}, {Shi}, {Liu},
  {Pan}, {Jiang}, {Ji}, {Jiang}, \& {Wang}}]{2017zhang2}
{Zhang}, S., {Zhou}, H., {Shi}, X., {et~al.} 2017{\natexlab{b}}, \apj, 845, 126

\bibitem[{{Zheng} {et~al.}(2002){Zheng}, {Xia}, {Mao}, {Wu}, \&
  {Deng}}]{2002zheng1}
{Zheng}, X.~Z., {Xia}, X.~Y., {Mao}, S., {Wu}, H., \& {Deng}, Z.~G. 2002, \aj,
  124, 18

\bibitem[{{Zhuang} {et~al.}(2018){Zhuang}, {Ho}, \& {Shangguan}}]{2018zhuang1}
{Zhuang}, M.-Y., {Ho}, L.~C., \& {Shangguan}, J. 2018, \apj, 862, 118

\end{thebibliography}

\newpage
\appendix

\onecolumn
\section{Tables}

\begin{table*}[!h]
\caption[]{Basic data of the sample.}
\centering
\begin{tabular}{l l l l l l l}
\hline\hline
Short name & NED Alias                & RA            & Dec           & $z$   & Scale   & Morph. \\ 
          &                          & (hh mm ss.ss) & (dd mm ss.ss) &       & (kpc/") & type  \\ \hline
J0347+0105 & IRAS 03450+0055          & 03 47 40.19    & 01 05 14.04  & 0.031 & 0.620   & E     \\
J0629-0545 & IRAS 06269-0543          & 06 29 24.77    & -05 45 26.34 & 0.117 & 2.116   & I     \\
J0632+6340 & UGC 3478                 & 06 32 47.17    & 63 40 25.29  & 0.013 & 0.266   & E     \\
J0706+3901 & FBQS J0706+3901          & 07 06 25.14    & 39 01 51.74  & 0.086 & 1.612   & I     \\
J0713+3820 & FBQS J0713+3820          & 07 13 40.28    & 38 20 39.92  & 0.123 & 2.210   & E     \\
J0804+3853 & FBQS J0804+3853          & 08 04 09.25    & 38 53 48.80  & 0.212 & 3.453   & E     \\
J0806+7248 & RGB J0806+728            & 08 06 38.96	   & 72 48 20.45  & 0.098 & 1.812   & I     \\
J0814+5609 & SDSS J081432.11+560956.6 & 08 14 32.12	   & 56 09 56.68  & 0.510 & 6.168   & I     \\
J0913+3658 & RX J0913.2+3658          & 09 13 13.73	   & 36 58 17.25  & 0.107 & 1.958   & I     \\
J0925+5217 & Mrk 110                  & 09 25 12.85	   & 52 17 10.40  & 0.035 & 0.697   & E     \\
J0926+1244 & Mrk 705                  & 09 26 03.27	   & 12 44 03.75  & 0.029 & 0.581   & I     \\
J0937+3615 & SDSS J093703.03+361537.2 & 09 37 03.03	   & 36 15 37.24  & 0.180 & 3.035   & E     \\
J0952-0136 & Mrk 1239                 & 09 52 19.09	   & -01 36 43.57 & 0.020 & 0.405   & I     \\
J1034+3938 & KUG 1031+398             & 10 34 38.59	   & 39 38 28.16  & 0.042 & 0.829   & I     \\
J1038+4227 & SDSS J103859.58+422742.3 & 10 38 59.58	   & 42 27 42.22  & 0.220 & 3.552   & E     \\
J1047+4725 & SDSS J104732.68+472532.0 & 10 47 32.67    & 47 25 32.04  & 0.799 & 7.505   & E     \\
J1048+2222 & SDSS J104816.57+222238.9 & 10 48 16.58    & 22 22 38.99  & 0.330 & 4.752   & I     \\
J1102+2239 & SDSS J110223.38+223920.7 & 11 02 23.38    & 22 39 20.70  & 0.453 & 5.781   & I     \\
J1110+3653 & SDSS J111005.03+365336.3 & 11 10 05.04	   & 36 53 36.27  & 0.629 & 6.830   & I     \\
J1121+5351 & SBS 1118+541             & 11 21 08.58	   & 53 51 21.03  & 0.103 & 1.893   & I     \\
J1138+3653 & SDSS J113824.54+365327.1 & 11 38 24.54    & 36 53 27.16  & 0.356 & 4.994   & I     \\
J1159+2838 & SDSS J115917.32+283814.5 & 11 59 17.32	   & 28 38 14.58  & 0.210 & 3.427   & I     \\
J1203+4431 & NGC 4051                 & 12 03 09.61	   & 44 31 52.72  & 0.002 & 0.041   & E     \\
J1209+3217 & RX J1209.7+3217          & 12 09 45.21	   & 32 17 01.13  & 0.144 & 2.527   & I     \\
J1215+5442 & SBS 1213+549A            & 12 15 49.44	   & 54 42 23.97  & 0.150 & 2.614   & I     \\
J1218+2948 & Mrk 766                  & 12 18 26.52	   & 29 48 46.51  & 0.013 & 0.266   & E     \\
J1227+3214 & SDSS J122749.14+321458.9 & 12 27 49.14	   & 32 14 58.96  & 0.136 & 2.408   & I     \\
J1242+3317 & WAS 61                   & 12 42 10.62	   & 33 17 02.73  & 0.044 & 0.866   & E     \\
J1246+0222 & PG 1244+026              & 12 46 35.25	   & 02 22 08.72  & 0.048 & 0.941   & E     \\
J1302+1624 & Mrk 783                  & 13 02 58.84    & 16 24 27.81  & 0.067 & 1.284   & E     \\
J1305+5116 & SDSS J130522.74+511640.2 & 13 05 22.74    & 51 16 40.15  & 0.788 & 7.469   & E     \\
J1317+6010 & SBS 1315+604             & 13 17 50.33    & 60 10 40.70  & 0.137 & 2.423   & E     \\
J1337+2423 & IRAS 13349+2438          & 13 37 18.71	   & 24 23 03.30  & 0.108 & 1.974   & I     \\
J1348+2622 & SDSS J134834.28+262205.9 & 13 48 34.28	   & 26 22 05.95  & 0.917 & 7.832   & I     \\
J1358+2658 & SDSS J135845.38+265808.4 & 13 58 45.38	   & 26 58 08.49  & 0.331 & 4.762   & I     \\
J1402+2159 & RX J1402.5+2159          & 14 02 34.47	   & 21 59 51.65  & 0.066 & 1.266   & I     \\
J1536+5433 & Mrk 486                  & 15 36 38.40	   & 54 33 33.23  & 0.039 & 0.772   & I     \\
J1555+1911 & Mrk 291                  & 15 55 07.92	   & 19 11 32.82  & 0.035 & 0.697   & E     \\
J1559+3501 & Mrk 493                  & 15 59 09.62	   & 35 01 47.54  & 0.031 & 0.620   & E     \\
J1633+4718 & SDSS J163323.58+471858.9 & 16 33 23.58	   & 47 18 58.96  & 0.116 & 2.101   & I     \\
J1703+4540 & SDSS J170330.38+454047.1 & 17 03 30.39	   & 45 40 47.19  & 0.060 & 1.159   & I     \\
J1713+3523 & FBQS J1713+3523          & 17 13 04.46	   & 35 23 33.55  & 0.083 & 1.561   & I     \\
J2242+2943 & Ark 564                  & 22 42 39.34	   & 29 43 31.09  & 0.025 & 0.504   & E     \\
J2314+2243 & RX J2314.9+2243          & 23 14 55.89	   & 22 43 22.79  & 0.169 & 2.884   & I     \\

\hline
\end{tabular}
\tablefoot{(1) Short name, (2) alias, (3), (4) right ascension and declination (J2000) from the JVLA radio maps, (5) redshift, (6) scale, (7) radio morphology type: I = intermediate, E = extended.}
\label{tab:basicdata}
\end{table*}

\begin{table*}[!h]
\caption[]{Basic data of the sample.}
\centering
\begin{tabular}{l l l l}
\hline\hline
Short name & log $M_{\textrm{BH}}$ & $L/L_{\textrm{Edd}}$ & log $ \nu L_{\nu, \textrm{int}}$ \\ 
           & ($M_{\odot}$)         & (erg s$^{-1}$)       & (erg s$^{-1}$)  \\ \hline
J0347+0105 &    7.26 & 	0.07 & 	39.17 $\pm$ 0.01  \\
J0629-0545 &    7.80 & 	0.11 & 	40.43 $\pm$ 0.01  \\
J0632+6340 & 	6.48 & 	0.02 & 	37.76 $\pm$ 0.01  \\
J0706+3901 & 	7.04 & 	0.04 & 	39.33 $\pm$ 0.01  \\
J0713+3820 & 	8.20 & 	0.17 & 	39.82 $\pm$ 0.01  \\
J0804+3853 & 	7.72 & 	0.14 & 	39.69 $\pm$ 0.02  \\
J0806+7248 & 	6.94 & 	0.08 & 	40.17 $\pm$ 0.01  \\
J0814+5609 & 	8.44 & 	0.11 & 	42.19 $\pm$ 0.01  \\
J0913+3658 & 	7.01 & 	0.06 & 	38.66 $\pm$ 0.06  \\
J0925+5217 & 	7.51 & 	0.10 & 	38.72 $\pm$ 0.01  \\
J0926+1244 & 	7.31 & 	0.04 & 	38.52 $\pm$ 0.01  \\
J0937+3615 & 	7.58 & 	0.05 & 	39.71 $\pm$ 0.02  \\
J0952-0136 & 	7.13 & 	0.05 & 	39.04 $\pm$ 0.01  \\
J1034+3938 & 	6.03 & 	0.17 & 	39.24 $\pm$ 0.01  \\
J1038+4227 & 	7.88 & 	0.08 & 	40.67 $\pm$ 0.01  \\
J1047+4725 & 	8.39 & 	0.08 & 	43.77 $\pm$ 0.01  \\
J1048+2222 & 	7.53 & 	0.11 & 	39.75 $\pm$ 0.05  \\
J1102+2239 & 	8.17 & 	0.06 & 	40.47 $\pm$ 0.04  \\
J1110+3653 & 	7.09 & 	0.23 & 	41.89 $\pm$ 0.01  \\
J1121+5351 & 	7.64 & 	0.13 & 	39.25 $\pm$ 0.01  \\
J1138+3653 & 	7.61 & 	0.09 & 	41.02 $\pm$ 0.01  \\
J1159+2838 & 	7.54 & 	0.09 & 	39.76 $\pm$ 0.02  \\
J1203+4431 & 	6.30 & 	0.01 & 	36.46 $\pm$ 0.01  \\
J1209+3217 & 	7.48 & 	0.10 & 	39.31 $\pm$ 0.02  \\
J1215+5442 & 	7.51 & 	0.10 & 	39.30 $\pm$ 0.01  \\
J1218+2948 & 	6.43 & 	0.05 & 	38.47 $\pm$ 0.01  \\
J1227+3214 & 	6.84 & 	0.16 & 	39.94 $\pm$ 0.01  \\
J1242+3317 & 	7.06 & 	0.06 & 	38.74 $\pm$ 0.01  \\
J1246+0222 & 	7.09 & 	0.05 & 	38.31 $\pm$ 0.02  \\
J1302+1624 & 	7.36 & 	0.12 & 	39.91 $\pm$ 0.01  \\
J1305+5116 & 	8.20 & 	0.67 & 	42.91 $\pm$ 0.01  \\
J1317+6010 & 	6.25 & 	0.15 & 	39.42 $\pm$ 0.02  \\
J1337+2423 & 	8.04 & 	0.22 & 	40.26 $\pm$ 0.01  \\
J1348+2622$^a$ & 	--	 & --	 &  40.84 $\pm$ 0.02  \\
J1358+2658 & 	7.84 & 	0.12 & 	40.05 $\pm$ 0.02  \\
J1402+2159 & 	7.15 & 	0.11 & 	38.34 $\pm$ 0.02  \\
J1536+5433 & 	7.34 & 	0.09 & 	37.98 $\pm$ 0.03  \\
J1555+1911 & 	6.50 & 	0.03 & 	37.85 $\pm$ 0.03  \\
J1559+3501 & 	6.86 & 	0.06 & 	38.19 $\pm$ 0.01  \\
J1633+4718 & 	6.91 & 	0.11 & 	40.67 $\pm$ 0.01  \\
J1703+4540 & 	7.73 & 	0.01 & 	40.20 $\pm$ 0.01  \\
J1713+3523 & 	6.69 & 	0.06 & 	39.54 $\pm$ 0.01  \\
J2242+2943 & 	6.67 & 	0.08 & 	38.88 $\pm$ 0.01  \\
J2314+2243 & 	8.00 &  0.17 & 	40.48 $\pm$ 0.01  \\
\hline
\end{tabular}
\tablefoot{(1) Short name, (2) logarithmic black hole mass estimate, (3), Eddington ratio estimate, (4) logarithm of the 5.2~GHz integrated radio luminosity. $^a$ The redshift of this source is too high to measure the H$\beta$ line and derive a black hole mass estimate, or an Eddington ratio.}
\label{tab:b18data}
\end{table*}

\begin{table*}[!h]
\caption[]{Measurements from the normal radio map, and 90 and 60k$\lambda$ tapered maps, and associated rms for our sample.}
\centering
\begin{tabular}{l l l l l l l l}
\hline\hline
Short name & rms       & $S_{\textrm{peak}}$ & $S_{\textrm{int}}$ & rms$_{90\textrm{k}\lambda}$ & $S_{\textrm{90k}\lambda, \textrm{int}}$ & rms$_{60\textrm{k}\lambda}$ & $S_{\textrm{60k}\lambda, \textrm{int}}$ \\
 & ($\mu$Jy beam$^{-1}$) & (mJy beam$^{-1}$) & (mJy)              & ($\mu$Jy beam$^{-1}$)                   & (mJy)                                    & ($\mu$Jy beam$^{-1}$)                   & (mJy)  \\ \hline
J0347+0105 & 14        & 7.54 $\pm$ 0.07   & 11.65 $\pm$ 0.05 & 25 & 12.10 $\pm$ 0.07 & 30 & 12.04 $\pm$	0.08    \\
J0629-0545 & 14        & 12.52	$\pm$ 0.06 & 14.58 $\pm$ 0.05 & 22 & 14.81 $\pm$ 0.06 & 26 & 14.82 $\pm$	0.06    \\
J0632+6340 & 10        & 1.69	$\pm$ 0.03 & 2.83 $\pm$ 0.04  & 14 & 3.02 $\pm$	0.04  & 16 & 3.02 $\pm$	0.04    \\
J0706+3901 & 13        & 1.67	$\pm$ 0.02 & 2.14 $\pm$ 0.03  & 17 & 2.18 $\pm$	0.04  & 21 & 2.21 $\pm$	0.04    \\
J0713+3820 & 13        & 2.32	$\pm$ 0.03 & 3.39 $\pm$ 0.04  & 19 & 3.71 $\pm$	0.06  & 23 & 3.80 $\pm$	0.06    \\
J0804+3853 & 10        & 0.37  $\pm$ 0.02  & 0.81 $\pm$ 0.03  & 18 & 0.74 $\pm$	0.03  & 23 & 0.72 $\pm$	0.04    \\
J0806+7248 & 10        & 10.92	$\pm$ 0.02 & 11.46 $\pm$ 0.03 & 11 & 11.33 $\pm$ 0.03 & 12 & 11.33 $\pm$	0.03    \\
J0814+5609 & 10        & 25.85	$\pm$ 0.02 & 29.27 $\pm$0.06  & 16 & 29.38 $\pm$ 0.06 & 19 & 30.09 $\pm$	0.08    \\
J0913+3658 & 11        & 0.27	$\pm$ 0.01 & 0.30 $\pm$ 0.02  & 17 & 0.41 $\pm$	0.03  & 26 & 0.36 $\pm$	0.04    \\
J0925+5217 & 12        & 1.70  $\pm$ 0.02  & 2.25 $\pm$ 0.03  & 16 & 3.10 $\pm$	0.06  & 21 & 3.11 $\pm$	0.06    \\
J0926+1244 & 11        & 2.81  $\pm$ 0.03  & 3.36 $\pm$ 0.04  & 17 & 3.46 $\pm$	0.05  & 22 & 3.37 $\pm$	0.05    \\
J0937+3615 & 12        & 0.79  $\pm$ 0.01  & 1.03 $\pm$ 0.03  & 15 & 0.94 $\pm$	0.03  & 19 & 0.90 $\pm$	0.03    \\
J0952-0136 & 10        & 17.49 $\pm$ 0.03  & 18.54 $\pm$ 0.04 & 12 & 18.92 $\pm$ 0.04 & 14 & 18.97 $\pm$	0.04    \\
J1034+3938 & 10        & 7.05  $\pm$ 0.02  & 7.37 $\pm$ 0.03  & 12 & 7.63 $\pm$	0.04  & 14 & 7.71 $\pm$	0.04 \\
J1038+4227 & 10        & 2.06  $\pm$ 0.03  & 3.75 $\pm$ 0.07  & 13 & 5.36 $\pm$	0.08  & 17 & 5.22 $\pm$	0.07    \\
J1047+4725 & 35      & 190.90 $\pm$ 5.70 & 381.75 $\pm$ 0.17 & 50  & 381.47 $\pm$ 0.15 & 80 & 381.34 $\pm$ 0.23     \\
J1048+2222 & 8         & 0.27 $\pm$ 0.01   & 0.24 $\pm$ 0.01  & 9  & 0.32 $\pm$	0.02  & 10 & 0.47 $\pm$	0.03     \\
J1102+2239 & 13        & 0.70 $\pm$ 0.01   & 0.70 $\pm$ 0.02  & 36 & 0.58 $\pm$	0.05  & 45 & 1.22 $\pm$	0.10     \\
J1110+3653 & 10        & 8.12 $\pm$ 0.03   & 9.34 $\pm$ 0.05  & 11 & 9.82 $\pm$	0.04  & 12 & 9.93 $\pm$	0.04    \\
J1121+5351 & 11        & 1.02 $\pm$ 0.02   & 1.22 $\pm$ 0.03  & 12 & 1.25 $\pm$	0.03  & 14 & 1.24 $\pm$	0.03    \\
J1138+3653 & 10        & 4.37 $\pm$ 0.01   & 4.64 $\pm$ 0.02  & 12 & 4.63 $\pm$	0.03  & 12 & 4.62 $\pm$	0.03    \\
J1159+2838 & 10        & 0.79 $\pm$ 0.01   & 0.84 $\pm$ 0.02  & 13 & 0.82 $\pm$	0.02  & 14 & 0.79 $\pm$	0.02    \\
J1203+4431 & 12        & 0.78 $\pm$ 0.04   & 5.47 $\pm$ 0.10  & 29 & 5.82 $\pm$	0.12  & 38 & 5.64 $\pm$	0.13    \\
J1209+3217 & 13        & 0.60 $\pm$ 0.03   & 0.72 $\pm$ 0.03  & 14 & 0.61 $\pm$	0.02  & 15 & 0.63 $\pm$	0.03     \\
J1215+5442 & 10        & 0.56 $\pm$ 0.02   & 0.64 $\pm$ 0.02  & 12 & 0.65 $\pm$	0.02  & 13 & 0.67 $\pm$	0.02     \\
J1218+2948 & 9         & 11.17 $\pm$ 0.08  & 15.46 $\pm$ 0.04 & 10 & 15.56 $\pm$ 0.03 & 12 & 15.55 $\pm$	0.03    \\
J1227+3214 & 11        & 2.96 $\pm$ 0.03   & 3.34 $\pm$ 0.03  & 17 & 3.63 $\pm$	0.05  & 23 & 3.60 $\pm$	0.06    \\
J1242+3317 & 12        & 1.32 $\pm$ 0.03   & 2.23 $\pm$ 0.04  & 16 & 2.21 $\pm$	0.04  & 21 & 2.09 $\pm$	0.04     \\
J1246+0222 & 10        & 0.43 $\pm$ 0.02   & 0.71 $\pm$ 0.03  & 11 & 0.85 $\pm$	0.03  & 15 & 0.82 $\pm$	0.03    \\
J1302+1624 & 12        & 3.28 $\pm$ 0.05   & 11.47 $\pm$ 0.13 & 28 & 11.87 $\pm$ 0.13 & 35 & 11.99 $\pm$	0.13    \\
J1305+5116 & 14        & 32.10 $\pm$ 1.60  & 53.41 $\pm$ 0.05 & 16 & 53.37 $\pm$ 0.05 & 20 & 53.35 $\pm$	0.06    \\
J1317+6010 & 16        & 0.51 $\pm$ 0.03   & 0.64 $\pm$ 0.03  & 24 & 0.80 $\pm$	0.04  & 29 & 0.79 $\pm$	0.05    \\
J1337+2423 & 13        & 9.64 $\pm$ 0.03   & 9.96 $\pm$ 0.04  & 14 & 9.84 $\pm$	0.04  & 16 & 9.83 $\pm$	0.04     \\
J1348+2622 & 12        & 0.38 $\pm$ 0.01   & 0.38 $\pm$ 0.02  & 14 & 0.38 $\pm$	0.02  & 15 & 0.37 $\pm$	0.02   \\
J1358+2658 & 11        & 0.52 $\pm$ 0.02   & 0.61 $\pm$ 0.02  & 13 & 0.62 $\pm$	0.03  & 15 & 0.60 $\pm$	0.03   \\
J1402+2159 & 16        & 0.28 $\pm$ 0.01   & 0.33 $\pm$ 0.03  & 22 & 0.46 $\pm$ 0.04  & 22 & 0.28 $\pm$	0.03    \\
J1536+5433 & 10        & 0.44 $\pm$ 0.01   & 0.50 $\pm$ 0.02  & 18 & 0.45 $\pm$	0.03  & 23 & 0.40 $\pm$	0.03   \\
J1555+1911 & 11        & 0.12 $\pm$ 0.01   & 0.36 $\pm$ 0.03  & 15 & 0.42 $\pm$	0.03  & 17 & 0.44 $\pm$	0.03      \\
J1559+3501 & 11        & 0.46 $\pm$ 0.04   & 1.41 $\pm$ 0.04  & 14 & 1.47 $\pm$	0.03  & 16 & 1.48 $\pm$	0.04   \\
J1633+4718 & 10        & 24.10 $\pm$ 0.03  & 25.27 $\pm$ 0.04 & 12 & 25.53 $\pm$ 0.04 & 14 & 25.59 $\pm$	0.05    \\
J1703+4540 & 11        & 32.26 $\pm$ 0.06  & 34.68 $\pm$ 0.04 & 17 & 35.97 $\pm$ 0.06 & 22 & 35.01 $\pm$	0.07    \\
J1713+3523 & 10        & 3.71 $\pm$ 0.01   & 3.81 $\pm$ 0.02  & 13 & 3.95 $\pm$	0.03  & 15 & 3.98 $\pm$	0.04    \\
J2242+2943 & 11        & 5.61 $\pm$ 0.18   & 10.33 $\pm$ 0.04 & 15 & 10.43 $\pm$ 0.04 & 25 & 10.41 $\pm$	0.06     \\
J2314+2243 & 10        & 6.17 $\pm$ 0.02   & 7.02 $\pm$ 0.03  & 14 & 6.96 $\pm$	0.03  & 15 & 6.99 $\pm$	0.03    \\

\hline
\end{tabular}
\tablefoot{(1) Short name, (2) map rms, (3), peak flux density, (4) integrated flux density, (5) rms of the 90k$\lambda$ tapered map, (6) integrated flux density of the 90k$\lambda$ tapered map, (7) rms of the 60k$\lambda$ tapered map, (8) integrated flux density of the 60k$\lambda$ tapered map.}
\label{tab:measurements}
\end{table*}

\begin{table*}[h!]
\caption[]{Weighted total spectral indices, core spectral indices, spectral indices of regions of interest for our sample, and the non-simultaneous 1.4 - 5.2~GHz spectral index.}
\centering
\begin{tabular}{l l l l l}
\hline\hline
Short name  & $\alpha_{\textrm{total}}$ & $\alpha_{\textrm{core}}$  & $\alpha_{\textrm{region}}$ & $\alpha_{\textrm{1.4-5.2~GHz}}$ \\ \hline
J0347+0105	&	-0.731	$\pm$	0.001	&	-0.806	$\pm$	0.001	&			& -0.87 $\pm$ 0.01 \\
J0629-0545	&	-0.923	$\pm$	0.001	&	-0.921	$\pm$	0.001	&			& -0.62 $\pm$ 0.04$^a$ \\
J0632+6340	&	-0.509	$\pm$	0.001	&	-0.626	$\pm$	0.003	&			& -1.16 $\pm$ 0.07$^a$ \\
J0706+3901	&	-0.807	$\pm$	0.002	&	-0.795	$\pm$	0.007	&			& -0.71 $\pm$ 0.03 \\
J0713+3820	&	-0.990	$\pm$	0.001	&	-1.004	$\pm$	0.004	&			& -0.88 $\pm$ 0.02 \\
J0804+3853	&	-0.979	$\pm$	0.005	&	-1.091	$\pm$	0.020	&			& -0.91 $\pm$ 0.07 \\
J0806+7248	&	-0.961	$\pm$	0.001	&	-0.981	$\pm$	0.001	&			& -1.12 $\pm$ 0.03$^a$ \\
J0814+5609	&	-0.104	$\pm$	0.001	&	-0.002	$\pm$	0.001	&	-0.448 $\pm$ 0.009 & -0.77 $\pm$ 0.01 \\
J0913+3658	&	-0.767	$\pm$	0.006	&	-0.992	$\pm$	0.036	&			& -0.78 $\pm$ 0.17 \\
J0925+5217	&	-0.341	$\pm$	0.001	&	-0.327	$\pm$	0.002	&			& -0.98 $\pm$ 0.03 \\
J0926+1244	&	-0.592	$\pm$	0.001	&	-0.746	$\pm$	0.004	&			& -0.71 $\pm$ 0.02 \\
J0937+3615	&	-0.765	$\pm$	0.002	&	-0.946	$\pm$	0.010	&			& -0.87 $\pm$ 0.05 \\
J0952-0136	&	-0.983	$\pm$	0.001	&	-1.073	$\pm$	0.002	&			& -0.89 $\pm$ 0.01 \\
J1034+3938	&	-1.170	$\pm$	0.001	&	-1.130	$\pm$	0.002	&			& -0.96 $\pm$ 0.01 \\
J1038+4227	&	-0.893	$\pm$	0.002	&	-0.767	$\pm$	0.004	&			& 0.33 $\pm$ 0.06 \\
J1047+4725	&	-0.747	$\pm$	0.001	&	-0.629	$\pm$	0.000	&			& -0.53 $\pm$ 0.01 \\
J1048+2222	&	-0.559	$\pm$	0.004	&	-0.710	$\pm$	0.015	&			& -1.39 $\pm$ 0.11 \\
J1102+2239	&	-0.761	$\pm$	0.004	&	-0.955	$\pm$	0.016	&			& -0.72 $\pm$ 0.09 \\
J1110+3653	&	-0.182	$\pm$	0.001	&	-0.099	$\pm$	0.001	&			& -0.61 $\pm$ 0.01  \\
J1121+5351	&	-0.612	$\pm$	0.002	&	-0.760	$\pm$	0.008	&			&  -0.43 $\pm$ 0.06 \\
J1138+3653	&	-0.588	$\pm$	0.001	&	-0.680	$\pm$	0.001	&			& -0.76 $\pm$ 0.01  \\
J1159+2838	&	-0.711	$\pm$	0.004	&	-0.926	$\pm$	0.015	&			& -0.66 $\pm$ 0.07  \\
J1203+4431	&	-0.616	$\pm$	0.002	&	-0.853	$\pm$	0.008	&			& -0.96 $\pm$ 0.02  \\
J1209+3217	&	-0.548	$\pm$	0.006	&	-0.621	$\pm$	0.016	&			& -0.76 $\pm$ 0.09  \\
J1215+5442	&	-1.012	$\pm$	0.005	&	-1.160	$\pm$	0.017	&			&  -0.99 $\pm$ 0.07 \\
J1218+2948	&	-0.830	$\pm$	0.001	&	-0.835	$\pm$	0.001	&			& -0.73 $\pm$ 0.01  \\
J1227+3214	&	-0.671	$\pm$	0.001	&	-0.737	$\pm$	0.002	&			& -0.50 $\pm$ 0.02  \\
J1242+3317	&	-0.739	$\pm$	0.003	&	-0.883	$\pm$	0.007	&			& -0.79 $\pm$ 0.03  \\
J1246+0222	&	0.199	$\pm$	0.001	&	0.053	$\pm$	0.004	&			& -0.87 $\pm$ 0.09  \\
J1302+1624	&	-1.500	$\pm$	0.003	&	-0.822	$\pm$	0.003	&	-0.666 $\pm$ 0.015 & -0.70 $\pm$ 0.01  \\
J1305+5116	&	-0.318	$\pm$	0.001	&	-0.072	$\pm$	0.000	&	-0.727 $\pm$ 0.001 & -0.37 $\pm$ 0.01  \\
J1317+6010	&	-0.191	$\pm$	0.005	&	-0.339	$\pm$	0.018	&			& -0.80 $\pm$ 0.10  \\
J1337+2423	&	-1.030	$\pm$	0.001	&	-1.009	$\pm$	0.003	&			& -0.53 $\pm$ 0.01  \\
J1348+2622	&	-1.055	$\pm$	0.004	&	-1.298	$\pm$	0.027	&			& -1.00 $\pm$ 0.11  \\
J1358+2658	&	-0.424	$\pm$	0.001	&	-0.509	$\pm$	0.008	&			& -0.50 $\pm$ 0.11  \\
J1402+2159	&	-0.954	$\pm$	0.004	&	-1.543	$\pm$	0.027	&			& -0.74 $\pm$ 0.17 \\
J1536+5433	&	-0.346	$\pm$	0.003	&	-0.497	$\pm$	0.011	&			&  -0.69 $\pm$ 0.12 \\
J1555+1911	&	-0.824	$\pm$	0.004	&	-1.128	$\pm$	0.018	&			&  -1.25 $\pm$ 0.12 \\
J1559+3501	&	-1.157	$\pm$	0.001	&	-1.103	$\pm$	0.008	&			& -0.67 $\pm$ 0.05  \\
J1633+4718	&	-0.543	$\pm$	0.001	&	-0.537	$\pm$	0.001	&	-0.694 $\pm$ 0.006 & -0.72 $\pm$ 0.01  \\
J1703+4540	&	-0.868	$\pm$	0.001	&	-0.912	$\pm$	0.001	&			&  -0.94 $\pm$ 0.01 \\
J1713+3523	&	-0.974	$\pm$	0.001	&	-1.050	$\pm$	0.003	&			& -0.82 $\pm$ 0.01  \\
J2242+2943	&	-0.812	$\pm$	0.001	&	-0.850	$\pm$	0.002	&			& -0.79 $\pm$ 0.03$^a$  \\
J2314+2243	&	-0.812	$\pm$	0.001	&	-0.850	$\pm$	0.003	&			& -0.75 $\pm$ 0.04$^a$  \\ \hline
\end{tabular}
\tablefoot{(1) Short name, (2) weighted average spectral index of the total emitting region, (3), weighted average spectral index of the centremost region within a circle with a radius of 2 pixels, (4) weighted average spectral index of a region of interest, (5) Non-simultaneous 1.4-5.2~GHz spectral index. 1.4~GHz data are from FIRST, except for the sources marked with $^a$ from NVSS. 5.2~GHz data are the intergrated flux densities in this paper.}
\label{tab:spinds}
\end{table*}

\begin{table*}[h!]
\caption[]{Parameters related to the star formation estimation for our sample.}
\centering
\begin{tabular}{l l l l l l l l}
\hline\hline
Short name & W3 & W4 & W3-W4 & $S_{\textrm{W3}}$ & $S_{\textrm{20cm}}$ CDFS & $S_{\textrm{20~cm}}$ ELAIS & $q22$ \\
 & (mag) & (mag) & (mag) & (mJy) & (mJy) & (mJy) & \\ \hline
J0347+0105 & 5.32 $\pm$ 0.02 & 3.03 $\pm$ 0.02 & 2.29 $\pm$ 0.05 & 234.80 $\pm$ 3.46 & 21.03 $\pm$ 1.43 & 20.01 $\pm$ 2.44 & 1.21\\
J0629-0545 & 5.48 $\pm$ 0.02 & 2.60 $\pm$ 0.02 & 2.88 $\pm$ 0.08 & 203.38 $\pm$ 2.44 & 31.36 $\pm$ 2.11 & 29.82 $\pm$ 3.61 & 1.19\\
J0632+6340 & 6.28 $\pm$ 0.02 & 3.76 $\pm$ 0.02 & 2.52 $\pm$ 0.06 & 97.43 $\pm$ 1.35 & 10.74 $\pm$ 0.71 & 10.23 $\pm$ 1.23 & 1.65\\
J0706+3901 & 8.60 $\pm$ 0.03 & 6.05 $\pm$ 0.05 & 2.55 $\pm$ 0.14 & 11.51 $\pm$ 0.28 & 1.31 $\pm$ 0.13 & 1.25 $\pm$ 0.19 & 0.70\\
J0713+3820 & 6.26 $\pm$ 0.02 & 3.98 $\pm$ 0.02 & 2.28 $\pm$ 0.05 & 99.15 $\pm$ 1.37 & 8.74 $\pm$ 0.60 & 8.32 $\pm$ 1.02 & 1.19\\
J0804+3853 & 8.08 $\pm$ 0.02 & 5.37 $\pm$ 0.03 & 2.70 $\pm$ 0.10 & 18.62 $\pm$ 0.38 & 2.44 $\pm$ 0.19 & 2.32 $\pm$ 0.31 & 1.36\\
J0806+7248 & 8.11 $\pm$ 0.02 & 5.54 $\pm$ 0.04 & 2.57 $\pm$ 0.11 & 17.99 $\pm$ 0.40 & 2.08 $\pm$ 0.18 & 1.99 $\pm$ 0.28 & 0.11\\
J0814+5609 & 10.41 $\pm$ 0.07 & 8.43 $\pm$ 0.45 & 1.98 $\pm$ 0.66 & 2.17 $\pm$ 0.15 & 0.15 $\pm$ 0.07 & 0.15 $\pm$ 0.08 & -0.99\\
J0913+3658 & 8.20 $\pm$ 0.02 & 5.65 $\pm$ 0.04 & 2.55 $\pm$ 0.11 & 16.67 $\pm$ 0.32 & 1.89 $\pm$ 0.16 & 1.80 $\pm$ 0.25 & 1.67\\
J0925+5217 & 6.72 $\pm$ 0.02 & 4.71 $\pm$ 0.03 & 2.01 $\pm$ 0.05 & 64.67 $\pm$ 1.01 & 4.46 $\pm$ 0.34 & 4.25 $\pm$ 0.56 & 1.35\\
J0926+1244 & 6.16 $\pm$ 0.02 & 3.80 $\pm$ 0.02 & 2.36 $\pm$ 0.05 & 108.92 $\pm$ 1.50 & 10.40 $\pm$ 0.75 & 9.90 $\pm$ 1.25 & 1.54\\
J0937+3615 & 8.14 $\pm$ 0.02 & 5.47 $\pm$ 0.04 & 2.66 $\pm$ 0.12 & 17.65 $\pm$ 0.34 & 2.22 $\pm$ 0.18 & 2.12 $\pm$ 0.29 & 1.34\\
J0952-0136 & 4.33 $\pm$ 0.01 & 2.31 $\pm$ 0.02 & 2.02 $\pm$ 0.04 & 589.25 $\pm$ 7.60 & 40.81 $\pm$ 2.56 & 38.83 $\pm$ 4.52 & 1.16\\
J1034+3938 & 6.65 $\pm$ 0.02 & 4.51 $\pm$ 0.03 & 2.14 $\pm$ 0.06 & 69.49 $\pm$ 0.96 & 5.40 $\pm$ 0.40 & 5.14 $\pm$ 0.66 & 0.57\\
J1038+4227 & 8.87 $\pm$ 0.03 & 6.58 $\pm$ 0.06 & 2.29 $\pm$ 0.13 & 8.97 $\pm$ 0.23 & 0.80 $\pm$ 0.08 & 0.77 $\pm$ 0.12 & 0.07\\
J1047+4725 & 10.44 $\pm$ 0.09 & 7.75 $\pm$ 0.19 & 2.69 $\pm$ 0.52 & 2.12 $\pm$ 0.17 & 0.27 $\pm$ 0.06 & 0.27 $\pm$ 0.08 & -2.18\\
J1048+2222 & 9.29 $\pm$ 0.04 & 7.08 $\pm$ 0.11 & 2.20 $\pm$ 0.21 & 6.08 $\pm$ 0.25 & 0.50 $\pm$ 0.08 & 0.48 $\pm$ 0.10 & 1.09\\
J1102+2239 & 9.15 $\pm$ 0.03 & 6.49 $\pm$ 0.07 & 2.66 $\pm$ 0.20 & 6.96 $\pm$ 0.22 & 0.87 $\pm$ 0.10 & 0.83 $\pm$ 0.14 & 0.81\\
J1110+3653 & nd              &              nd & N/A             &            N/A  & N/A             & N/A             & N/A \\
J1121+5351 & 7.32 $\pm$ 0.02 & 4.99 $\pm$ 0.03 & 2.33 $\pm$ 0.07 & 37.39 $\pm$ 0.62 & 3.46 $\pm$ 0.27 & 3.30 $\pm$ 0.44 & 1.48\\
J1138+3653 & 10.53 $\pm$ 0.08 & 8.01 $\pm$ 0.21 & 2.52 $\pm$ 0.50 & 1.94 $\pm$ 0.15 & 0.22 $\pm$ 0.05 & 0.21 $\pm$ 0.06 & -0.28\\
J1159+2838 & 9.02 $\pm$ 0.03 & 5.98 $\pm$ 0.04 & 3.03 $\pm$ 0.17 & 7.85 $\pm$ 0.25 & 1.39 $\pm$ 0.12 & 1.33 $\pm$ 0.19 & 1.22\\
J1203+4431 & 4.58 $\pm$ 0.02 & 2.16 $\pm$ 0.02 & 2.41 $\pm$ 0.05 & 468.06 $\pm$ 6.47 & 46.81 $\pm$ 2.97 & 44.53 $\pm$ 5.23 & 1.96\\
J1209+3217 & 8.18 $\pm$ 0.02 & 5.30 $\pm$ 0.03 & 2.88 $\pm$ 0.11 & 16.93 $\pm$ 0.33 & 2.60 $\pm$ 0.20 & 2.48 $\pm$ 0.32 & 1.75\\
J1215+5442 & 7.48 $\pm$ 0.02 & 5.01 $\pm$ 0.04 & 2.47 $\pm$ 0.10 & 32.17 $\pm$ 0.53 & 3.40 $\pm$ 0.28 & 3.24 $\pm$ 0.44 & 1.52\\
J1218+2948 & 5.00 $\pm$ 0.01 & 2.13 $\pm$ 0.02 & 2.87 $\pm$ 0.07 & 316.16 $\pm$ 4.08 & 48.08 $\pm$ 3.23 & 45.74 $\pm$ 5.54 & 1.40\\
J1227+3214 & 7.42 $\pm$ 0.02 & 4.80 $\pm$ 0.02 & 2.62 $\pm$ 0.06 & 34.10 $\pm$ 0.57 & 4.12 $\pm$ 0.29 & 3.92 $\pm$ 0.49 & 1.06\\
J1242+3317 & 6.70 $\pm$ 0.02 & 3.98 $\pm$ 0.02 & 2.72 $\pm$ 0.07 & 66.30 $\pm$ 0.98 & 8.82 $\pm$ 0.62 & 8.39 $\pm$ 1.04 & 1.59\\
J1246+0222 & 7.22 $\pm$ 0.02 & 4.66 $\pm$ 0.03 & 2.56 $\pm$ 0.08 & 40.95 $\pm$ 0.64 & 4.69 $\pm$ 0.34 & 4.46 $\pm$ 0.57 & 2.26\\
J1302+1624 & 7.88 $\pm$ 0.02 & 5.26 $\pm$ 0.04 & 2.62 $\pm$ 0.12 & 22.28 $\pm$ 0.41 & 2.70 $\pm$ 0.22 & 2.58 $\pm$ 0.35 & -0.11\\
J1305+5116 & 8.91 $\pm$ 0.03 & 6.57 $\pm$ 0.07 & 2.34 $\pm$ 0.16 & 8.63 $\pm$ 0.21 & 0.81 $\pm$ 0.09 & 0.78 $\pm$ 0.13 & -0.61\\
J1317+6010 & 8.40 $\pm$ 0.02 & 5.80 $\pm$ 0.04 & 2.60 $\pm$ 0.11 & 13.79 $\pm$ 0.27 & 1.64 $\pm$ 0.14 & 1.57 $\pm$ 0.22 & 1.60\\
J1337+2423 & 4.59 $\pm$ 0.01 & 2.66 $\pm$ 0.02 & 1.95 $\pm$ 0.03 & 462.06 $\pm$ 5.96 & 30.03 $\pm$ 1.91 & 28.57 $\pm$ 3.35 & 1.29\\
J1348+2622 & 10.40 $\pm$ 0.06 & 8.57 $\pm$ 0.29 & 1.83 $\pm$ 0.40 & 2.19 $\pm$ 0.13 & 0.13 $\pm$ 0.04 & 0.13 $\pm$ 0.05 & 0.32\\
J1358+2658 & 9.22 $\pm$ 0.03 & 6.53 $\pm$ 0.06 & 2.69 $\pm$ 0.18 & 6.47 $\pm$ 0.18 & 0.84 $\pm$ 0.09 & 0.80 $\pm$ 0.13 & 1.30\\
J1402+2159 & 7.18 $\pm$ 0.02 & 4.34 $\pm$ 0.03 & 2.75 $\pm$ 0.11 & 42.41 $\pm$ 0.63 & 5.78 $\pm$ 0.43 & 5.50 $\pm$ 0.70 & 2.16\\
J1536+5433 & 6.95 $\pm$ 0.02 & 5.02 $\pm$ 0.02 & 1.92 $\pm$ 0.04 & 52.76 $\pm$ 0.78 & 3.36 $\pm$ 0.23 & 3.20 $\pm$ 0.40 & 2.12\\
J1555+1911 & 7.97 $\pm$ 0.02 & 5.35 $\pm$ 0.04 & 2.62 $\pm$ 0.12 & 20.47 $\pm$ 0.38 & 2.49 $\pm$ 0.20 & 2.37 $\pm$ 0.32 & 1.67\\
J1559+3501 & 6.62 $\pm$ 0.02 & 4.28 $\pm$ 0.02 & 2.34 $\pm$ 0.05 & 71.04 $\pm$ 0.98 & 6.64 $\pm$ 0.47 & 6.33 $\pm$ 0.78 & 1.38\\
J1633+4718 & 7.64 $\pm$ 0.02 & 4.97 $\pm$ 0.02 & 2.67 $\pm$ 0.06 & 27.79 $\pm$ 0.44 & 3.51 $\pm$ 0.25 & 3.35 $\pm$ 0.42 & 0.22\\
J1703+4540 & 6.01 $\pm$ 0.02 & 3.42 $\pm$ 0.02 & 2.59 $\pm$ 0.06 & 125.17 $\pm$ 1.73 & 14.75 $\pm$ 0.98 & 14.04 $\pm$ 1.69 & 0.52\\
J1713+3523 & 7.48 $\pm$ 0.02 & 5.38 $\pm$ 0.03 & 2.11 $\pm$ 0.06 & 32.11 $\pm$ 0.53 & 2.43 $\pm$ 0.20 & 2.32 $\pm$ 0.31 & 0.62\\
J2242+2943 & 5.78 $\pm$ 0.01 & 3.21 $\pm$ 0.02 & 2.57 $\pm$ 0.06 & 153.99 $\pm$ 1.84 & 17.80 $\pm$ 1.16 & 16.94 $\pm$ 2.02 & 1.16\\
J2314+2243 & 6.74 $\pm$ 0.02 & 4.40 $\pm$ 0.03 & 2.33 $\pm$ 0.07 & 64.08 $\pm$ 0.94 & 5.93 $\pm$ 0.43 & 5.65 $\pm$ 0.72 & 0.85\\
\hline
\end{tabular}
\tablefoot{ (1) Short name, (2) WISE W3-band magnitude, (3) WISE W4-band magnitude, (4) WISE W3-W4 magnitude colour, (5) WISE W3-band flux density in mJy (6) 20~cm flux density in mJy estimated from the WISE W4-band flux density, using a relation based on CDFS data \citep{2007boyle1}, (7) 20~cm flux density in mJy estimated from the WISE W4-band flux density, using a relation based on ELAIS data \citep{2007boyle1}, (8) $q22$ parameter = log ($S_{22 \mu m}$ /$S_{\textrm{1.4~GHz}}$).}
\label{tab:sf}
\end{table*}

\begin{table*}[h!]
\caption[]{Summary of the radio and mid-infrared results.}
\centering
\begin{tabular}{l l l l l l l l l}
\hline\hline
Short name & Morph.        & Spectral & Other radio & W3-W4  & $q22$ & $R_{\textrm{CDFS}}$             & $R_{\textrm{W3}}$           & Verdict \\
           & type          & type     & properties  & colour &       &                                 &                             &   \\ \hline
J0347+0105 & E             & S        &             & nSF    & SF    & ?                               & SF                          & SF \\
J0629-0545 & I             & S        & J?          & SF     & SF    & ?                               & SF                          & mix \\
J0632+6340 & E             & S        & O?          & SF     & SF    & SF                              & SF                          & mix \\
J0706+3901 & I             & S        &             & SF     & nSF   & AGN                             & SF?                         & ? \\
J0713+3820 & E             & S        &             & nSF    & SF    & ?                               & SF                          & SF \\
J0804+3853 & E             & S        & O?          & SF     & SF    & SF                              & SF                          & mix \\
J0806+7248 & I             & S        & CSS         & SF     & nSF   & AGN                             & AGN?                        & AGN \\
J0814+5609 & I             & F        & J           & nSF    & nSF   & AGN                             & AGN?                        & AGN \\
J0913+3658 & I             & S        &             & SF     & SF    & SF                              & SF                          & SF \\
J0925+5217 & E             & F        &             & nSF    & SF    & AGN?                            & SF?                         & mix \\
J0926+1244 & I             & S        & J           & nSF    & SF    & SF                              & SF                          & mix \\
J0937+3615 & E             & S        &             & SF     & SF    & SF                              & SF                          & SF \\
J0952-0136 & I             & S        & J           & nSF    & SF    & ?                               & SF                          & AGN/mix \\
J1034+3938 & I             & S        &             & nSF    & nSF   & AGN                             & SF?                         & mix \\
J1038+4227 & E             & S        &             & nSF    & nSF   & AGN                             & ?                           & AGN \\
J1047+4725 & E             & S        & CSS         & SF     & nSF   & AGN                             & AGN?                        & AGN \\
J1048+2222 & I             & S        & O           & nSF    & SF    & AGN                             & SF                          & AGN/mix\\
J1102+2239 & I             & S        &             & nSF    & nSF   & AGN                             & SF                          & mix \\
J1110+3653 & I             & F        & J           & N/A    & N/A   & N/A                             & N/A                         & AGN \\
J1121+5351 & I             & S        &             & nSF    & SF    & SF                              & SF                          & SF/mix \\
J1138+3653 & I             & S        & CSS         & SF     & nSF   & AGN                             & AGN?                        & AGN \\
J1159+2838 & I             & S        &             & SF     & SF    & ?                               & SF                          & SF \\
J1203+4431 & E             & S        & J           & nSF    & SF    & SF                              & SF                          & AGN \\
J1209+3217 & I             & S        &             & SF     & SF    & SF                              & SF                          & SF \\
J1215+5442 & I             & S        &             & nSF    & SF    & SF                              & SF                          & SF \\
J1218+2948 & E             & S        &             & SF     & SF    & SF                              & SF                          & mix \\
J1227+3214 & I             & S        &             & SF     & SF    & AGN                             & SF                          & mix \\
J1242+3317 & E             & S        & J/O         & SF     & SF    & SF                              & SF                          & mix \\
J1246+0222 & E             & F        &             & SF     & SF    & SF                              & SF                          & mix \\
J1302+1624 & E             & S        & J           & SF     & nSF   & AGN                             & AGN?                        & AGN \\
J1305+5116 & E             & F        & J           & nSF    & nSF   & AGN                             & AGN?                        & AGN \\
J1317+6010 & E             & S        &             & SF     & SF    & ?                               & SF                          & mix? \\
J1337+2423$^a$ & I         & S        &             & nSF    & SF    & SF                              & SF                          & mix \\
J1348+2622 & I             & S        & CSS         & nSF    & nSF   & AGN                             & ?                           & AGN \\
J1358+2658 & I             & F        &             & SF     & SF    & AGN?                            & SF                          & mix \\
J1402+2159 & I             & S        &             & SF     & SF    & SF                              & SF                          & SF \\
J1536+5433 & I             & F        &             & nSF    & SF    & SF                              & SF                          & ? \\
J1555+1911 & E             & S        &             & SF     & SF    & SF                              & SF                          & SF \\
J1559+3501 & E             & S        &             & nSF    & SF    & SF                              & SF                          & SF \\
J1633+4718 & I             & S        & J           & SF     & nSF   & AGN                             & ?                           & AGN \\
J1703+4540 & I             & S        & J           & SF     & nSF   & AGN                             & ?                           & AGN/mix \\
J1713+3523 & I             & S        &             & nSF    & nSF   & AGN                             & SF                          & mix? \\
J2242+2943 & E             & S        &             & SF     & SF    & ?                               & SF                          & mix \\
J2314+2243 & I             & S        & CSS         & nSF    & nSF   & AGN                             & SF                          & AGN \\

\hline
\end{tabular}
\tablefoot{ (1) Short name, (2) radio morphology type: I = intermediate, E = extended, same as in Table~\ref{tab:basicdata}, (3) radio spectral type: F = flat, S = steep, (4) other radio properties: CSS = CSS-like, J = jet, O = outflow, (5), W3-W4 colour: SF = supports enhanced star formation, nSF = does not require enhanced star formation, (6) $q22$ parameter, SF and nSF as before, (7) $S_{\textrm{5.2~GHz}}$ CDFS vs. $S_{\textrm{5.2~GHz}}$ JVLA: SF = $S_{\textrm{5.2~GHz}}$ JVLA could be explained by star formation, AGN = $S_{\textrm{5.2~GHz}}$ JVLA cannot be explained by star formation, ? = neither of the options is preferred, (8) $S_{\textrm{W3}}$ WISE vs. $S_{\textrm{1.4~GHz}}$ JVLA: SF = star formation dominated, AGN = AGN dominated, ? = combination, or no preferred option, (8) the origin of radio emission based on all data and diagnostic tools: SF = star formation, AGN = AGN activity, mix = combination of star formation and AGN activity, ? = inconclusive.}
\label{tab:summary}
\end{table*}

\section{$\alpha$, $\Delta \alpha$, and tapered maps}
\label{app:maps}

\begin{figure*}
     \centering
     \begin{subfigure}[b]{0.47\textwidth}
         \centering
         \includegraphics[width=\textwidth]{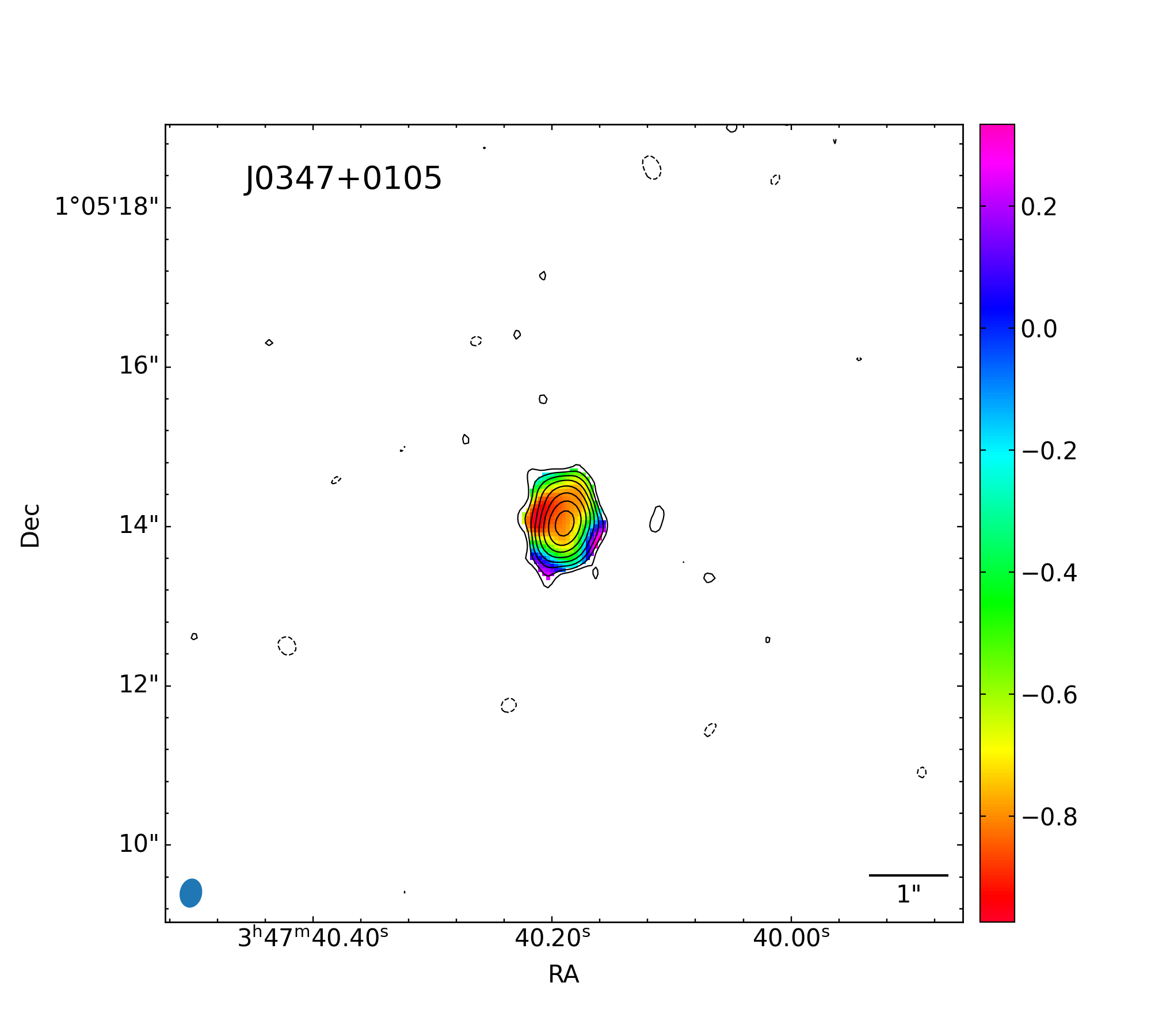}
         \caption{Spectral index map, rms = 14$\mu$Jy beam$^{-1}$, contour levels at -3, 3 $\times$ 2$^n$, $n \in$ [0, 7], beam size 0.23 $\times$ 0.17~kpc. } \label{fig:J0347spind}
     \end{subfigure}
     \hfill
     \begin{subfigure}[b]{0.47\textwidth}
         \centering
         \includegraphics[width=\textwidth]{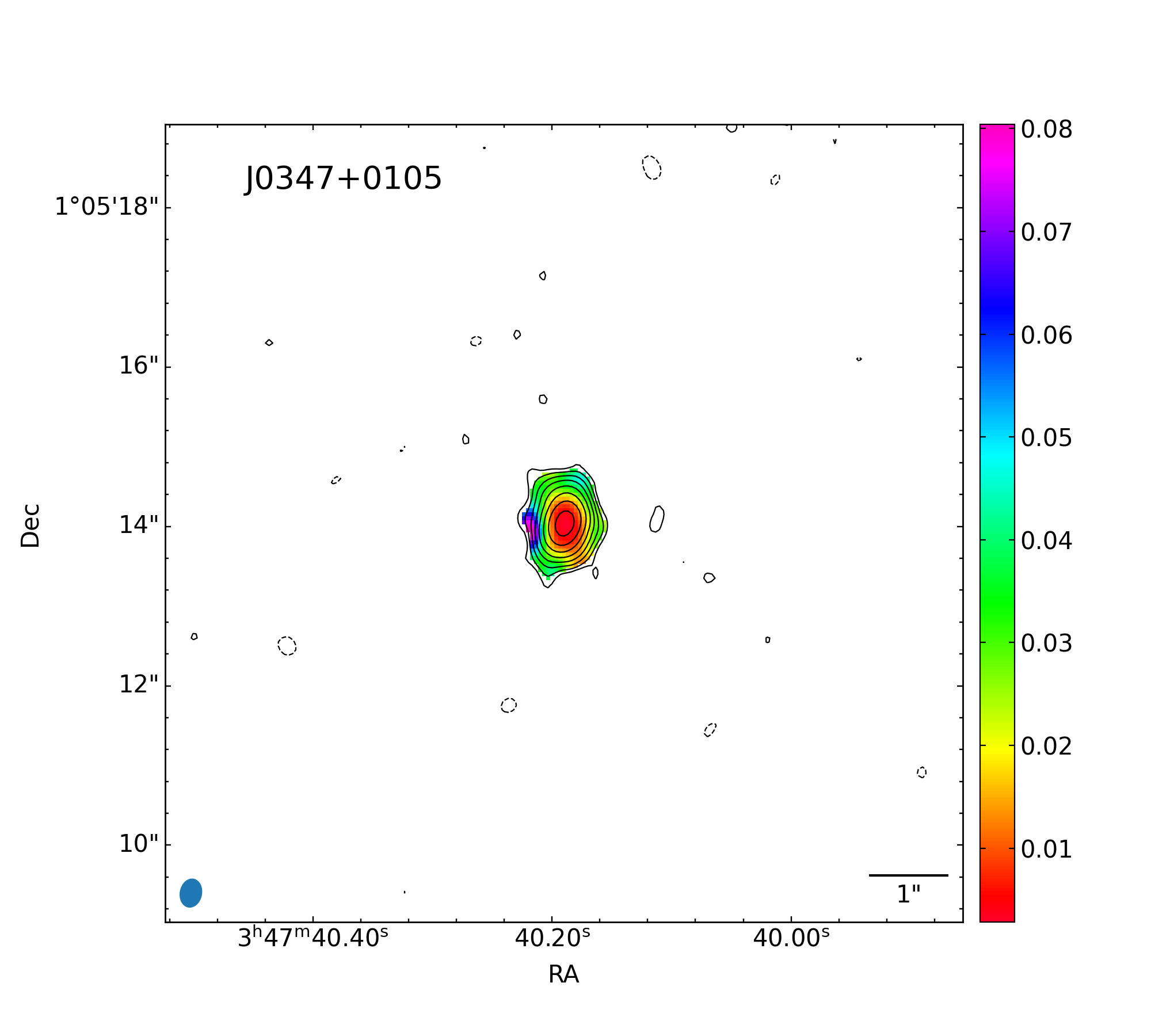}
         \caption{Spectral index error map, rms, contour levels, and beam size as in Fig.~\ref{fig:J0347spind}.} \label{fig:J0347spinderr}
     \end{subfigure}
     \hfill
     \\
     \begin{subfigure}[b]{0.47\textwidth}
         \centering
         \includegraphics[width=\textwidth]{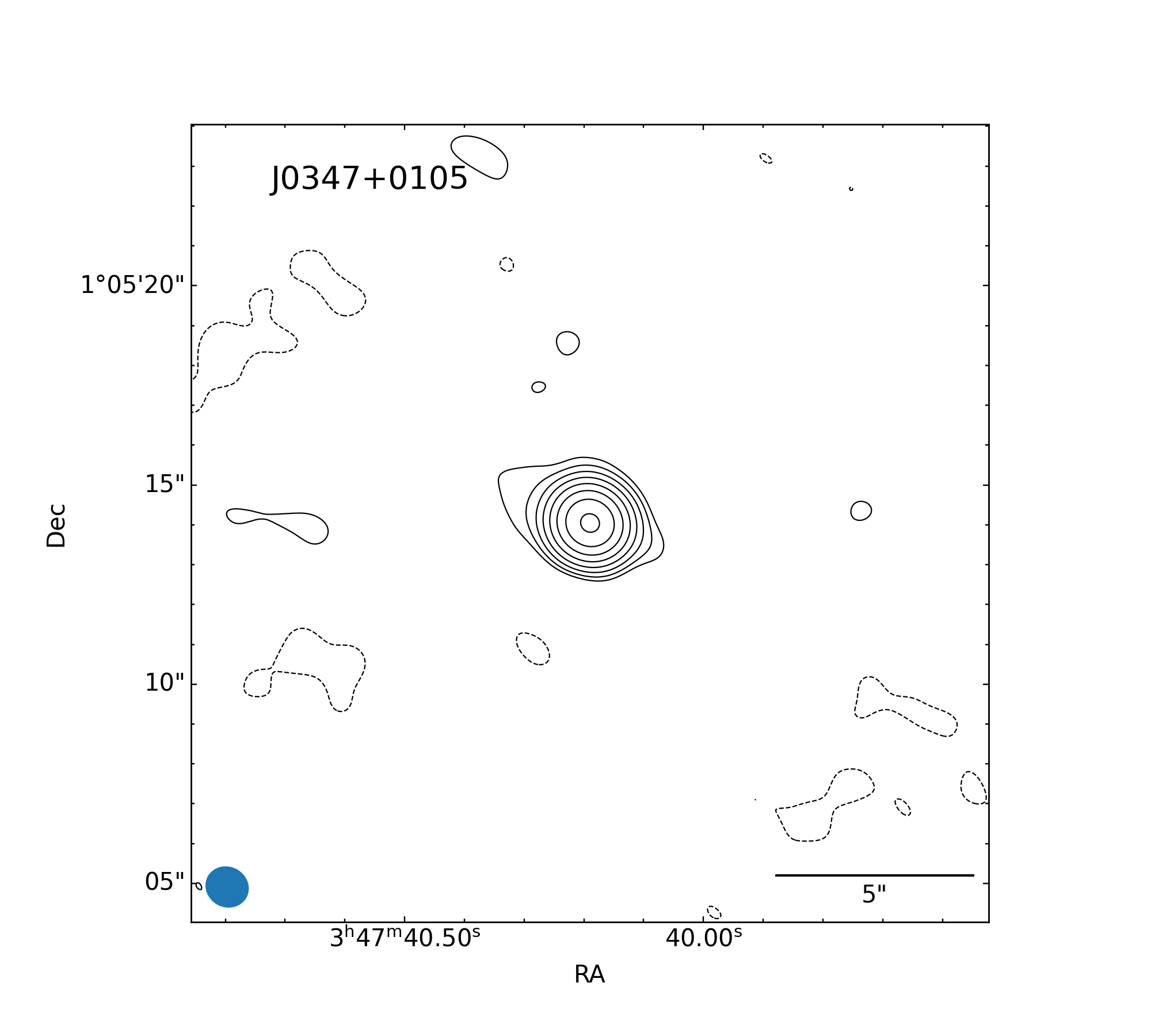}
         \caption{Tapered map with \texttt{uvtaper} = 90k$\lambda$, rms = 25$\mu$Jy beam$^{-1}$, contour levels at -3, 3 $\times$ 2$^n$, $n \in$ [0, 7], beam size 0.69 $\times$ 0.63~kpc.} \label{fig:J0347-90k}
     \end{subfigure}
          \hfill
     \begin{subfigure}[b]{0.47\textwidth}
         \centering
         \includegraphics[width=\textwidth]{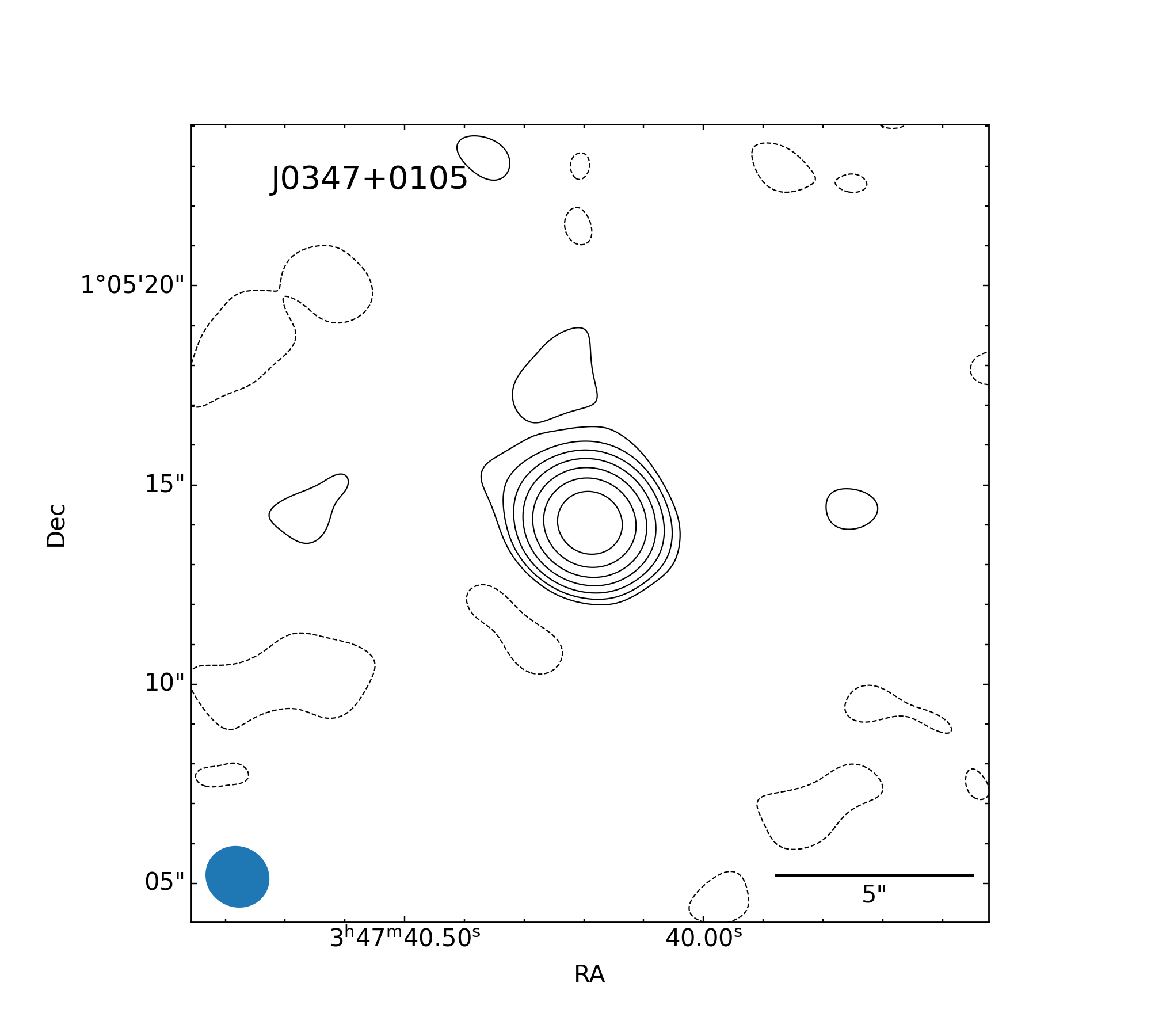}
         \caption{Tapered map with \texttt{uvtaper} = 60k$\lambda$, rms = 33$\mu$Jy beam$^{-1}$, contour levels at -3, 3 $\times$ 2$^n$, $n \in$ [0, 6], beam size 1.01 $\times$ 0.94~kpc.} \label{fig:J0347-60k}
     \end{subfigure}
          \hfill
     \\
     \begin{subfigure}[b]{0.47\textwidth}
         \centering
         \includegraphics[width=\textwidth]{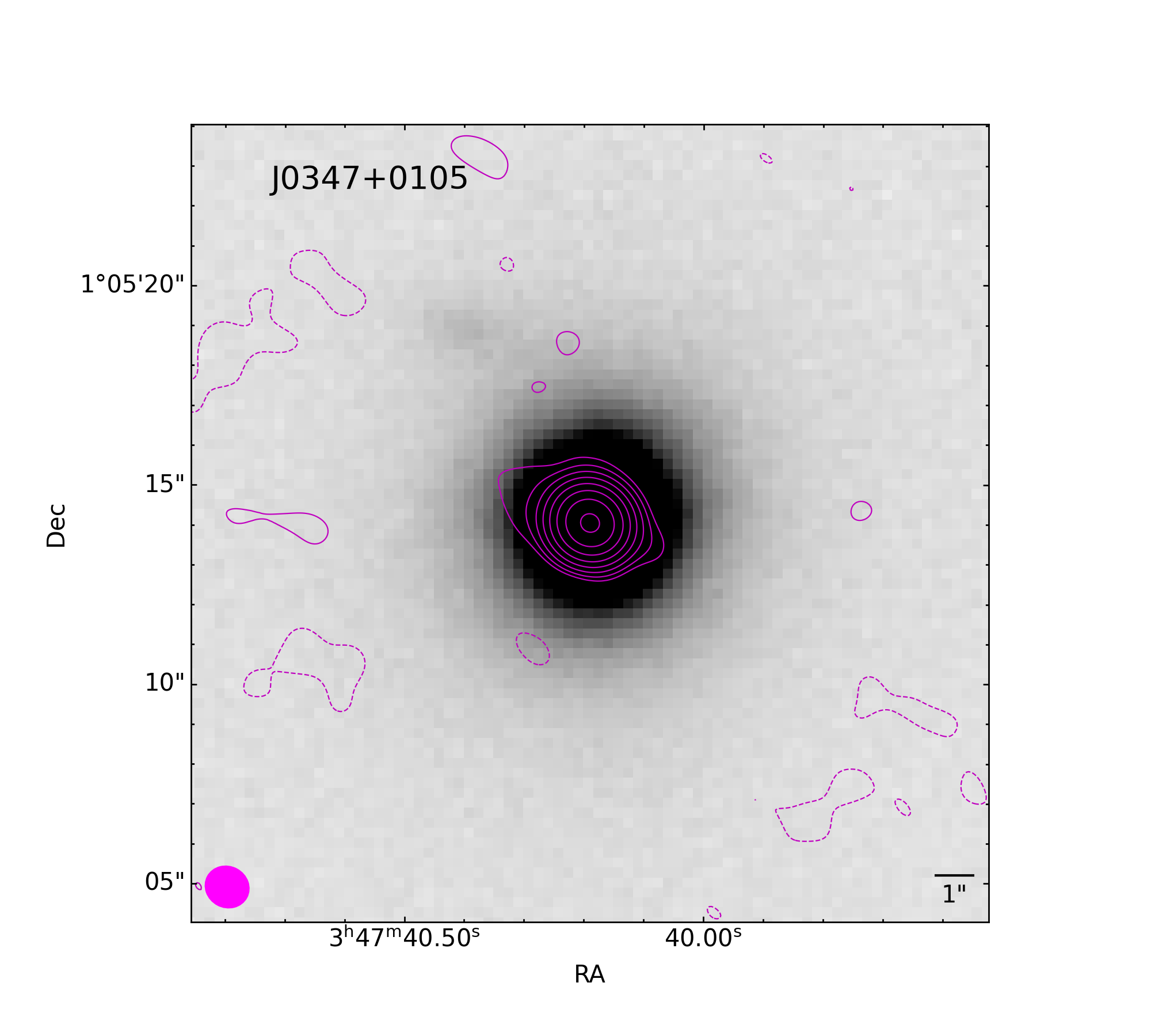}
         \caption{PanSTARRS $i$ band image of the host galaxy overlaid with the 90k$\lambda$ tapered map. Radio map properties as in Fig.~\ref{fig:J0347-90k}}. \label{fig:J0347-host}
     \end{subfigure}
        \caption{} \label{fig:J0347}
\end{figure*}


\begin{figure*}
     \centering
     \begin{subfigure}[b]{0.47\textwidth}
         \centering
         \includegraphics[width=\textwidth]{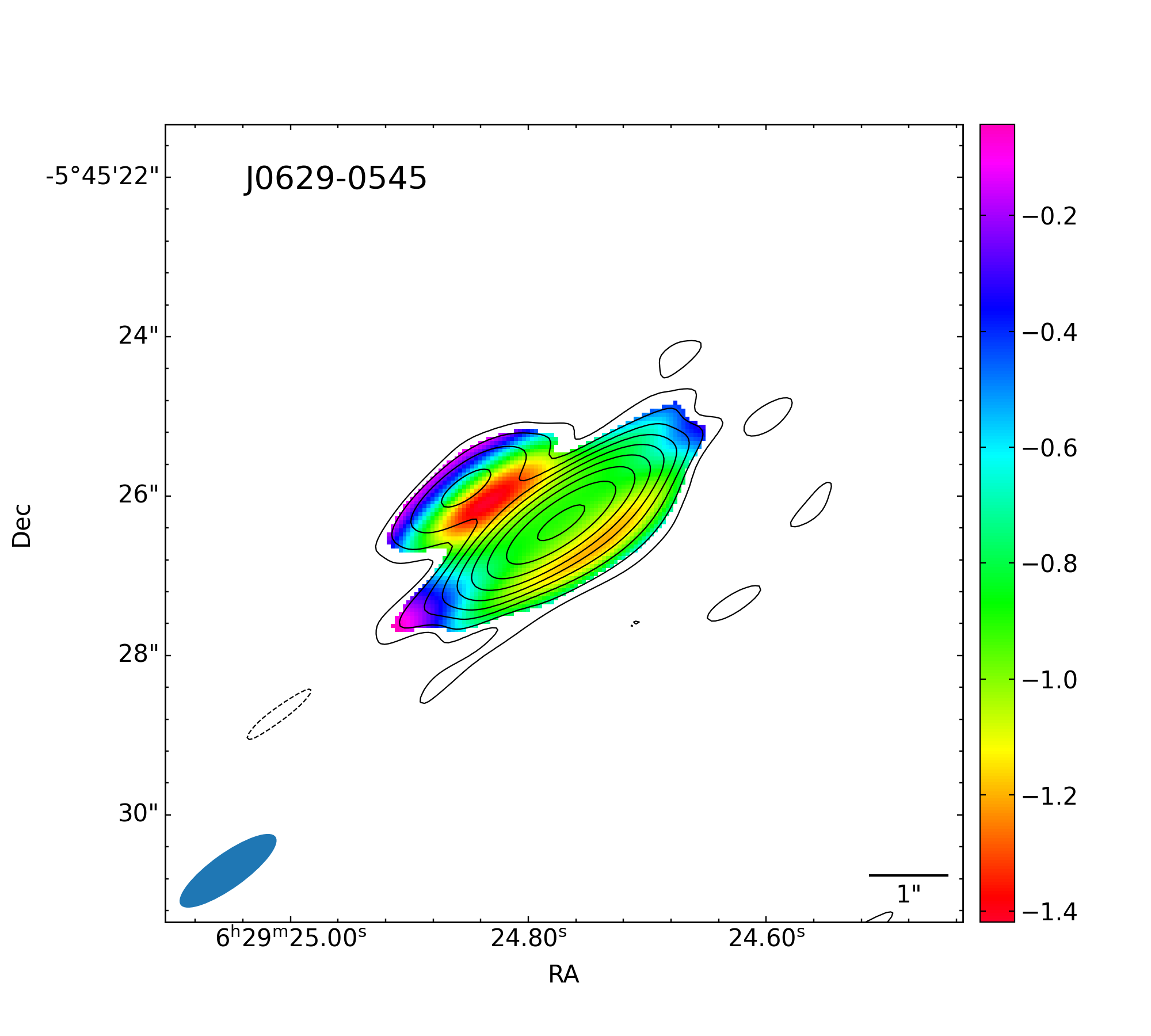}
         \caption{Spectral index map, rms = 14$\mu$Jy beam$^{-1}$, contour levels at -3, 3 $\times$ 2$^n$, $n \in$ [0, 8], beam size 3.09 $\times$ 0.93~kpc. } \label{fig:J0629spind}
     \end{subfigure}
     \hfill
     \begin{subfigure}[b]{0.47\textwidth}
         \centering
         \includegraphics[width=\textwidth]{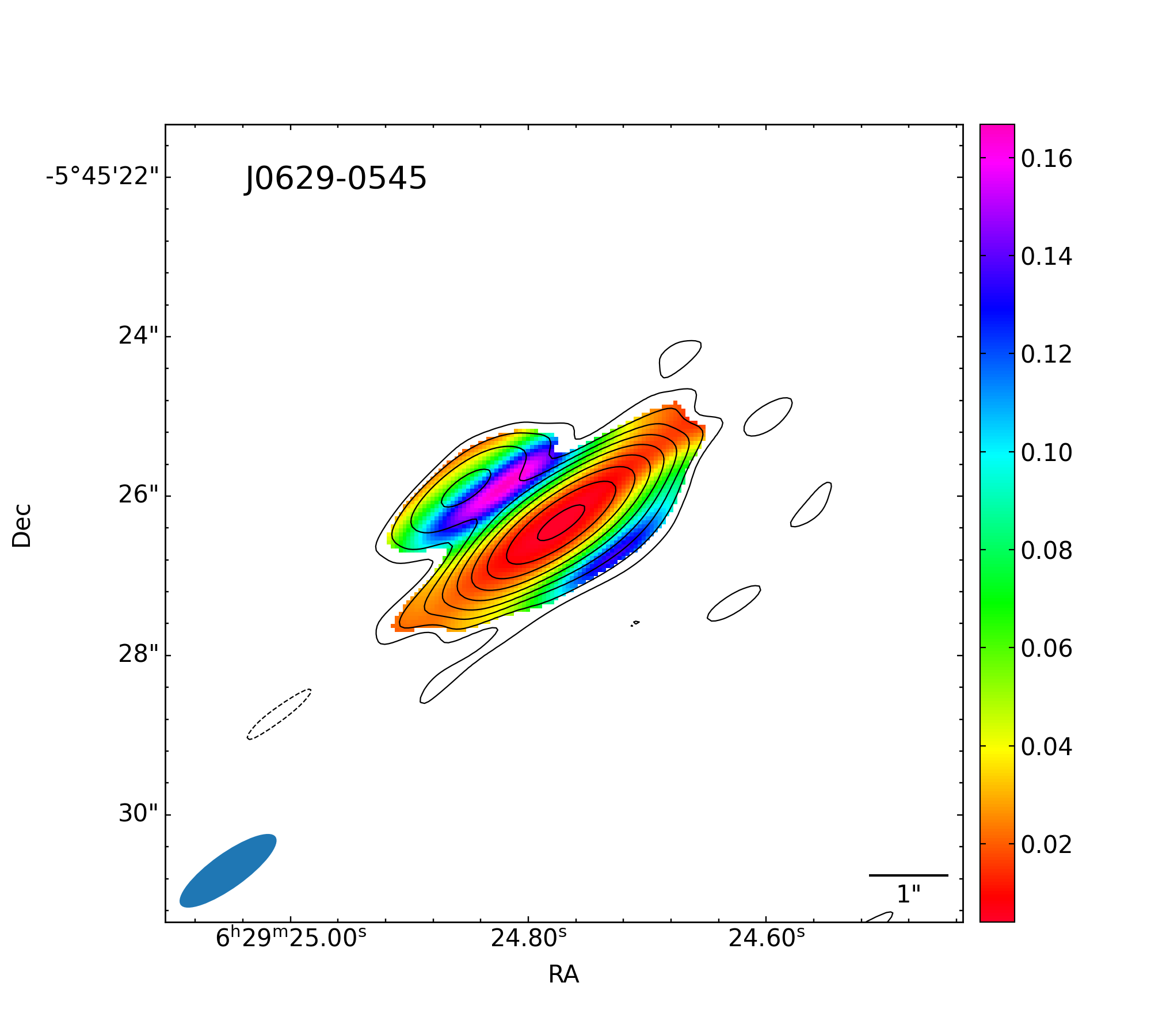}
         \caption{Spectral index error map, rms, contour levels, and beam size as in Fig.~\ref{fig:J0629spind}.} \label{fig:J0629spinderr}
     \end{subfigure}
     \hfill
     \\
     \begin{subfigure}[b]{0.47\textwidth}
         \centering
         \includegraphics[width=\textwidth]{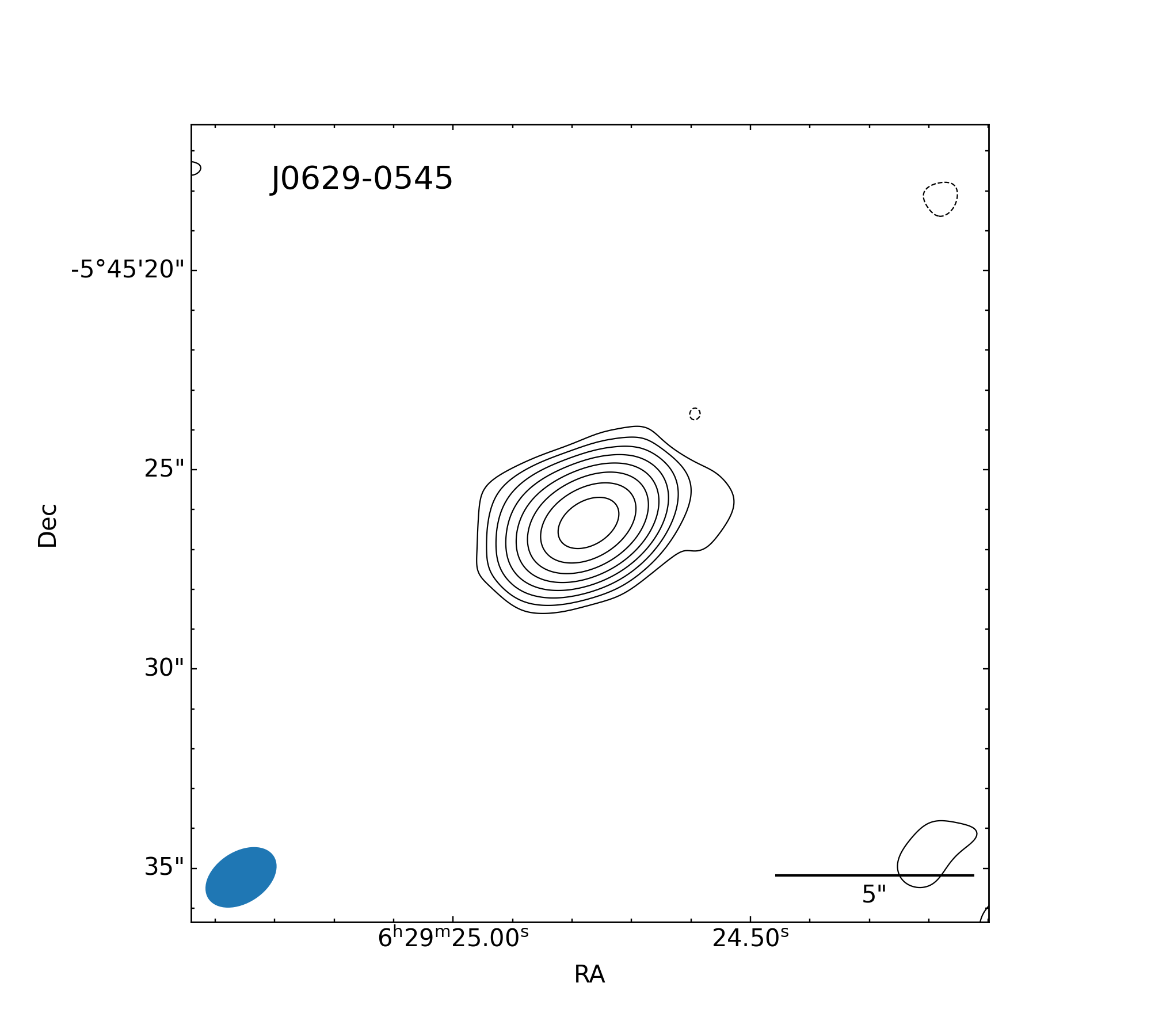}
         \caption{Tapered map with \texttt{uvtaper} = 90k$\lambda$, rms = 22$\mu$Jy beam$^{-1}$, contour levels at -3, 3 $\times$ 2$^n$, $n \in$ [0, 7], beam size 4.15 $\times$ 2.73~kpc.} \label{fig:J0629-90k}
     \end{subfigure}
          \hfill
     \begin{subfigure}[b]{0.47\textwidth}
         \centering
         \includegraphics[width=\textwidth]{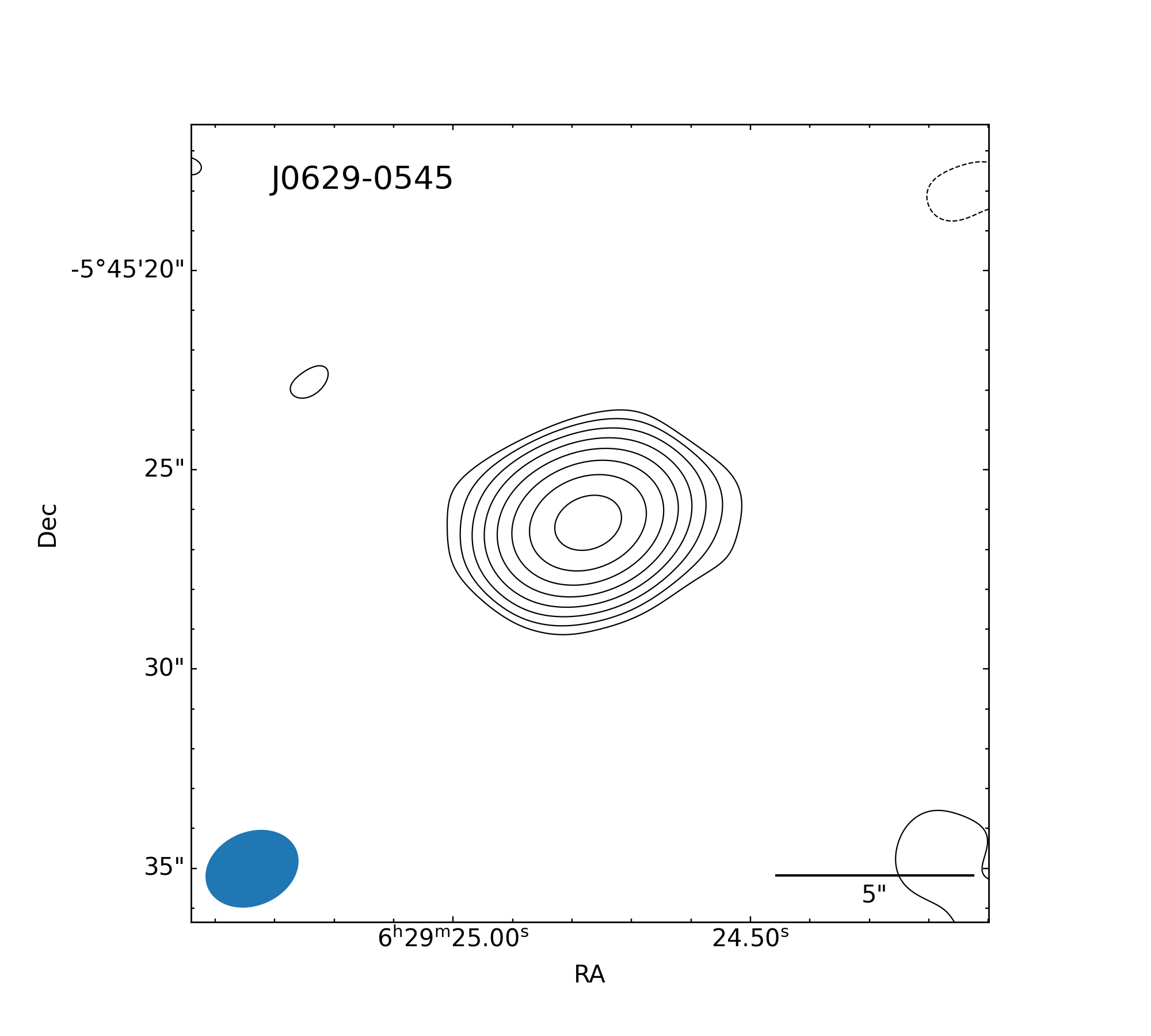}
         \caption{Tapered map with \texttt{uvtaper} = 60k$\lambda$, rms = 26$\mu$Jy beam$^{-1}$, contour levels at -3, 3 $\times$ 2$^n$, $n \in$ [0, 7], beam size 5.12 $\times$ 3.91~kpc.} \label{fig:J0629-60k}
     \end{subfigure}
     \hfill
     \\
     \begin{subfigure}[b]{0.47\textwidth}
         \centering
         \includegraphics[width=\textwidth]{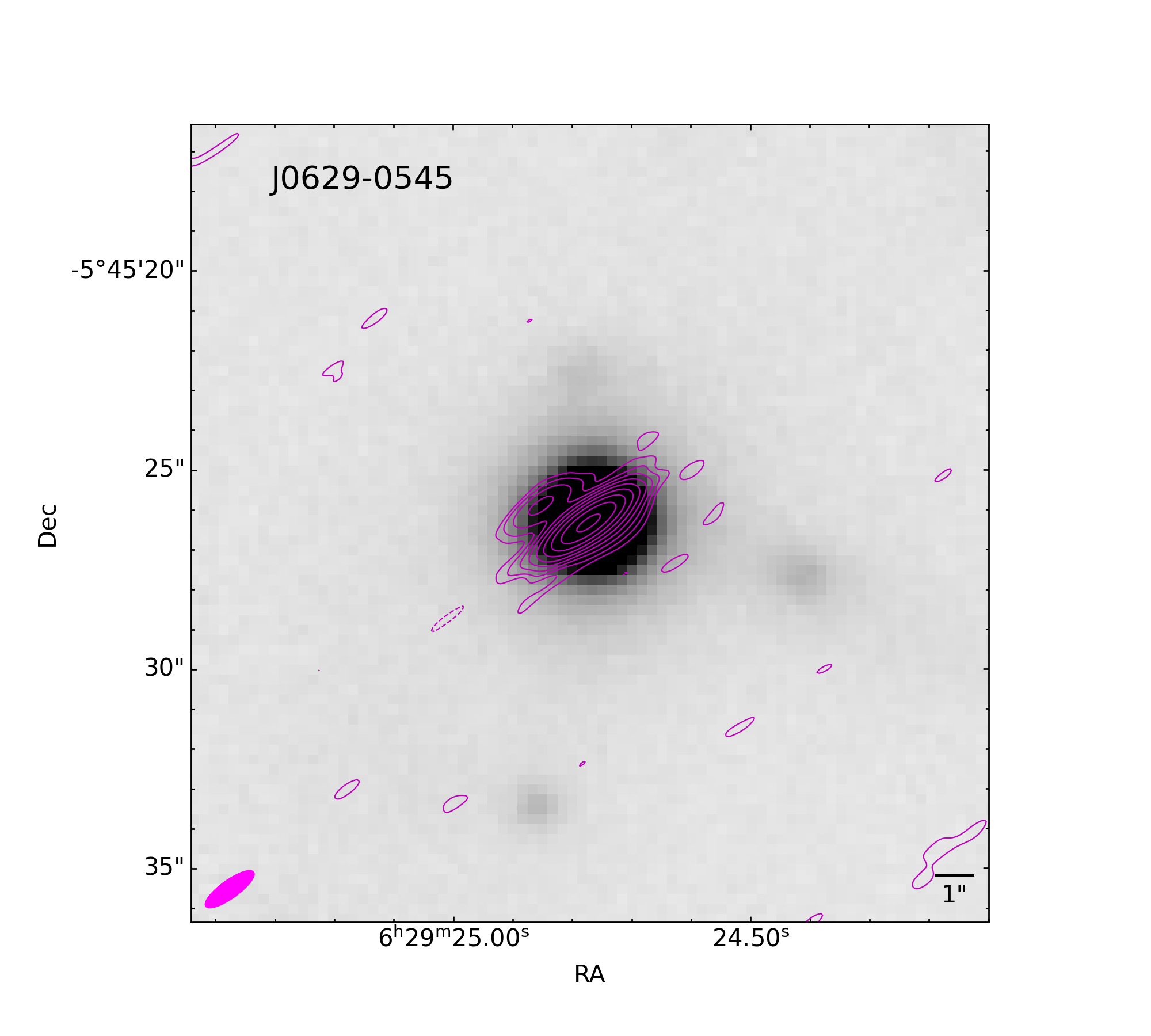}
         \caption{PanSTARRS $i$ band image of the host galaxy overlaid with the normal map. Radio map properties as in Fig.~\ref{fig:J0629spind}}. \label{fig:J0629-host-zoom}
     \end{subfigure}
     \hfill
     \begin{subfigure}[b]{0.47\textwidth}
         \centering
         \includegraphics[width=\textwidth]{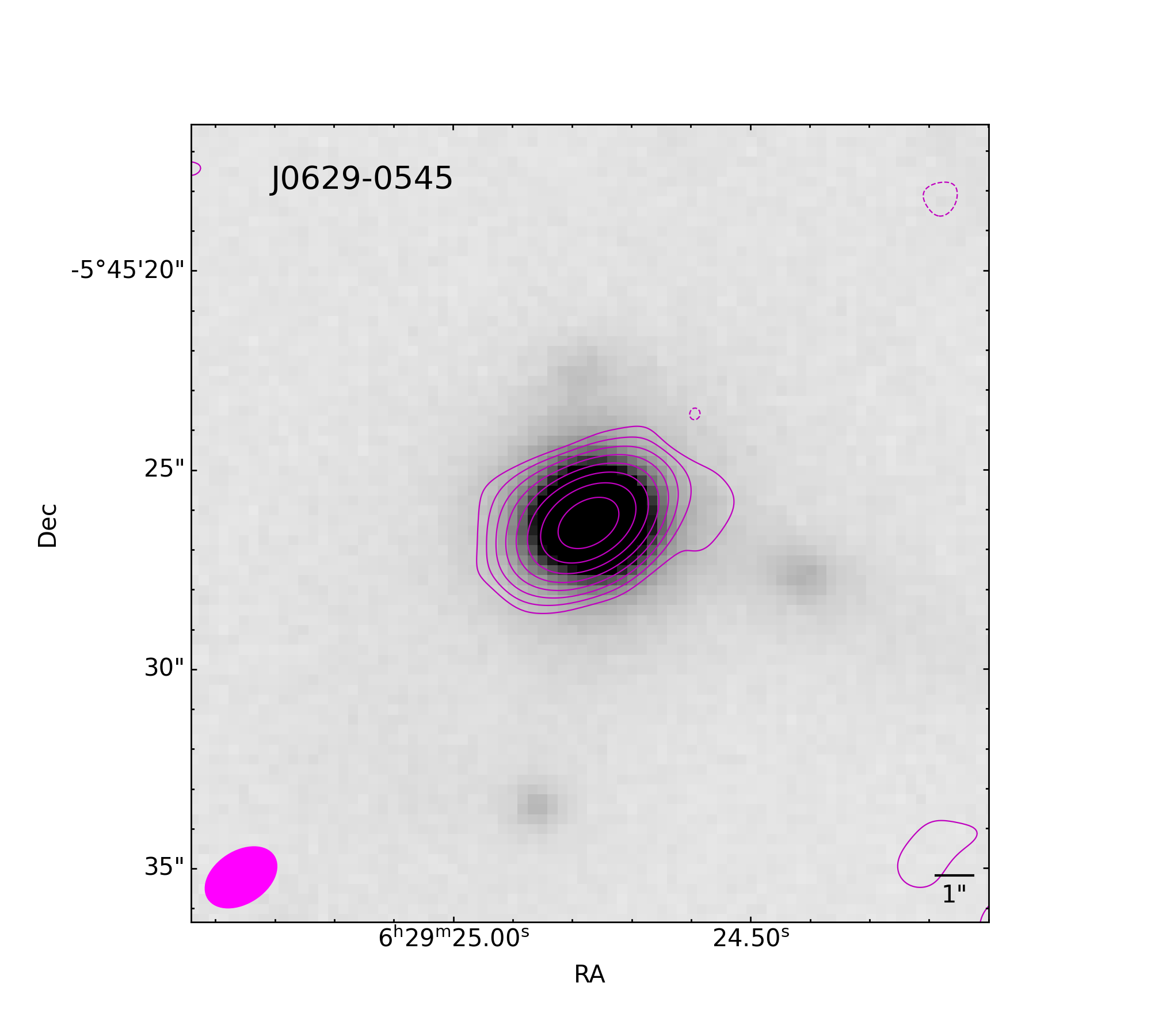}
         \caption{PanSTARRS $i$ band image of the host galaxy overlaid with the 90k$\lambda$ tapered map. Radio map properties as in Fig.~\ref{fig:J0629-90k}}. \label{fig:J0629-host}
     \end{subfigure}
        \caption{}
        \label{fig:J0629}
\end{figure*}


\begin{figure*}
     \centering
     \begin{subfigure}[b]{0.47\textwidth}
         \centering
         \includegraphics[width=\textwidth]{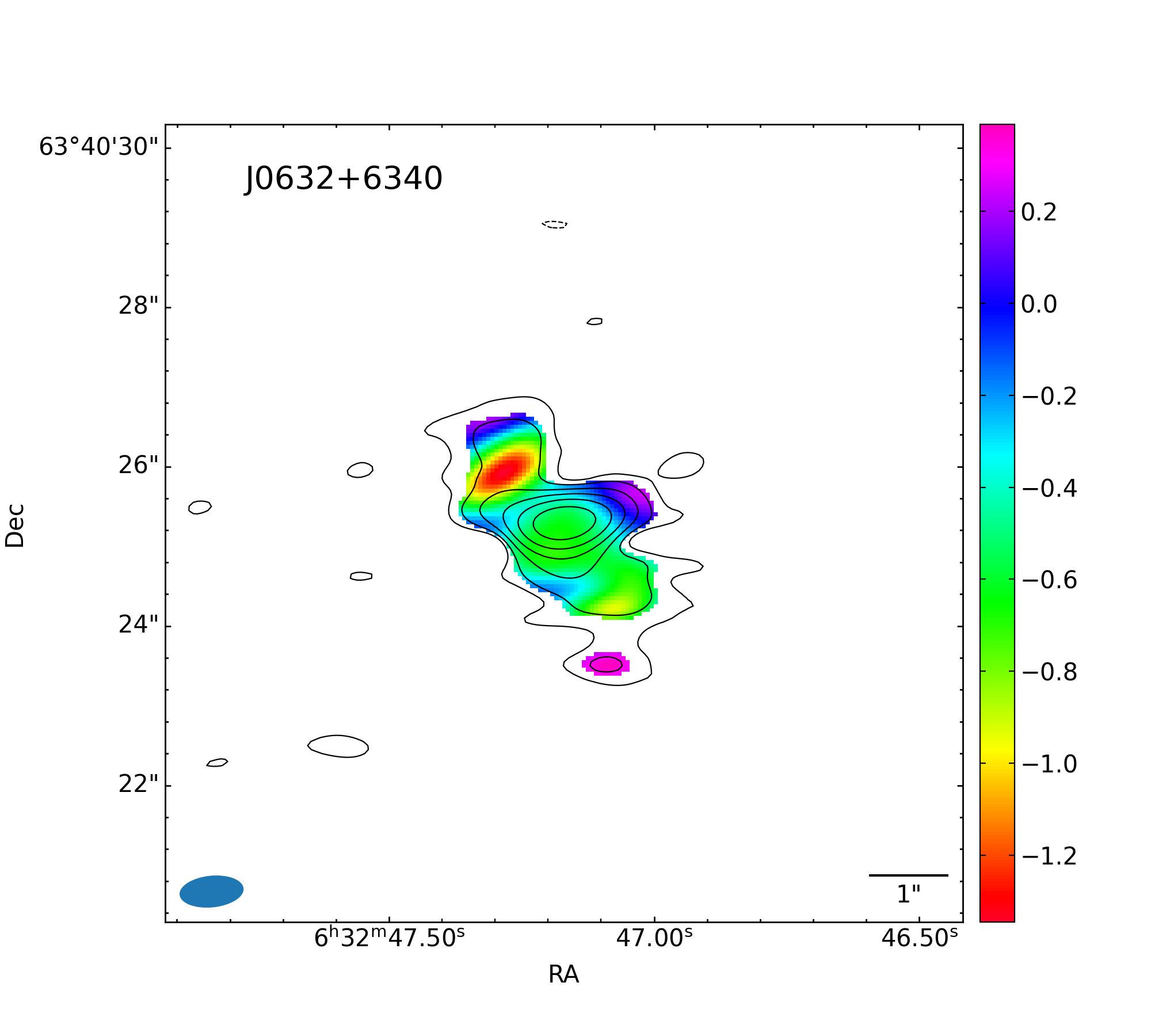}
         \caption{Spectral index map, rms = 10 $\mu$Jy beam$^{-1}$, contour levels at -3, 3 $\times$ 2$^n$, $n \in$ [0, 5], beam size 0.22 $\times$ 0.11~kpc. } \label{fig:J0632spind}
     \end{subfigure}
     \hfill
     \begin{subfigure}[b]{0.47\textwidth}
         \centering
         \includegraphics[width=\textwidth]{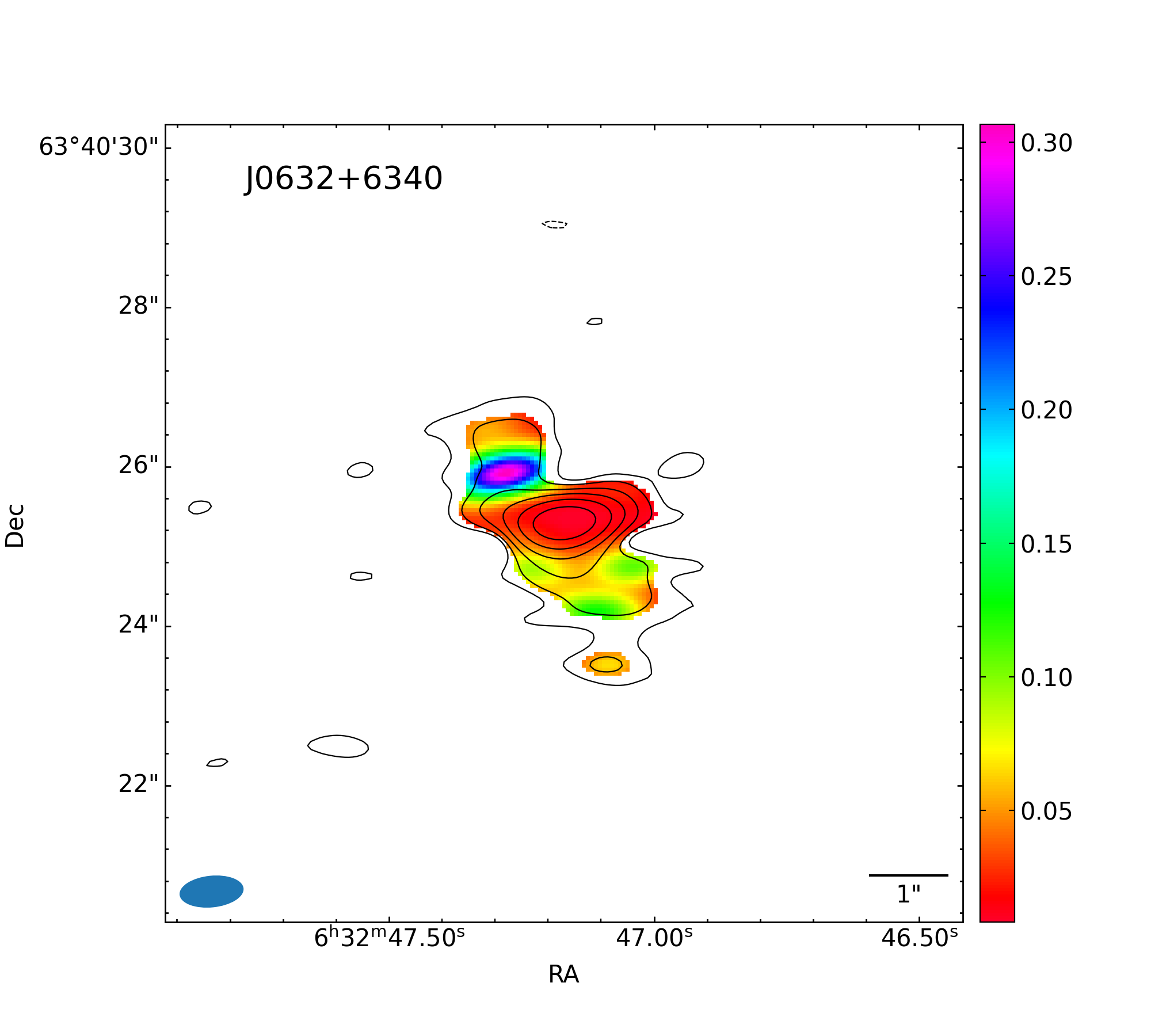}
         \caption{Spectral index error map, rms, contour levels, and beam size as in Fig.~\ref{fig:J0632spind}.} \label{fig:J0632spinderr}
     \end{subfigure}
     \hfill
     \\
     \begin{subfigure}[b]{0.47\textwidth}
         \centering
         \includegraphics[width=\textwidth]{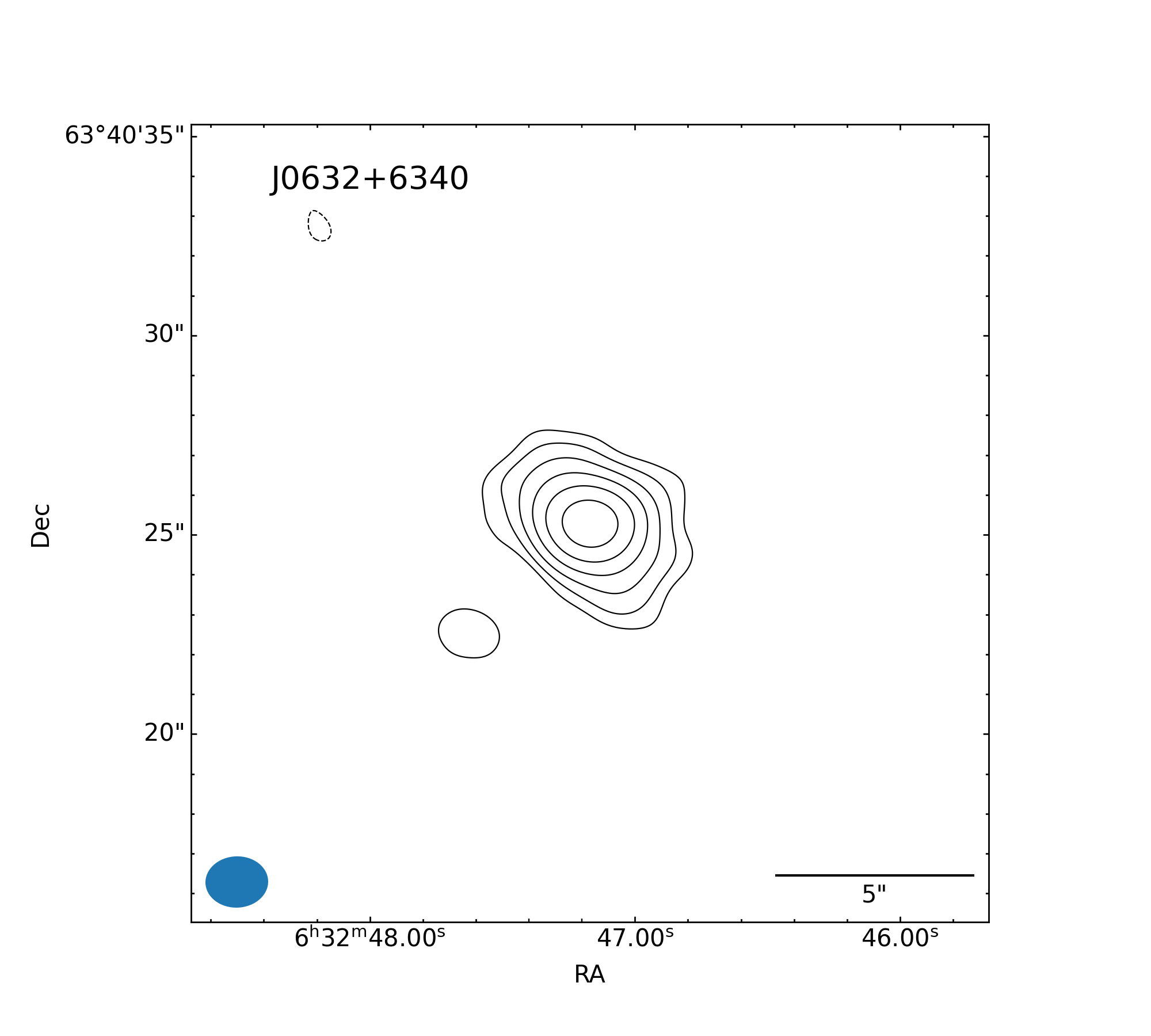}
         \caption{Tapered map with \texttt{uvtaper} = 90k$\lambda$, rms = 14$\mu$Jy beam$^{-1}$, contour levels at -3, 3 $\times$ 2$^n$, $n \in$ [0, 5], beam size 0.42 $\times$ 0.34~kpc.} \label{fig:J0632-90k}
     \end{subfigure}
          \hfill
     \begin{subfigure}[b]{0.47\textwidth}
         \centering
         \includegraphics[width=\textwidth]{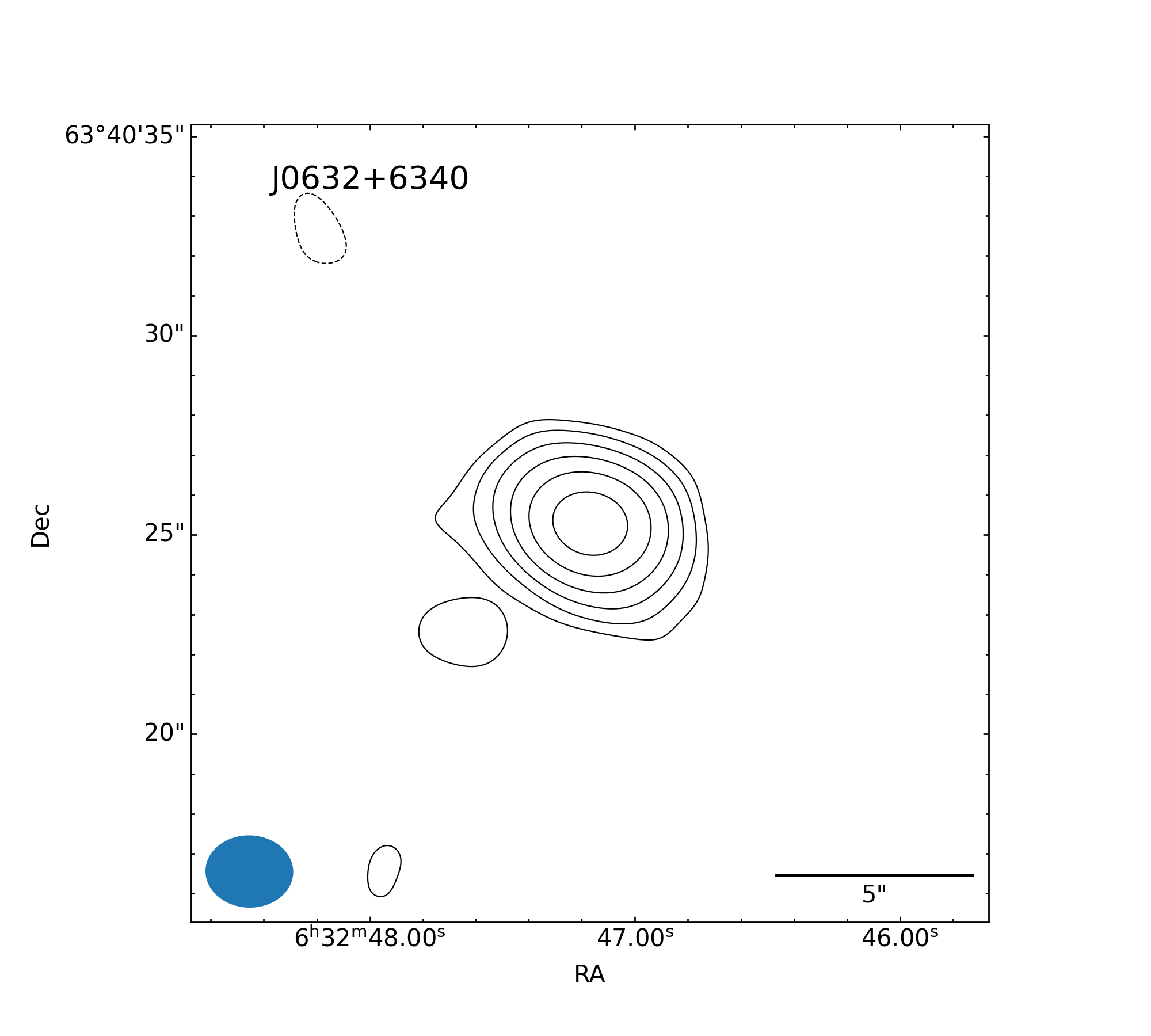}
         \caption{Tapered map with \texttt{uvtaper} = 60k$\lambda$, rms = 16$\mu$Jy beam$^{-1}$, contour levels at -3, 3 $\times$ 2$^n$, $n \in$ [0, 5], beam size 0.59 $\times$ 0.48~kpc.} \label{fig:J0632-60k}
     \end{subfigure}
  \hfill
     \\
     \begin{subfigure}[b]{0.47\textwidth}
         \centering
         \includegraphics[width=\textwidth]{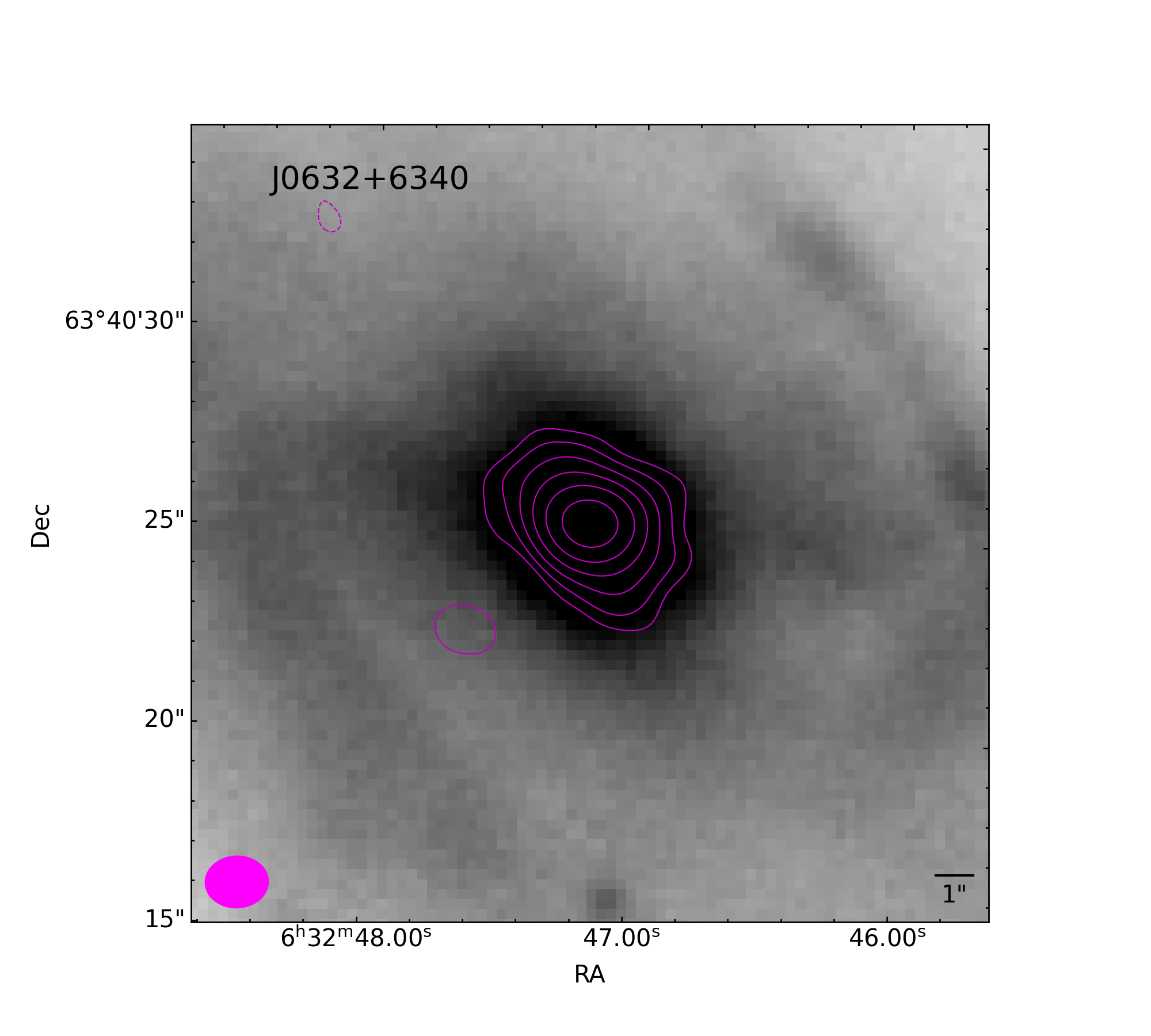}
         \caption{PanSTARRS $i$ band image of the host galaxy overlaid with the 90k$\lambda$ tapered map. Radio map properties as in Fig.~\ref{fig:J0632-90k}}. \label{fig:J0632-host}
     \end{subfigure}
        \hfill
     \begin{subfigure}[b]{0.47\textwidth}
         \centering
         \includegraphics[width=\textwidth]{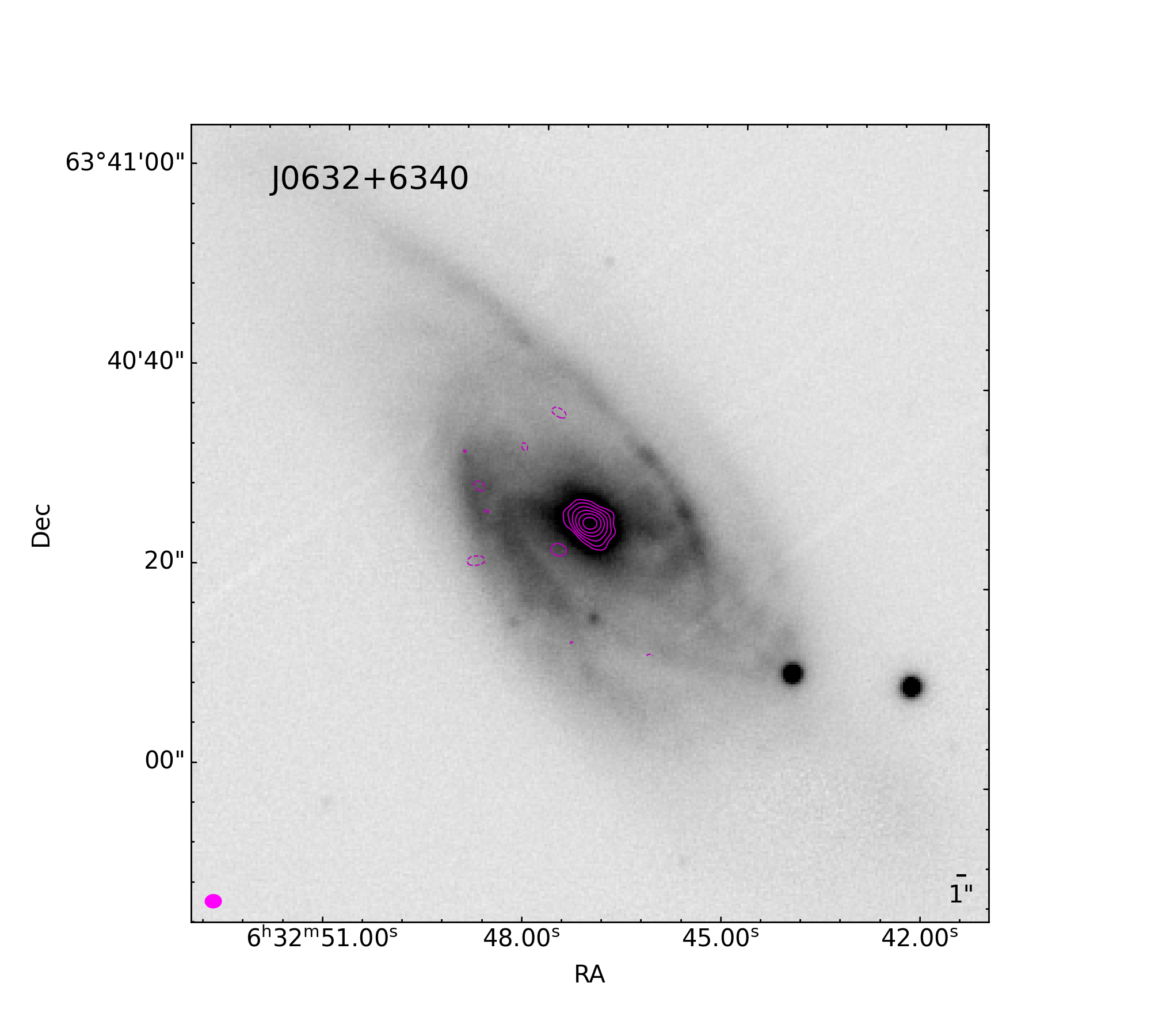}
         \caption{PanSTARRS $i$ band image of the host galaxy overlaid with the 90k$\lambda$ tapered map. Radio map properties as in Fig.~\ref{fig:J0632-90k}}. \label{fig:J0632-host-zoom}
     \end{subfigure}
        \caption{}
        \label{fig:J0632}
\end{figure*}


\begin{figure*}
     \centering
     \begin{subfigure}[b]{0.47\textwidth}
         \centering
         \includegraphics[width=\textwidth]{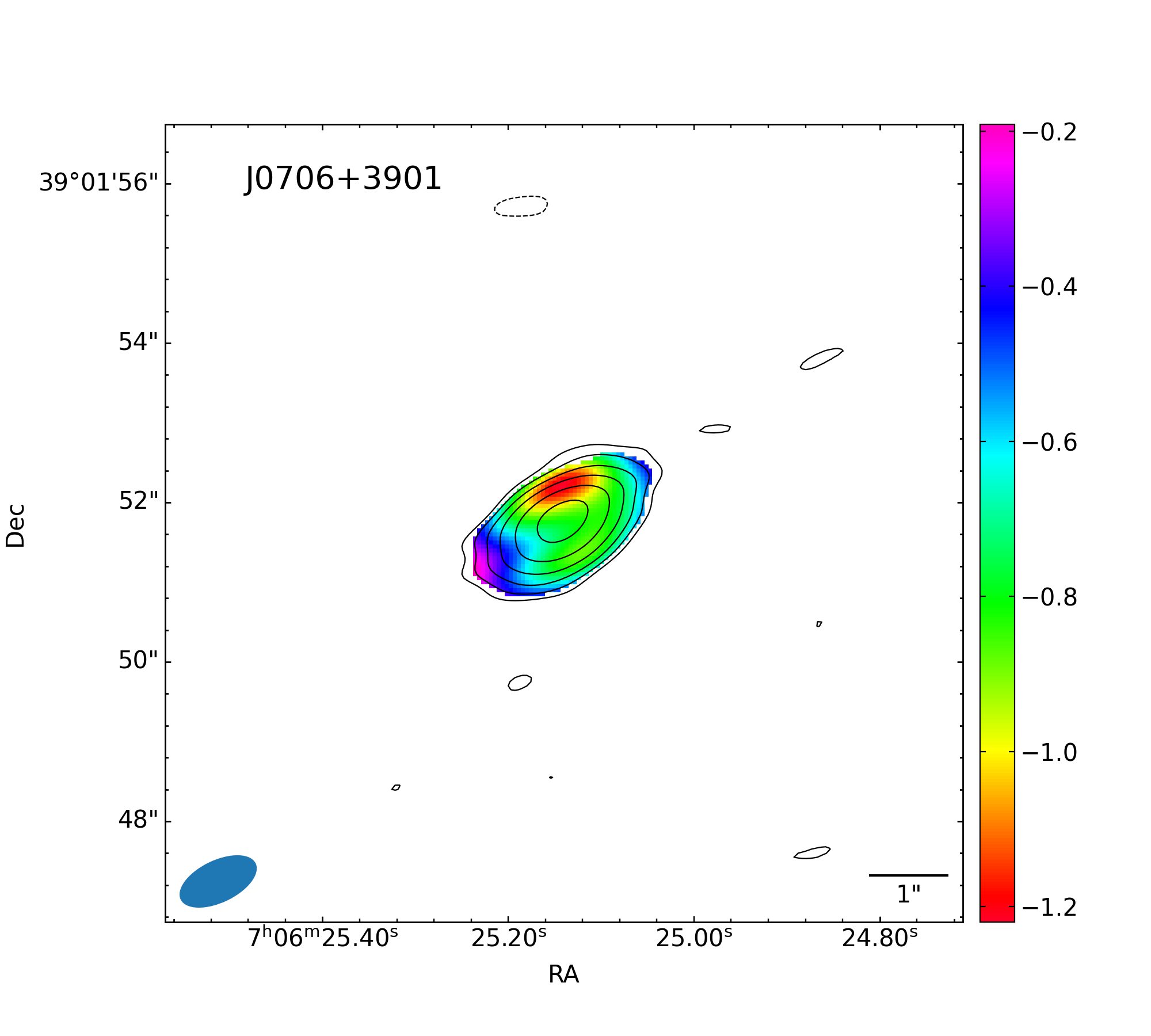}
         \caption{Spectral index map, rms = 13$\mu$Jy beam$^{-1}$, contour levels at -3, 3 $\times$ 2$^n$, $n \in$ [0, 5], beam size 1.68 $\times$ 0.85~kpc. } \label{fig:J0706spind}
     \end{subfigure}
     \hfill
     \begin{subfigure}[b]{0.47\textwidth}
         \centering
         \includegraphics[width=\textwidth]{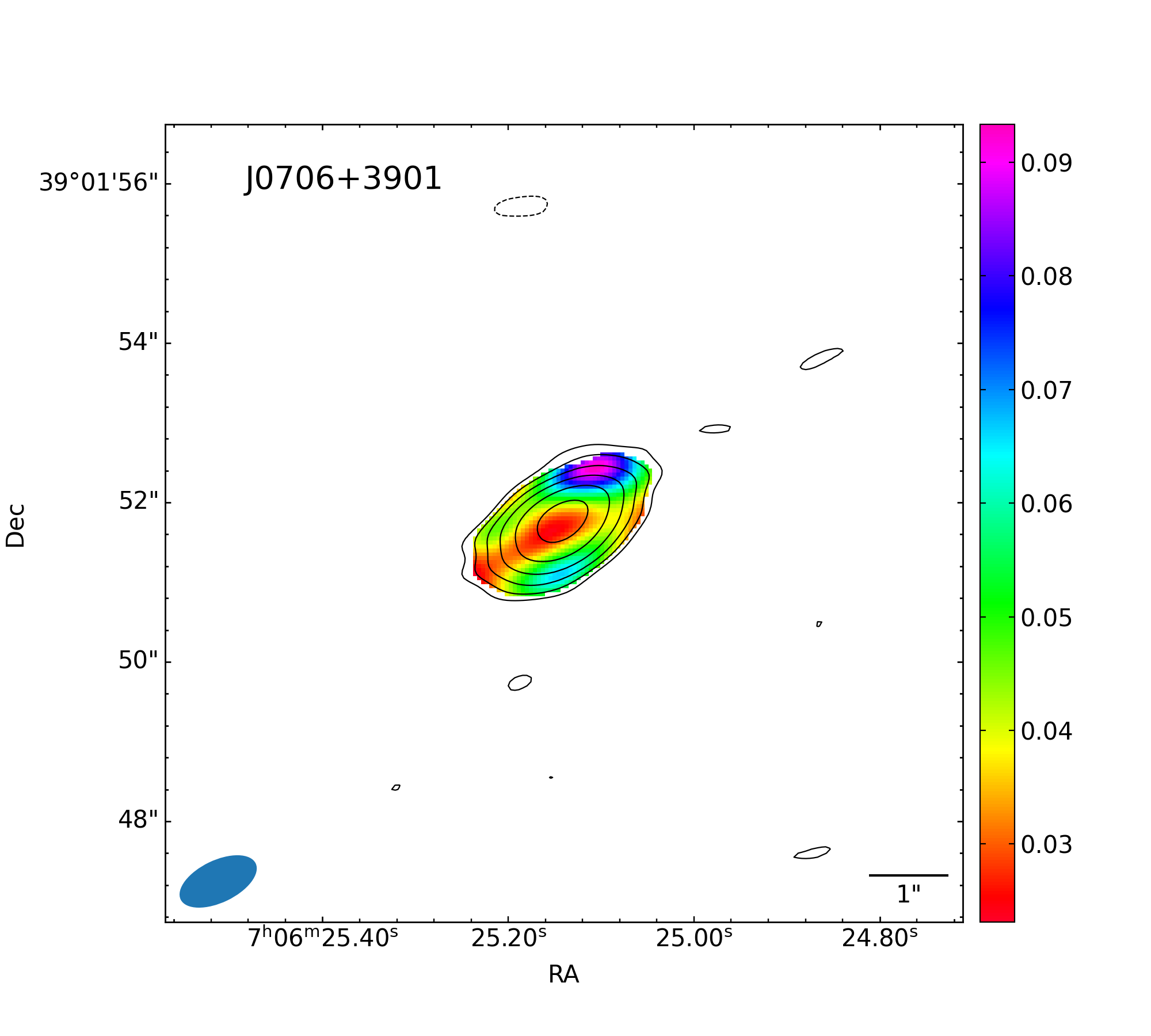}
         \caption{Spectral index error map, rms, contour levels, and beam size as in Fig.~\ref{fig:J0706spind}.} \label{fig:J0706spinderr}
     \end{subfigure}
     \hfill
     \\
     \begin{subfigure}[b]{0.47\textwidth}
         \centering
         \includegraphics[width=\textwidth]{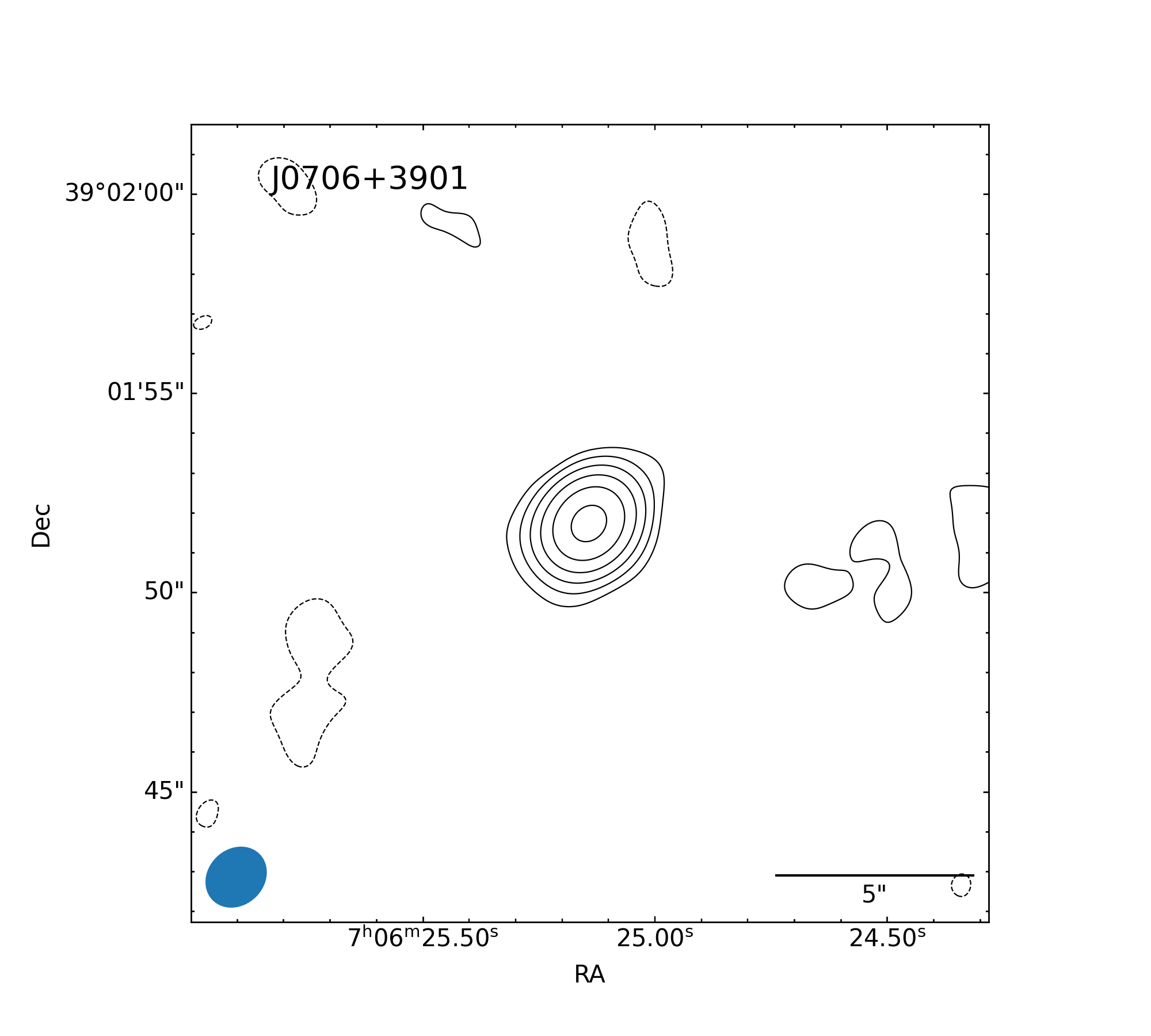}
         \caption{Tapered map with \texttt{uvtaper} = 90k$\lambda$, rms = 17$\mu$Jy beam$^{-1}$, contour levels at -3, 3 $\times$ 2$^n$, $n \in$ [0, 5], beam size 2.66 $\times$ 2.27~kpc.} \label{fig:J0706-90k}
     \end{subfigure}
          \hfill
     \begin{subfigure}[b]{0.47\textwidth}
         \centering
         \includegraphics[width=\textwidth]{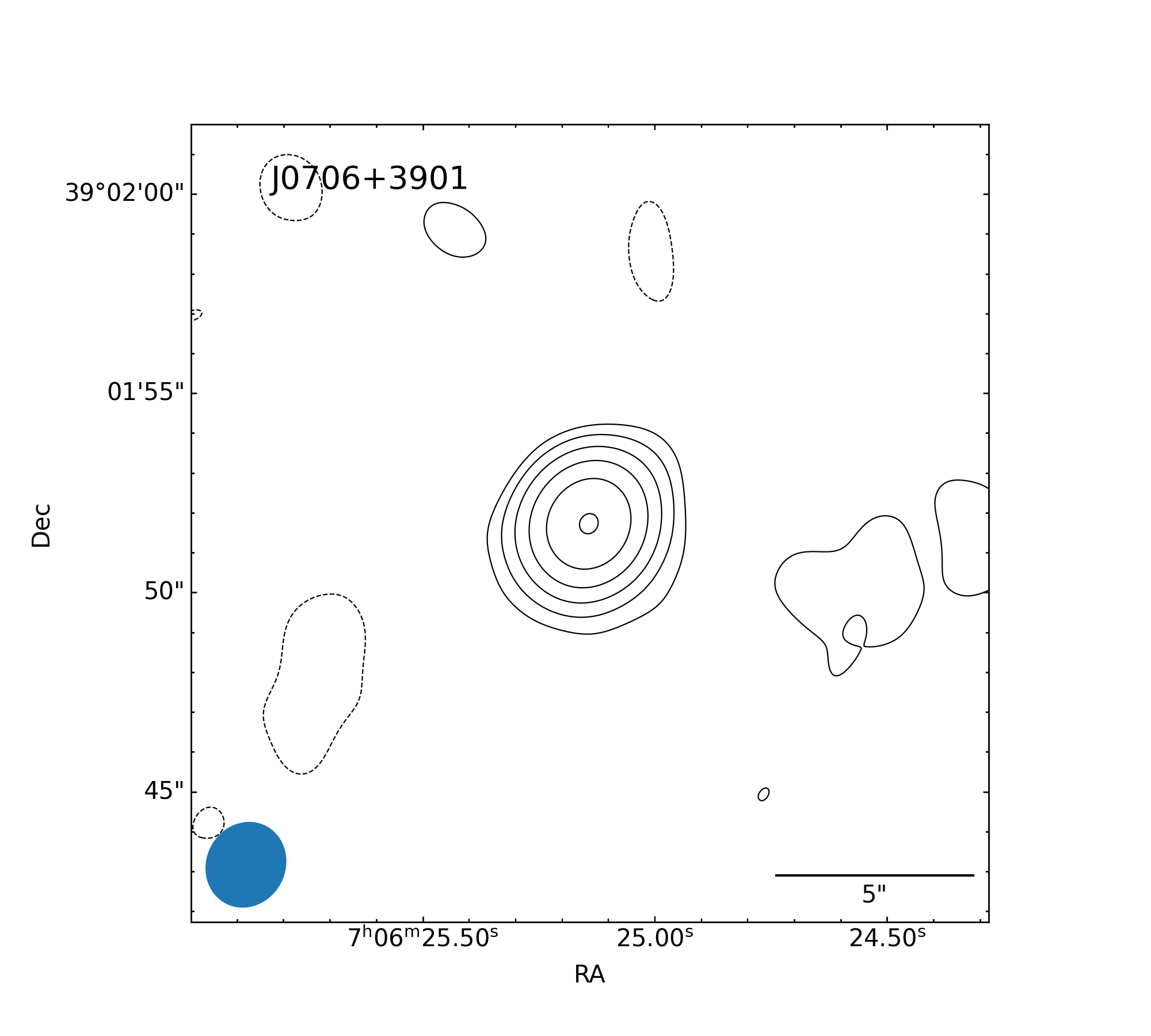}
         \caption{Tapered map with \texttt{uvtaper} = 60k$\lambda$, rms = 21$\mu$Jy beam$^{-1}$, contour levels at -3, 3 $\times$ 2$^n$, $n \in$ [0, 5], beam size 3.53 $\times$ 3.21~kpc.} \label{fig:J0706-60k}
     \end{subfigure}
    \hfill
     \\
     \begin{subfigure}[b]{0.47\textwidth}
         \centering
         \includegraphics[width=\textwidth]{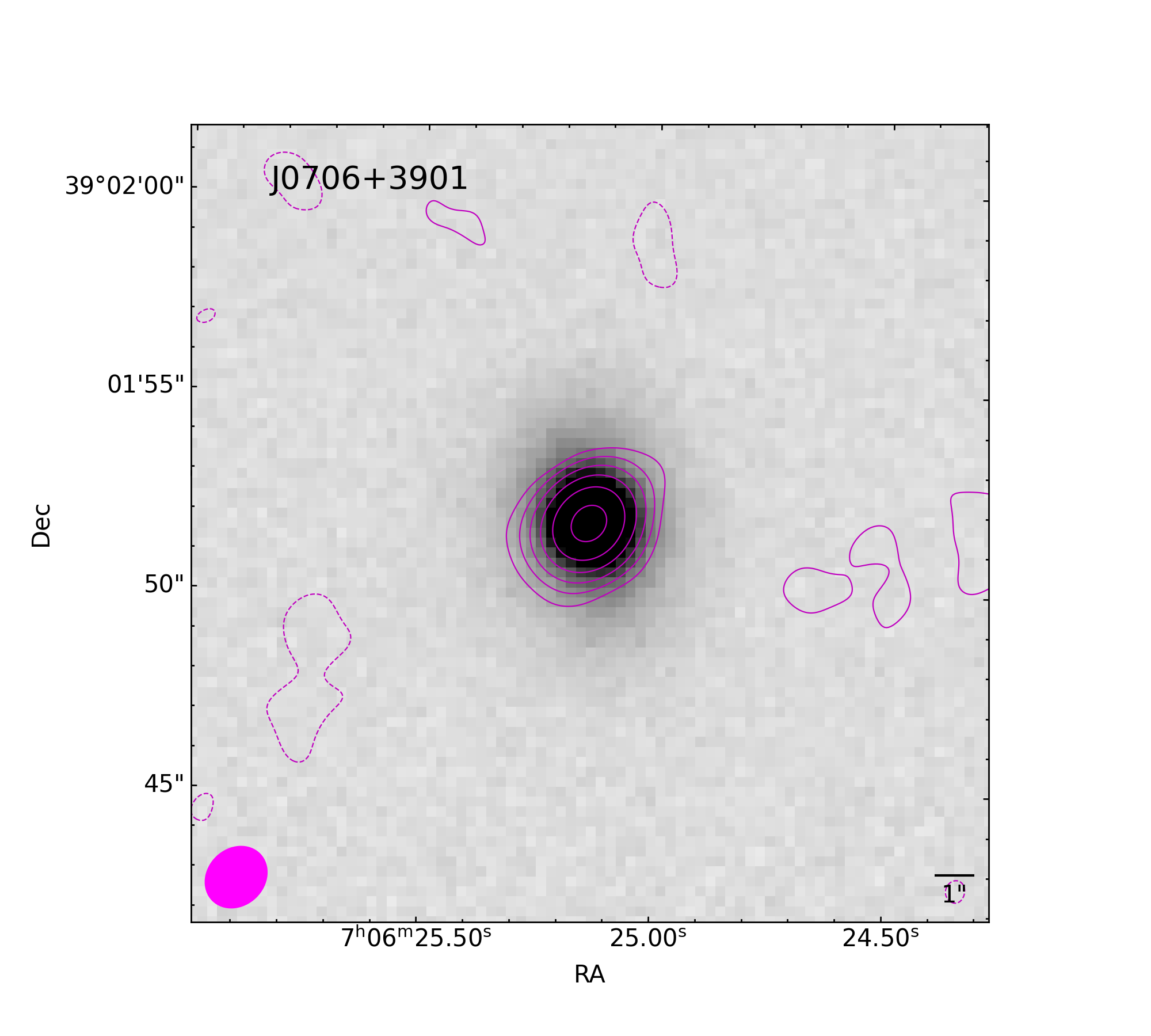}
         \caption{PanSTARRS $i$ band image of the host galaxy overlaid with the 90k$\lambda$ tapered map. Radio map properties as in Fig.~\ref{fig:J0706-90k}}. \label{fig:J0706-host}
     \end{subfigure}
        \caption{}
        \label{fig:J0706}
\end{figure*}


\begin{figure*}
     \centering
     \begin{subfigure}[b]{0.47\textwidth}
         \centering
         \includegraphics[width=\textwidth]{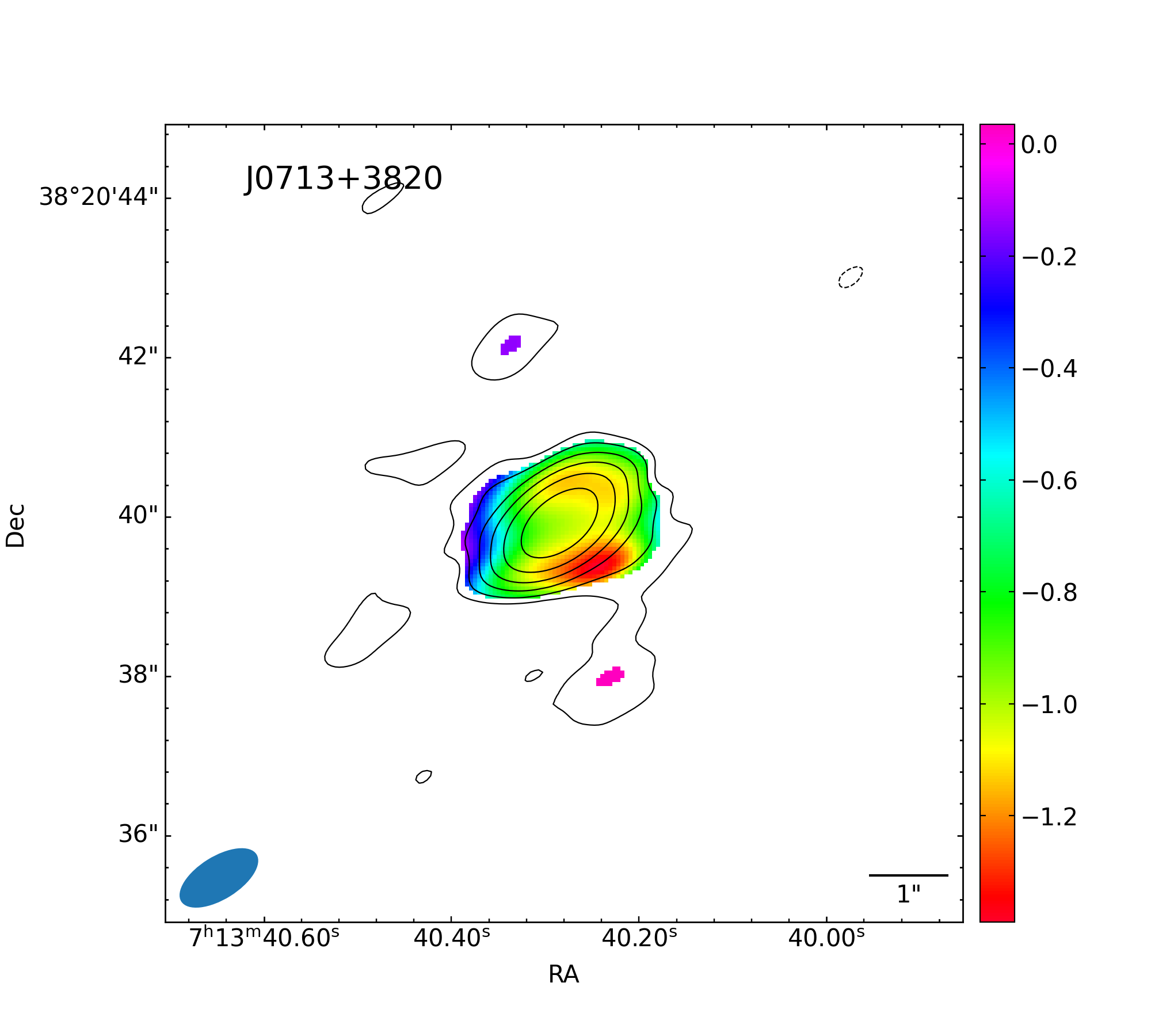}
         \caption{Spectral index map, rms = 13$\mu$Jy beam$^{-1}$, contour levels at -3, 3 $\times$ 2$^n$, $n \in$ [0, 5], beam size 2.45 $\times$ 1.19~kpc. } \label{fig:J0713spind}
     \end{subfigure}
     \hfill
     \begin{subfigure}[b]{0.47\textwidth}
         \centering
         \includegraphics[width=\textwidth]{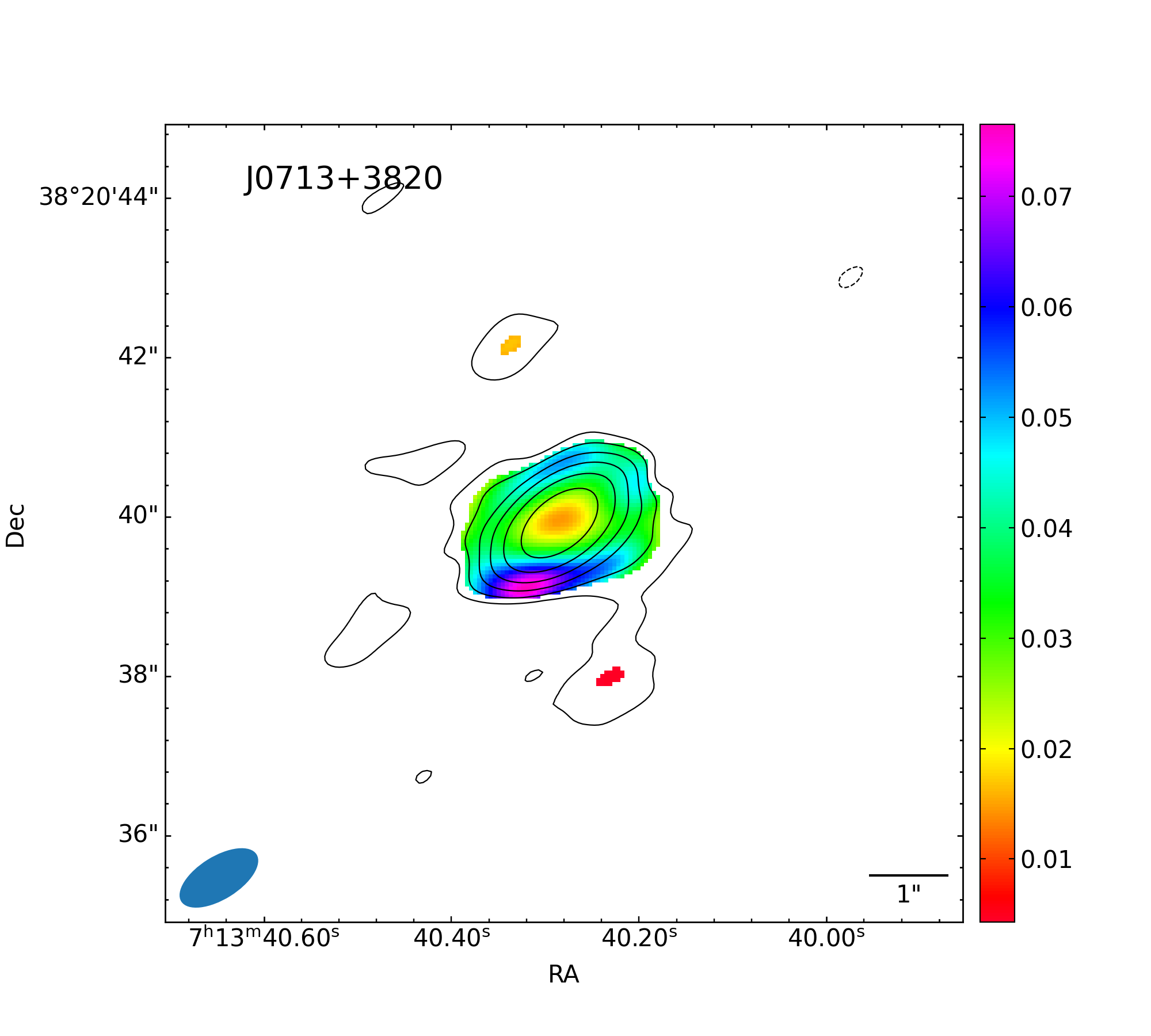}
         \caption{Spectral index error map, rms, contour levels, and beam size as in Fig.~\ref{fig:J0713spind}.} \label{fig:J0713spinderr}
     \end{subfigure}
     \hfill
     \\
     \begin{subfigure}[b]{0.47\textwidth}
         \centering
         \includegraphics[width=\textwidth]{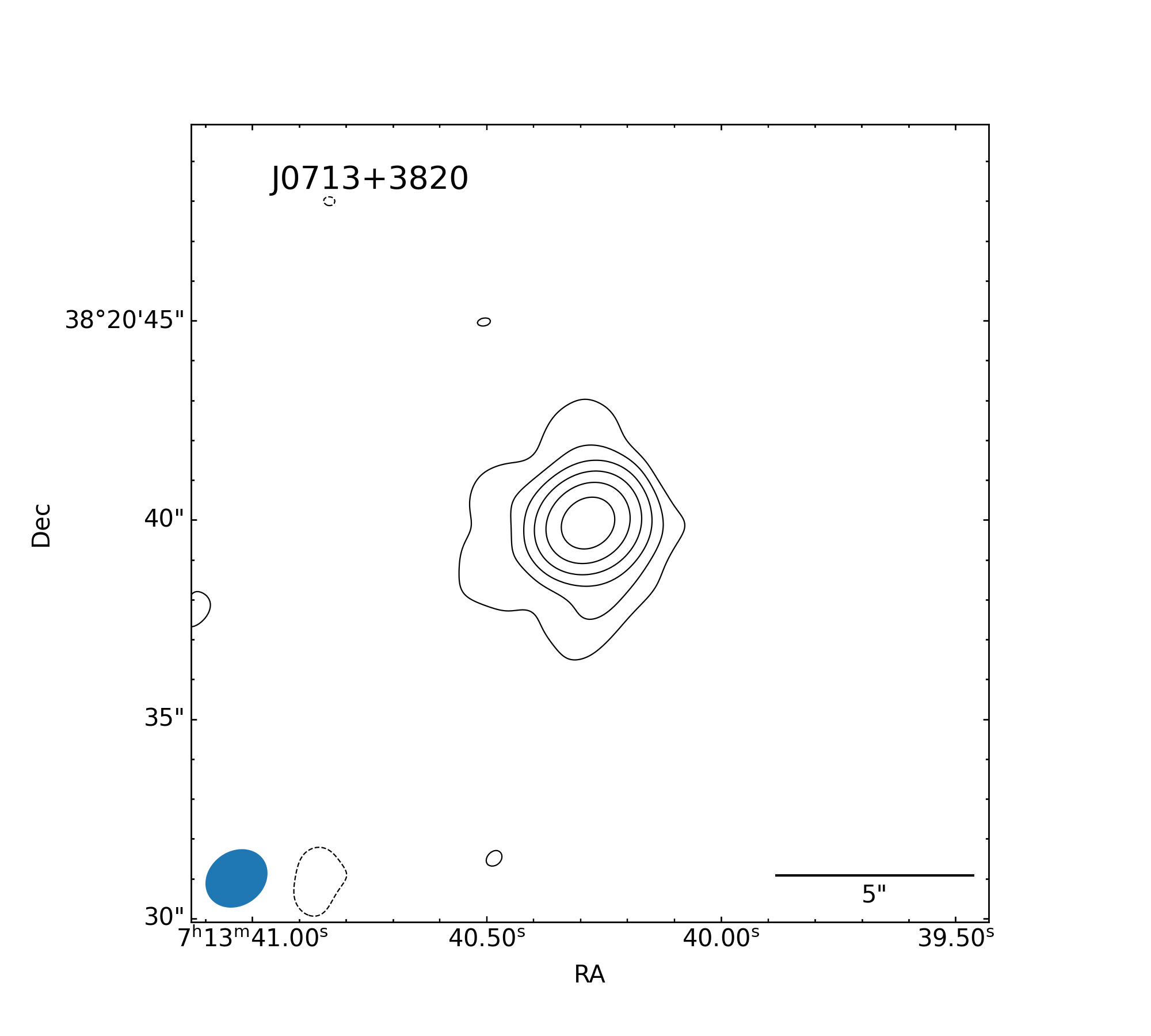}
         \caption{Tapered map with \texttt{uvtaper} = 90k$\lambda$, rms = 19$\mu$Jy beam$^{-1}$, contour levels at -3, 3 $\times$ 2$^n$, $n \in$ [0, 5], beam size 3.65 $\times$ 3.01~kpc.} \label{fig:J0713-90k}
     \end{subfigure}
          \hfill
     \begin{subfigure}[b]{0.47\textwidth}
         \centering
         \includegraphics[width=\textwidth]{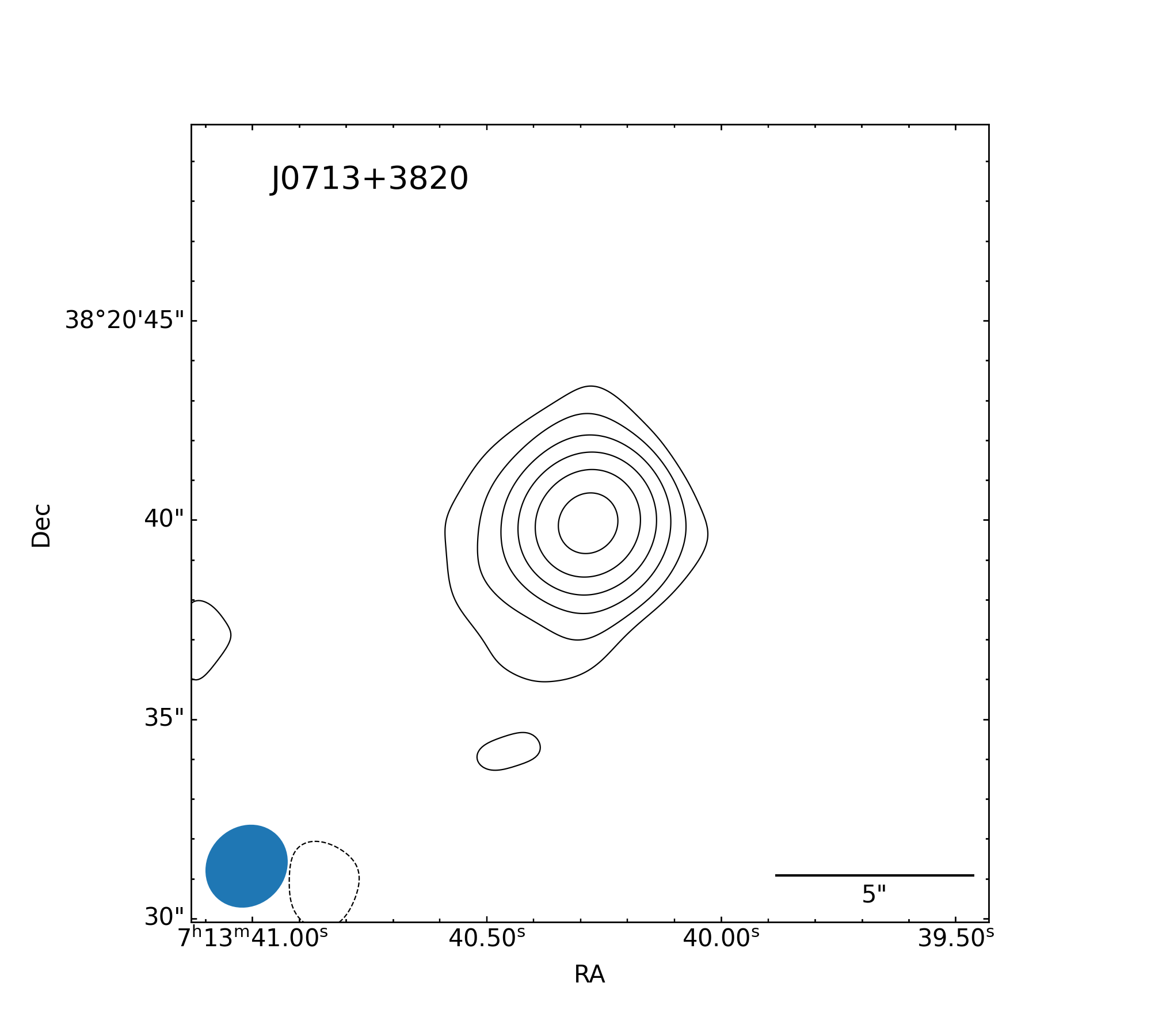}
         \caption{Tapered map with \texttt{uvtaper} = 60k$\lambda$, rms = 23$\mu$Jy beam$^{-1}$, contour levels at -3, 3 $\times$ 2$^n$, $n \in$ [0, 5], beam size 4.82 $\times$ 4.33~kpc.} \label{fig:J0713-60k}
     \end{subfigure}
          \hfill
     \\
     \begin{subfigure}[b]{0.47\textwidth}
         \centering
         \includegraphics[width=\textwidth]{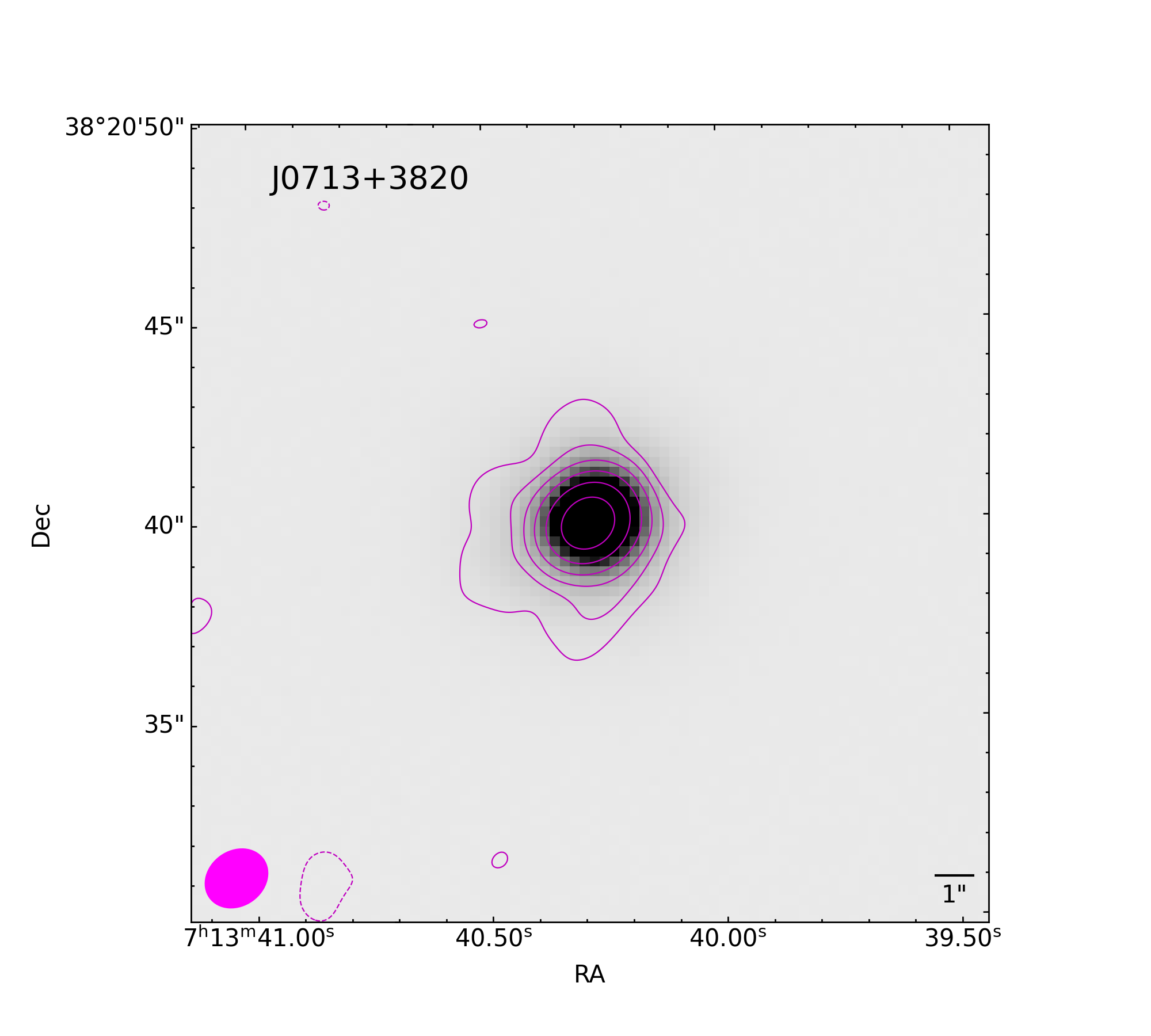}
         \caption{PanSTARRS $i$ band image of the host galaxy overlaid with the 90k$\lambda$ tapered map. Radio map properties as in Fig.~\ref{fig:J0713-90k}}. \label{fig:J0713-host}
     \end{subfigure}
        \caption{}
        \label{fig:J0713}
\end{figure*}


\begin{figure*}
     \centering
     \begin{subfigure}[b]{0.47\textwidth}
         \centering
         \includegraphics[width=\textwidth]{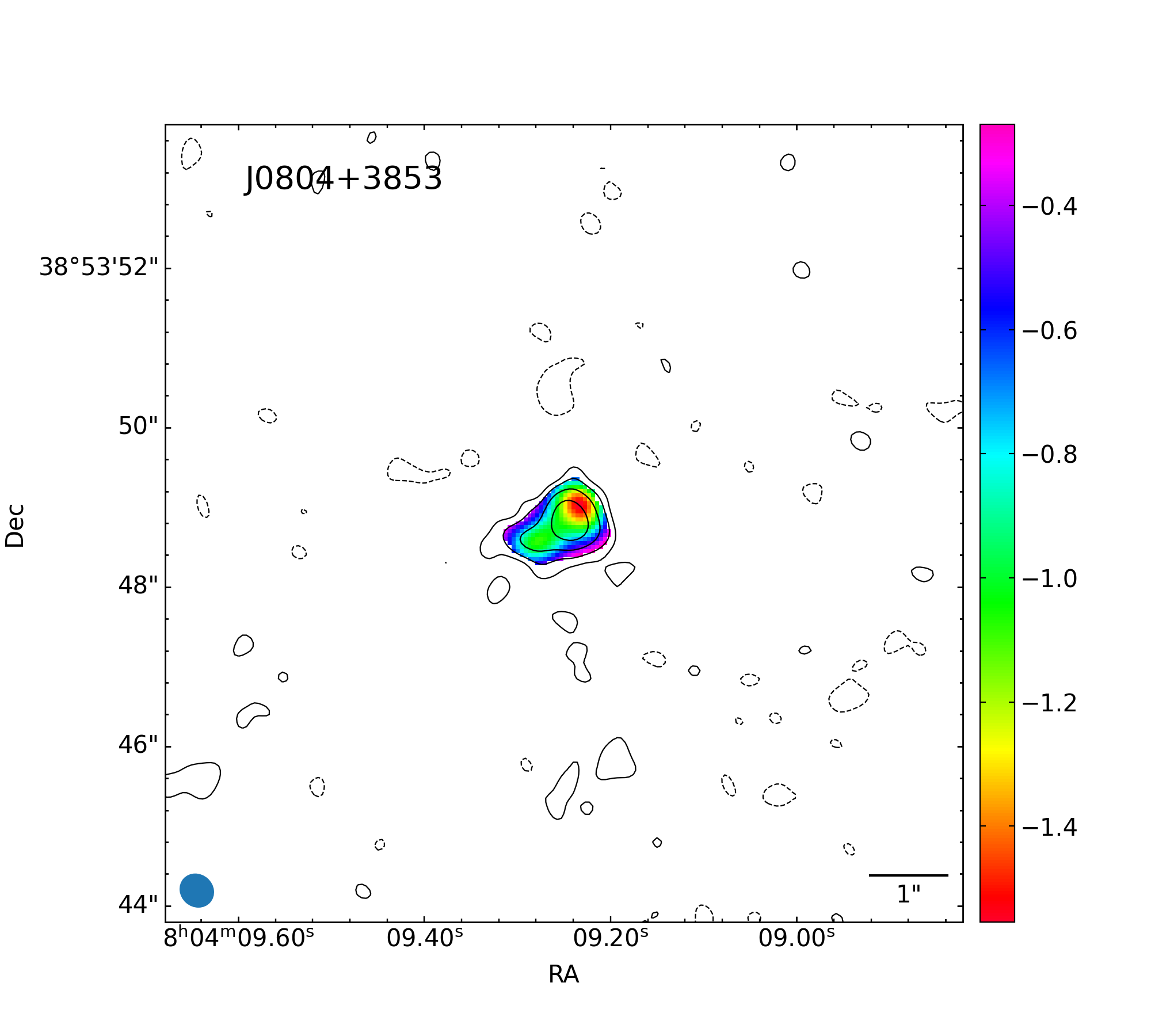}
         \caption{Spectral index map, rms = 10$\mu$Jy beam$^{-1}$, contour levels at -3, 3 $\times$ 2$^n$, $n \in$ [0, 3], beam size 1.55 $\times$ 1.42~kpc. } \label{fig:J0804spind}
     \end{subfigure}
     \hfill
     \begin{subfigure}[b]{0.47\textwidth}
         \centering
         \includegraphics[width=\textwidth]{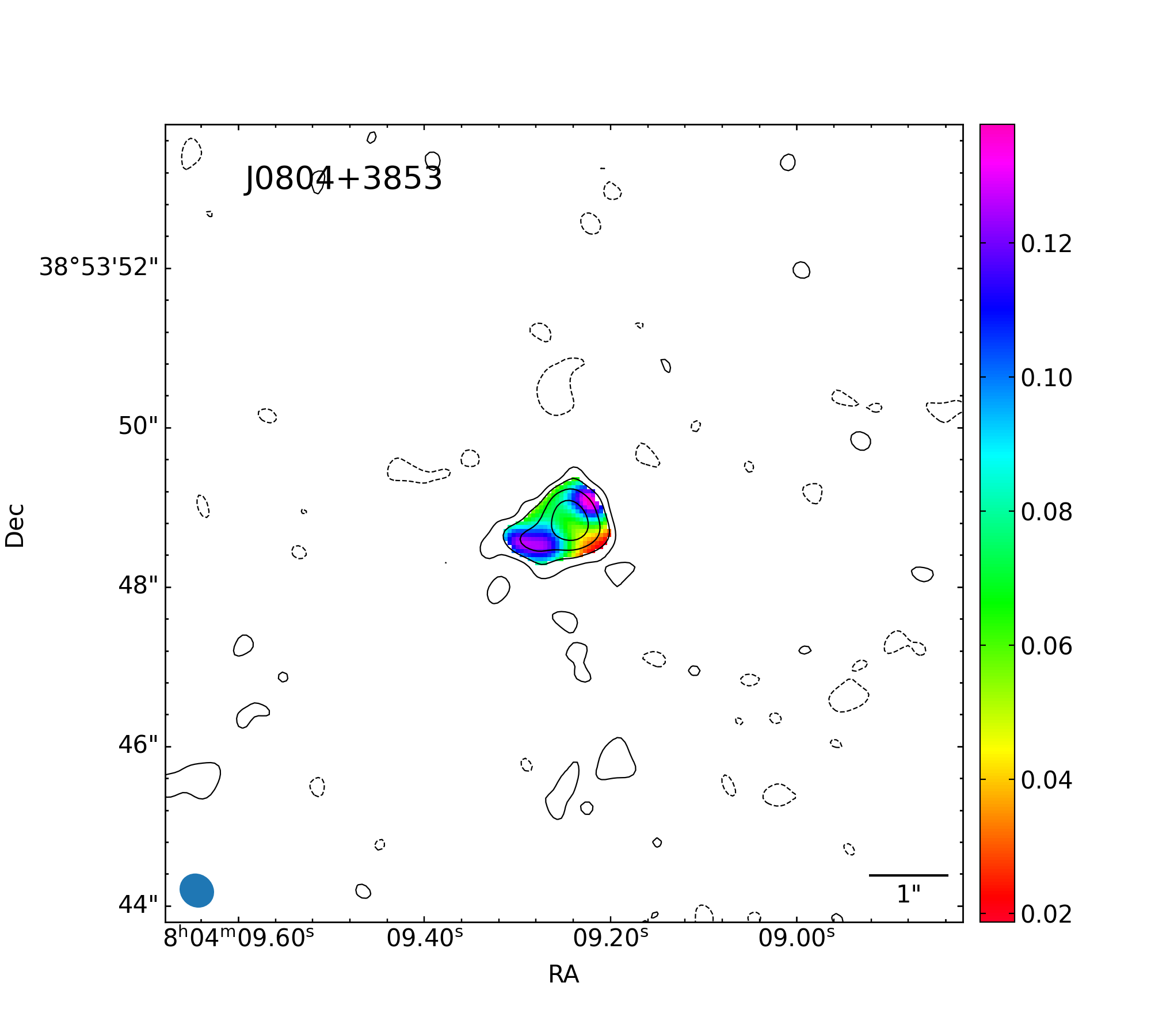}
         \caption{Spectral index error map, rms, contour levels, and beam size as in Fig.~\ref{fig:J0804spind}.} \label{fig:J0804spinderr}
     \end{subfigure}
     \hfill
     \\
     \begin{subfigure}[b]{0.47\textwidth}
         \centering
         \includegraphics[width=\textwidth]{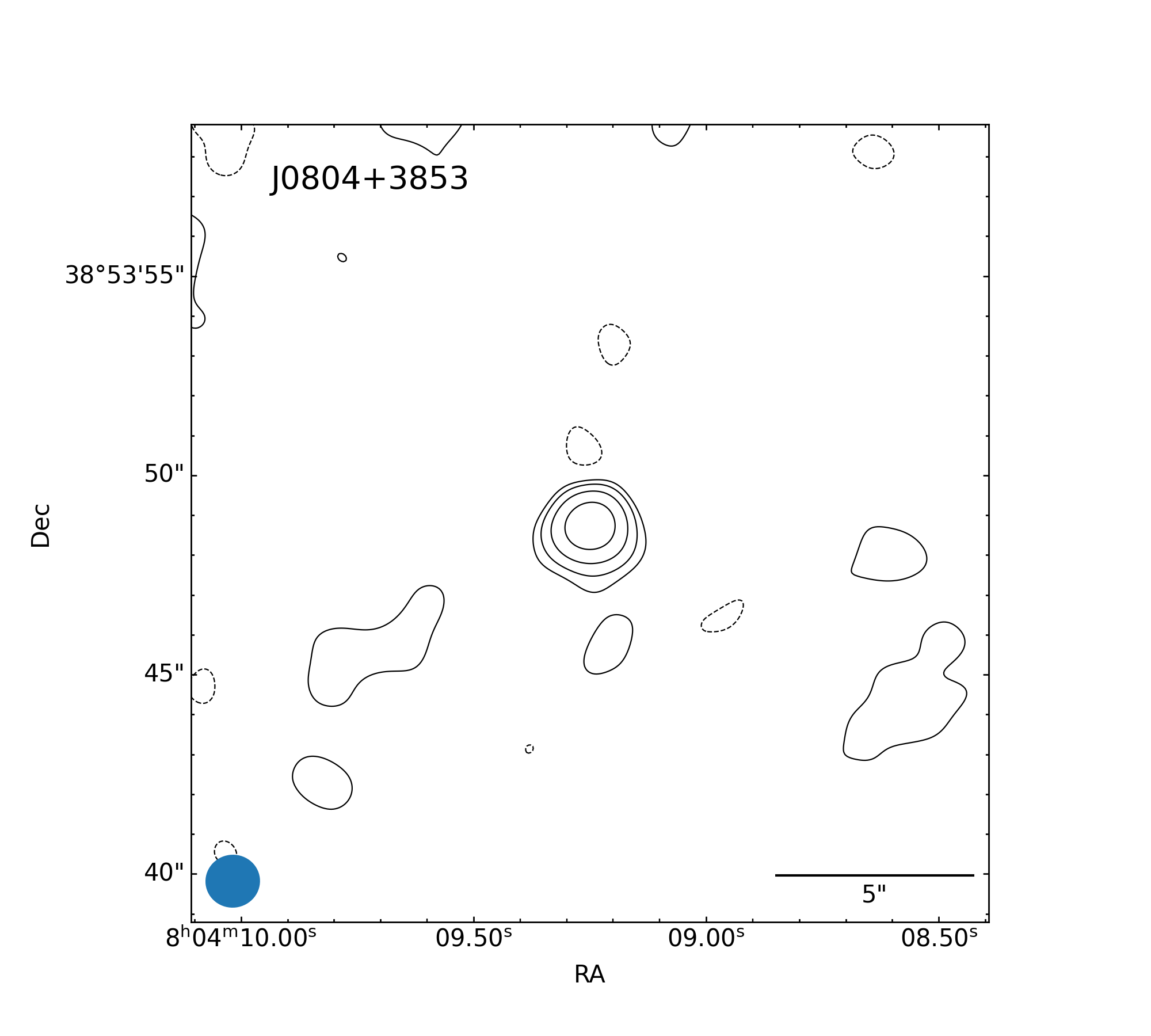}
         \caption{Tapered map with \texttt{uvtaper} = 90k$\lambda$, rms = 18$\mu$Jy beam$^{-1}$, contour levels at -3, 3 $\times$ 2$^n$, $n \in$ [0, 3], beam size 4.73 $\times$ 4.59~kpc.} \label{fig:J0804-90k}
     \end{subfigure}
          \hfill
     \begin{subfigure}[b]{0.47\textwidth}
         \centering
         \includegraphics[width=\textwidth]{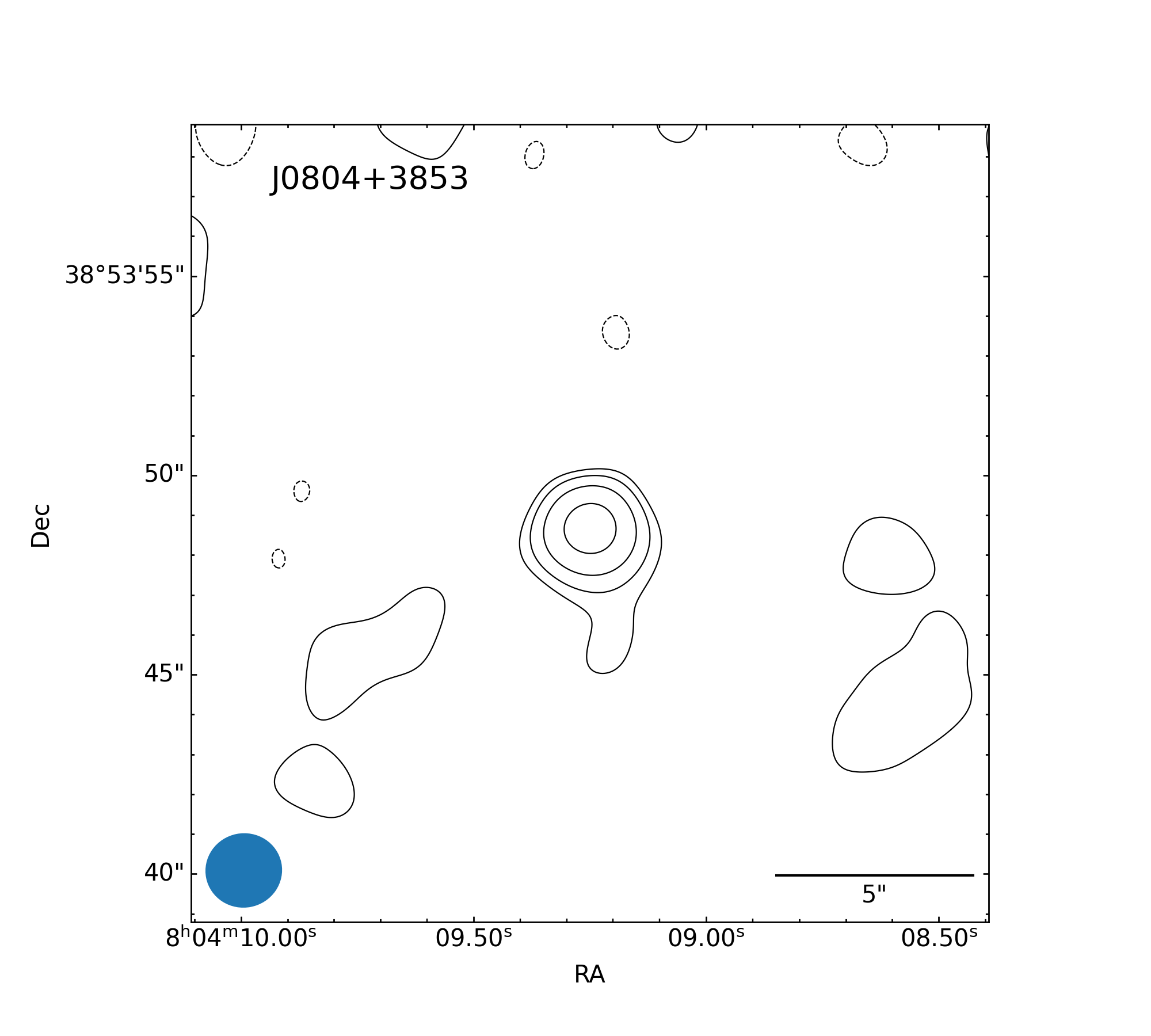}
         \caption{Tapered map with \texttt{uvtaper} = 60k$\lambda$, rms = 23$\mu$Jy beam$^{-1}$, contour levels at -3, 3 $\times$ 2$^n$, $n \in$ [0, 3], beam size 6.66 $\times$ 6.46~kpc.} \label{fig:J0804-60k}
     \end{subfigure}
          \hfill
     \\
     \begin{subfigure}[b]{0.47\textwidth}
         \centering
         \includegraphics[width=\textwidth]{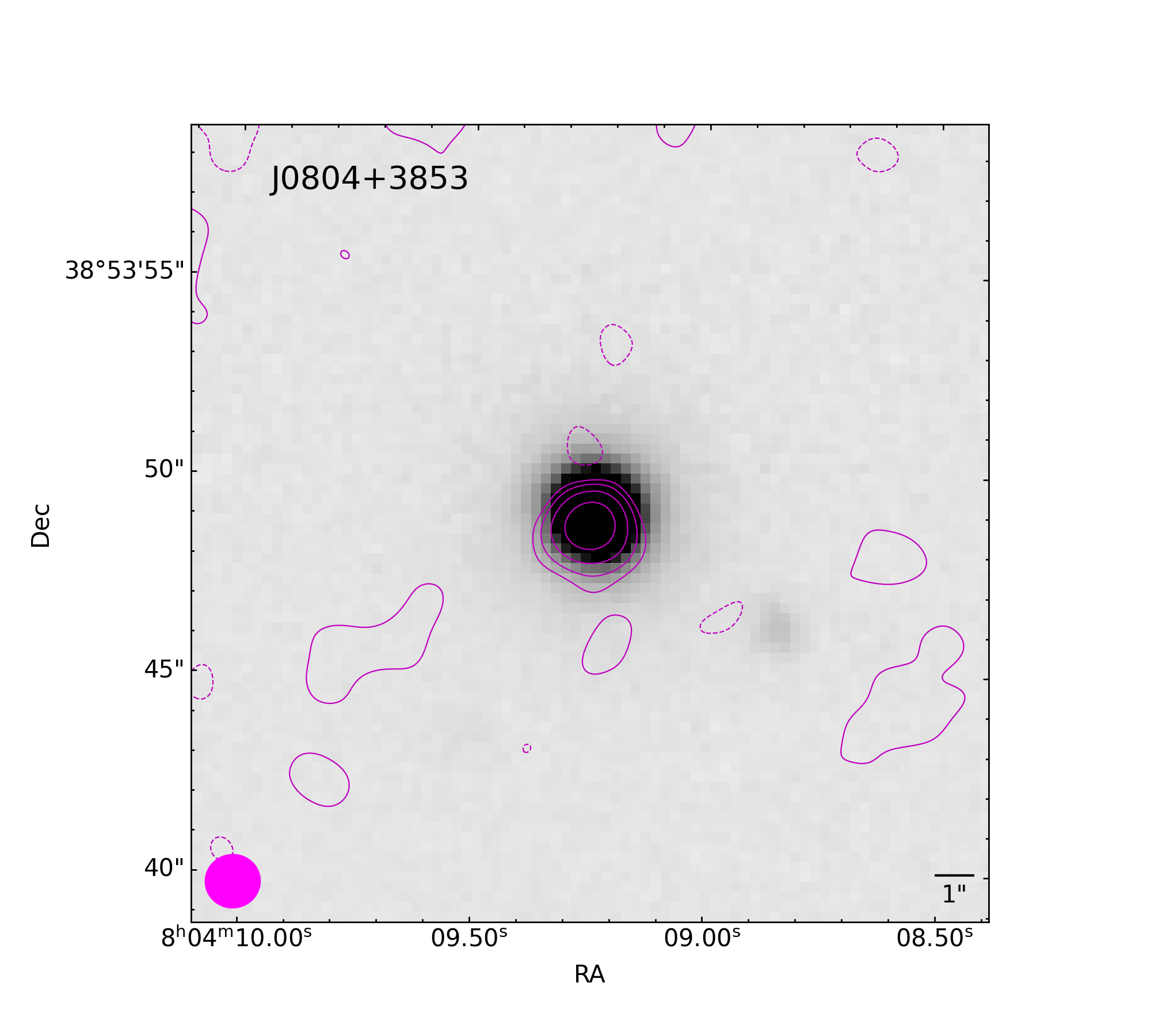}
         \caption{PanSTARRS $i$ band image of the host galaxy overlaid with the 90k$\lambda$ tapered map. Radio map properties as in Fig.~\ref{fig:J0804-90k}}. \label{fig:J0804-host}
     \end{subfigure}
        \caption{}
        \label{fig:J0804}
\end{figure*}


\begin{figure*}
     \centering
     \begin{subfigure}[b]{0.47\textwidth}
         \centering
         \includegraphics[width=\textwidth]{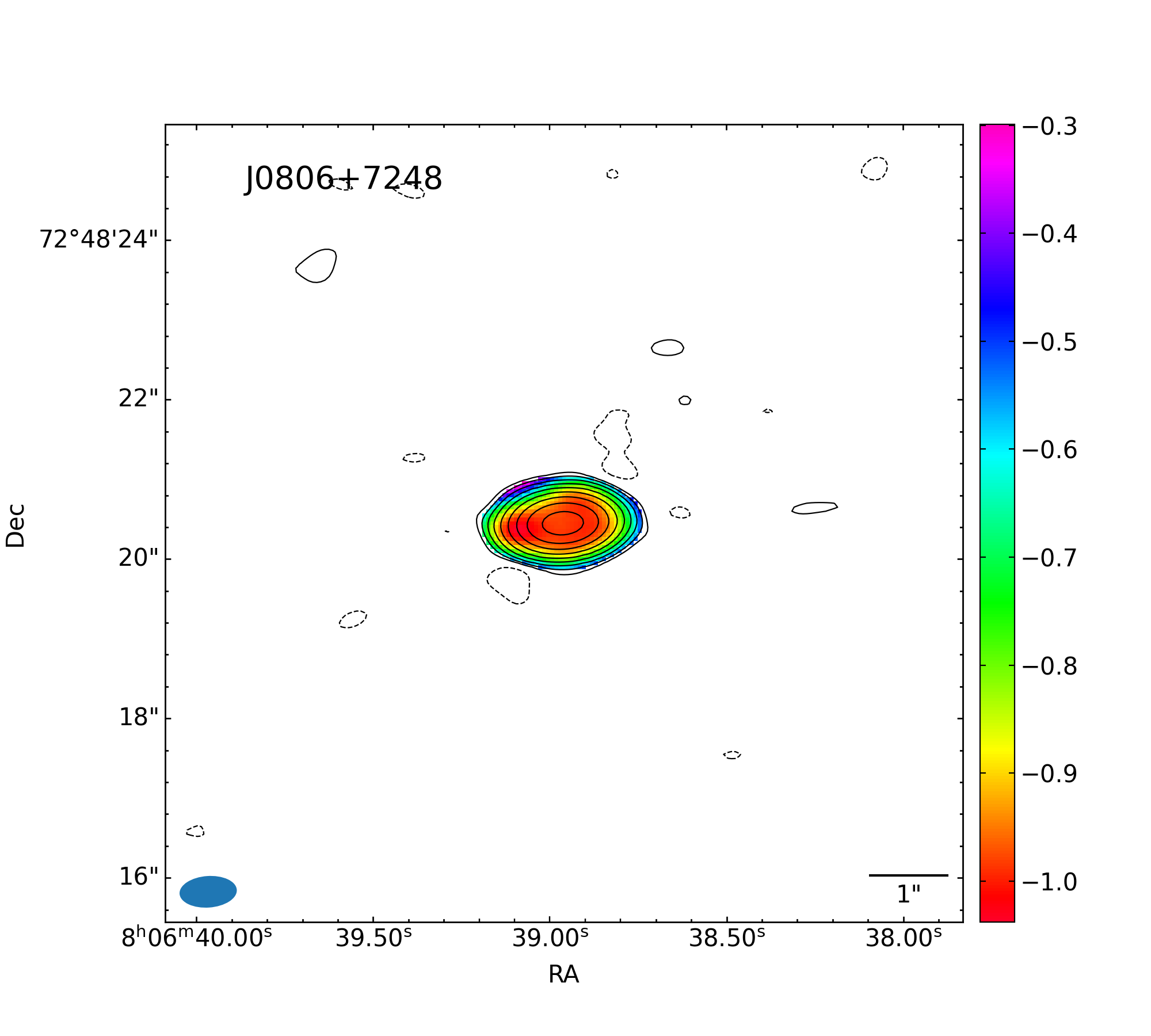}
         \caption{Spectral index map, rms = 10$\mu$Jy beam$^{-1}$, contour levels at -3, 3 $\times$ 2$^n$, $n \in$ [0, 8], beam size 1.30 $\times$ 0.72~kpc. } \label{fig:J0806spind}
     \end{subfigure}
     \hfill
     \begin{subfigure}[b]{0.47\textwidth}
         \centering
         \includegraphics[width=\textwidth]{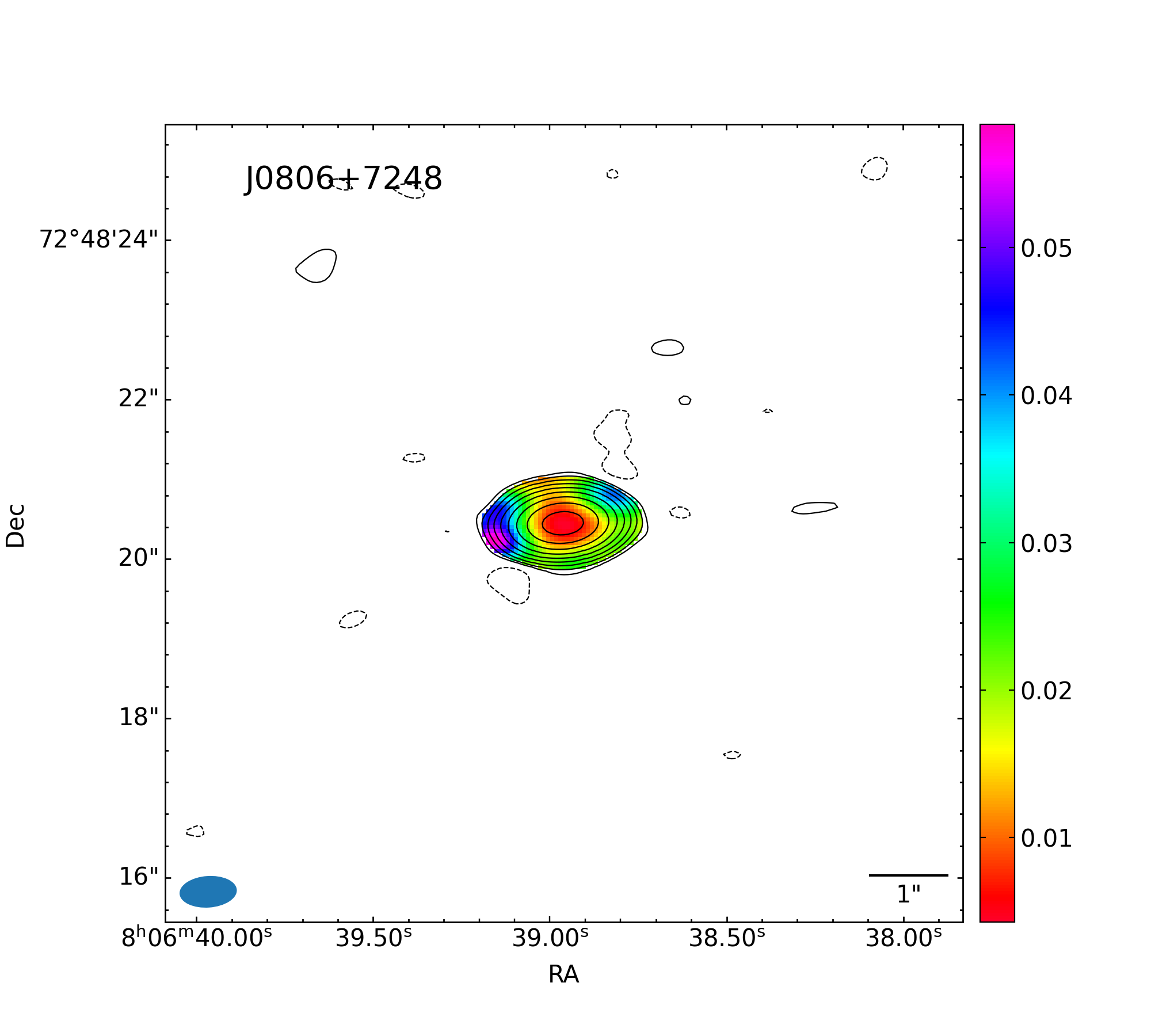}
         \caption{Spectral index error map, rms, contour levels, and beam size as in Fig.~\ref{fig:J0806spind}.} \label{fig:J0806spinderr}
     \end{subfigure}
     \hfill
     \\
     \begin{subfigure}[b]{0.47\textwidth}
         \centering
         \includegraphics[width=\textwidth]{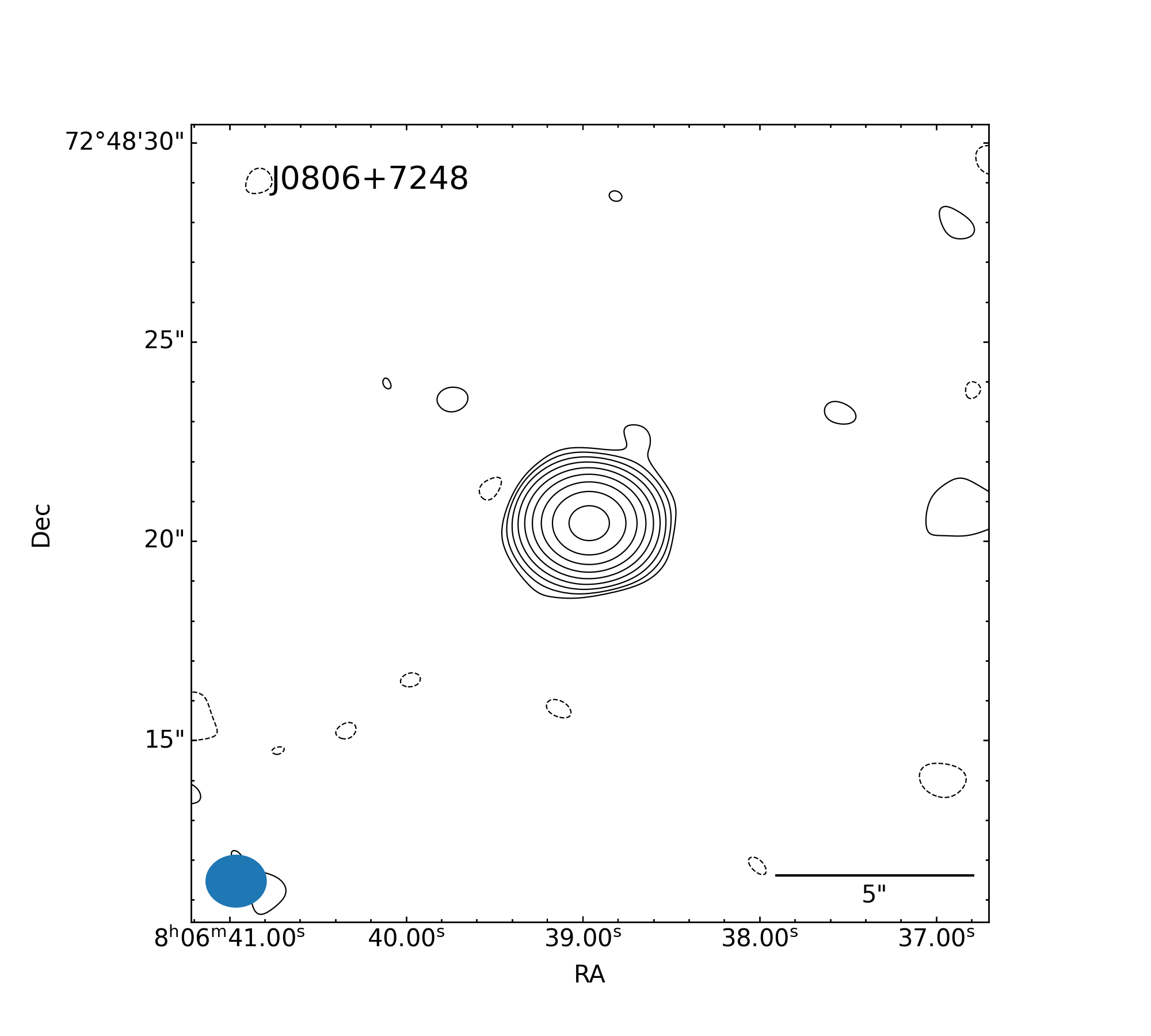}
         \caption{Tapered map with \texttt{uvtaper} = 90k$\lambda$, rms = 11$\mu$Jy beam$^{-1}$, contour levels at -3, 3 $\times$ 2$^n$, $n \in$ [0, 8], beam size 2.79 $\times$ 2.41~kpc.} \label{fig:J0806-90k}
     \end{subfigure}
          \hfill
     \begin{subfigure}[b]{0.47\textwidth}
         \centering
         \includegraphics[width=\textwidth]{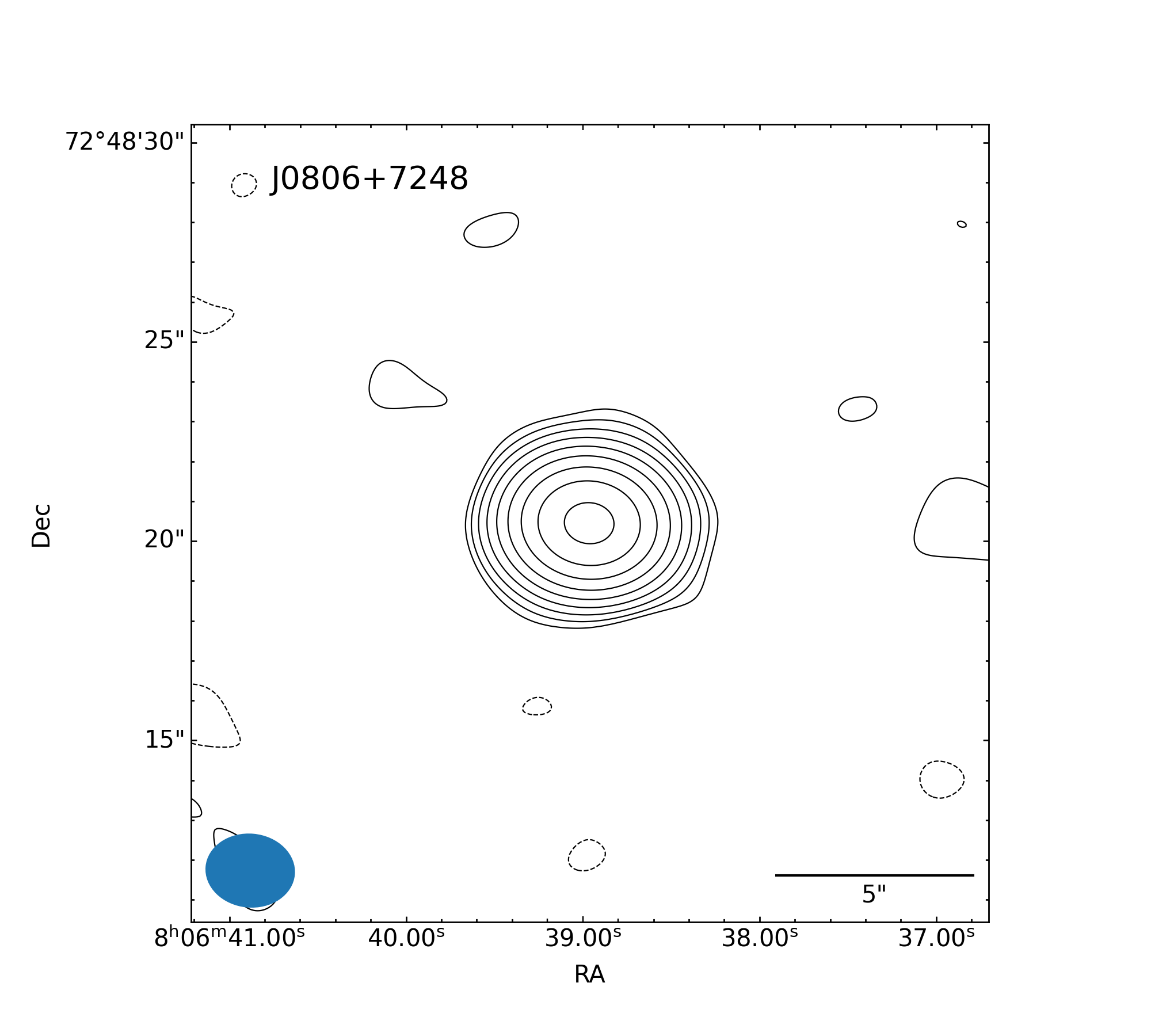}
         \caption{Tapered map with \texttt{uvtaper} = 60k$\lambda$, rms = 12$\mu$Jy beam$^{-1}$, contour levels at -3, 3 $\times$ 2$^n$, $n \in$ [0, 8], beam size 4.08 $\times$ 3.37~kpc.} \label{fig:J0806-60k}
     \end{subfigure}
          \hfill
     \\
     \begin{subfigure}[b]{0.47\textwidth}
         \centering
         \includegraphics[width=\textwidth]{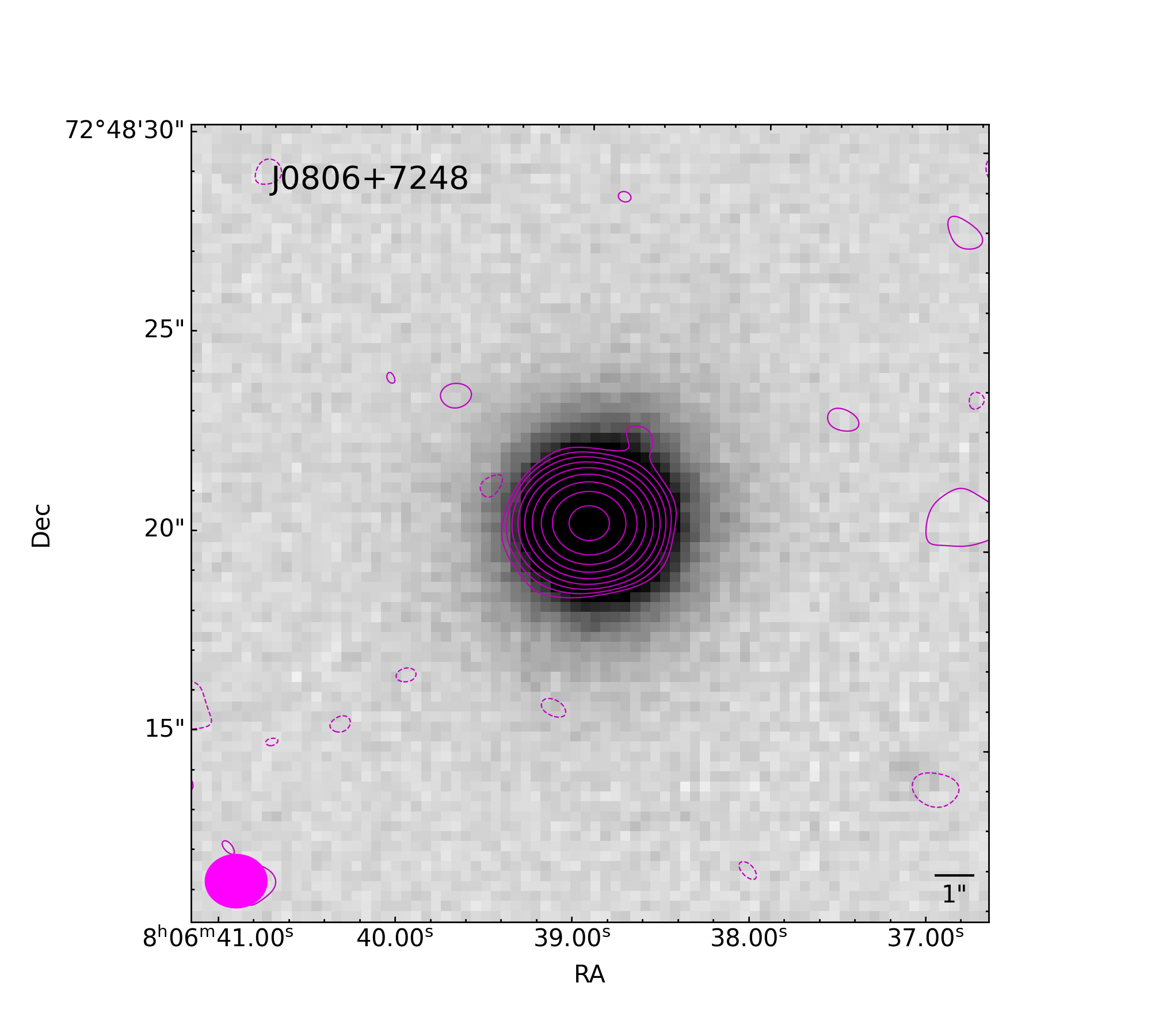}
         \caption{PanSTARRS $i$ band image of the host galaxy overlaid with the 90k$\lambda$ tapered map. Radio map properties as in Fig.~\ref{fig:J0806-90k}}. \label{fig:J0806-host}
     \end{subfigure}
        \caption{}
        \label{fig:J0806}
\end{figure*}


\begin{figure*}
     \centering
     \begin{subfigure}[b]{0.47\textwidth}
         \centering
         \includegraphics[width=\textwidth]{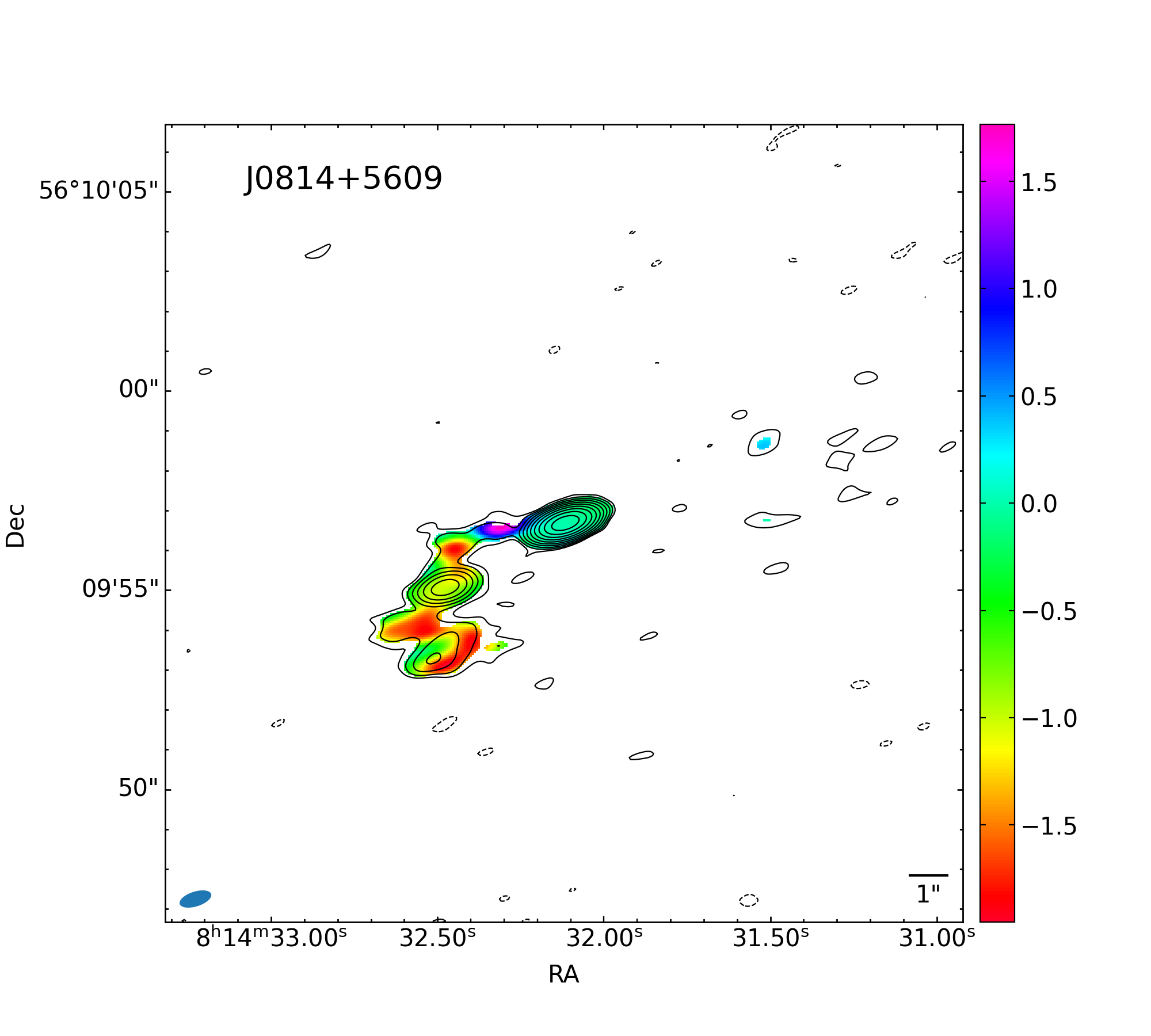}
         \caption{Spectral index map, rms = 10$\mu$Jy beam$^{-1}$, contour levels at -3, 3 $\times$ 2$^n$, $n \in$ [0, 9], beam size 5.12 $\times$ 2.28~kpc. } \label{fig:J0814spind}
     \end{subfigure}
     \hfill
     \begin{subfigure}[b]{0.47\textwidth}
         \centering
         \includegraphics[width=\textwidth]{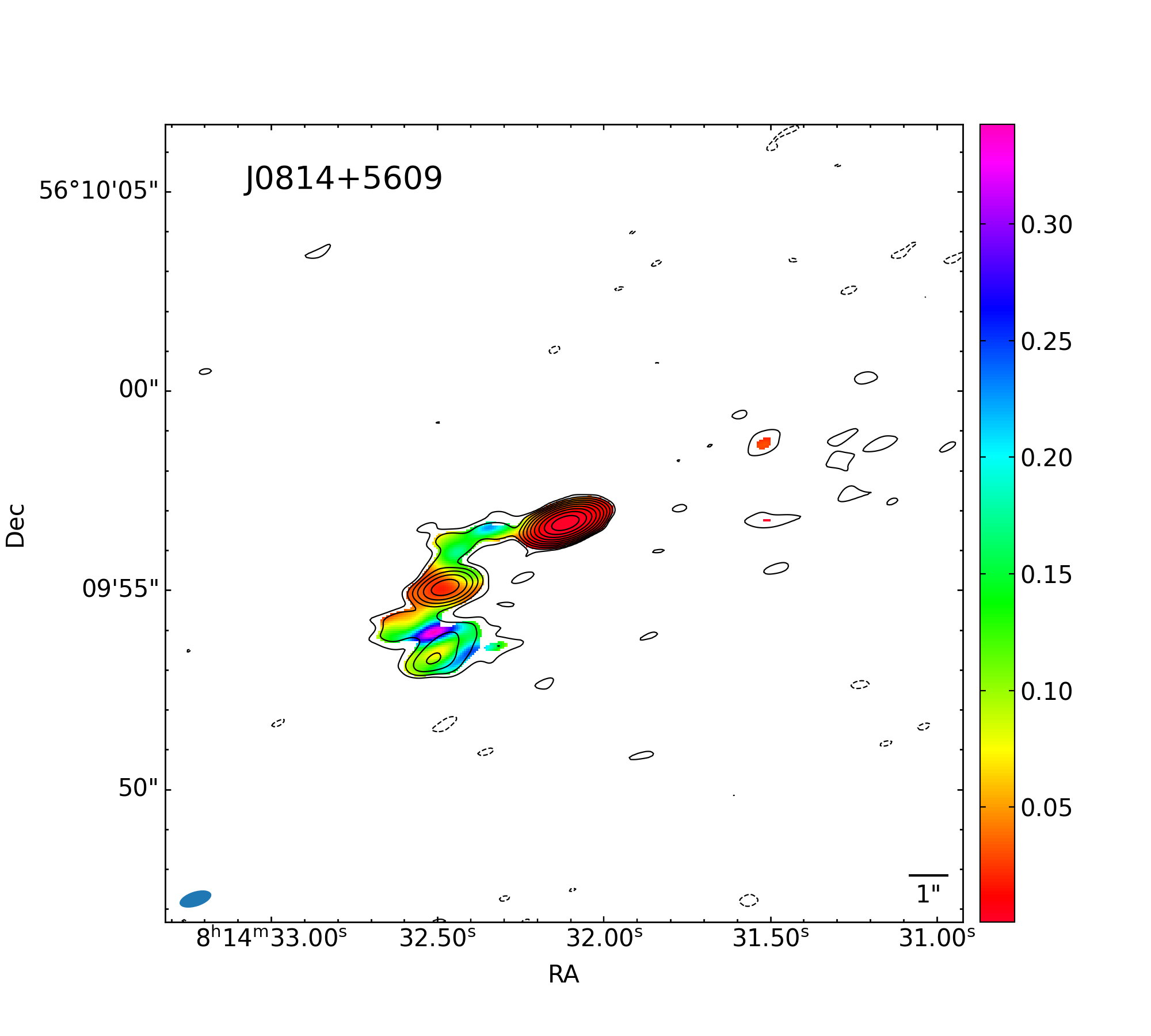}
         \caption{Spectral index error map, rms, contour levels, and beam size as in Fig.~\ref{fig:J0814spind}.} \label{fig:J0814spinderr}
     \end{subfigure}
     \hfill
     \\
     \begin{subfigure}[b]{0.47\textwidth}
         \centering
         \includegraphics[width=\textwidth]{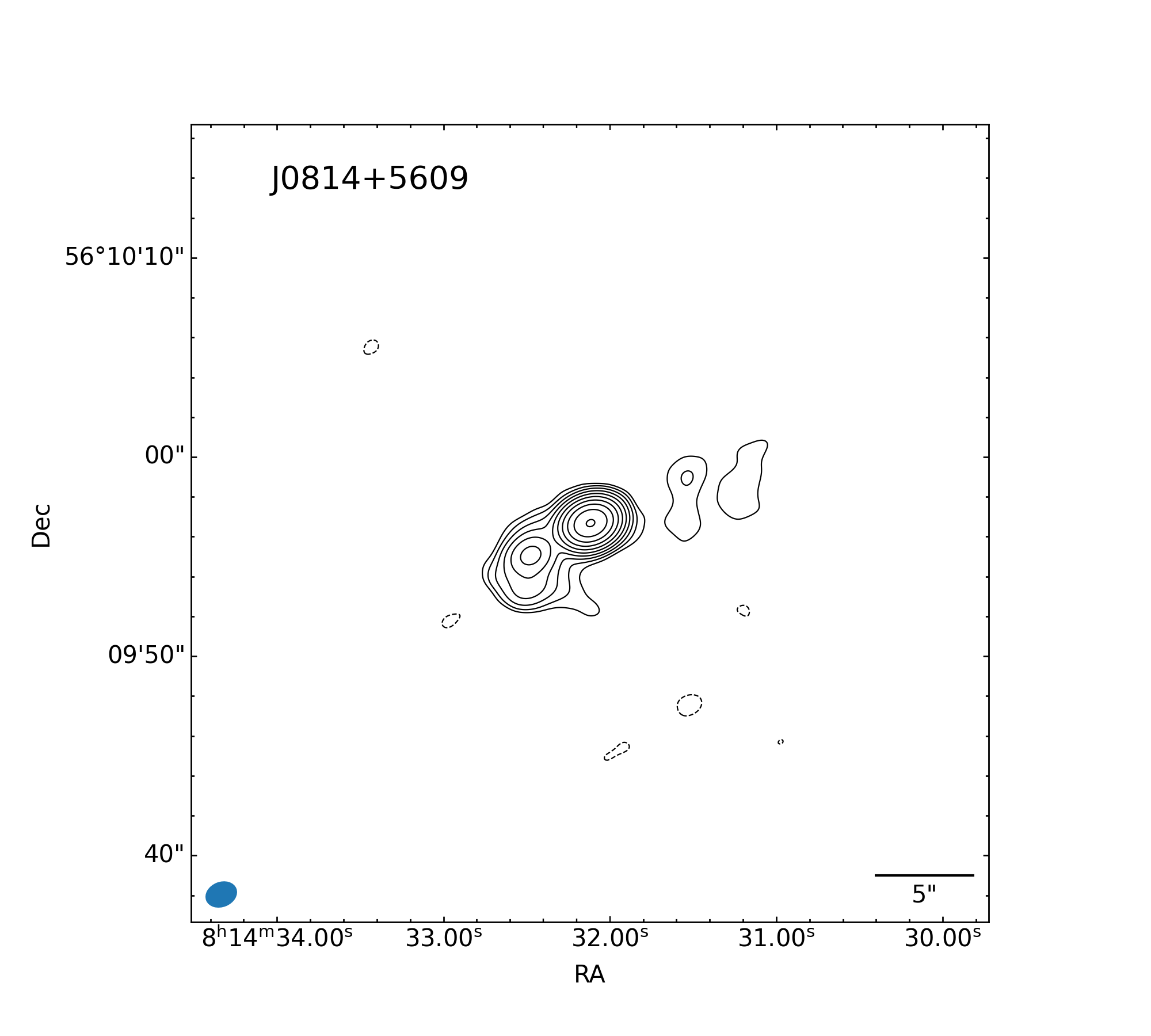}
         \caption{Tapered map with \texttt{uvtaper} = 90k$\lambda$, rms = 16$\mu$Jy beam$^{-1}$, contour levels at -3, 3 $\times$ 2$^n$, $n \in$ [0, 9], beam size 10.05 $\times$ 7.77~kpc.} \label{fig:J0814-90k}
     \end{subfigure}
          \hfill
     \begin{subfigure}[b]{0.47\textwidth}
         \centering
         \includegraphics[width=\textwidth]{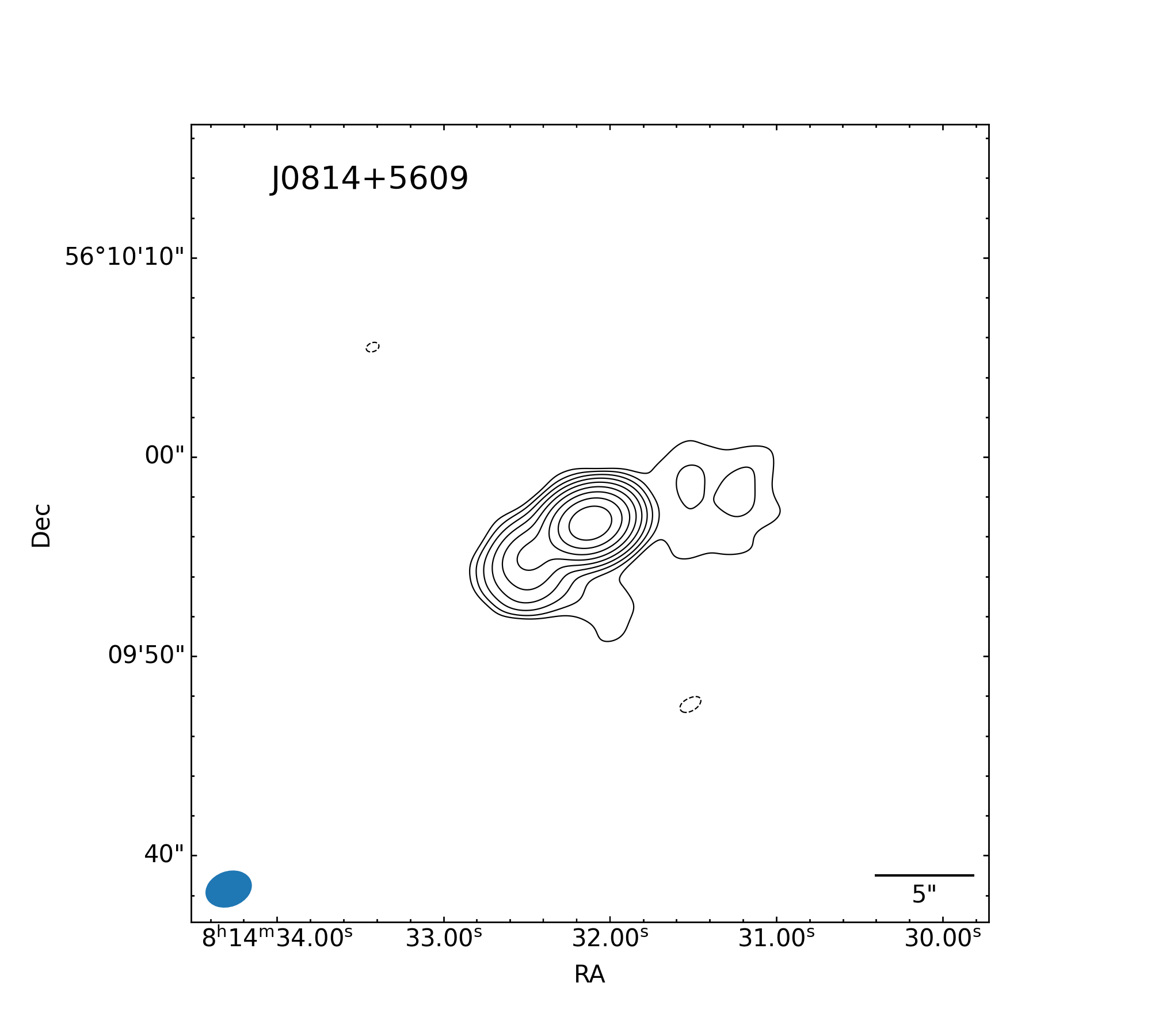}
         \caption{Tapered map with \texttt{uvtaper} = 60k$\lambda$, rms = 19$\mu$Jy beam$^{-1}$, contour levels at -3, 3 $\times$ 2$^n$, $n \in$ [0, 8], beam size 14.74 $\times$ 10.98~kpc.} \label{fig:J0814-60k}
     \end{subfigure}
      \hfill
     \\
          \begin{subfigure}[b]{0.47\textwidth}
         \centering
         \includegraphics[width=\textwidth]{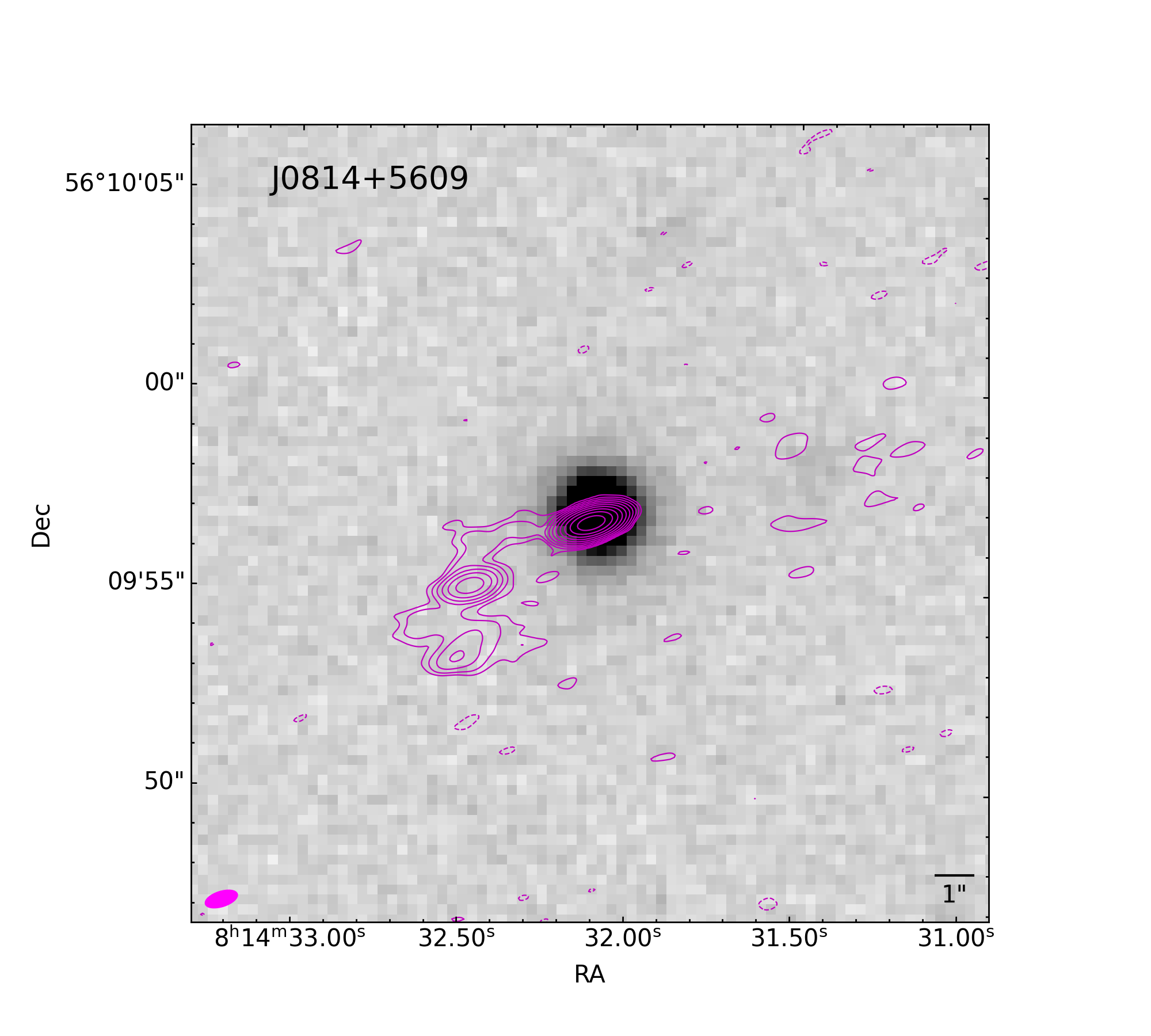}
         \caption{PanSTARRS $i$ band image of the host galaxy overlaid with the normal map. Radio map properties as in Fig.~\ref{fig:J0814spind}}. \label{fig:J0814-host-zoom}
     \end{subfigure}
     \hfill
     \begin{subfigure}[b]{0.47\textwidth}
         \centering
         \includegraphics[width=\textwidth]{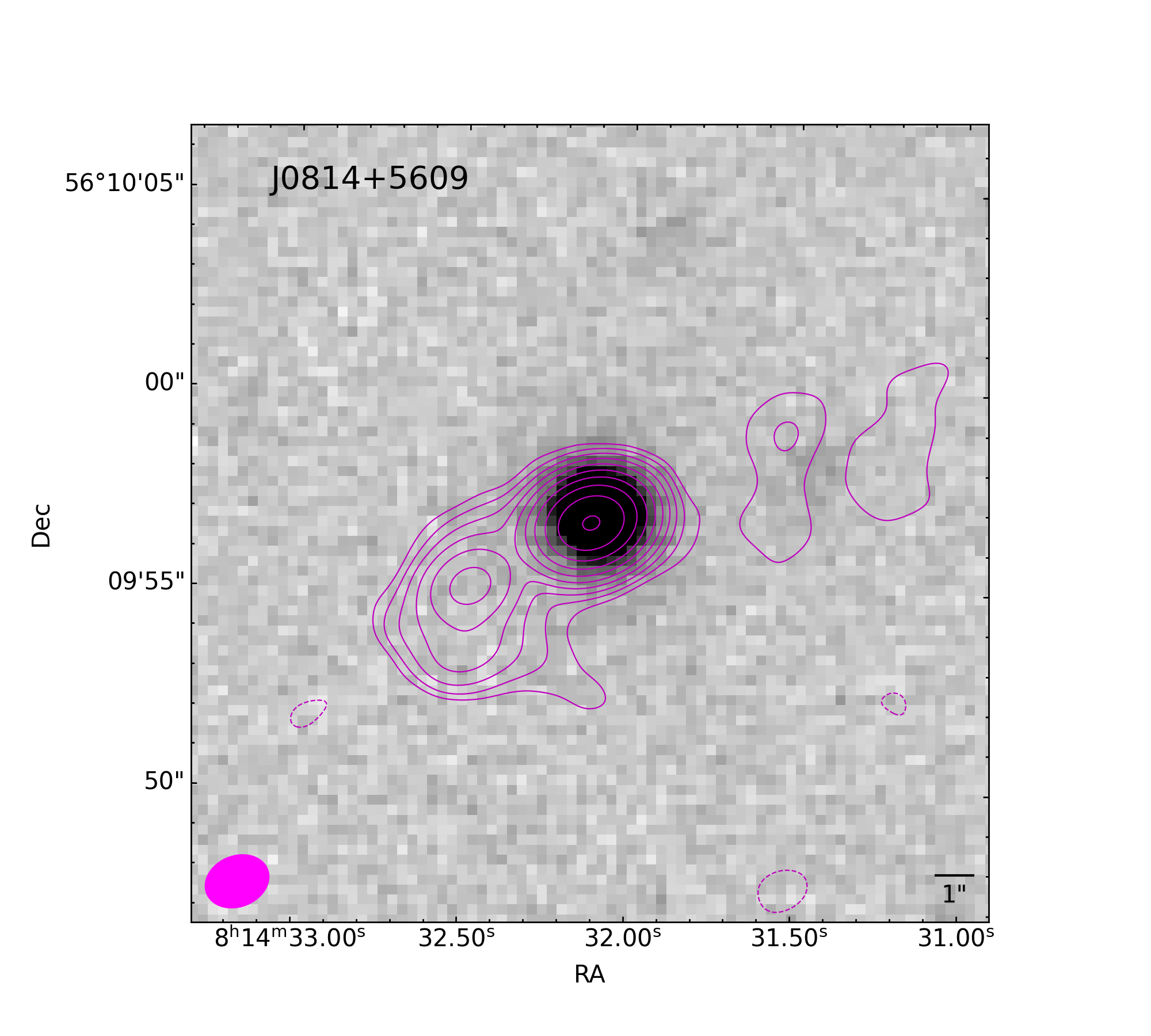}
         \caption{PanSTARRS $i$ band image of the host galaxy overlaid with the 90k$\lambda$ tapered map. Radio map properties as in Fig.~\ref{fig:J0814-90k}}. \label{fig:J0814-host}
     \end{subfigure}
        \caption{}
        \label{fig:J0814}
\end{figure*}

\clearpage
\begin{figure*}
     \centering
     \begin{subfigure}[b]{0.47\textwidth}
         \centering
         \includegraphics[width=\textwidth]{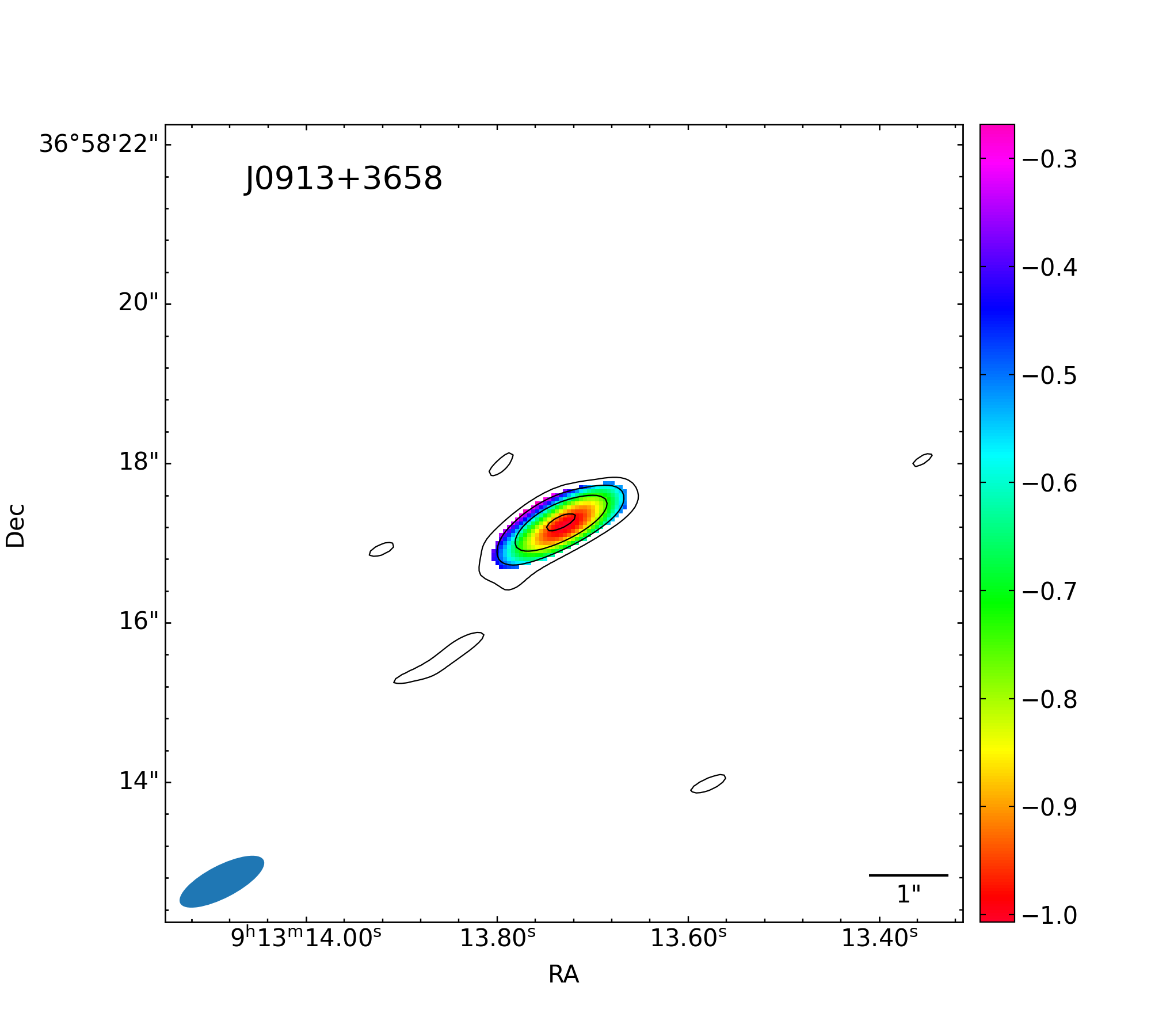}
         \caption{Spectral index map, rms = 11$\mu$Jy beam$^{-1}$, contour levels at -3, 3 $\times$ 2$^n$, $n \in$ [0, 3], beam size 2.31 $\times$ 0.80~kpc. } \label{fig:J0913spind}
     \end{subfigure}
     \hfill
     \begin{subfigure}[b]{0.47\textwidth}
         \centering
         \includegraphics[width=\textwidth]{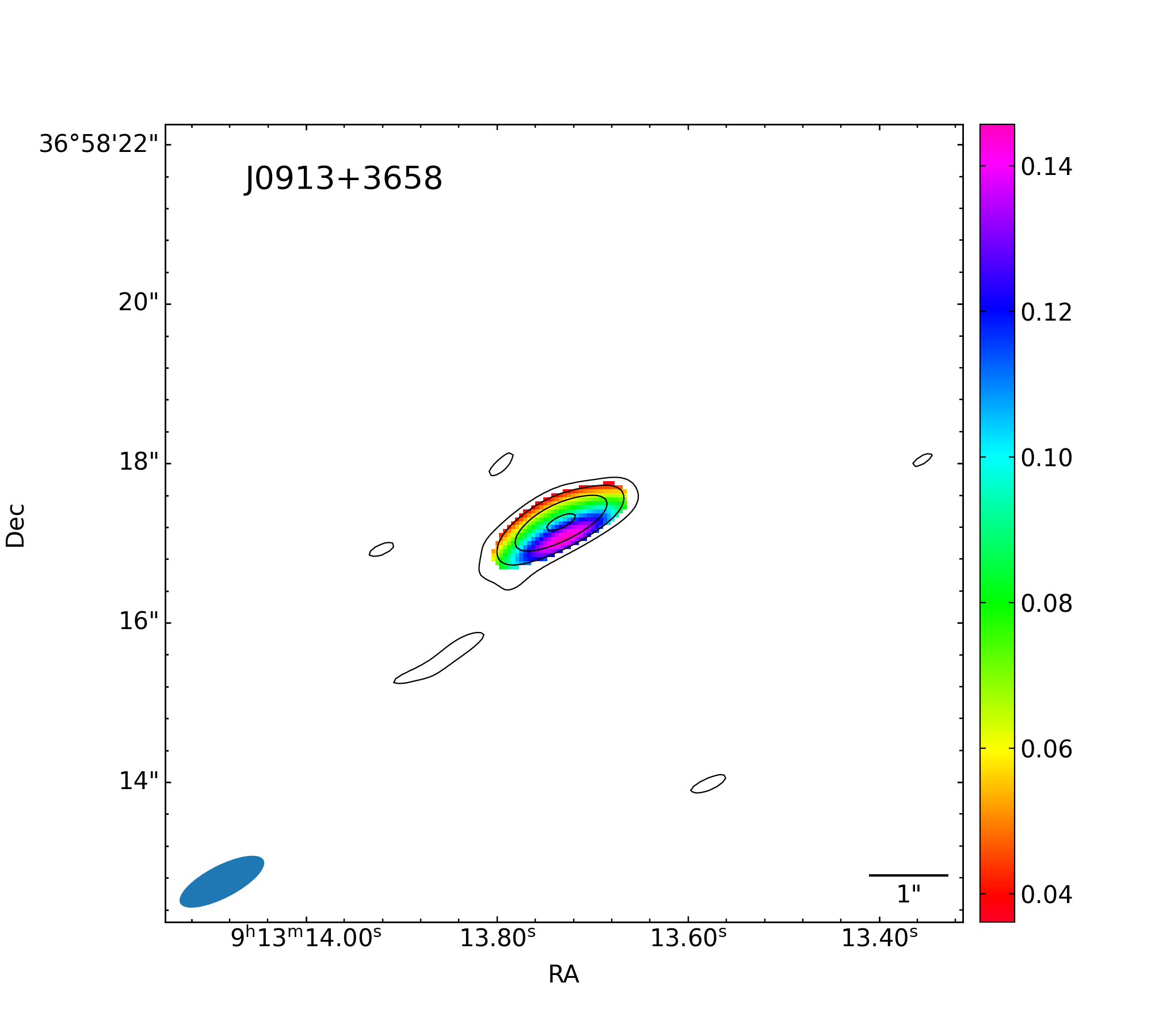}
         \caption{Spectral index error map, rms, contour levels, and beam size as in Fig.~\ref{fig:J0913spind}.} \label{fig:J0913spinderr}
     \end{subfigure}
     \hfill
     \\
     \begin{subfigure}[b]{0.47\textwidth}
         \centering
         \includegraphics[width=\textwidth]{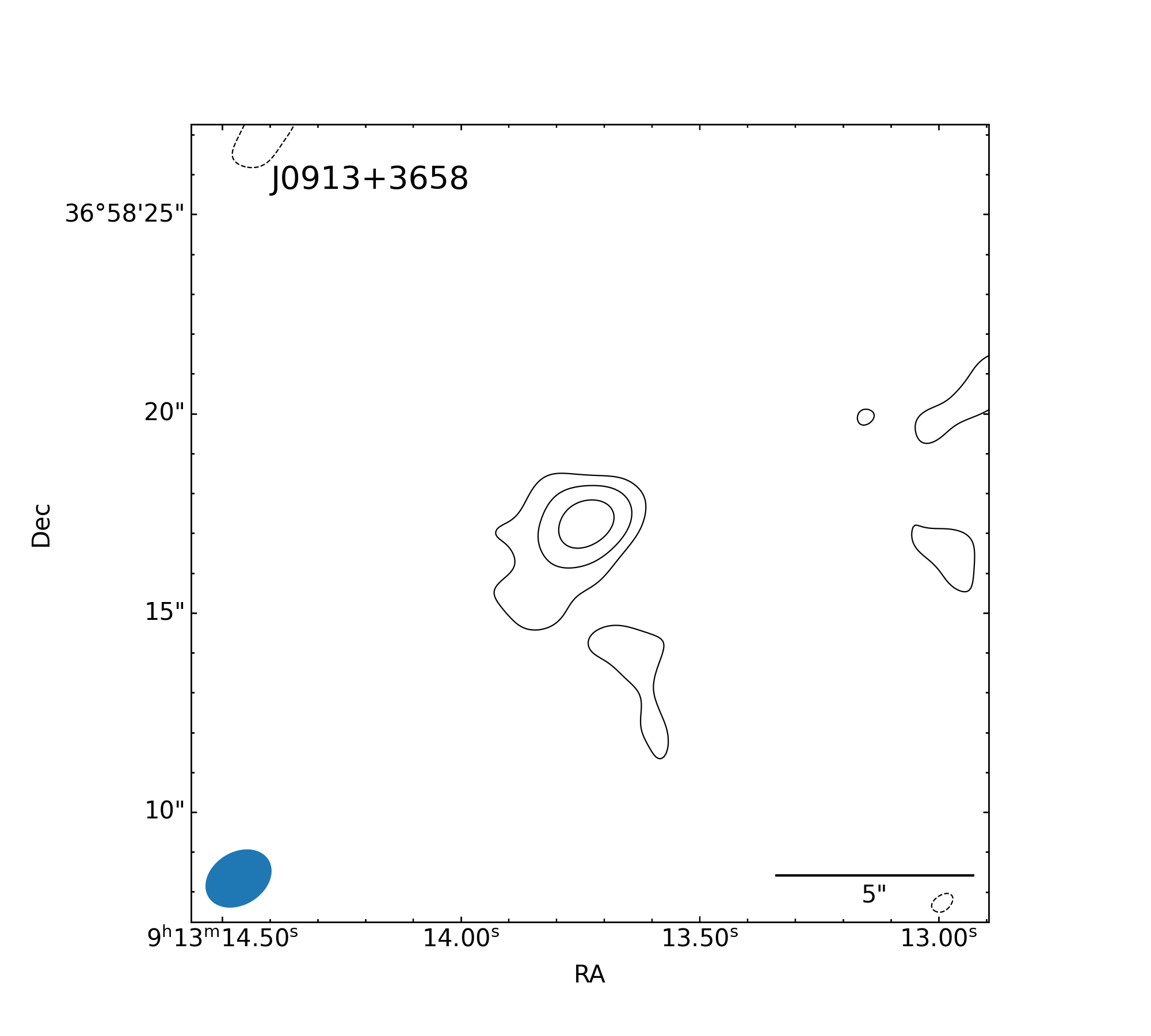}
         \caption{Tapered map with \texttt{uvtaper} = 90k$\lambda$, rms = 17$\mu$Jy beam$^{-1}$, contour levels at -3, 3 $\times$ 2$^n$, $n \in$ [0, 2], beam size 3.47 $\times$ 2.58~kpc.} \label{fig:J0913-90k}
     \end{subfigure}
          \hfill
     \begin{subfigure}[b]{0.47\textwidth}
         \centering
         \includegraphics[width=\textwidth]{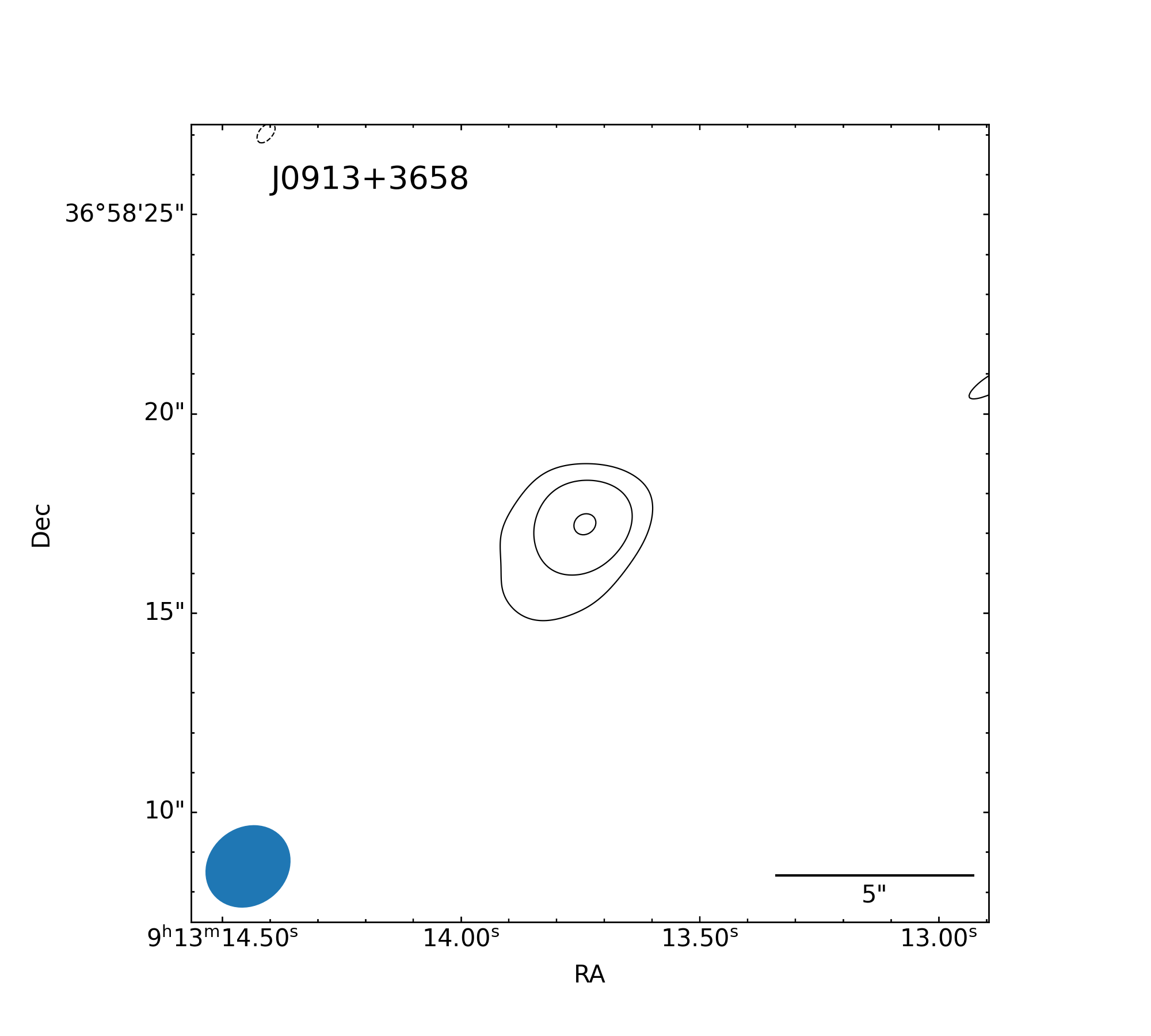}
         \caption{Tapered map with \texttt{uvtaper} = 60k$\lambda$, rms = 26$\mu$Jy beam$^{-1}$, contour levels at -3, 3 $\times$ 2$^n$, $n \in$ [0, 2], beam size 4.41 $\times$ 3.82~kpc.} \label{fig:J0913-60k}
     \end{subfigure}
          \hfill
     \\
     \begin{subfigure}[b]{0.47\textwidth}
         \centering
         \includegraphics[width=\textwidth]{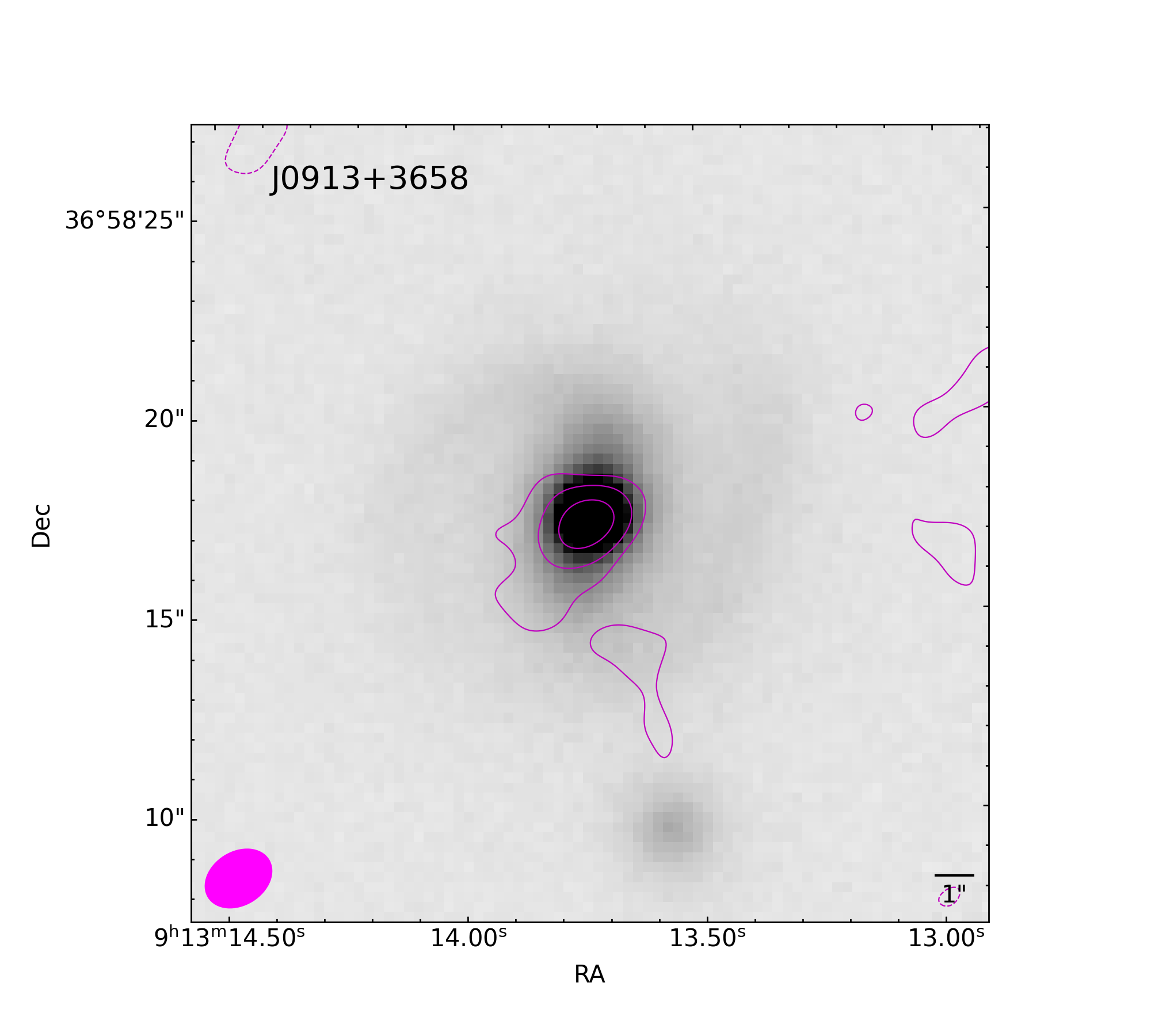}
         \caption{PanSTARRS $i$ band image of the host galaxy overlaid with the 90k$\lambda$ tapered map. Radio map properties as in Fig.~\ref{fig:J0913-90k}}. \label{fig:J0913-host}
     \end{subfigure}
        \caption{}
        \label{fig:J0913}
\end{figure*}


\begin{figure*}
     \centering
     \begin{subfigure}[b]{0.47\textwidth}
         \centering
         \includegraphics[width=\textwidth]{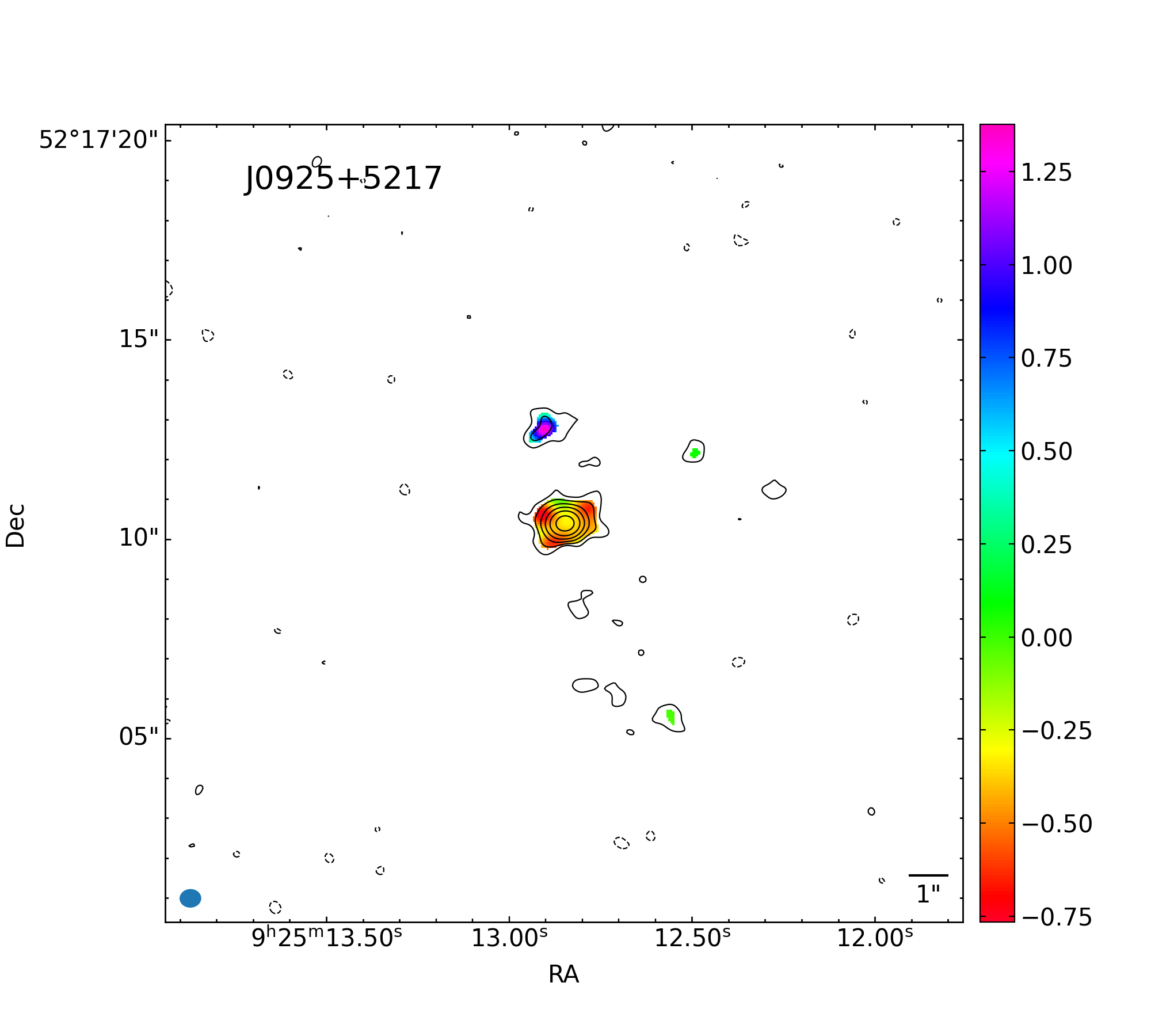}
         \caption{Spectral index map, rms = 12$\mu$Jy beam$^{-1}$, contour levels at -3, 3 $\times$ 2$^n$, $n \in$ [0, 5], beam size 0.38 $\times$ 0.33~kpc. } \label{fig:J0925spind}
     \end{subfigure}
     \hfill
     \begin{subfigure}[b]{0.47\textwidth}
         \centering
         \includegraphics[width=\textwidth]{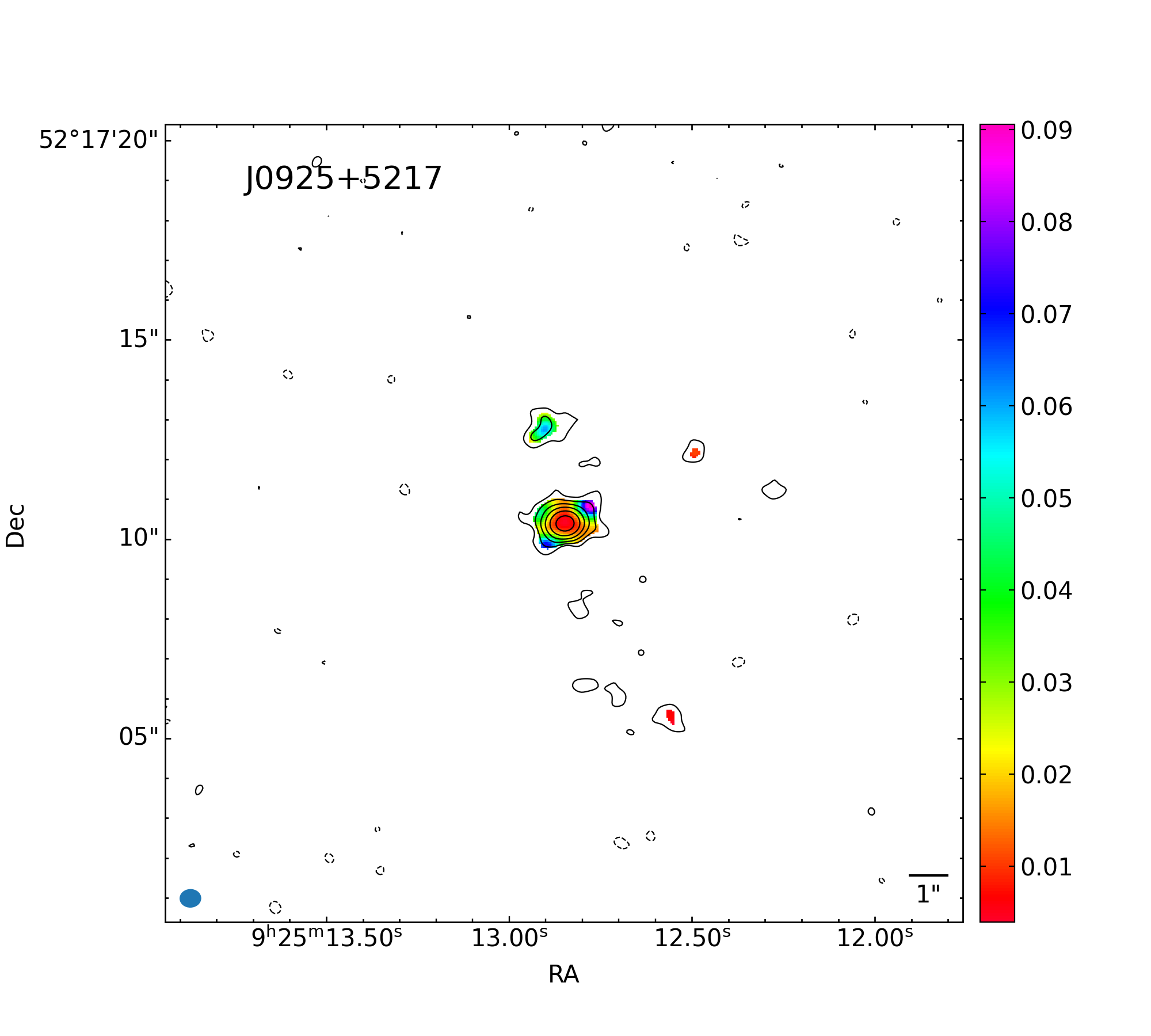}
         \caption{Spectral index error map, rms, contour levels, and beam size as in Fig.~\ref{fig:J0925spind}.} \label{fig:J0925spinderr}
     \end{subfigure}
     \hfill
     \\
     \begin{subfigure}[b]{0.47\textwidth}
         \centering
         \includegraphics[width=\textwidth]{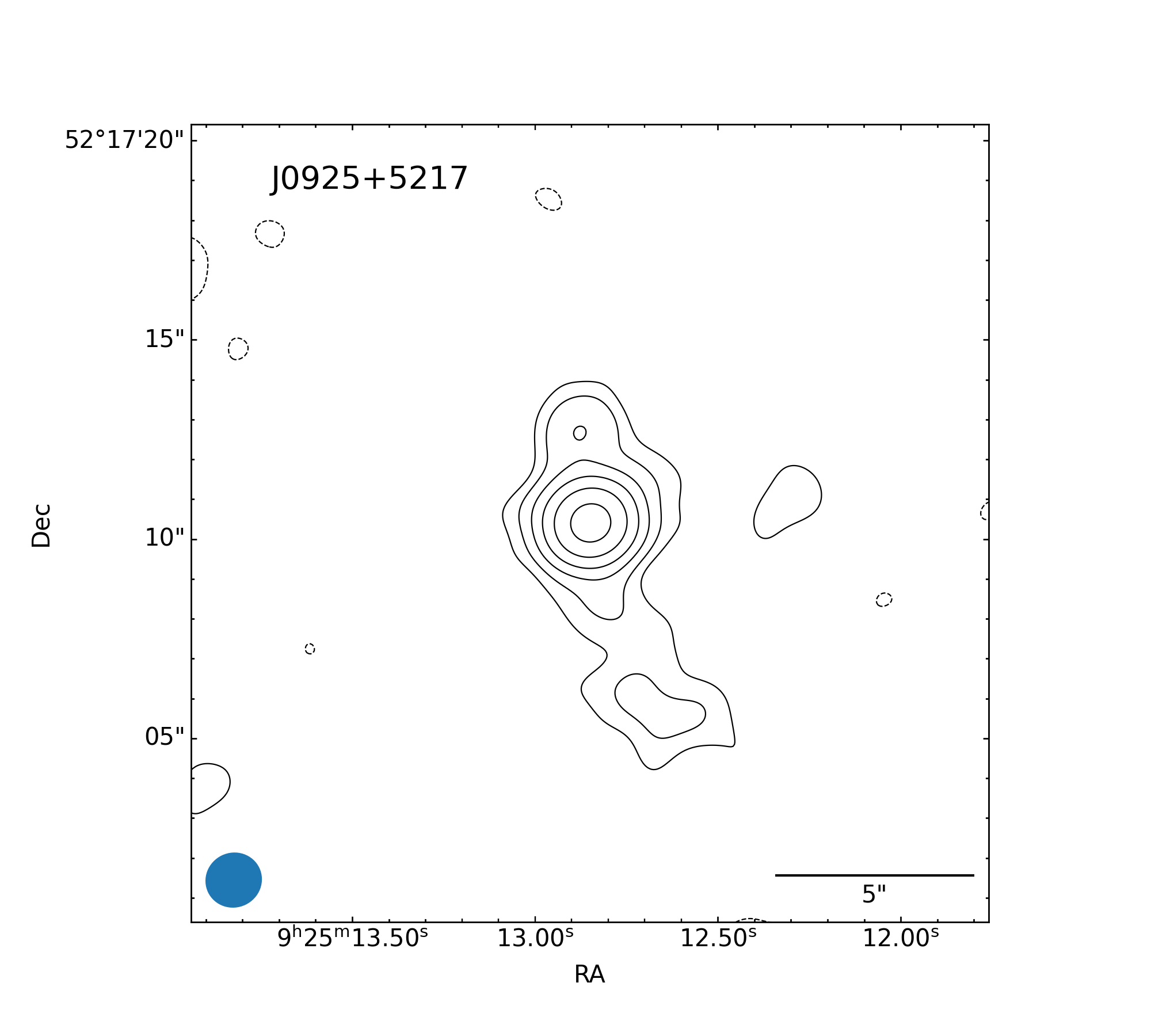}
         \caption{Tapered map with \texttt{uvtaper} = 90k$\lambda$, rms = 16$\mu$Jy beam$^{-1}$, contour levels at -3, 3 $\times$ 2$^n$, $n \in$ [0, 5], beam size 1.00 $\times$ 0.95~kpc.} \label{fig:J0925-90k}
     \end{subfigure}
          \hfill
     \begin{subfigure}[b]{0.47\textwidth}
         \centering
         \includegraphics[width=\textwidth]{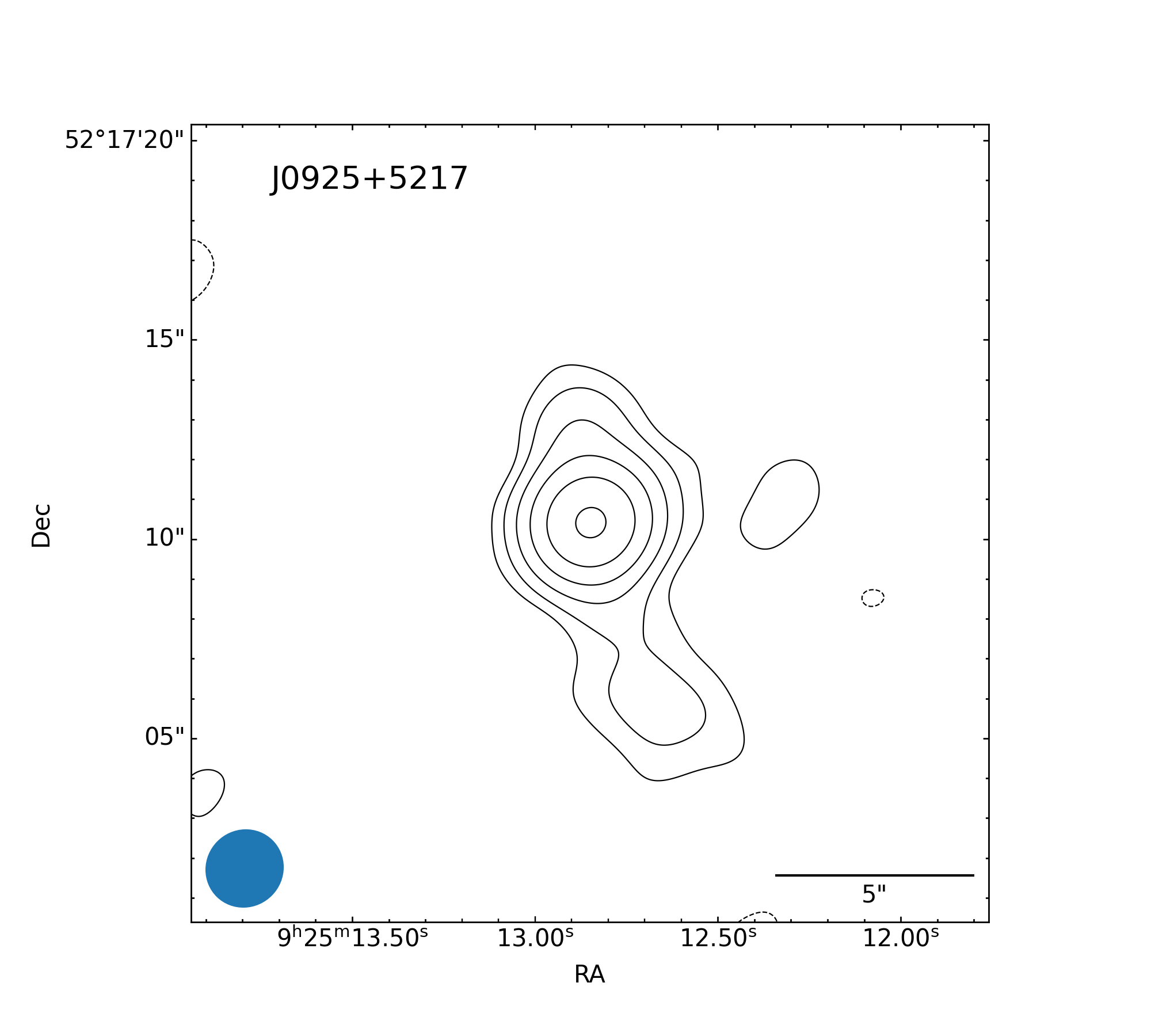}
         \caption{Tapered map with \texttt{uvtaper} = 60k$\lambda$, rms = 21$\mu$Jy beam$^{-1}$, contour levels at -3, 3 $\times$ 2$^n$, $n \in$ [0, 5], beam size 1.39 $\times$ 1.35~kpc.} \label{fig:J0925-60k}
     \end{subfigure}
          \hfill
     \\
     \begin{subfigure}[b]{0.47\textwidth}
         \centering
         \includegraphics[width=\textwidth]{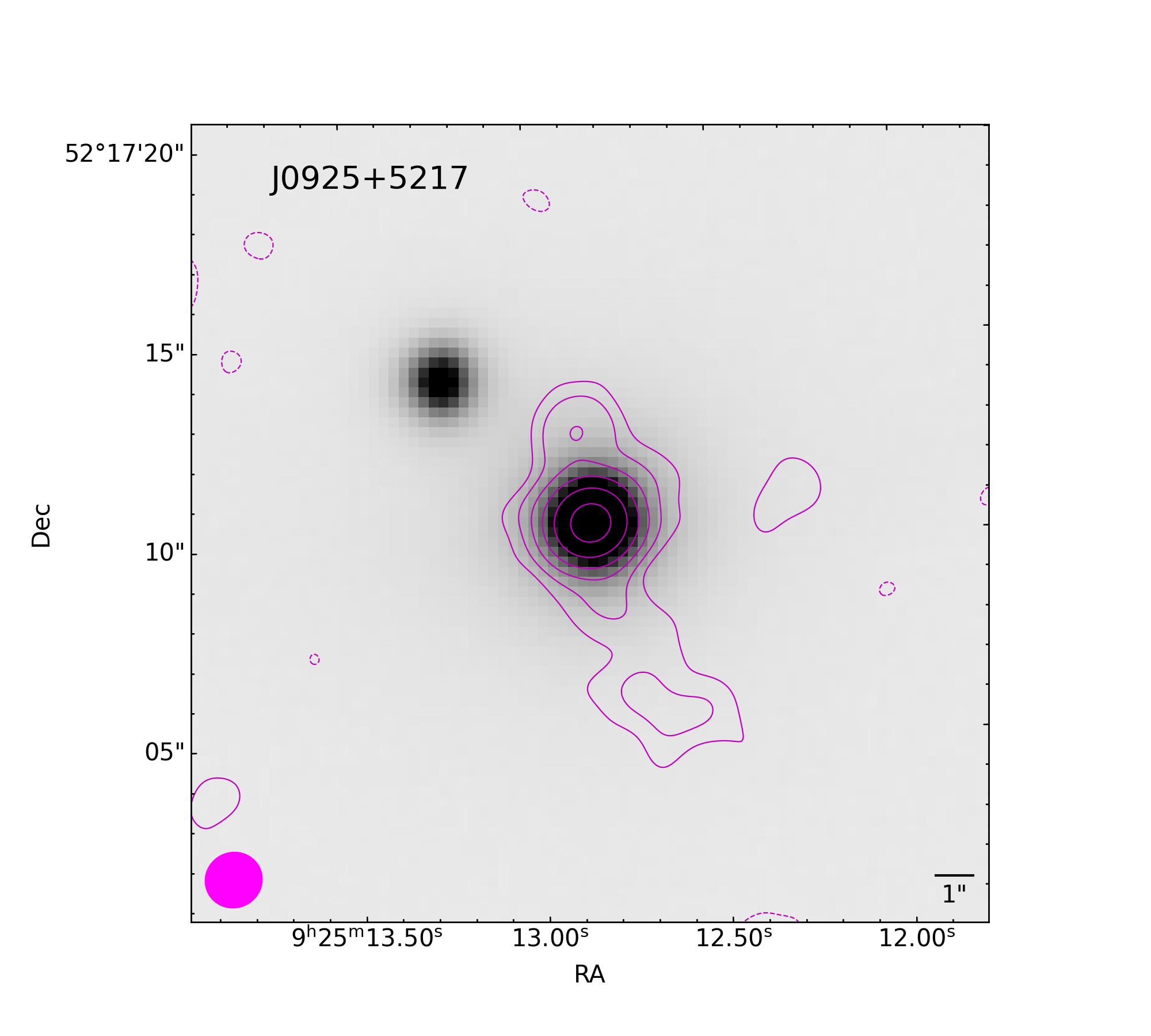}
         \caption{PanSTARRS $i$ band image of the host galaxy overlaid with the 90k$\lambda$ tapered map. Radio map properties as in Fig.~\ref{fig:J0925-90k}}. \label{fig:J0925-host}
     \end{subfigure}
        \caption{}
        \label{fig:J0925}
\end{figure*}


\begin{figure*}
     \centering
     \begin{subfigure}[b]{0.47\textwidth}
         \centering
         \includegraphics[width=\textwidth]{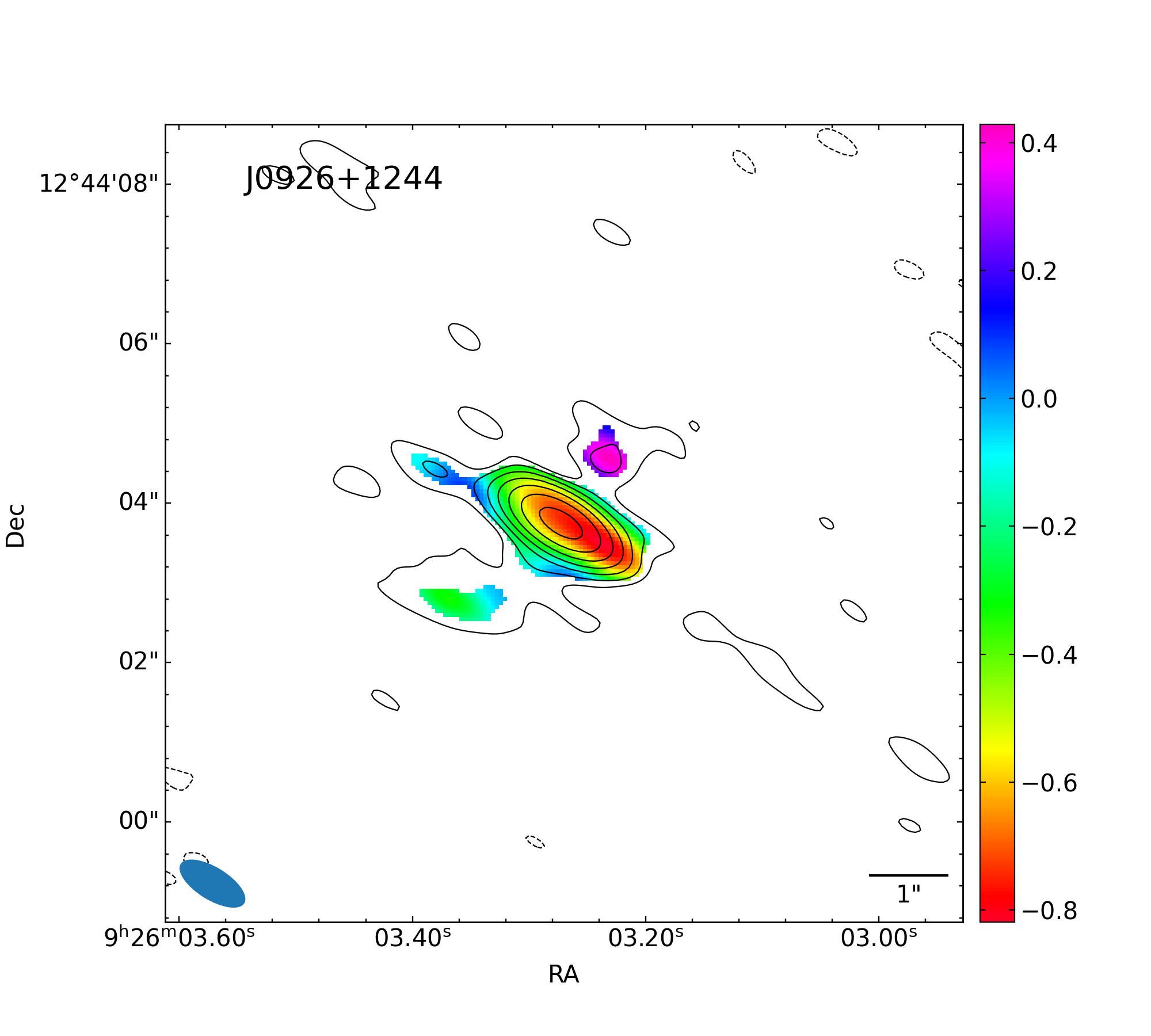}
         \caption{Spectral index map, rms = 11$\mu$Jy beam$^{-1}$, contour levels at -3, 3 $\times$ 2$^n$, $n \in$ [0, 6], beam size 0.55 $\times$ 0.24~kpc. } \label{fig:J0926spind}
     \end{subfigure}
     \hfill
     \begin{subfigure}[b]{0.47\textwidth}
         \centering
         \includegraphics[width=\textwidth]{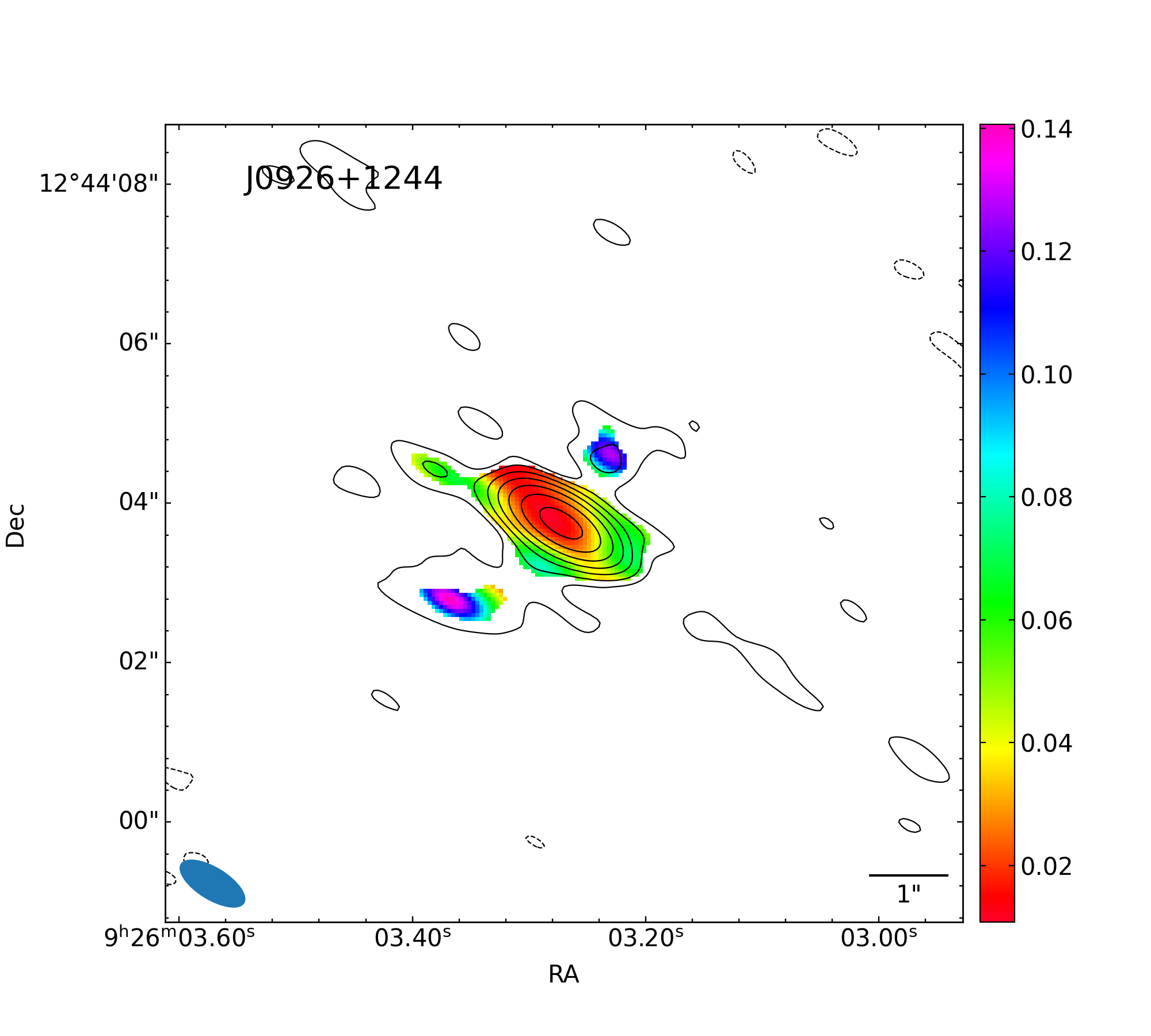}
         \caption{Spectral index error map, rms, contour levels, and beam size as in Fig.~\ref{fig:J0926spind}.} \label{fig:J0926spinderr}
     \end{subfigure}
     \hfill
     \\
     \begin{subfigure}[b]{0.47\textwidth}
         \centering
         \includegraphics[width=\textwidth]{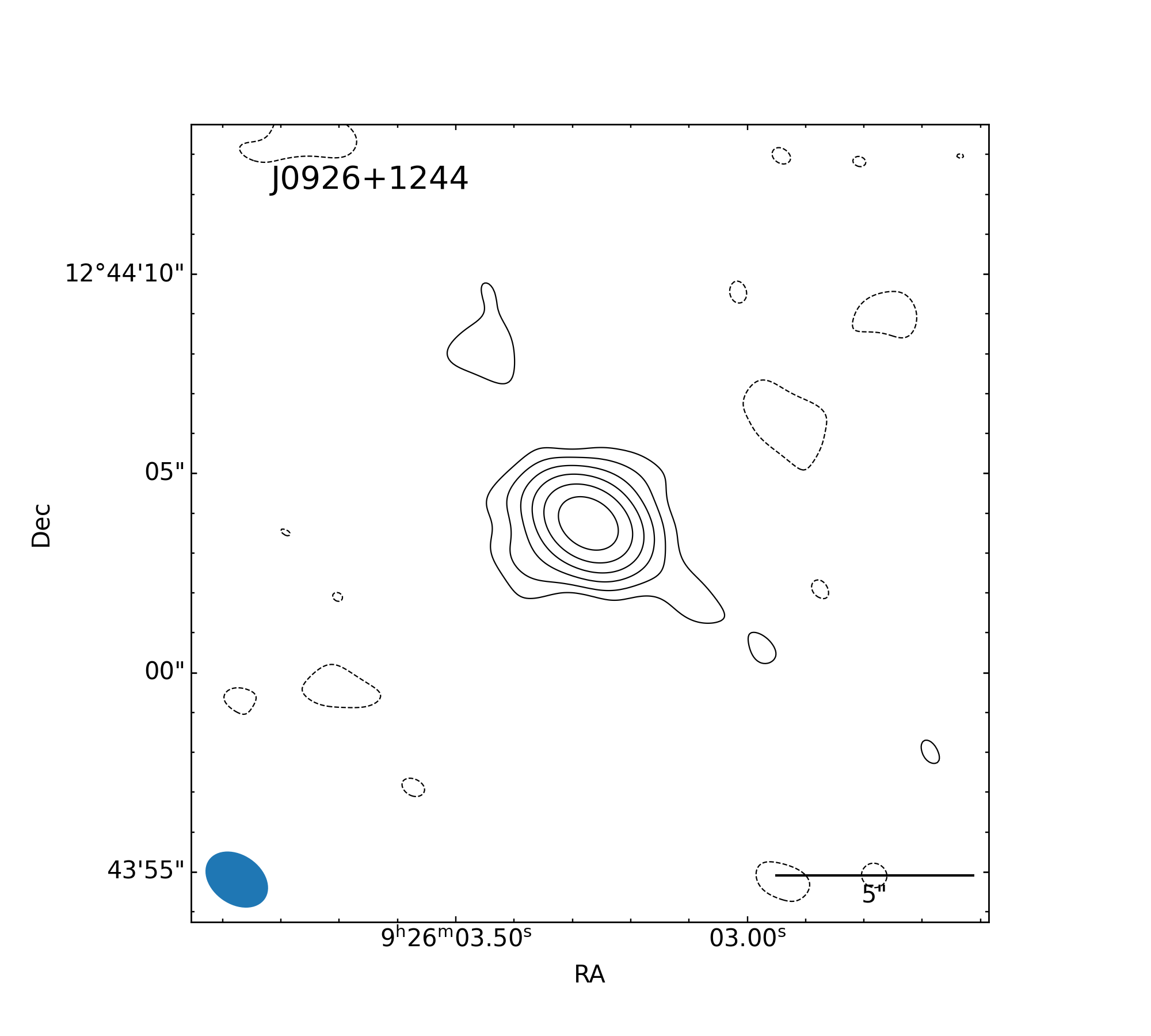}
         \caption{Tapered map with \texttt{uvtaper} = 90k$\lambda$, rms = 17$\mu$Jy beam$^{-1}$, contour levels at -3, 3 $\times$ 2$^n$, $n \in$ [0, 5], beam size 0.99 $\times$ 0.72~kpc.} \label{fig:J0926-90k}
     \end{subfigure}
          \hfill
     \begin{subfigure}[b]{0.47\textwidth}
         \centering
         \includegraphics[width=\textwidth]{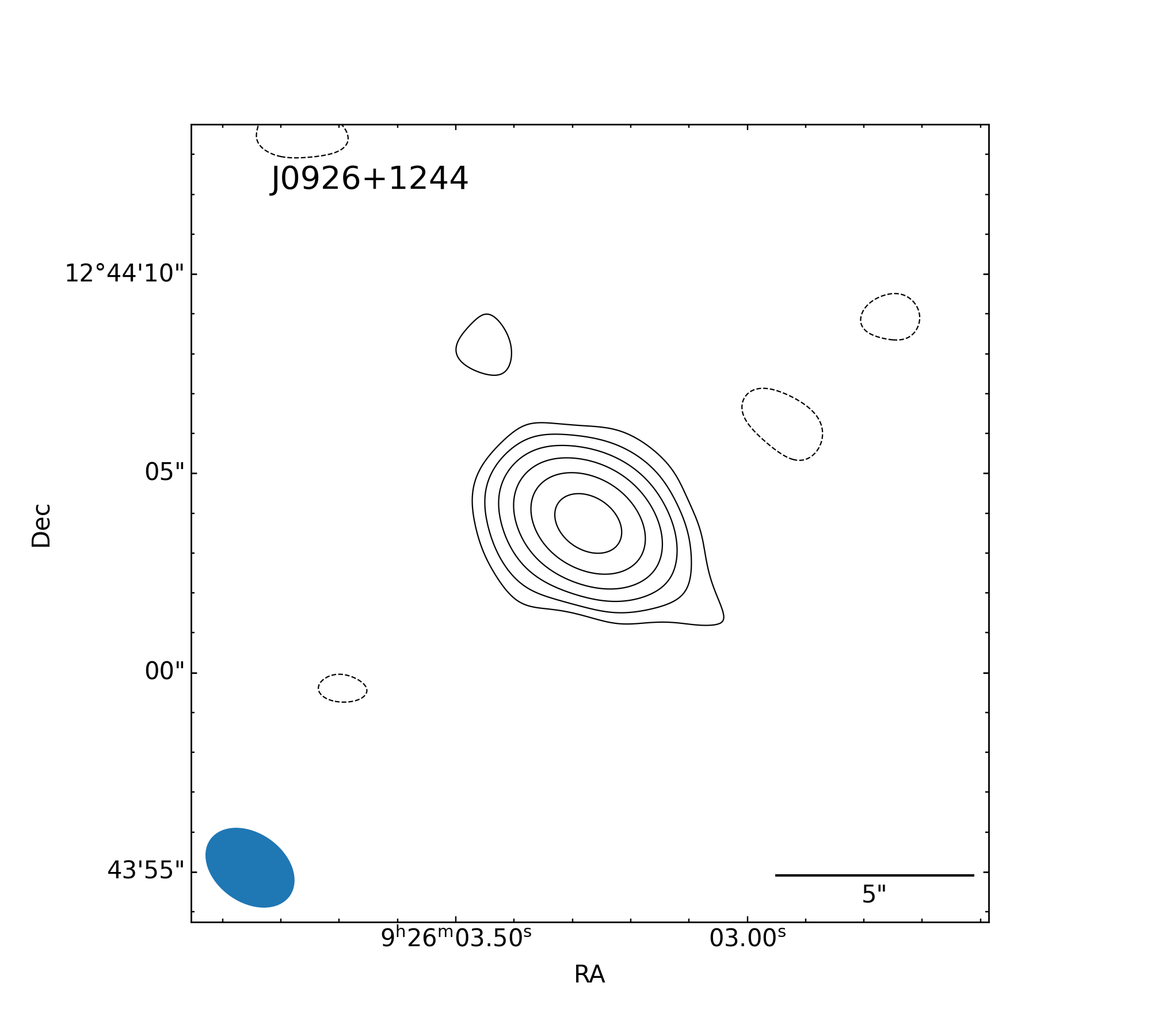}
         \caption{Tapered map with \texttt{uvtaper} = 60k$\lambda$, rms = 22$\mu$Jy beam$^{-1}$, contour levels at -3, 3 $\times$ 2$^n$, $n \in$ [0, 5], beam size 1.42 $\times$ 1.01~kpc.} \label{fig:J0926-60k}
     \end{subfigure}
          \hfill
     \\
     \begin{subfigure}[b]{0.47\textwidth}
         \centering
         \includegraphics[width=\textwidth]{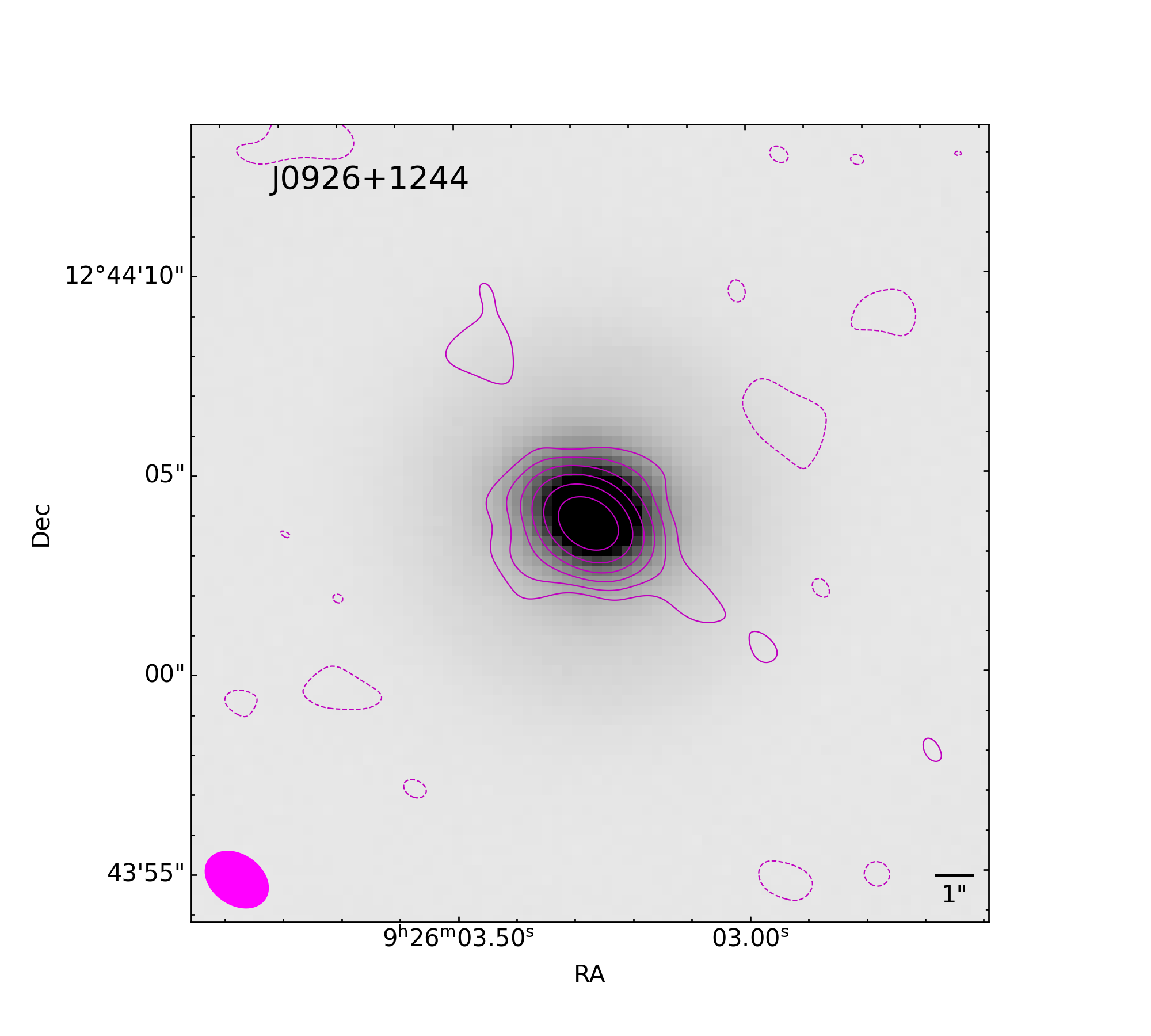}
         \caption{PanSTARRS $i$ band image of the host galaxy overlaid with the 90k$\lambda$ tapered map. Radio map properties as in Fig.~\ref{fig:J0926-90k}}. \label{fig:J0926-host}
     \end{subfigure}
        \caption{}
        \label{fig:J0926}
\end{figure*}


\begin{figure*}
     \centering
     \begin{subfigure}[b]{0.47\textwidth}
         \centering
         \includegraphics[width=\textwidth]{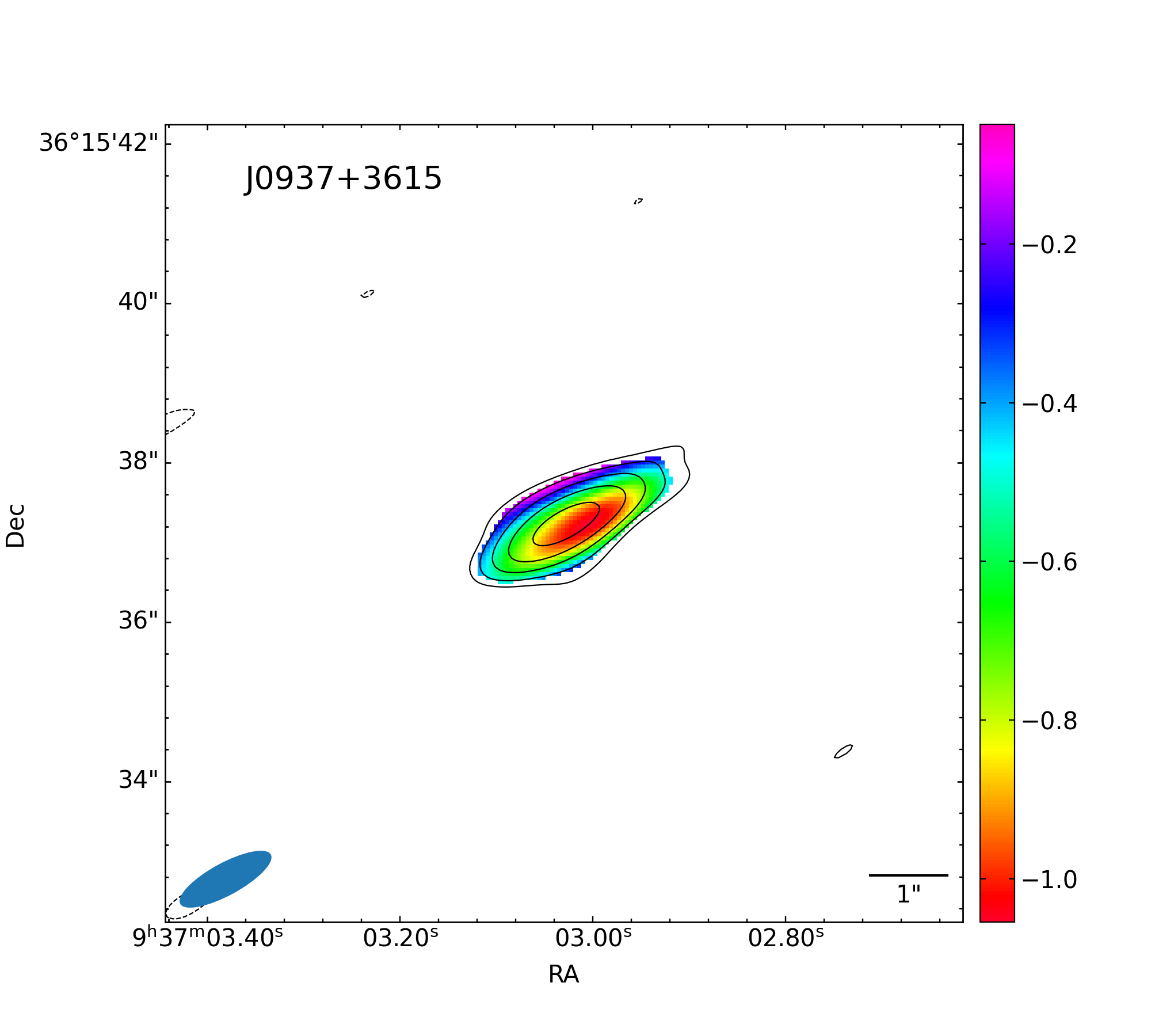}
         \caption{Spectral index map, rms = 12$\mu$Jy beam$^{-1}$, contour levels at -3, 3 $\times$ 2$^n$, $n \in$ [0, 4], beam size 3.92 $\times$ 1.24~kpc. } \label{fig:J0937spind}
     \end{subfigure}
     \hfill
     \begin{subfigure}[b]{0.47\textwidth}
         \centering
         \includegraphics[width=\textwidth]{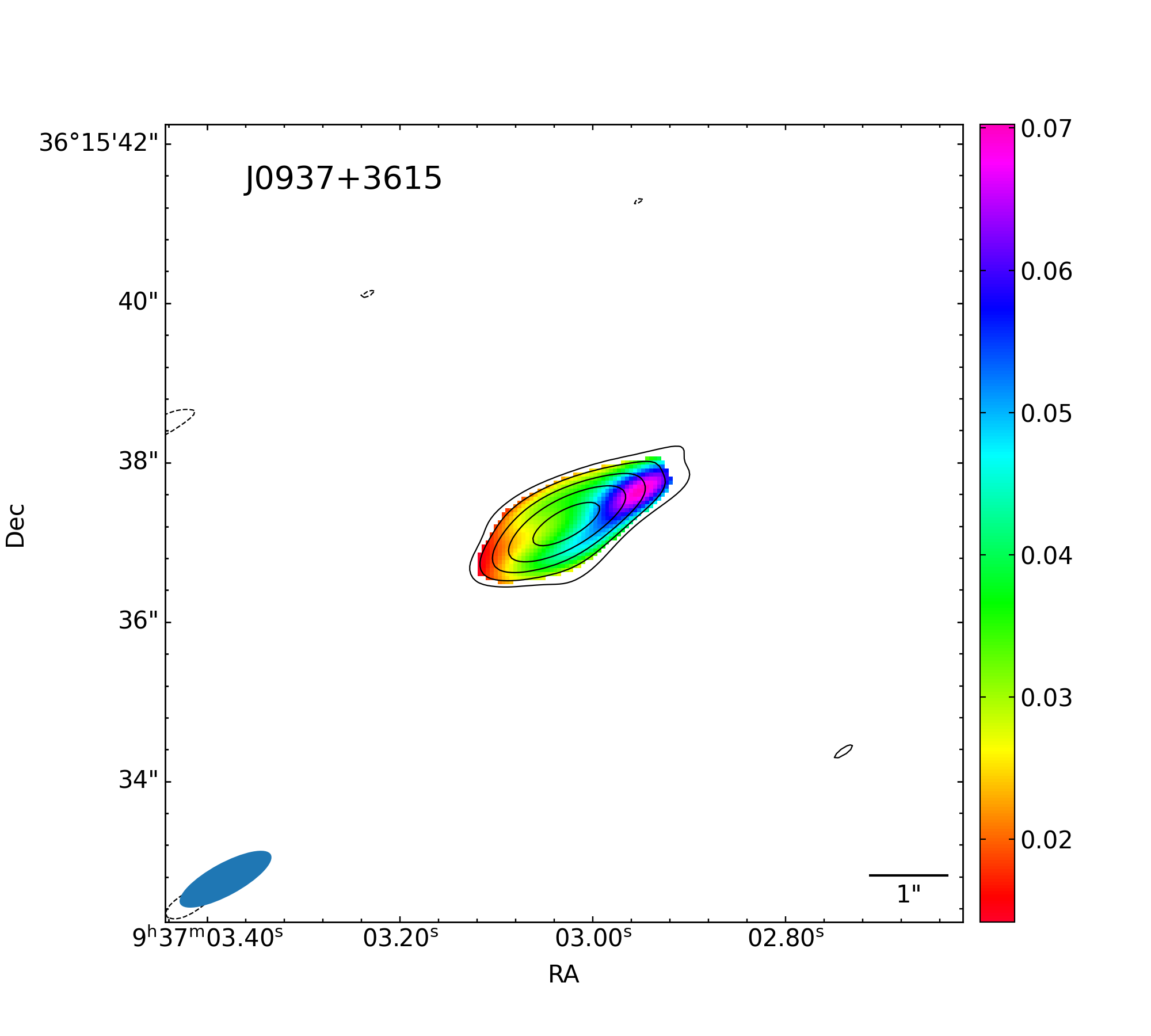}
         \caption{Spectral index error map, rms, contour levels, and beam size as in Fig.~\ref{fig:J0937spind}.} \label{fig:J0937spinderr}
     \end{subfigure}
     \hfill
     \\
     \begin{subfigure}[b]{0.47\textwidth}
         \centering
         \includegraphics[width=\textwidth]{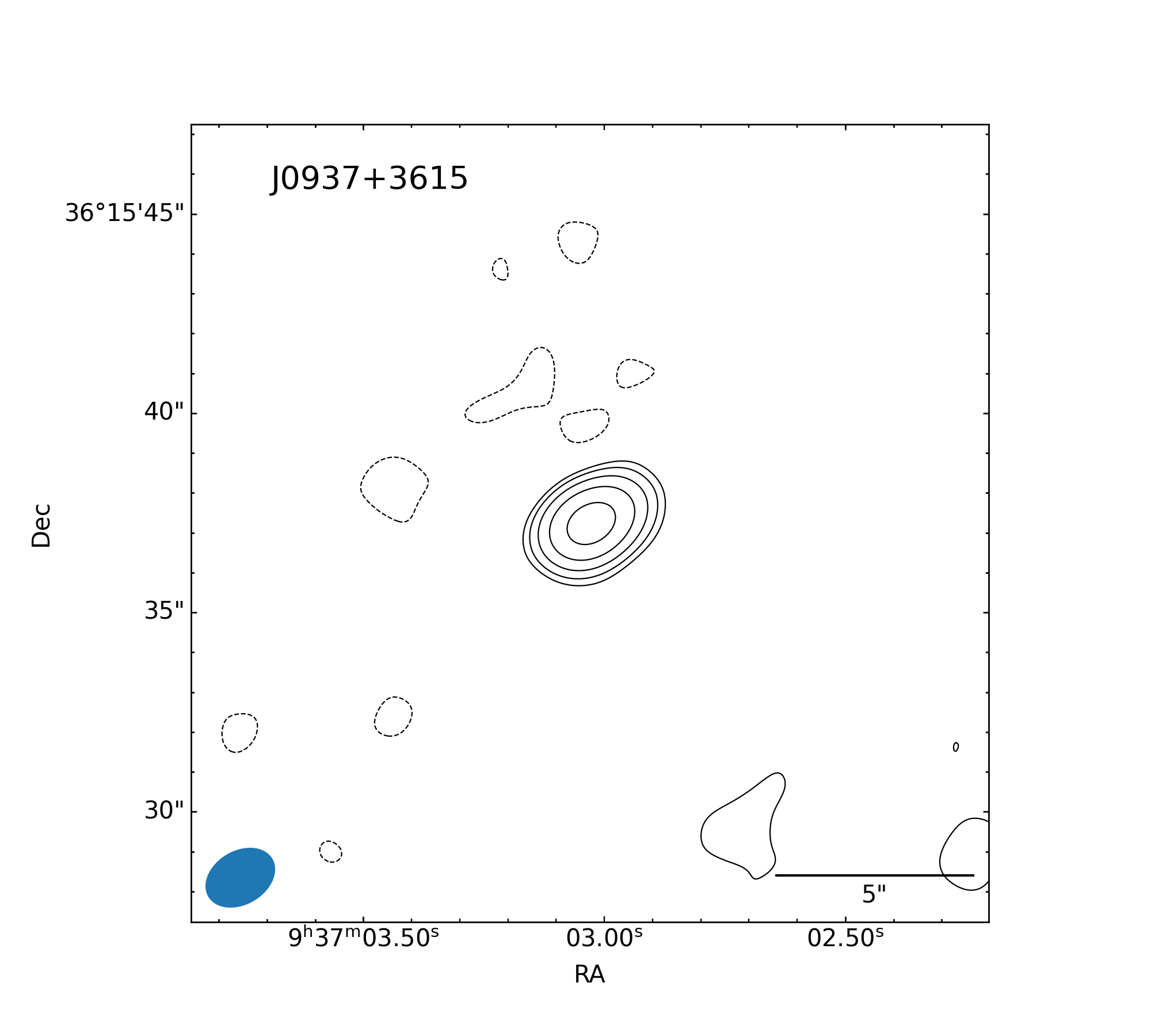}
         \caption{Tapered map with \texttt{uvtaper} = 90k$\lambda$, rms = 15$\mu$Jy beam$^{-1}$, contour levels at -3, 3 $\times$ 2$^n$, $n \in$ [0, 4], beam size 5.68 $\times$ 4.10~kpc.} \label{fig:J0937-90k}
     \end{subfigure}
          \hfill
     \begin{subfigure}[b]{0.47\textwidth}
         \centering
         \includegraphics[width=\textwidth]{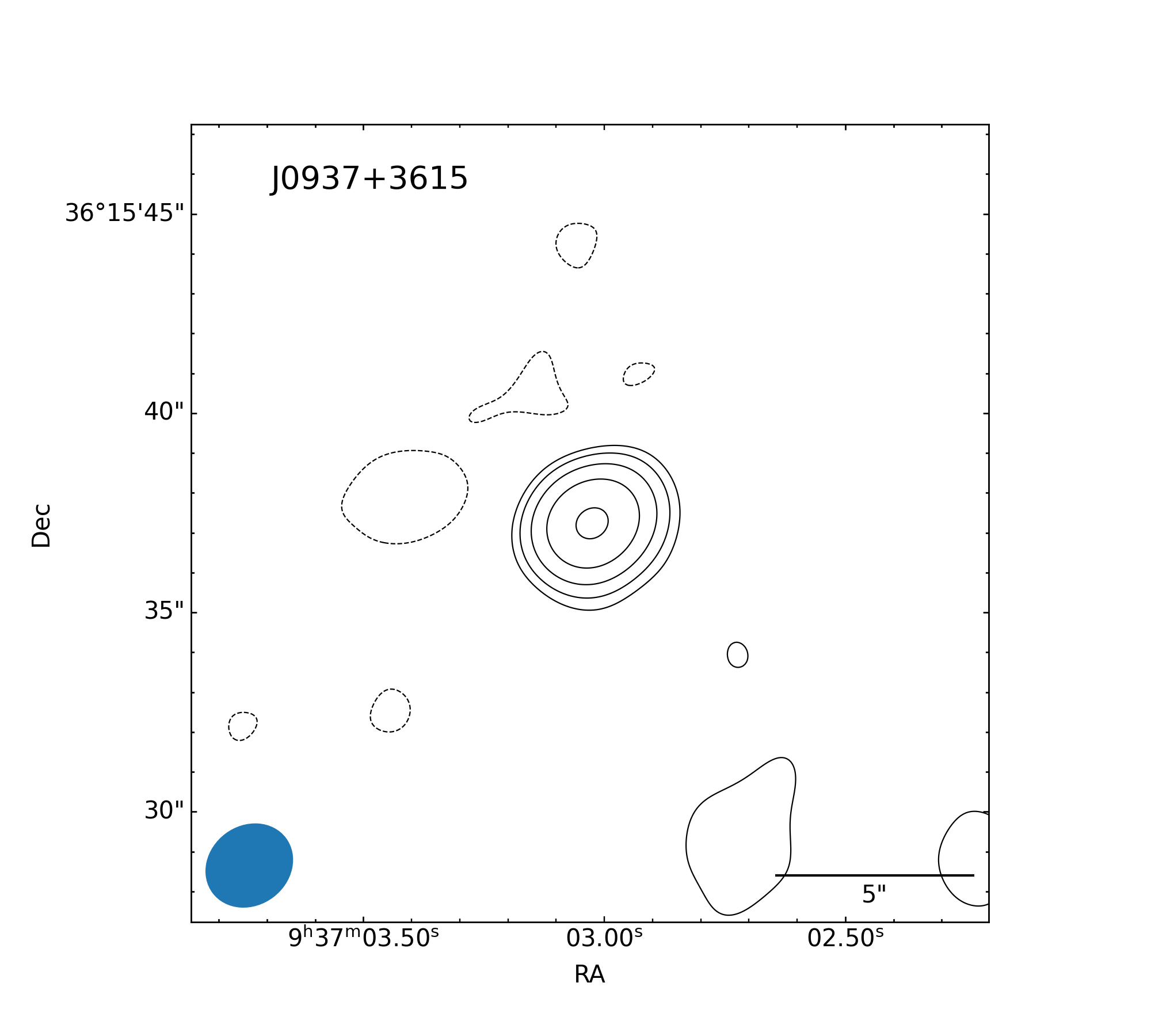}
         \caption{Tapered map with \texttt{uvtaper} = 60k$\lambda$, rms = 19$\mu$Jy beam$^{-1}$, contour levels at -3, 3 $\times$ 2$^n$, $n \in$ [0, 4], beam size 6.98 $\times$ 6.07~kpc.} \label{fig:J0937-60k}
     \end{subfigure}
          \hfill
     \\
     \begin{subfigure}[b]{0.47\textwidth}
         \centering
         \includegraphics[width=\textwidth]{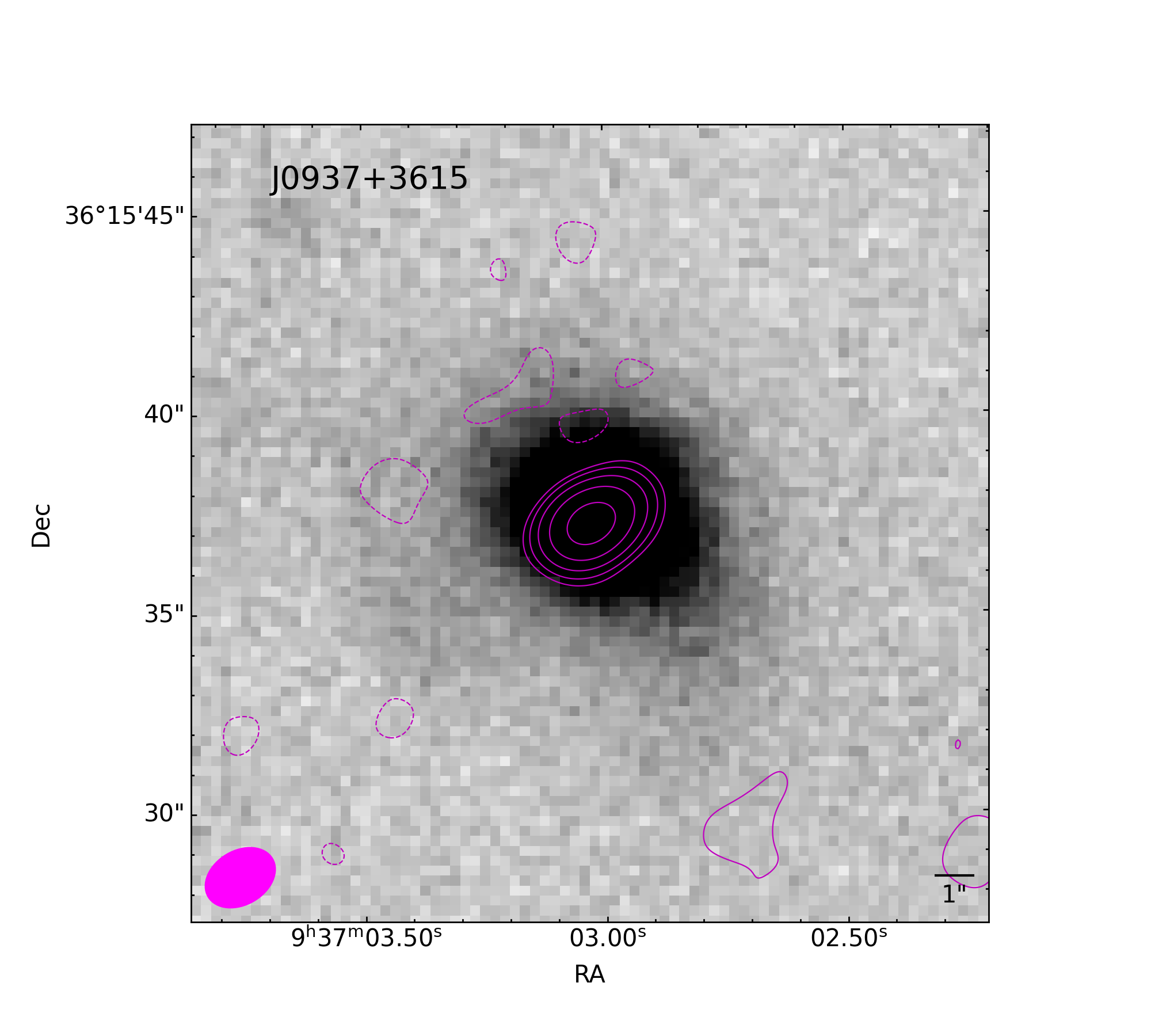}
         \caption{PanSTARRS $i$ band image of the host galaxy overlaid with the 90k$\lambda$ tapered map. Radio map properties as in Fig.~\ref{fig:J0937-90k}}. \label{fig:J0937-host}
     \end{subfigure}
        \caption{}
        \label{fig:J0937}
\end{figure*}


\begin{figure*}
     \centering
     \begin{subfigure}[b]{0.47\textwidth}
         \centering
         \includegraphics[width=\textwidth]{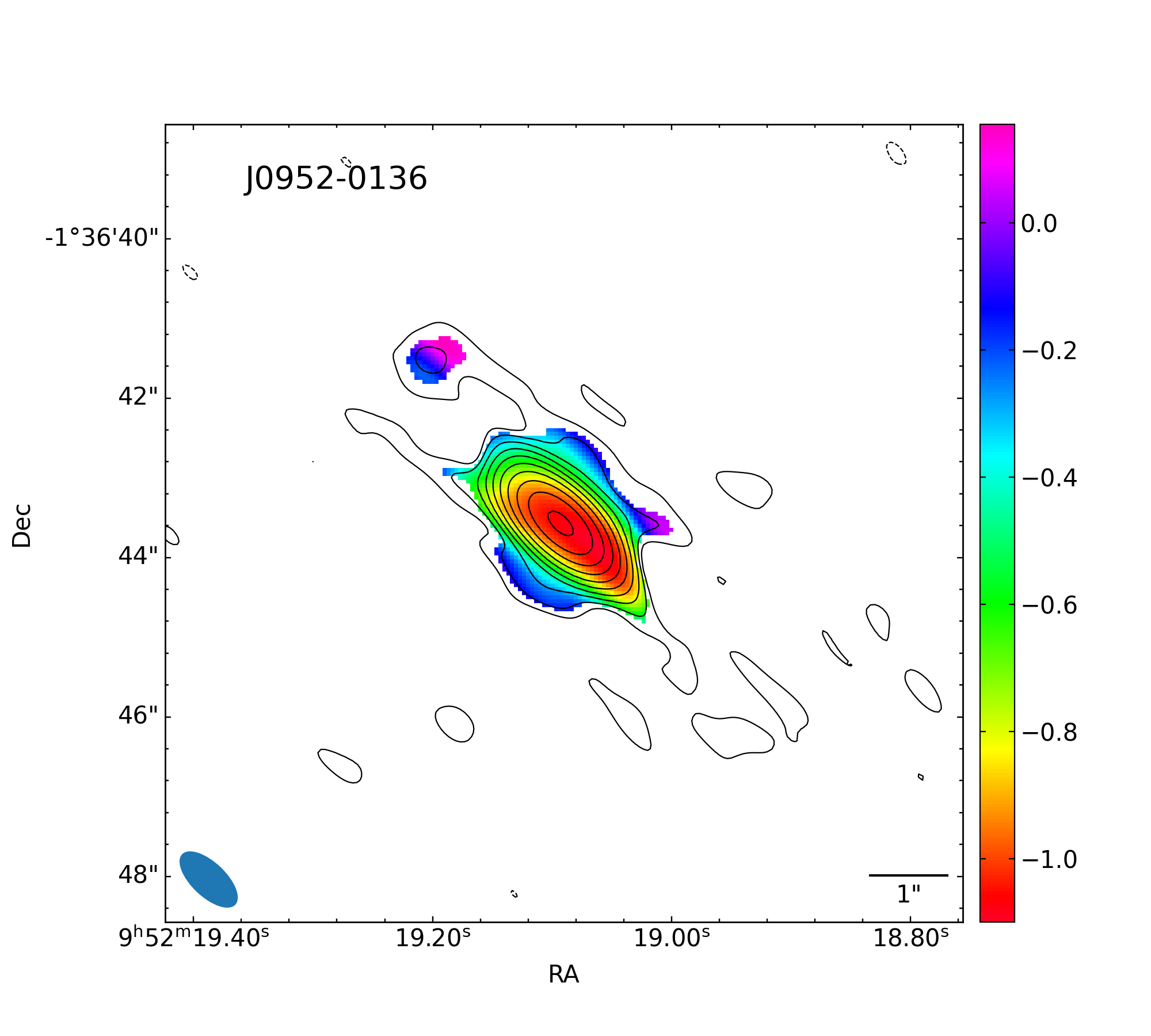}
         \caption{Spectral index map, rms = 10$\mu$Jy beam$^{-1}$, contour levels at -3, 3 $\times$ 2$^n$, $n \in$ [0, 9], beam size 0.37 $\times$ 0.17~kpc. } \label{fig:J0952spind}
     \end{subfigure}
     \hfill
     \begin{subfigure}[b]{0.47\textwidth}
         \centering
         \includegraphics[width=\textwidth]{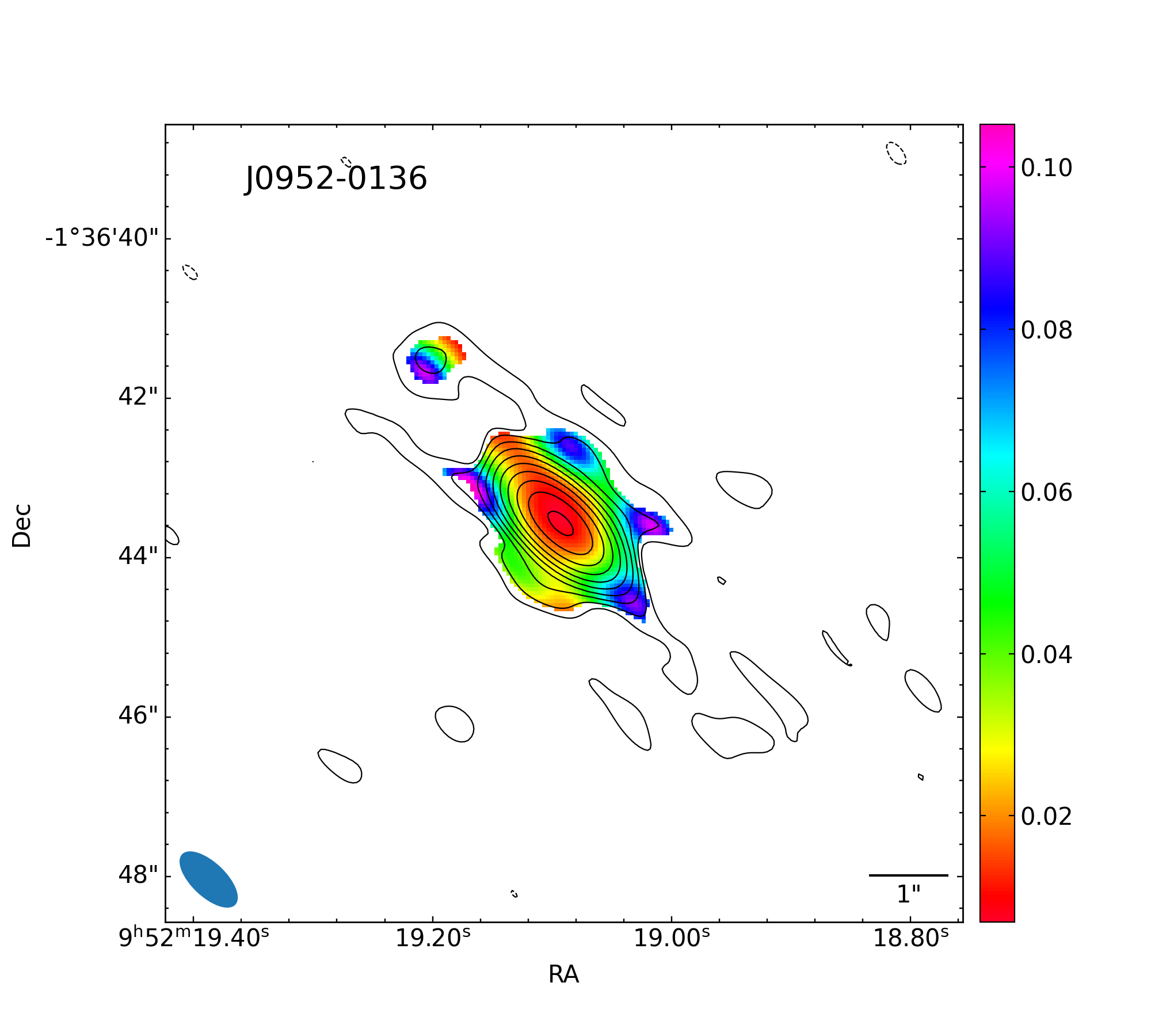}
         \caption{Spectral index error map, rms, contour levels, and beam size as in Fig.~\ref{fig:J0952spind}.} \label{fig:J0952spinderr}
     \end{subfigure}
     \hfill
     \\
     \begin{subfigure}[b]{0.47\textwidth}
         \centering
         \includegraphics[width=\textwidth]{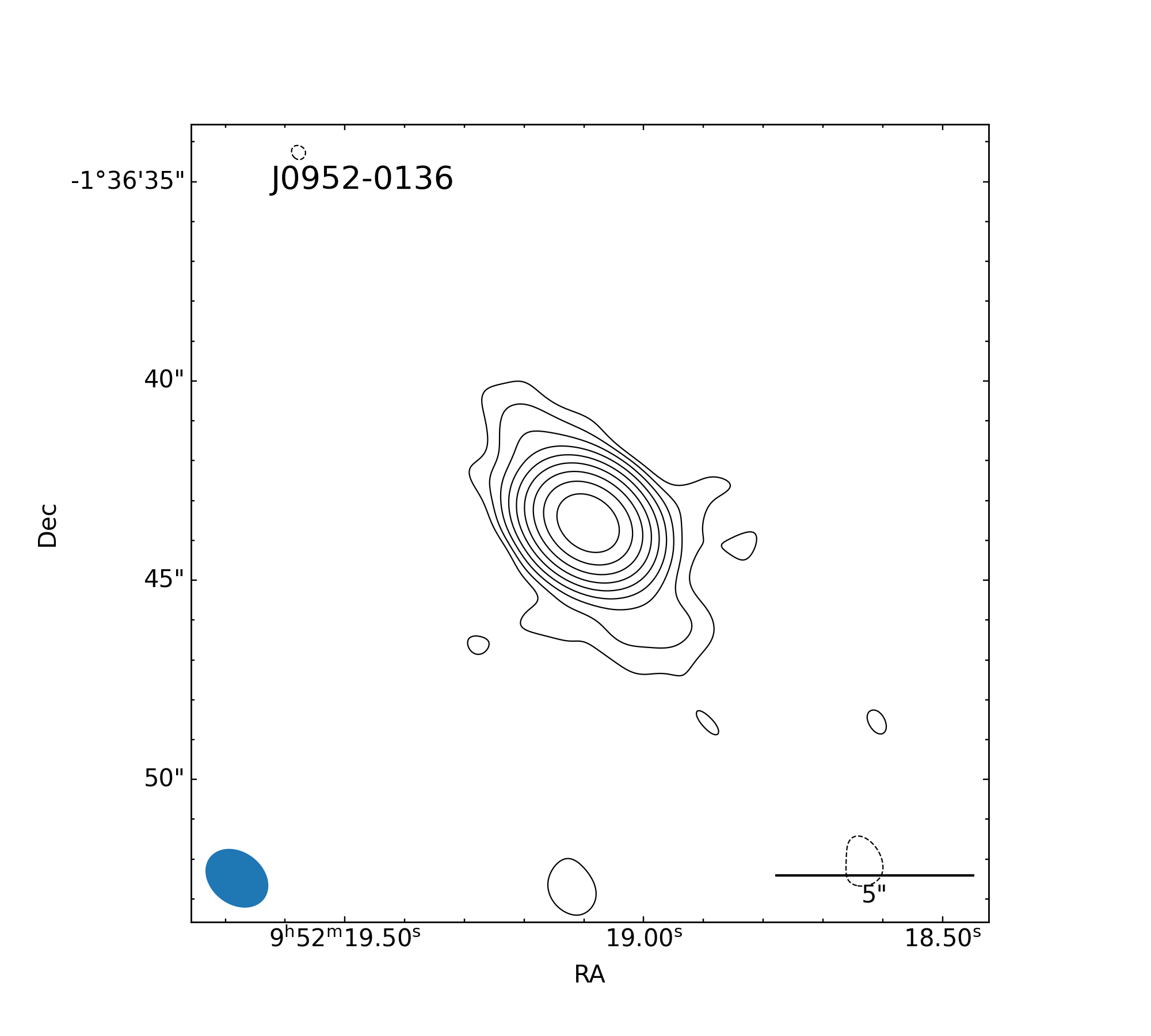}
         \caption{Tapered map with \texttt{uvtaper} = 90k$\lambda$, rms = 12$\mu$Jy beam$^{-1}$, contour levels at -3, 3 $\times$ 2$^n$, $n \in$ [0, 8], beam size 0.69 $\times$ 0.53~kpc.} \label{fig:J0952-90k}
     \end{subfigure}
          \hfill
     \begin{subfigure}[b]{0.47\textwidth}
         \centering
         \includegraphics[width=\textwidth]{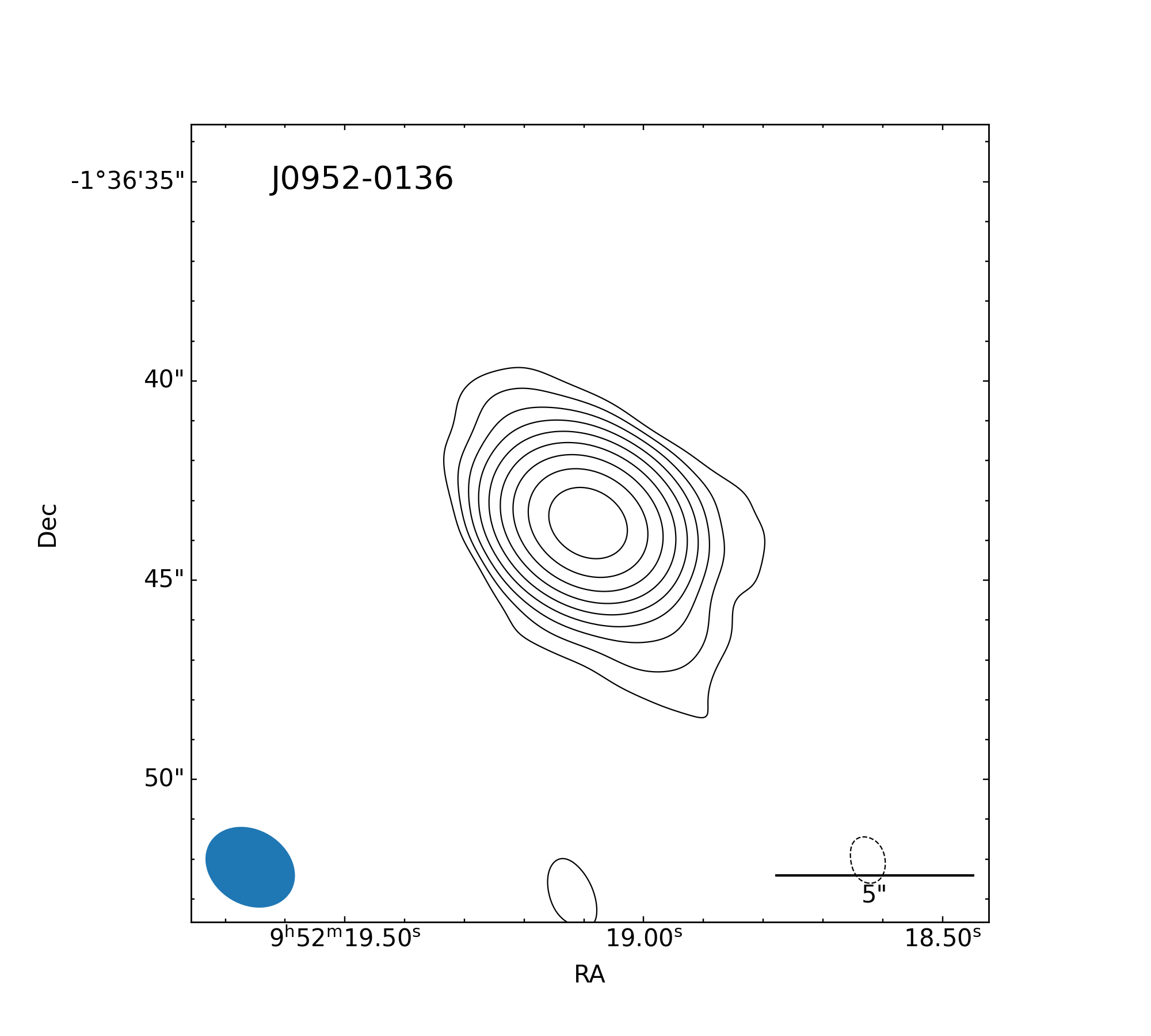}
         \caption{Tapered map with \texttt{uvtaper} = 60k$\lambda$, rms = 14$\mu$Jy beam$^{-1}$, contour levels at -3, 3 $\times$ 2$^n$, $n \in$ [0, 8], beam size 0.96 $\times$ 0.76~kpc.} \label{fig:J0952-60k}
     \end{subfigure}
          \hfill
     \\
     \begin{subfigure}[b]{0.47\textwidth}
         \centering
         \includegraphics[width=\textwidth]{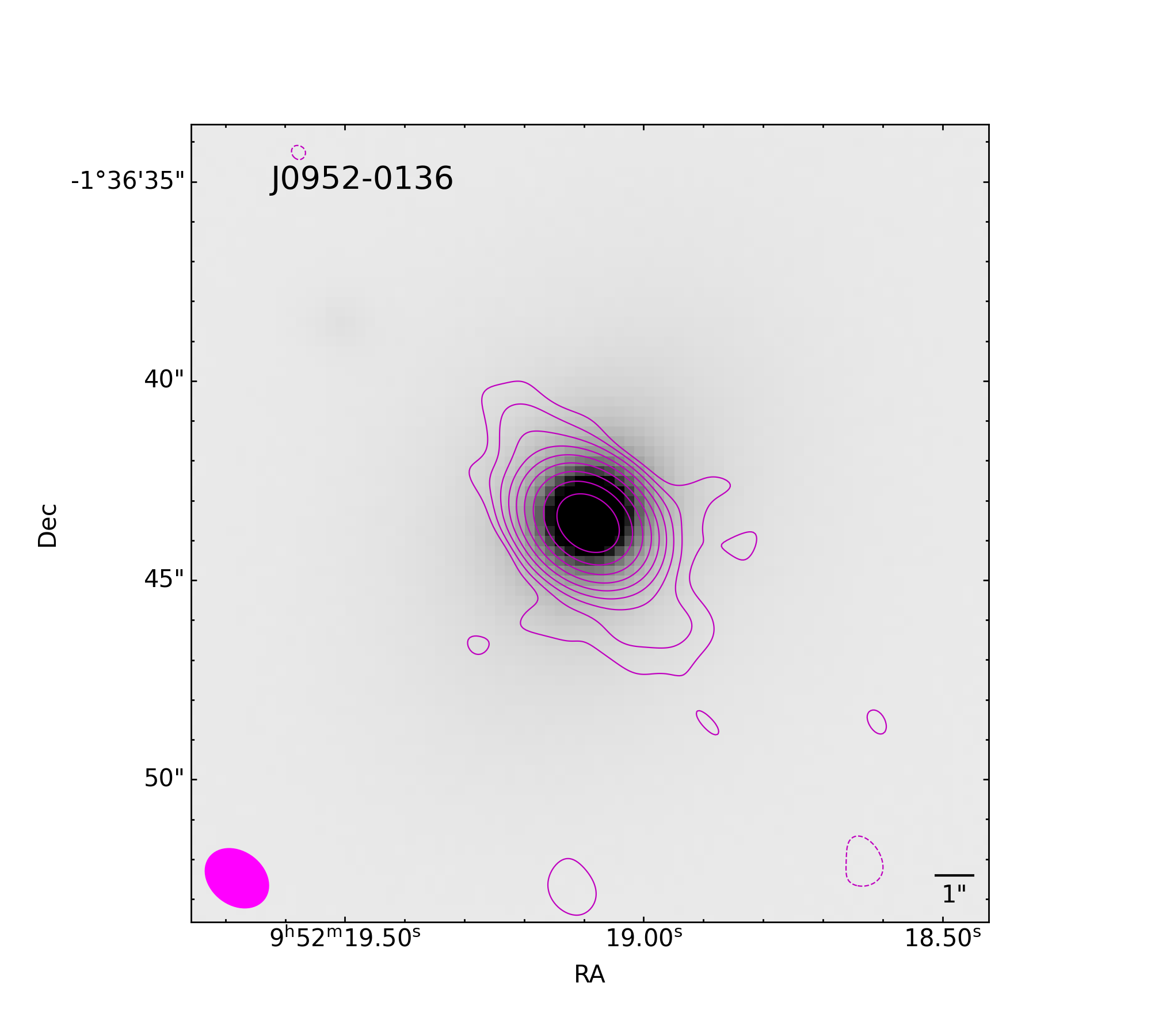}
         \caption{PanSTARRS $i$ band image of the host galaxy overlaid with the 90k$\lambda$ tapered map. Radio map properties as in Fig.~\ref{fig:J0952-90k}}. \label{fig:J0952-host}
     \end{subfigure}
        \caption{}
        \label{fig:J0952}
\end{figure*}


\begin{figure*}
     \centering
     \begin{subfigure}[b]{0.47\textwidth}
         \centering
         \includegraphics[width=\textwidth]{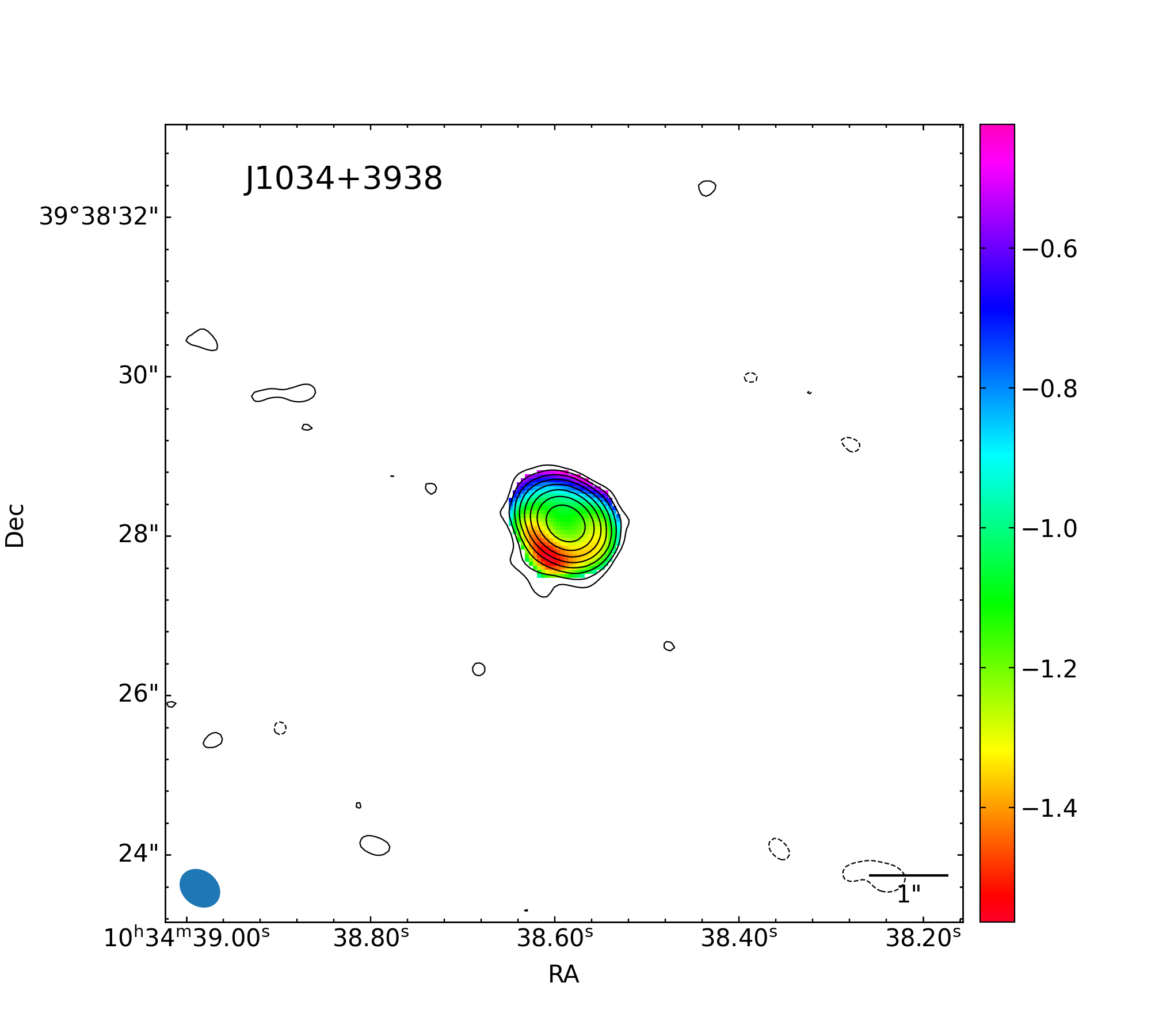}
         \caption{Spectral index map, rms = 10$\mu$Jy, contour levels at -3, 3 $\times$ 2$^n$, $n \in$ [0, 7], beam size 0.46 $\times$ 0.36~kpc. } \label{fig:J1034spind}
     \end{subfigure}
     \hfill
     \begin{subfigure}[b]{0.47\textwidth}
         \centering
         \includegraphics[width=\textwidth]{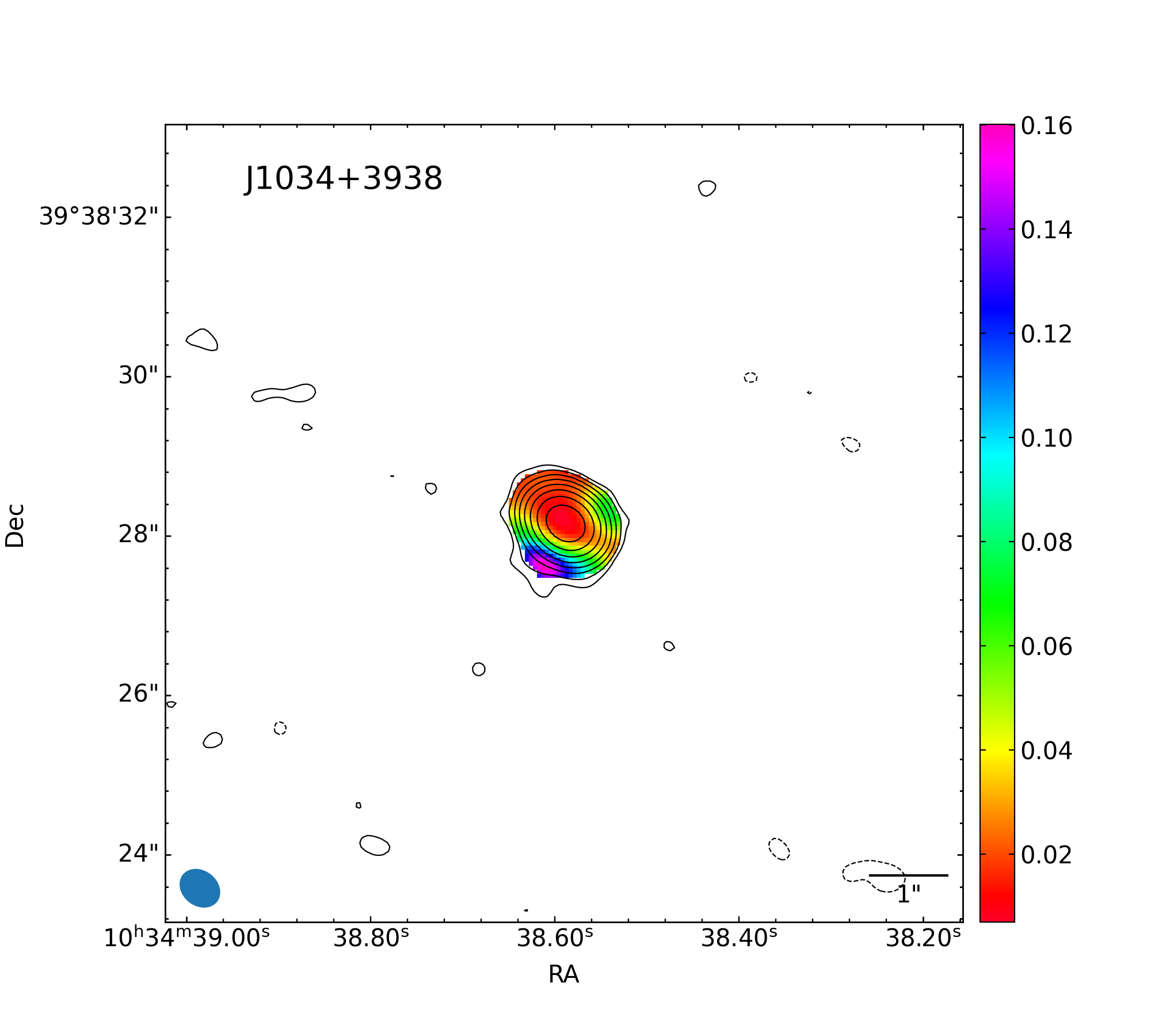}
         \caption{Spectral index error map, rms, contour levels, and beam size as in Fig.~\ref{fig:J1034spind}.} \label{fig:J1034spinderr}
     \end{subfigure}
     \hfill
     \\
     \begin{subfigure}[b]{0.47\textwidth}
         \centering
         \includegraphics[width=\textwidth]{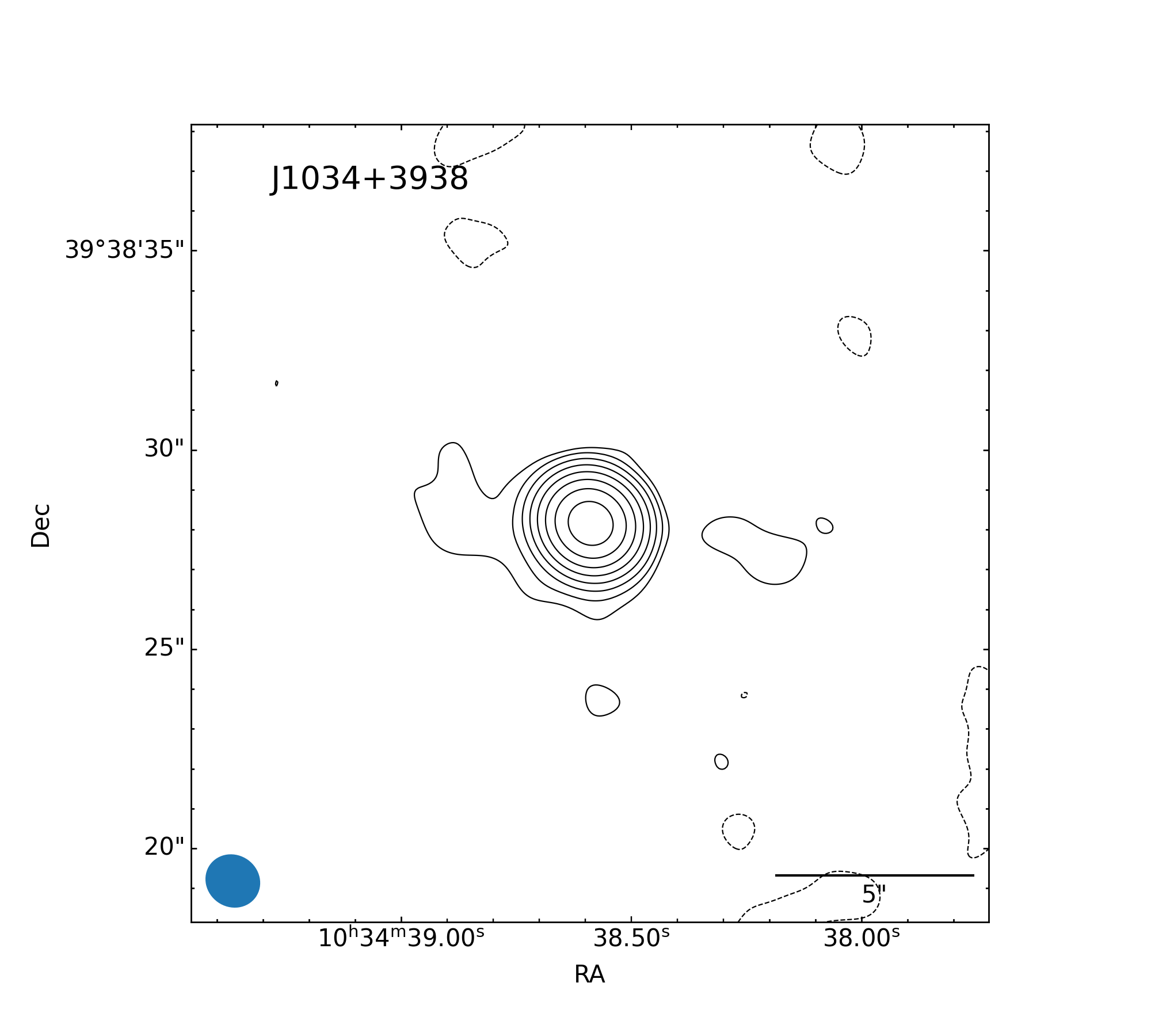}
         \caption{Tapered map with \texttt{uvtaper} = 90k$\lambda$, rms = 12$\mu$Jy beam$^{-1}$, contour levels at -3, 3 $\times$ 2$^n$, $n \in$ [0, 7], beam size 1.17 $\times$ 1.08~kpc.} \label{fig:J1034-90k}
     \end{subfigure}
          \hfill
     \begin{subfigure}[b]{0.47\textwidth}
         \centering
         \includegraphics[width=\textwidth]{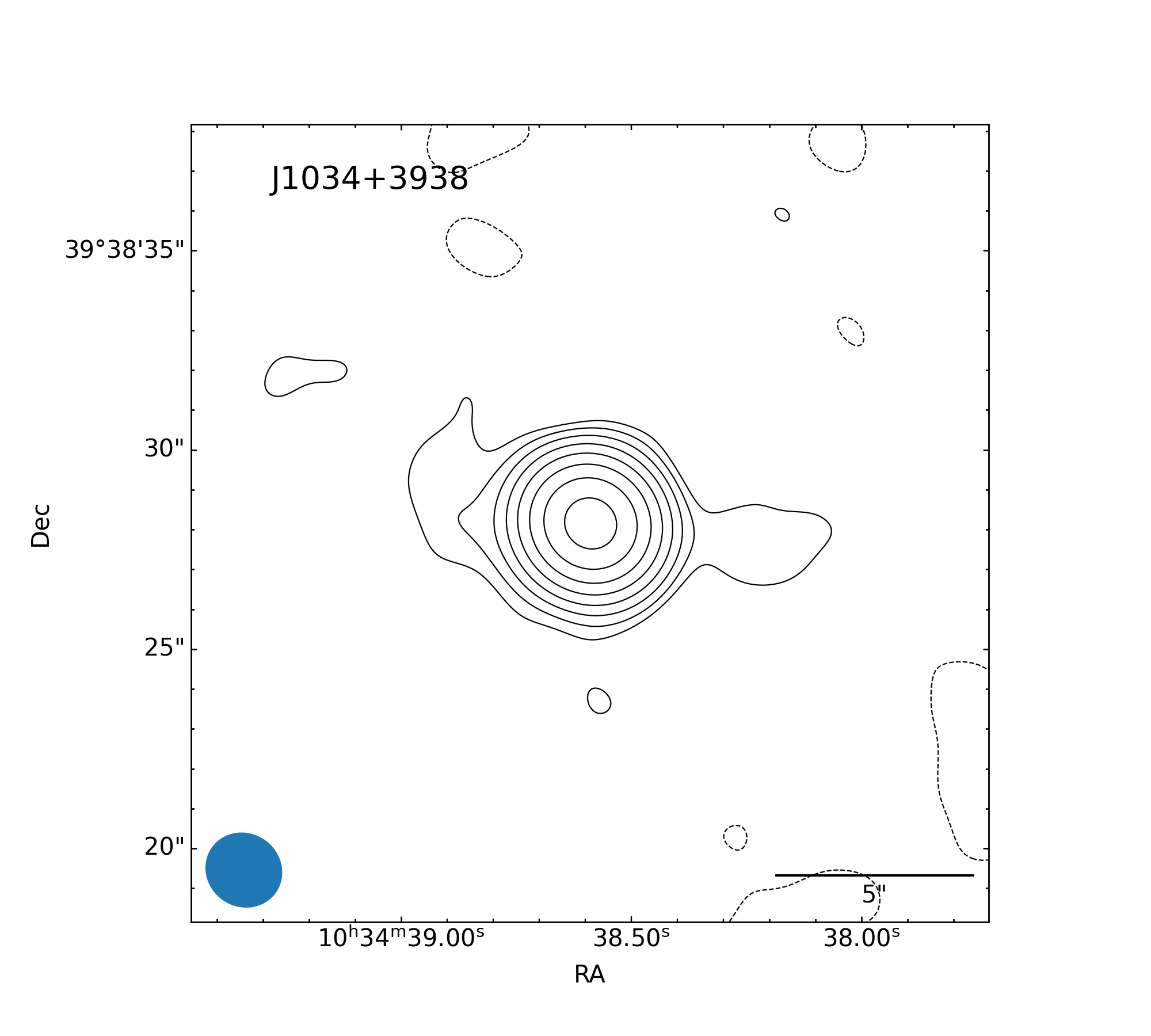}
         \caption{Tapered map with \texttt{uvtaper} = 60k$\lambda$, rms = 14$\mu$Jy beam$^{-1}$, contour levels at -3, 3 $\times$ 2$^n$, $n \in$ [0, 7], beam size 1.63 $\times$ 1.53~kpc.} \label{fig:J1034-60k}
     \end{subfigure}
          \hfill
     \\
     \begin{subfigure}[b]{0.47\textwidth}
         \centering
         \includegraphics[width=\textwidth]{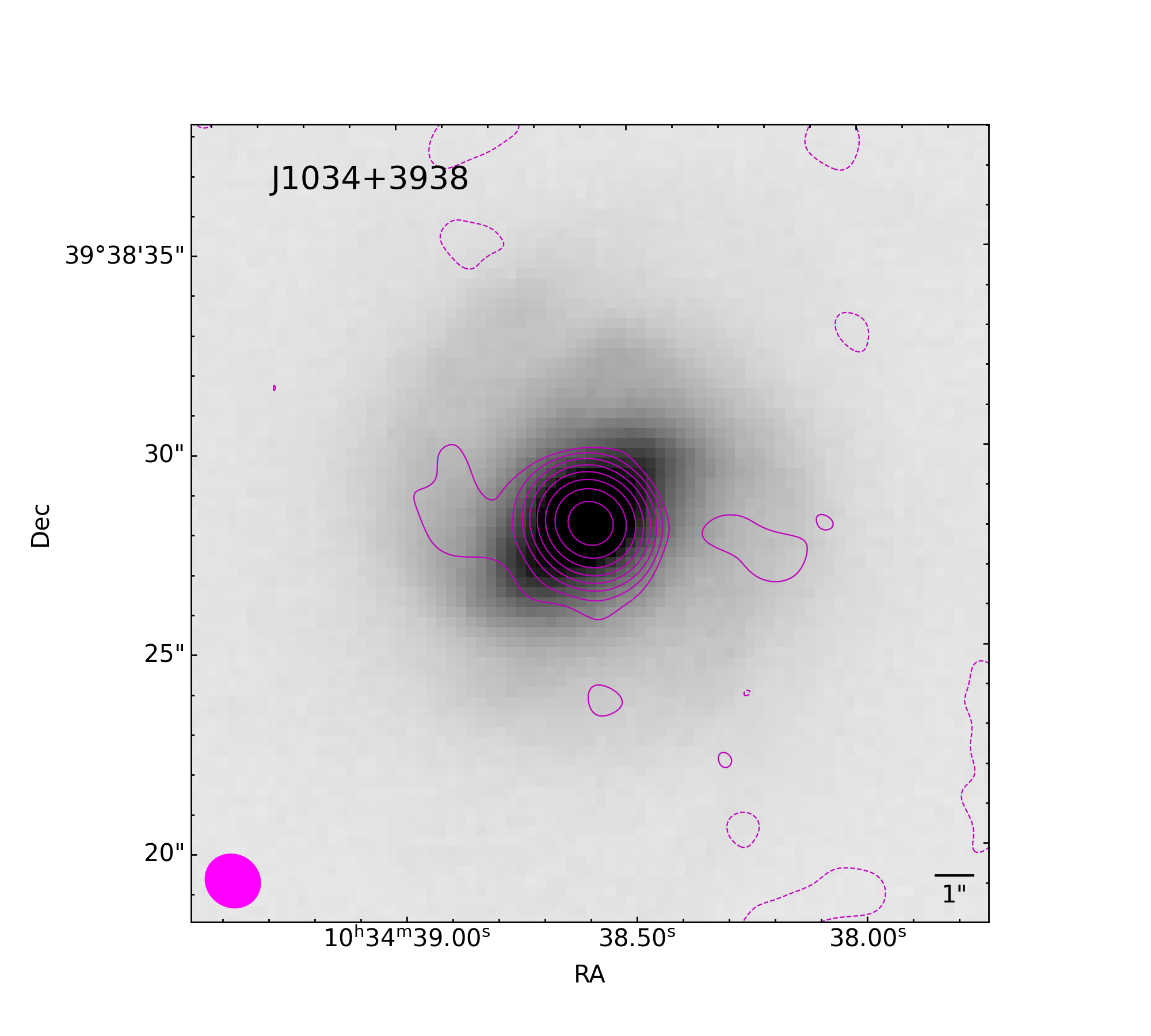}
         \caption{PanSTARRS $i$ band image of the host galaxy overlaid with the 90k$\lambda$ tapered map. Radio map properties as in Fig.~\ref{fig:J1034-90k}}. \label{fig:J1034-host}
     \end{subfigure}
        \caption{}
        \label{fig:J1034}
\end{figure*}


\begin{figure*}
     \centering
     \begin{subfigure}[b]{0.47\textwidth}
         \centering
         \includegraphics[width=\textwidth]{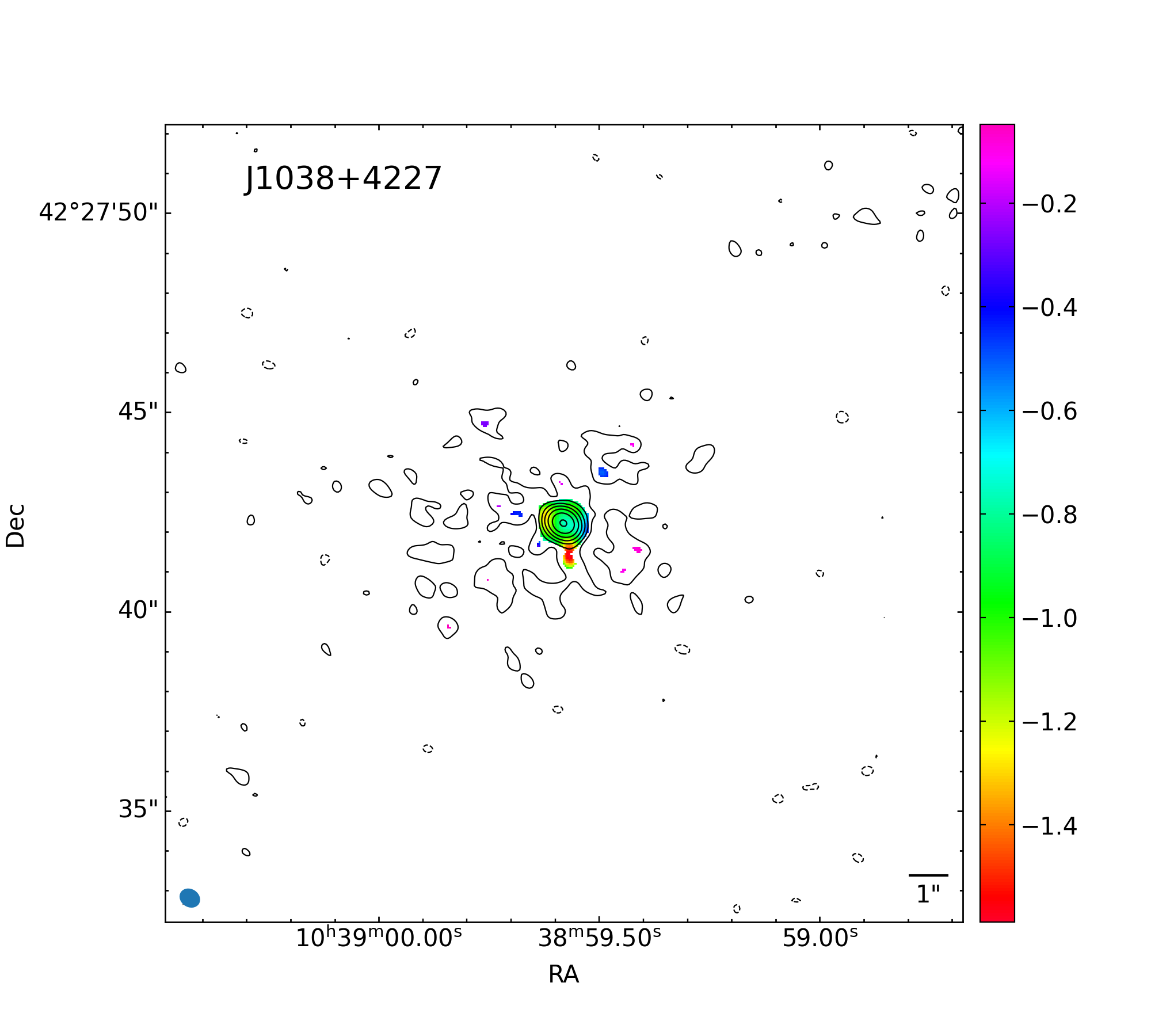}
         \caption{Spectral index map, rms = 10$\mu$Jy beam$^{-1}$, contour levels at -3, 3 $\times$ 2$^n$, $n \in$ [0, 6], beam size 1.92 $\times$ 1.60~kpc. } \label{fig:J1038spind}
     \end{subfigure}
     \hfill
     \begin{subfigure}[b]{0.47\textwidth}
         \centering
         \includegraphics[width=\textwidth]{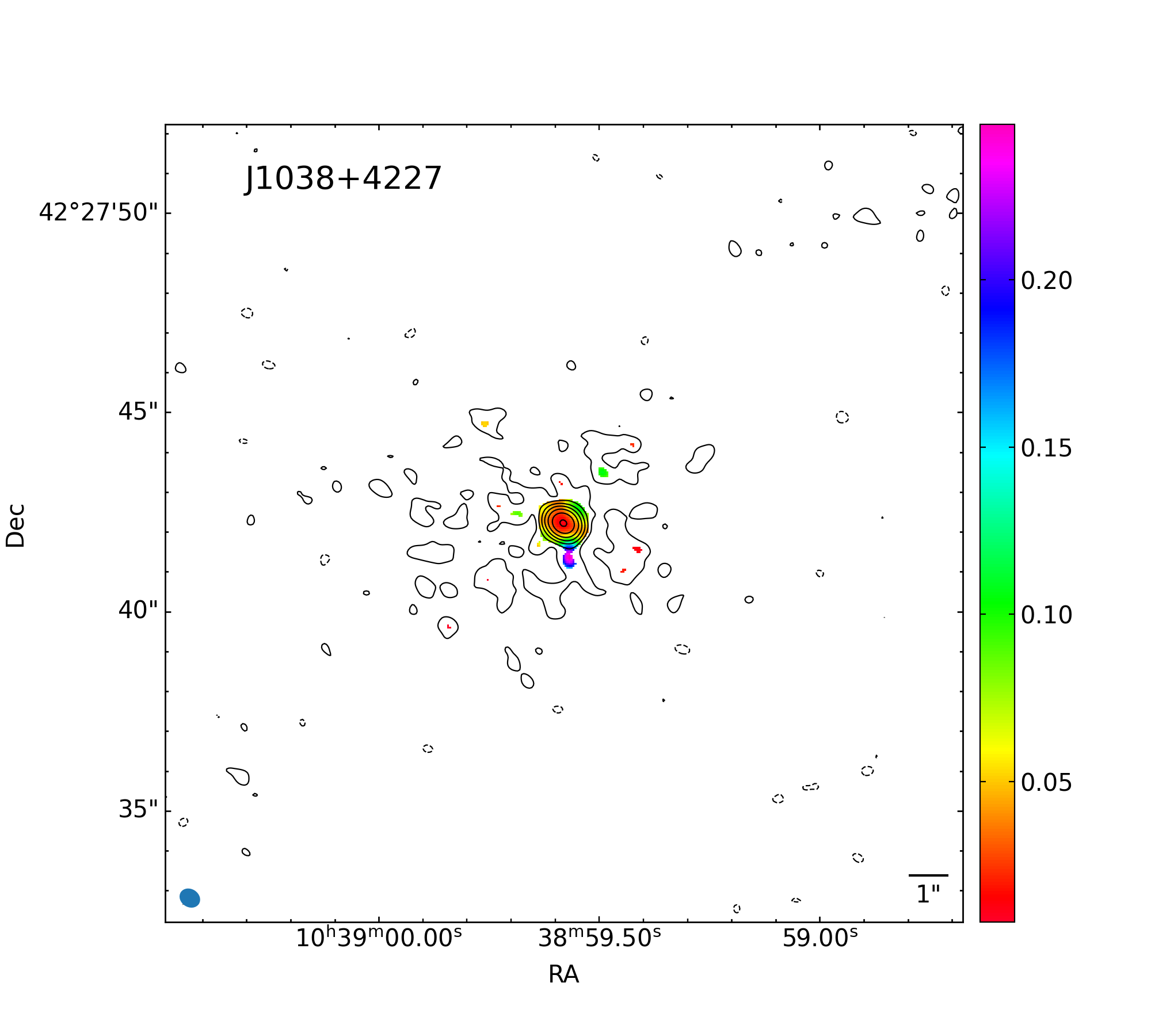}
         \caption{Spectral index error map, rms, contour levels, and beam size as in Fig.~\ref{fig:J1038spind}.} \label{fig:J1038spinderr}
     \end{subfigure}
     \hfill
     \\
     \begin{subfigure}[b]{0.47\textwidth}
         \centering
         \includegraphics[width=\textwidth]{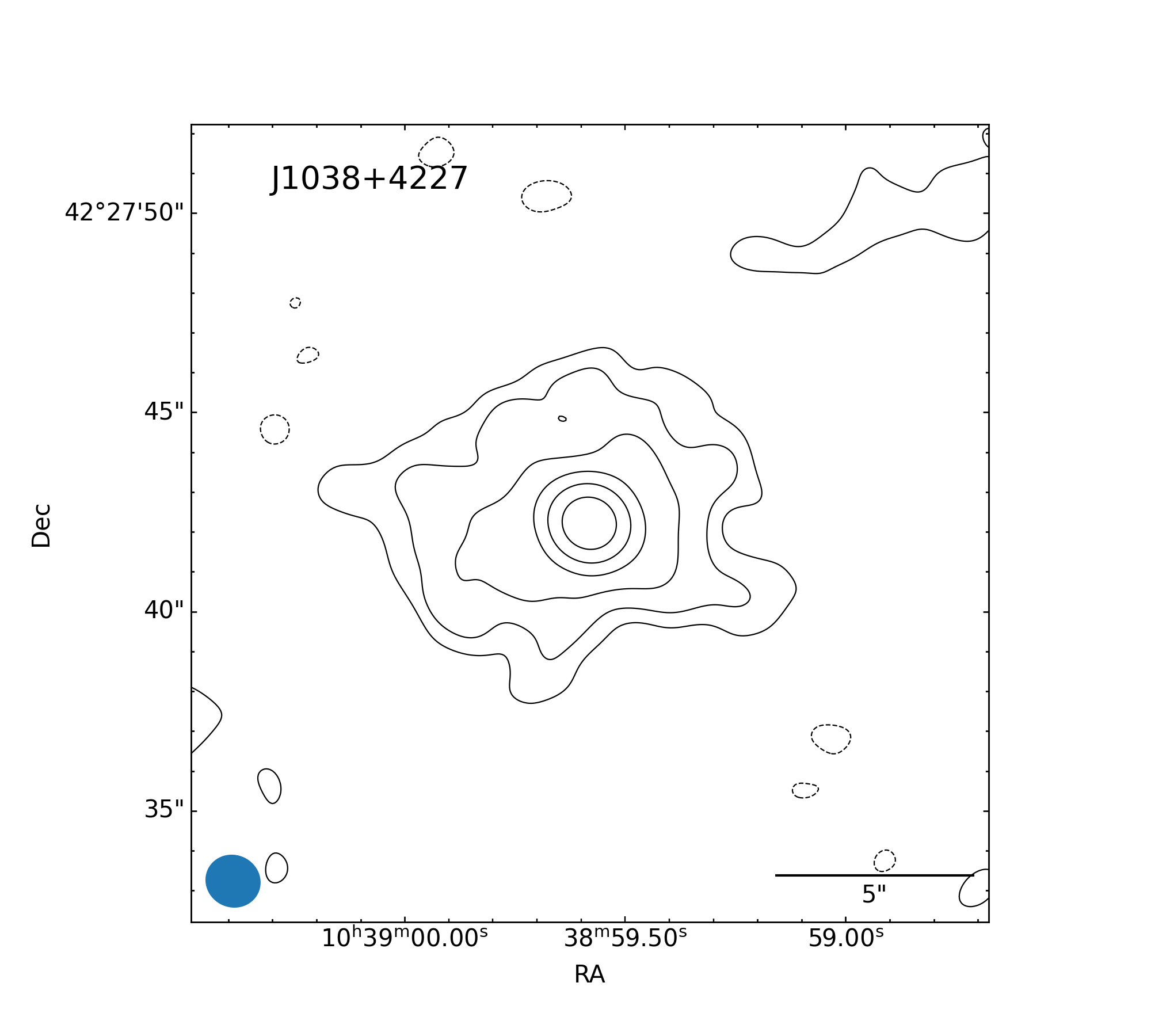}
         \caption{Tapered map with \texttt{uvtaper} = 90k$\lambda$, rms = 13$\mu$Jy beam$^{-1}$, contour levels at -3, 3 $\times$ 2$^n$, $n \in$ [0, 5], beam size 4.97 $\times$ 4.65~kpc.} \label{fig:J1038-90k}
     \end{subfigure}
          \hfill
     \begin{subfigure}[b]{0.47\textwidth}
         \centering
         \includegraphics[width=\textwidth]{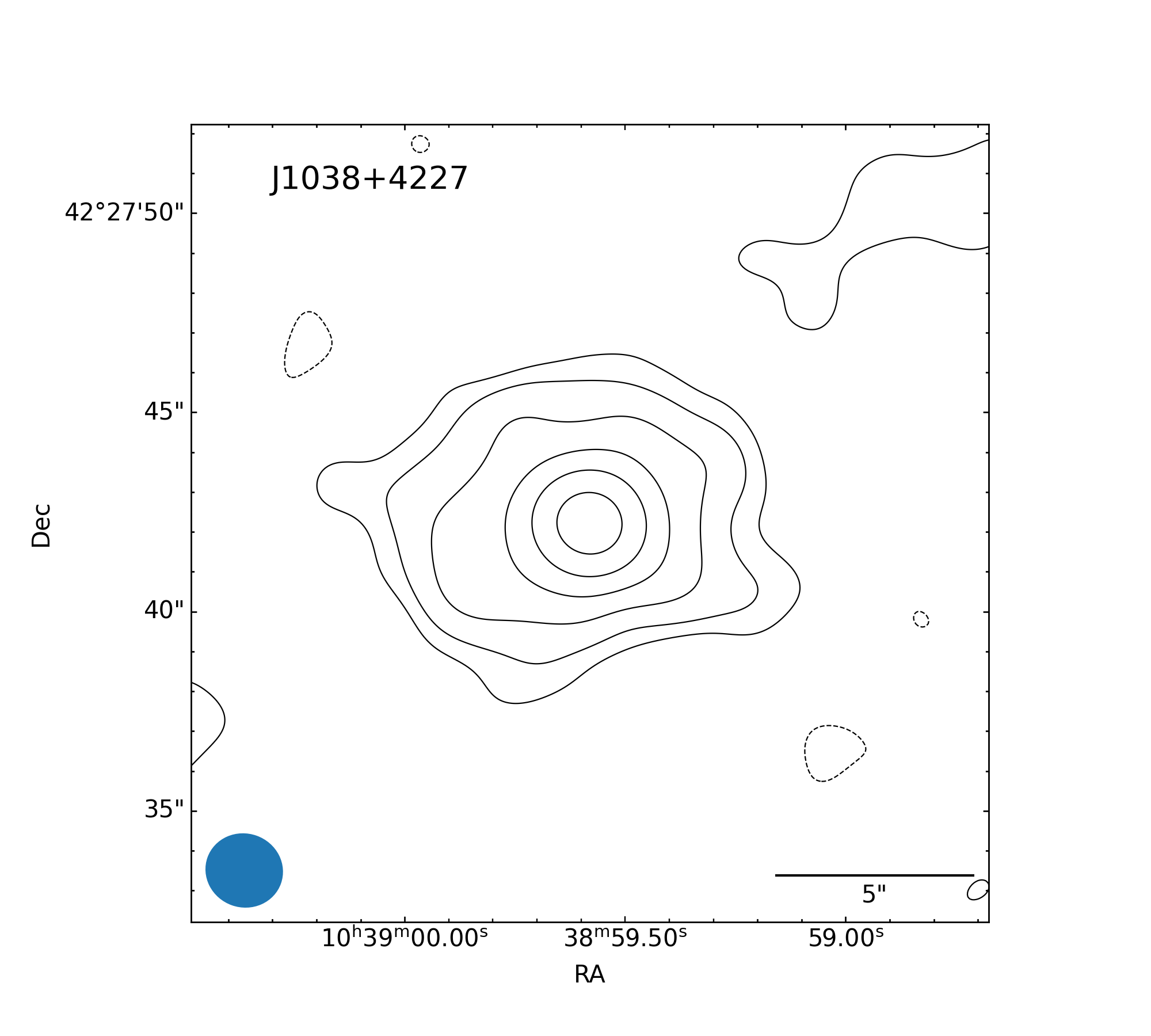}
         \caption{Tapered map with \texttt{uvtaper} = 60k$\lambda$, rms = 17$\mu$Jy beam$^{-1}$, contour levels at -3, 3 $\times$ 2$^n$, $n \in$ [0, 5], beam size 6.96 $\times$ 6.57~kpc.} \label{fig:J1038-60k}
     \end{subfigure}
     \hfill
     \\
     \begin{subfigure}[b]{0.47\textwidth}
         \centering
         \includegraphics[width=\textwidth]{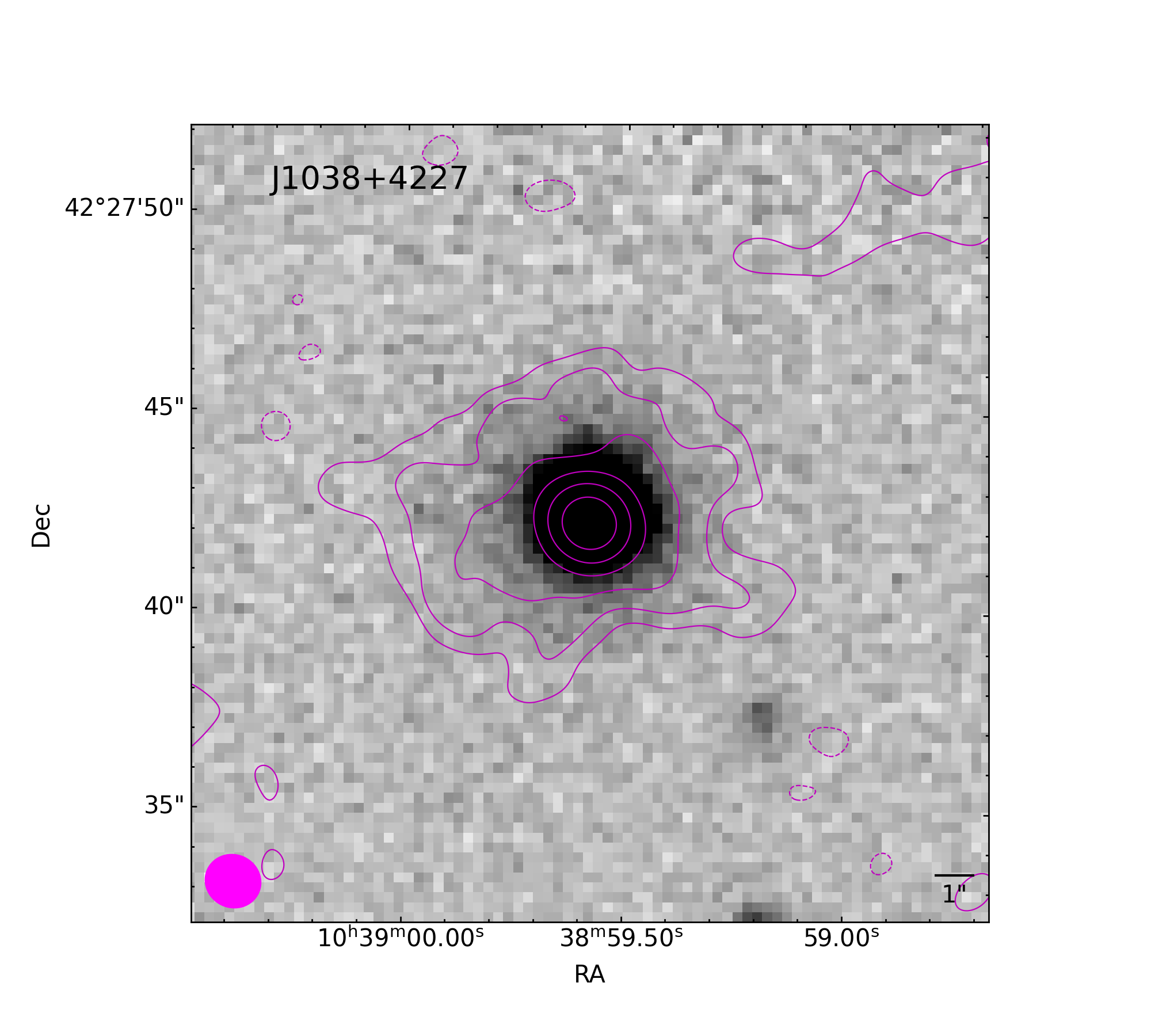}
         \caption{PanSTARRS $i$ band image of the host galaxy overlaid with the 90k$\lambda$ tapered map. Radio map properties as in Fig.~\ref{fig:J1038-90k}}. \label{fig:J1038-host}
     \end{subfigure}
        \caption{}
        \label{fig:J1038}
\end{figure*}


\begin{figure*}
     \centering
     \begin{subfigure}[b]{0.47\textwidth}
         \centering
         \includegraphics[width=\textwidth]{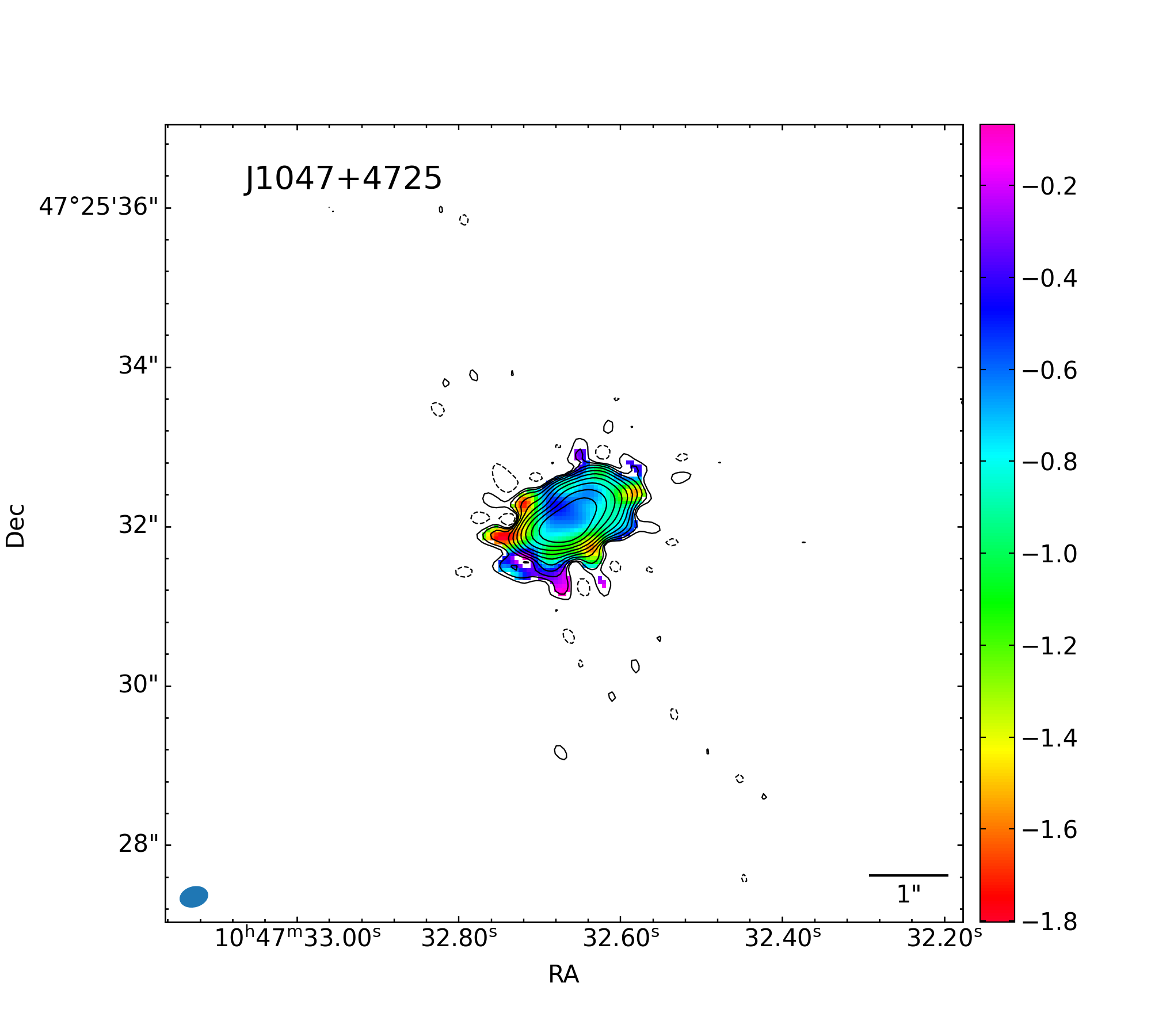}
         \caption{Spectral index map, rms = 35$\mu$Jy beam$^{-1}$, contour levels at -3, 3 $\times$ 2$^n$, $n \in$ [0, 10], beam size 2.78 $\times$ 1.95~kpc. } \label{fig:J1047spind} 
     \end{subfigure}
     \hfill
     \begin{subfigure}[b]{0.47\textwidth}
         \centering
         \includegraphics[width=\textwidth]{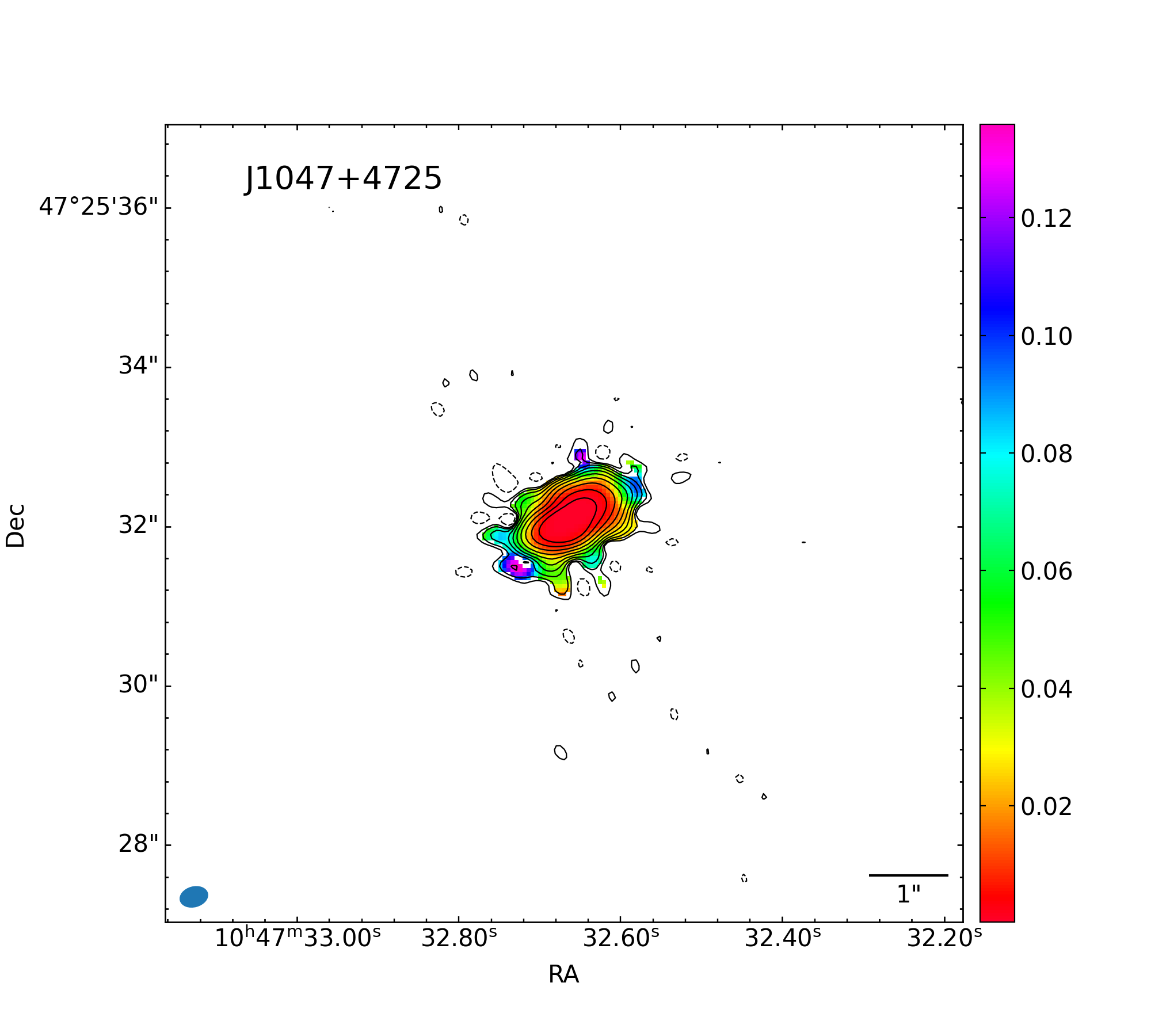}
         \caption{Spectral index error map, rms, contour levels, and beam size as in Fig.~\ref{fig:J1047spind}.} \label{fig:J1047spinderr}
     \end{subfigure}
     \hfill
     \\
     \begin{subfigure}[b]{0.47\textwidth}
         \centering
         \includegraphics[width=\textwidth]{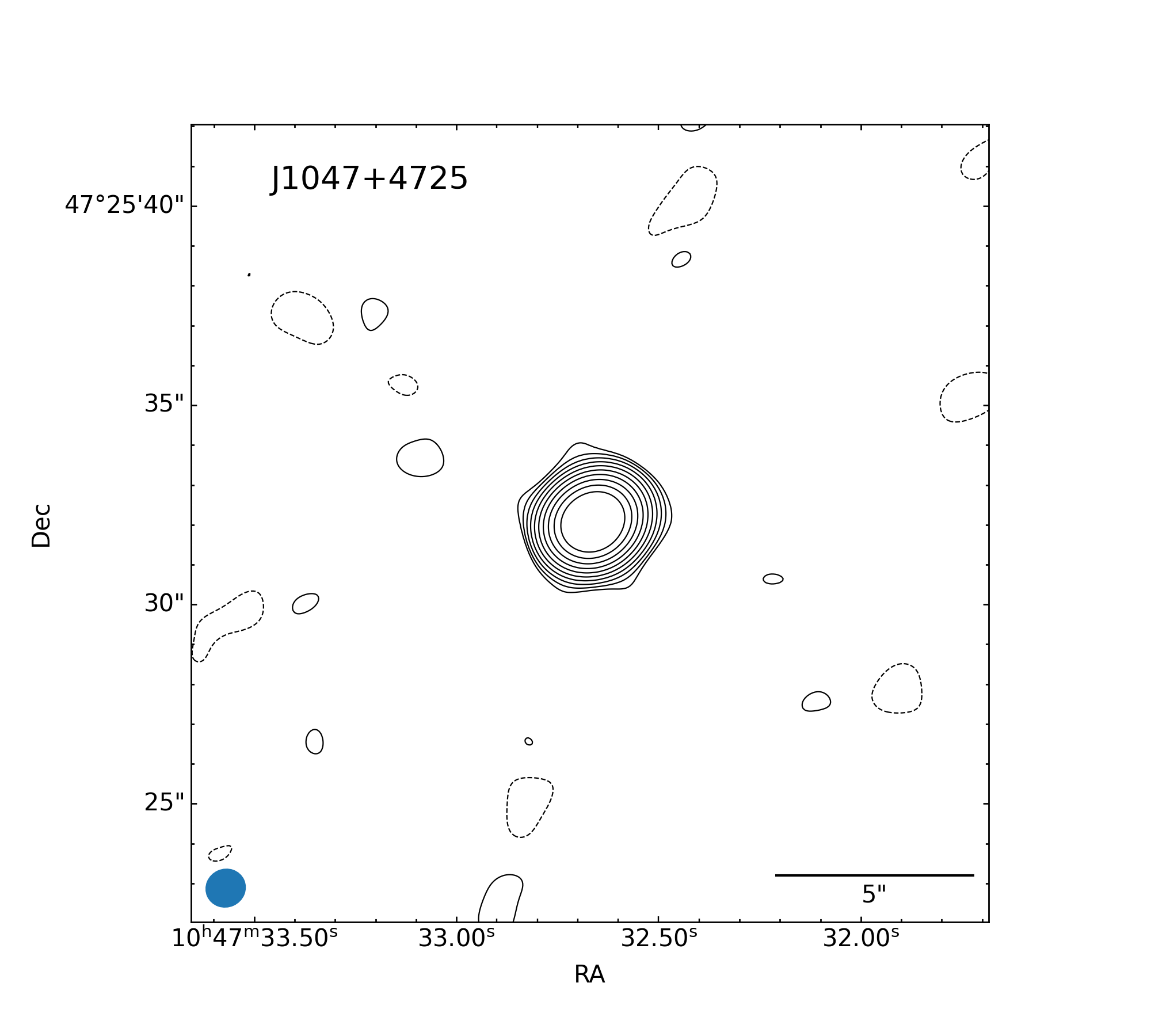}
         \caption{Tapered map with \texttt{uvtaper} = 90k$\lambda$, rms = 50$\mu$Jy beam$^{-1}$, contour levels at -3, 3 $\times$ 2$^n$, $n \in$ [0, 11], beam size 7.66 $\times$ 7.28~kpc.} \label{fig:J1047-90k}
     \end{subfigure}
          \hfill
     \begin{subfigure}[b]{0.47\textwidth}
         \centering
         \includegraphics[width=\textwidth]{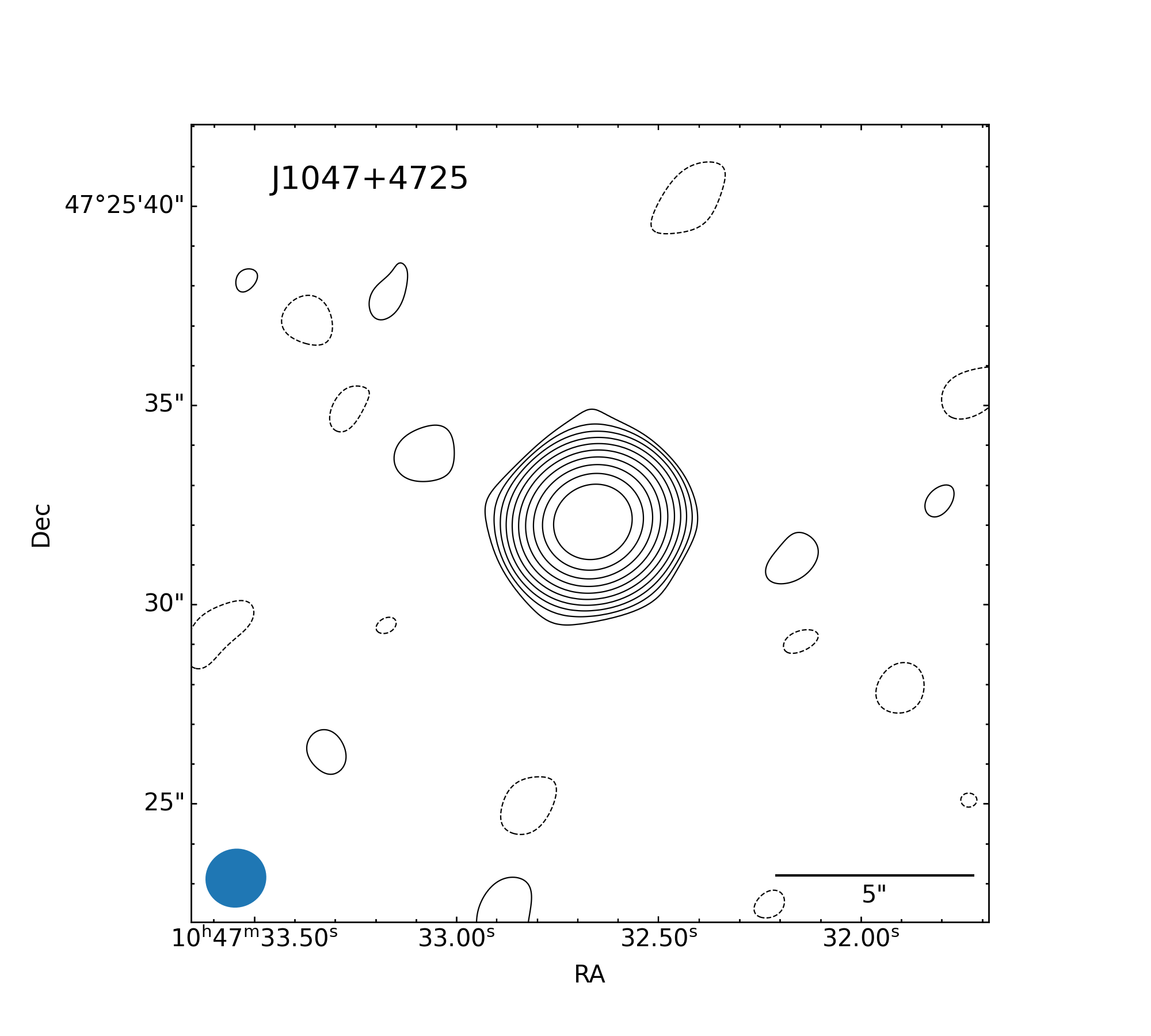}
         \caption{Tapered map with \texttt{uvtaper} = 60k$\lambda$, rms = 80$\mu$Jy beam$^{-1}$, contour levels at -3, 3 $\times$ 2$^n$, $n \in$ [0, 10], beam size 11.56 $\times$ 11.03~kpc.} \label{fig:J1047-60k} 
     \end{subfigure}
          \hfill
     \\
     \begin{subfigure}[b]{0.47\textwidth}
         \centering
         \includegraphics[width=\textwidth]{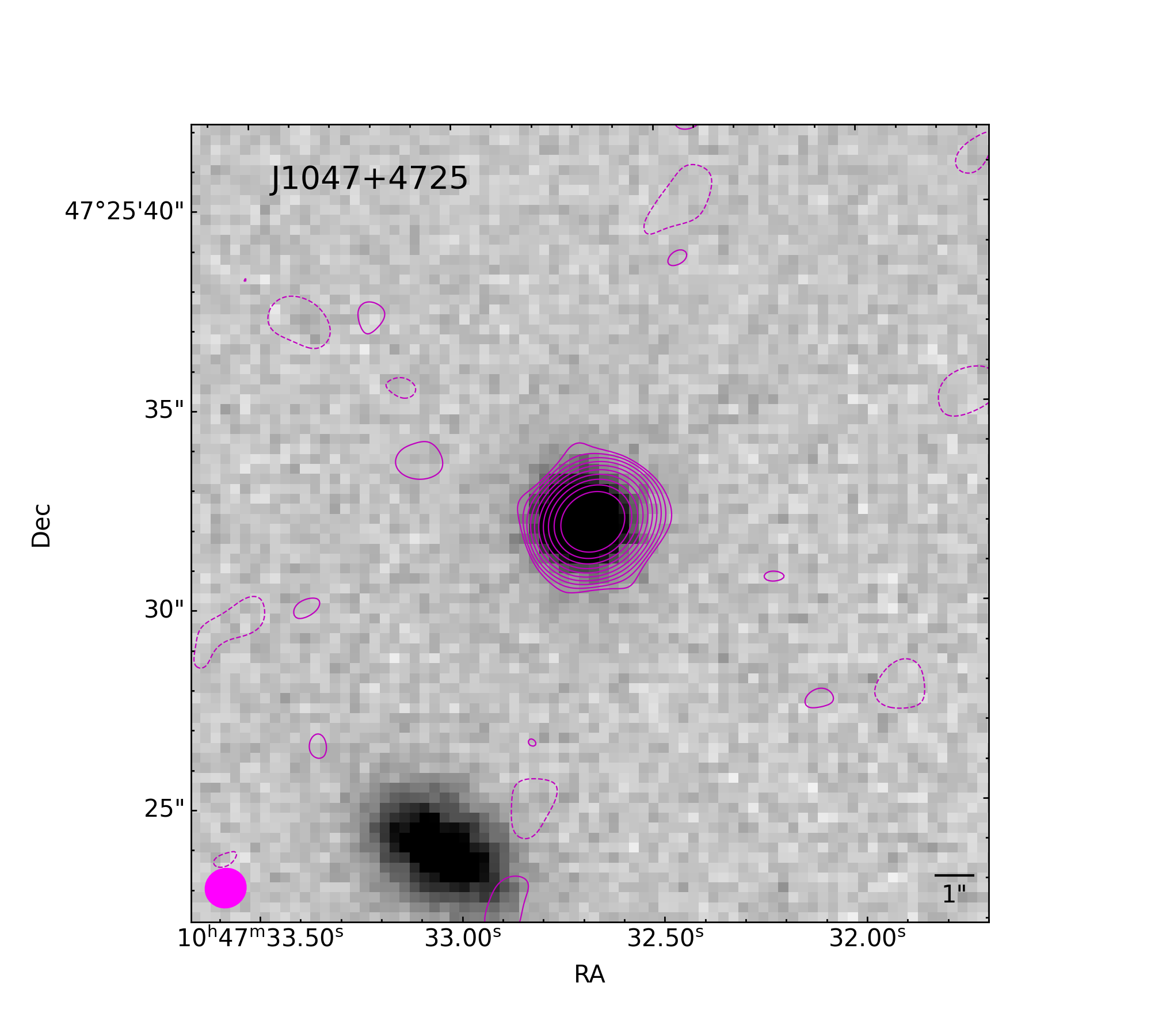}
         \caption{PanSTARRS $i$ band image of the host galaxy overlaid with the 90k$\lambda$ tapered map. Radio map properties as in Fig.~\ref{fig:J1047-90k}}. \label{fig:J1047-host}
     \end{subfigure}
        \caption{}
        \label{fig:J1047}
\end{figure*}

\clearpage
\begin{figure*}
     \centering
     \begin{subfigure}[b]{0.47\textwidth}
         \centering
         \includegraphics[width=\textwidth]{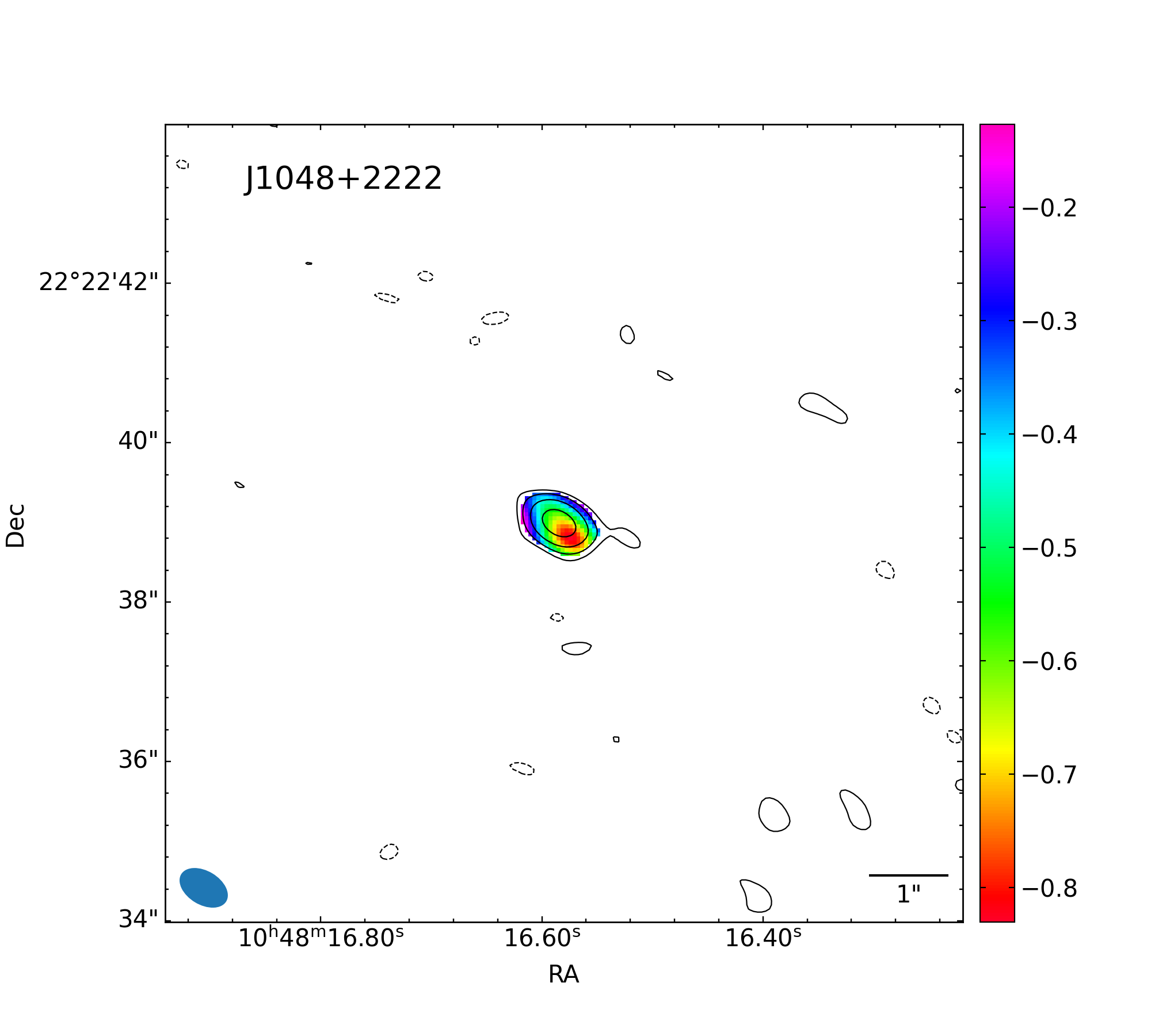}
         \caption{Spectral index map, rms = 8$\mu$Jy beam$^{-1}$, contour levels at -3, 3 $\times$ 2$^n$, $n \in$ [0, 3], beam size 3.14 $\times$ 2.00~kpc. } \label{fig:J1048spind}
     \end{subfigure}
     \hfill
     \begin{subfigure}[b]{0.47\textwidth}
         \centering
         \includegraphics[width=\textwidth]{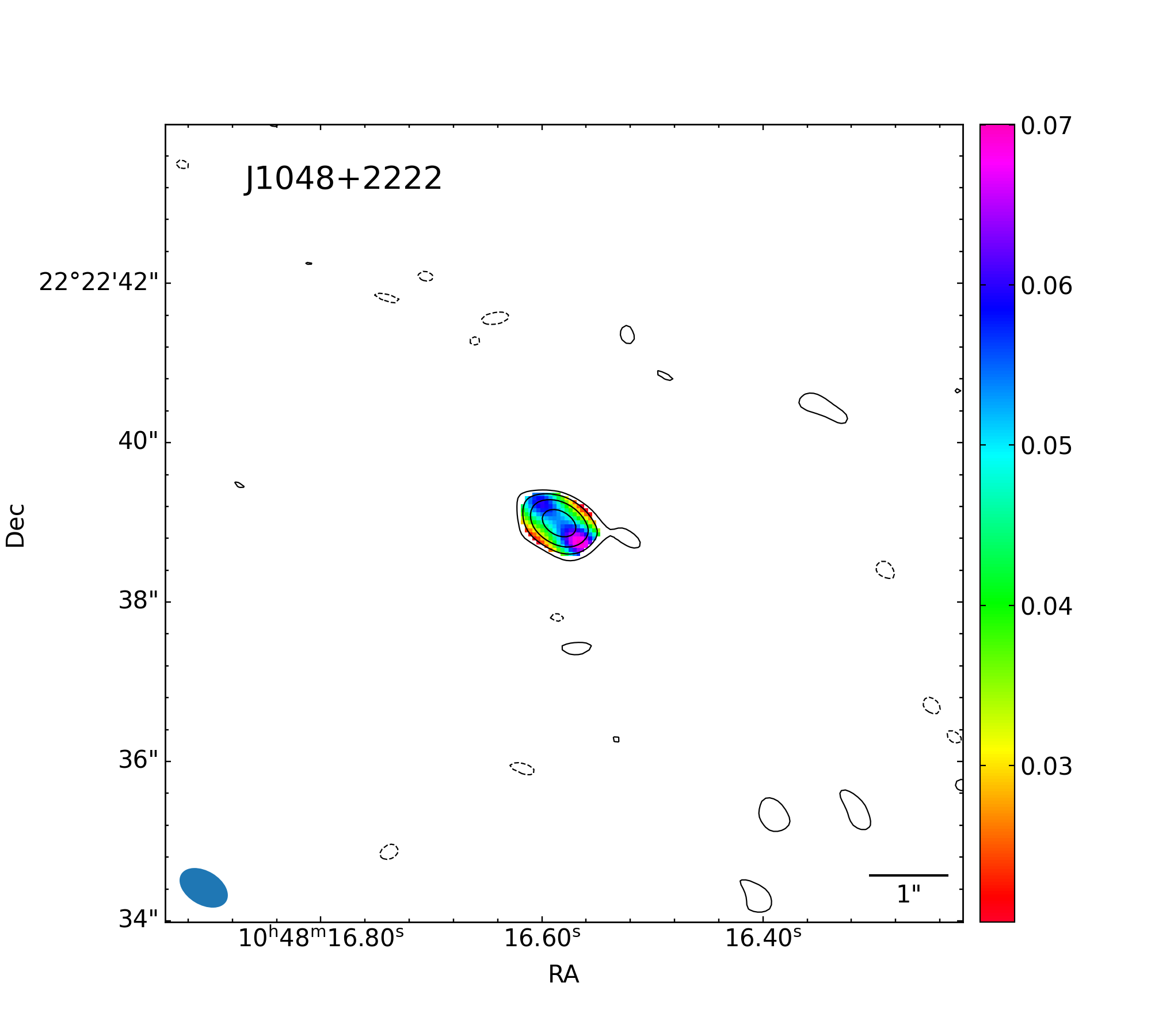}
         \caption{Spectral index error map, rms, contour levels, and beam size as in Fig.~\ref{fig:J1048spind}.} \label{fig:J1048spinderr}
     \end{subfigure}
     \hfill
     \\
     \begin{subfigure}[b]{0.47\textwidth}
         \centering
         \includegraphics[width=\textwidth]{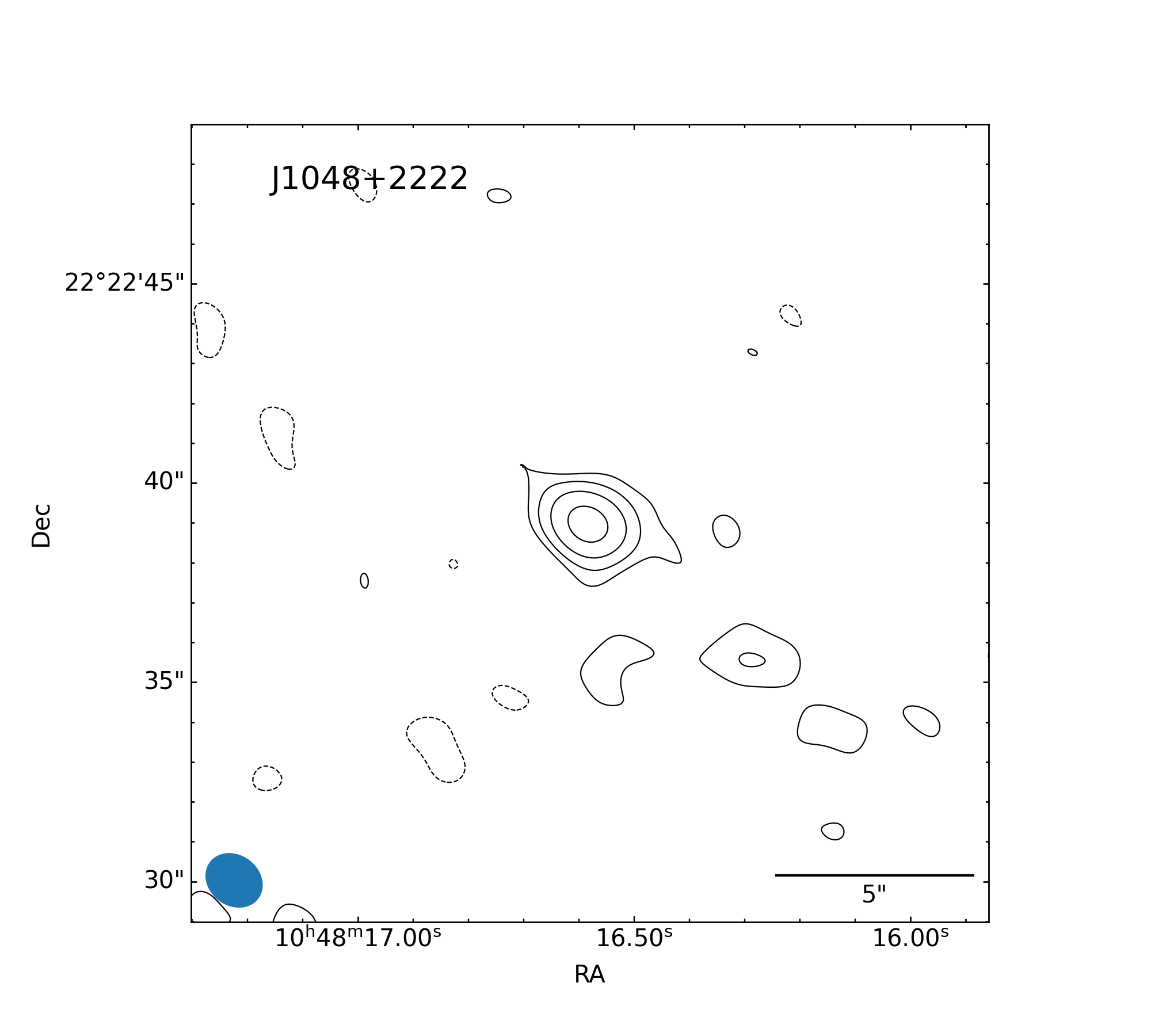}
         \caption{Tapered map with \texttt{uvtaper} = 90k$\lambda$, rms = 9$\mu$Jy beam$^{-1}$, contour levels at -3, 3 $\times$ 2$^n$, $n \in$ [0, 3], beam size 7.27 $\times$ 6.04~kpc.} \label{fig:J1048-90k}
     \end{subfigure}
          \hfill
     \begin{subfigure}[b]{0.47\textwidth}
         \centering
         \includegraphics[width=\textwidth]{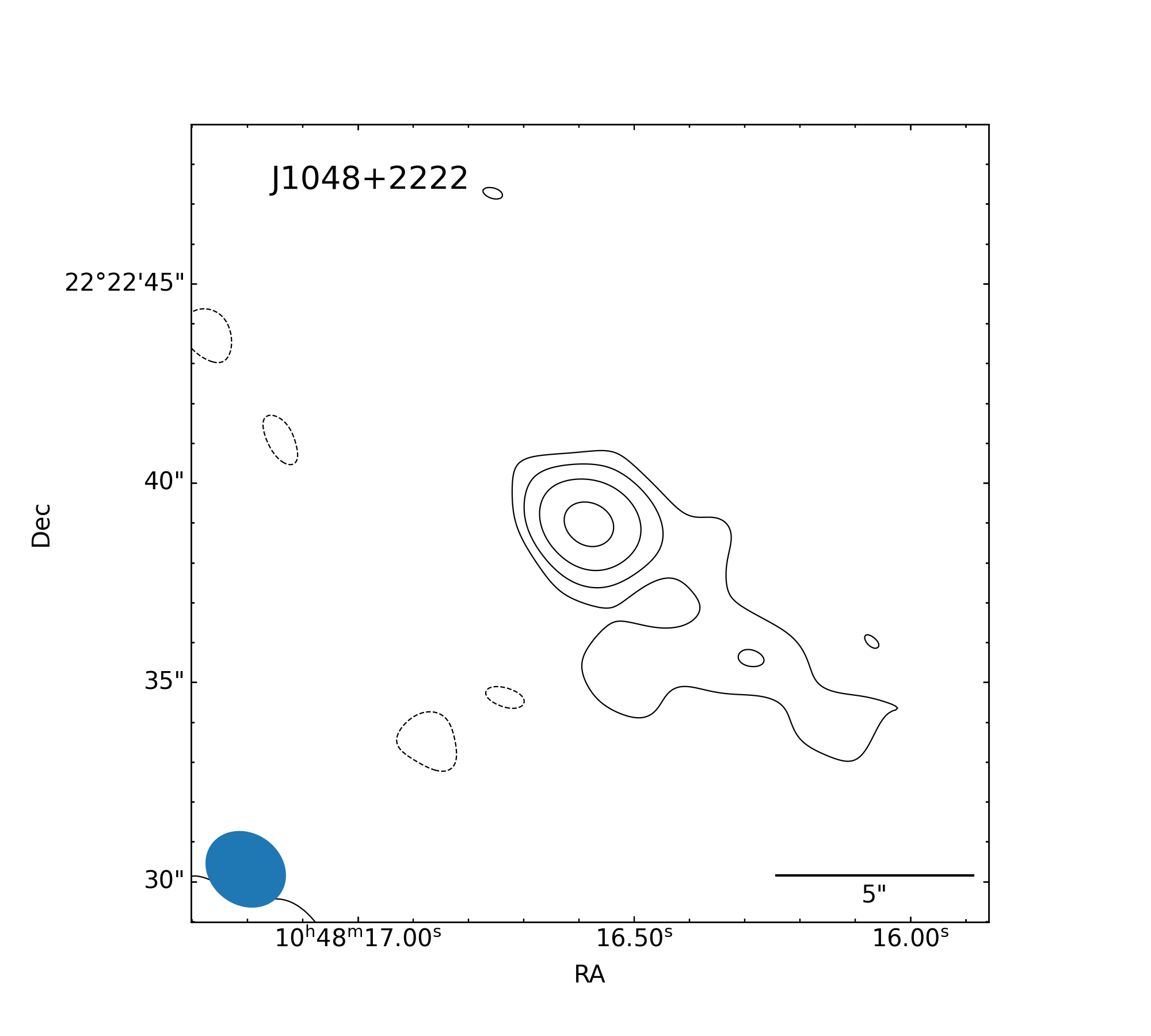}
         \caption{Tapered map with \texttt{uvtaper} = 60k$\lambda$, rms = 10$\mu$Jy beam$^{-1}$, contour levels at -3, 3 $\times$ 2$^n$, $n \in$ [0, 3], beam size 10.12 $\times$ 8.55~kpc.} \label{fig:J1048-60k}
     \end{subfigure}
          \hfill
     \\
     \begin{subfigure}[b]{0.47\textwidth}
         \centering
         \includegraphics[width=\textwidth]{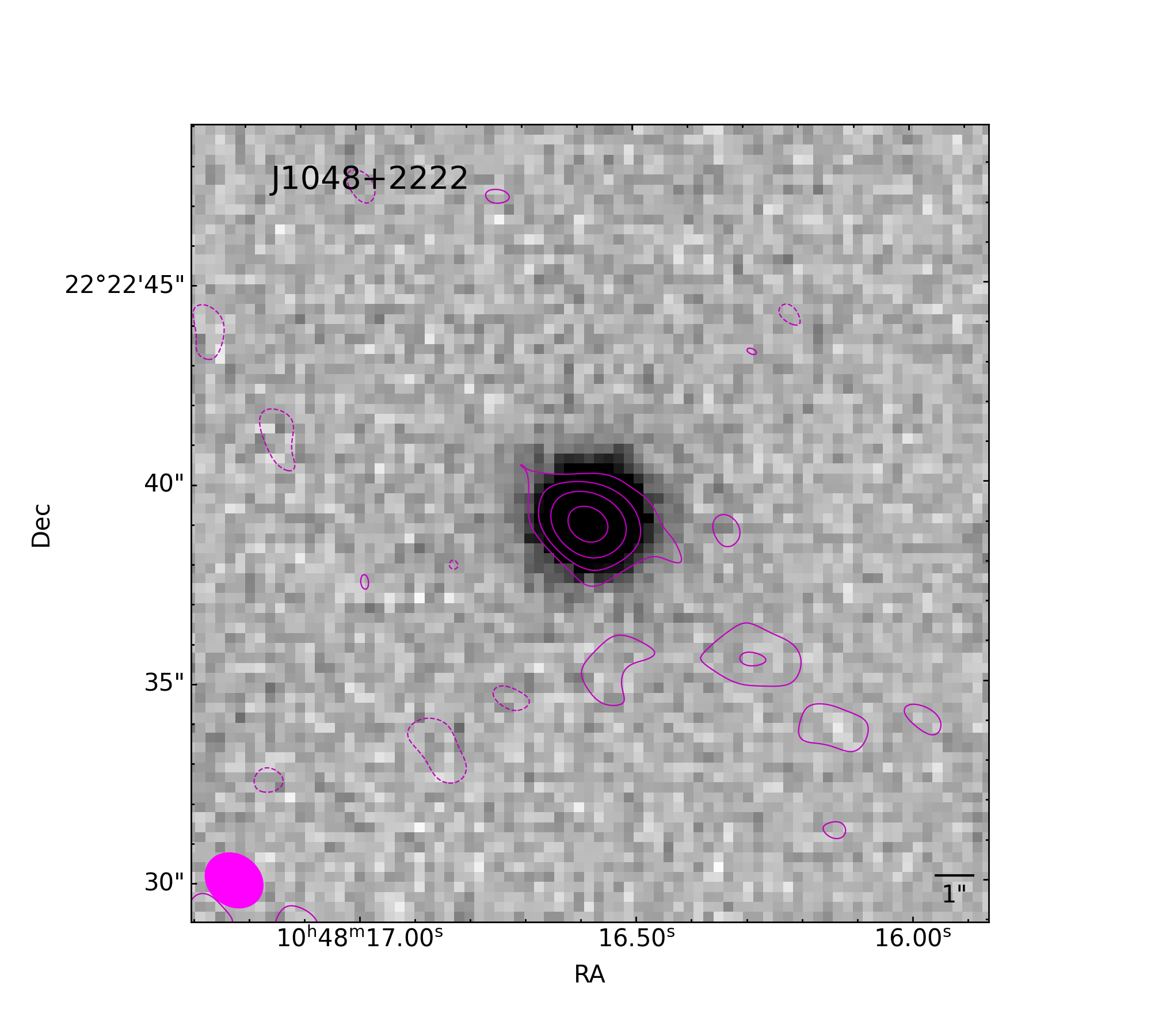}
         \caption{PanSTARRS $i$ band image of the host galaxy overlaid with the 90k$\lambda$ tapered map. Radio map properties as in Fig.~\ref{fig:J1048-90k}}. \label{fig:J1048-host}
     \end{subfigure}
        \caption{}
        \label{fig:J1048}
\end{figure*}


\begin{figure*}
     \centering
     \begin{subfigure}[b]{0.47\textwidth}
         \centering
         \includegraphics[width=\textwidth]{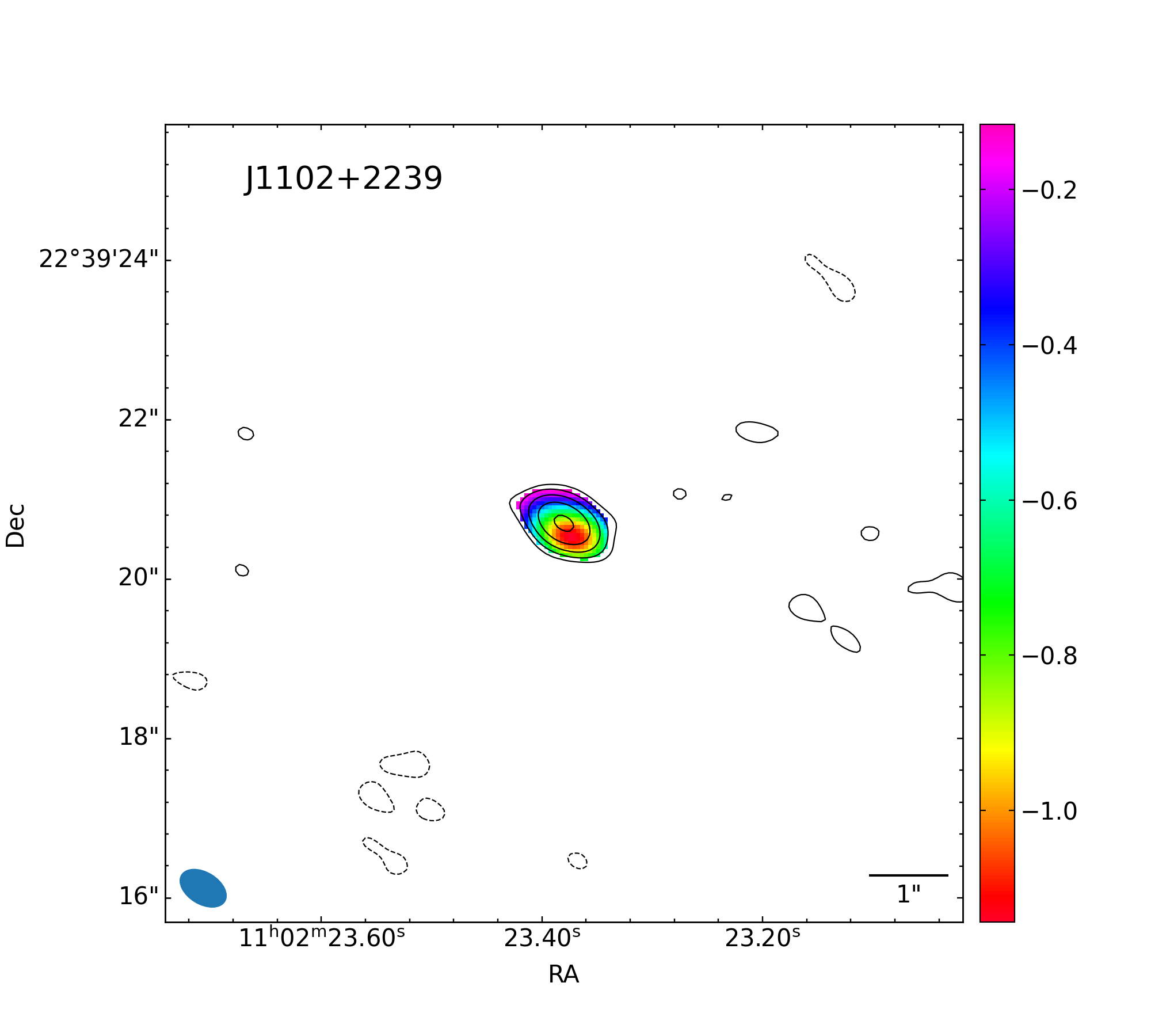}
         \caption{Spectral index map, rms = 13$\mu$Jy beam$^{-1}$, contour levels at -3, 3 $\times$ 2$^n$, $n \in$ [0, 4], beam size 3.76 $\times$ 2.37~kpc. } \label{fig:J1102spind}
     \end{subfigure}
     \hfill
     \begin{subfigure}[b]{0.47\textwidth}
         \centering
         \includegraphics[width=\textwidth]{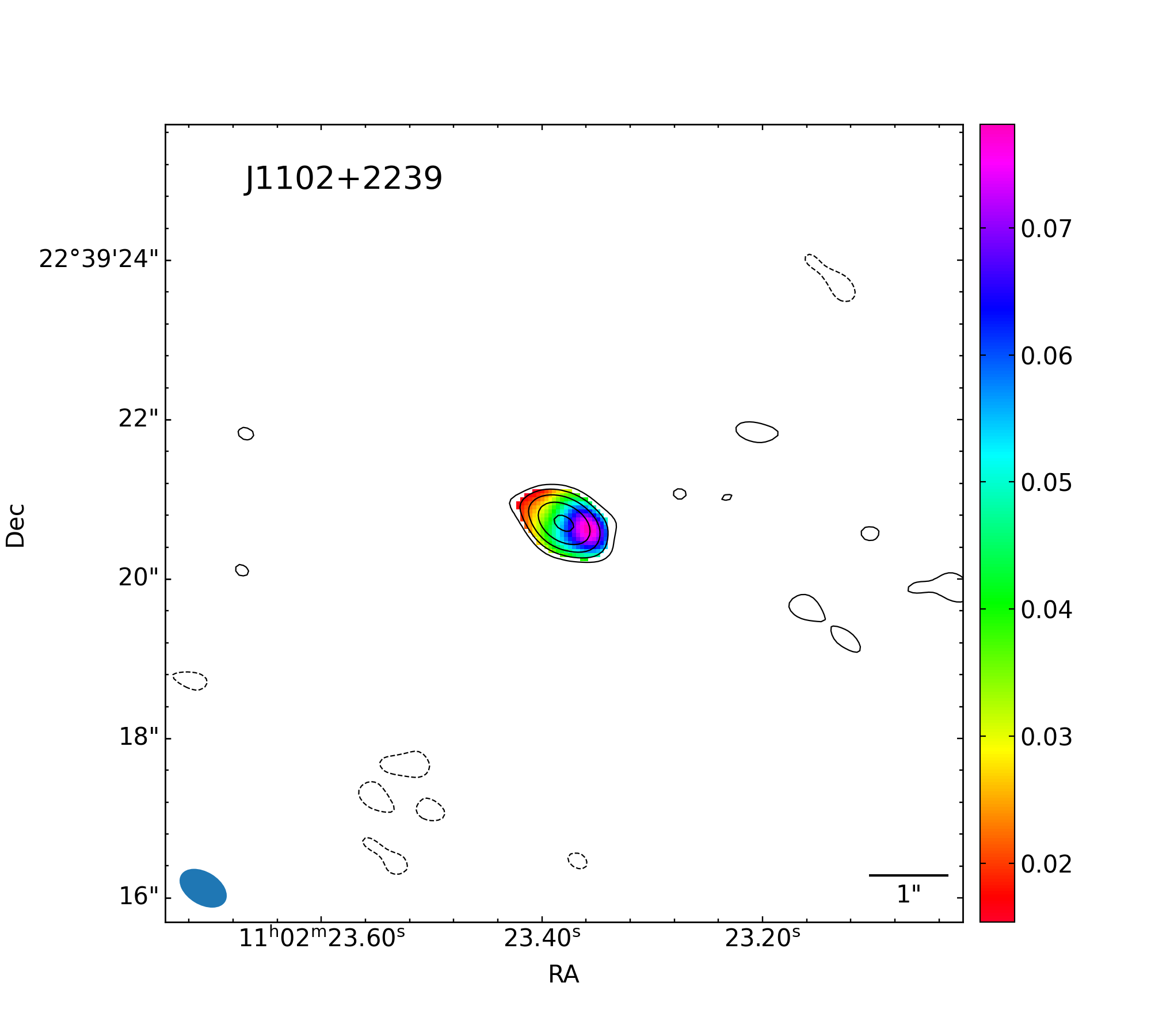}
         \caption{Spectral index error map, rms, contour levels, and beam size as in Fig.~\ref{fig:J1102spind}.} \label{fig:J1102spinderr}
     \end{subfigure}
     \hfill
     \\
     \begin{subfigure}[b]{0.47\textwidth}
         \centering
         \includegraphics[width=\textwidth]{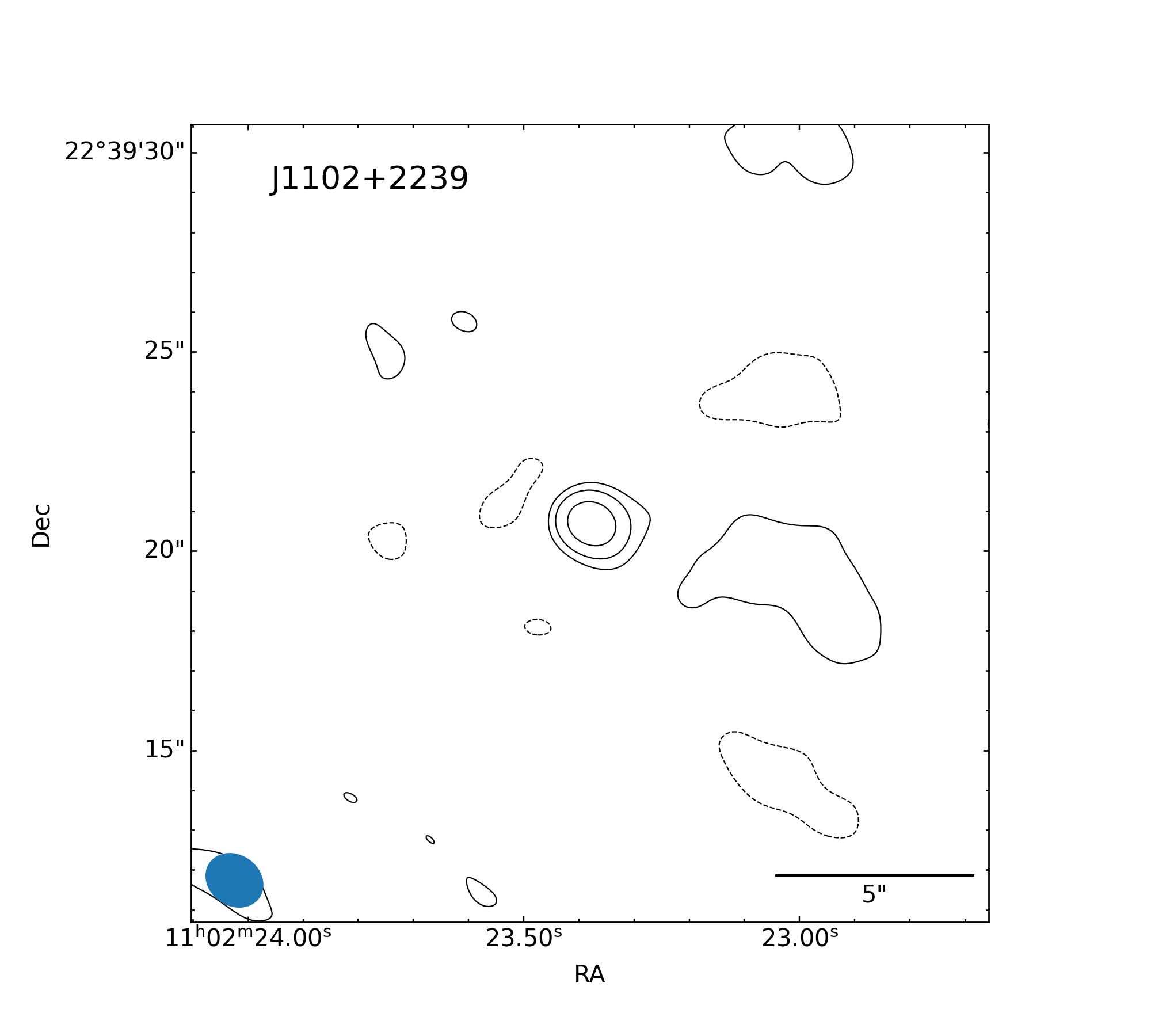}
         \caption{Tapered map with \texttt{uvtaper} = 90k$\lambda$, rms = 36$\mu$Jy beam$^{-1}$, contour levels at -3, 3 $\times$ 2$^n$, $n \in$ [0, 2], beam size 8.84 $\times$ 7.40~kpc.} \label{fig:J1102-90k}
     \end{subfigure}
          \hfill
     \begin{subfigure}[b]{0.47\textwidth}
         \centering
         \includegraphics[width=\textwidth]{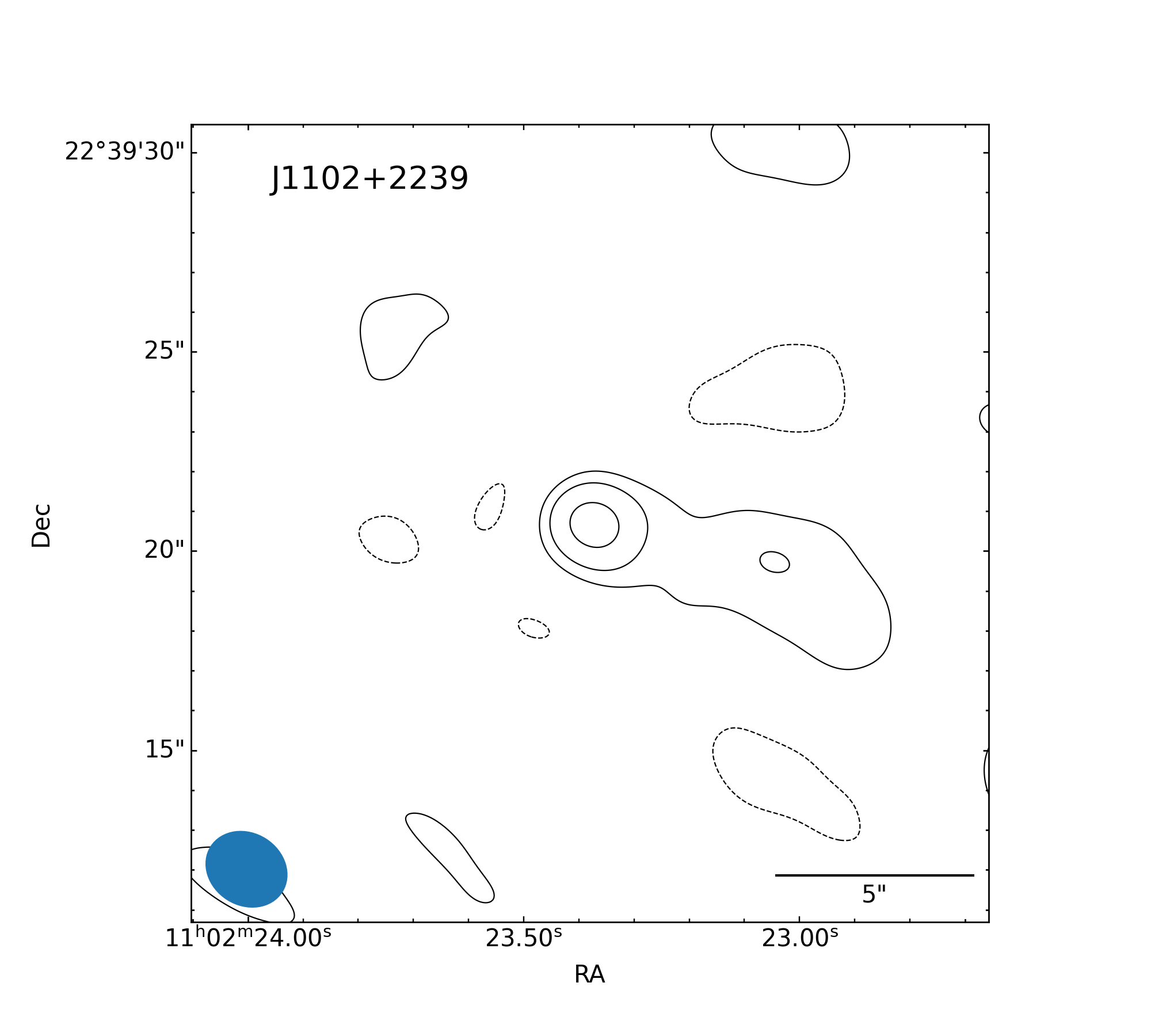}
         \caption{Tapered map with \texttt{uvtaper} = 60k$\lambda$, rms = 45$\mu$Jy beam$^{-1}$, contour levels at -3, 3 $\times$ 2$^n$, $n \in$ [0, 2], beam size 12.43 $\times$ 10.58~kpc.} \label{fig:J1102-60k}
     \end{subfigure}
          \hfill
     \\
     \begin{subfigure}[b]{0.47\textwidth}
         \centering
         \includegraphics[width=\textwidth]{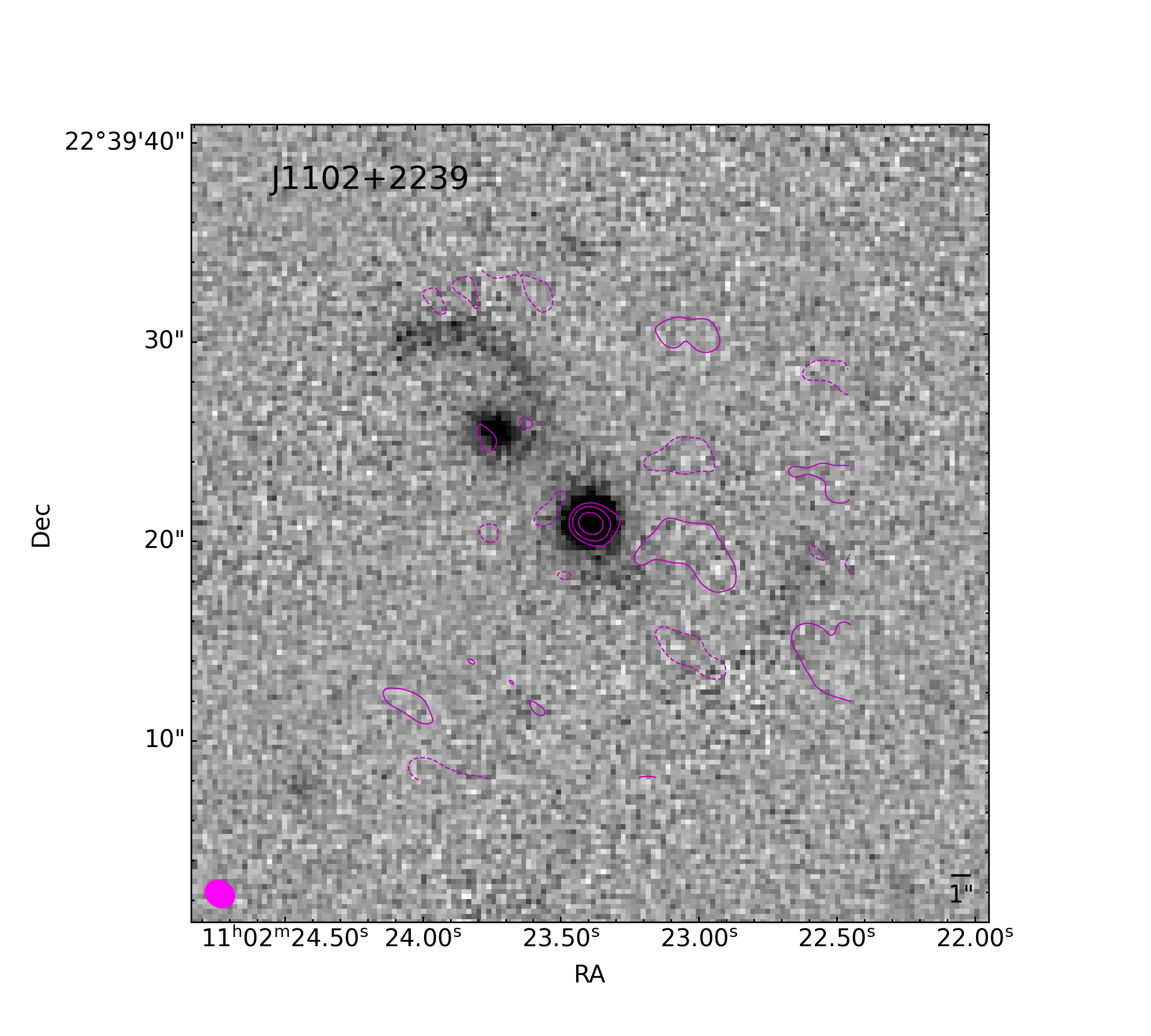}
         \caption{PanSTARRS $i$ band image of the host galaxy overlaid with the 90k$\lambda$ tapered map. Radio map properties as in Fig.~\ref{fig:J1102-90k}}. \label{fig:J1102-host}
     \end{subfigure}
        \caption{}
        \label{fig:J1102}
\end{figure*}


\begin{figure*}
     \centering
     \begin{subfigure}[b]{0.47\textwidth}
         \centering
         \includegraphics[width=\textwidth]{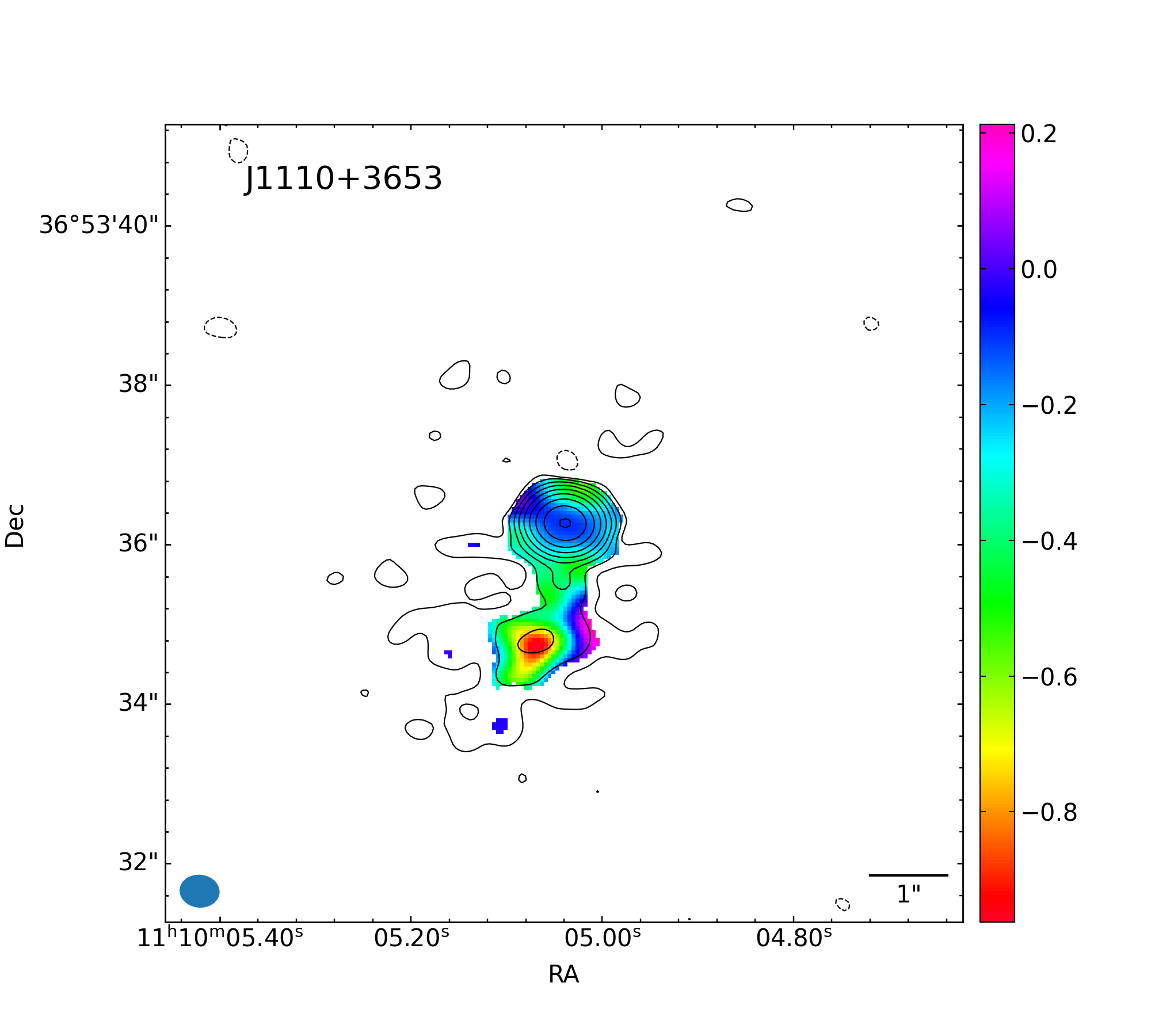}
         \caption{Spectral index map, rms = 10$\mu$Jy beam$^{-1}$, contour levels at -3, 3 $\times$ 2$^n$, $n \in$ [0, 8], beam size 3.42 $\times$ 2.80~kpc. } \label{fig:J1110spind}
     \end{subfigure}
     \hfill
     \begin{subfigure}[b]{0.47\textwidth}
         \centering
         \includegraphics[width=\textwidth]{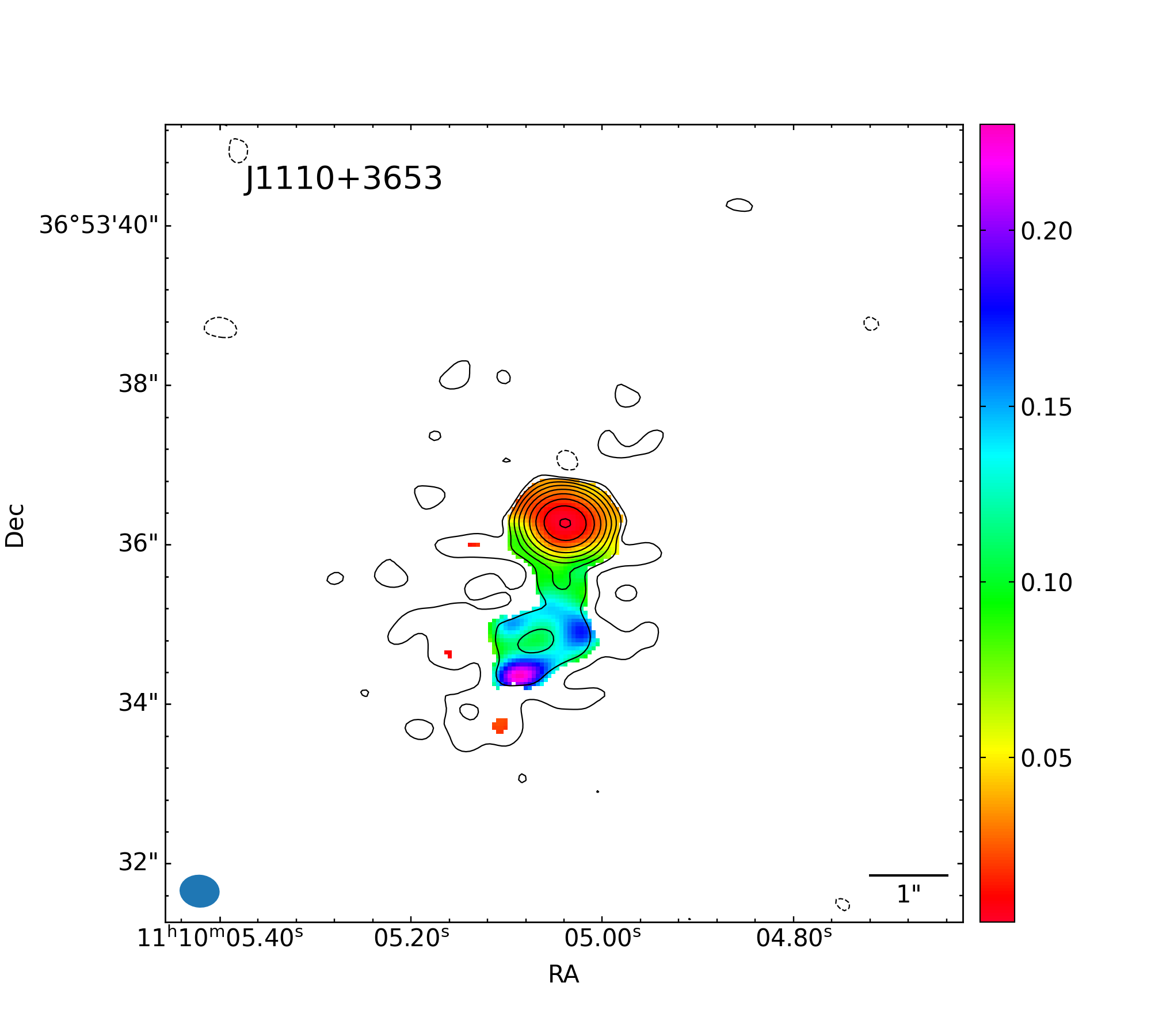}
         \caption{Spectral index error map, rms, contour levels, and beam size as in Fig.~\ref{fig:J1110spind}.} \label{fig:J1110spinderr}
     \end{subfigure}
     \hfill
     \\
     \begin{subfigure}[b]{0.47\textwidth}
         \centering
         \includegraphics[width=\textwidth]{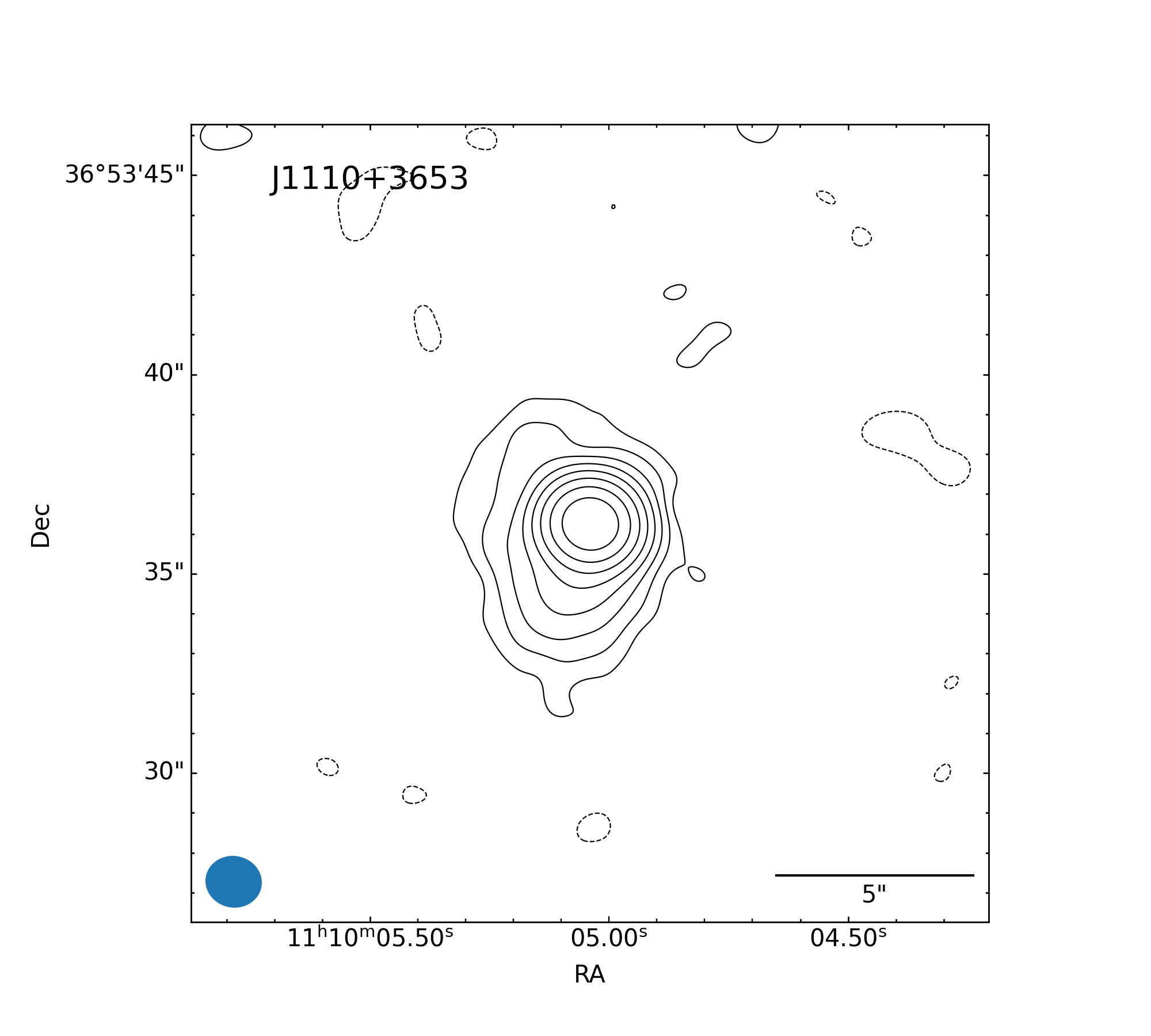}
         \caption{Tapered map with \texttt{uvtaper} = 90k$\lambda$, rms = 11$\mu$Jy beam$^{-1}$, contour levels at -3, 3 $\times$ 2$^n$, $n \in$ [0, 7], beam size 9.70 $\times$ 8.81~kpc.} \label{fig:J1110-90k}
     \end{subfigure}
          \hfill
     \begin{subfigure}[b]{0.47\textwidth}
         \centering
         \includegraphics[width=\textwidth]{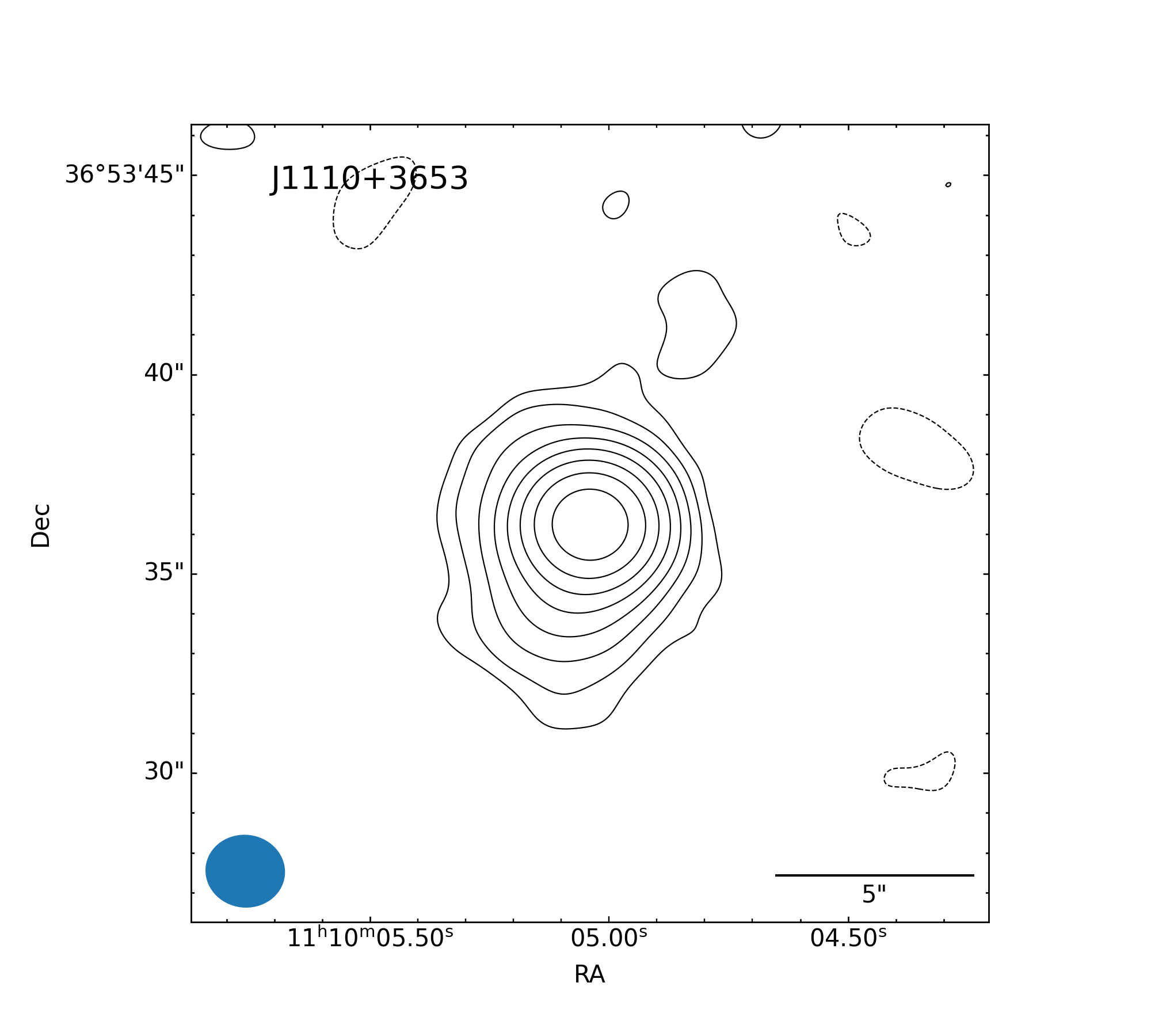}
         \caption{Tapered map with \texttt{uvtaper} = 60k$\lambda$, rms = 12$\mu$Jy beam$^{-1}$, contour levels at -3, 3 $\times$ 2$^n$, $n \in$ [0, 7], beam size 13.66 $\times$ 12.50~kpc.} \label{fig:J1110-60k}
     \end{subfigure}
          \hfill
     \\
     \begin{subfigure}[b]{0.47\textwidth}
         \centering
         \includegraphics[width=\textwidth]{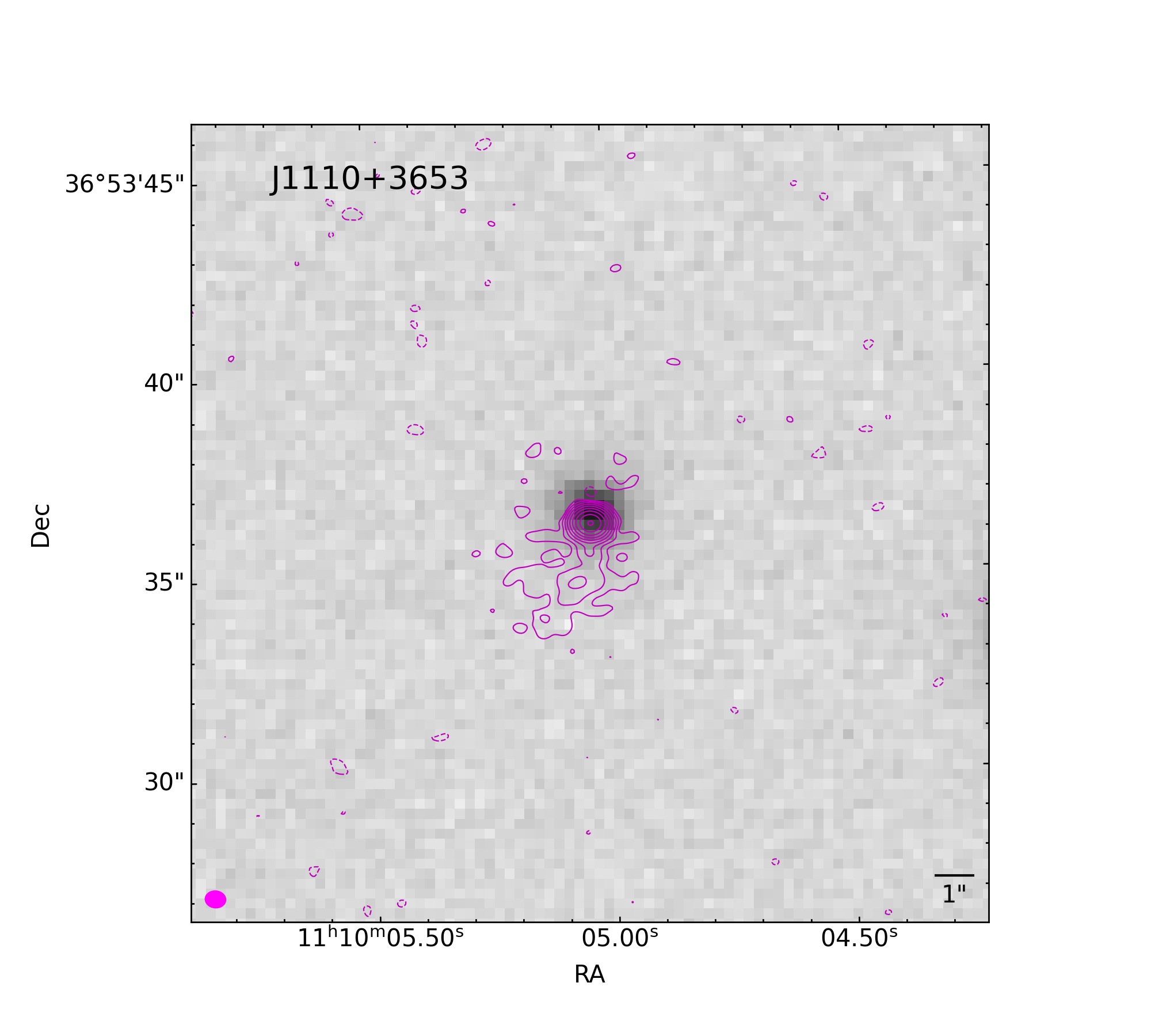}
         \caption{PanSTARRS $i$ band image of the host galaxy overlaid with the normal map. Radio map properties as in Fig.~\ref{fig:J1110spind}}. \label{fig:J1110-host-zoom}
     \end{subfigure}
     \hfill     
     \begin{subfigure}[b]{0.47\textwidth}
         \centering
         \includegraphics[width=\textwidth]{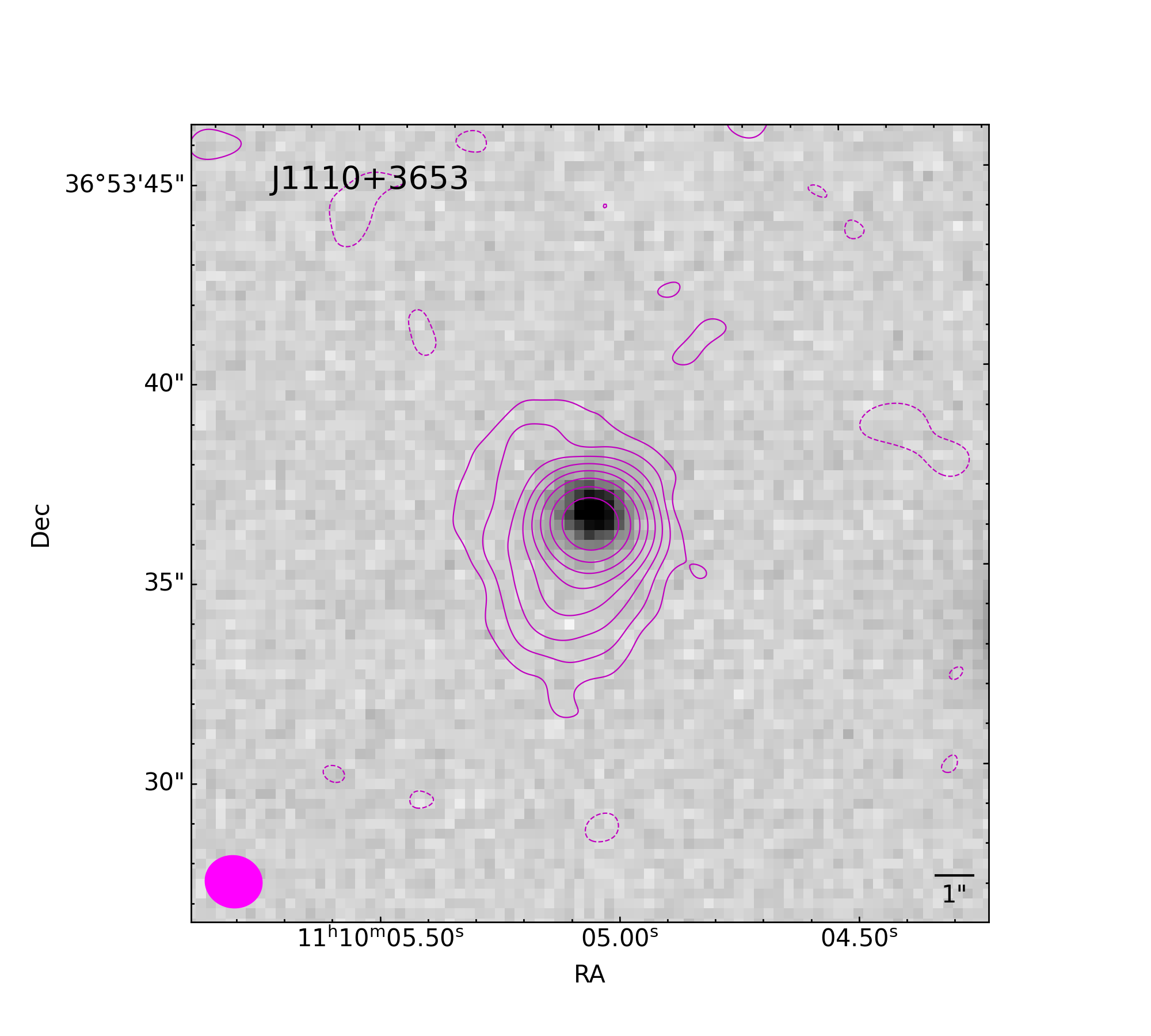}
         \caption{PanSTARRS $i$ band image of the host galaxy overlaid with the 90k$\lambda$ tapered map. Radio map properties as in Fig.~\ref{fig:J1110-90k}}. \label{fig:J1110-host}
     \end{subfigure}
        \caption{}
        \label{fig:J1110}
\end{figure*}


\begin{figure*}
     \centering
     \begin{subfigure}[b]{0.47\textwidth}
         \centering
         \includegraphics[width=\textwidth]{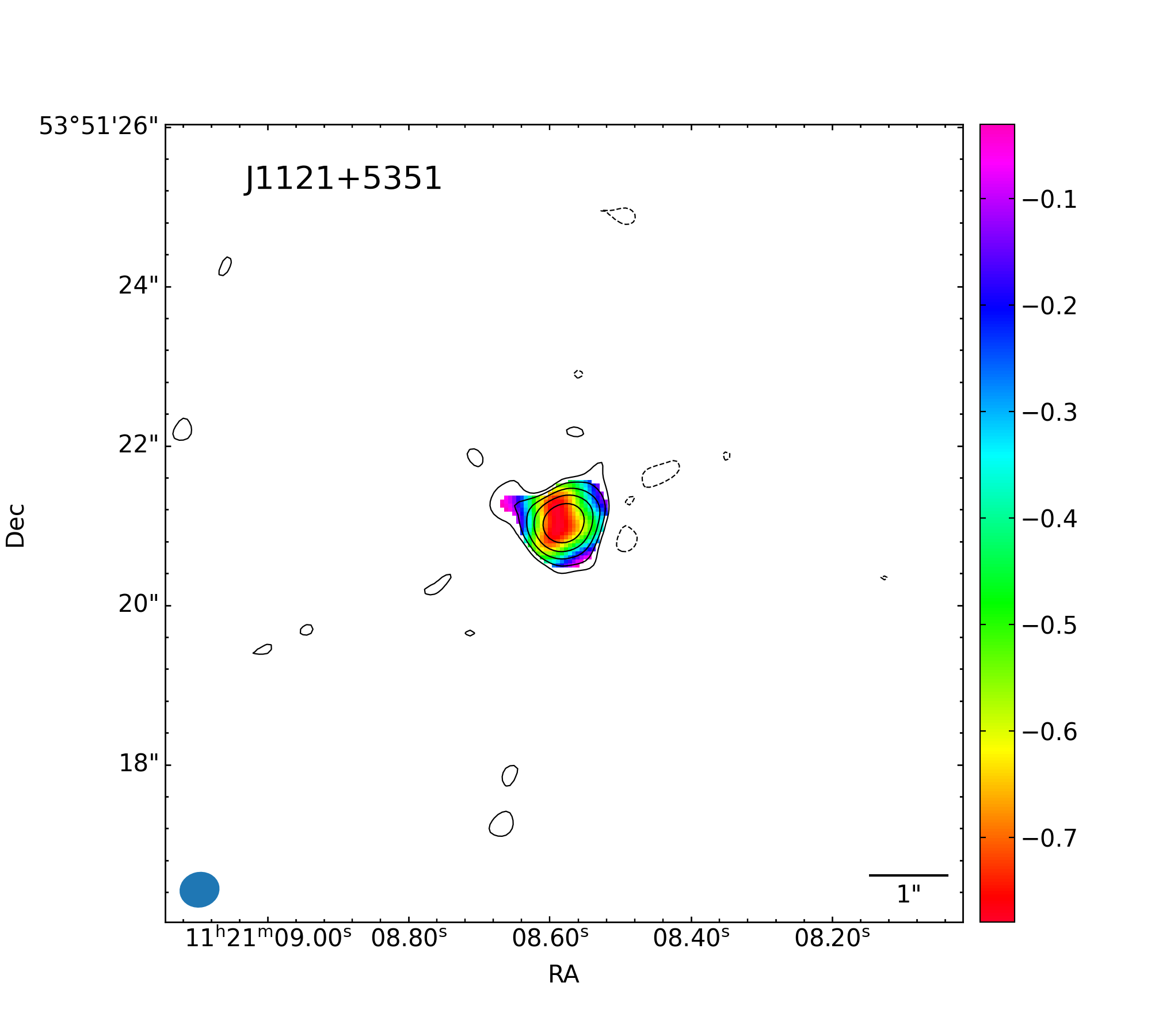}
         \caption{Spectral index map, rms = 11$\mu$Jy beam$^{-1}$, contour levels at -3, 3 $\times$ 2$^n$, $n \in$ [0, 4], beam size 0.97 $\times$ 0.85~kpc. } \label{fig:J1121spind}
     \end{subfigure}
     \hfill
     \begin{subfigure}[b]{0.47\textwidth}
         \centering
         \includegraphics[width=\textwidth]{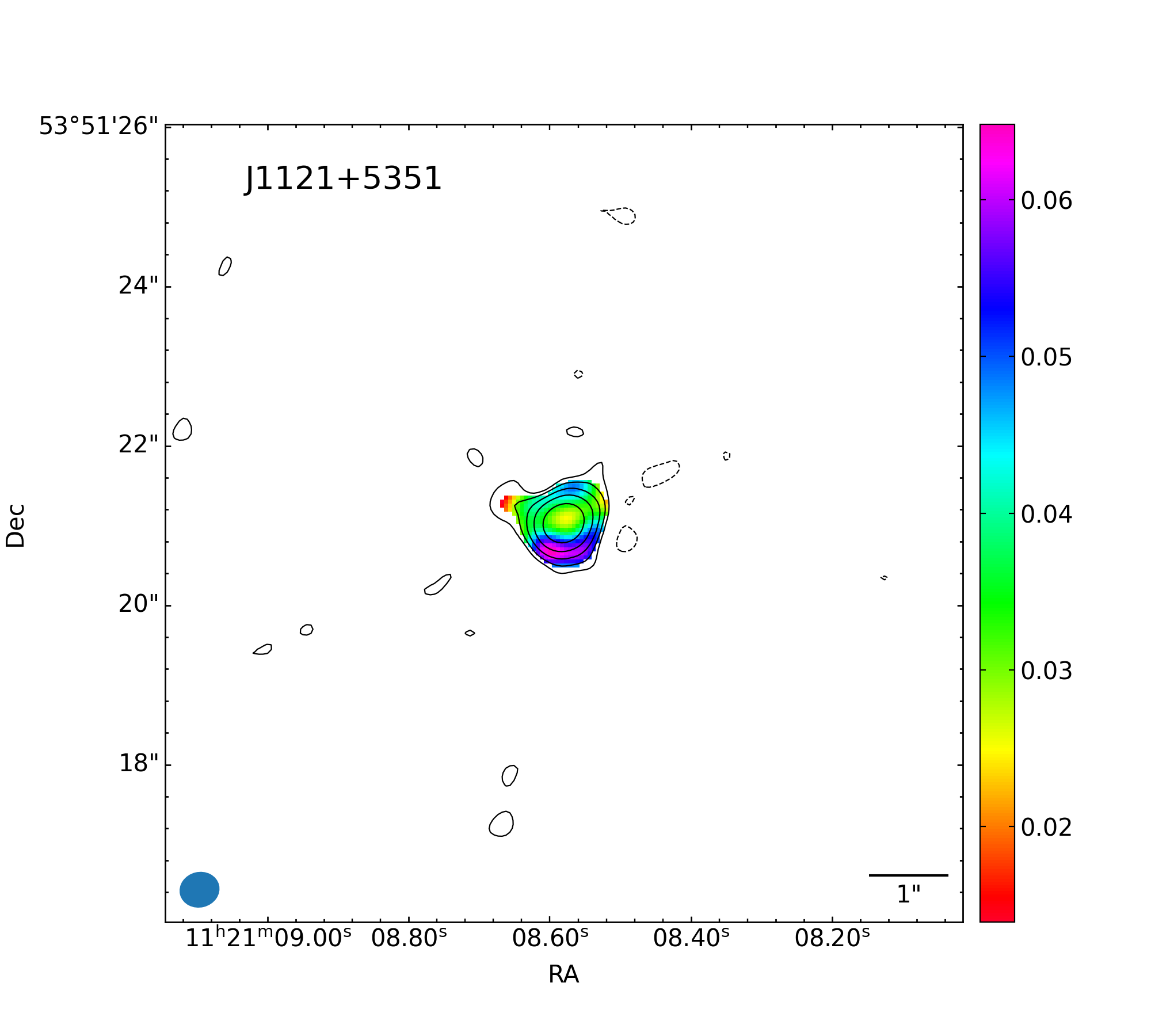}
         \caption{Spectral index error map, rms, contour levels, and beam size as in Fig.~\ref{fig:J1121spind}.} \label{fig:J1121spinderr}
     \end{subfigure}
     \hfill
     \\
     \begin{subfigure}[b]{0.47\textwidth}
         \centering
         \includegraphics[width=\textwidth]{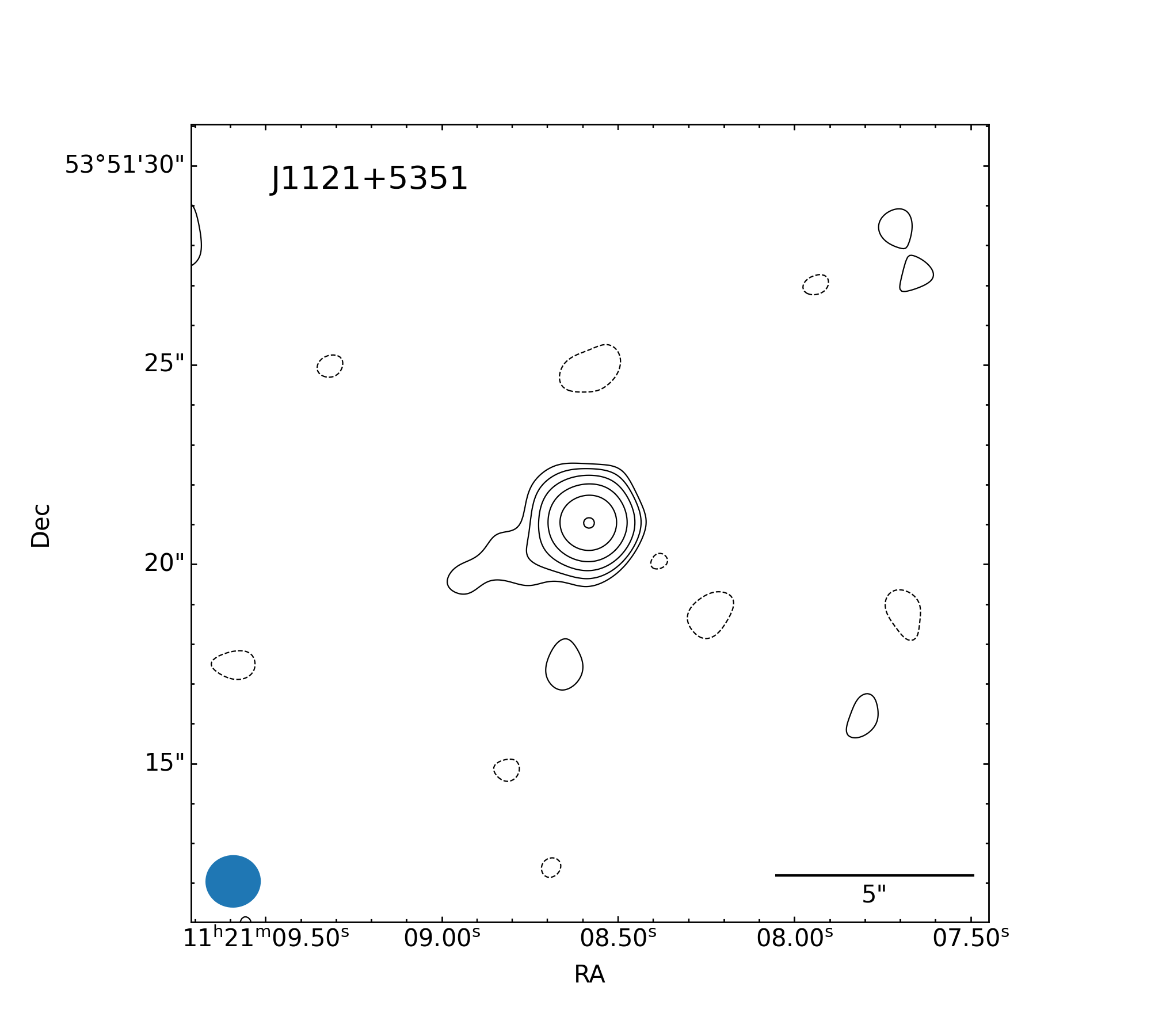}
         \caption{Tapered map with \texttt{uvtaper} = 90k$\lambda$, rms = 12$\mu$Jy beam$^{-1}$, contour levels at -3, 3 $\times$ 2$^n$, $n \in$ [0, 5], beam size 2.63 $\times$ 2.50~kpc.} \label{fig:J1121-90k}
     \end{subfigure}
          \hfill
     \begin{subfigure}[b]{0.47\textwidth}
         \centering
         \includegraphics[width=\textwidth]{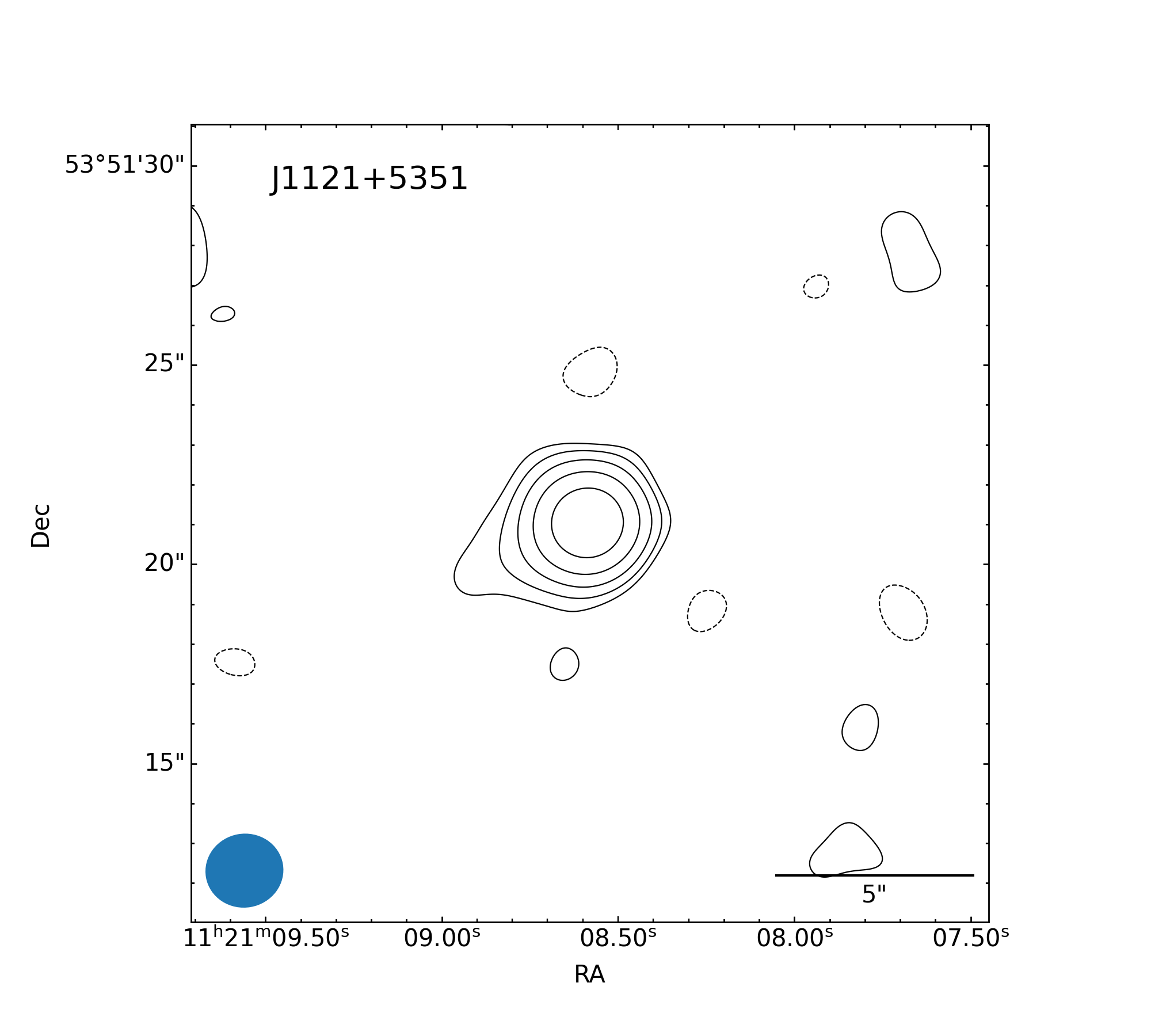}
         \caption{Tapered map with \texttt{uvtaper} = 60k$\lambda$, rms = 14$\mu$Jy beam$^{-1}$, contour levels at -3, 3 $\times$ 2$^n$, $n \in$ [0, 4], beam size 3.71 $\times$ 3.50~kpc.} \label{fig:J1121-60k}
     \end{subfigure}
          \hfill
     \\
     \begin{subfigure}[b]{0.47\textwidth}
         \centering
         \includegraphics[width=\textwidth]{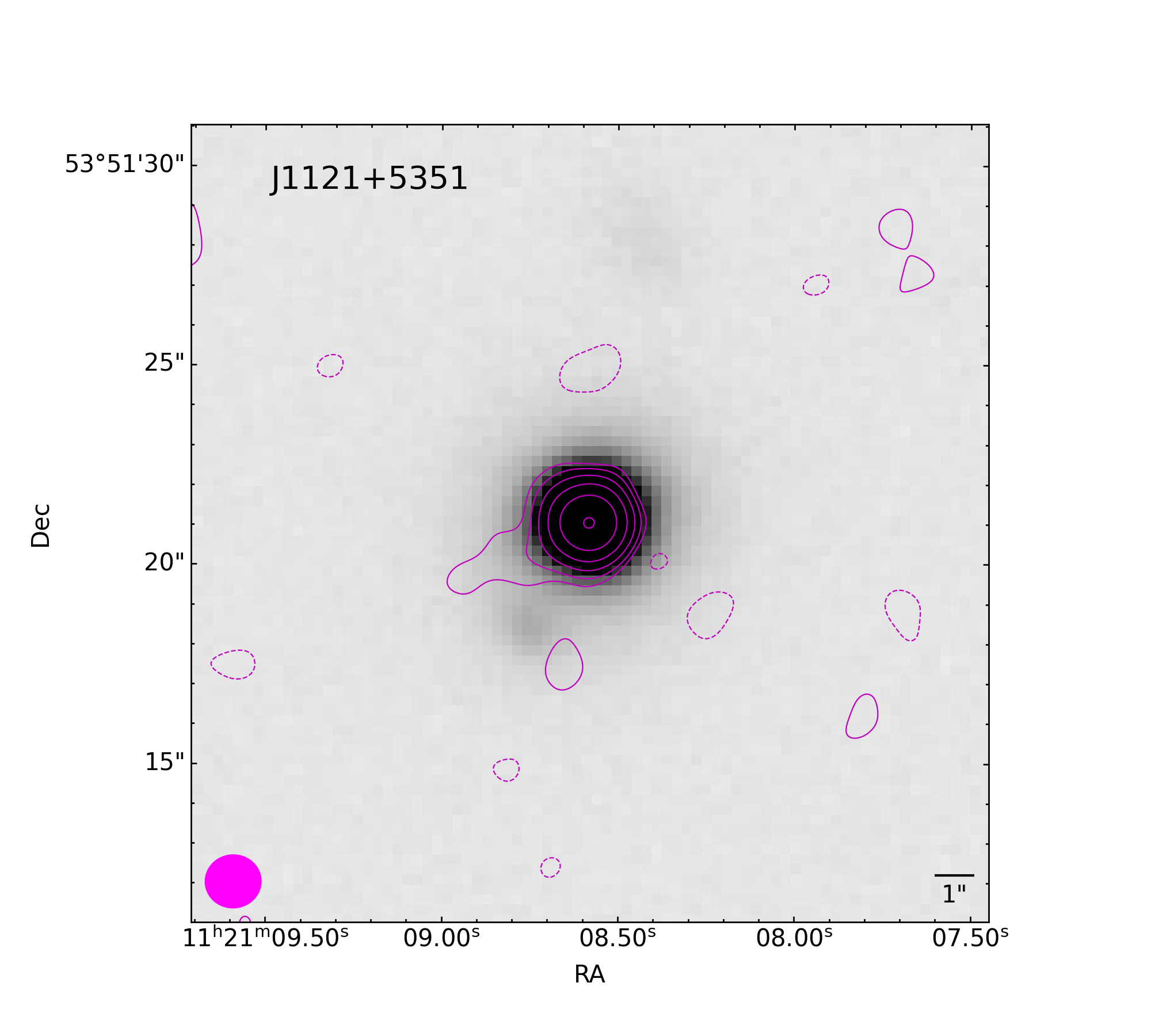}
         \caption{PanSTARRS $i$ band image of the host galaxy overlaid with the 90k$\lambda$ tapered map. Radio map properties as in Fig.~\ref{fig:J1121-90k}}. \label{fig:J1121-host}
     \end{subfigure}
        \caption{}
        \label{fig:J1121}
\end{figure*}


\begin{figure*}
     \centering
     \begin{subfigure}[b]{0.47\textwidth}
         \centering
         \includegraphics[width=\textwidth]{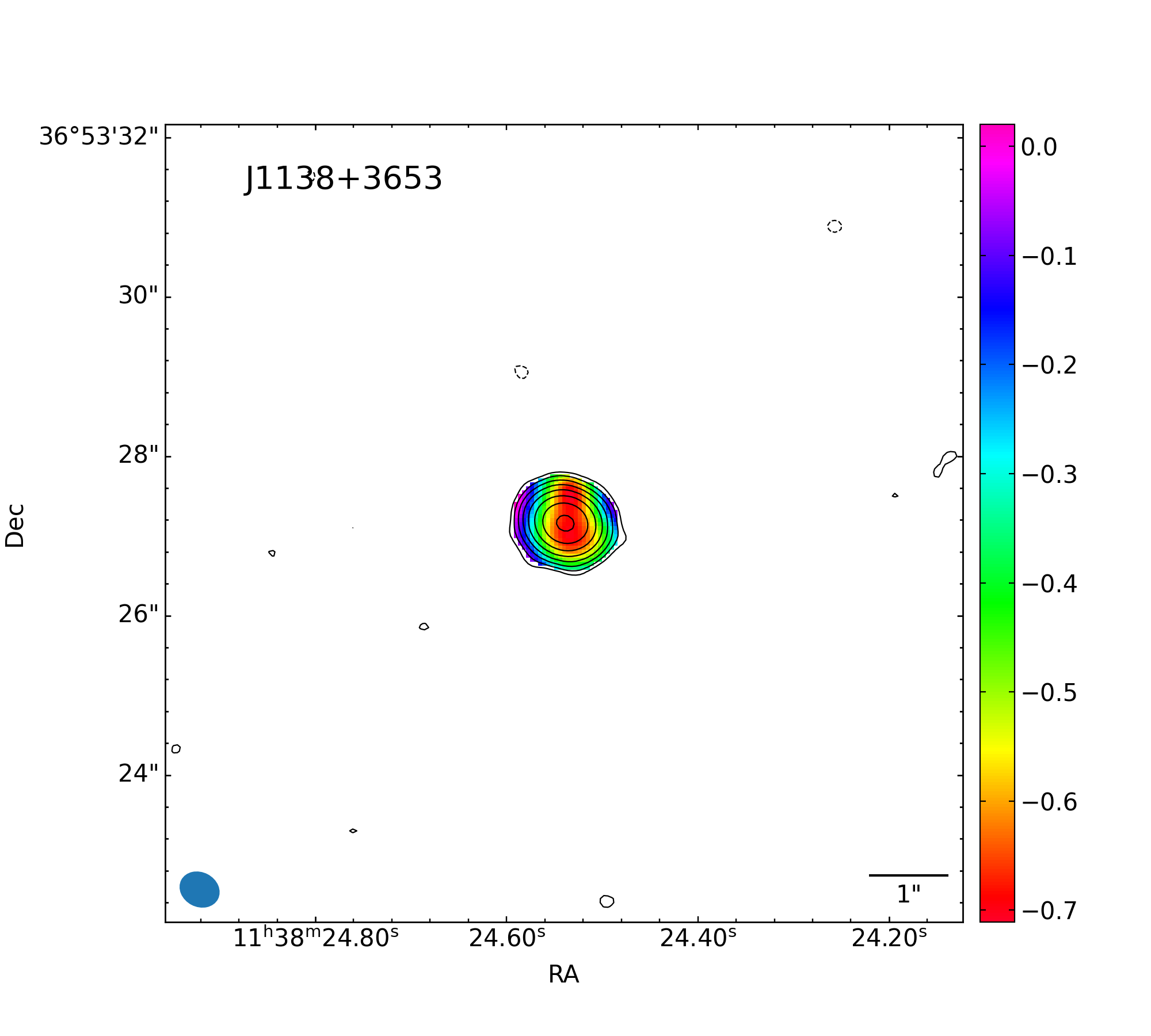}
         \caption{Spectral index map, rms = 10$\mu$Jy beam$^{-1}$, contour levels at -3, 3 $\times$ 2$^n$, $n \in$ [0, 7], beam size 2.60 $\times$ 2.20~kpc. } \label{fig:J1138spind}
     \end{subfigure}
     \hfill
     \begin{subfigure}[b]{0.47\textwidth}
         \centering
         \includegraphics[width=\textwidth]{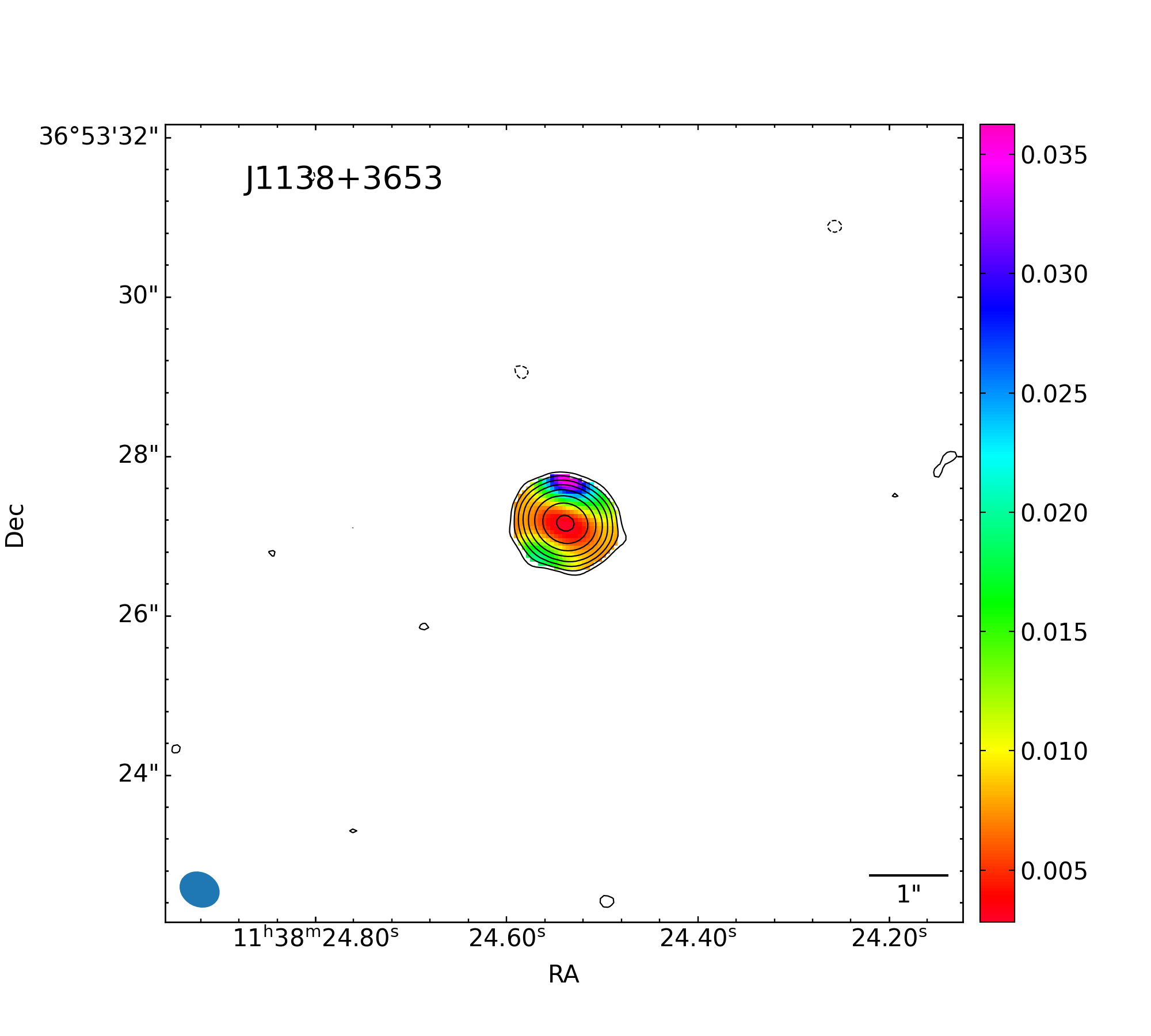}
         \caption{Spectral index error map, rms, contour levels, and beam size as in Fig.~\ref{fig:J1138spind}.} \label{fig:J1138spinderr}
     \end{subfigure}
     \hfill
     \\
     \begin{subfigure}[b]{0.47\textwidth}
         \centering
         \includegraphics[width=\textwidth]{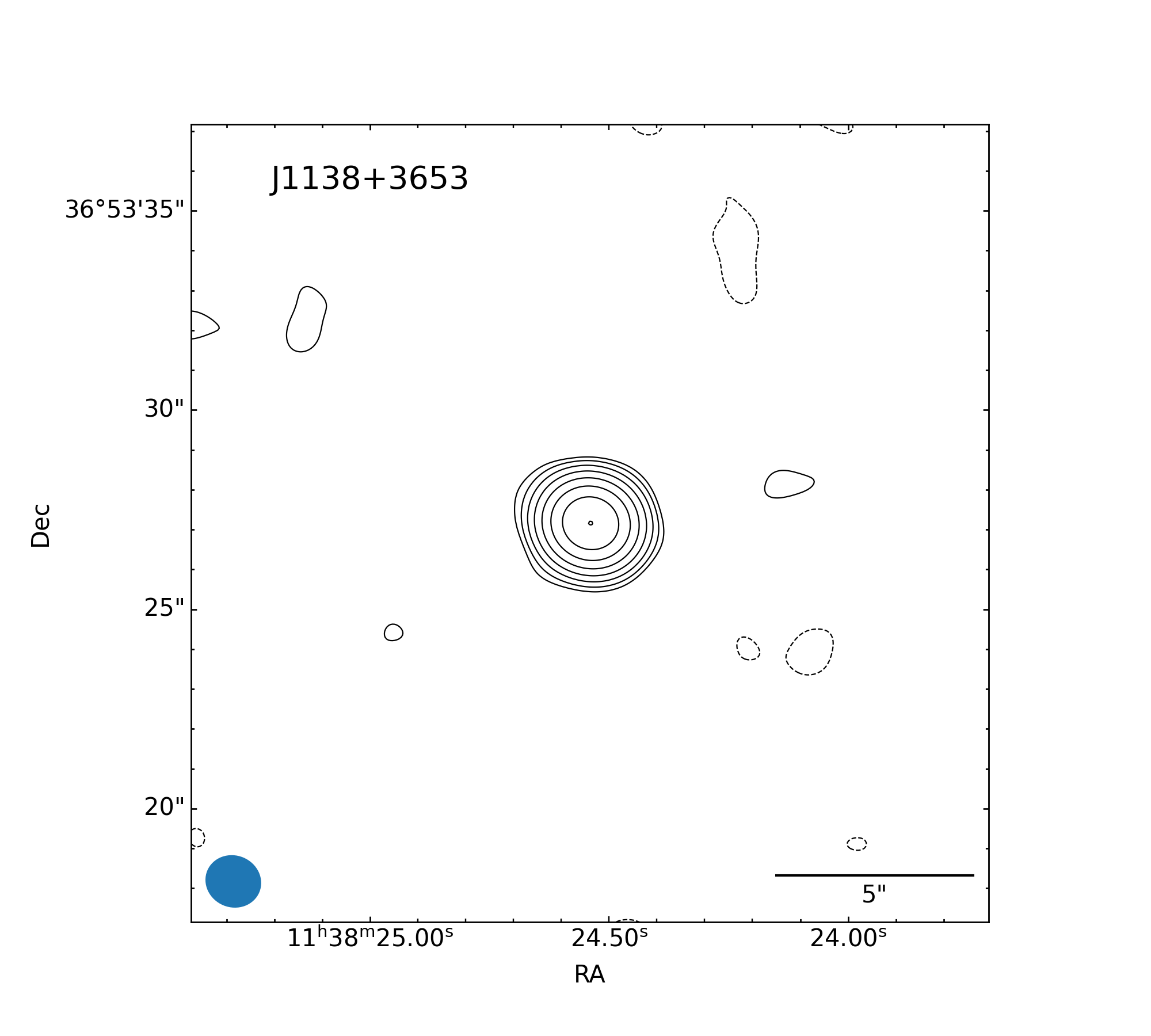}
         \caption{Tapered map with \texttt{uvtaper} = 90k$\lambda$, rms = 12$\mu$Jy beam$^{-1}$, contour levels at -3, 3 $\times$ 2$^n$, $n \in$ [0, 7], beam size 7.04 $\times$ 6.49~kpc.} \label{fig:J1138-90k}
     \end{subfigure}
          \hfill
     \begin{subfigure}[b]{0.47\textwidth}
         \centering
         \includegraphics[width=\textwidth]{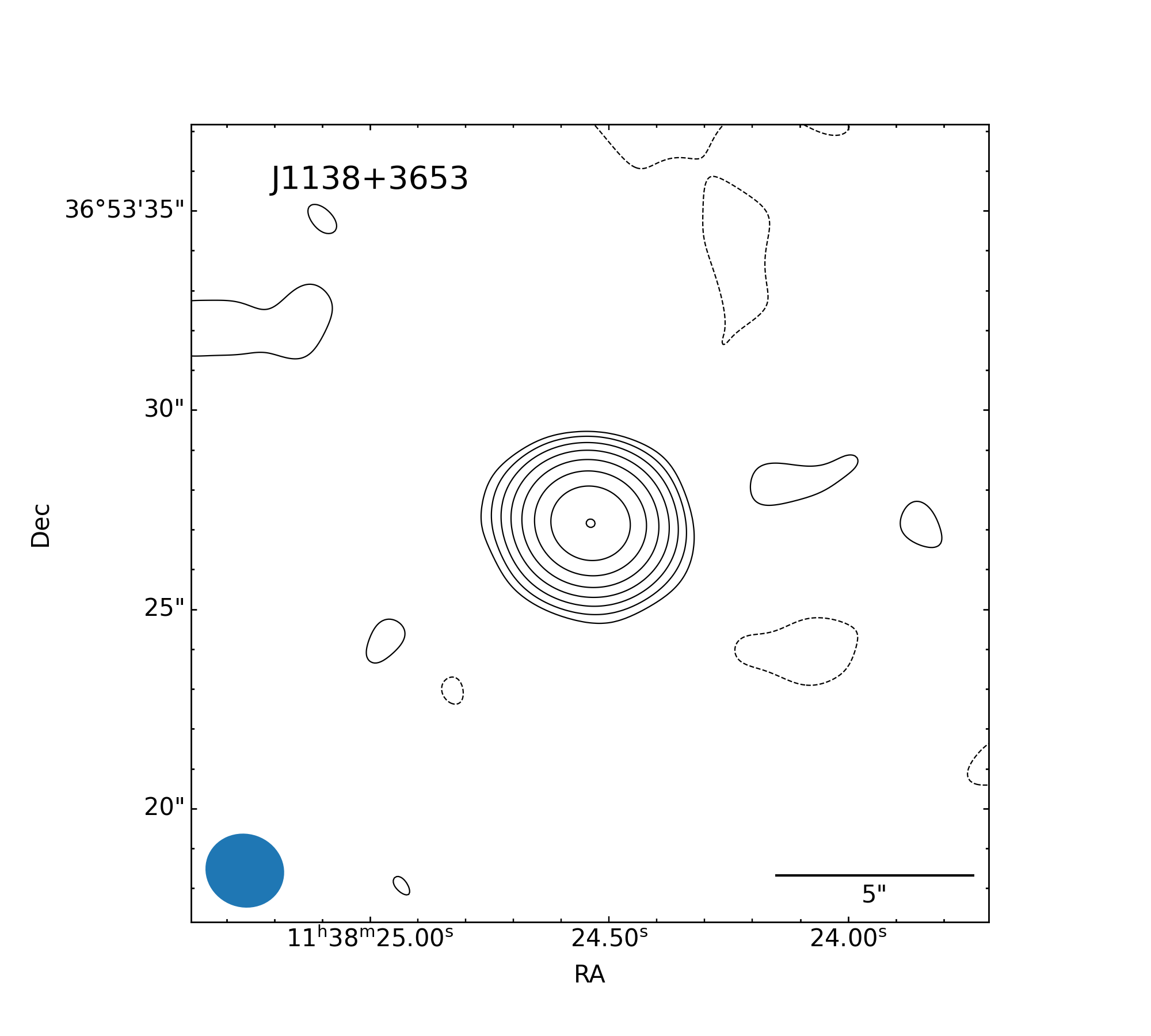}
         \caption{Tapered map with \texttt{uvtaper} = 60k$\lambda$, rms = 12$\mu$Jy beam$^{-1}$, contour levels at -3, 3 $\times$ 2$^n$, $n \in$ [0, 7], beam size 9.94 $\times$ 9.19~kpc.} \label{fig:J1138-60k}
     \end{subfigure}
          \hfill
     \\
     \begin{subfigure}[b]{0.47\textwidth}
         \centering
         \includegraphics[width=\textwidth]{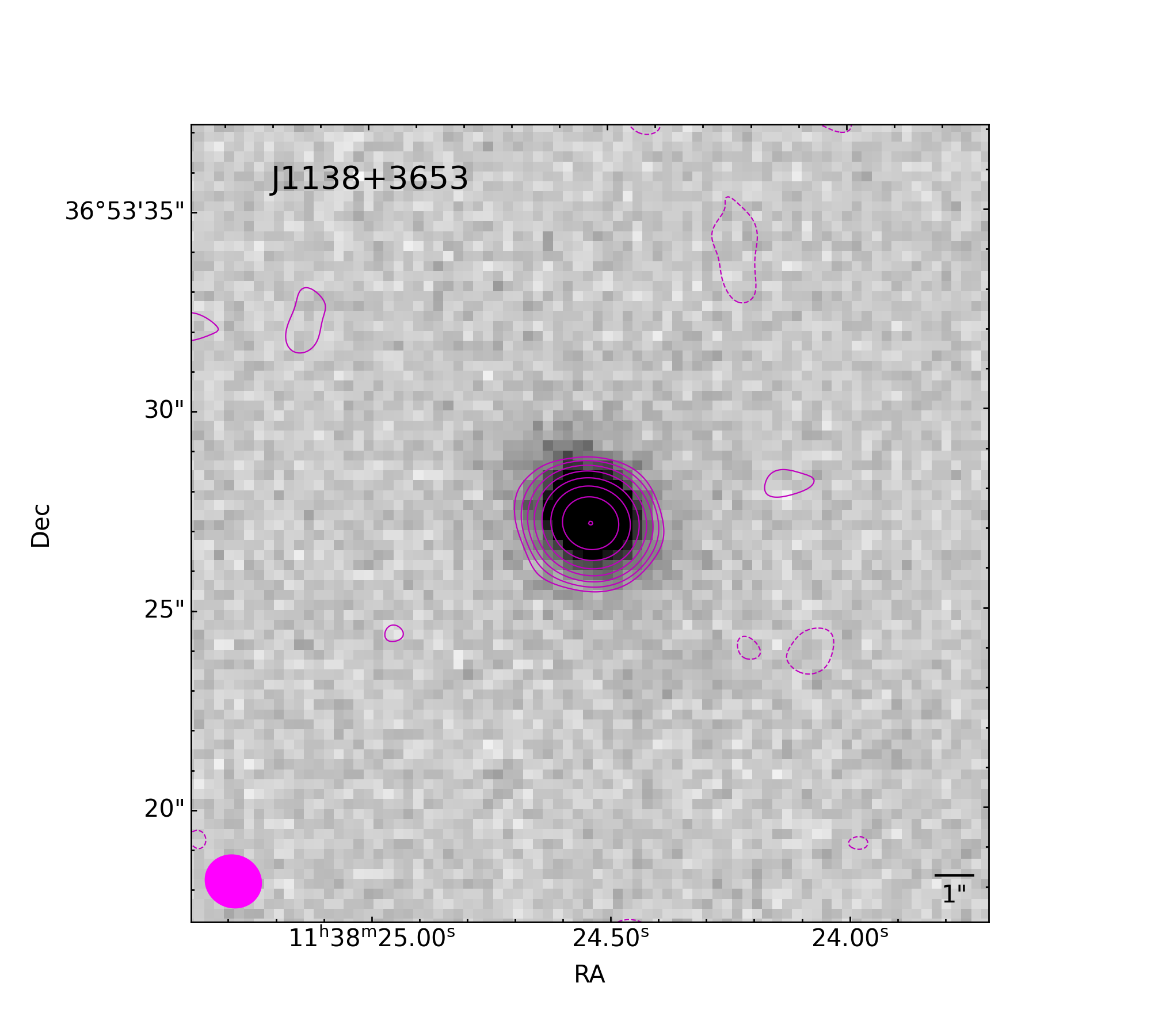}
         \caption{PanSTARRS $i$ band image of the host galaxy overlaid with the 90k$\lambda$ tapered map. Radio map properties as in Fig.~\ref{fig:J1138-90k}}. \label{fig:J1138-host}
     \end{subfigure}
        \caption{}
        \label{fig:J1138}
\end{figure*}


\begin{figure*}
     \centering
     \begin{subfigure}[b]{0.47\textwidth}
         \centering
         \includegraphics[width=\textwidth]{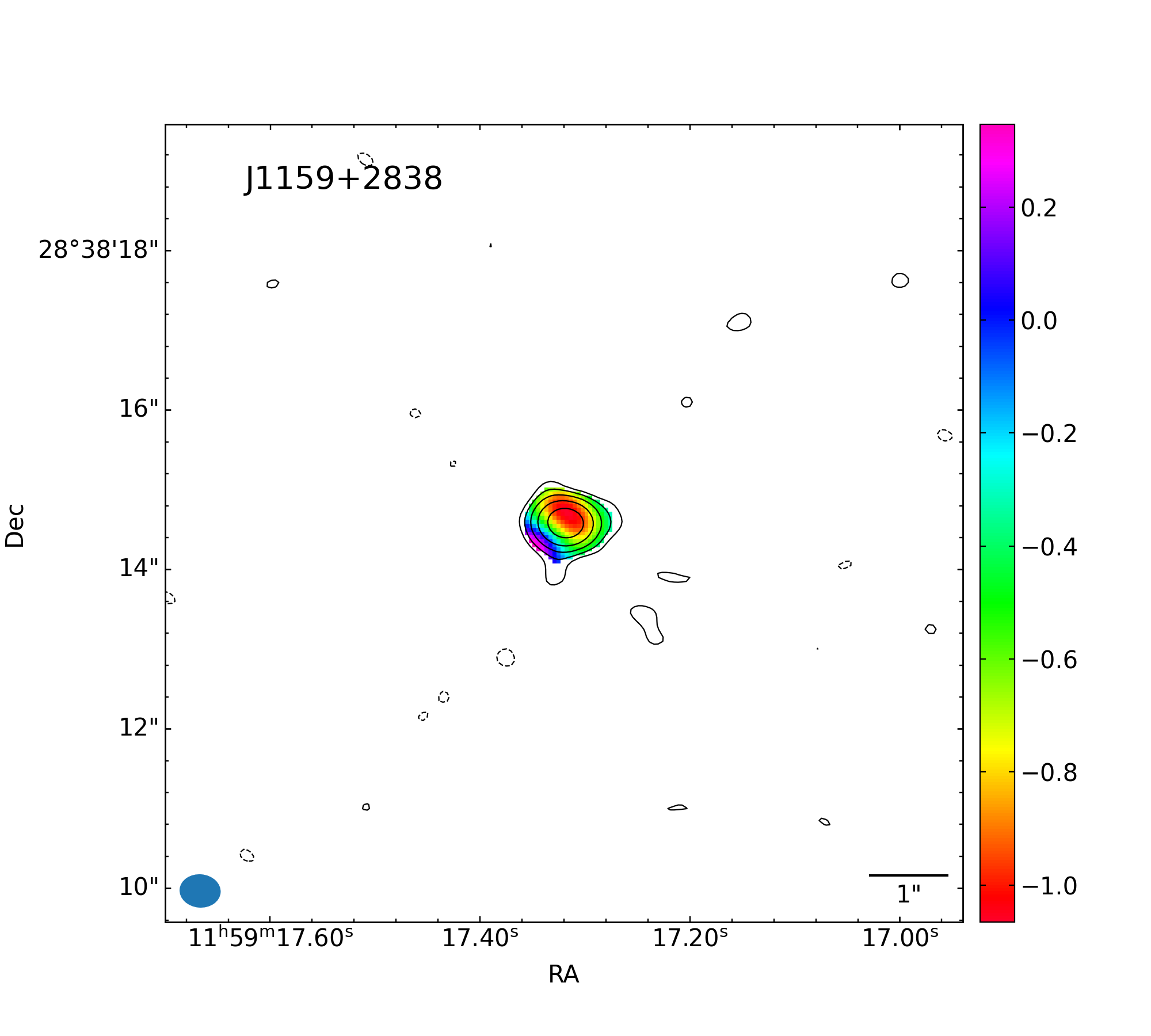}
         \caption{Spectral index map, rms = 10$\mu$Jy beam$^{-1}$, contour levels at -3, 3 $\times$ 2$^n$, $n \in$ [0, 4], beam size 1.78 $\times$ 1.44~kpc. } \label{fig:J1159spind}
     \end{subfigure}
     \hfill
     \begin{subfigure}[b]{0.47\textwidth}
         \centering
         \includegraphics[width=\textwidth]{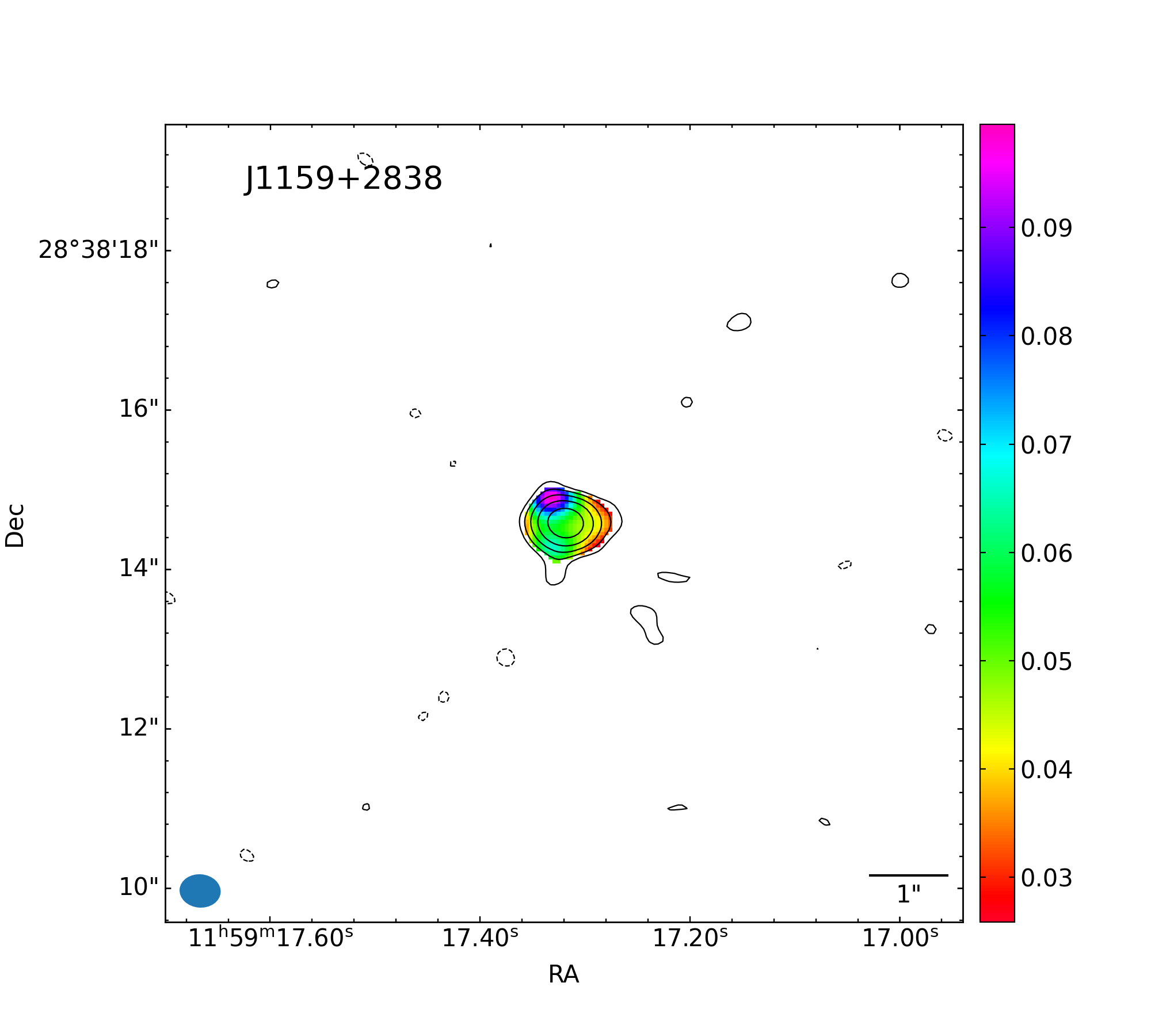}
         \caption{Spectral index error map, rms, contour levels, and beam size as in Fig.~\ref{fig:J1159spind}.} \label{fig:J1159spinderr}
     \end{subfigure}
     \hfill
     \\
     \begin{subfigure}[b]{0.47\textwidth}
         \centering
         \includegraphics[width=\textwidth]{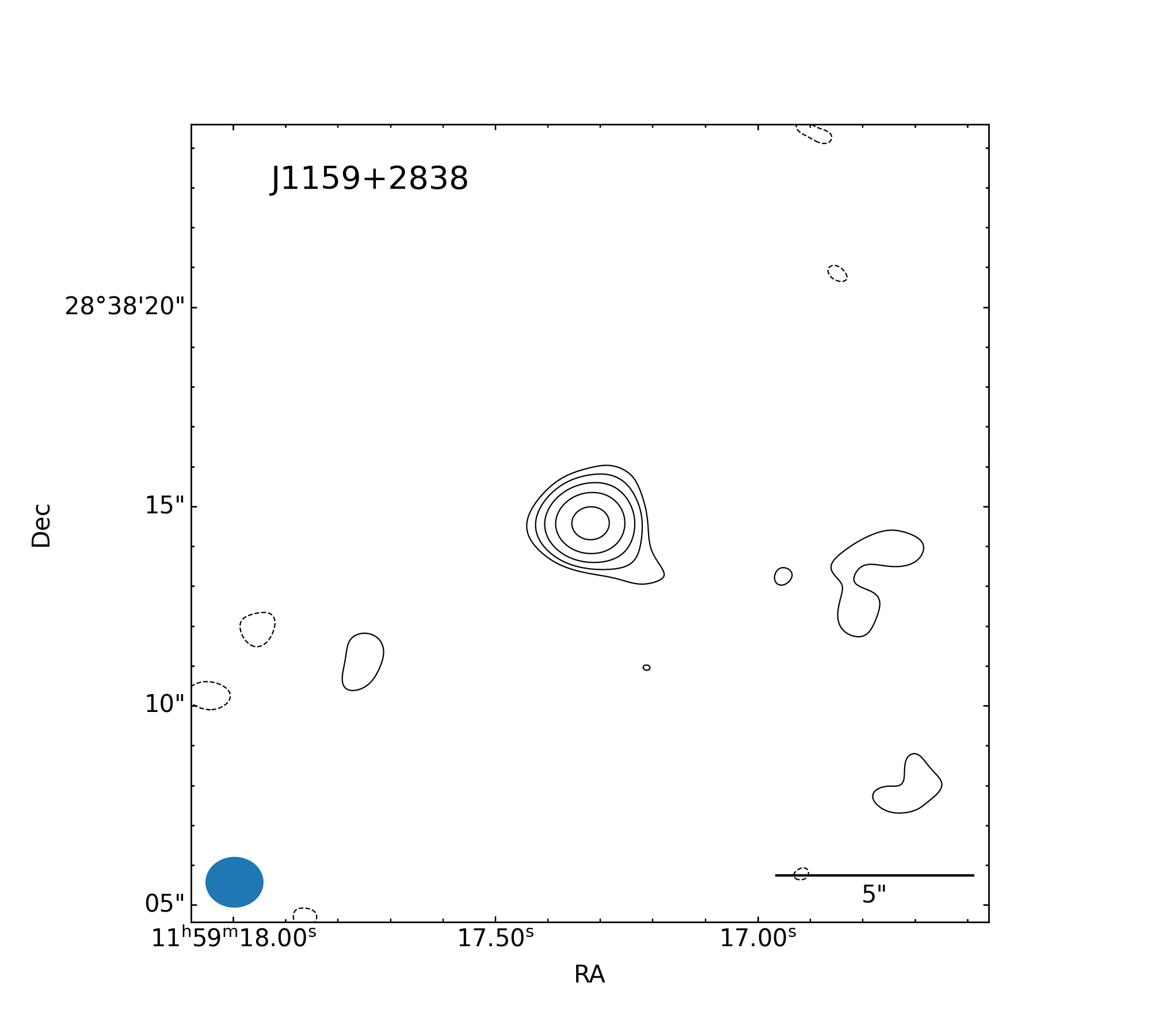}
         \caption{Tapered map with \texttt{uvtaper} = 90k$\lambda$, rms = 13$\mu$Jy beam$^{-1}$, contour levels at -3, 3 $\times$ 2$^n$, $n \in$ [0, 4], beam size 5.00 $\times$ 4.35~kpc.} \label{fig:J1159-90k}
     \end{subfigure}
          \hfill
     \begin{subfigure}[b]{0.47\textwidth}
         \centering
         \includegraphics[width=\textwidth]{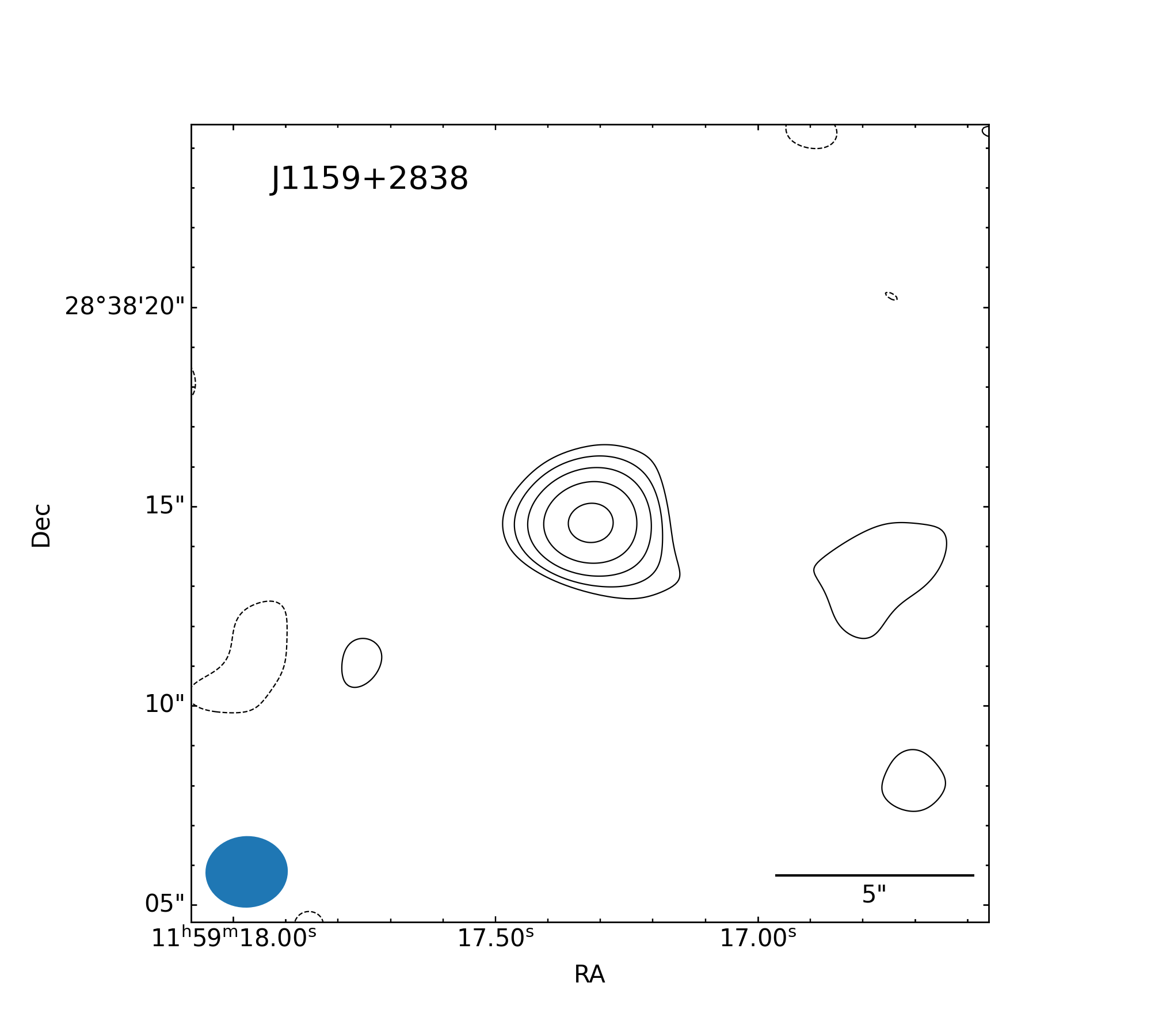}
         \caption{Tapered map with \texttt{uvtaper} = 60k$\lambda$, rms = 14$\mu$Jy beam$^{-1}$, contour levels at -3, 3 $\times$ 2$^n$, $n \in$ [0, 4], beam size 7.09 $\times$ 6.13~kpc.} \label{fig:J1159-60k}
     \end{subfigure}
          \hfill
     \\
     \begin{subfigure}[b]{0.47\textwidth}
         \centering
         \includegraphics[width=\textwidth]{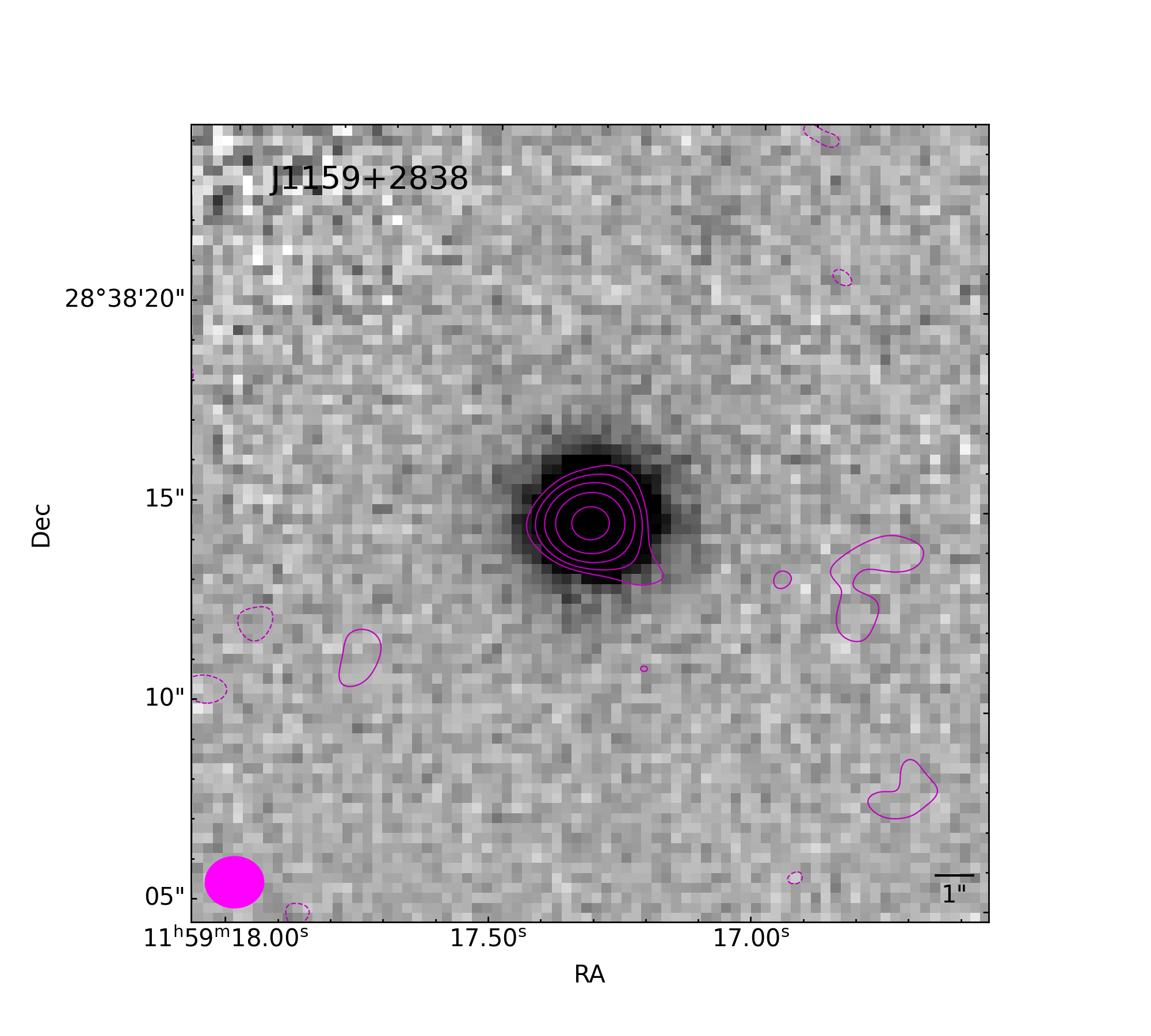}
         \caption{PanSTARRS $i$ band image of the host galaxy overlaid with the 90k$\lambda$ tapered map. Radio map properties as in Fig.~\ref{fig:J1159-90k}}. \label{fig:J1159-host}
     \end{subfigure}
        \caption{}
        \label{fig:J1159}
\end{figure*}


\begin{figure*}
     \centering
     \begin{subfigure}[b]{0.47\textwidth}
         \centering
         \includegraphics[width=\textwidth]{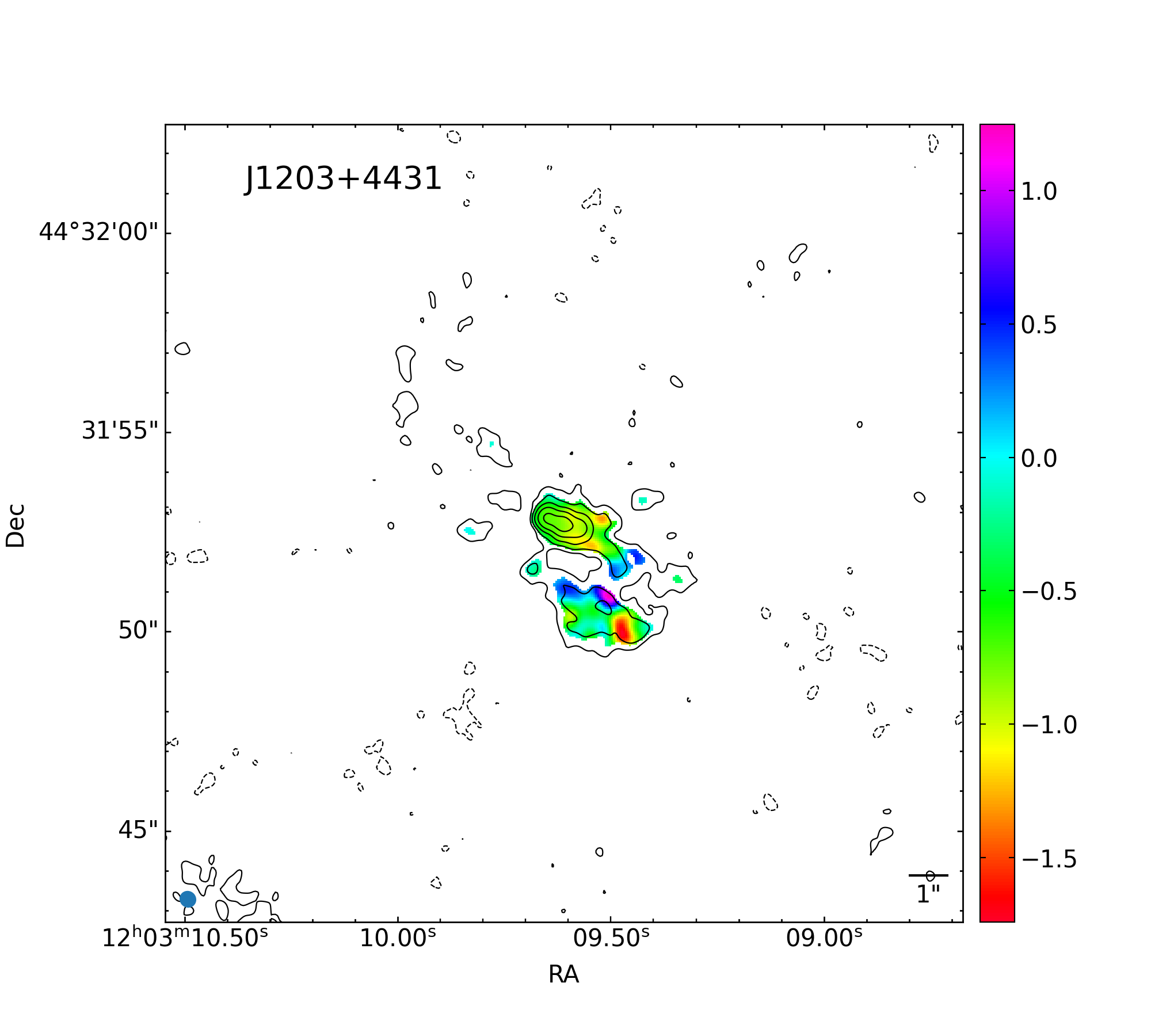}
         \caption{Spectral index map, rms = 12$\mu$Jy beam$^{-1}$, contour levels at -3, 3 $\times$ 2$^n$, $n \in$ [0, 4], beam size 0.02 $\times$ 0.02~kpc. } \label{fig:J1203spind}
     \end{subfigure}
     \hfill
     \begin{subfigure}[b]{0.47\textwidth}
         \centering
         \includegraphics[width=\textwidth]{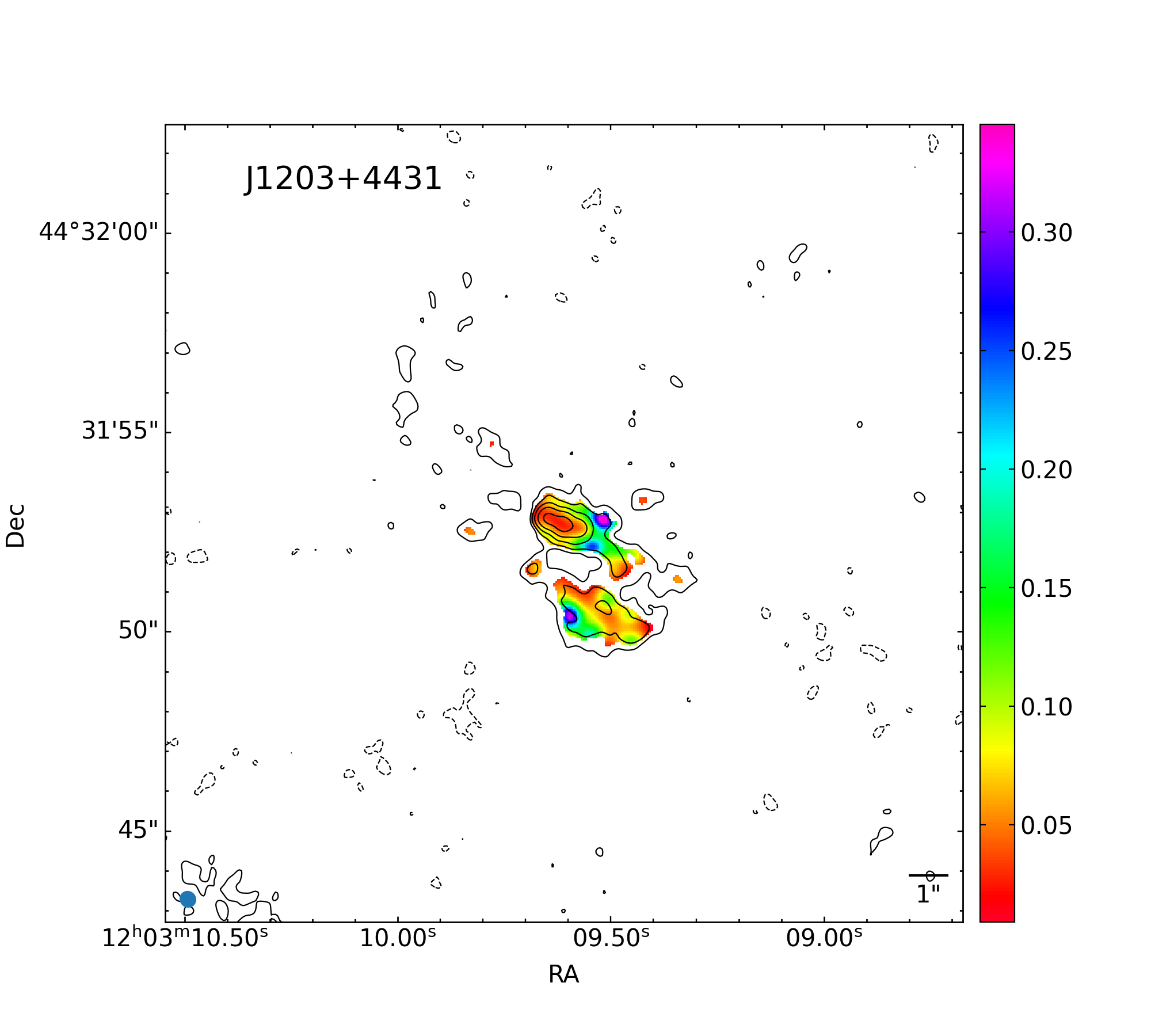}
         \caption{Spectral index error map, rms, contour levels, and beam size as in Fig.~\ref{fig:J1203spind}.} \label{fig:J1203spinderr}
     \end{subfigure}
     \hfill
     \\
     \begin{subfigure}[b]{0.47\textwidth}
         \centering
         \includegraphics[width=\textwidth]{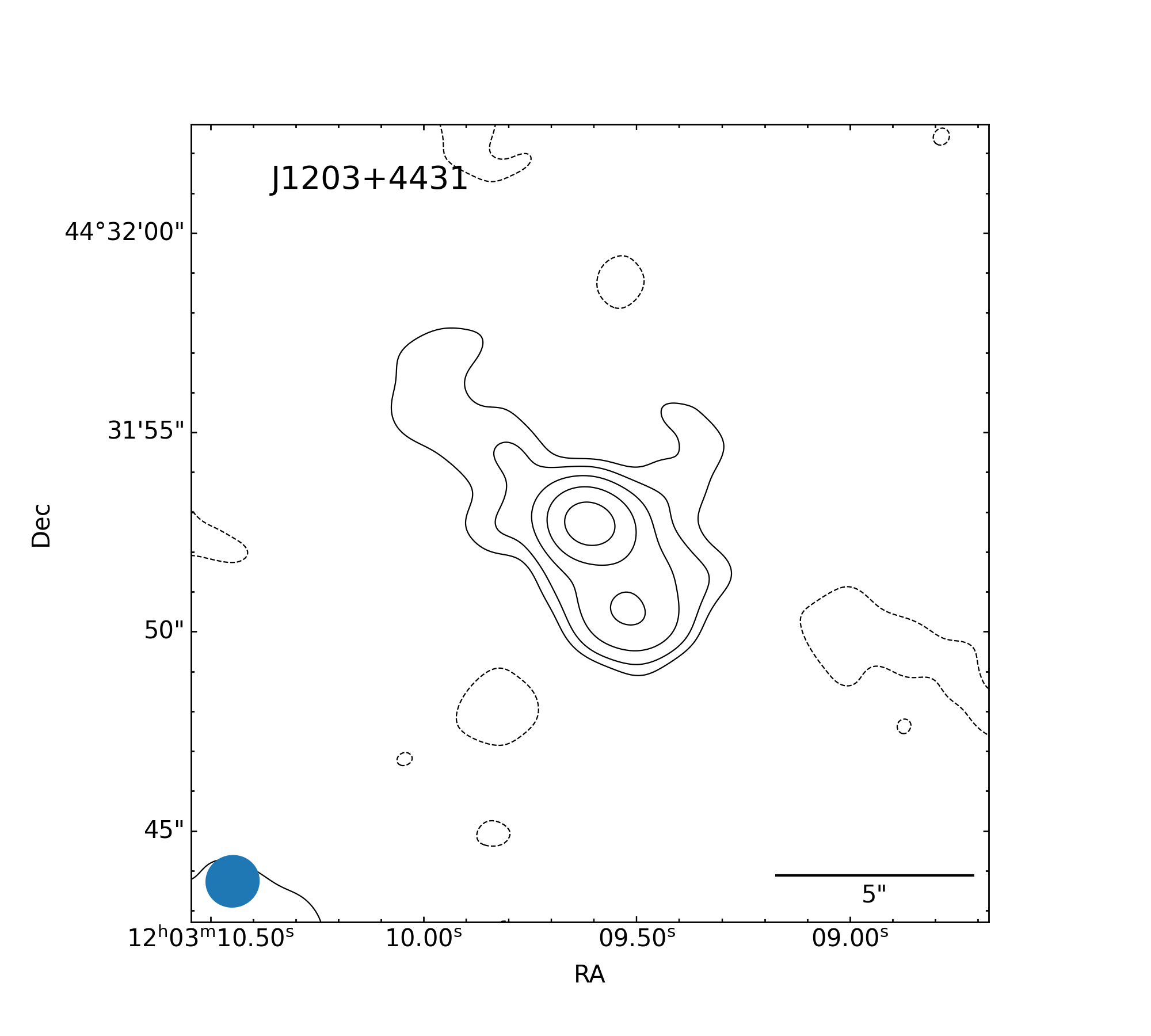}
         \caption{Tapered map with \texttt{uvtaper} = 90k$\lambda$, rms = 29$\mu$Jy beam$^{-1}$, contour levels at -3, 3 $\times$ 2$^n$, $n \in$ [0, 4], beam size 0.06 $\times$ 0.05~kpc.} \label{fig:J1203-90k}
     \end{subfigure}
          \hfill
     \begin{subfigure}[b]{0.47\textwidth}
         \centering
         \includegraphics[width=\textwidth]{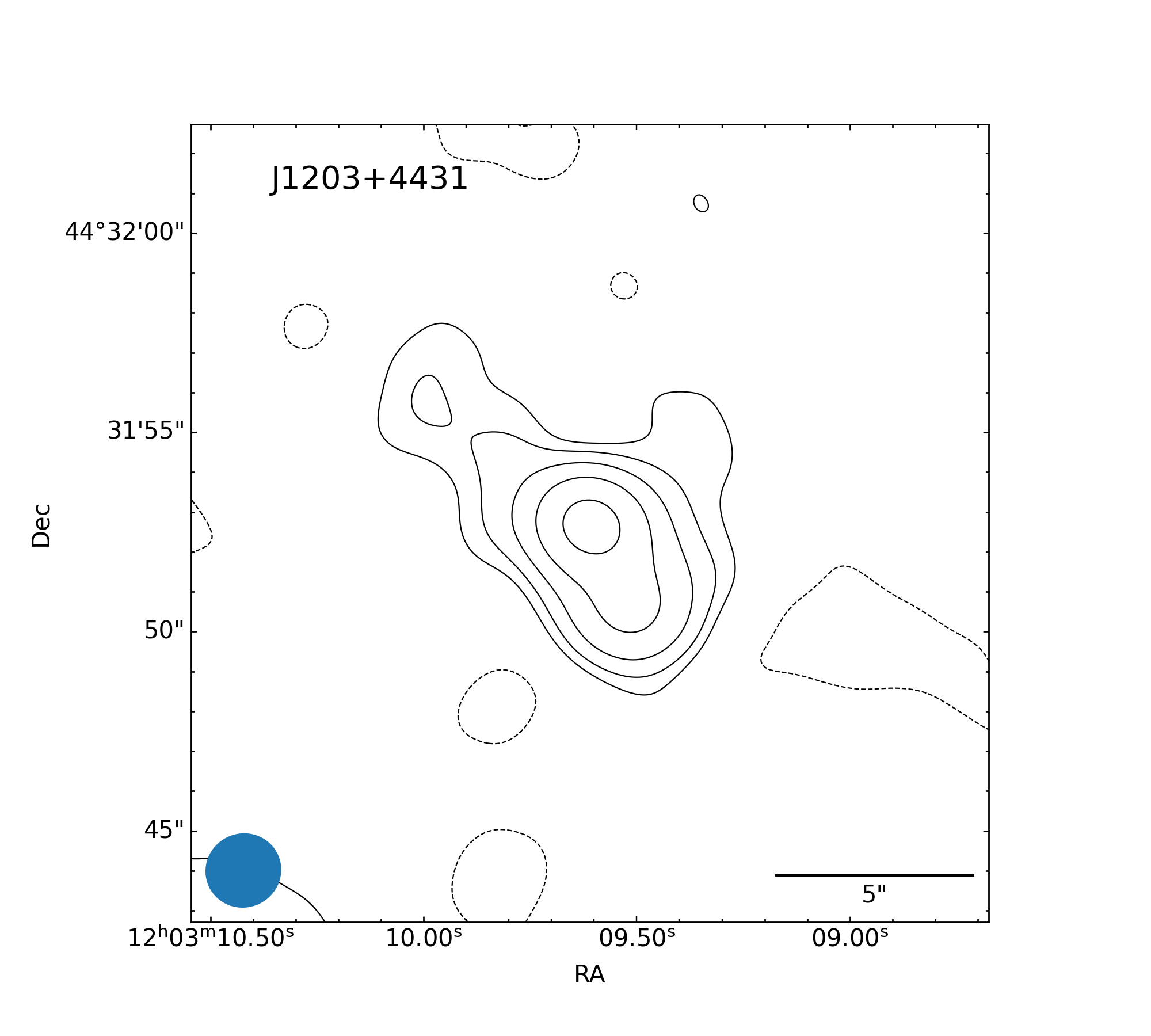}
         \caption{Tapered map with \texttt{uvtaper} = 60k$\lambda$, rms = 38$\mu$Jy beam$^{-1}$, contour levels at -3, 3 $\times$ 2$^n$, $n \in$ [0, 4], beam size 0.08 $\times$ 0.08~kpc.} \label{fig:J1203-60k}
     \end{subfigure}
          \hfill
     \\
     \begin{subfigure}[b]{0.47\textwidth}
         \centering
         \includegraphics[width=\textwidth]{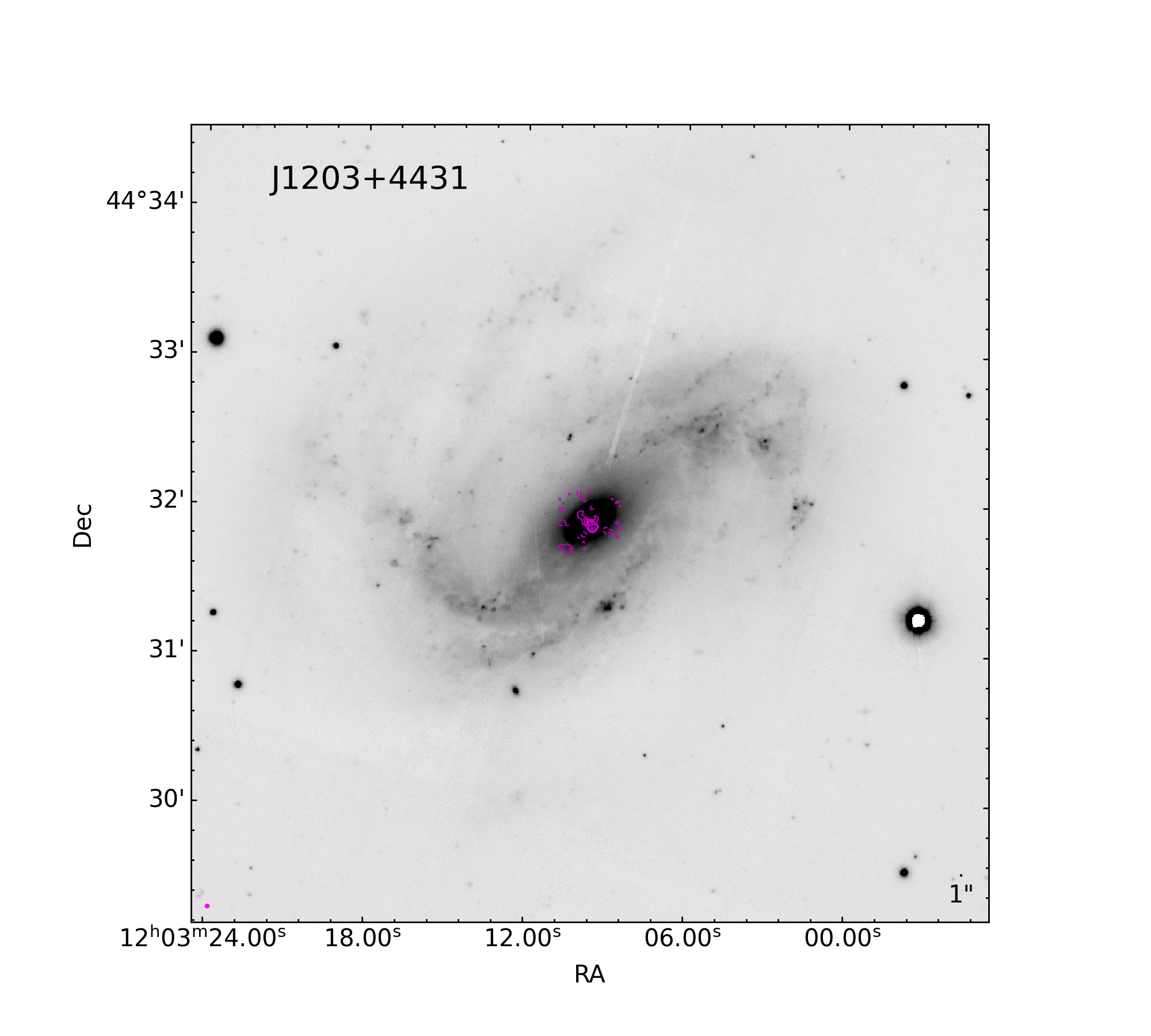}
         \caption{PanSTARRS $i$ band image of the host galaxy overlaid with the 90k$\lambda$ tapered map. Radio map properties as in Fig.~\ref{fig:J1203-90k}}. \label{fig:J1203-host}
     \end{subfigure}
          \hfill
     \begin{subfigure}[b]{0.47\textwidth}
         \centering
         \includegraphics[width=\textwidth]{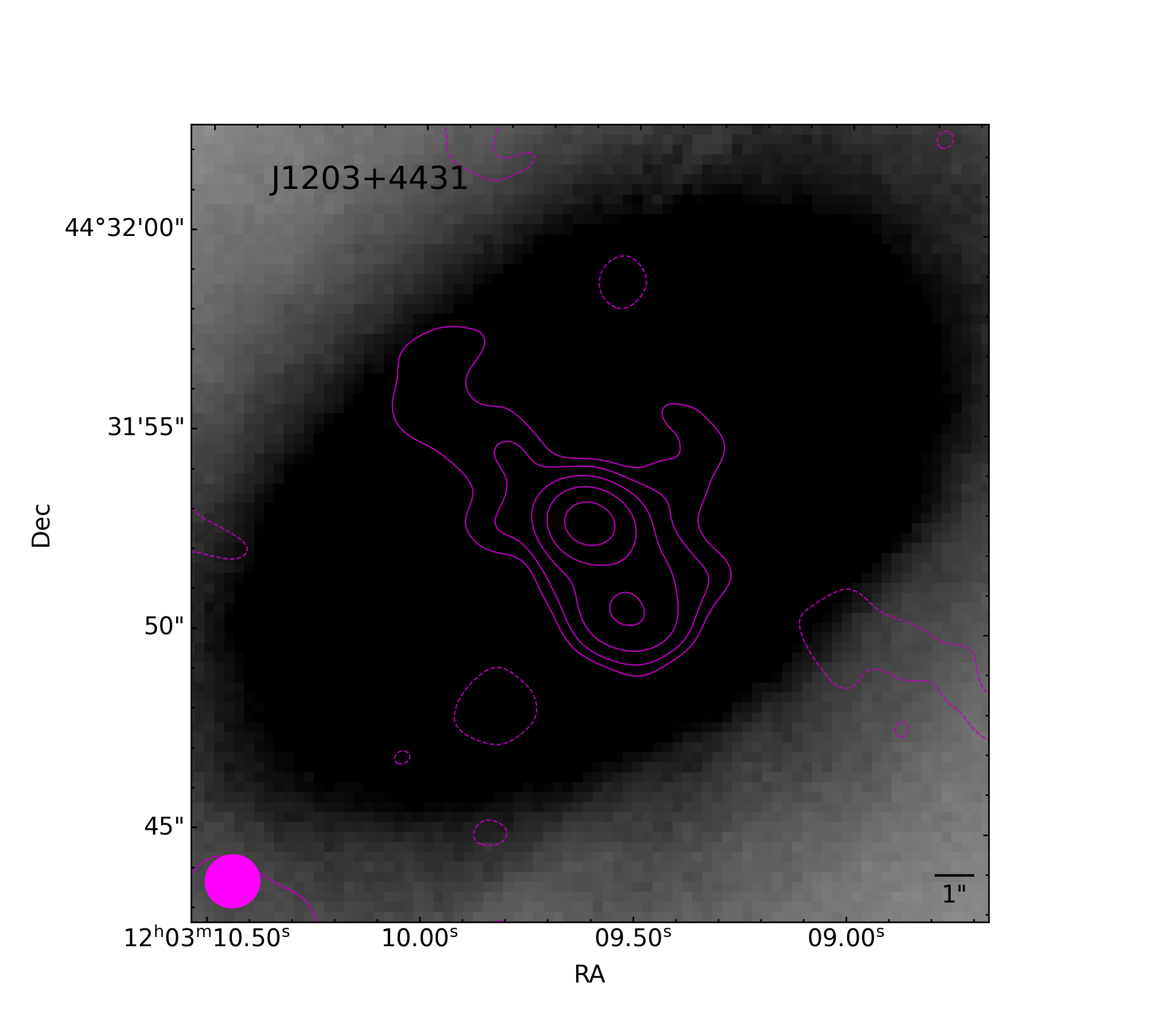}
         \caption{PanSTARRS $i$ band image of the host galaxy overlaid with the 90k$\lambda$ tapered map. Radio map properties as in Fig.~\ref{fig:J1203-90k}}. \label{fig:J1203-host-zoom}
     \end{subfigure}
        \caption{}
        \label{fig:J1203}
\end{figure*}


\begin{figure*}
     \centering
     \begin{subfigure}[b]{0.47\textwidth}
         \centering
         \includegraphics[width=\textwidth]{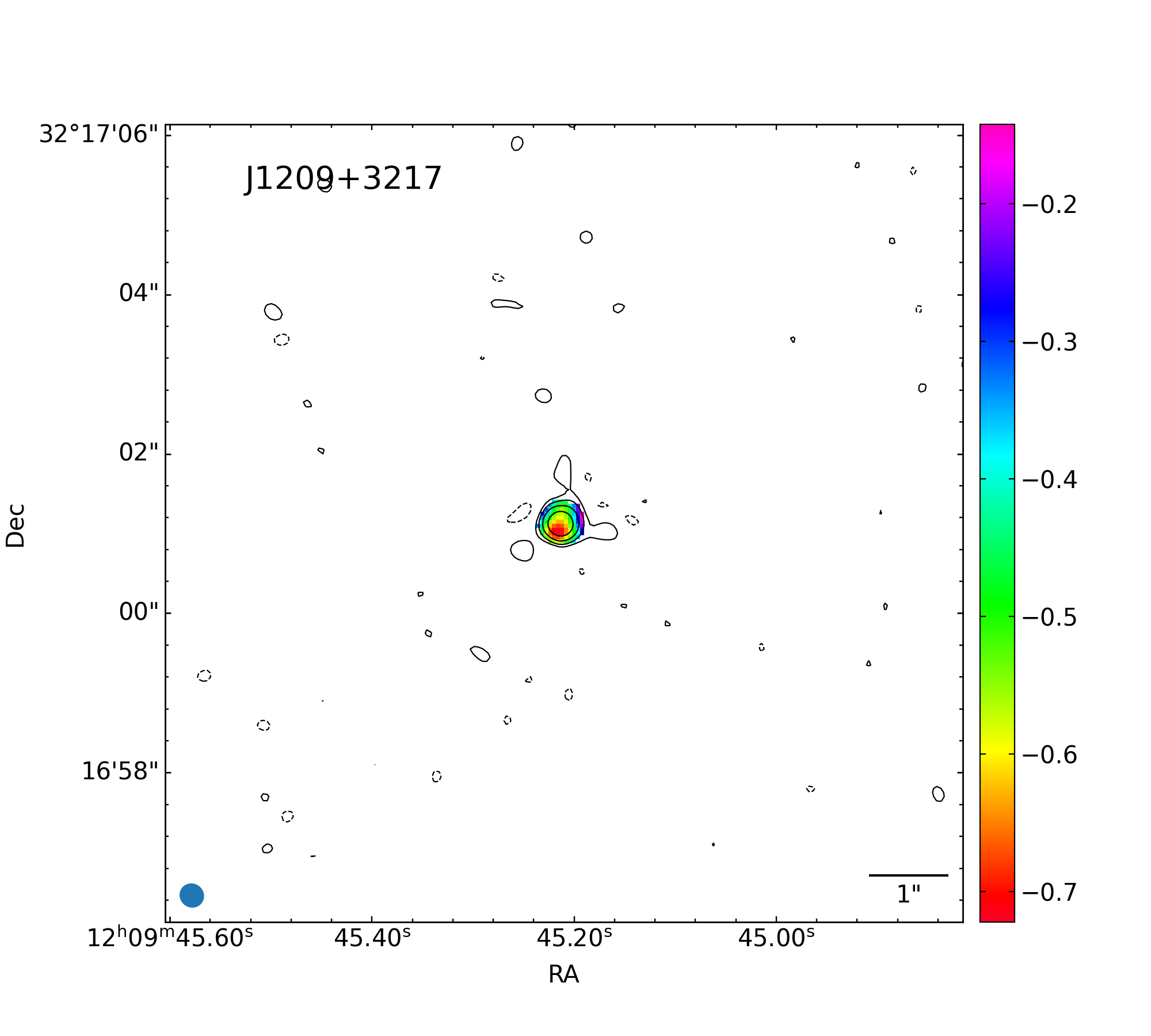}
         \caption{Spectral index map, rms = 13$\mu$Jy beam$^{-1}$, contour levels at -3, 3 $\times$ 2$^n$, $n \in$ [0, 3], beam size 0.78 $\times$ 0.76~kpc. } \label{fig:J1209spind}
     \end{subfigure}
     \hfill
     \begin{subfigure}[b]{0.47\textwidth}
         \centering
         \includegraphics[width=\textwidth]{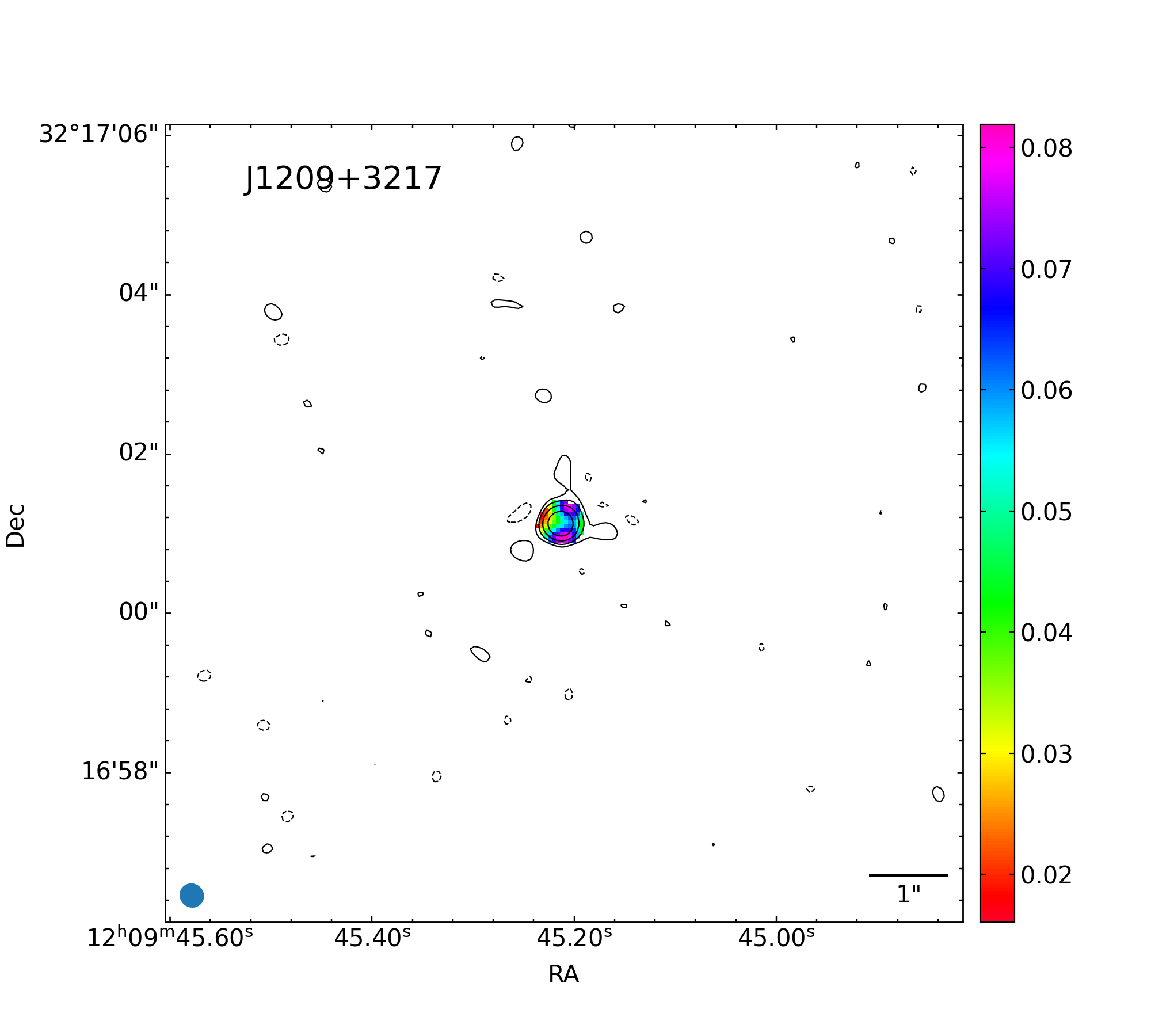}
         \caption{Spectral index error map, rms, contour levels, and beam size as in Fig.~\ref{fig:J1209spind}.} \label{fig:J1209spinderr}
     \end{subfigure}
     \hfill
     \\
     \begin{subfigure}[b]{0.47\textwidth}
         \centering
         \includegraphics[width=\textwidth]{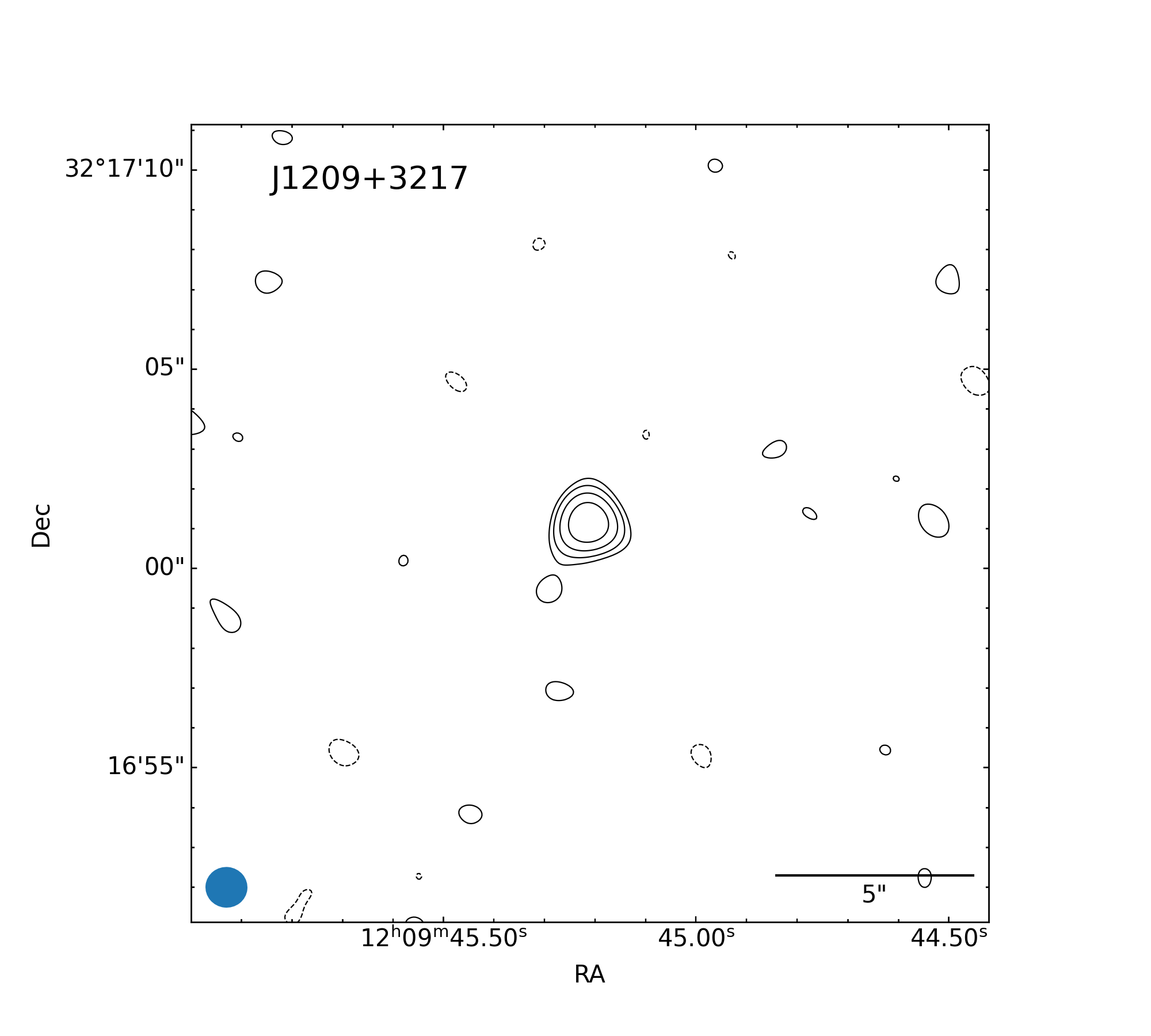}
         \caption{Tapered map with \texttt{uvtaper} = 90k$\lambda$, rms = 14$\mu$Jy beam$^{-1}$, contour levels at -3, 3 $\times$ 2$^n$, $n \in$ [0, 3], beam size 2.65 $\times$ 2.58~kpc.} \label{fig:J1209-90k}
     \end{subfigure}
          \hfill
     \begin{subfigure}[b]{0.47\textwidth}
         \centering
         \includegraphics[width=\textwidth]{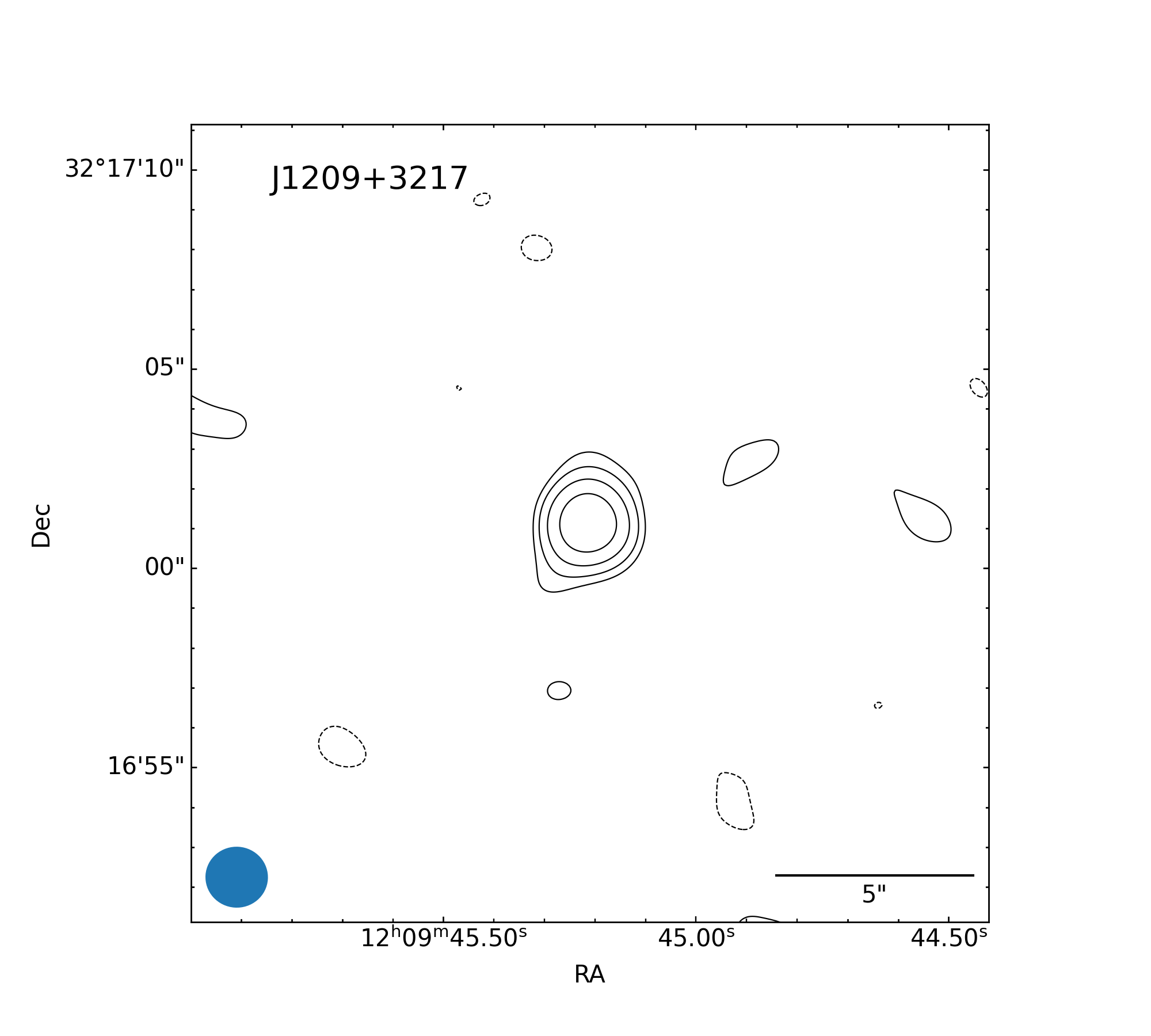}
         \caption{Tapered map with \texttt{uvtaper} = 60k$\lambda$, rms = 15$\mu$Jy beam$^{-1}$, contour levels at -3, 3 $\times$ 2$^n$, $n \in$ [0, 3], beam size 3.97 $\times$ 3.87~kpc.} \label{fig:J1209-60k}
     \end{subfigure}
          \hfill
     \\
     \begin{subfigure}[b]{0.47\textwidth}
         \centering
         \includegraphics[width=\textwidth]{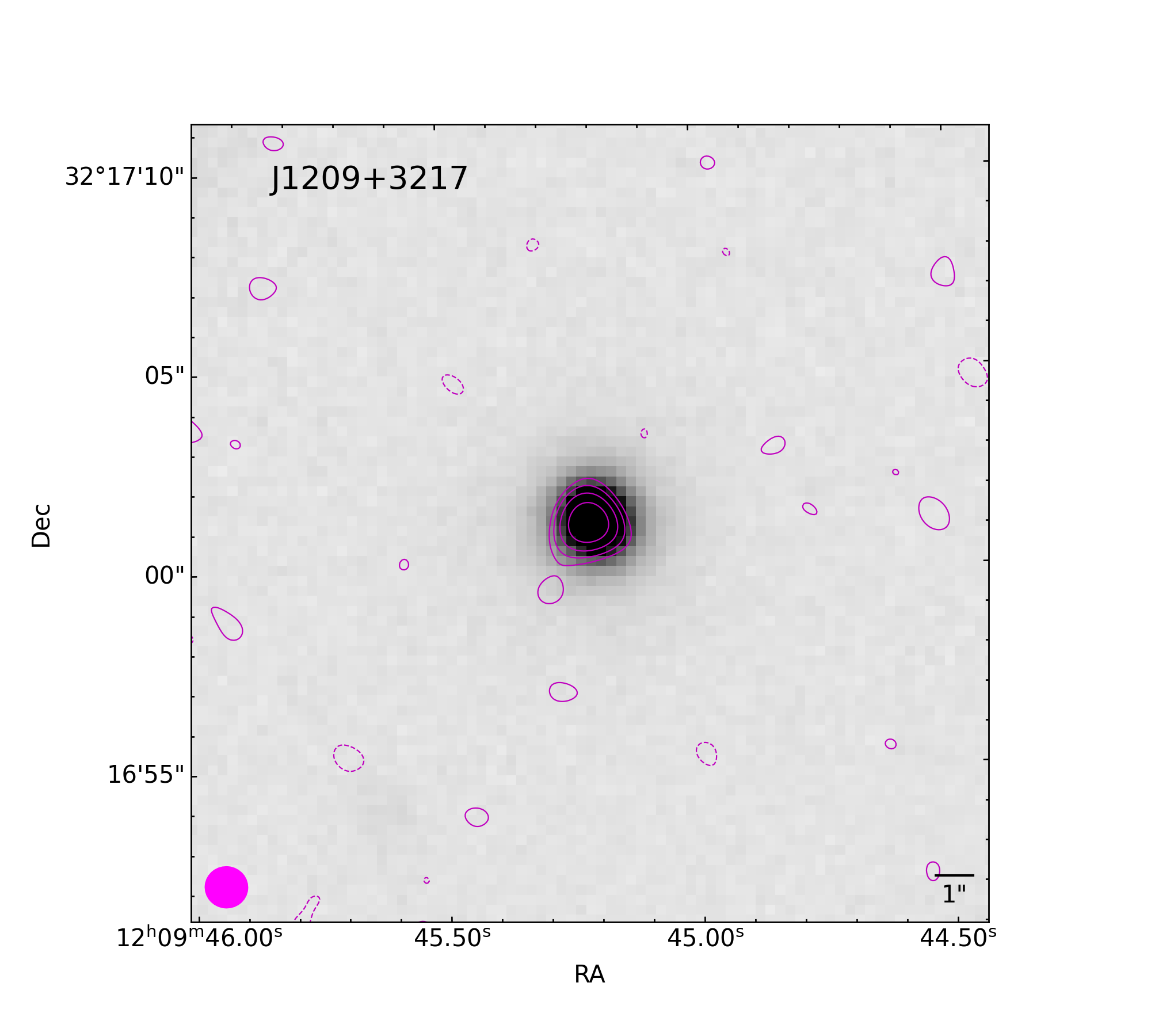}
         \caption{PanSTARRS $i$ band image of the host galaxy overlaid with the 90k$\lambda$ tapered map. Radio map properties as in Fig.~\ref{fig:J1209-90k}}. \label{fig:J1209-host}
     \end{subfigure}
        \caption{}
        \label{fig:J1209}
\end{figure*}


\begin{figure*}
     \centering
     \begin{subfigure}[b]{0.47\textwidth}
         \centering
         \includegraphics[width=\textwidth]{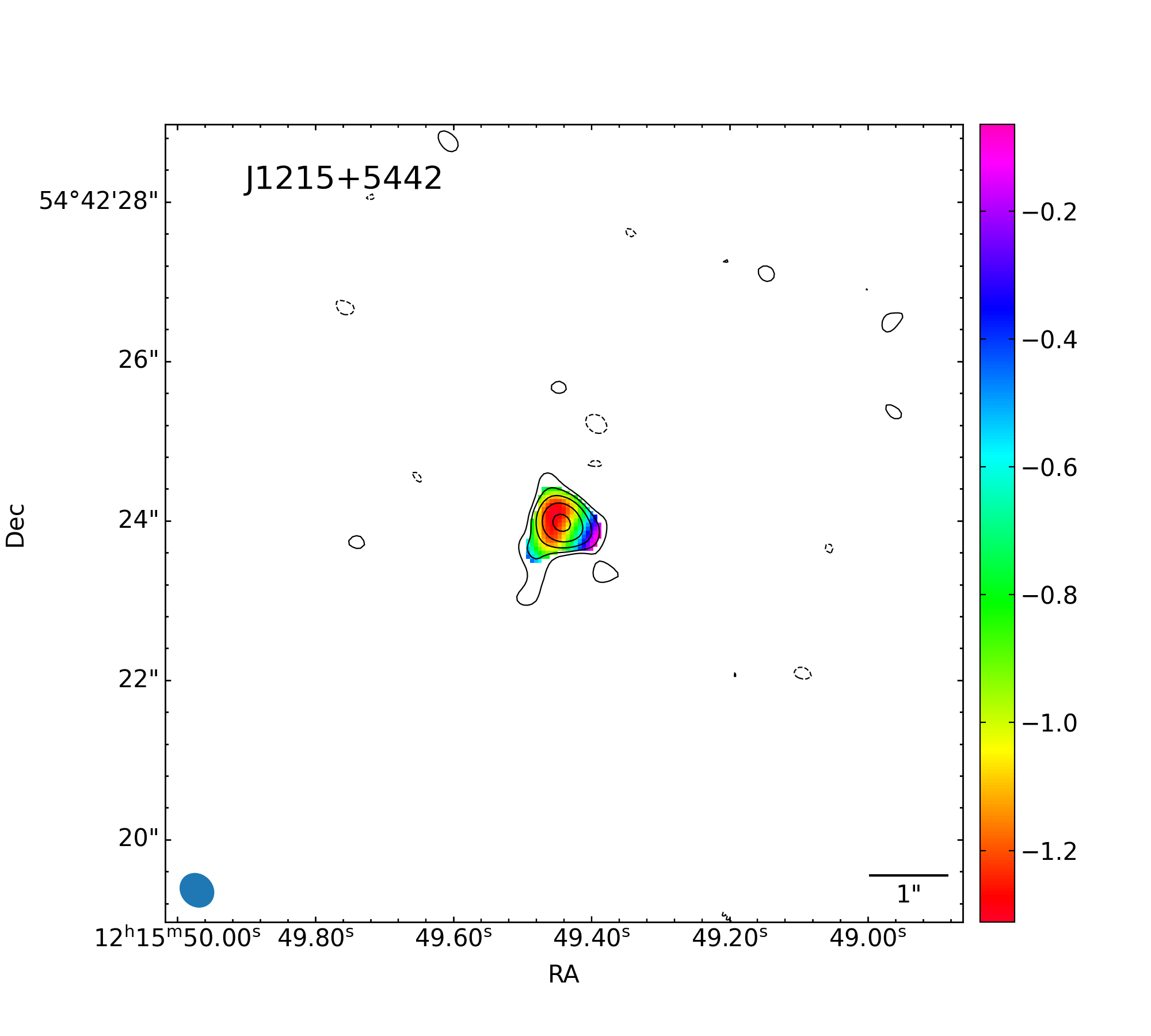}
         \caption{Spectral index map, rms = 10$\mu$Jy beam$^{-1}$, contour levels at -3, 3 $\times$ 2$^n$, $n \in$ [0, 4], beam size 1.20 $\times$ 1.07~kpc. } \label{fig:J1215spind}
     \end{subfigure}
     \hfill
     \begin{subfigure}[b]{0.47\textwidth}
         \centering
         \includegraphics[width=\textwidth]{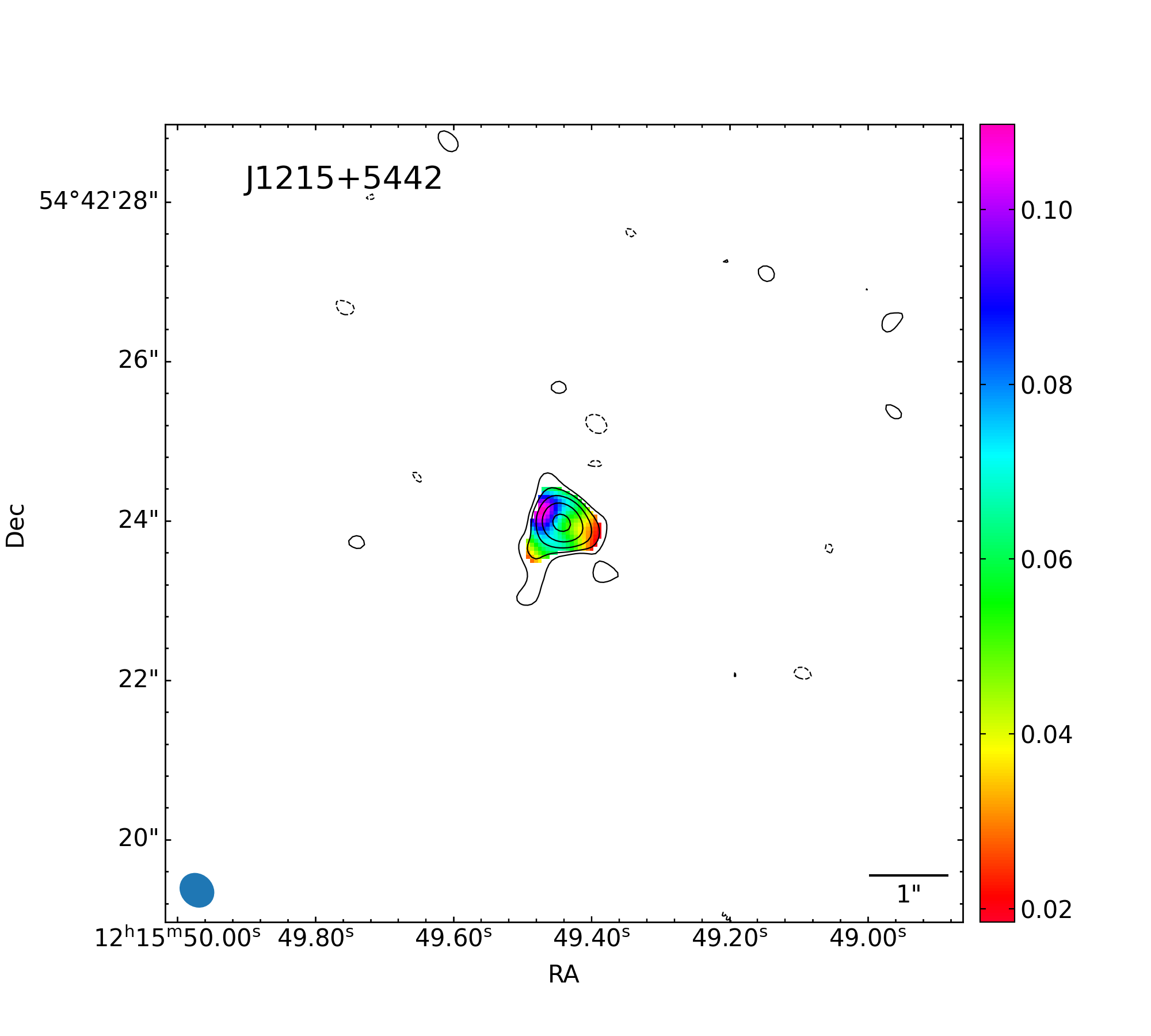}
         \caption{Spectral index error map, rms, contour levels, and beam size as in Fig.~\ref{fig:J1215spind}.} \label{fig:J1215spinderr}
     \end{subfigure}
     \hfill
     \\
     \begin{subfigure}[b]{0.47\textwidth}
         \centering
         \includegraphics[width=\textwidth]{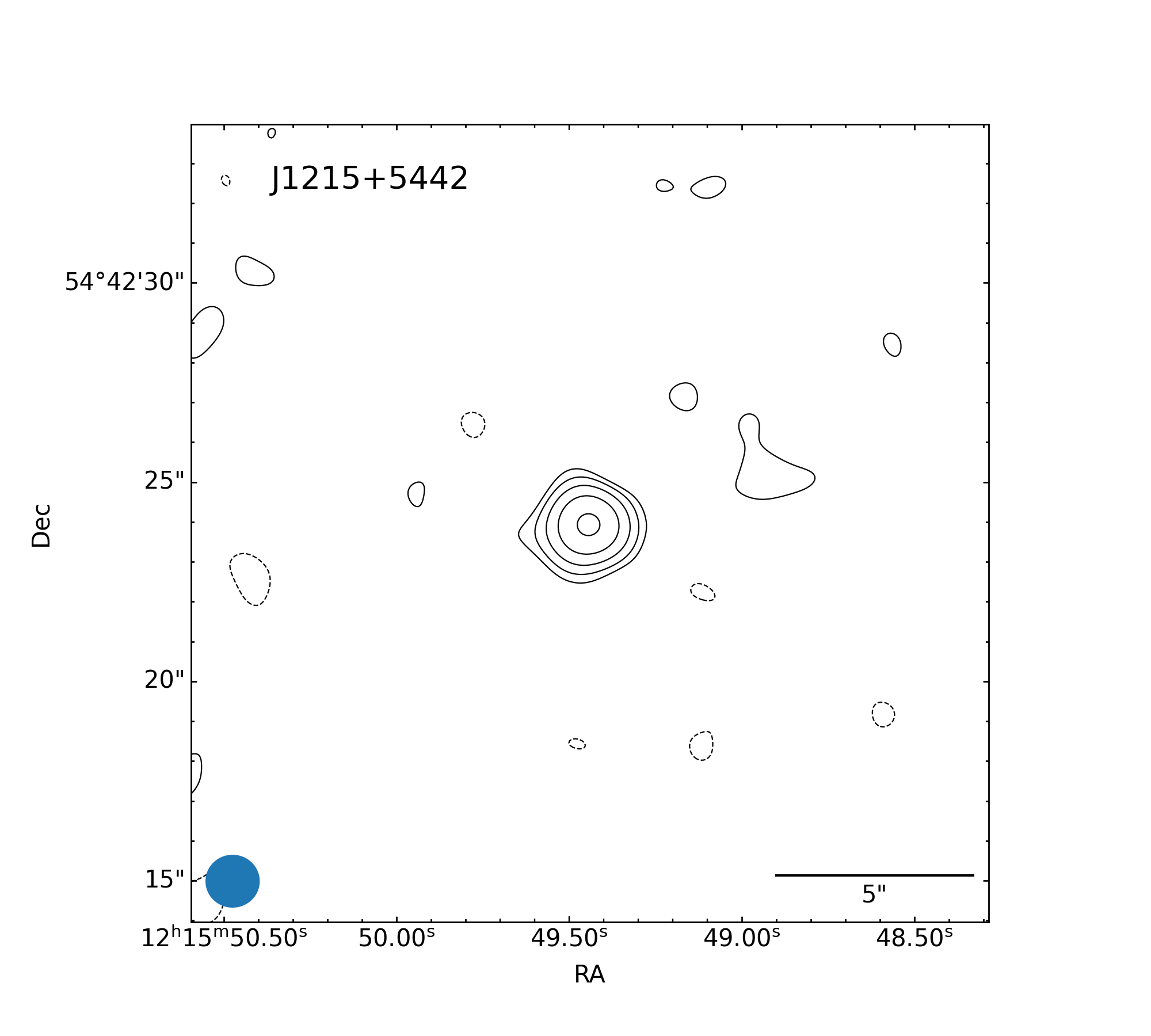}
         \caption{Tapered map with \texttt{uvtaper} = 90k$\lambda$, rms = 12$\mu$Jy beam$^{-1}$, contour levels at -3, 3 $\times$ 2$^n$, $n \in$ [0, 4], beam size 3.56 $\times$ 3.48~kpc.} \label{fig:J1215-90k}
     \end{subfigure}
          \hfill
     \begin{subfigure}[b]{0.47\textwidth}
         \centering
         \includegraphics[width=\textwidth]{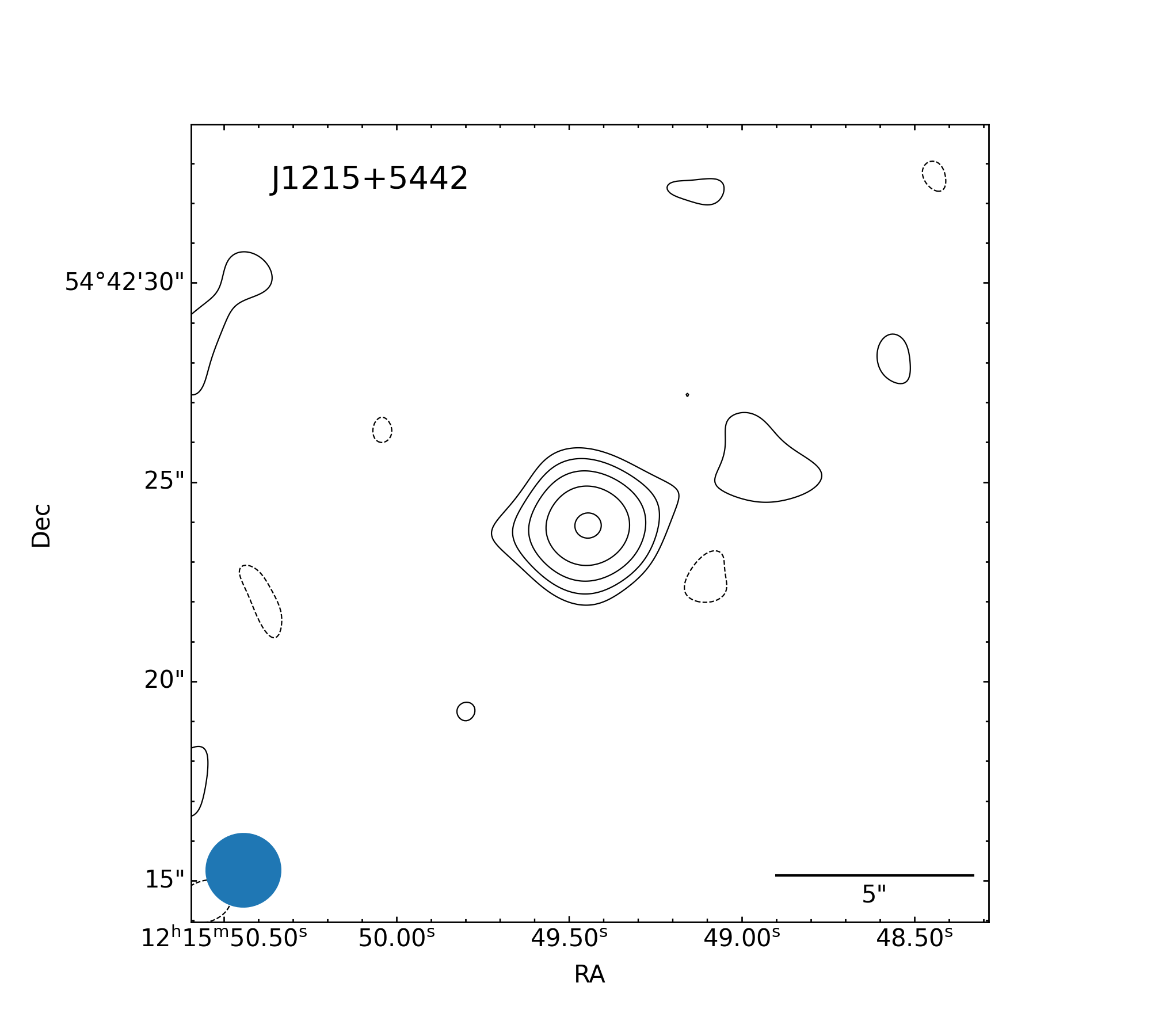}
         \caption{Tapered map with \texttt{uvtaper} = 60k$\lambda$, rms = 13$\mu$Jy beam$^{-1}$, contour levels at -3, 3 $\times$ 2$^n$, $n \in$ [0, 4], beam size 4.99 $\times$ 4.91~kpc.} \label{fig:J1215-60k}
     \end{subfigure}
          \hfill
     \\
     \begin{subfigure}[b]{0.47\textwidth}
         \centering
         \includegraphics[width=\textwidth]{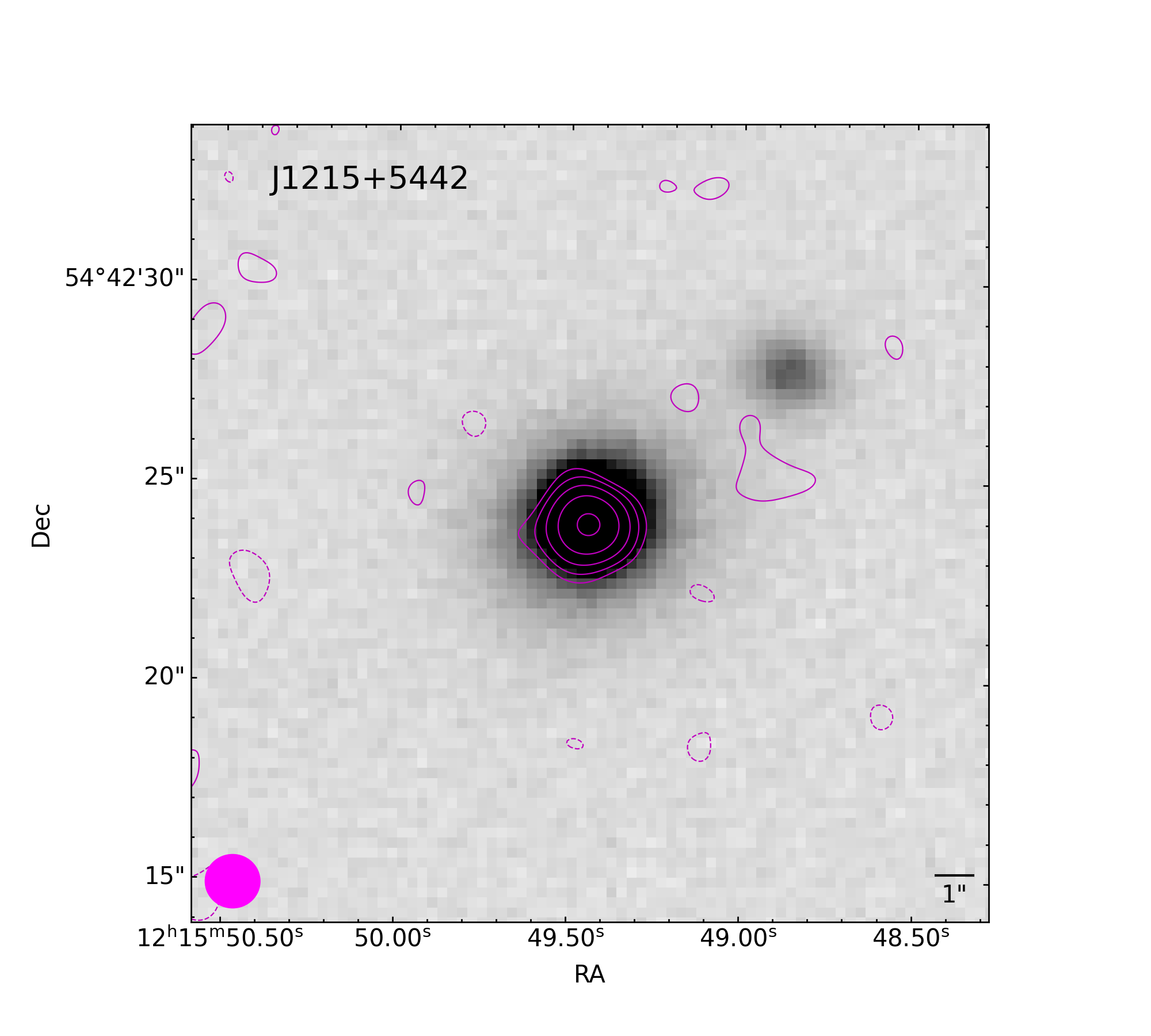}
         \caption{PanSTARRS $i$ band image of the host galaxy overlaid with the 90k$\lambda$ tapered map. Radio map properties as in Fig.~\ref{fig:J1215-90k}}. \label{fig:J1215-host}
     \end{subfigure}
        \caption{}
        \label{fig:J1215}
\end{figure*}


\begin{figure*}
     \centering
     \begin{subfigure}[b]{0.47\textwidth}
         \centering
         \includegraphics[width=\textwidth]{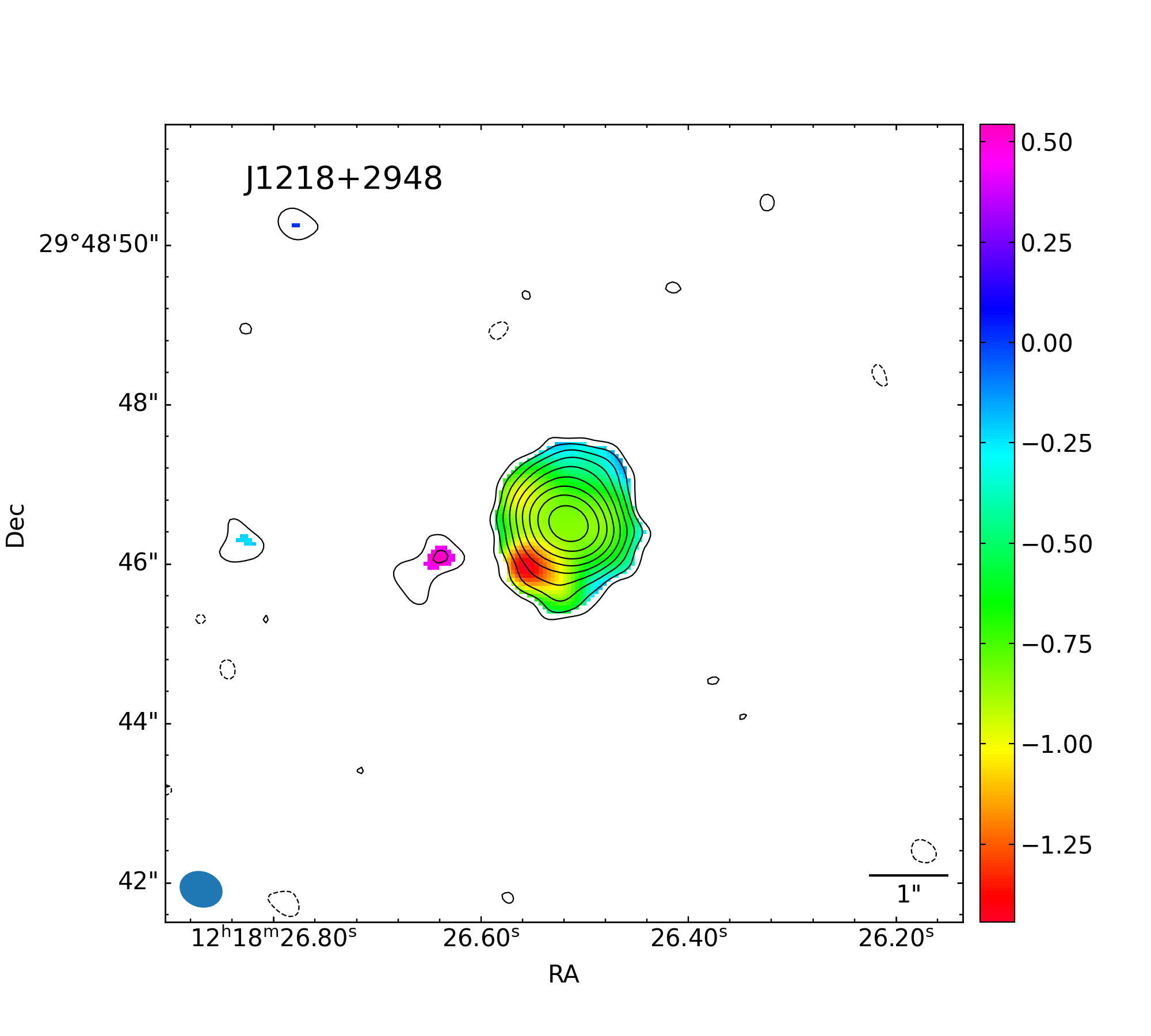}
         \caption{Spectral index map, rms = 9$\mu$Jy beam$^{-1}$, contour levels at -3, 3 $\times$ 2$^n$, $n \in$ [0, 8], beam size 0.15 $\times$ 0.12~kpc. } \label{fig:J1218spind}
     \end{subfigure}
     \hfill
     \begin{subfigure}[b]{0.47\textwidth}
         \centering
         \includegraphics[width=\textwidth]{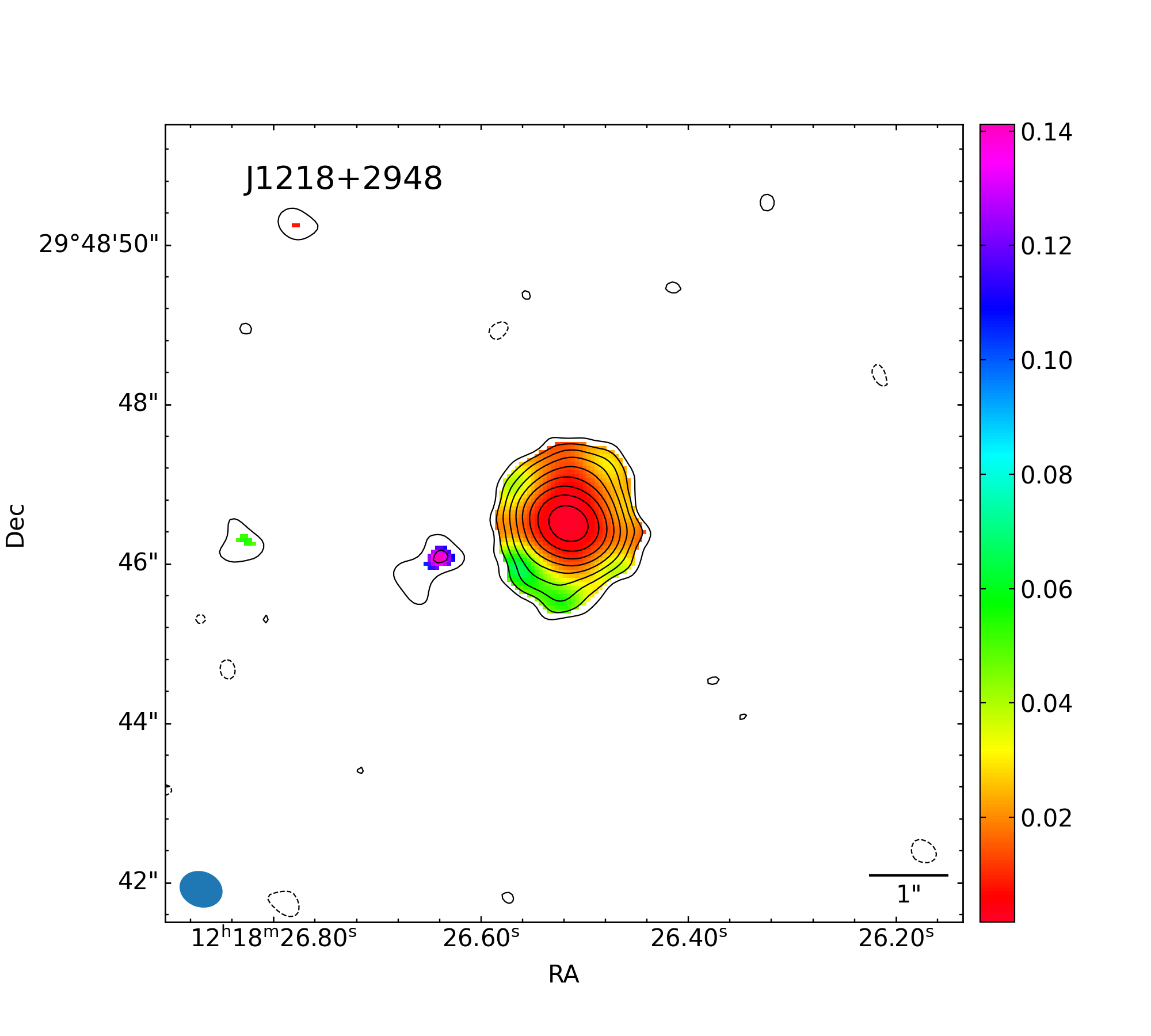}
         \caption{Spectral index error map, rms, contour levels, and beam size as in Fig.~\ref{fig:J1218spind}.} \label{fig:J1218spinderr}
     \end{subfigure}
     \hfill
     \\
     \begin{subfigure}[b]{0.47\textwidth}
         \centering
         \includegraphics[width=\textwidth]{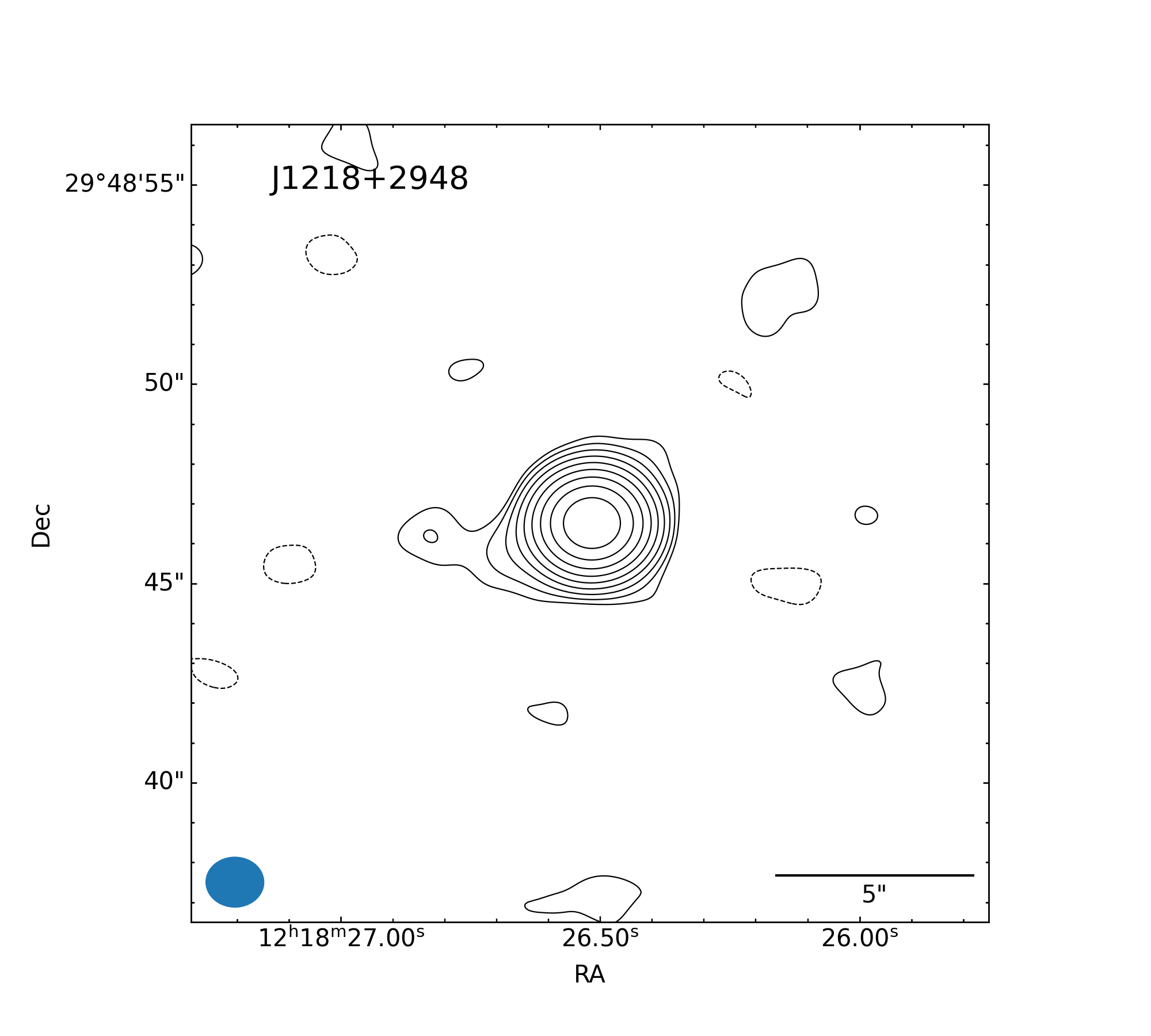}
         \caption{Tapered map with \texttt{uvtaper} = 90k$\lambda$, rms = 10$\mu$Jy beam$^{-1}$, contour levels at -3, 3 $\times$ 2$^n$, $n \in$ [0, 8], beam size 0.39 $\times$ 0.34~kpc.} \label{fig:J1218-90k}
     \end{subfigure}
          \hfill
     \begin{subfigure}[b]{0.47\textwidth}
         \centering
         \includegraphics[width=\textwidth]{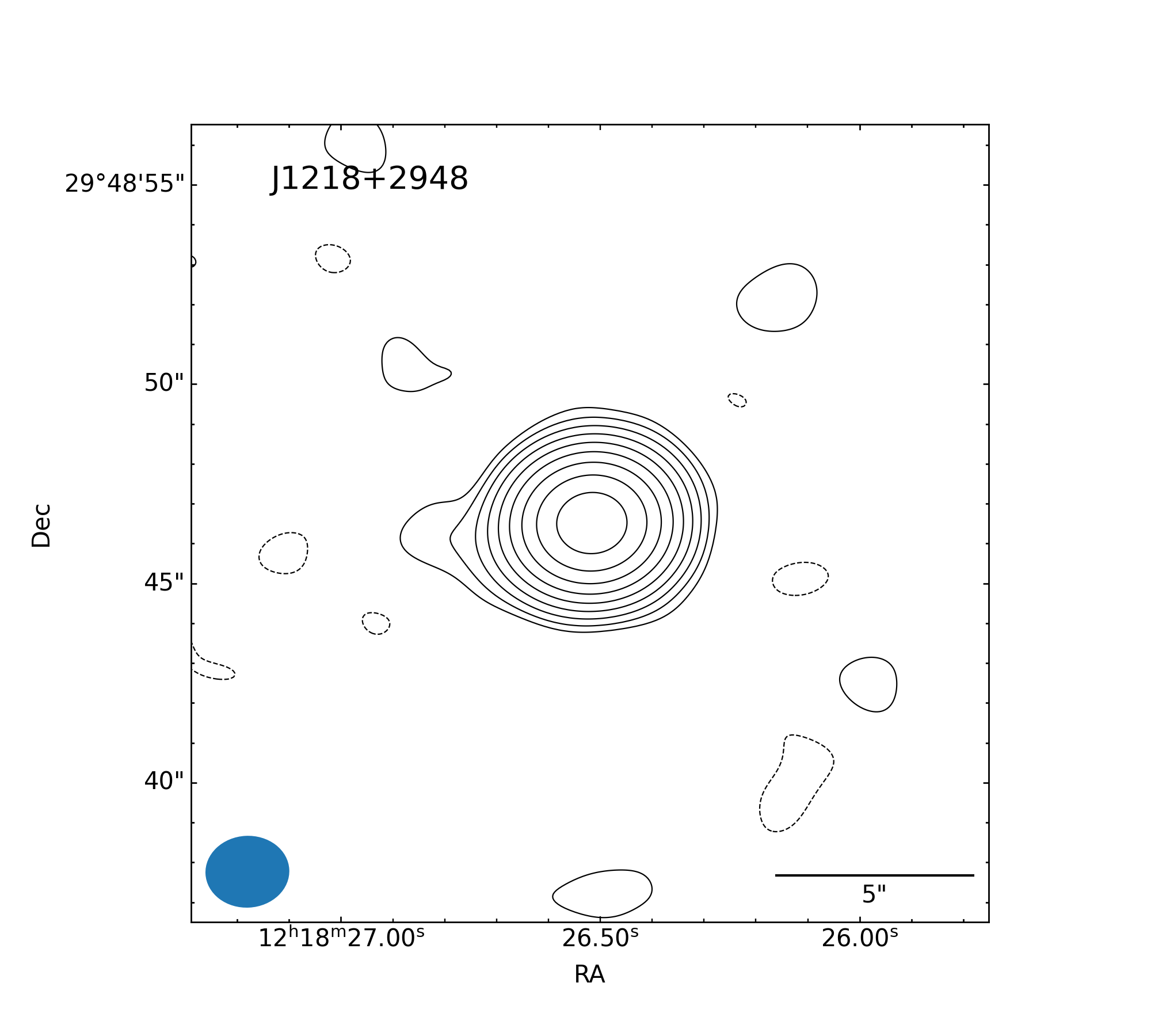}
         \caption{Tapered map with \texttt{uvtaper} = 60k$\lambda$, rms = 12$\mu$Jy beam$^{-1}$, contour levels at -3, 3 $\times$ 2$^n$, $n \in$ [0, 8], beam size 0.56 $\times$ 0.48~kpc.} \label{fig:J1218-60k}
     \end{subfigure}
          \hfill
     \\
     \begin{subfigure}[b]{0.47\textwidth}
         \centering
         \includegraphics[width=\textwidth]{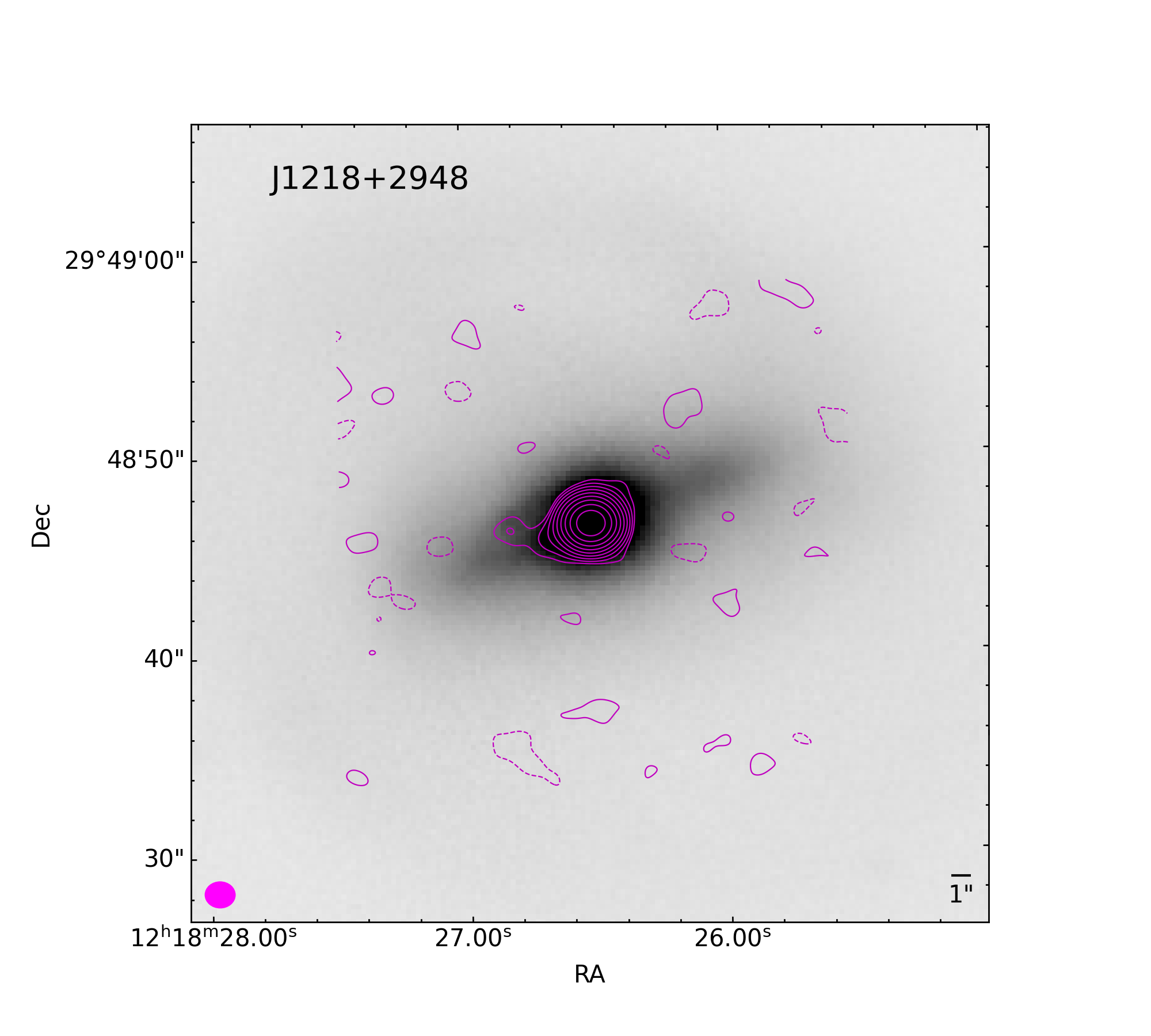}
         \caption{PanSTARRS $i$ band image of the host galaxy overlaid with the 90k$\lambda$ tapered map. Radio map properties as in Fig.~\ref{fig:J1218-90k}}. \label{fig:J1218-host}
     \end{subfigure}
              \hfill
     \begin{subfigure}[b]{0.47\textwidth}
         \centering
         \includegraphics[width=\textwidth]{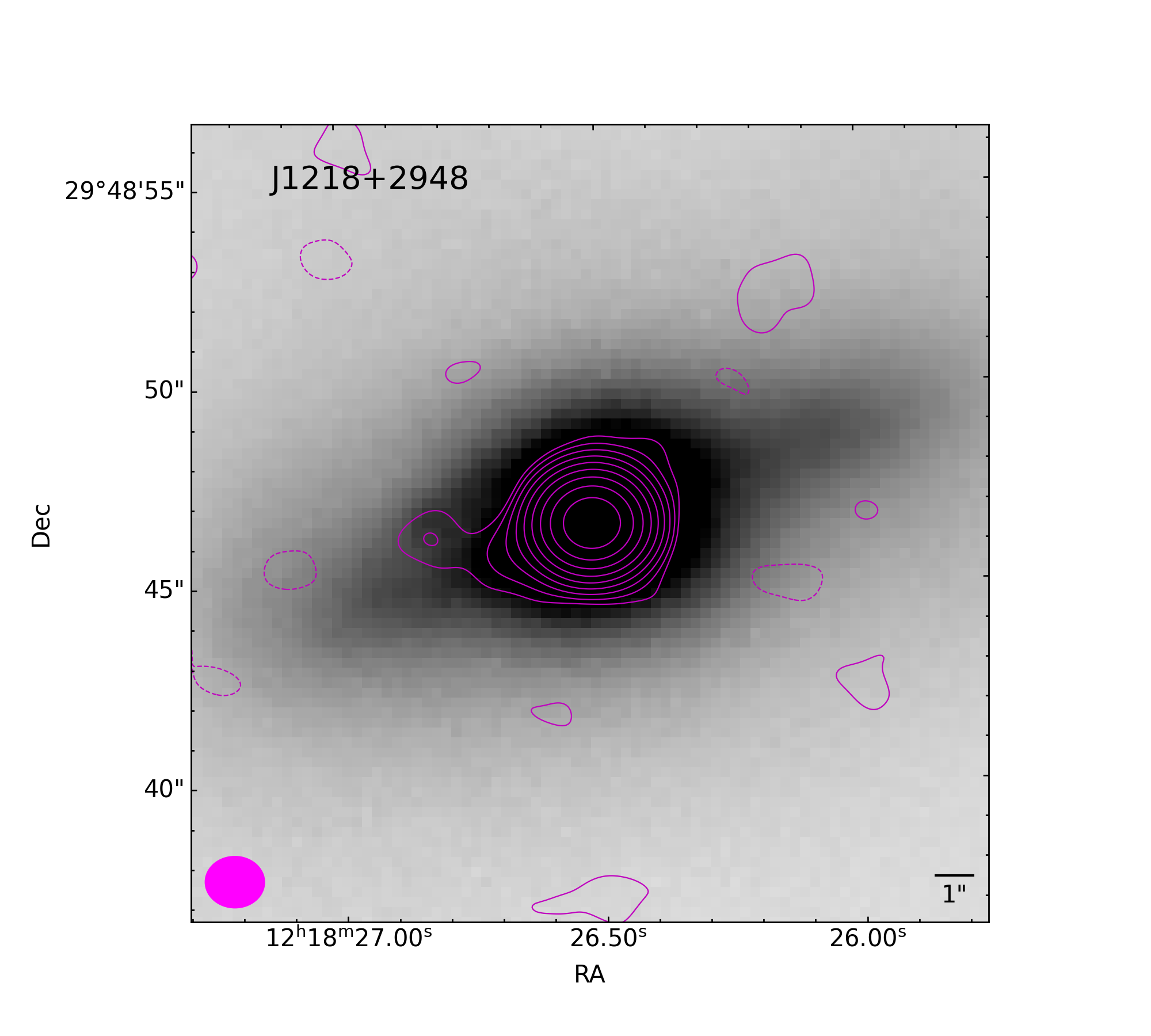}
         \caption{PanSTARRS $i$ band image of the host galaxy overlaid with the 90k$\lambda$ tapered map. Radio map properties as in Fig.~\ref{fig:J1218-90k}}. \label{fig:J1218-host-zoom}
     \end{subfigure}
        \caption{}
        \label{fig:J1218}
\end{figure*}


\begin{figure*}
     \centering
     \begin{subfigure}[b]{0.47\textwidth}
         \centering
         \includegraphics[width=\textwidth]{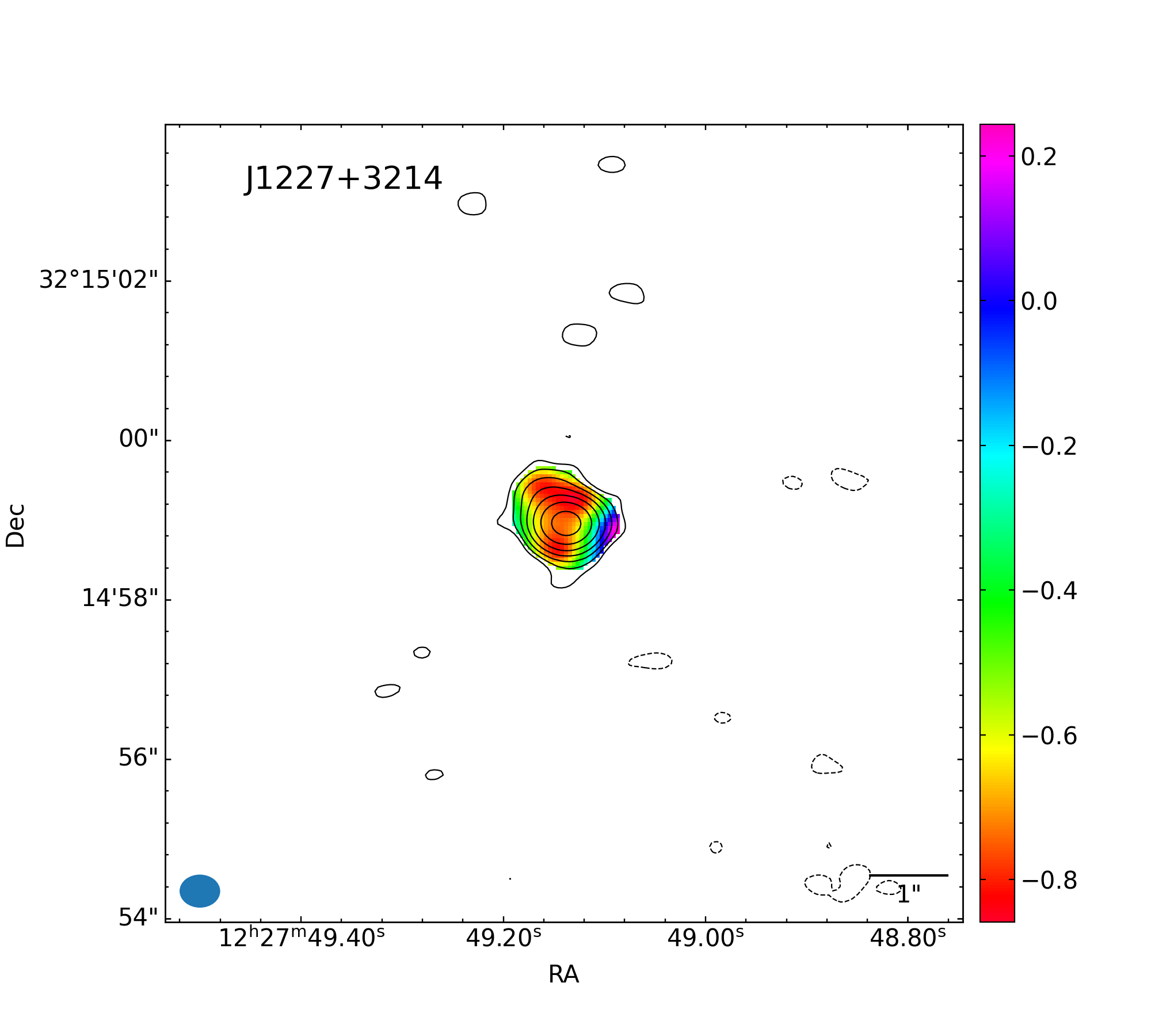}
         \caption{Spectral index map, rms = 11$\mu$Jy beam$^{-1}$, contour levels at -3, 3 $\times$ 2$^n$, $n \in$ [0, 6], beam size 1.23 $\times$ 1.01~kpc. } \label{fig:J1227spind}
     \end{subfigure}
     \hfill
     \begin{subfigure}[b]{0.47\textwidth}
         \centering
         \includegraphics[width=\textwidth]{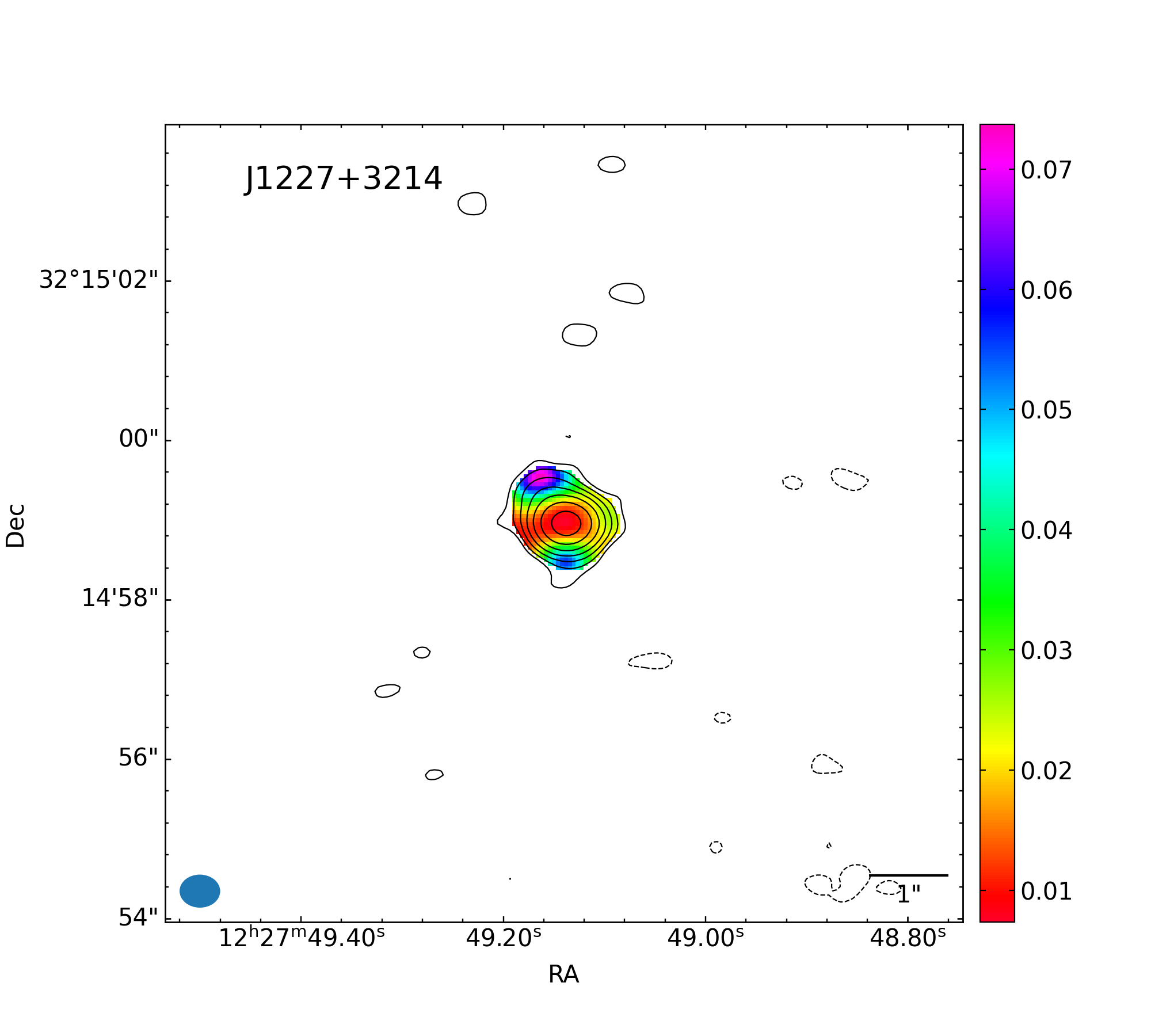}
         \caption{Spectral index error map, rms, contour levels, and beam size as in Fig.~\ref{fig:J1227spind}.} \label{fig:J1227spinderr}
     \end{subfigure}
     \hfill
     \\
     \begin{subfigure}[b]{0.47\textwidth}
         \centering
         \includegraphics[width=\textwidth]{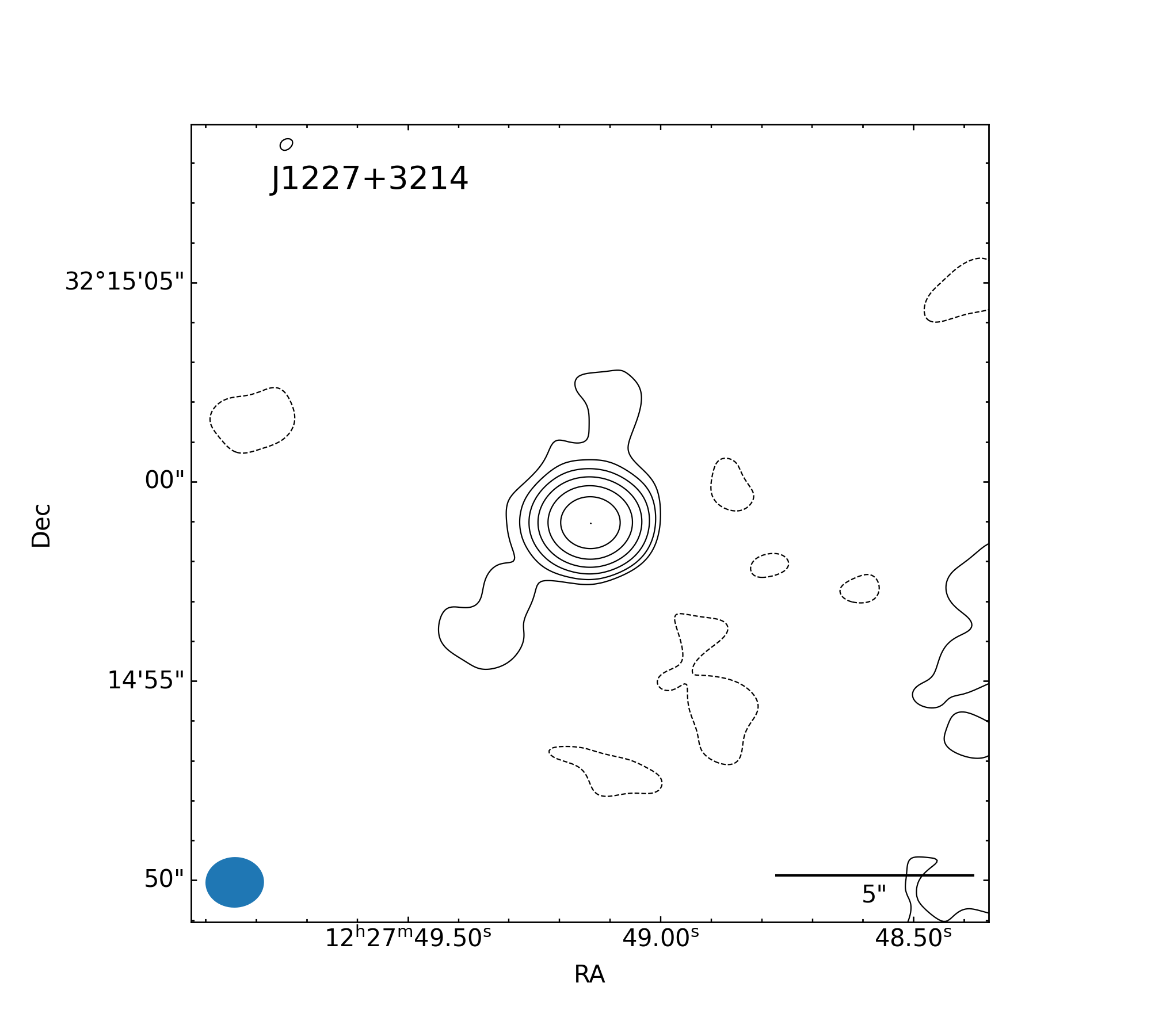}
         \caption{Tapered map with \texttt{uvtaper} = 90k$\lambda$, rms = 17$\mu$Jy beam$^{-1}$, contour levels at -3, 3 $\times$ 2$^n$, $n \in$ [0, 5], beam size 3.54 $\times$ 3.06~kpc.} \label{fig:J1227-90k}
     \end{subfigure}
          \hfill
     \begin{subfigure}[b]{0.47\textwidth}
         \centering
         \includegraphics[width=\textwidth]{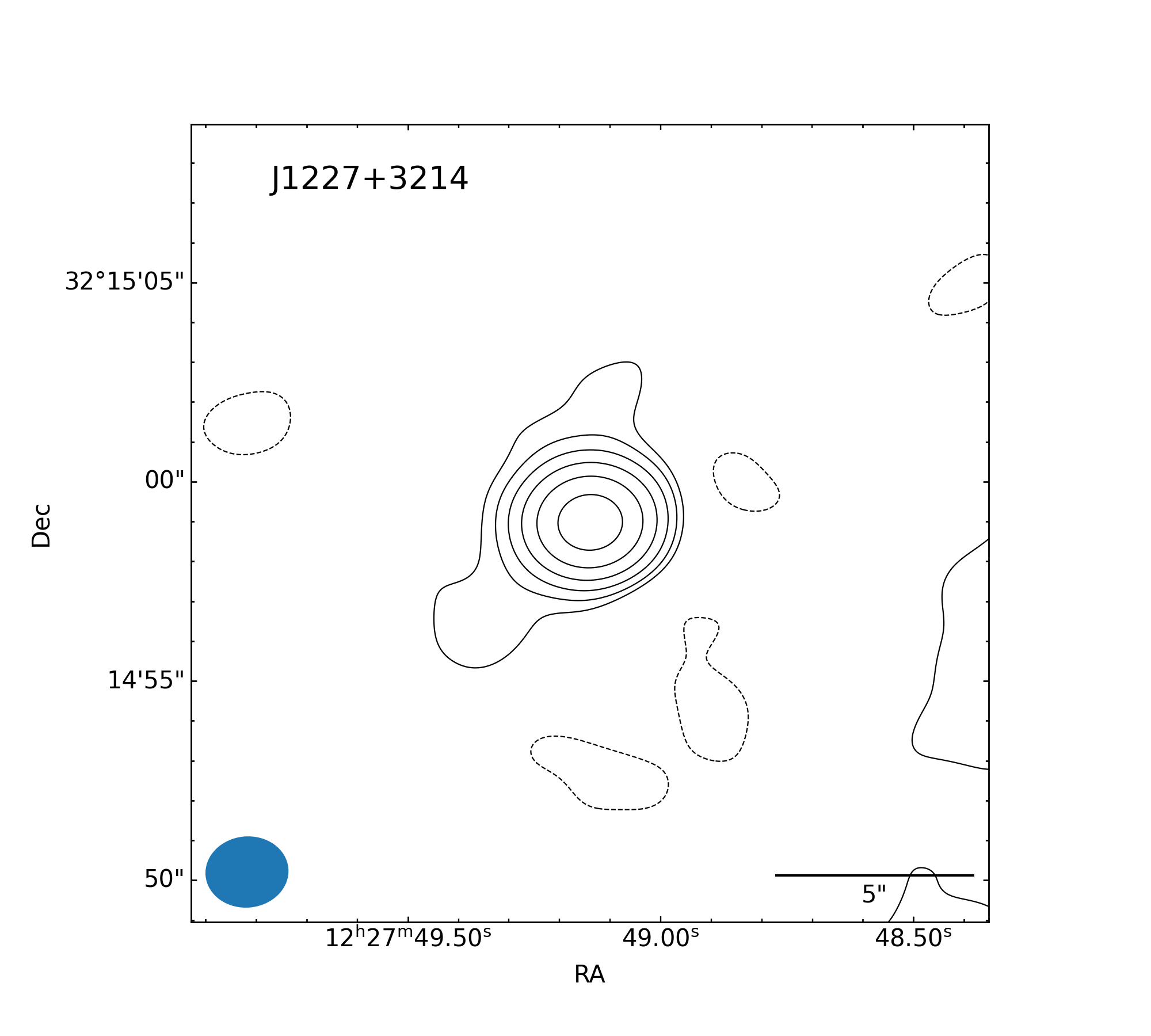}
         \caption{Tapered map with \texttt{uvtaper} = 60k$\lambda$, rms = 23$\mu$Jy beam$^{-1}$, contour levels at -3, 3 $\times$ 2$^n$, $n \in$ [0, 5], beam size 5.03 $\times$ 4.31~kpc.} \label{fig:J1227-60k}
     \end{subfigure}
    \hfill
     \\
     \begin{subfigure}[b]{0.47\textwidth}
         \centering
         \includegraphics[width=\textwidth]{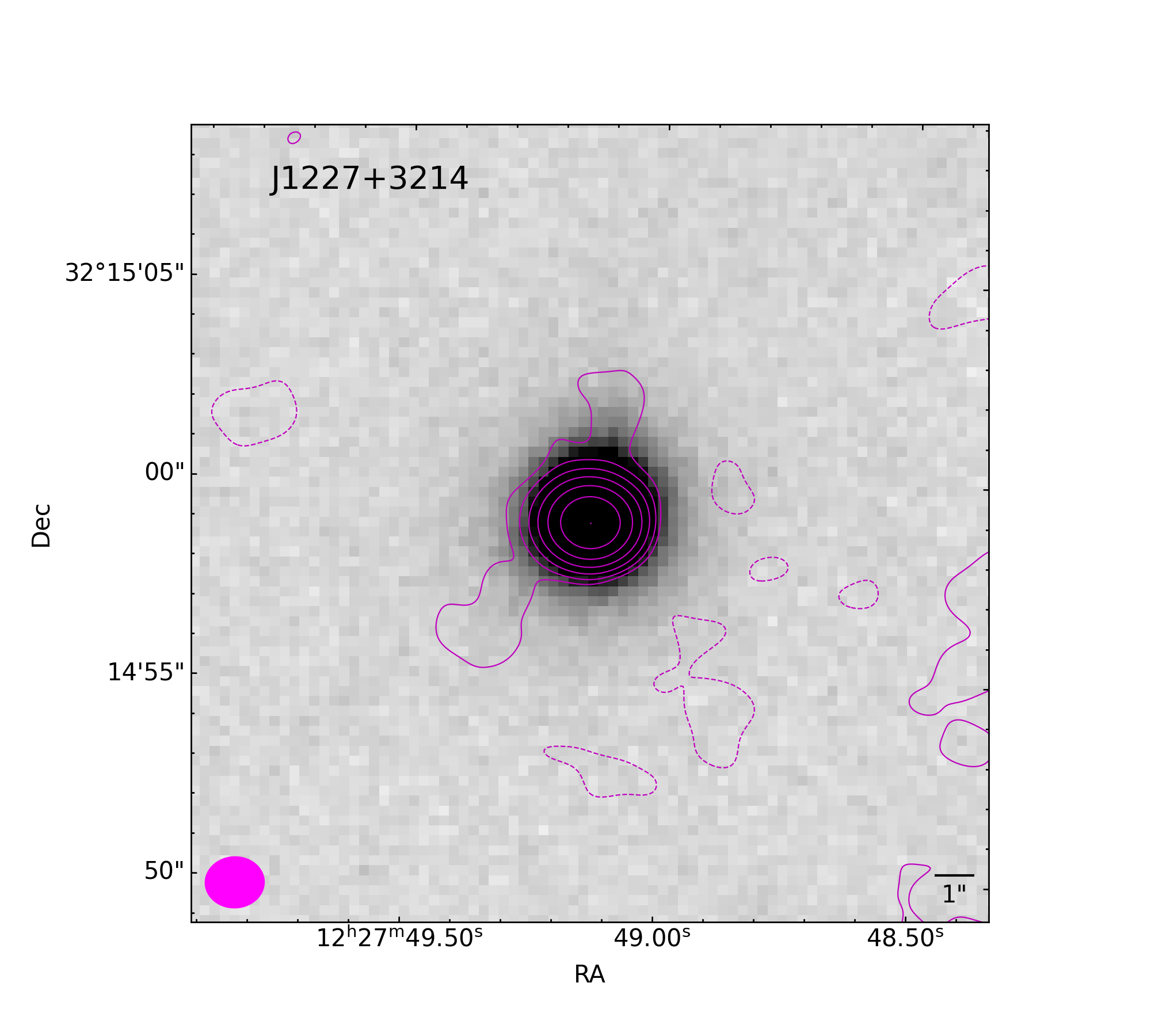}
         \caption{PanSTARRS $i$ band image of the host galaxy overlaid with the 90k$\lambda$ tapered map. Radio map properties as in Fig.~\ref{fig:J1227-90k}}. \label{fig:J1227-host}
     \end{subfigure}
        \caption{}
        \label{fig:J1227}
\end{figure*}


\begin{figure*}
     \centering
     \begin{subfigure}[b]{0.47\textwidth}
         \centering
         \includegraphics[width=\textwidth]{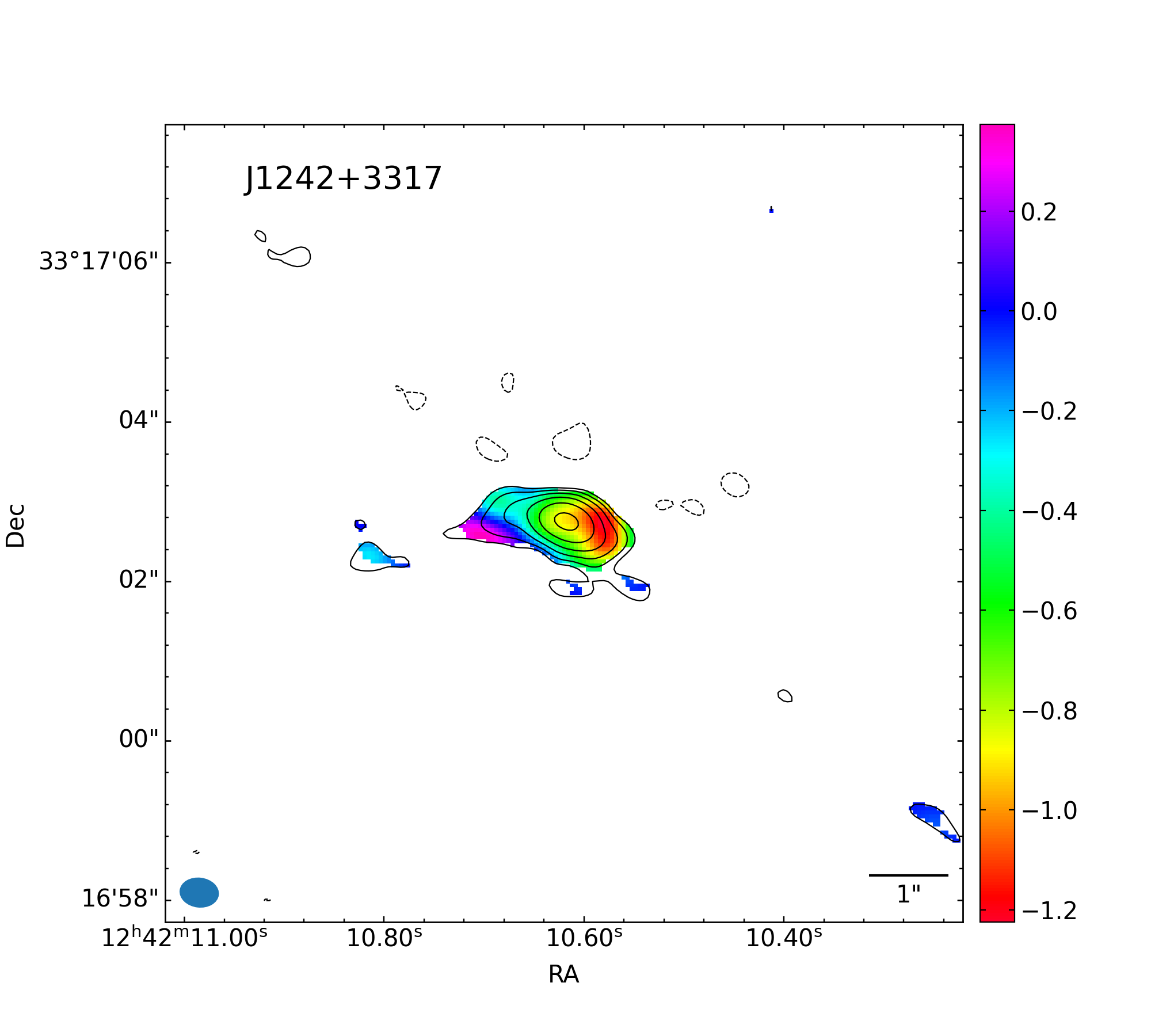}
         \caption{Spectral index map, rms = 12$\mu$Jy beam$^{-1}$, contour levels at -3, 3 $\times$ 2$^n$, $n \in$ [0, 5], beam size 0.43 $\times$ 0.33~kpc. } \label{fig:J1242spind}
     \end{subfigure}
     \hfill
     \begin{subfigure}[b]{0.47\textwidth}
         \centering
         \includegraphics[width=\textwidth]{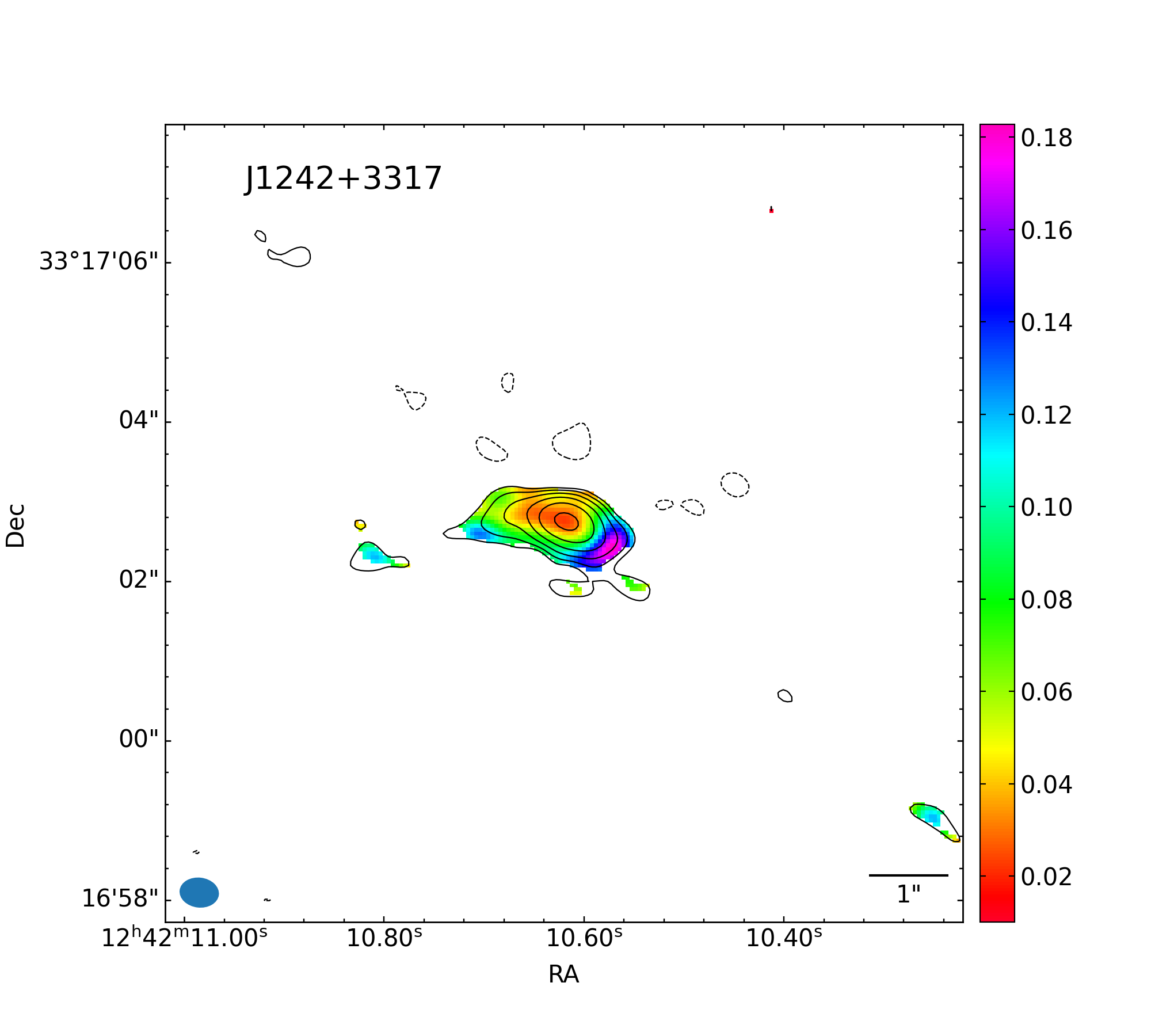}
         \caption{Spectral index error map, rms, contour levels, and beam size as in Fig.~\ref{fig:J1242spind}.} \label{fig:J1242spinderr}
     \end{subfigure}
     \hfill
     \\
     \begin{subfigure}[b]{0.47\textwidth}
         \centering
         \includegraphics[width=\textwidth]{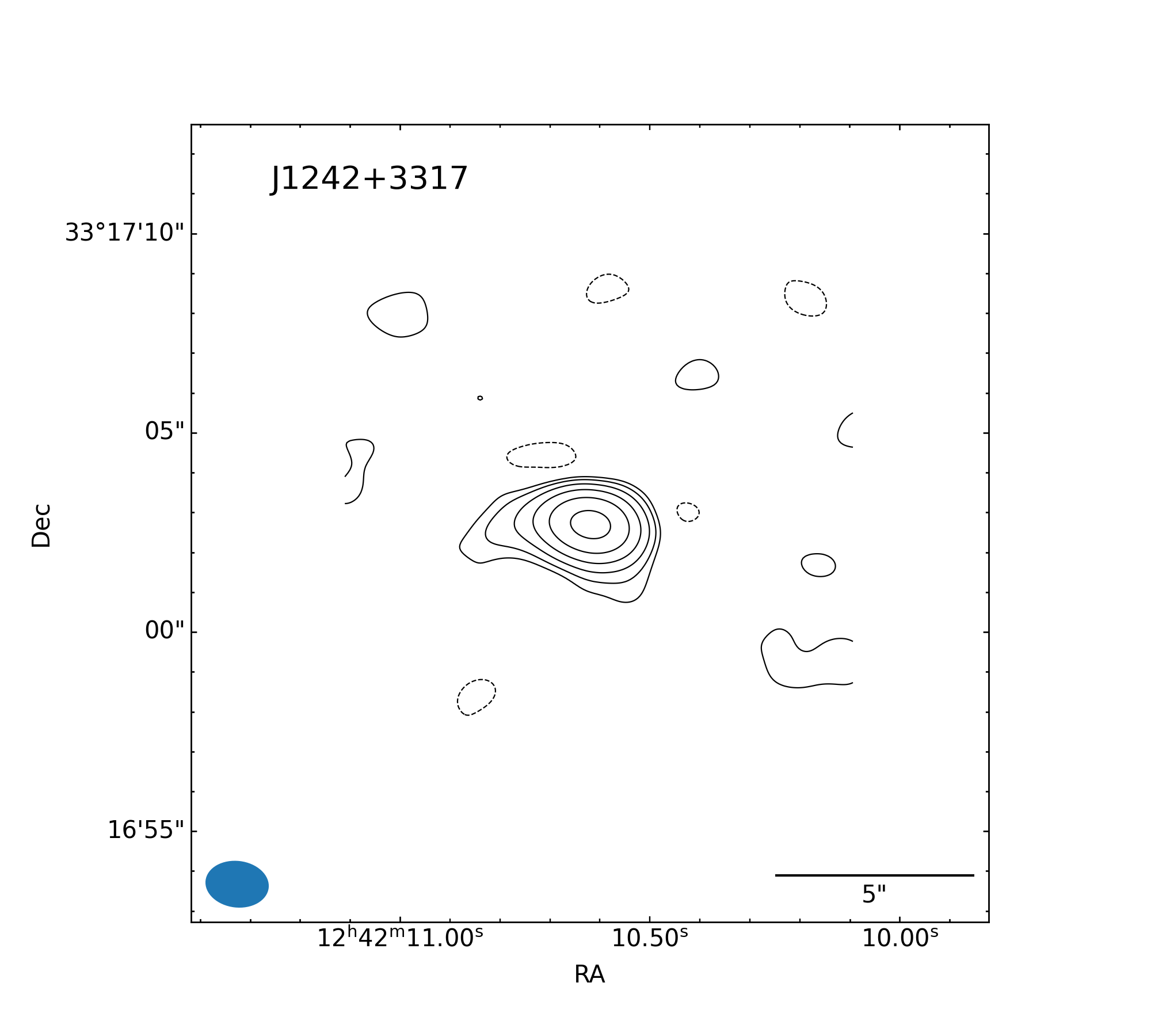}
         \caption{Tapered map with \texttt{uvtaper} = 90k$\lambda$, rms = 16$\mu$Jy beam$^{-1}$, contour levels at -3, 3 $\times$ 2$^n$, $n \in$ [0, 5], beam size 1.38 $\times$ 1.01~kpc.} \label{fig:J1242-90k}
     \end{subfigure}
          \hfill
     \begin{subfigure}[b]{0.47\textwidth}
         \centering
         \includegraphics[width=\textwidth]{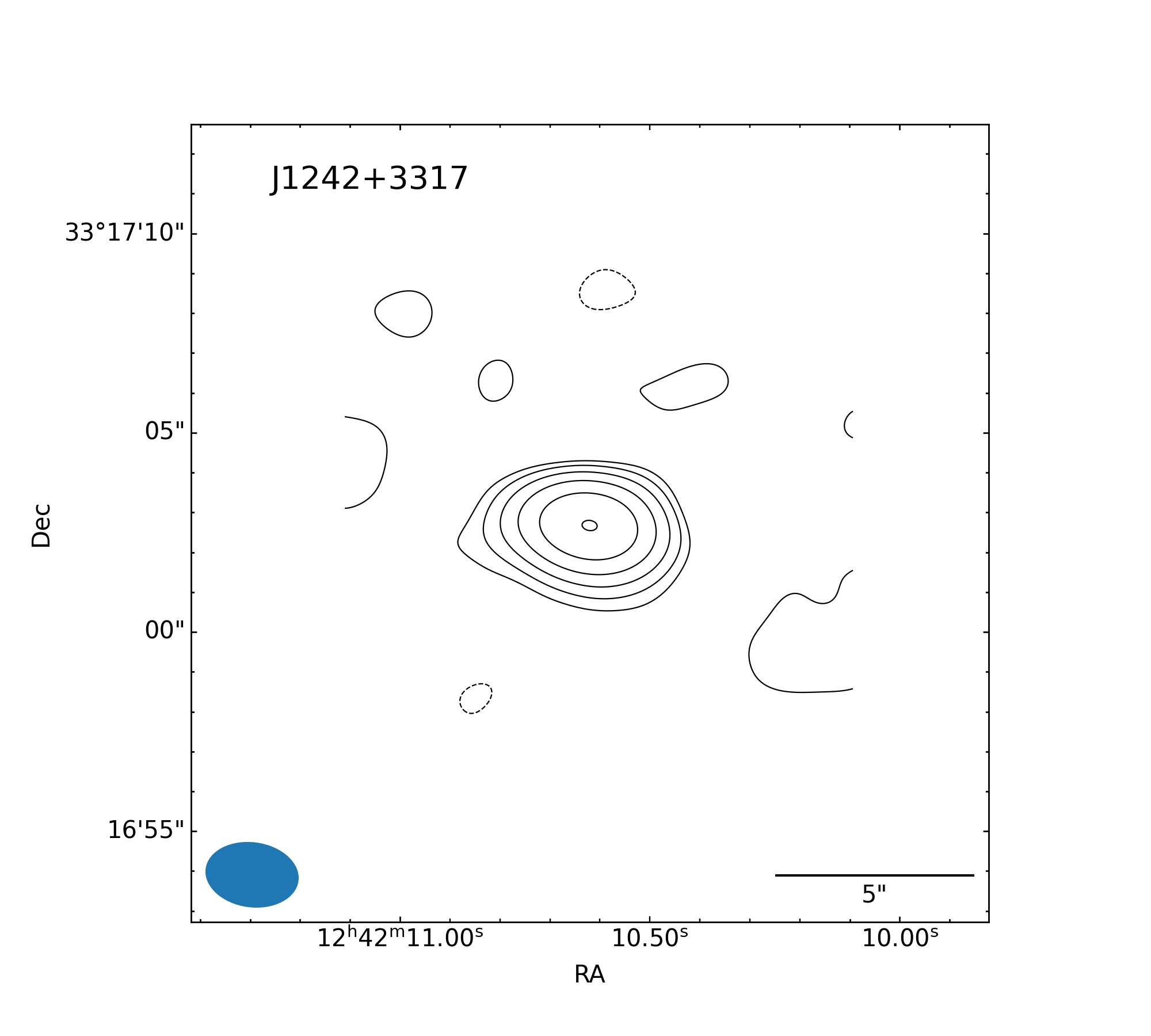}
         \caption{Tapered map with \texttt{uvtaper} = 60k$\lambda$, rms = 21$\mu$Jy beam$^{-1}$, contour levels at -3, 3 $\times$ 2$^n$, $n \in$ [0, 5], beam size 2.04 $\times$ 1.41~kpc.} \label{fig:J1242-60k}
     \end{subfigure}
          \hfill
     \\
     \begin{subfigure}[b]{0.47\textwidth}
         \centering
         \includegraphics[width=\textwidth]{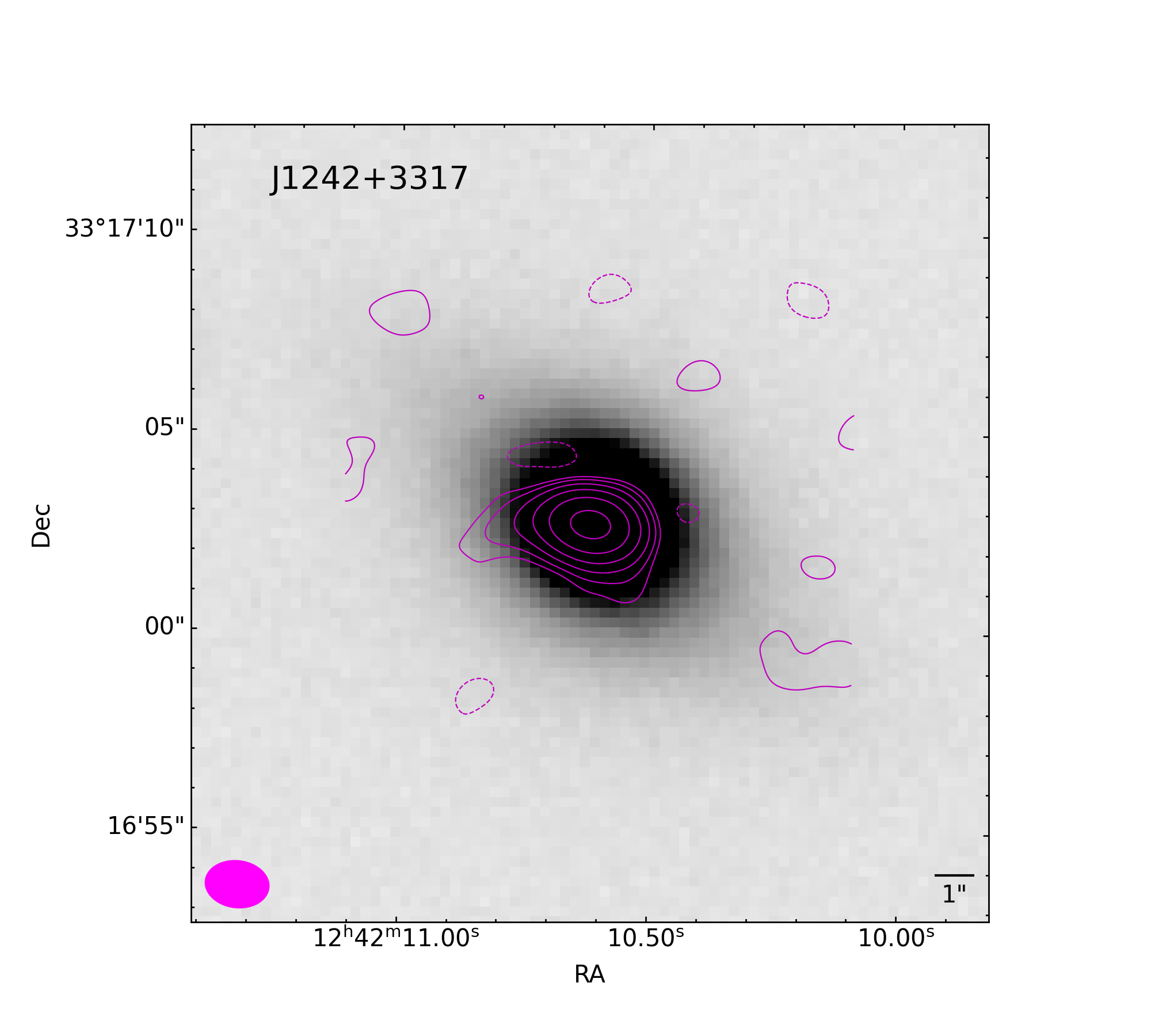}
         \caption{PanSTARRS $i$ band image of the host galaxy overlaid with the 90k$\lambda$ tapered map. Radio map properties as in Fig.~\ref{fig:J1242-90k}}. \label{fig:J1242-host}
     \end{subfigure}
        \caption{}
        \label{fig:J1242}
\end{figure*}


\begin{figure*}
     \centering
     \begin{subfigure}[b]{0.47\textwidth}
         \centering
         \includegraphics[width=\textwidth]{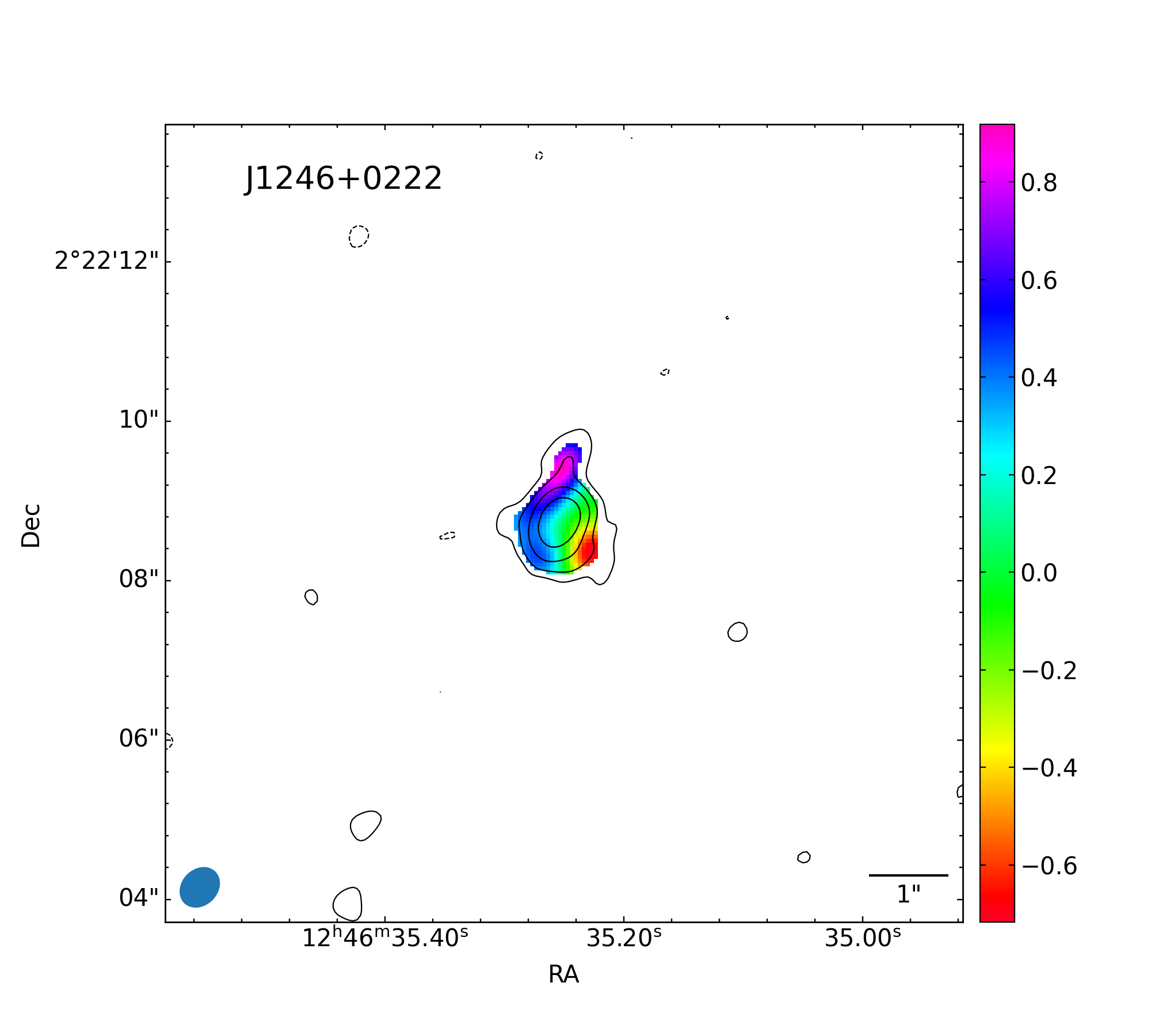}
         \caption{Spectral index map, rms = 10$\mu$Jy beam$^{-1}$, contour levels at -3, 3 $\times$ 2$^n$, $n \in$ [0, 3], beam size 0.53 $\times$ 0.42~kpc. } \label{fig:J1246spind}
     \end{subfigure}
     \hfill
     \begin{subfigure}[b]{0.47\textwidth}
         \centering
         \includegraphics[width=\textwidth]{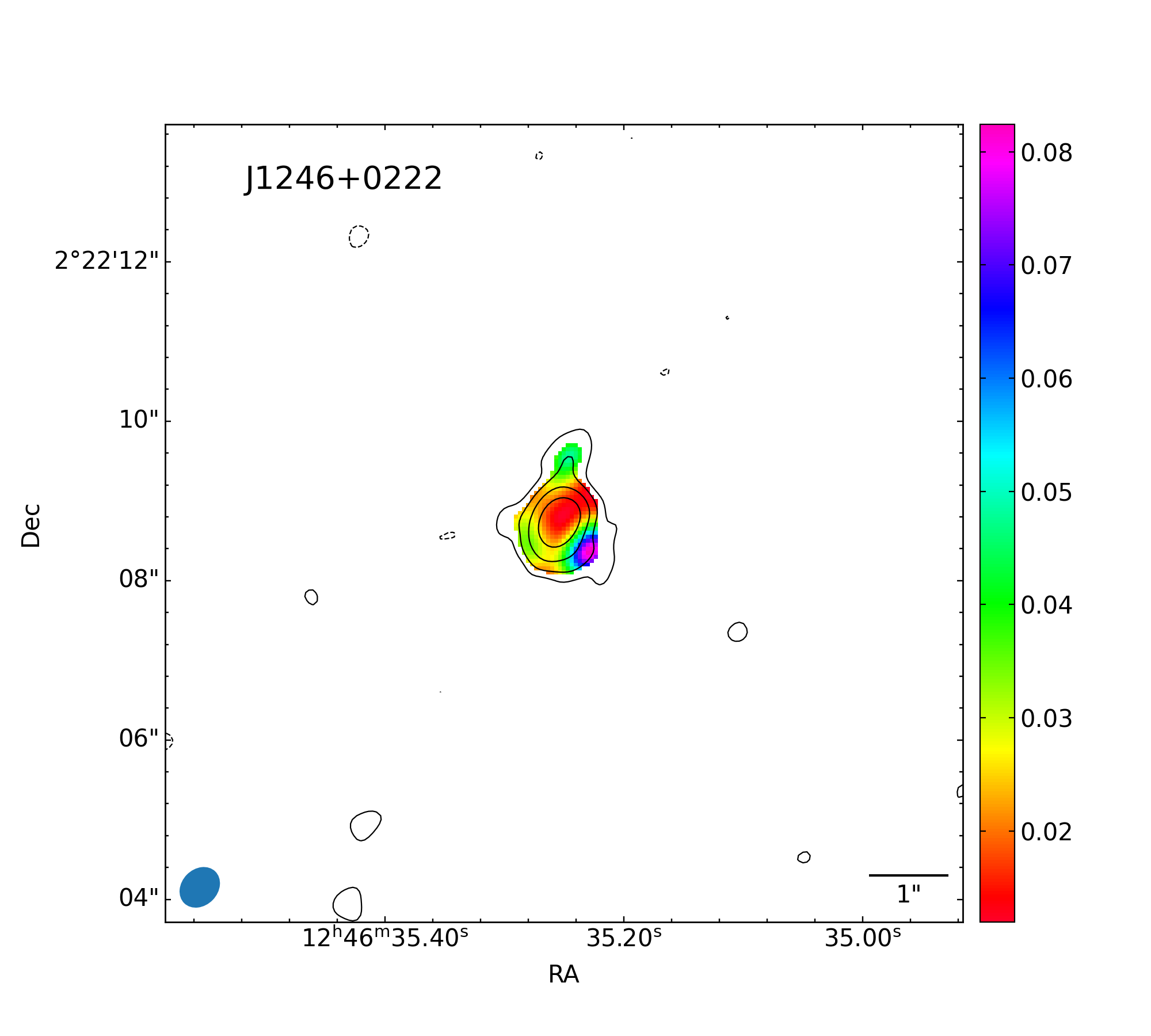}
         \caption{Spectral index error map, rms, contour levels, and beam size as in Fig.~\ref{fig:J1246spind}.} \label{fig:J1246spinderr}
     \end{subfigure}
     \hfill
     \\
     \begin{subfigure}[b]{0.47\textwidth}
         \centering
         \includegraphics[width=\textwidth]{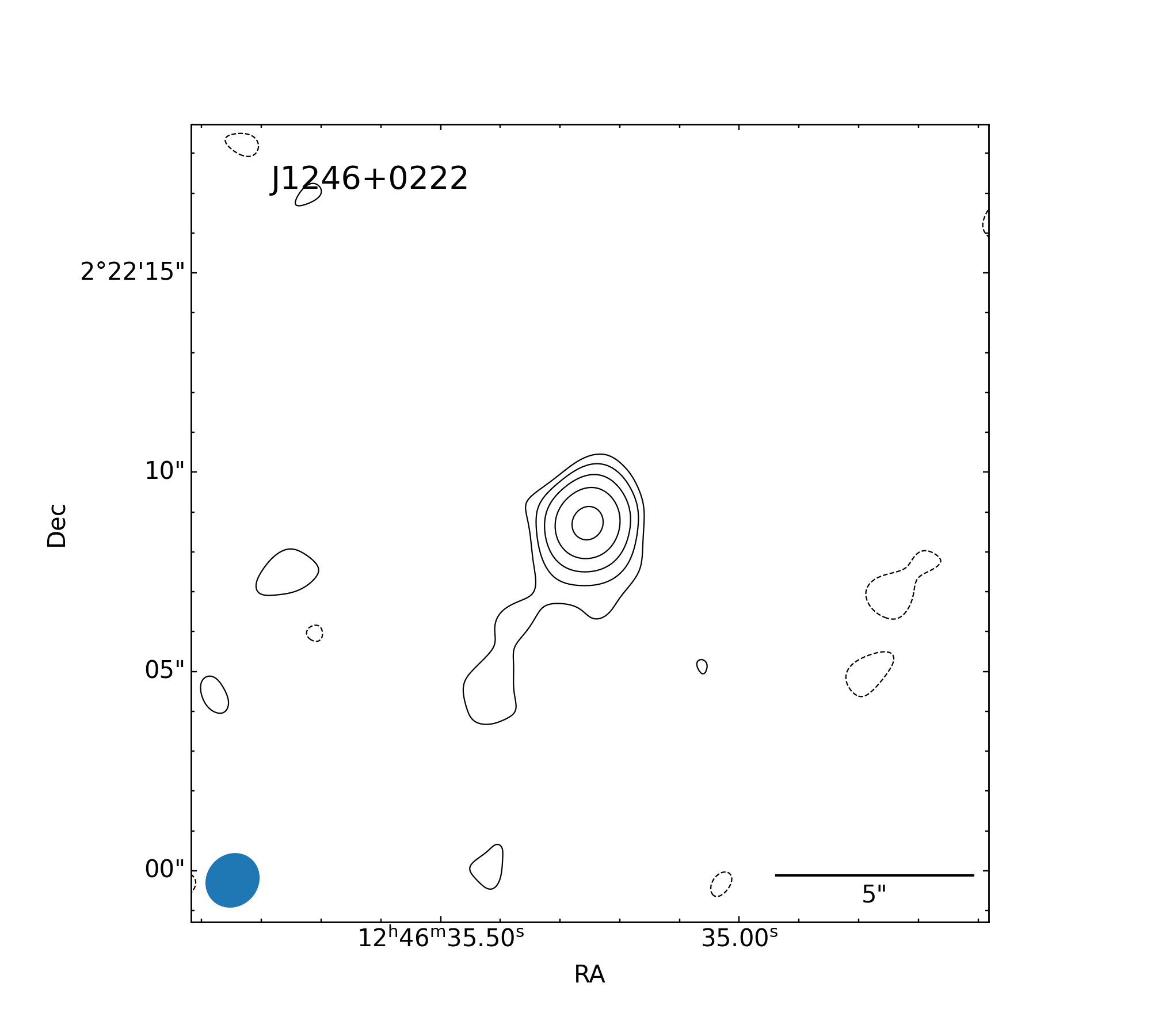}
         \caption{Tapered map with \texttt{uvtaper} = 90k$\lambda$, rms = 11$\mu$Jy beam$^{-1}$, contour levels at -3, 3 $\times$ 2$^n$, $n \in$ [0, 4], beam size 1.35 $\times$ 1.22~kpc.} \label{fig:J1246-90k}
     \end{subfigure}
          \hfill
     \begin{subfigure}[b]{0.47\textwidth}
         \centering
         \includegraphics[width=\textwidth]{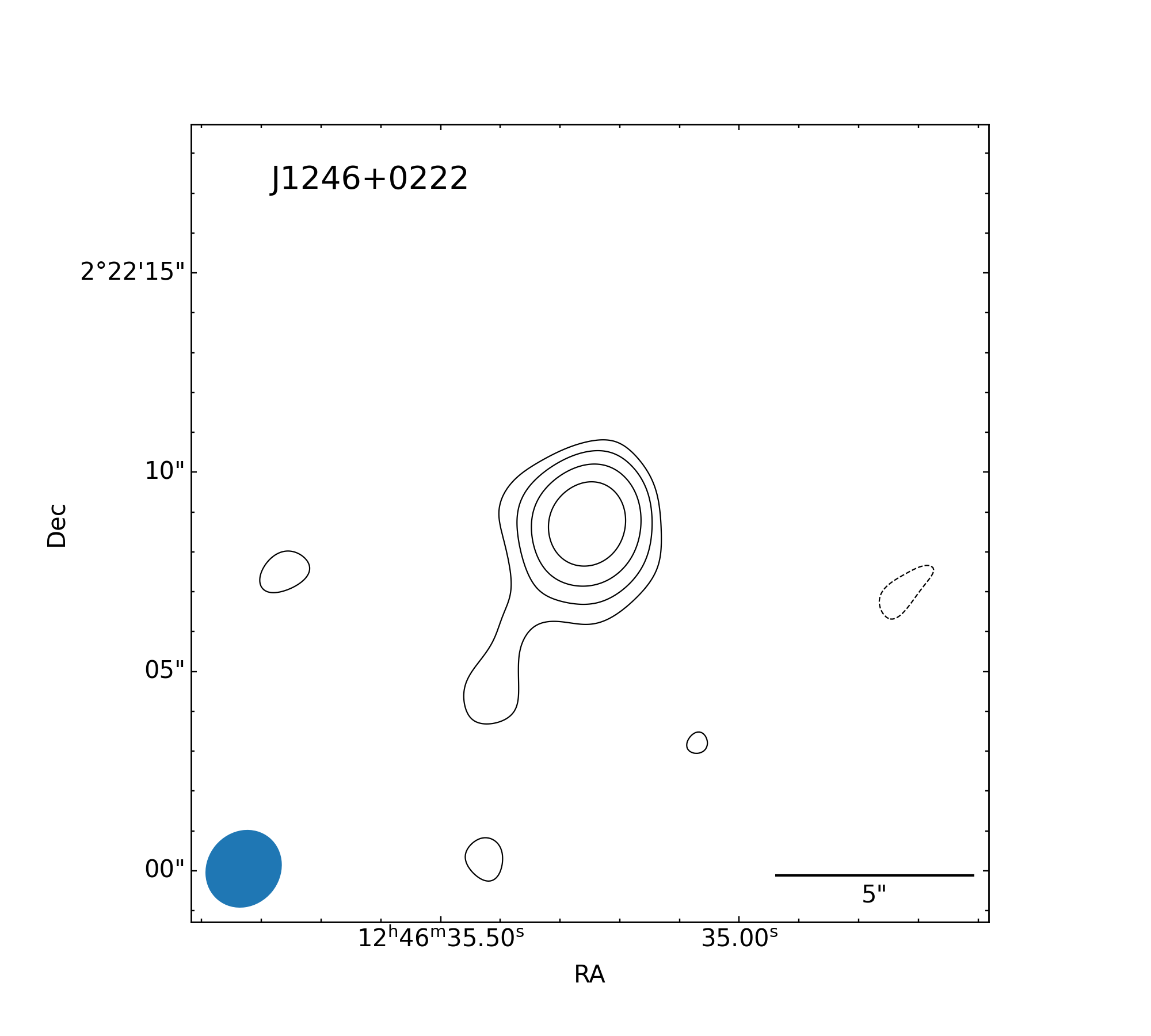}
         \caption{Tapered map with \texttt{uvtaper} = 60k$\lambda$, rms = 15$\mu$Jy beam$^{-1}$, contour levels at -3, 3 $\times$ 2$^n$, $n \in$ [0, 3], beam size 1.91 $\times$ 1.72~kpc.} \label{fig:J1246-60k}
     \end{subfigure}
          \hfill
     \\
     \begin{subfigure}[b]{0.47\textwidth}
         \centering
         \includegraphics[width=\textwidth]{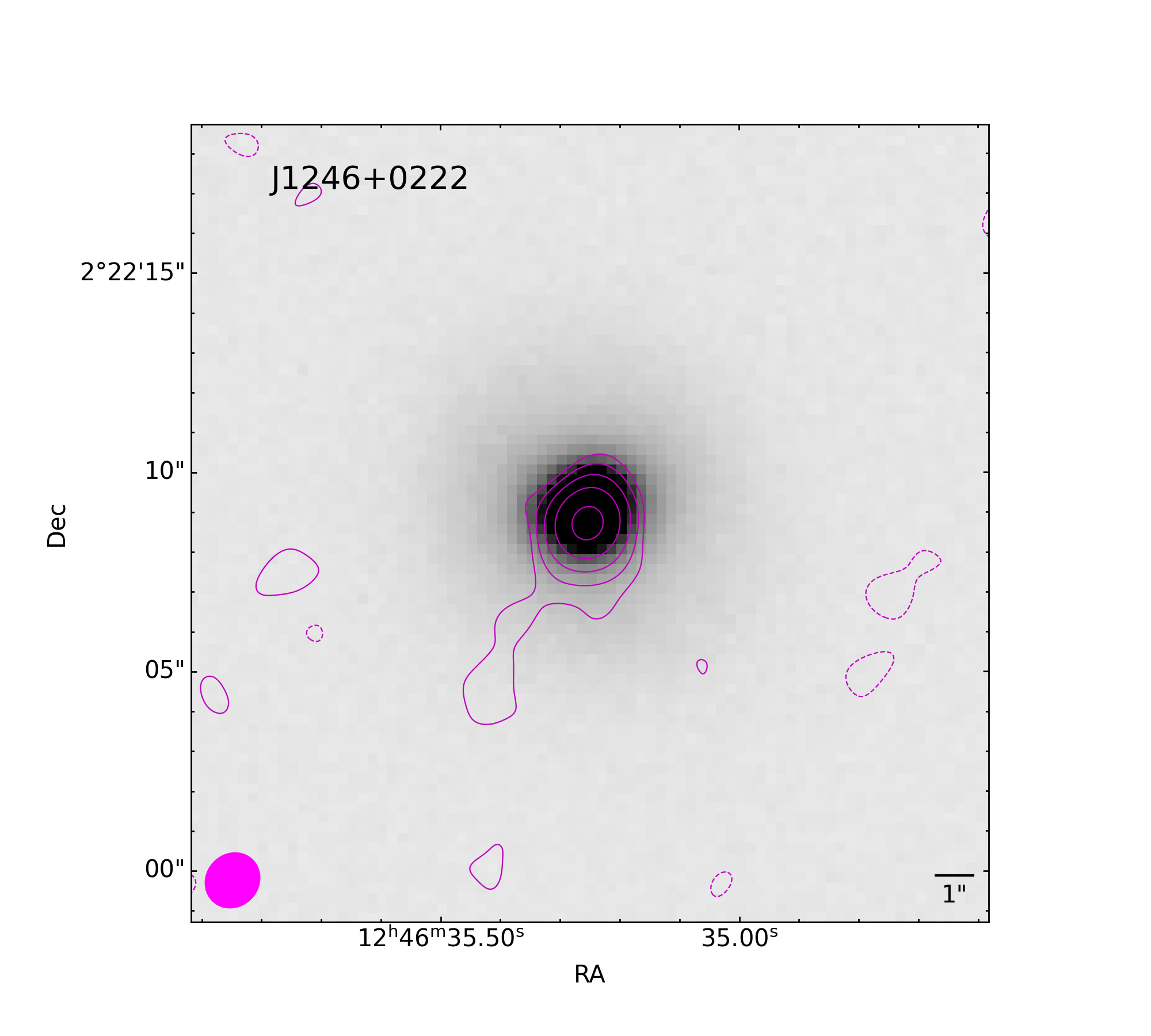}
         \caption{PanSTARRS $i$ band image of the host galaxy overlaid with the 90k$\lambda$ tapered map. Radio map properties as in Fig.~\ref{fig:J1246-90k}}. \label{fig:J1246-host}
     \end{subfigure}
        \caption{}
        \label{fig:J1246}
\end{figure*}


\begin{figure*}
     \begin{subfigure}[b]{0.47\textwidth}
         \centering
         \includegraphics[width=\textwidth]{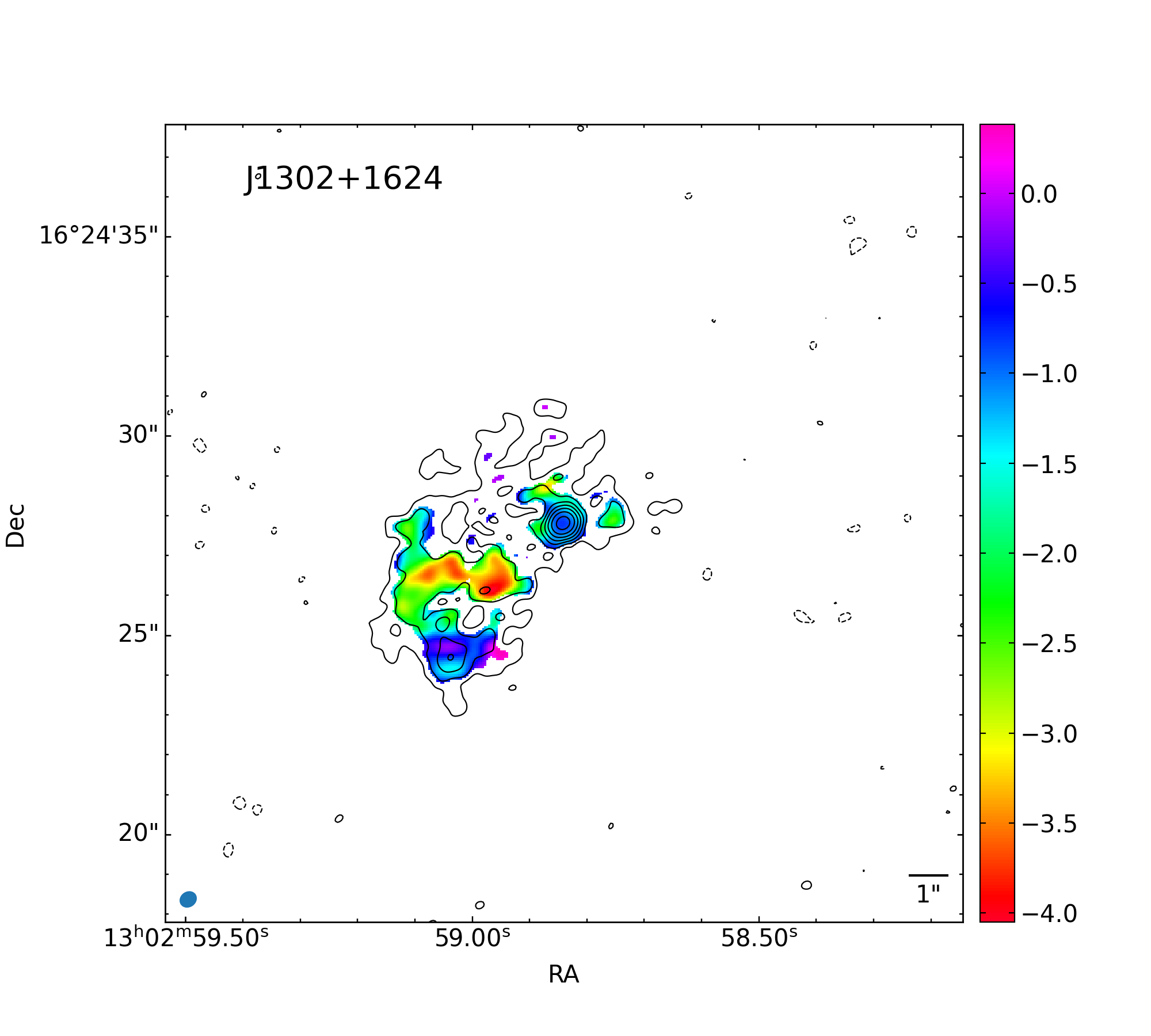}
         \caption{Spectral index map, rms = 12$\mu$Jy beam$^{-1}$, contour levels at -3, 3 $\times$ 2$^n$, $n \in$ [0, 6], beam size 0.58 $\times$ 0.50~kpc. Non-common uvrange.} \label{fig:J1302spindnoncommon}
     \end{subfigure}
     \hfill
     \begin{subfigure}[b]{0.47\textwidth}
         \centering
         \includegraphics[width=\textwidth]{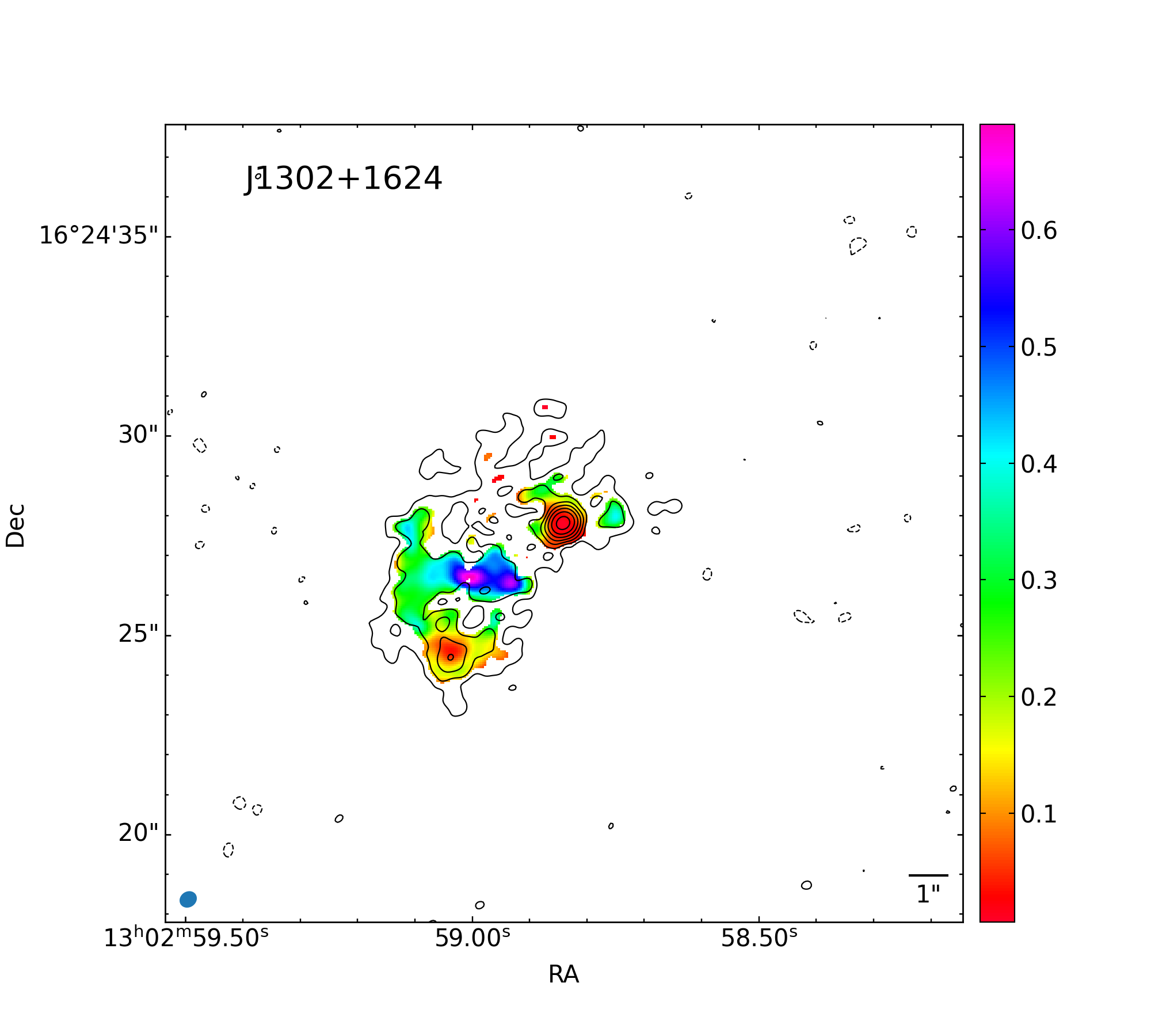}
         \caption{Spectral index error map, rms, contour levels, and beam size as in Fig.~\ref{fig:J1302spindnoncommon}.} \label{fig:J1302spinderrnoncommon}
     \end{subfigure}
     \hfill
         \\
     \begin{subfigure}[b]{0.47\textwidth}
         \centering
         \includegraphics[width=\textwidth]{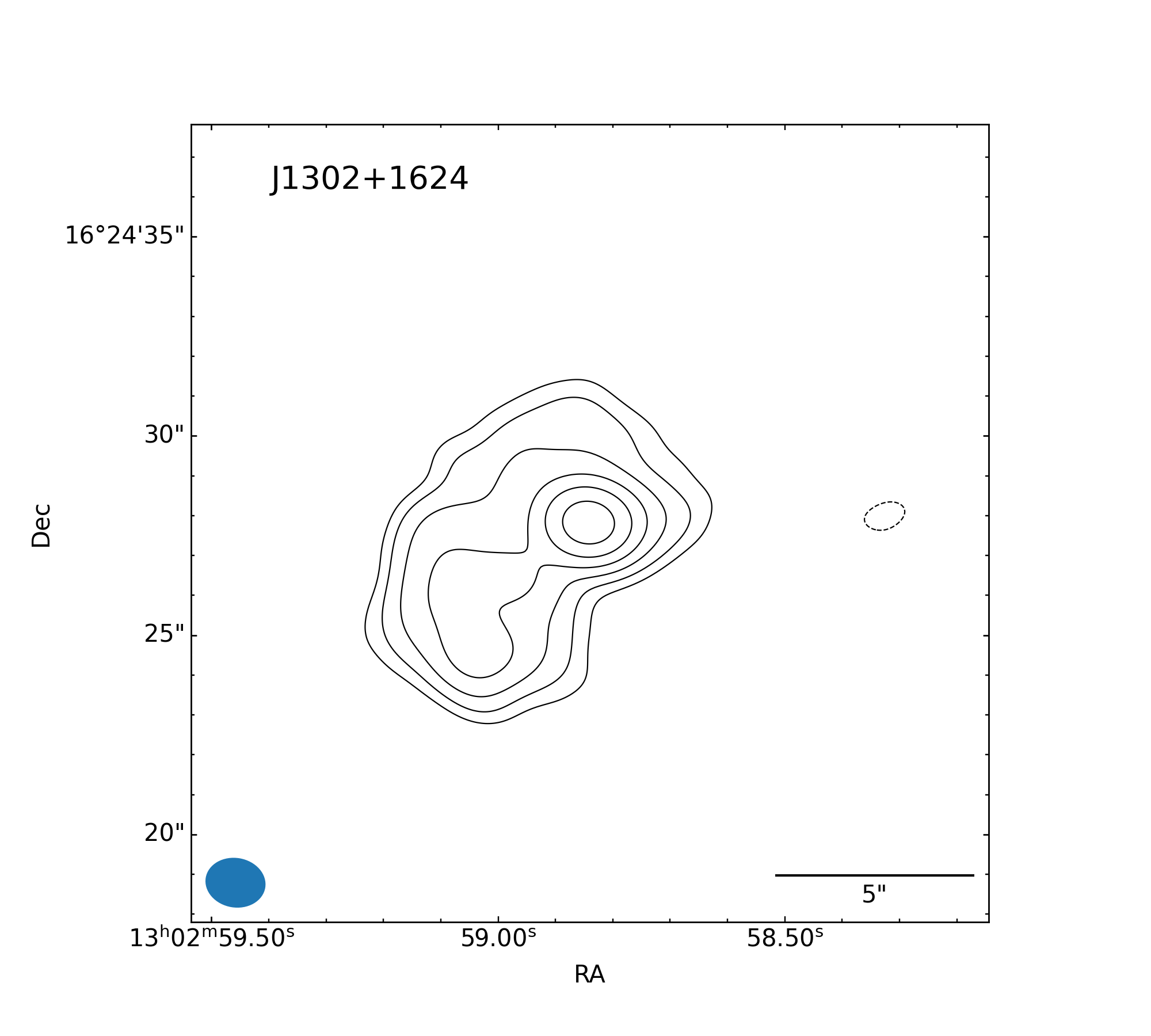}
         \caption{Tapered map with \texttt{uvtaper} = 90k$\lambda$, rms = 28$\mu$Jy beam$^{-1}$, contour levels at -3, 3 $\times$ 2$^n$, $n \in$ [0, 5], beam size 1.94 $\times$ 1.59~kpc.} \label{fig:J1302-90k}
     \end{subfigure}
          \hfill
     \begin{subfigure}[b]{0.47\textwidth}
         \centering
         \includegraphics[width=\textwidth]{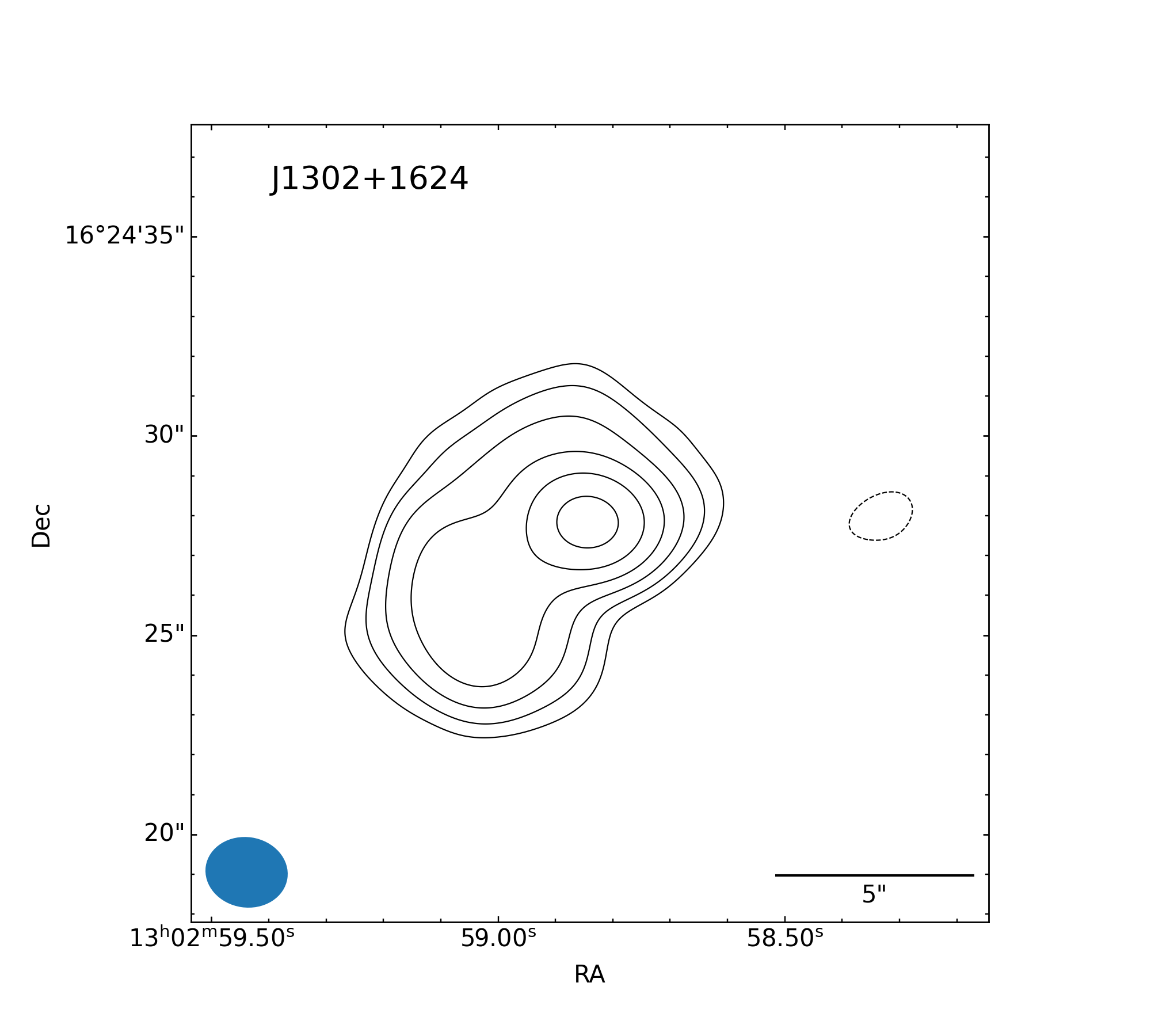}
         \caption{Tapered map with \texttt{uvtaper} = 60k$\lambda$, rms = 35$\mu$Jy beam$^{-1}$, contour levels at -3, 3 $\times$ 2$^n$, $n \in$ [0, 5], beam size 2.66 $\times$ 2.27~kpc.} \label{fig:J1302-60k}
     \end{subfigure}
          \hfill
          \\
     \begin{subfigure}[b]{0.47\textwidth}
         \centering
         \includegraphics[width=\textwidth]{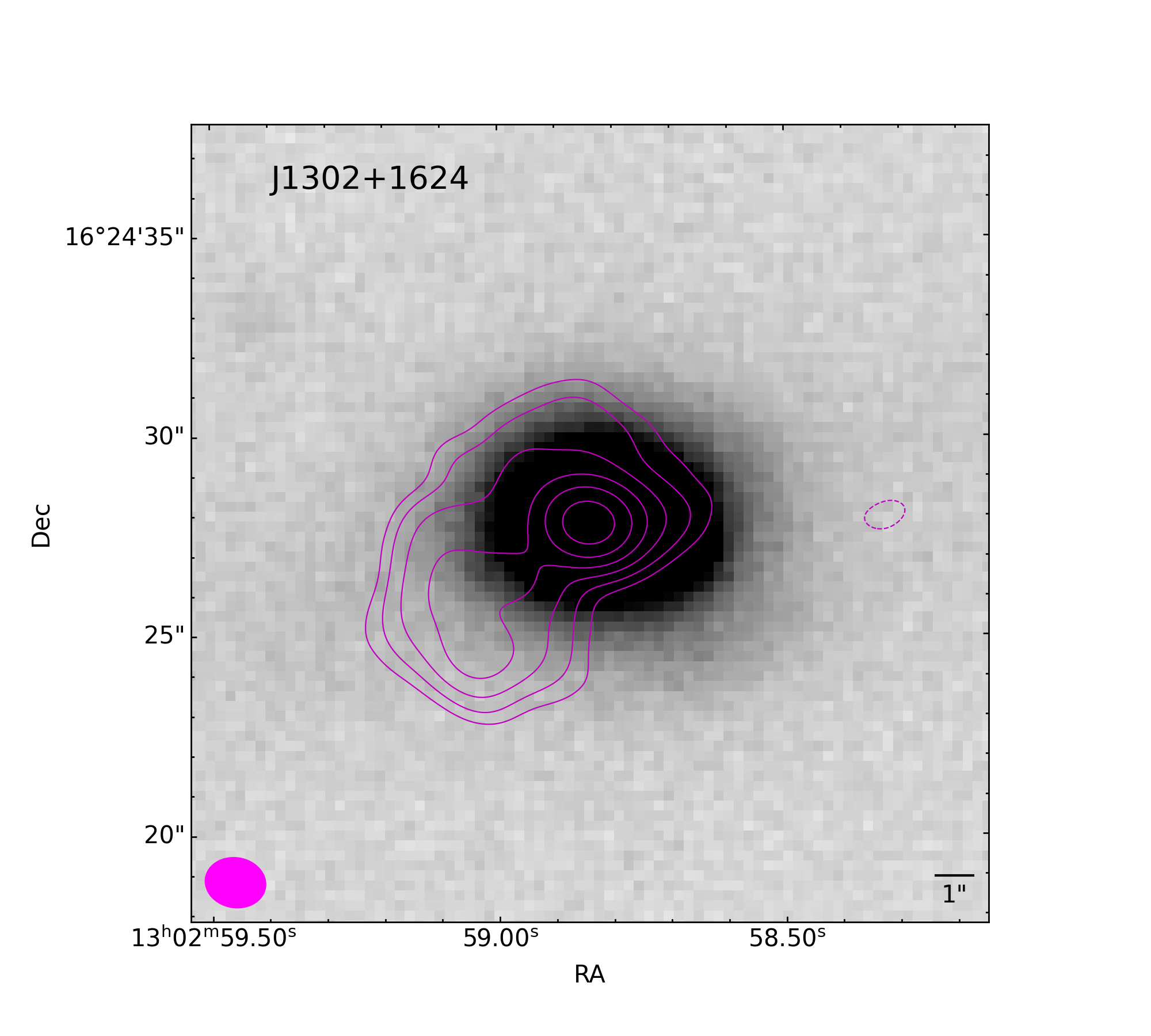}
         \caption{PanSTARRS $i$ band image of the host galaxy overlaid with the 90k$\lambda$ tapered map. Radio map properties as in Fig.~\ref{fig:J1302-90k}}. \label{fig:J1302-host}
     \end{subfigure}
      \caption{}
        \label{fig:J1302}
\end{figure*}

\begin{figure*}
     \centering
     \begin{subfigure}[b]{0.47\textwidth}
         \centering
         \includegraphics[width=\textwidth]{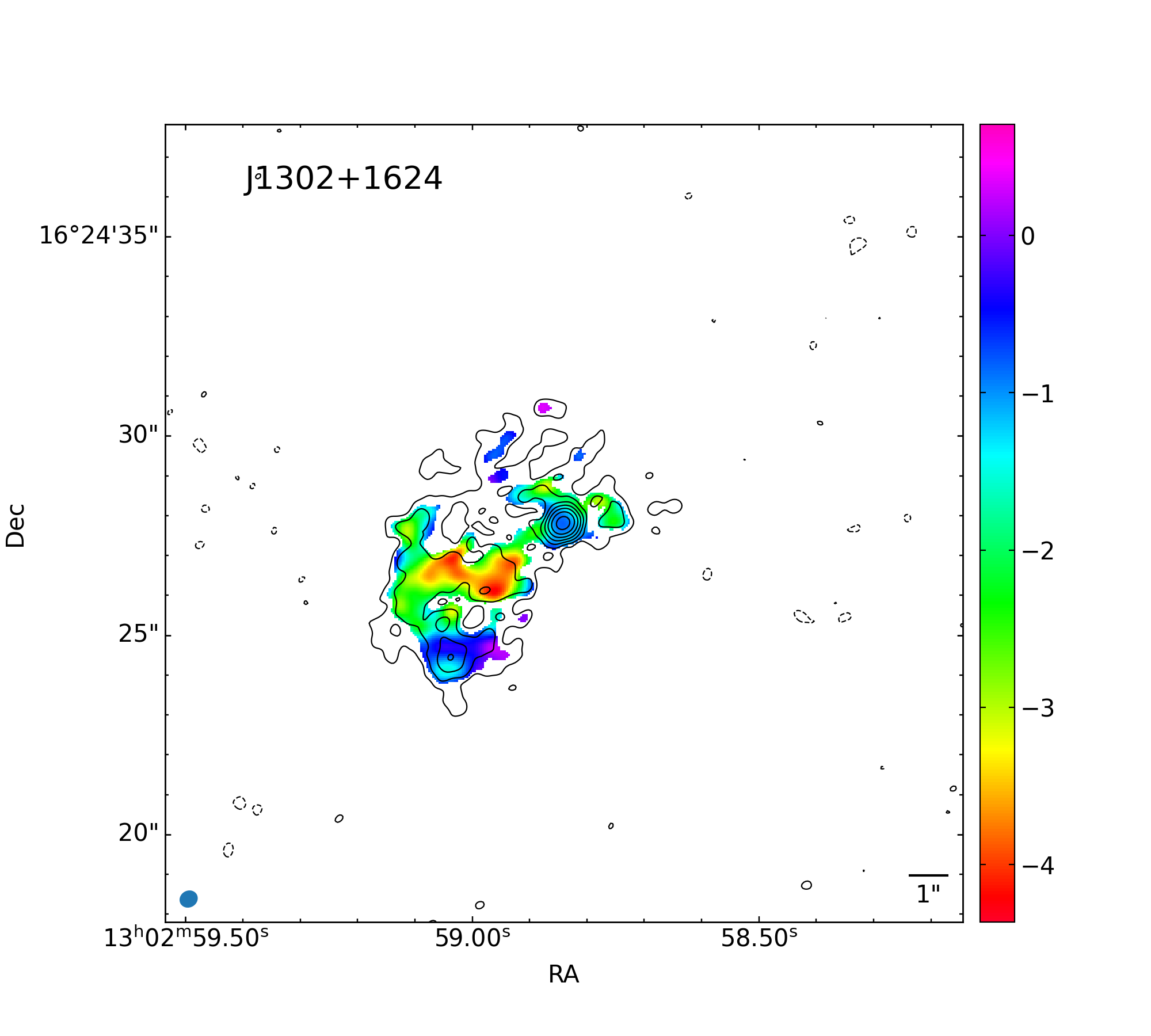}
         \caption{Spectral index map, rms = 12$\mu$Jy beam$^{-1}$, contour levels at -3, 3 $\times$ 2$^n$, $n \in$ [0, 6], beam size 0.60 $\times$ 0.54~kpc. Common uvrange. } \label{fig:J1302spindcommon} 0.47 x 0.42
     \end{subfigure}
     \hfill
     \begin{subfigure}[b]{0.47\textwidth}
         \centering
         \includegraphics[width=\textwidth]{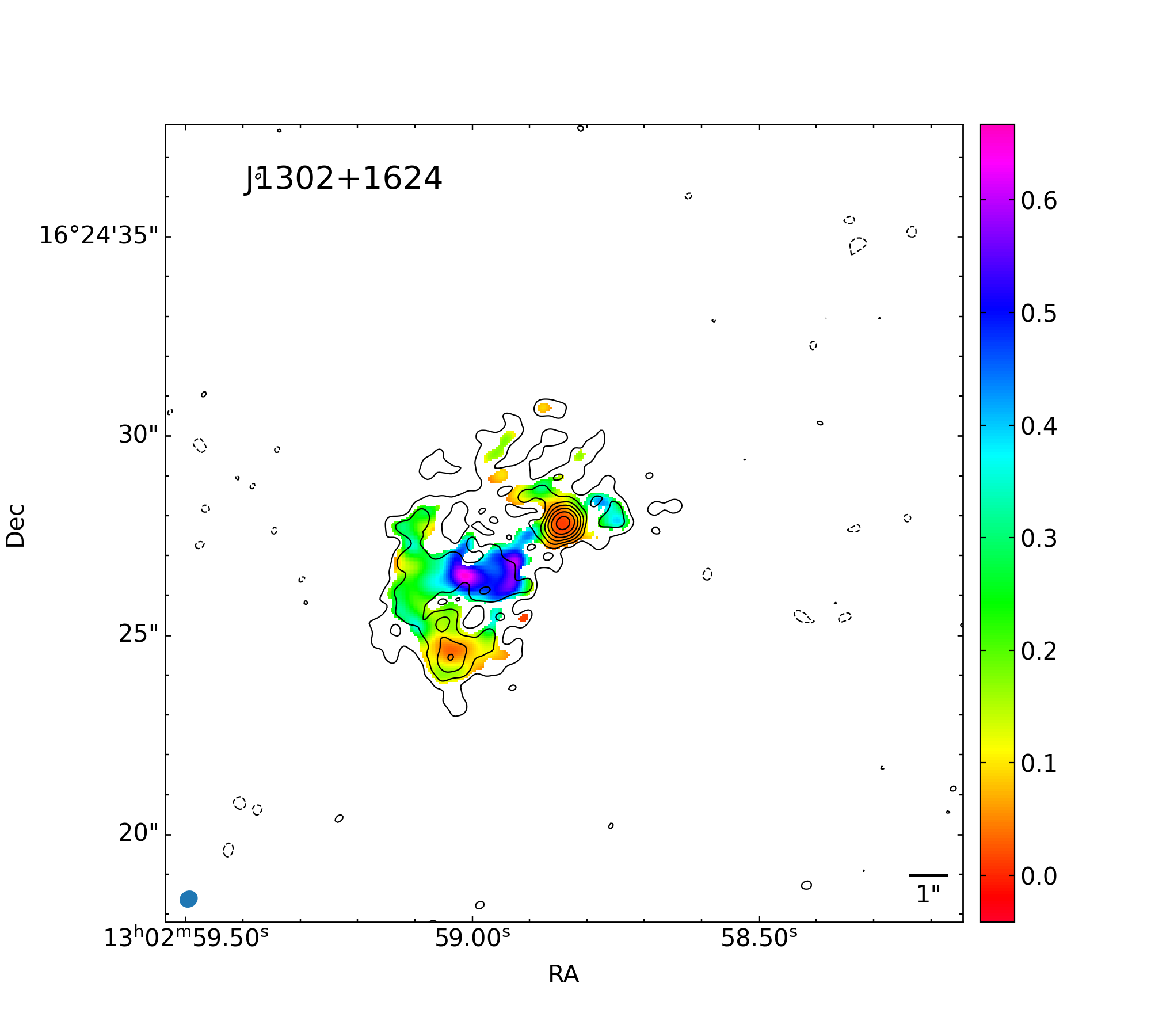}
         \caption{Spectral index error map, rms, contour levels, and beam size as in Fig.~\ref{fig:J1302spindcommon}.} \label{fig:J1302spinderrcommon}
     \end{subfigure}
      \caption{}
        \label{fig:J1302-common}
\end{figure*}


\begin{figure*}
     \centering
     \begin{subfigure}[b]{0.47\textwidth}
         \centering
         \includegraphics[width=\textwidth]{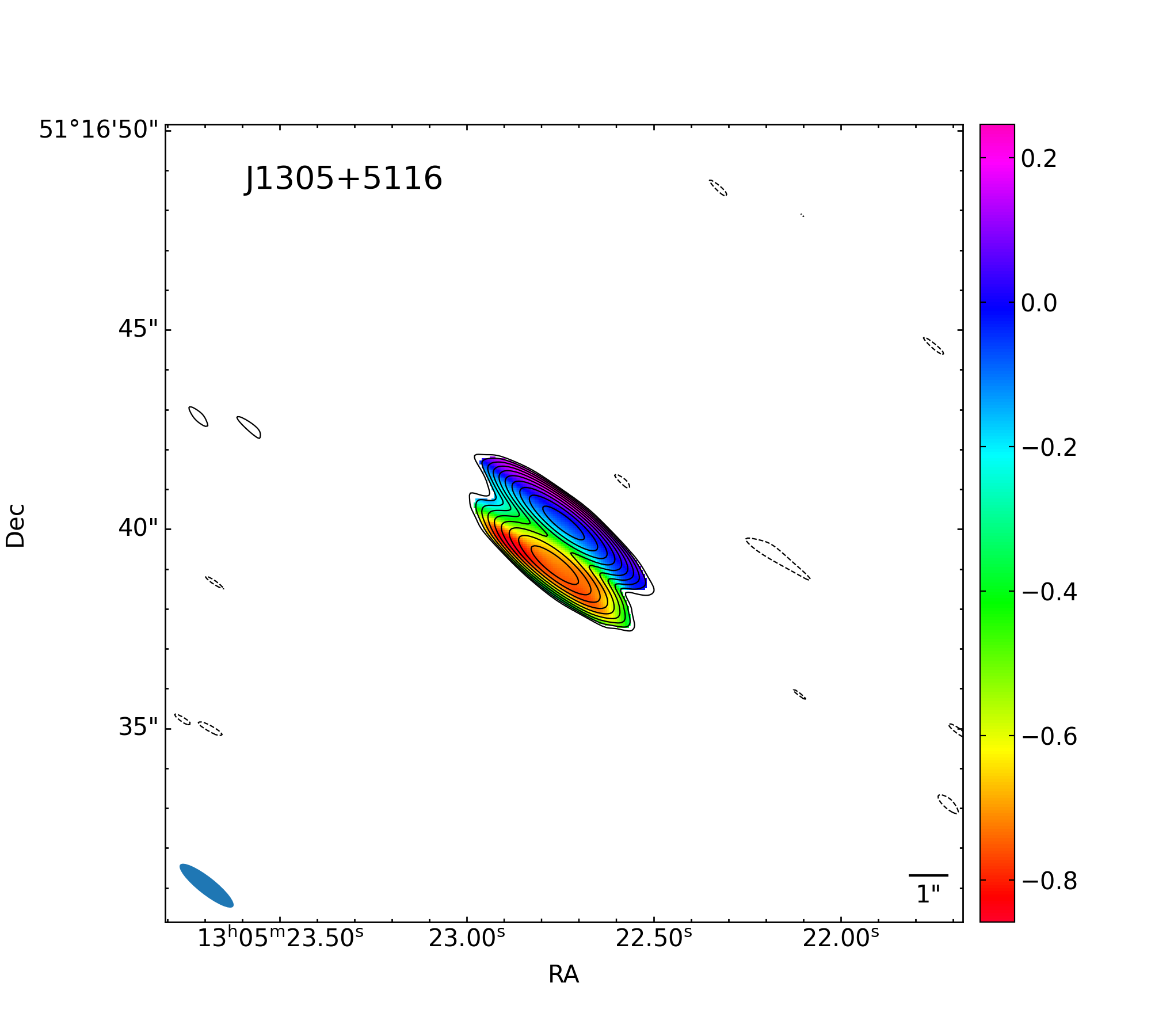}
         \caption{Spectral index map, rms = 14$\mu$Jy beam$^{-1}$, contour levels at -3, 3 $\times$ 2$^n$, $n \in$ [0, 9], beam size 12.70 $\times$ 3.21~kpc. } \label{fig:J1305spind}
     \end{subfigure}
     \hfill
     \begin{subfigure}[b]{0.47\textwidth}
         \centering
         \includegraphics[width=\textwidth]{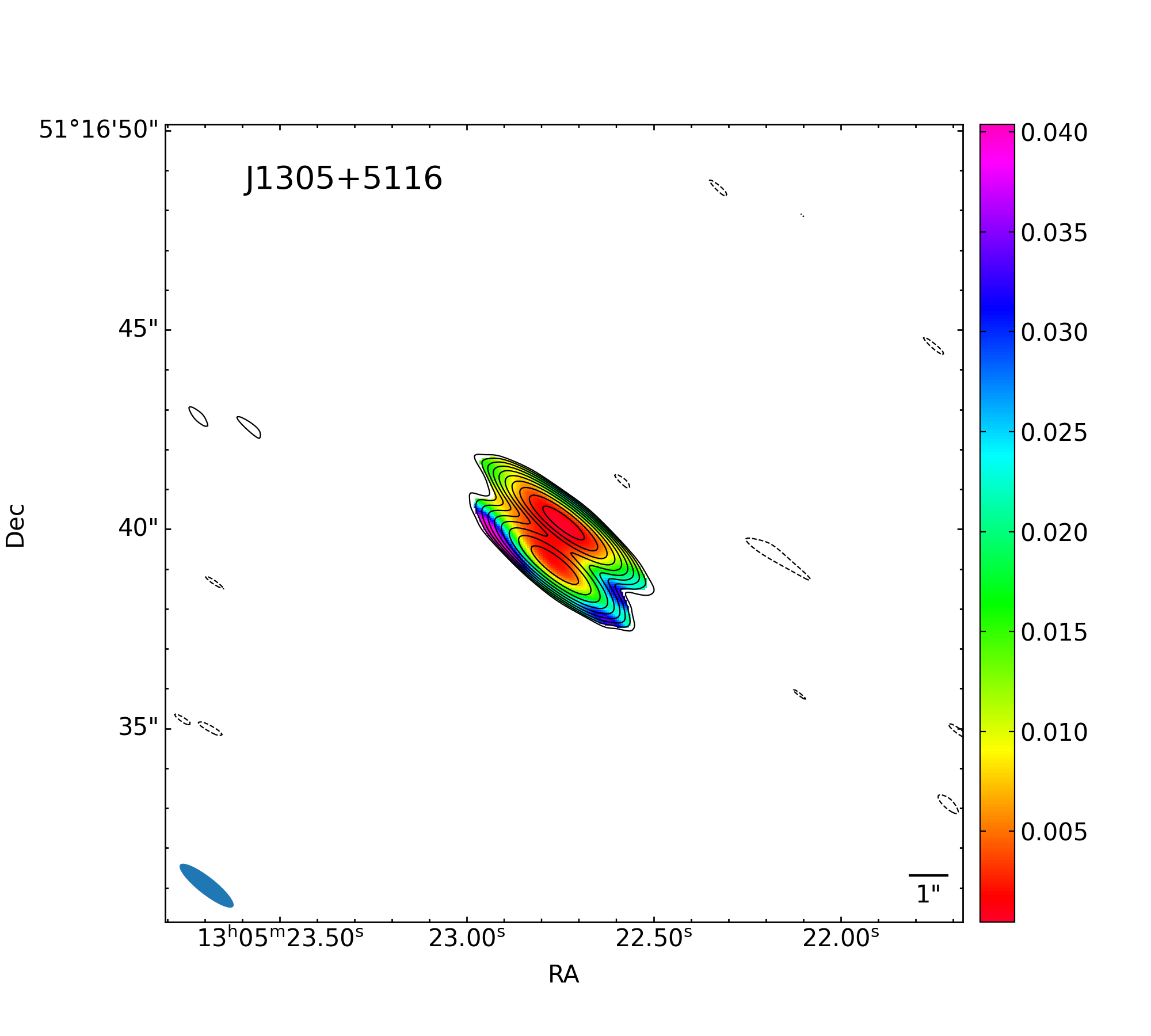}
         \caption{Spectral index error map, rms, contour levels, and beam size as in Fig.~\ref{fig:J1305spind}.} \label{fig:J1305spinderr}
     \end{subfigure}
     \hfill
     \\
     \begin{subfigure}[b]{0.47\textwidth}
         \centering
         \includegraphics[width=\textwidth]{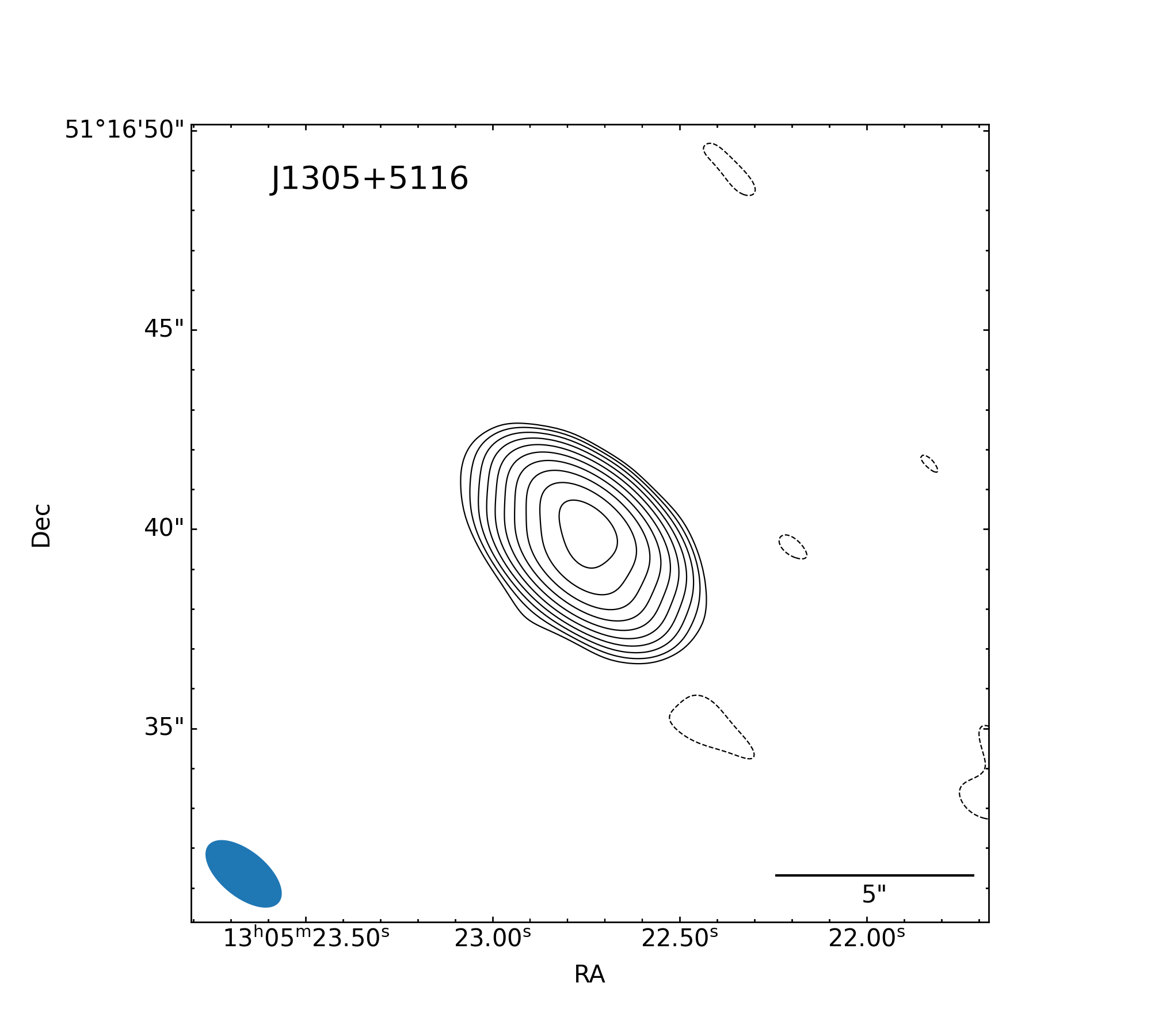}
         \caption{Tapered map with \texttt{uvtaper} = 90k$\lambda$, rms = 16$\mu$Jy beam$^{-1}$, contour levels at -3, 3 $\times$ 2$^n$, $n \in$ [0, 9], beam size 17.03 $\times$ 8.66~kpc.} \label{fig:J1305-90k}
     \end{subfigure}
          \hfill
     \begin{subfigure}[b]{0.47\textwidth}
         \centering
         \includegraphics[width=\textwidth]{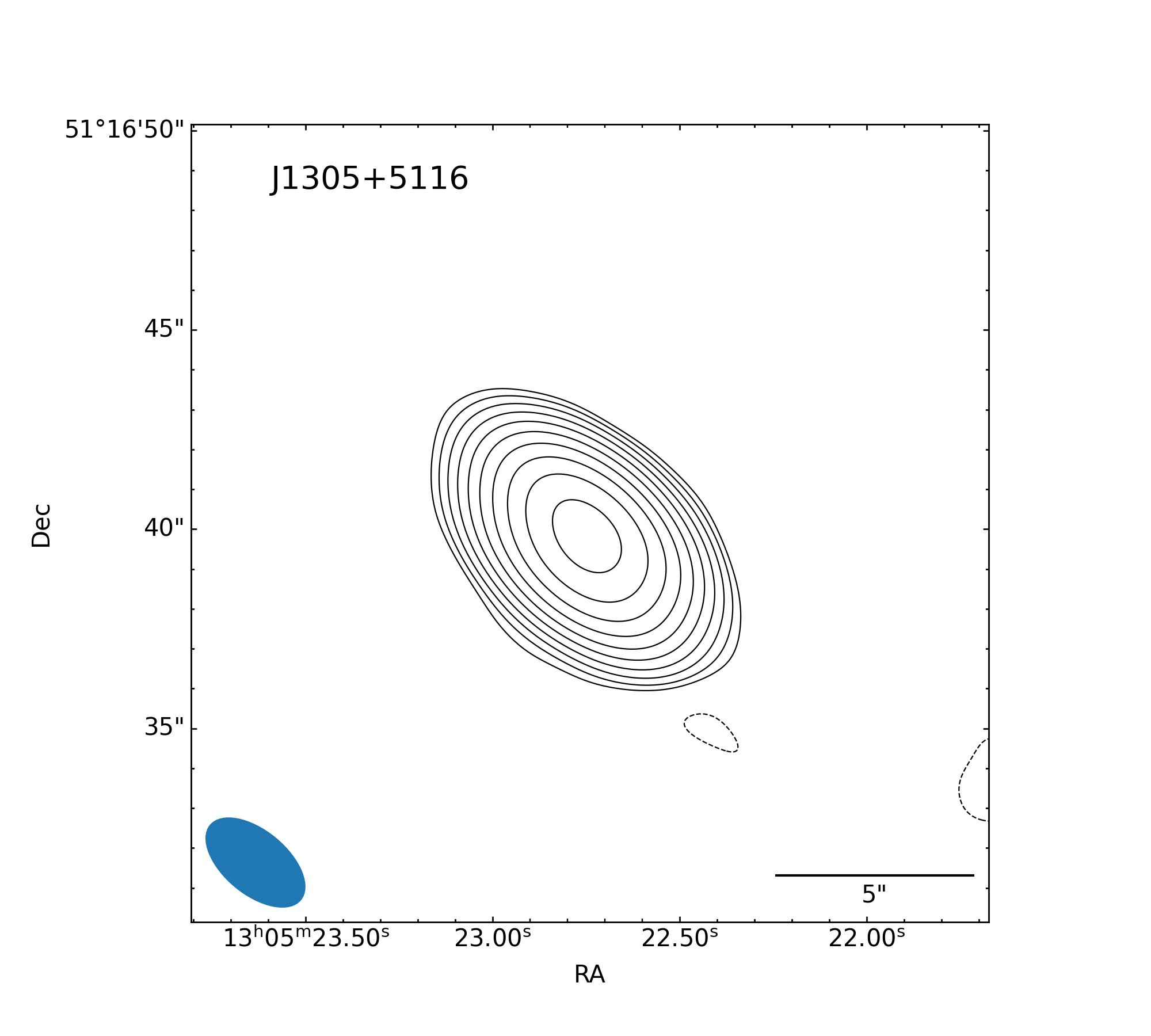}
         \caption{Tapered map with \texttt{uvtaper} = 60k$\lambda$, rms = 20$\mu$Jy beam$^{-1}$, contour levels at -3, 3 $\times$ 2$^n$, $n \in$ [0, 9], beam size 22.18 $\times$ 12.10~kpc.} \label{fig:J1305-60k}
     \end{subfigure}
          \hfill
     \\
     \begin{subfigure}[b]{0.47\textwidth}
         \centering
         \includegraphics[width=\textwidth]{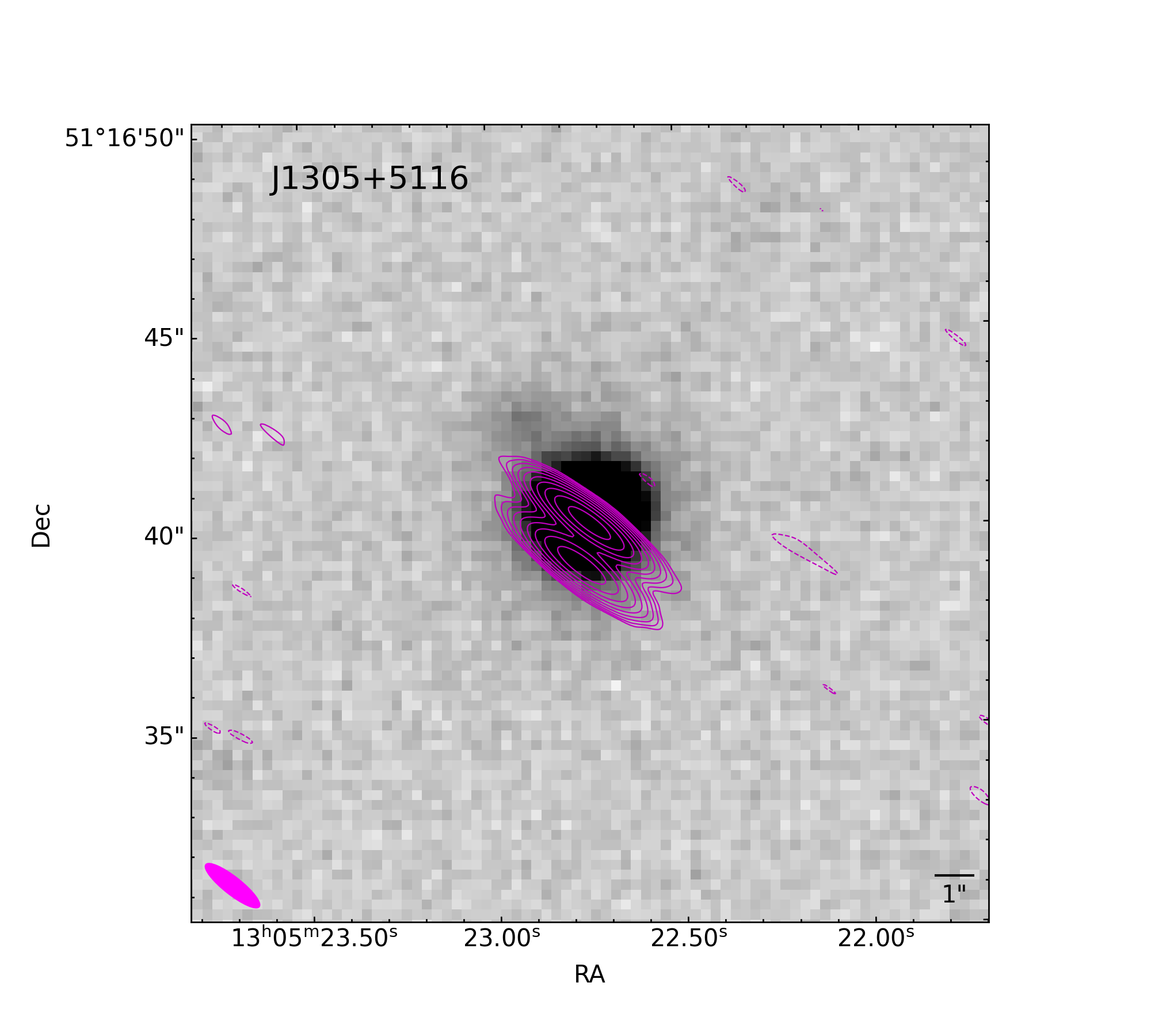}
         \caption{PanSTARRS $i$ band image of the host galaxy overlaid with the normal radio map. Radio map properties as in Fig.~\ref{fig:J1305spind}}. \label{fig:J1305-host-zoom}
     \end{subfigure}
          \hfill
     \begin{subfigure}[b]{0.47\textwidth}
         \centering
         \includegraphics[width=\textwidth]{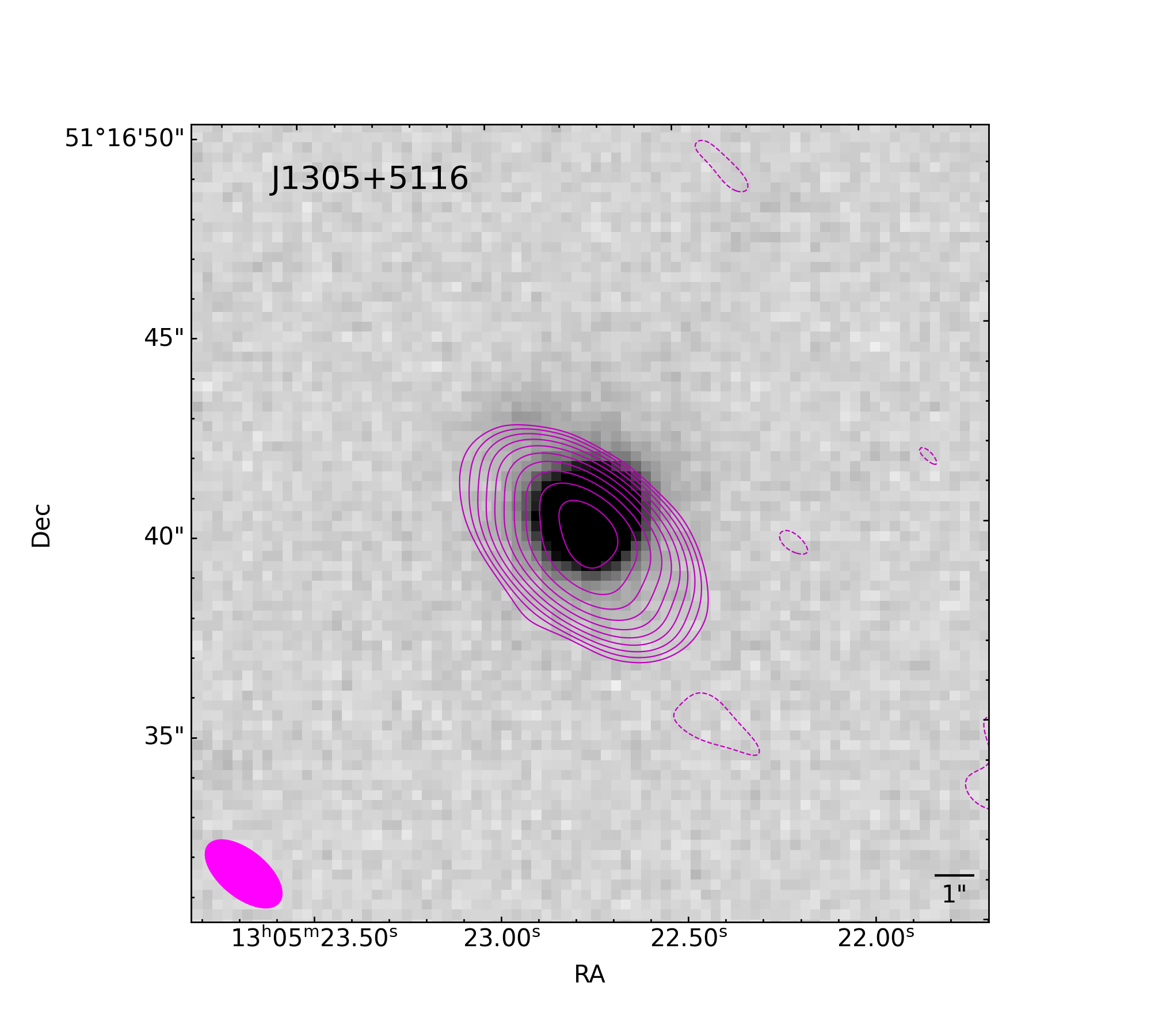}
         \caption{PanSTARRS $i$ band image of the host galaxy overlaid with the 90k$\lambda$ tapered map. Radio map properties as in Fig.~\ref{fig:J1305-90k}}. \label{fig:J1305-host}
     \end{subfigure}
        \caption{}
        \label{fig:J1305}
\end{figure*}


\begin{figure*}
     \centering
     \begin{subfigure}[b]{0.47\textwidth}
         \centering
         \includegraphics[width=\textwidth]{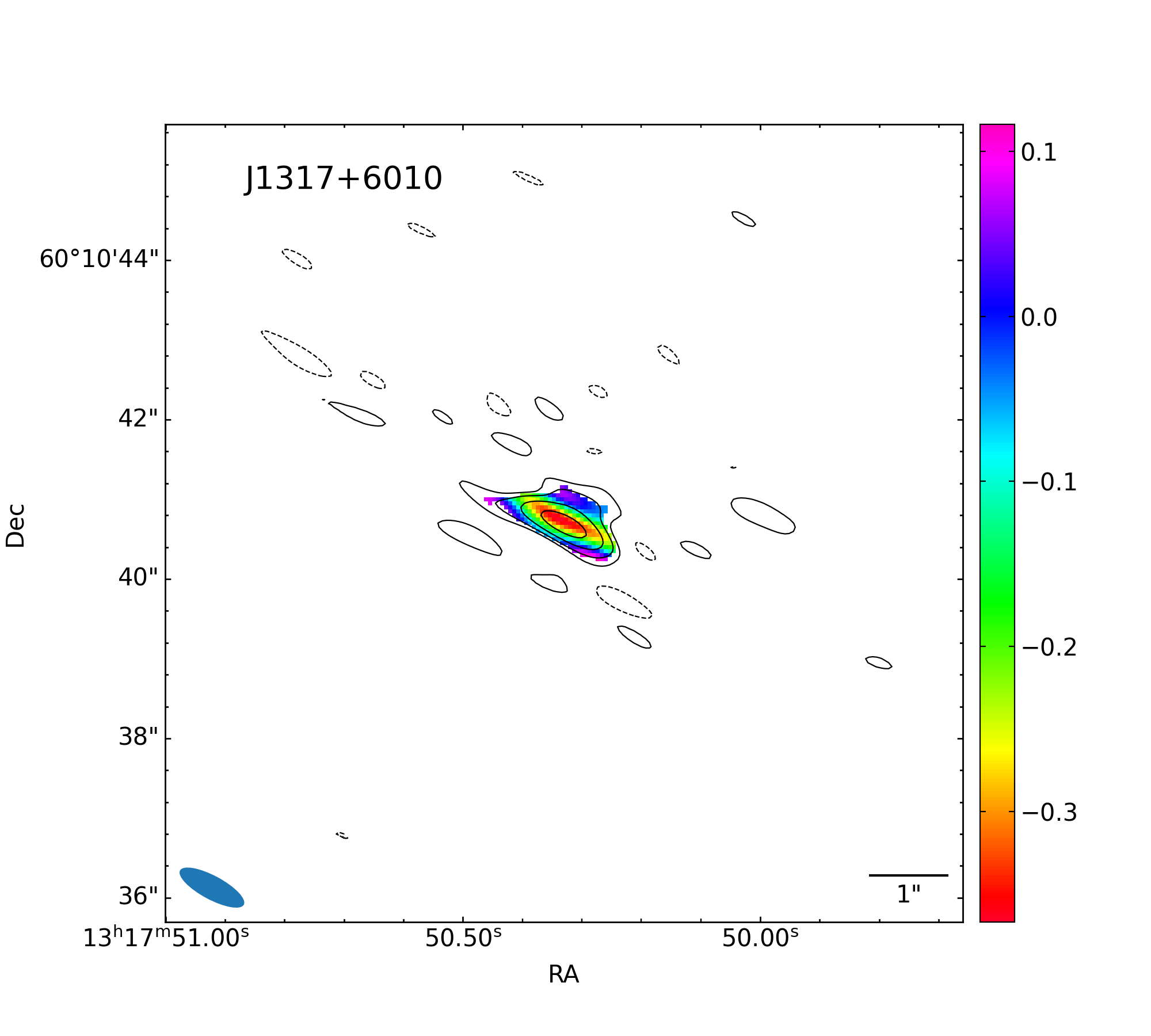}
         \caption{Spectral index map, rms = 16$\mu$Jy beam$^{-1}$, contour levels at -3, 3 $\times$ 2$^n$, $n \in$ [0, 3], beam size 2.20 $\times$ 0.73~kpc. } \label{fig:J1317spind}
     \end{subfigure}
     \hfill
     \begin{subfigure}[b]{0.47\textwidth}
         \centering
         \includegraphics[width=\textwidth]{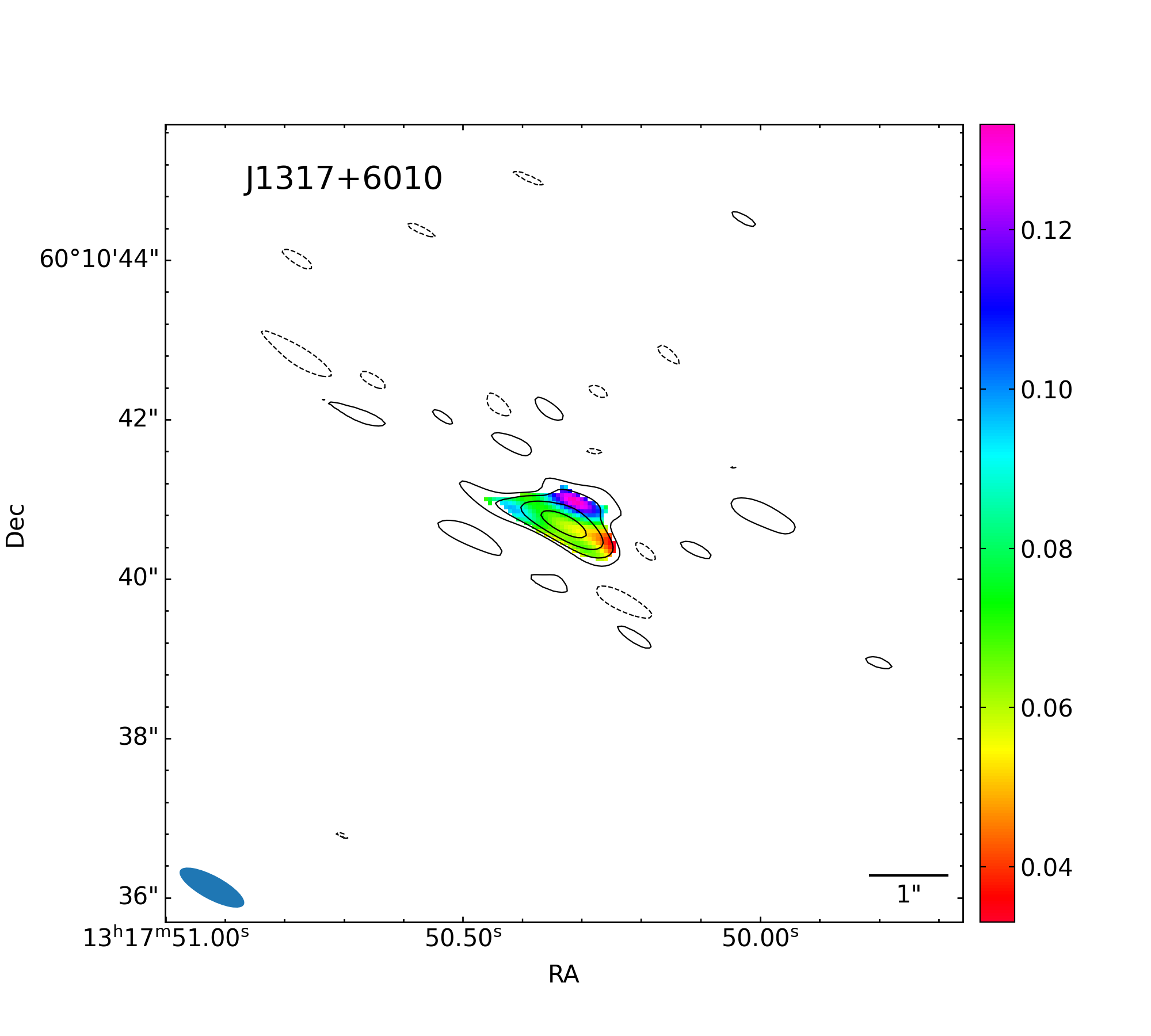}
         \caption{Spectral index error map, rms, contour levels, and beam size as in Fig.~\ref{fig:J1317spind}.} \label{fig:J1317spinderr}
     \end{subfigure}
     \hfill
     \\
     \begin{subfigure}[b]{0.47\textwidth}
         \centering
         \includegraphics[width=\textwidth]{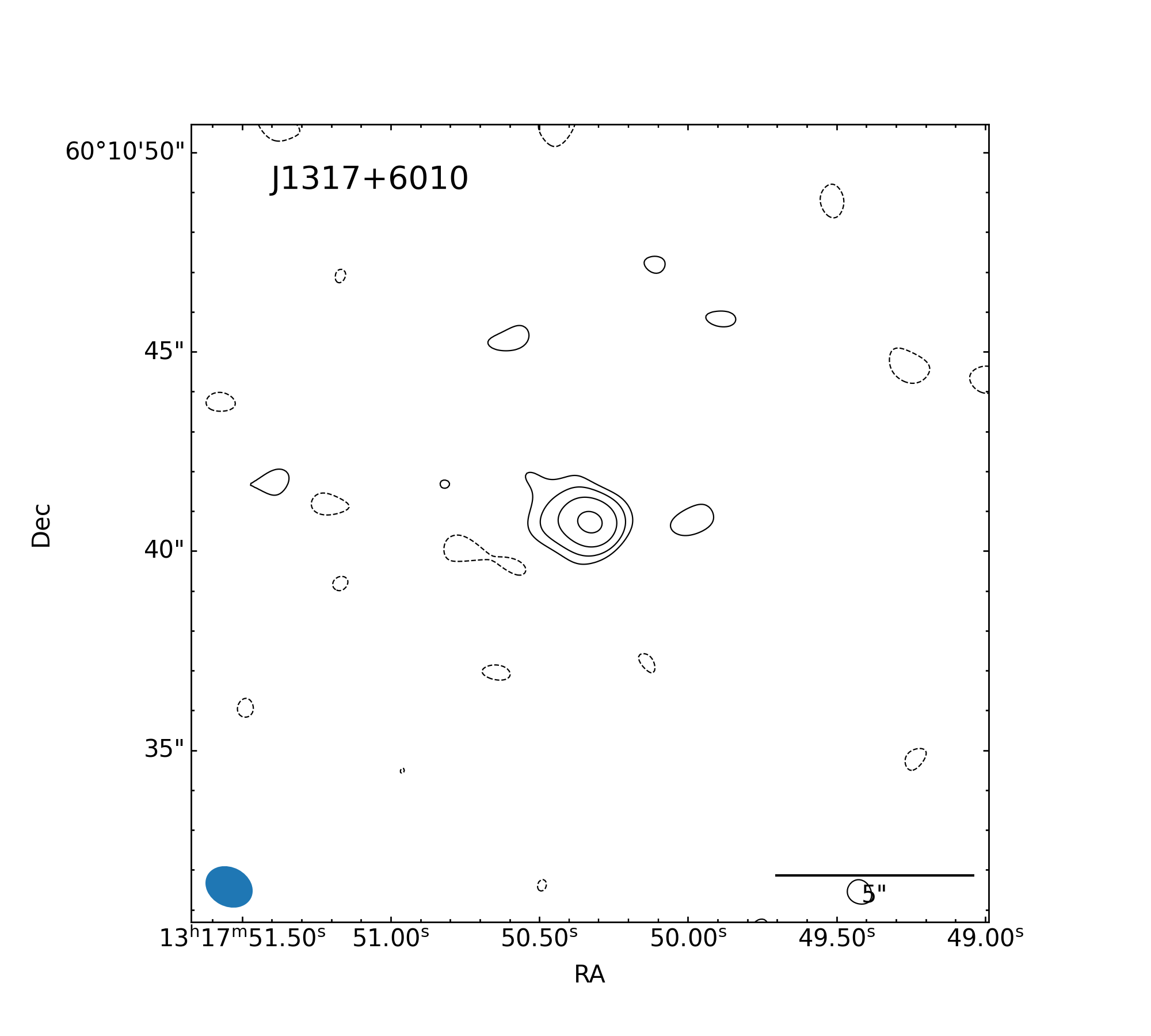}
         \caption{Tapered map with \texttt{uvtaper} = 90k$\lambda$, rms = 24$\mu$Jy beam$^{-1}$, contour levels at -3, 3 $\times$ 2$^n$, $n \in$ [0, 3], beam size 2.98 $\times$ 2.37~kpc.} \label{fig:J1317-90k}
     \end{subfigure}
          \hfill
     \begin{subfigure}[b]{0.47\textwidth}
         \centering
         \includegraphics[width=\textwidth]{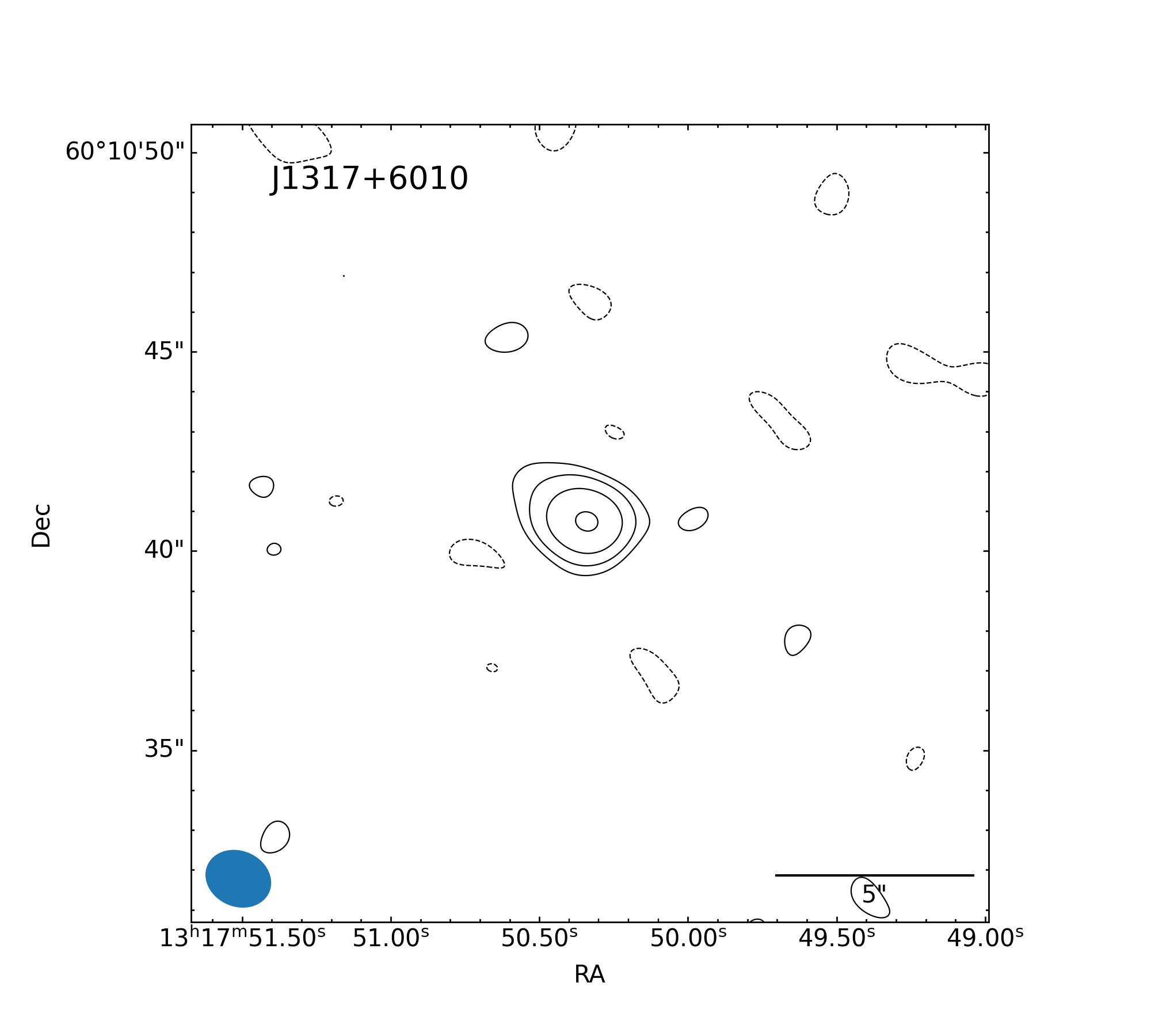}
         \caption{Tapered map with \texttt{uvtaper} = 60k$\lambda$, rms = 29$\mu$Jy beam$^{-1}$, contour levels at -3, 3 $\times$ 2$^n$, $n \in$ [0, 3], beam size 4.09 $\times$ 3.39~kpc.} \label{fig:J1317-60k}
     \end{subfigure}
          \hfill
     \\
     \begin{subfigure}[b]{0.47\textwidth}
         \centering
         \includegraphics[width=\textwidth]{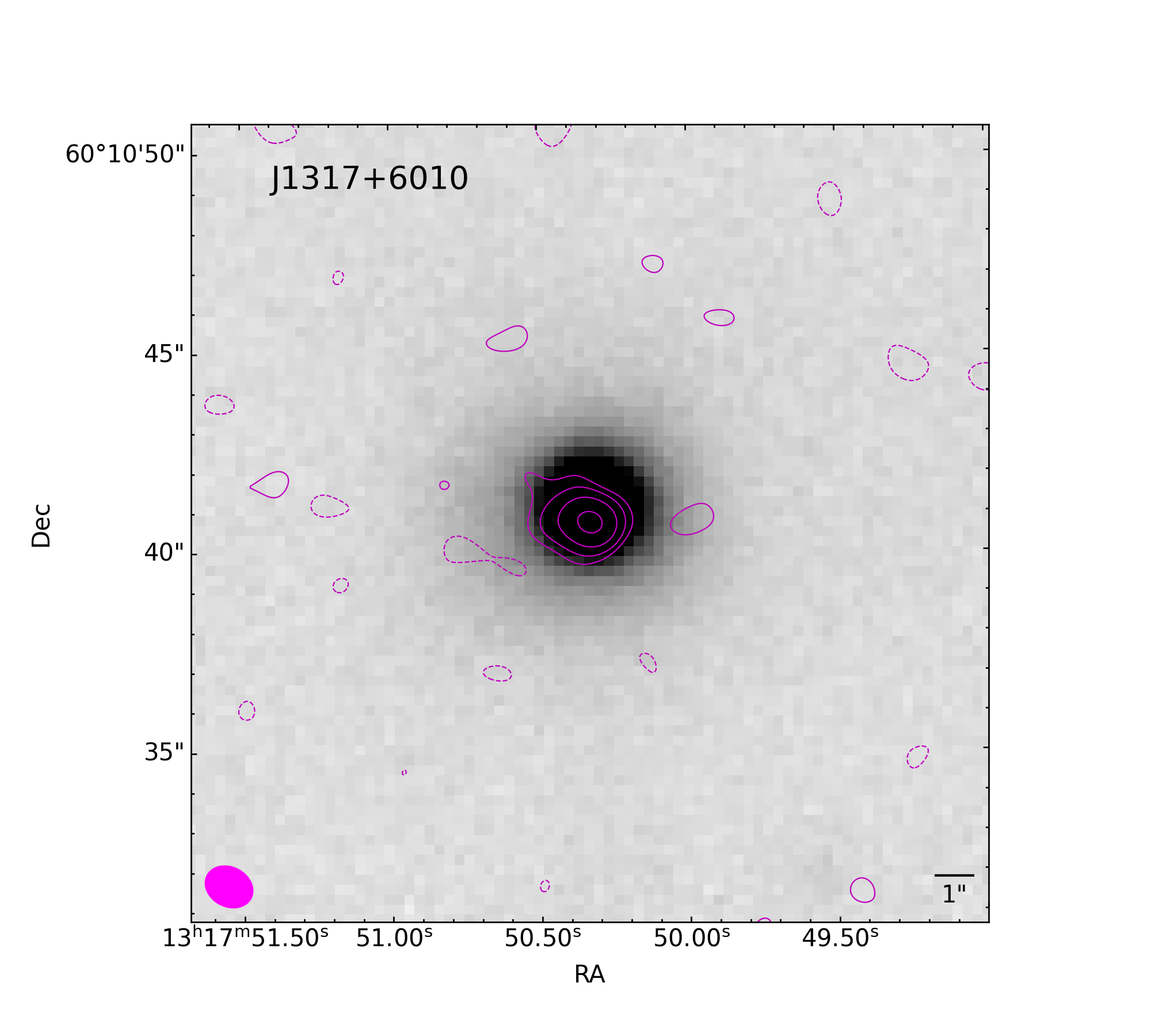}
         \caption{PanSTARRS $i$ band image of the host galaxy overlaid with the 90k$\lambda$ tapered map. Radio map properties as in Fig.~\ref{fig:J1317-90k}}. \label{fig:J1317-host}
     \end{subfigure}
        \caption{}
        \label{fig:J1317}
\end{figure*}


\begin{figure*}
     \centering
     \begin{subfigure}[b]{0.47\textwidth}
         \centering
         \includegraphics[width=\textwidth]{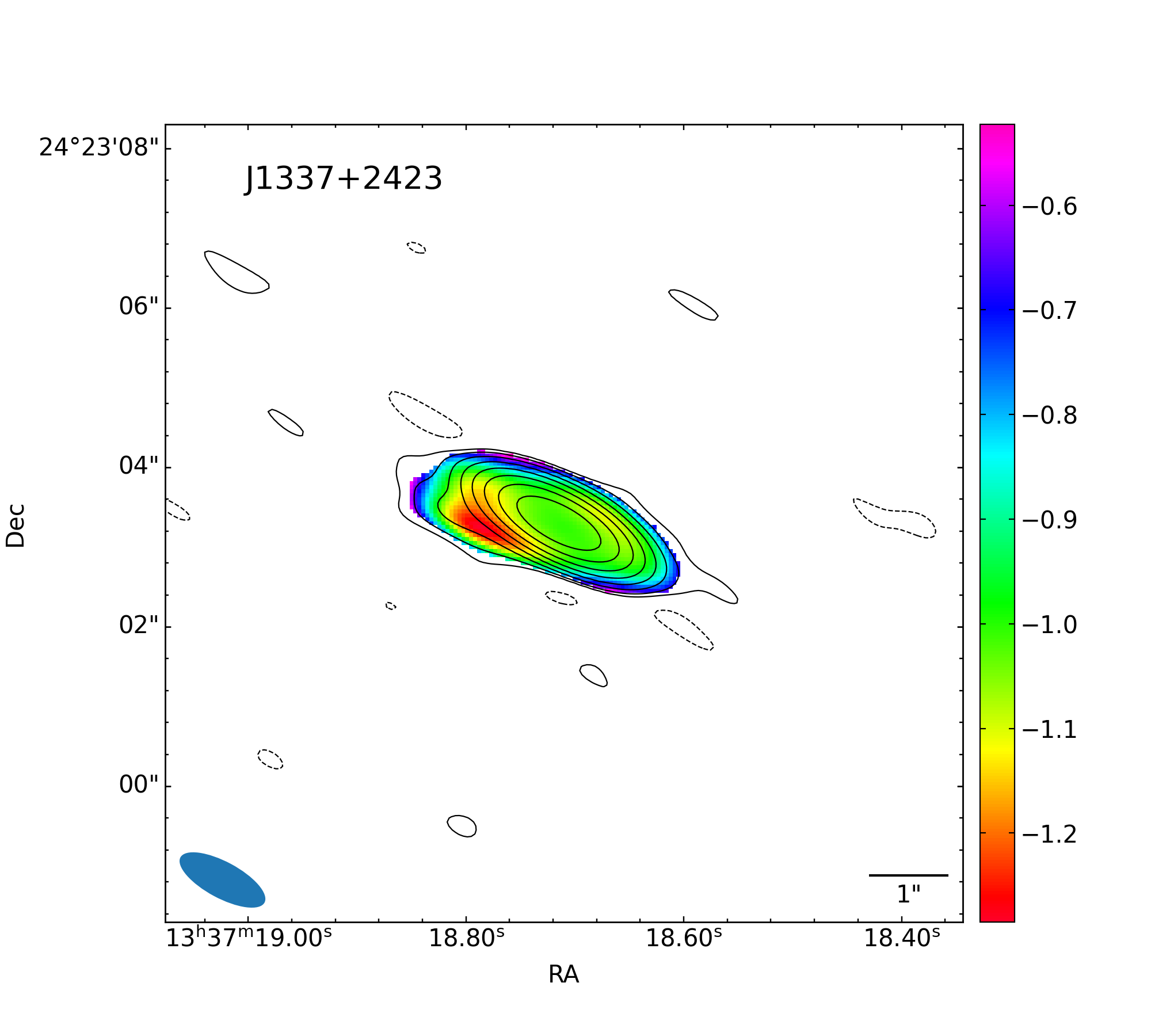}
         \caption{Spectral index map, rms = 13$\mu$Jy beam$^{-1}$, contour levels at -3, 3 $\times$ 2$^n$, $n \in$ [0, 7], beam size 2.37 $\times$ 0.91~kpc. } \label{fig:J1337spind}
     \end{subfigure}
     \hfill
     \begin{subfigure}[b]{0.47\textwidth}
         \centering
         \includegraphics[width=\textwidth]{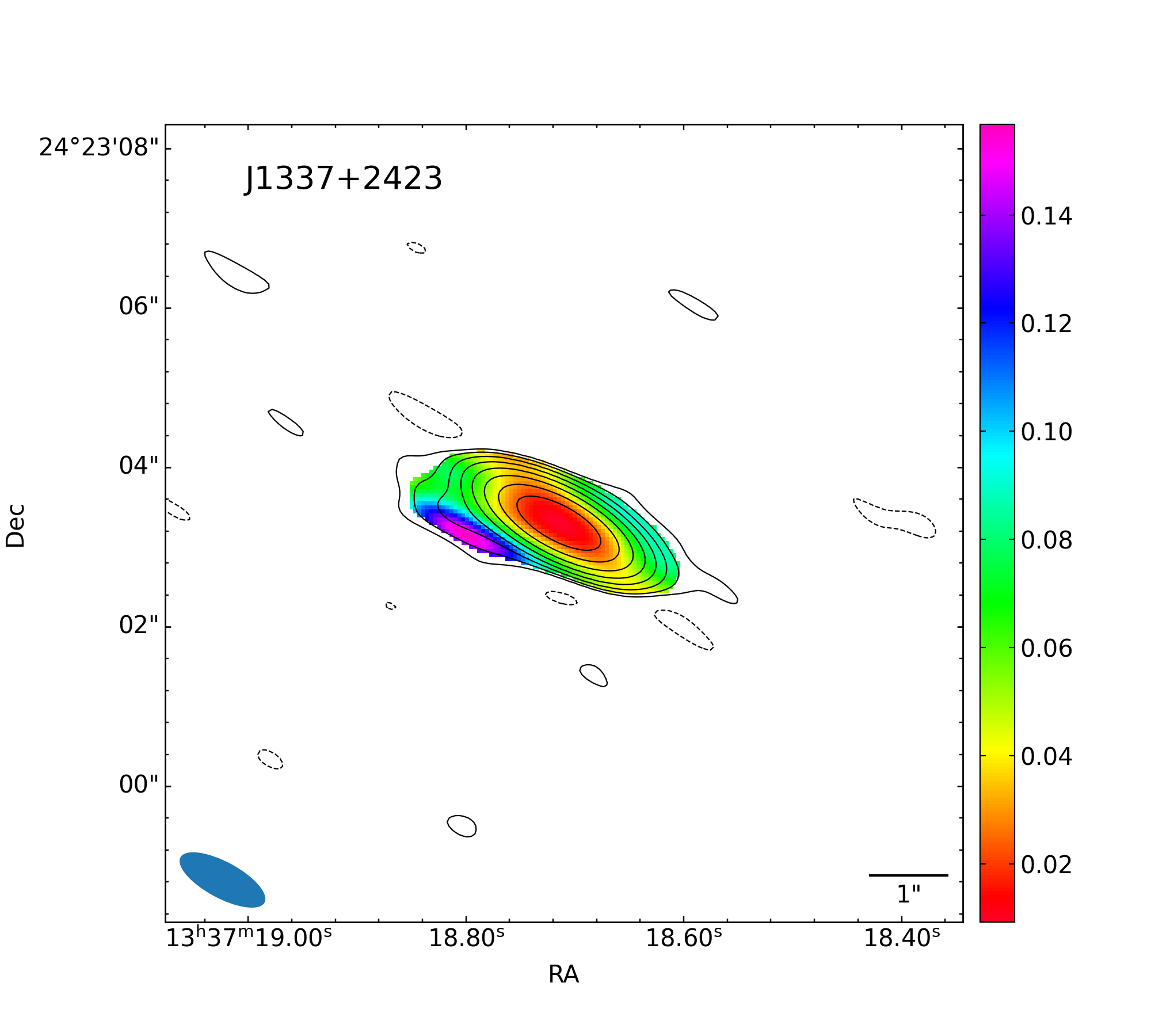}
         \caption{Spectral index error map, rms, contour levels, and beam size as in Fig.~\ref{fig:J1337spind}.} \label{fig:J1337spinderr}
     \end{subfigure}
     \hfill
     \\
     \begin{subfigure}[b]{0.47\textwidth}
         \centering
         \includegraphics[width=\textwidth]{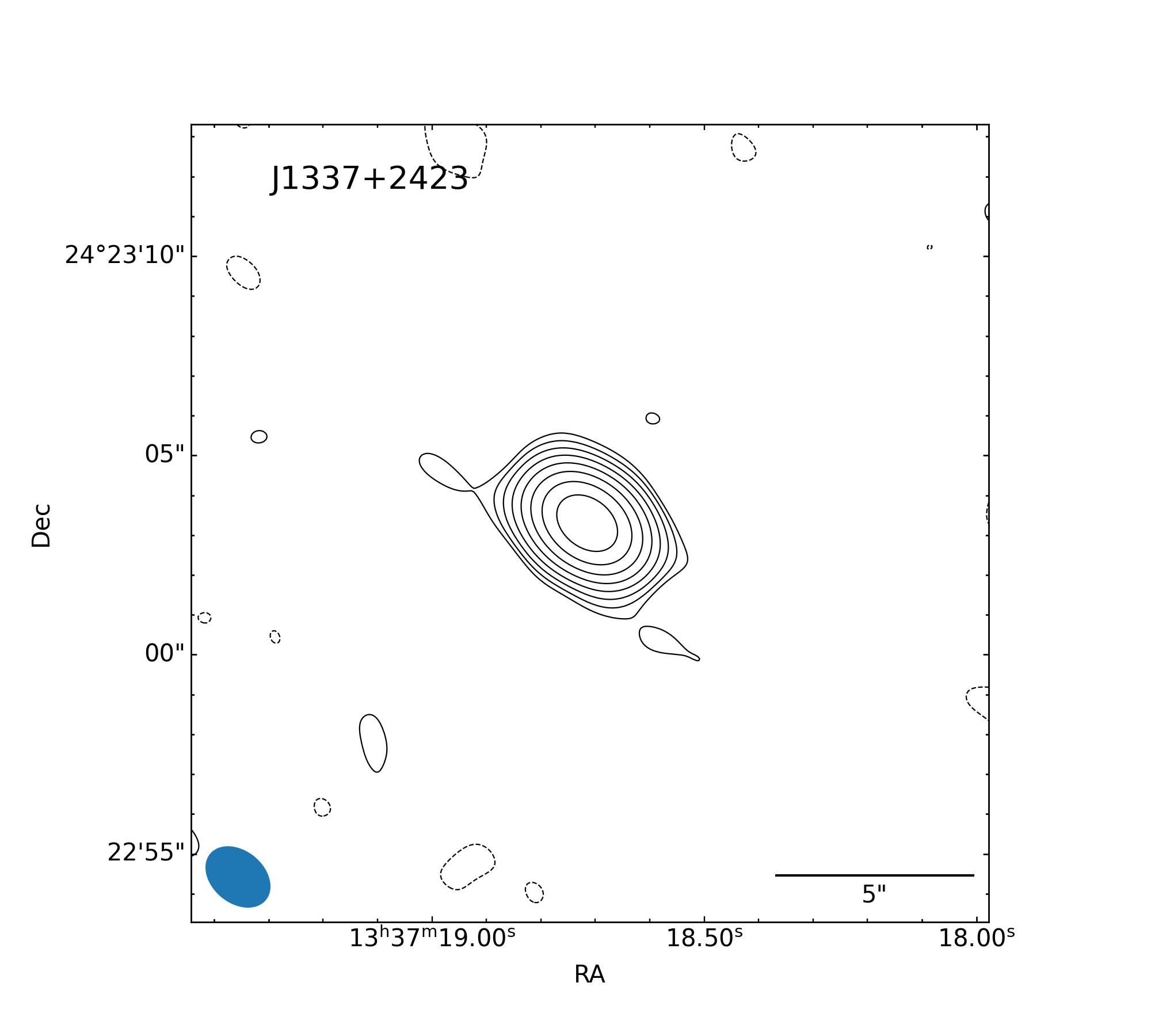}
         \caption{Tapered map with \texttt{uvtaper} = 90k$\lambda$, rms = 14$\mu$Jy beam$^{-1}$, contour levels at -3, 3 $\times$ 2$^n$, $n \in$ [0, 7], beam size 3.59 $\times$ 2.59~kpc.} \label{fig:J1337-90k}
     \end{subfigure}
          \hfill
     \begin{subfigure}[b]{0.47\textwidth}
         \centering
         \includegraphics[width=\textwidth]{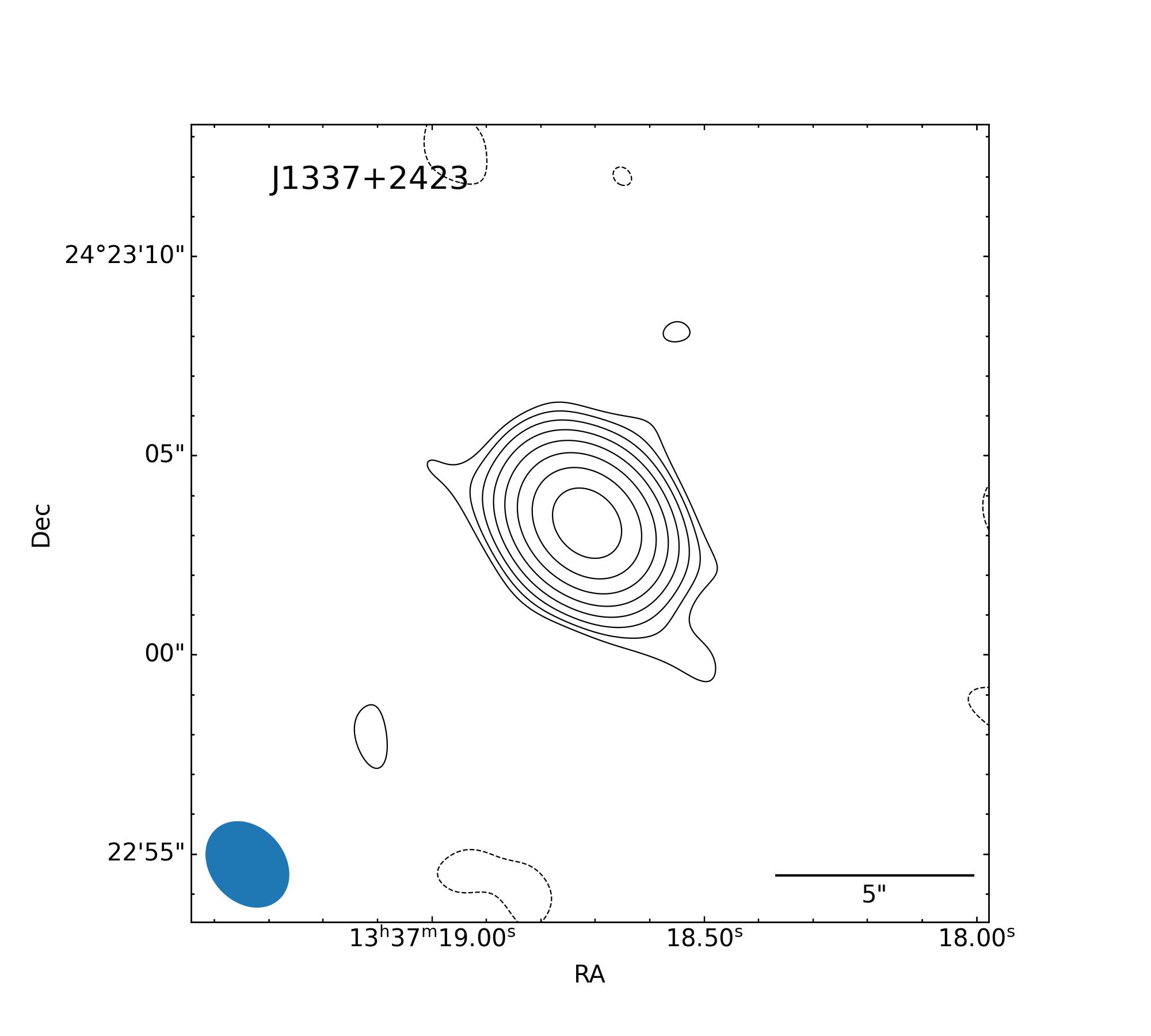}
         \caption{Tapered map with \texttt{uvtaper} = 60k$\lambda$, rms = 16$\mu$Jy beam$^{-1}$, contour levels at -3, 3 $\times$ 2$^n$, $n \in$ [0, 7], beam size 4.70 $\times$ 3.69~kpc.} \label{fig:J1337-60k}
     \end{subfigure}
          \hfill
     \\
     \begin{subfigure}[b]{0.47\textwidth}
         \centering
         \includegraphics[width=\textwidth]{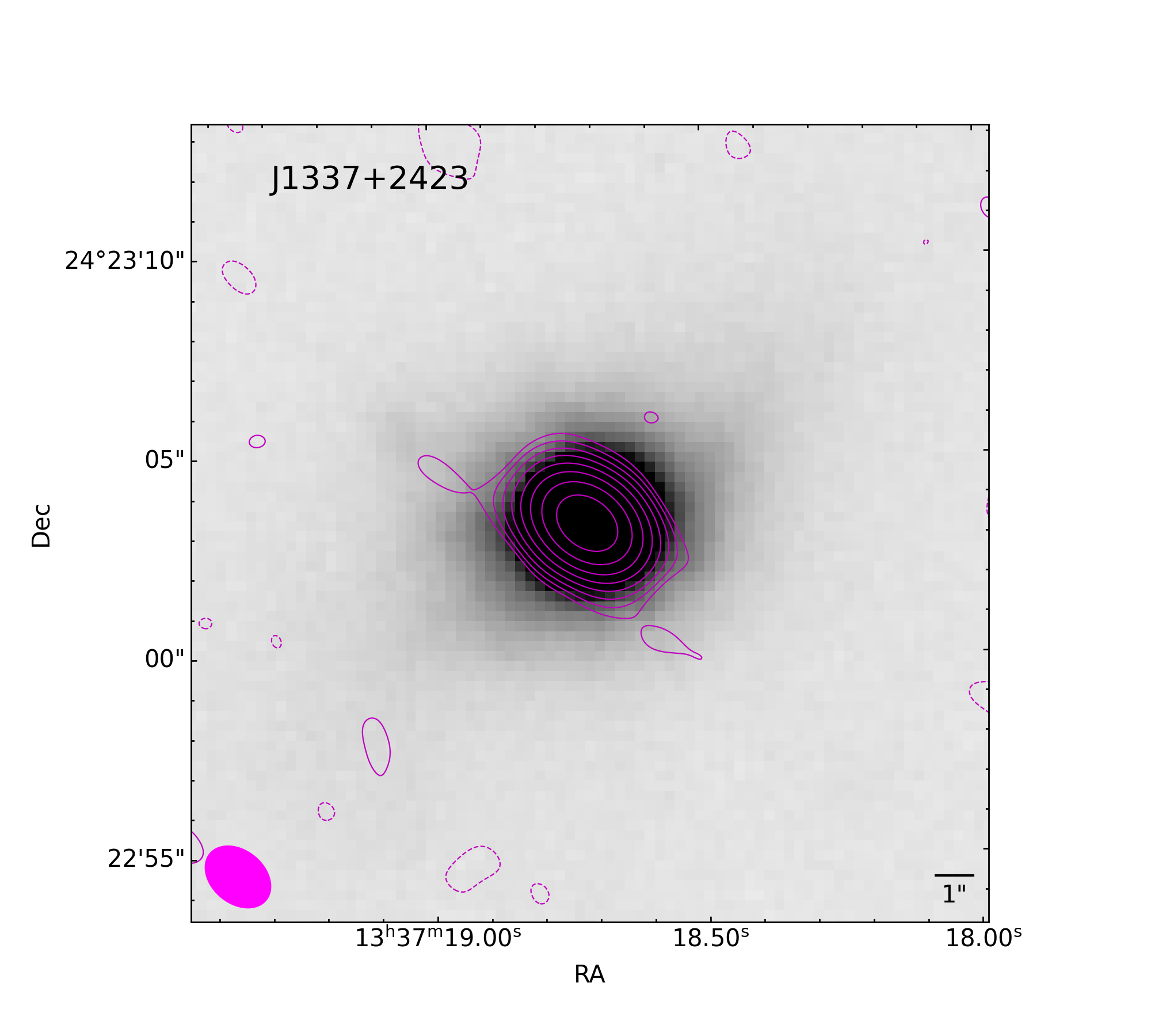}
         \caption{PanSTARRS $i$ band image of the host galaxy overlaid with the 90k$\lambda$ tapered map. Radio map properties as in Fig.~\ref{fig:J1337-90k}}. \label{fig:J1337-host}
     \end{subfigure}
        \caption{}
        \label{fig:J1337}
\end{figure*}


\begin{figure*}
     \centering
     \begin{subfigure}[b]{0.47\textwidth}
         \centering
         \includegraphics[width=\textwidth]{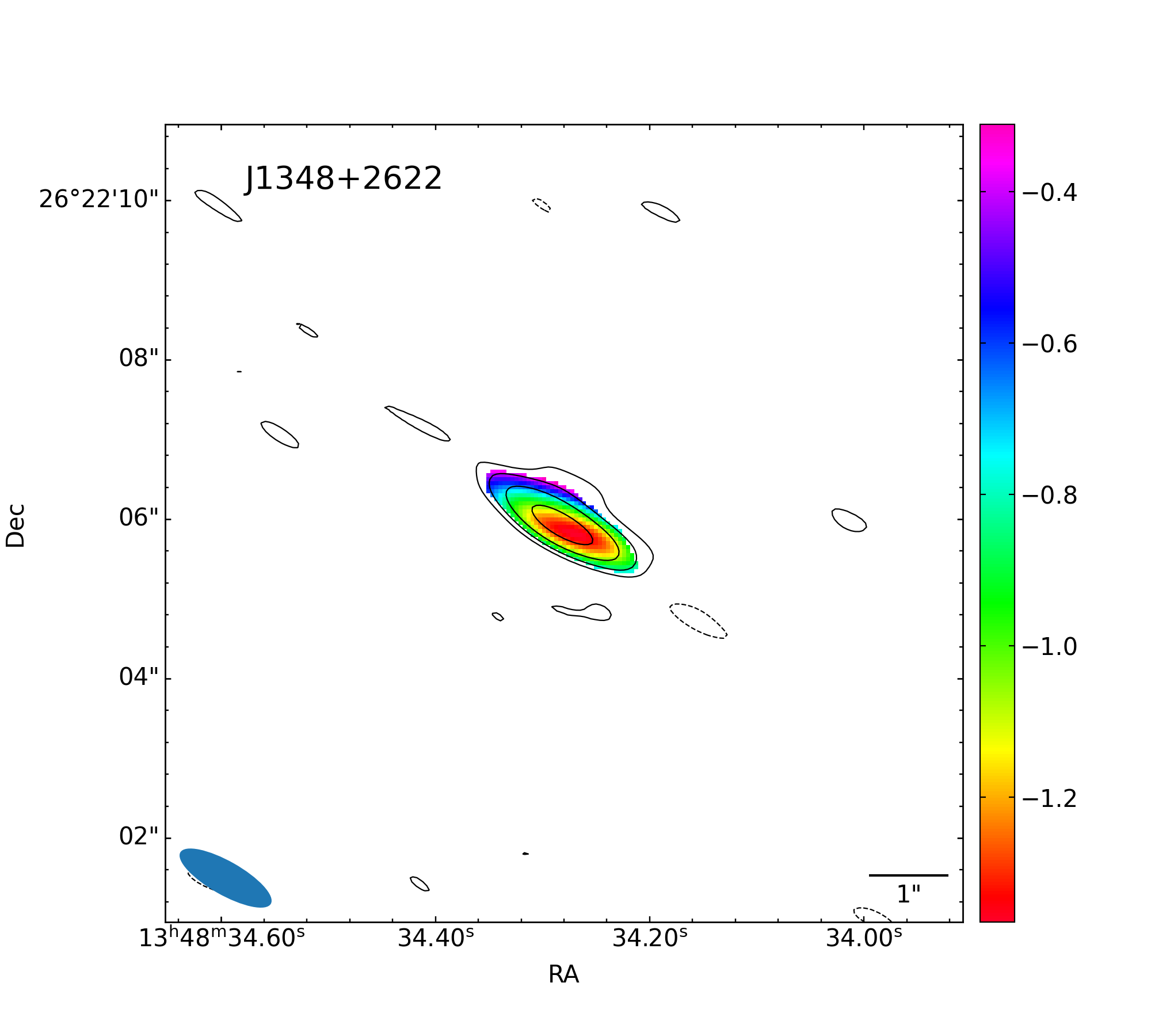}
         \caption{Spectral index map, rms = 12$\mu$Jy beam$^{-1}$, contour levels at -3, 3 $\times$ 2$^n$, $n \in$ [0, 3], beam size 10.26 $\times$ 3.29~kpc. } \label{fig:J1348spind}
     \end{subfigure}
     \hfill
     \begin{subfigure}[b]{0.47\textwidth}
         \centering
         \includegraphics[width=\textwidth]{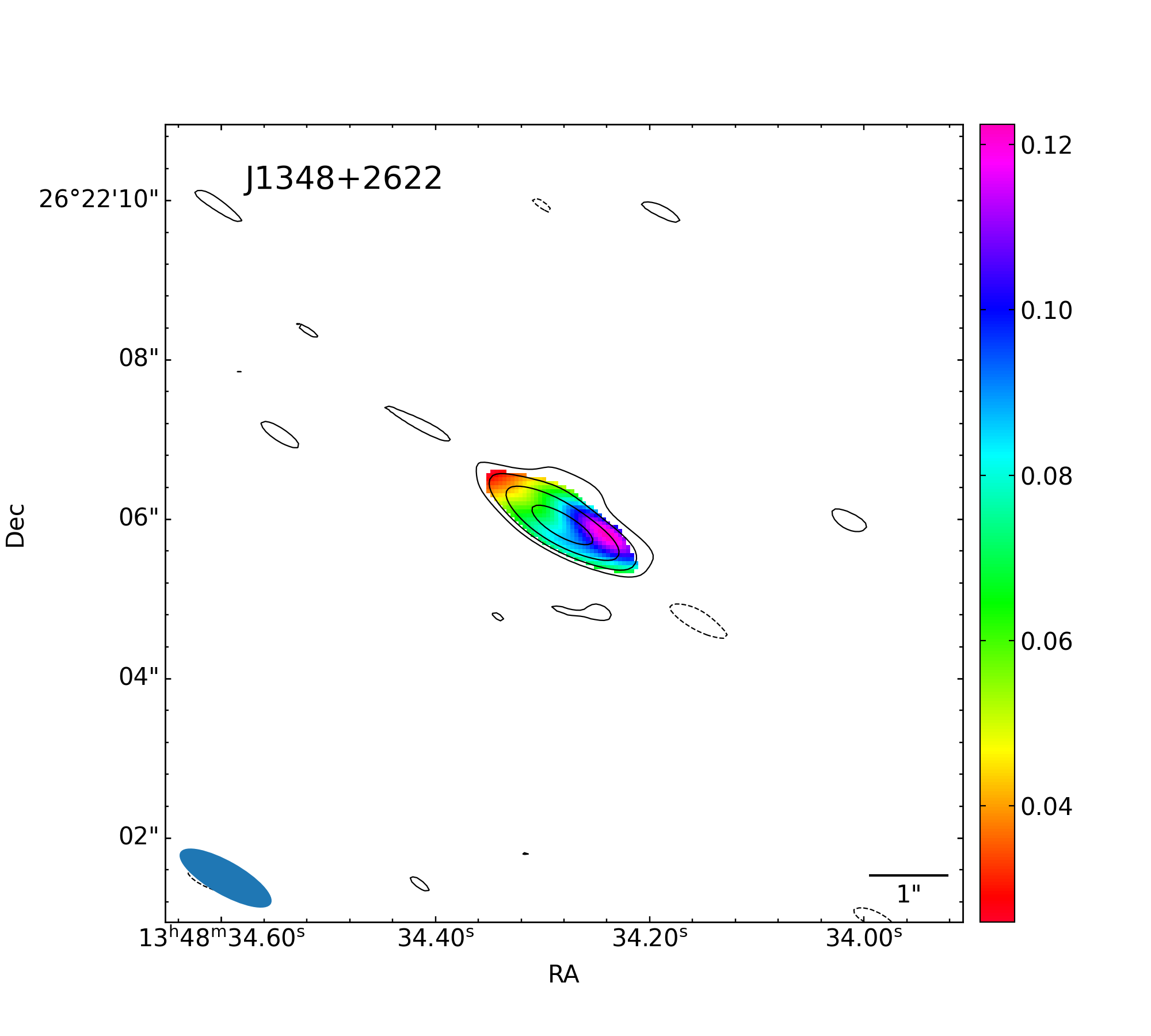}
         \caption{Spectral index error map, rms, contour levels, and beam size as in Fig.~\ref{fig:J1348spind}.} \label{fig:J1348spinderr}
     \end{subfigure}
     \hfill
     \\
     \begin{subfigure}[b]{0.47\textwidth}
         \centering
         \includegraphics[width=\textwidth]{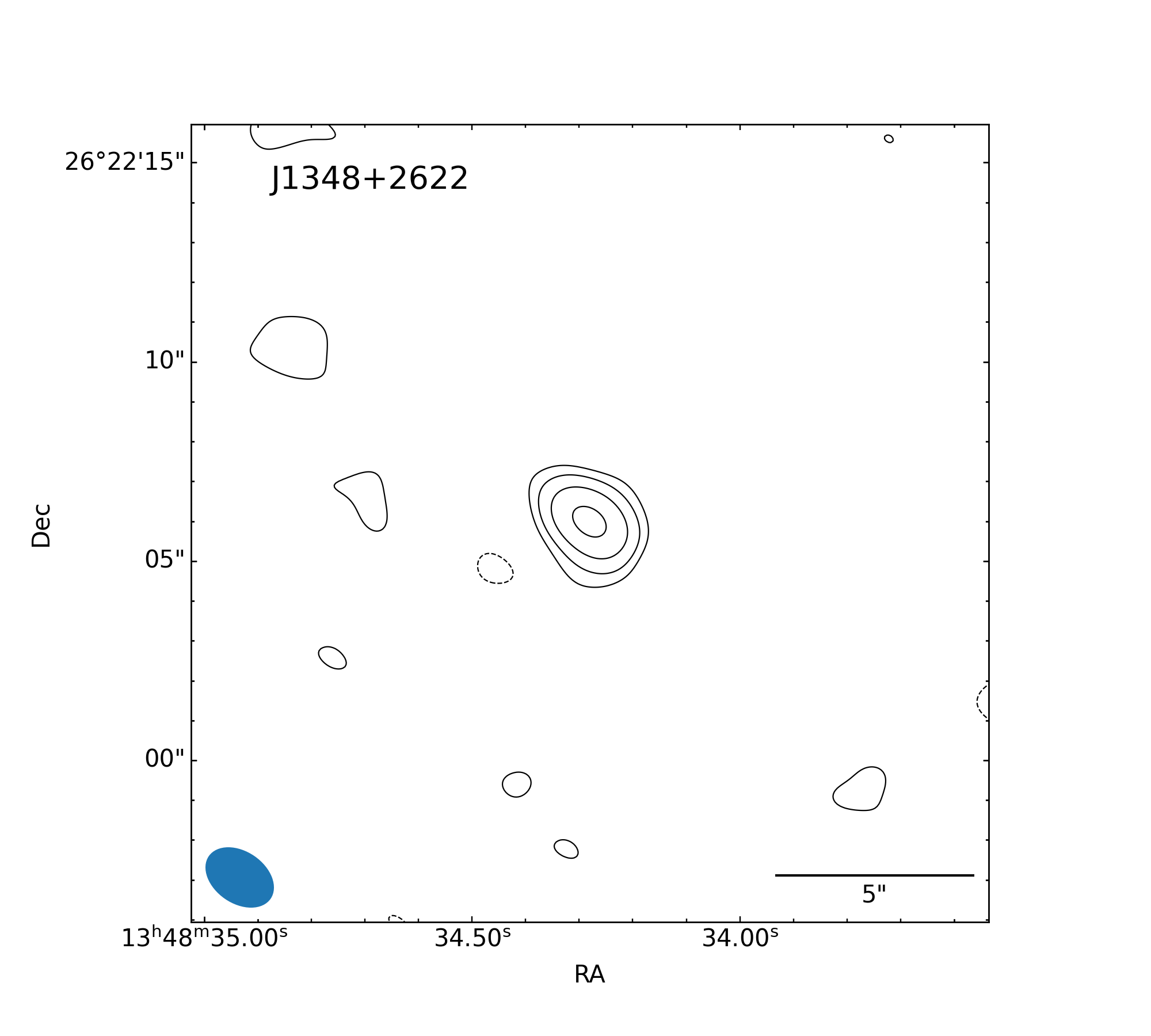}
         \caption{Tapered map with \texttt{uvtaper} = 90k$\lambda$, rms = 14$\mu$Jy beam$^{-1}$, contour levels at -3, 3 $\times$ 2$^n$, $n \in$ [0, 3], beam size 14.72 $\times$ 10.18~kpc.} \label{fig:J1348-90k}
     \end{subfigure}
          \hfill
     \begin{subfigure}[b]{0.47\textwidth}
         \centering
         \includegraphics[width=\textwidth]{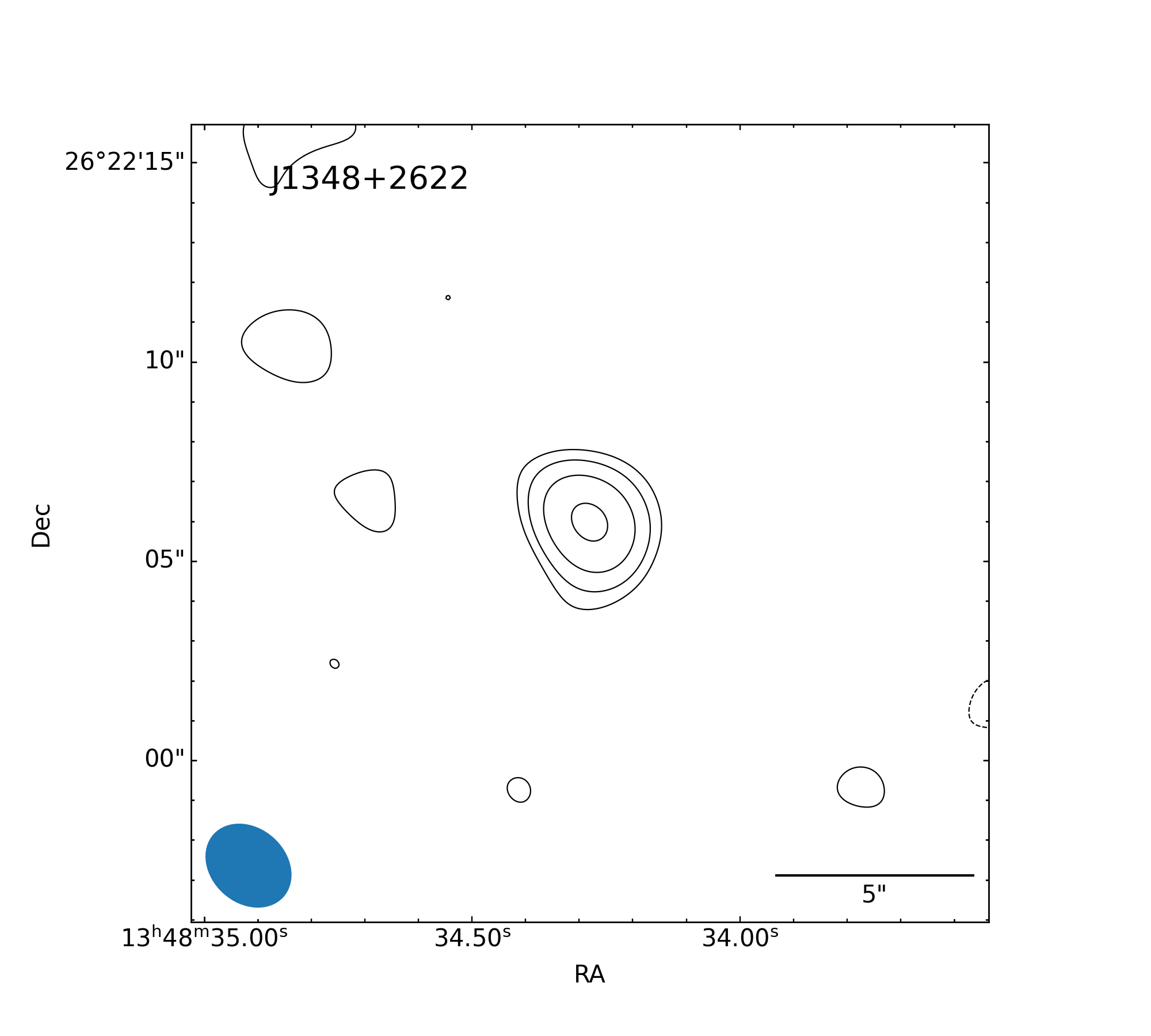}
         \caption{Tapered map with \texttt{uvtaper} = 60k$\lambda$, rms = 15$\mu$Jy beam$^{-1}$, contour levels at -3, 3 $\times$ 2$^n$, $n \in$ [0, 3], beam size 18.48 $\times$ 14.72~kpc.} \label{fig:J1348-60k}
     \end{subfigure}
          \hfill
     \\
     \begin{subfigure}[b]{0.47\textwidth}
         \centering
         \includegraphics[width=\textwidth]{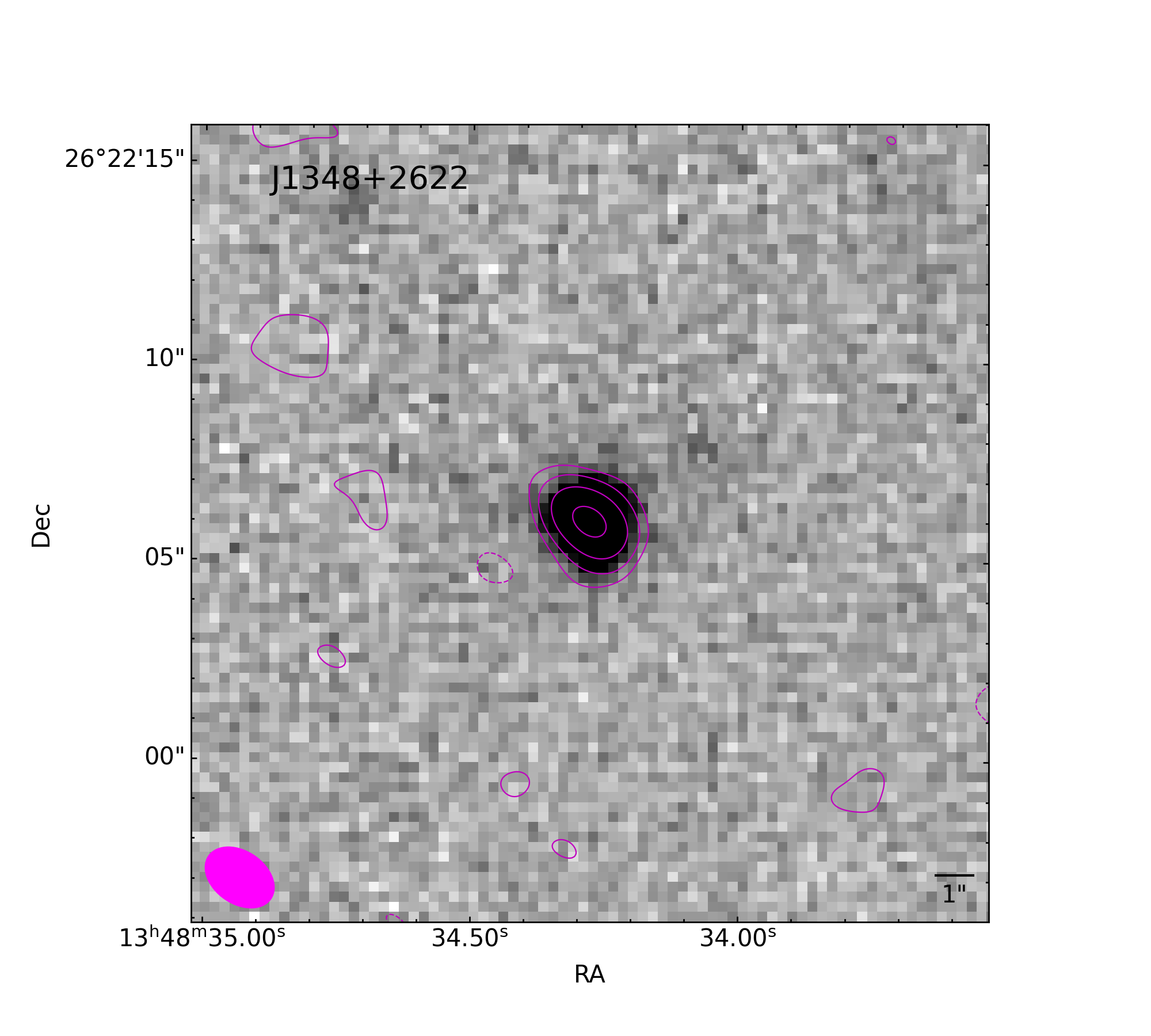}
         \caption{PanSTARRS $i$ band image of the host galaxy overlaid with the 90k$\lambda$ tapered map. Radio map properties as in Fig.~\ref{fig:J1348-90k}}. \label{fig:J1348-host}
     \end{subfigure}
        \caption{}
        \label{fig:J1348}
\end{figure*}


\begin{figure*}
     \centering
     \begin{subfigure}[b]{0.47\textwidth}
         \centering
         \includegraphics[width=\textwidth]{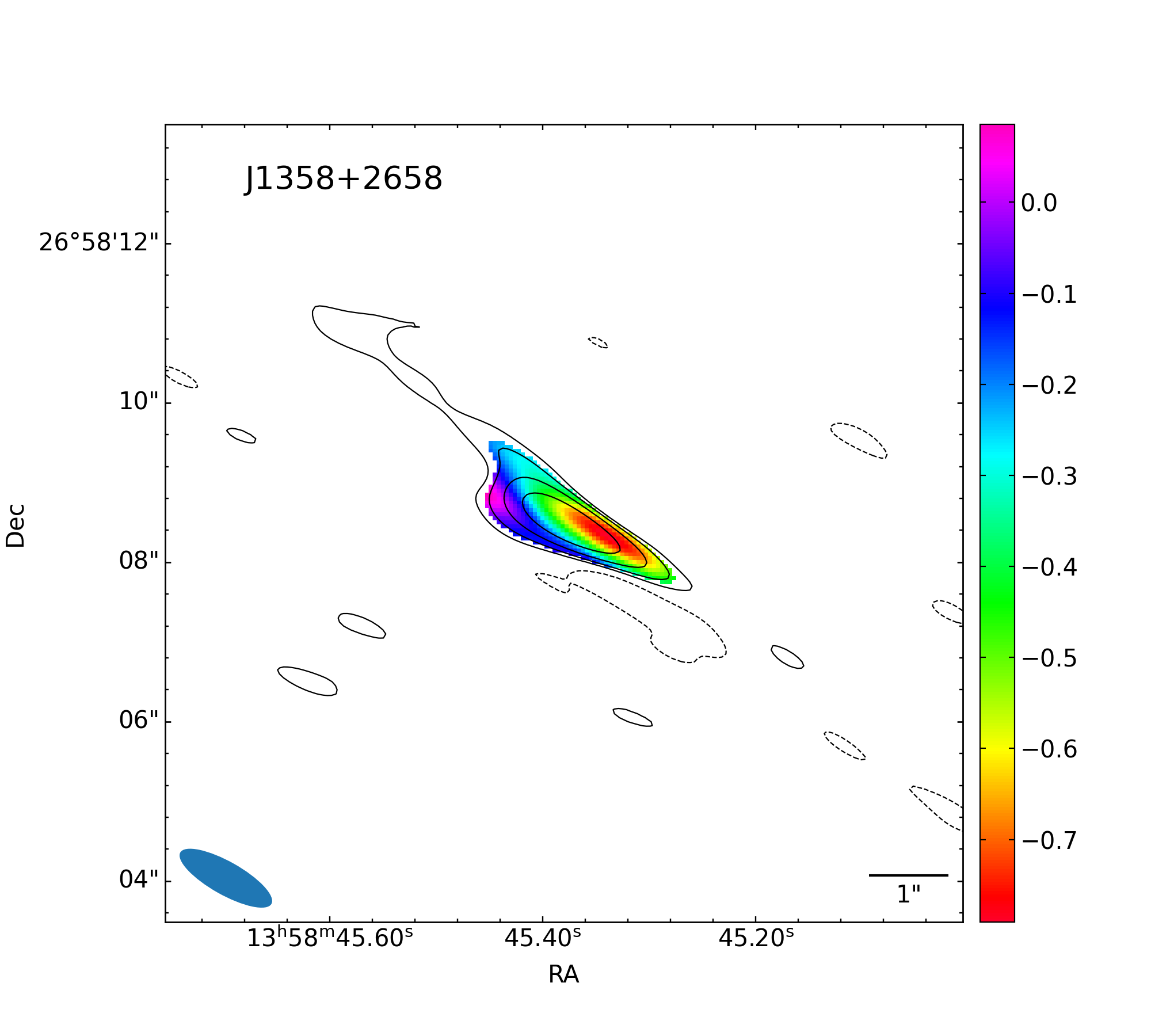}
         \caption{Spectral index map, rms = 11$\mu$Jy beam$^{-1}$, contour levels at -3, 3 $\times$ 2$^n$, $n \in$ [0, 3], beam size 6.24 $\times$ 2.00~kpc. } \label{fig:J1358spind}
     \end{subfigure}
     \hfill
     \begin{subfigure}[b]{0.47\textwidth}
         \centering
         \includegraphics[width=\textwidth]{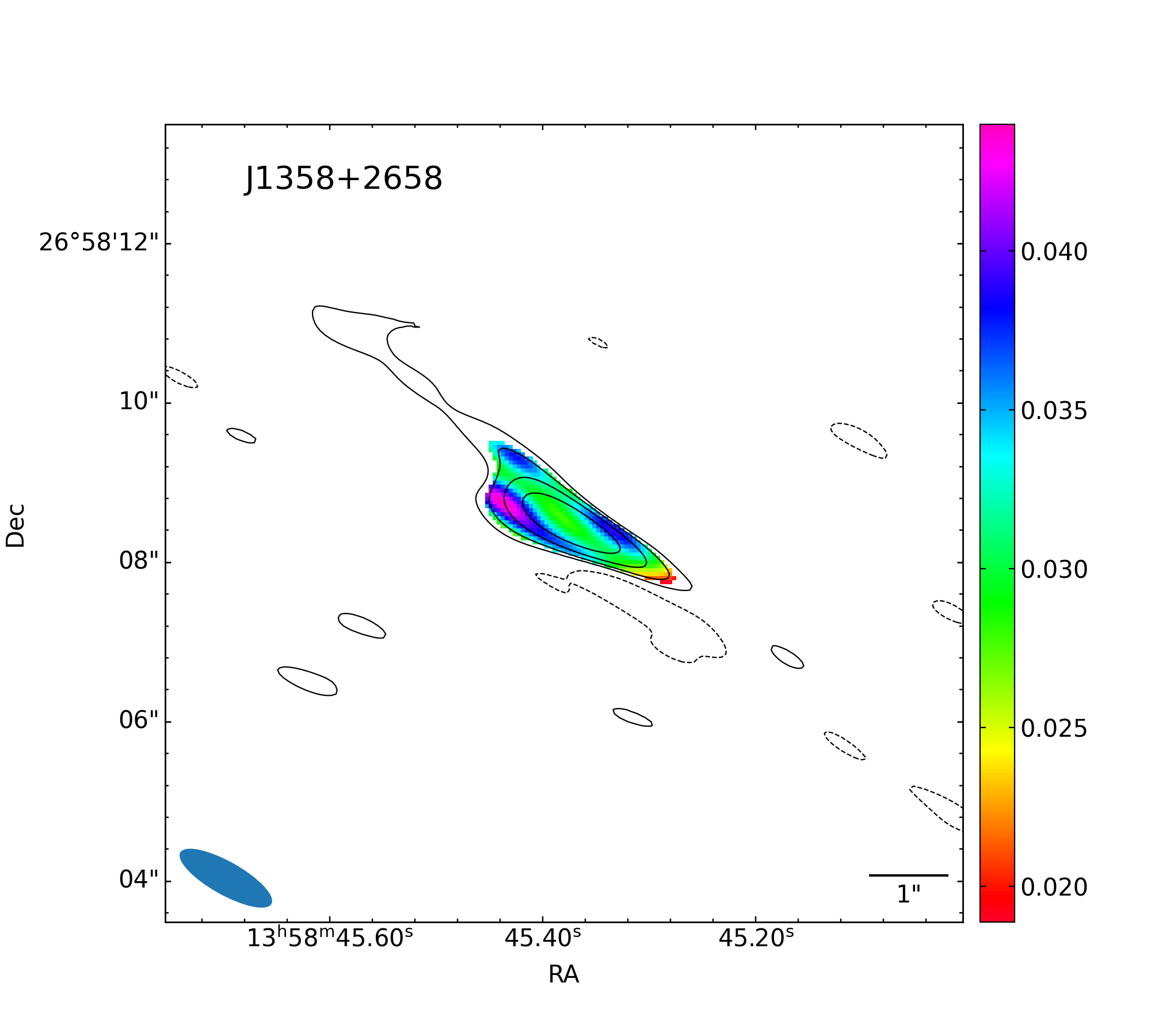}
         \caption{Spectral index error map, rms, contour levels, and beam size as in Fig.~\ref{fig:J1358spind}.} \label{fig:J1358spinderr}
     \end{subfigure}
     \hfill
     \\
     \begin{subfigure}[b]{0.47\textwidth}
         \centering
         \includegraphics[width=\textwidth]{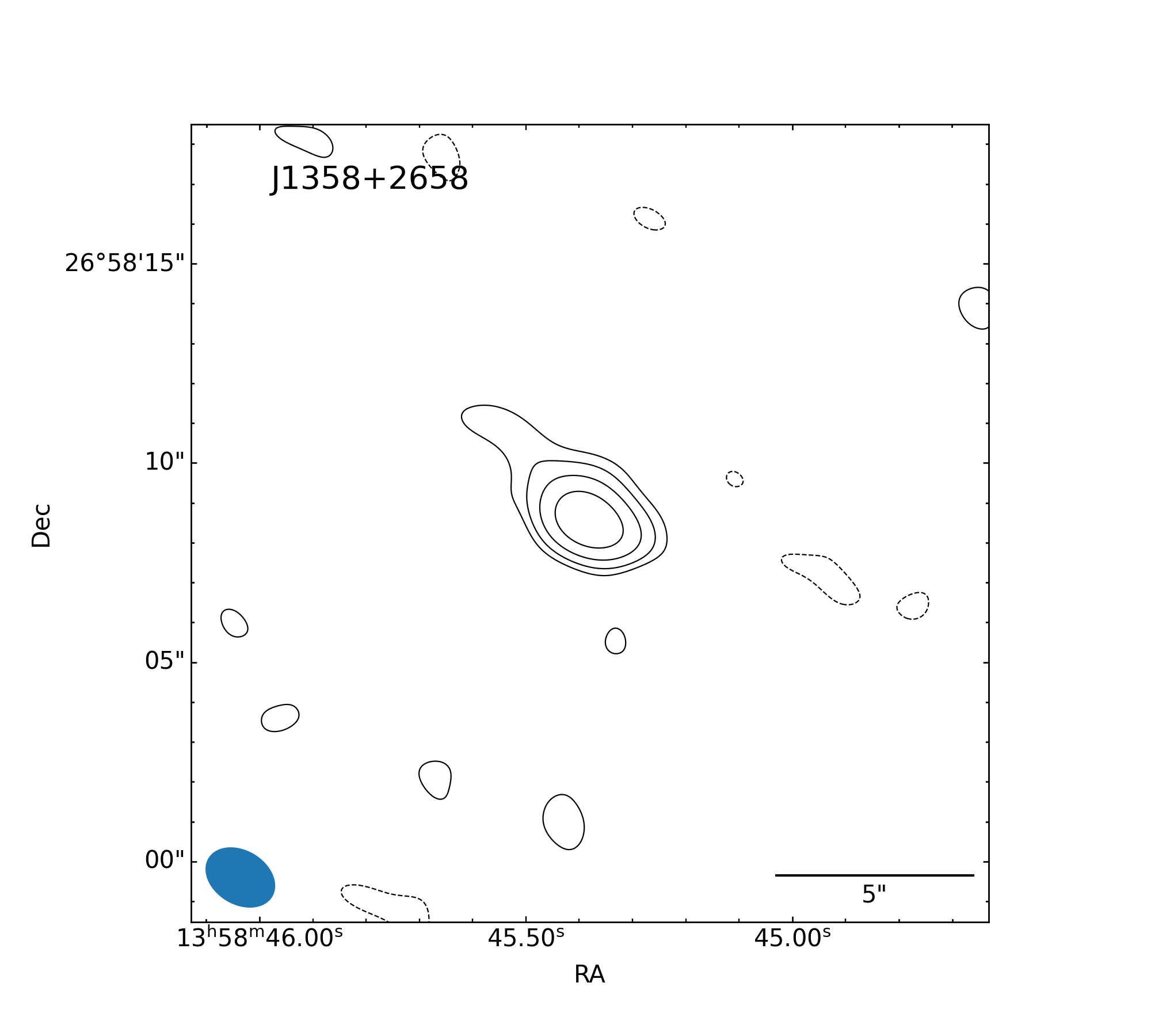}
         \caption{Tapered map with \texttt{uvtaper} = 90k$\lambda$, rms = 13$\mu$Jy beam$^{-1}$, contour levels at -3, 3 $\times$ 2$^n$, $n \in$ [0, 3], beam size 8.90 $\times$ 6.48~kpc.} \label{fig:J1358-90k}
     \end{subfigure}
          \hfill
     \begin{subfigure}[b]{0.47\textwidth}
         \centering
         \includegraphics[width=\textwidth]{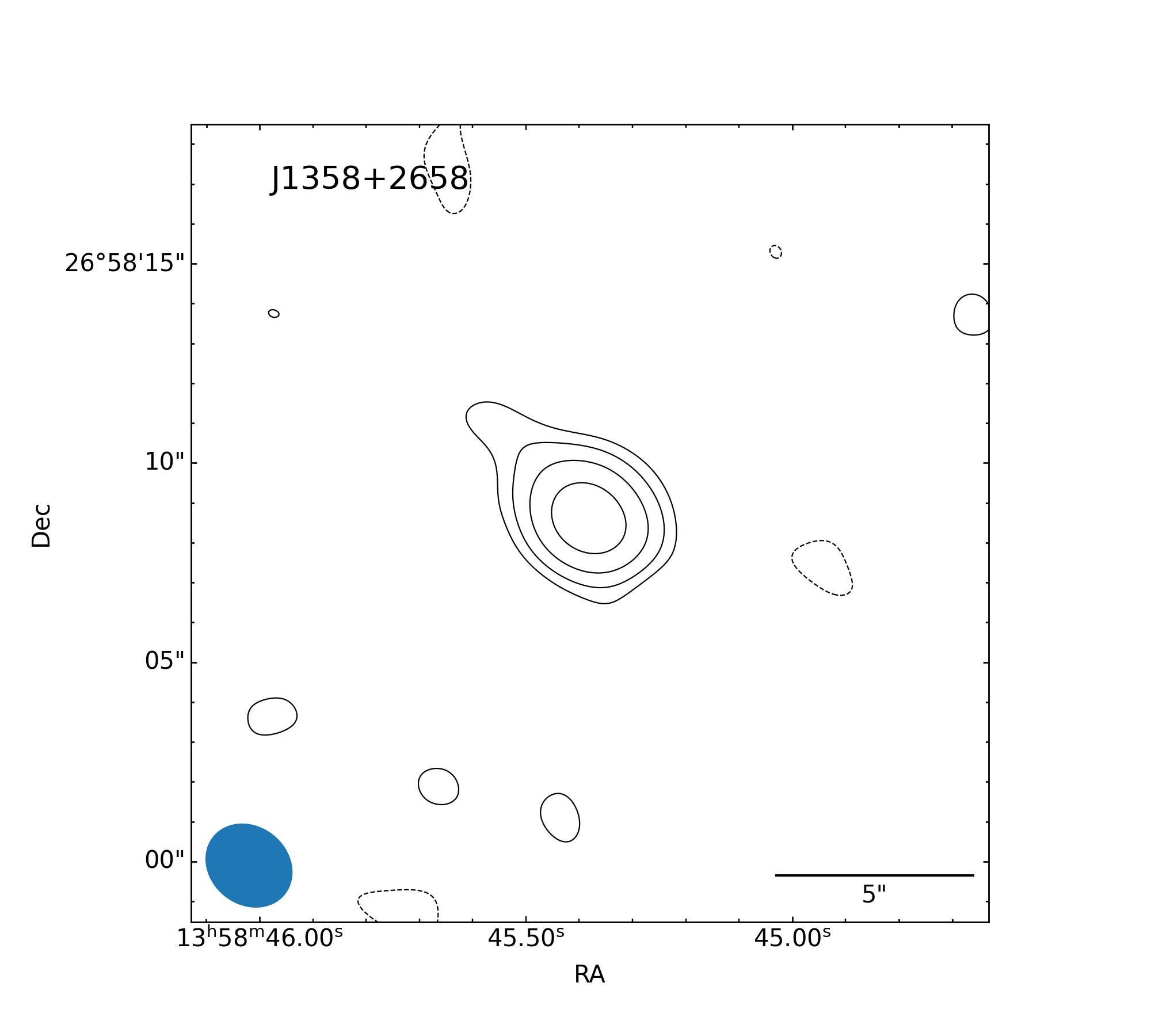}
         \caption{Tapered map with \texttt{uvtaper} = 60k$\lambda$, rms = 15$\mu$Jy beam$^{-1}$, contour levels at -3, 3 $\times$ 2$^n$, $n \in$ [0, 3], beam size 11.05 $\times$ 9.38~kpc.} \label{fig:J1358-60k}
     \end{subfigure}
          \hfill
     \\
     \begin{subfigure}[b]{0.47\textwidth}
         \centering
         \includegraphics[width=\textwidth]{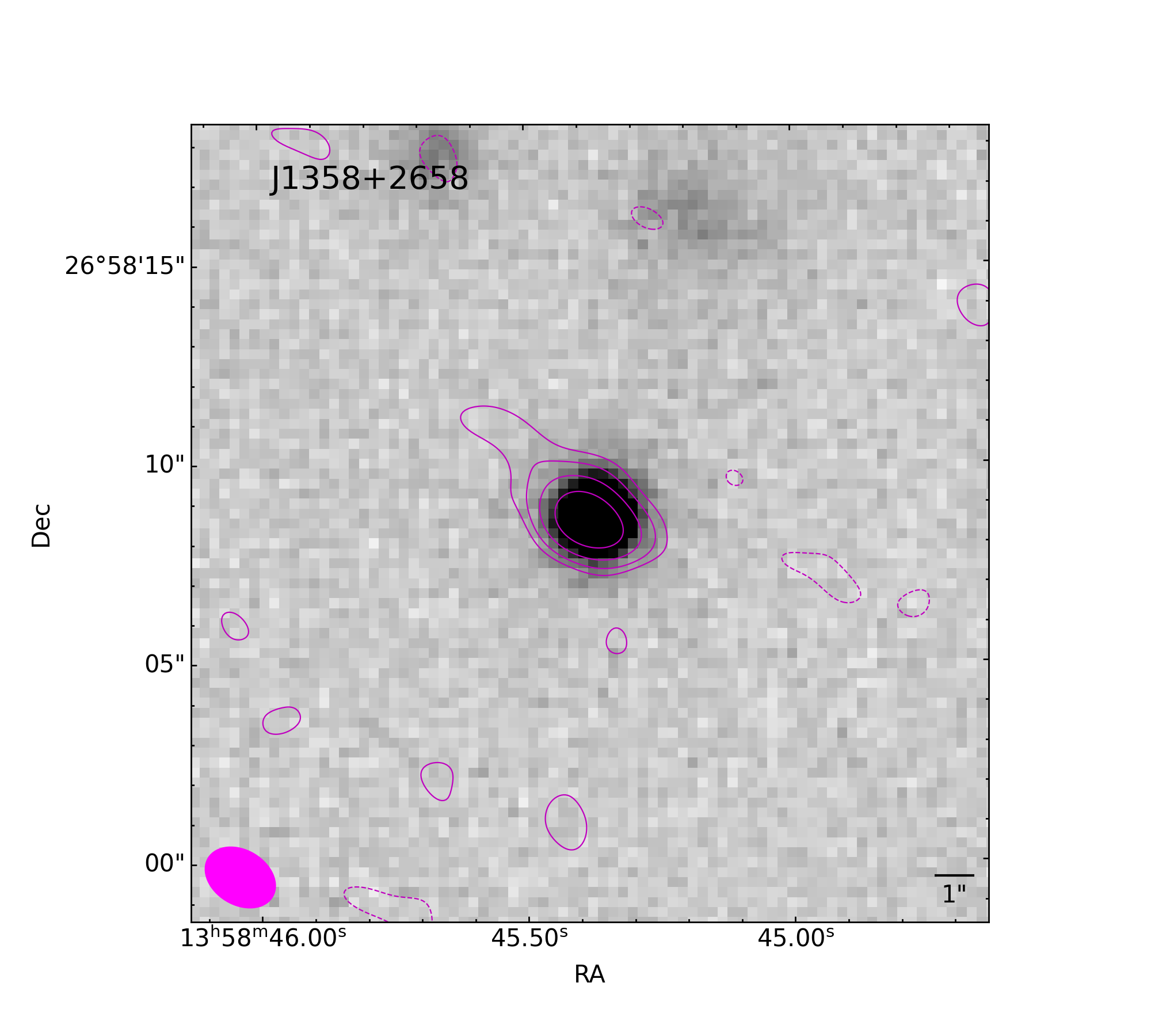}
         \caption{PanSTARRS $i$ band image of the host galaxy overlaid with the 90k$\lambda$ tapered map. Radio map properties as in Fig.~\ref{fig:J1358-90k}}. \label{fig:J1358-host}
     \end{subfigure}
        \caption{}
        \label{fig:J1358}
\end{figure*}


\begin{figure*}
     \centering
     \begin{subfigure}[b]{0.47\textwidth}
         \centering
         \includegraphics[width=\textwidth]{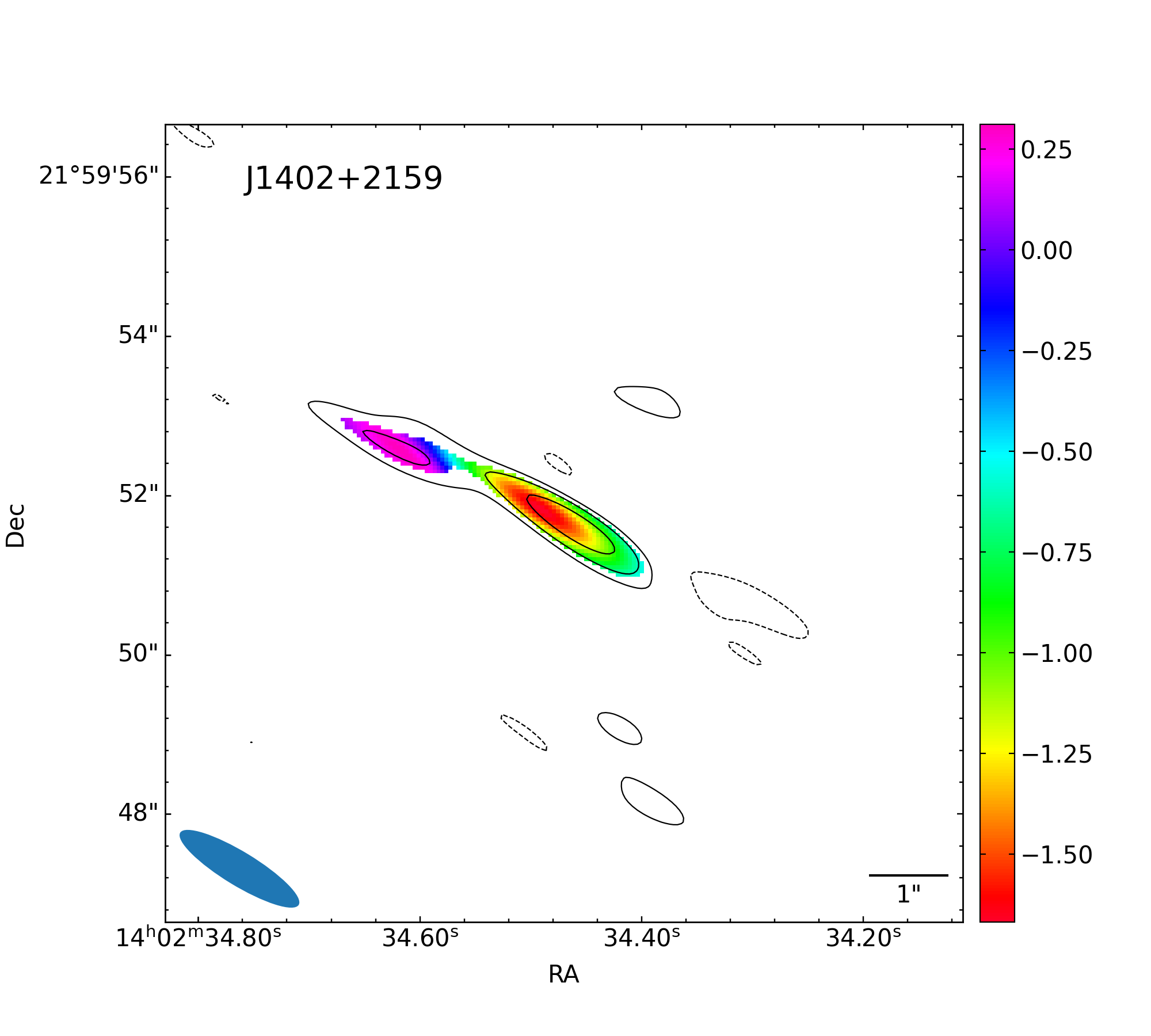}
         \caption{Spectral index map, rms = 16$\mu$Jy beam$^{-1}$, contour levels at -3, 3 $\times$ 2$^n$, $n \in$ [0, 2], beam size 2.20 $\times$ 0.54~kpc. } \label{fig:J1402spind}
     \end{subfigure}
     \hfill
     \begin{subfigure}[b]{0.47\textwidth}
         \centering
         \includegraphics[width=\textwidth]{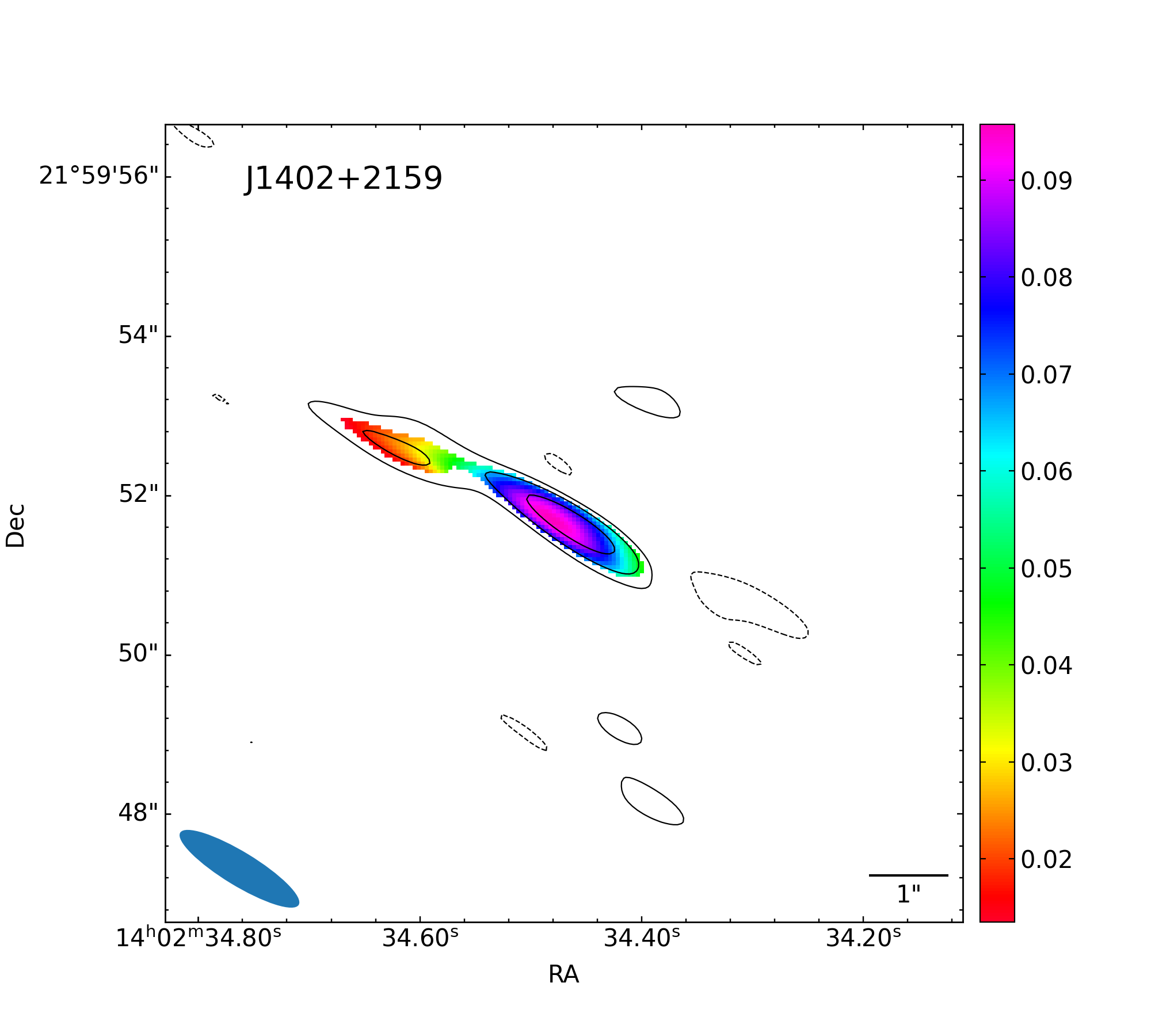}
         \caption{Spectral index error map, rms, contour levels, and beam size as in Fig.~\ref{fig:J1402spind}.} \label{fig:J1402spinderr}
     \end{subfigure}
     \hfill
     \\
     \begin{subfigure}[b]{0.47\textwidth}
         \centering
         \includegraphics[width=\textwidth]{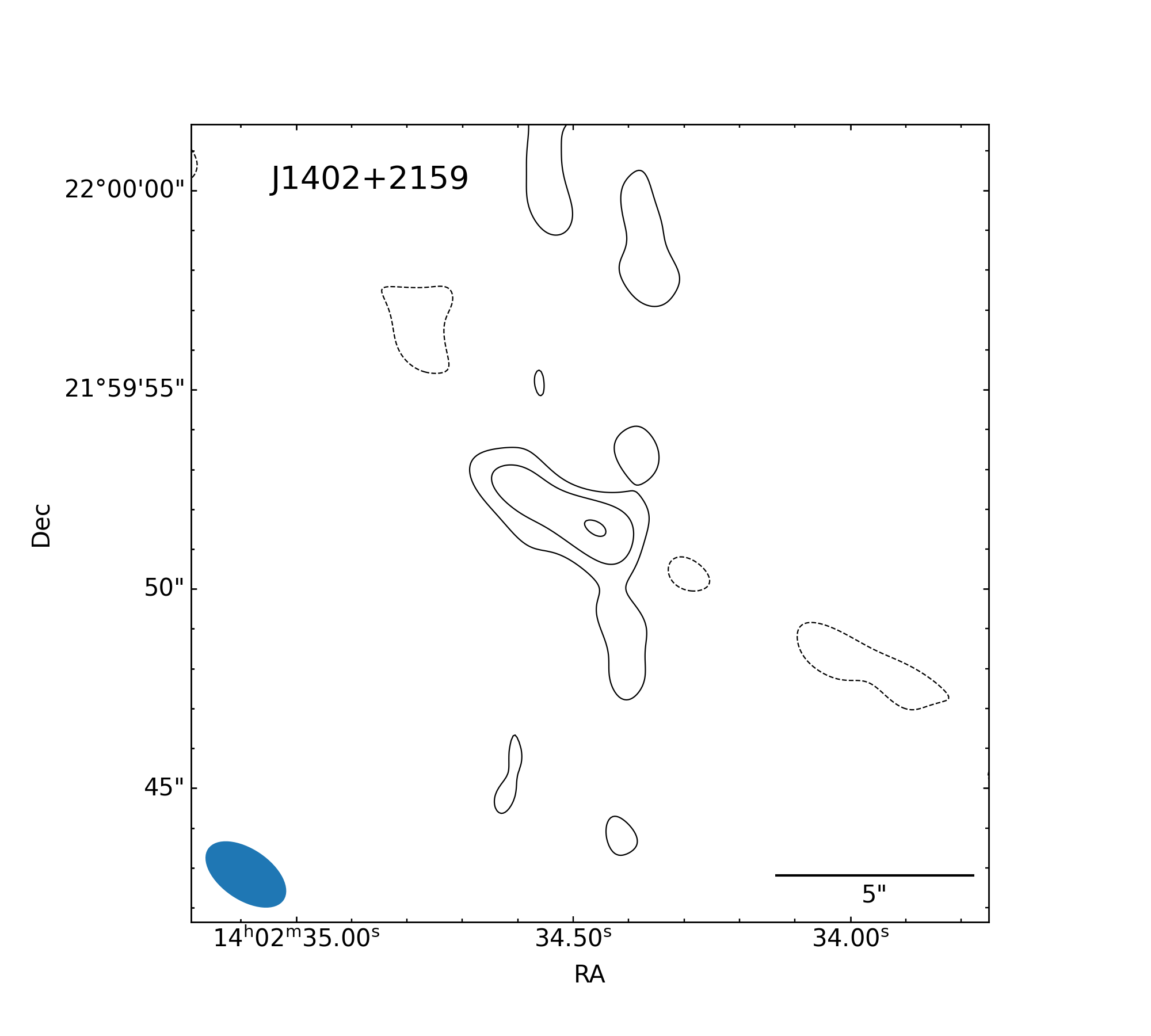}
         \caption{Tapered map with \texttt{uvtaper} = 90k$\lambda$, rms = 22$\mu$Jy beam$^{-1}$, contour levels at -3, 3 $\times$ 2$^n$, $n \in$ [0, 2], beam size 2.90 $\times$ 1.62~kpc.} \label{fig:J1402-90k}
     \end{subfigure}
          \hfill
     \begin{subfigure}[b]{0.47\textwidth}
         \centering
         \includegraphics[width=\textwidth]{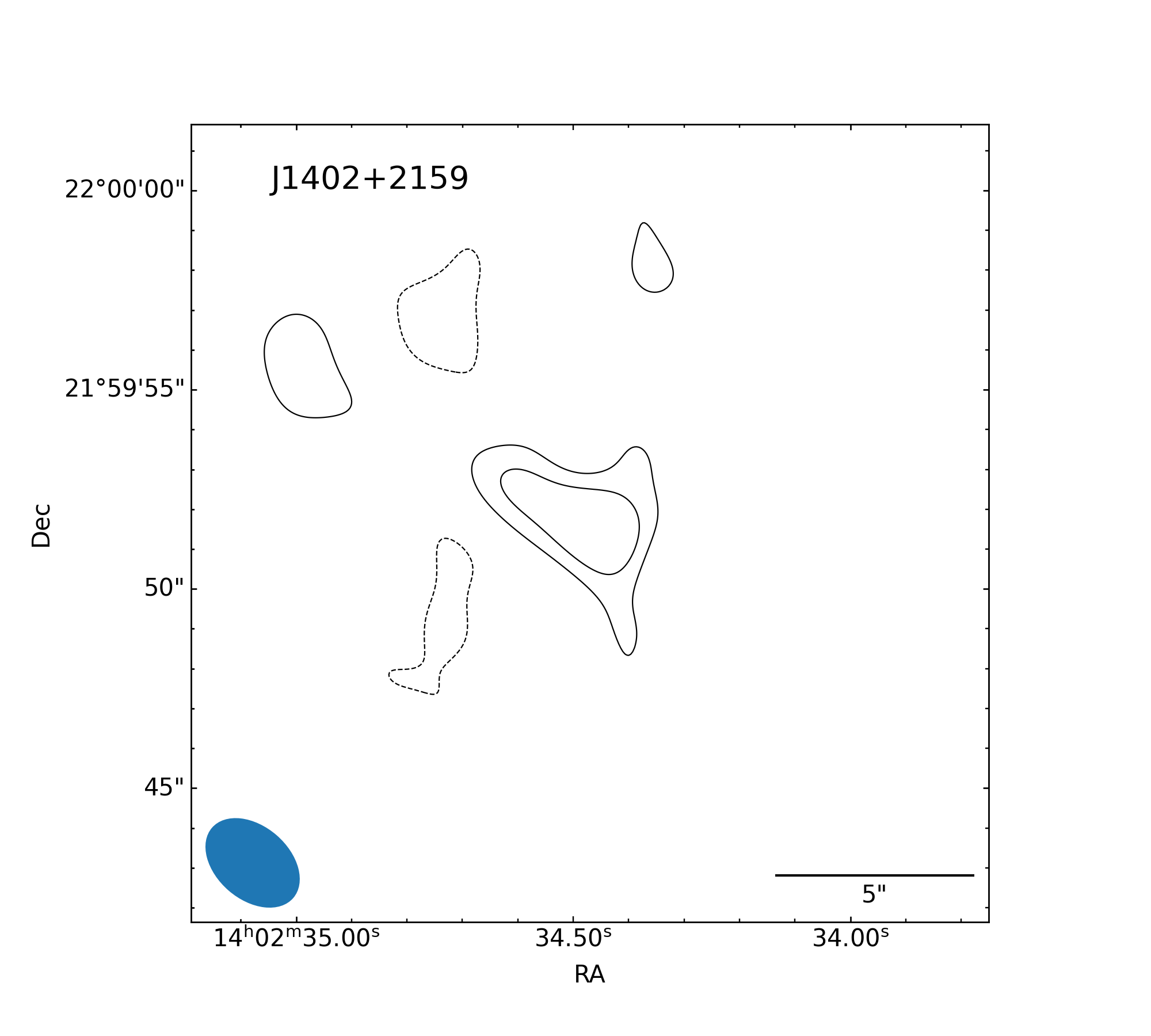}
         \caption{Tapered map with \texttt{uvtaper} = 60k$\lambda$, rms = 22$\mu$Jy beam$^{-1}$, contour levels at -3, 3, 6, beam size 3.42 $\times$ 2.33~kpc.} \label{fig:J1402-60k}
     \end{subfigure}
          \hfill
     \\
     \begin{subfigure}[b]{0.47\textwidth}
         \centering
         \includegraphics[width=\textwidth]{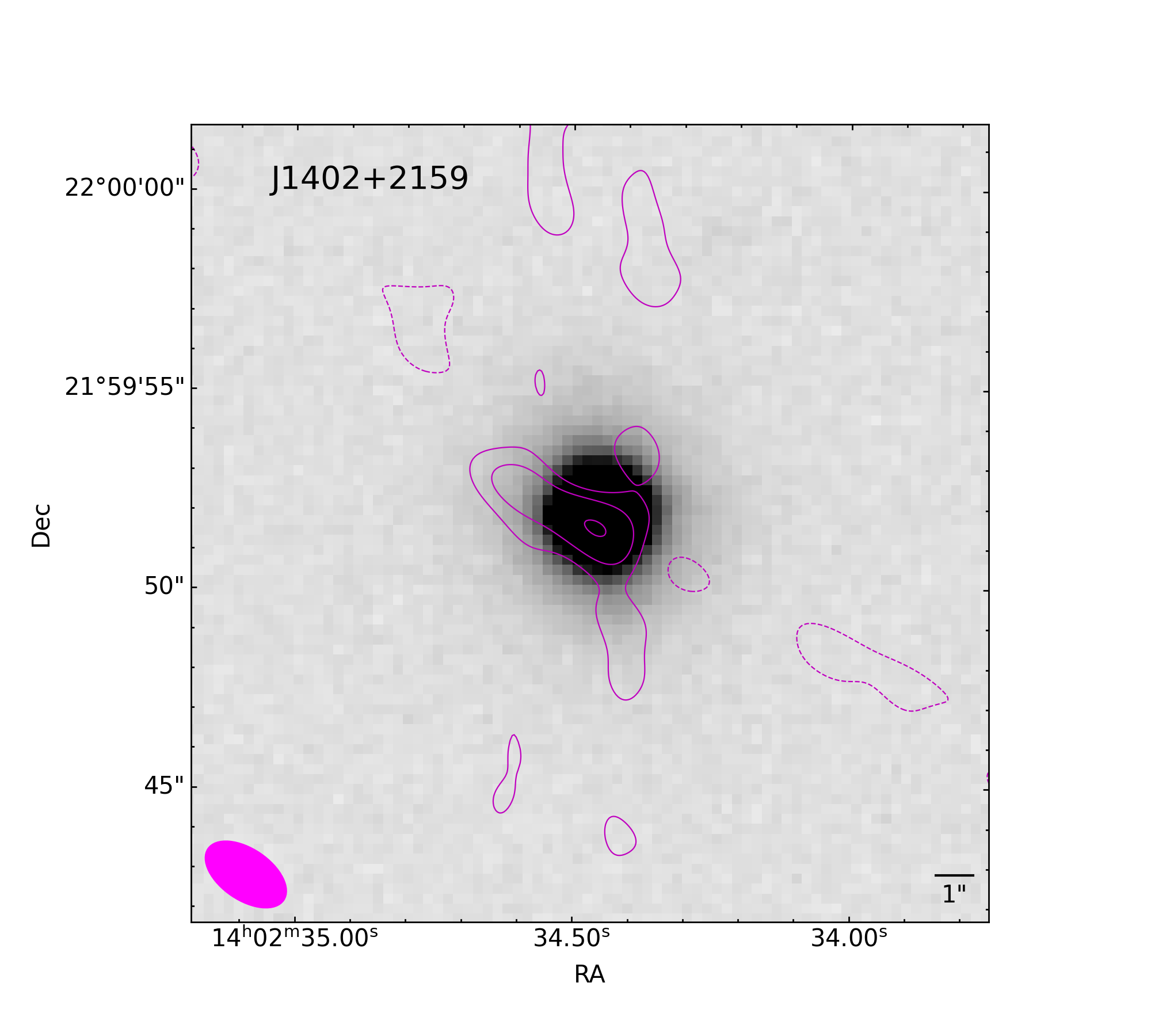}
         \caption{PanSTARRS $i$ band image of the host galaxy overlaid with the 90k$\lambda$ tapered map. Radio map properties as in Fig.~\ref{fig:J1402-90k}}. \label{fig:J1402-host}
     \end{subfigure}
        \caption{}
        \label{fig:J1402}
\end{figure*}


\begin{figure*}
     \centering
     \begin{subfigure}[b]{0.47\textwidth}
         \centering
         \includegraphics[width=\textwidth]{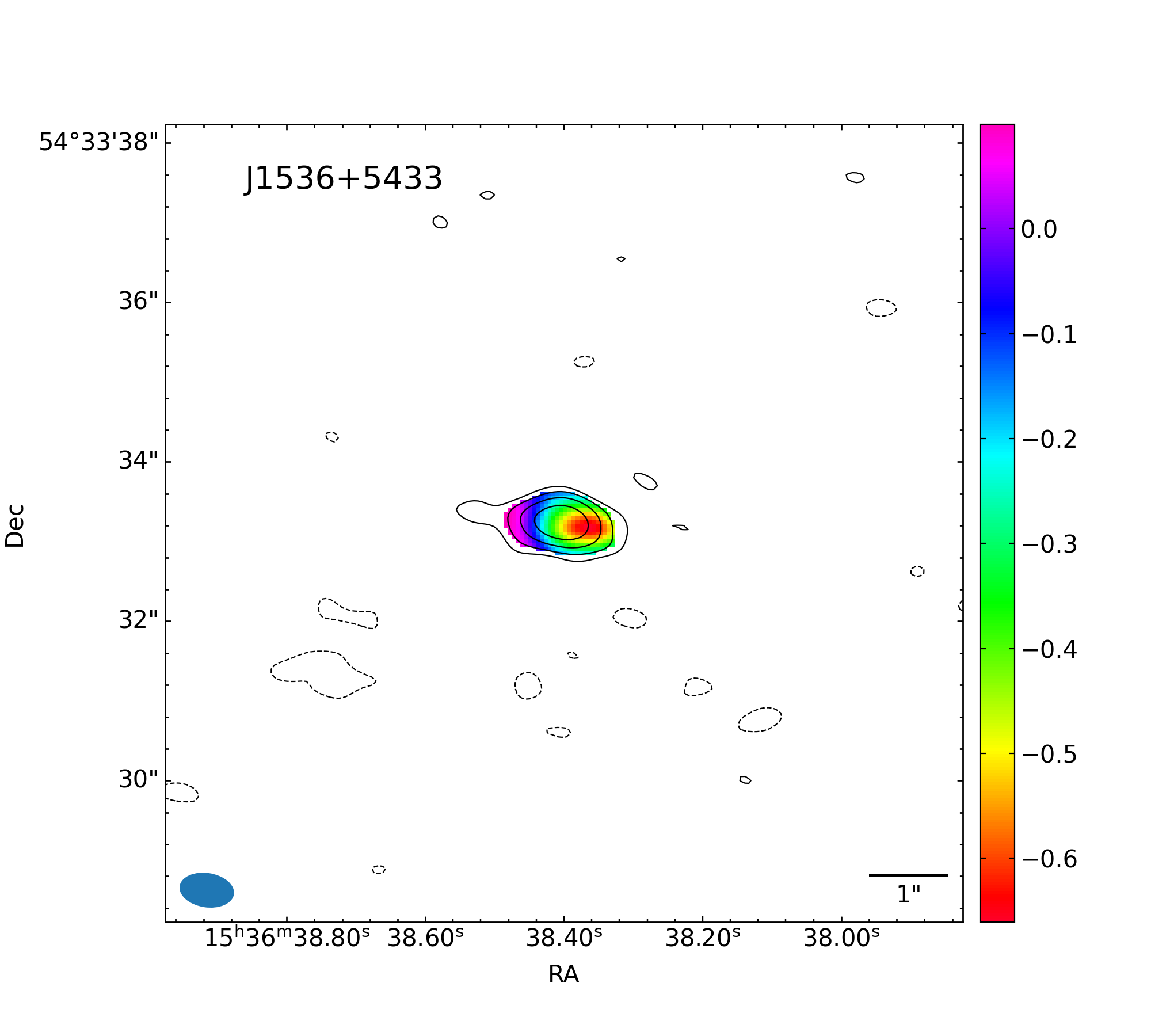}
         \caption{Spectral index map, rms = 10$\mu$Jy beam$^{-1}$, contour levels at -3, 3 $\times$ 2$^n$, $n \in$ [0, 3], beam size 0.53 $\times$ 0.33~kpc. } \label{fig:J1536spind}
     \end{subfigure}
     \hfill
     \begin{subfigure}[b]{0.47\textwidth}
         \centering
         \includegraphics[width=\textwidth]{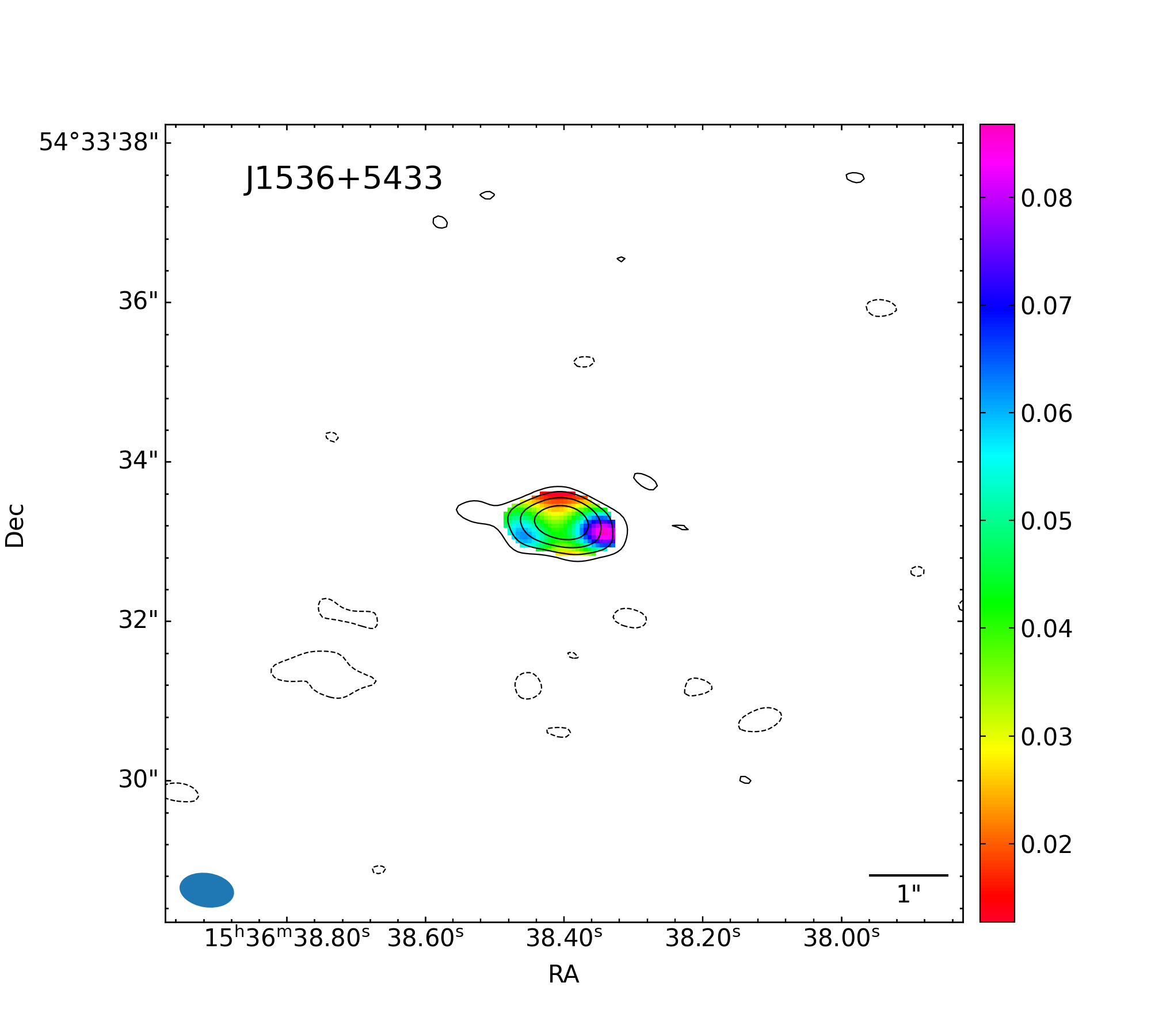}
         \caption{Spectral index error map, rms, contour levels, and beam size as in Fig.~\ref{fig:J1536spind}.} \label{fig:J1536spinderr}
     \end{subfigure}
     \hfill
     \\
     \begin{subfigure}[b]{0.47\textwidth}
         \centering
         \includegraphics[width=\textwidth]{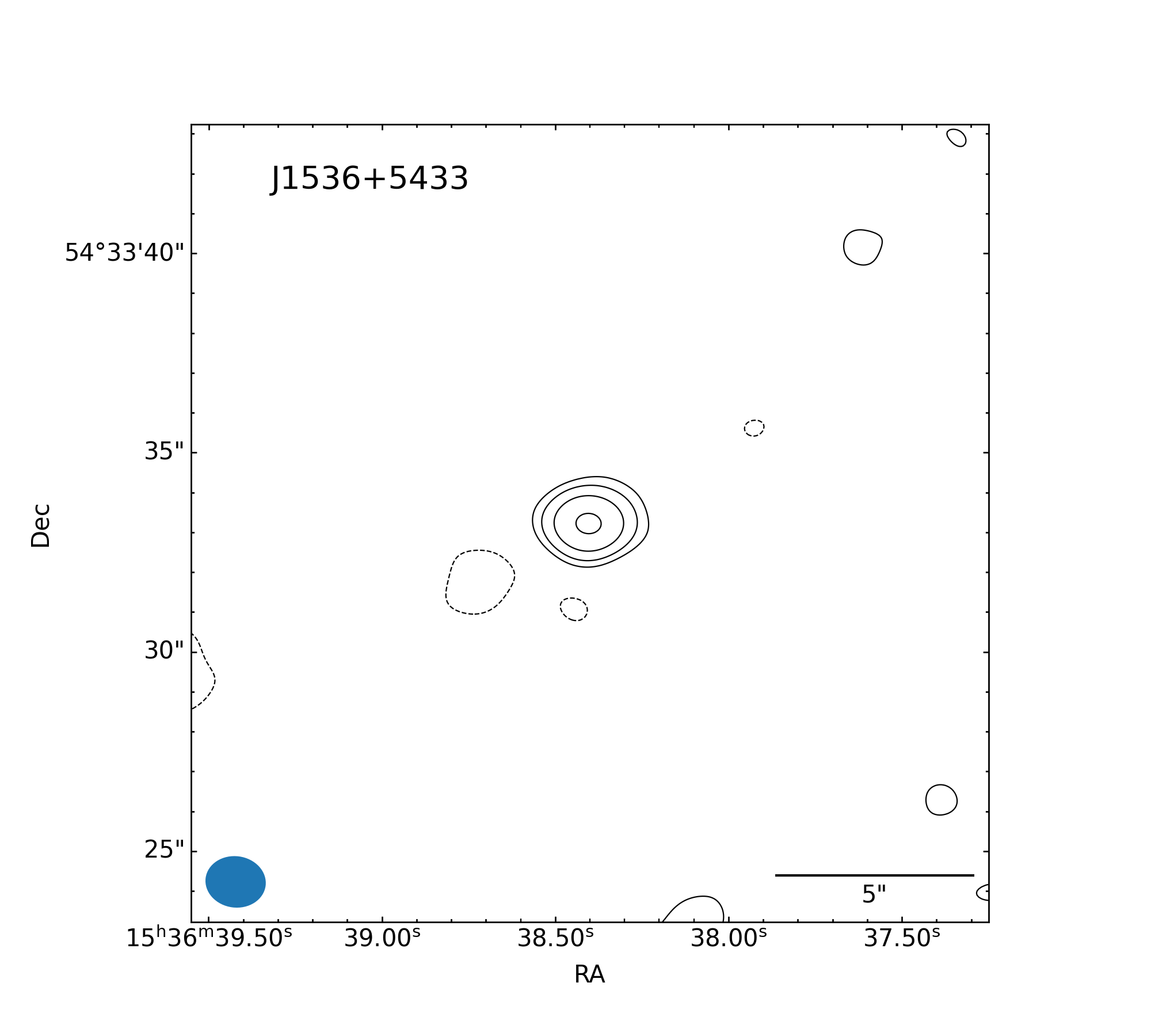}
         \caption{Tapered map with \texttt{uvtaper} = 90k$\lambda$, rms = 18$\mu$Jy beam$^{-1}$, contour levels at -3, 3 $\times$ 2$^n$, $n \in$ [0, 3], beam size 1.17 $\times$ 1.00~kpc.} \label{fig:J1536-90k}
     \end{subfigure}
          \hfill
     \begin{subfigure}[b]{0.47\textwidth}
         \centering
         \includegraphics[width=\textwidth]{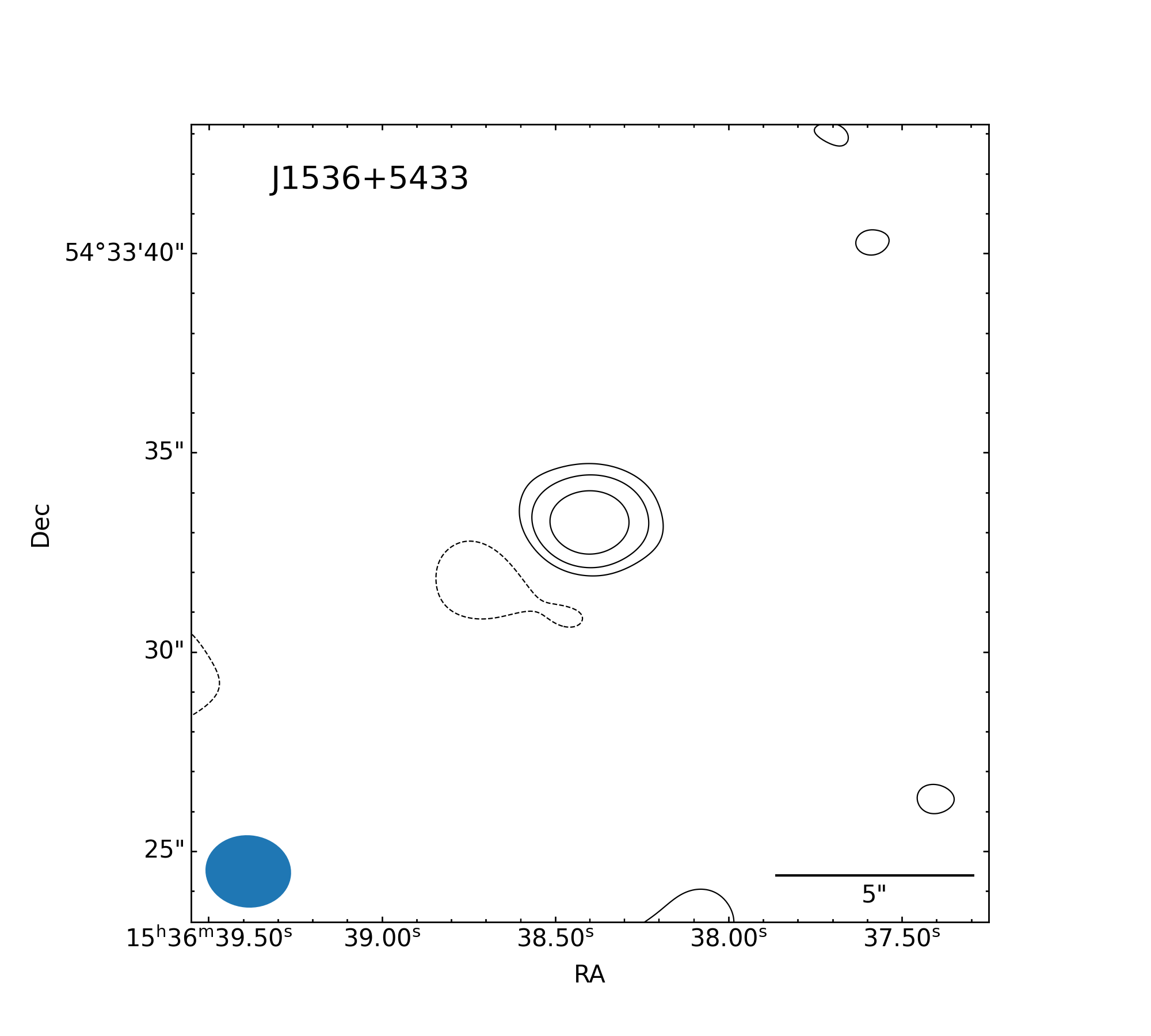}
         \caption{Tapered map with \texttt{uvtaper} = 60k$\lambda$, rms = 23$\mu$Jy beam$^{-1}$, contour levels at -3, 3 $\times$ 2$^n$, $n \in$ [0, 2], beam size 1.66 $\times$ 1.40~kpc.} \label{fig:J1536-60k}
     \end{subfigure}
          \hfill
     \\
     \begin{subfigure}[b]{0.47\textwidth}
         \centering
         \includegraphics[width=\textwidth]{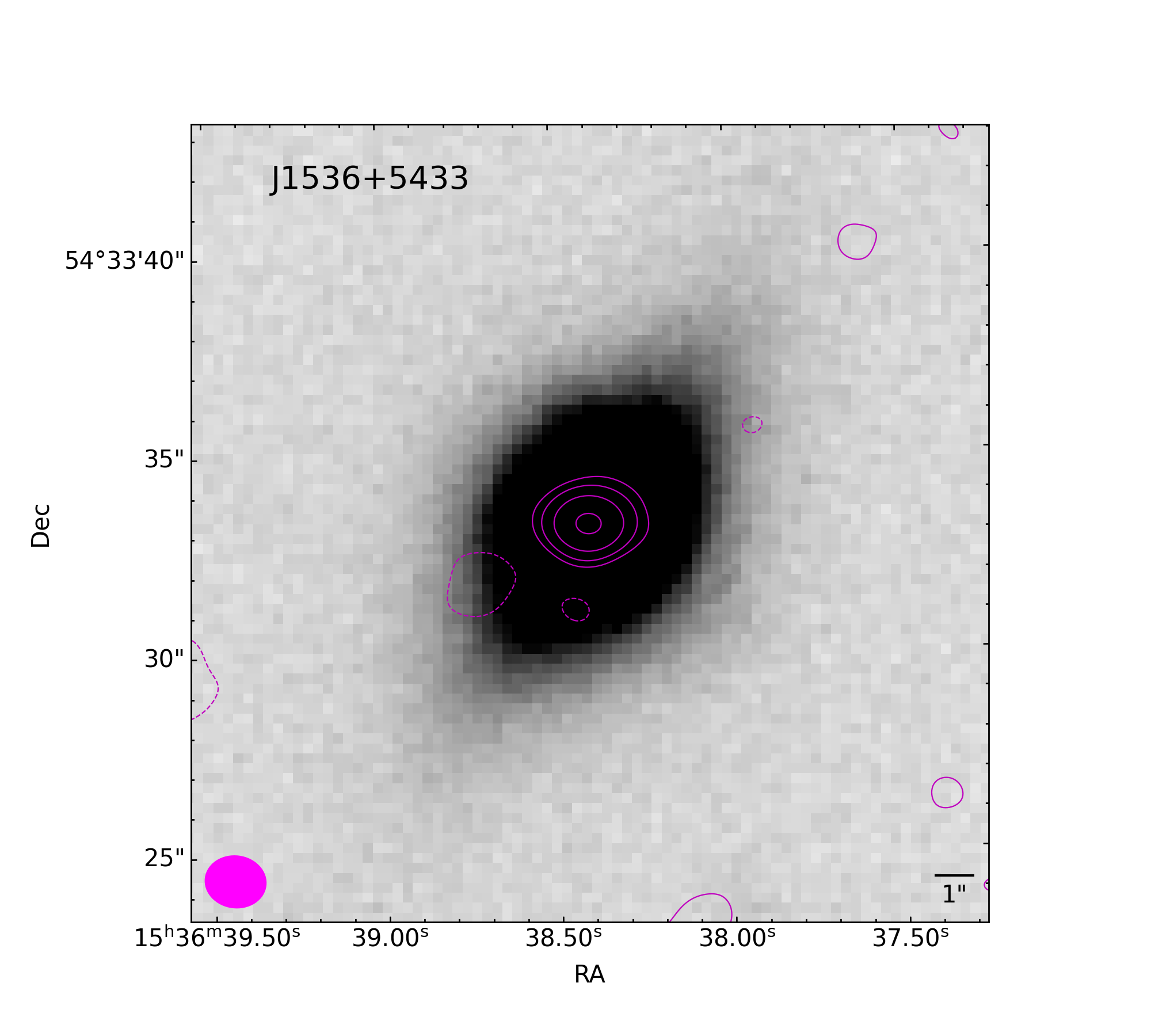}
         \caption{PanSTARRS $i$ band image of the host galaxy overlaid with the 90k$\lambda$ tapered map. Radio map properties as in Fig.~\ref{fig:J1536-90k}}. \label{fig:J1536-host}
     \end{subfigure}
        \caption{}
        \label{fig:J1536}
\end{figure*}


\begin{figure*}
     \centering
     \begin{subfigure}[b]{0.47\textwidth}
         \centering
         \includegraphics[width=\textwidth]{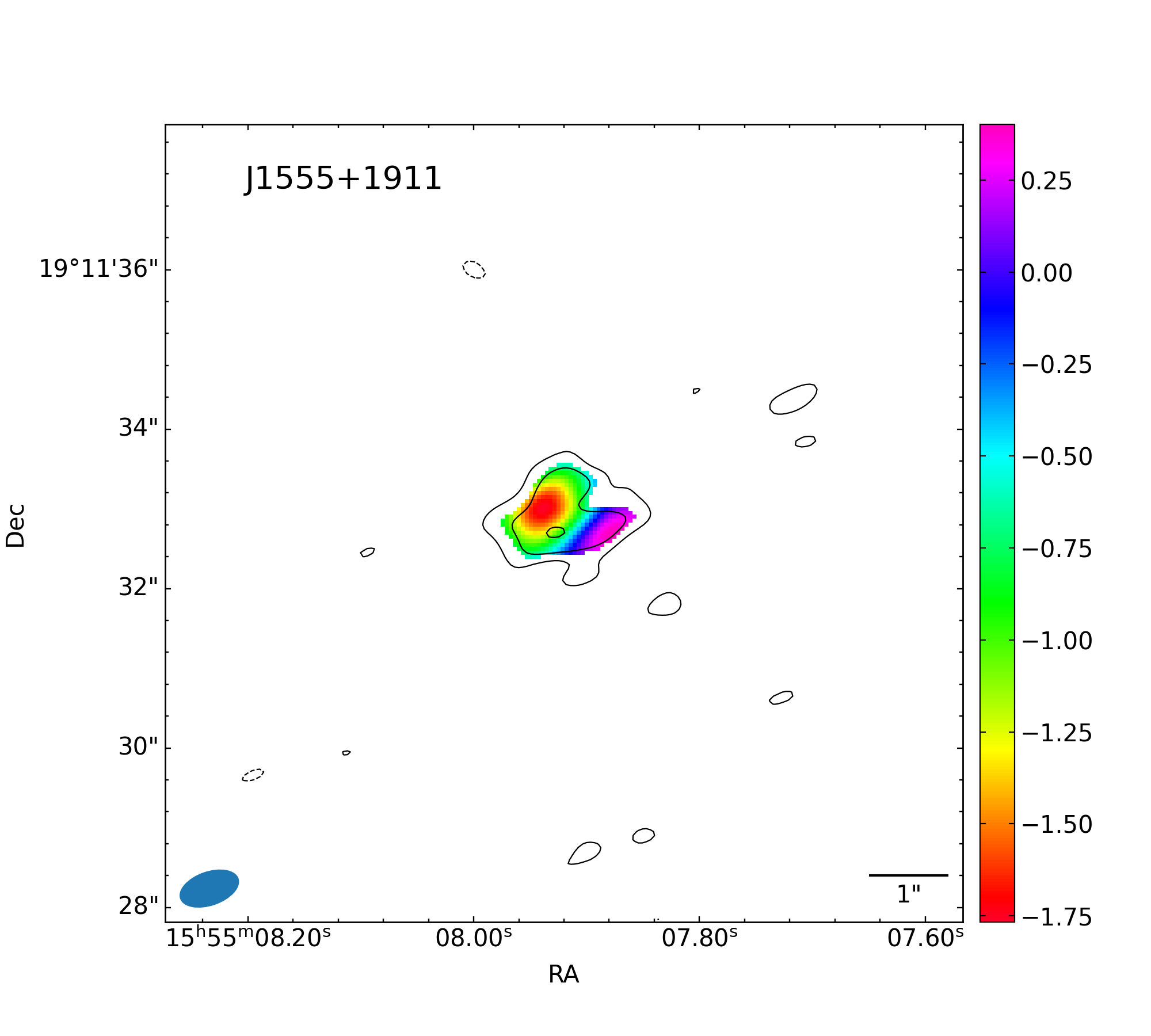}
         \caption{Spectral index map, rms = 11$\mu$Jy beam$^{-1}$, contour levels at -3, 3 $\times$ 2$^n$, $n \in$ [0, 2], beam size 0.54 $\times$ 0.30~kpc. } \label{fig:J1555spind}
     \end{subfigure}
     \hfill
     \begin{subfigure}[b]{0.47\textwidth}
         \centering
         \includegraphics[width=\textwidth]{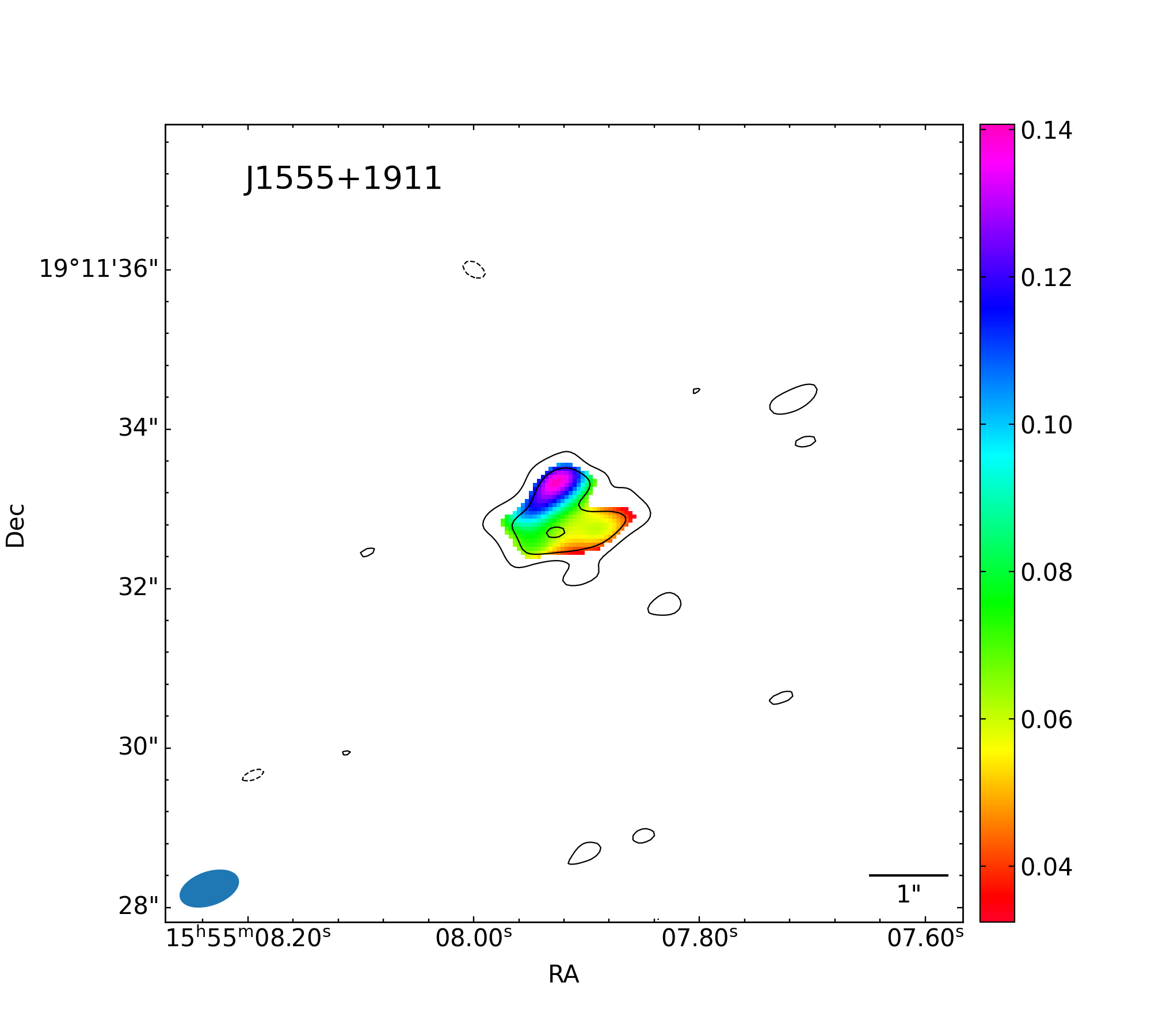}
         \caption{Spectral index error map, rms, contour levels, and beam size as in Fig.~\ref{fig:J1555spind}.} \label{fig:J1555spinderr}
     \end{subfigure}
     \hfill
     \\
     \begin{subfigure}[b]{0.47\textwidth}
         \centering
         \includegraphics[width=\textwidth]{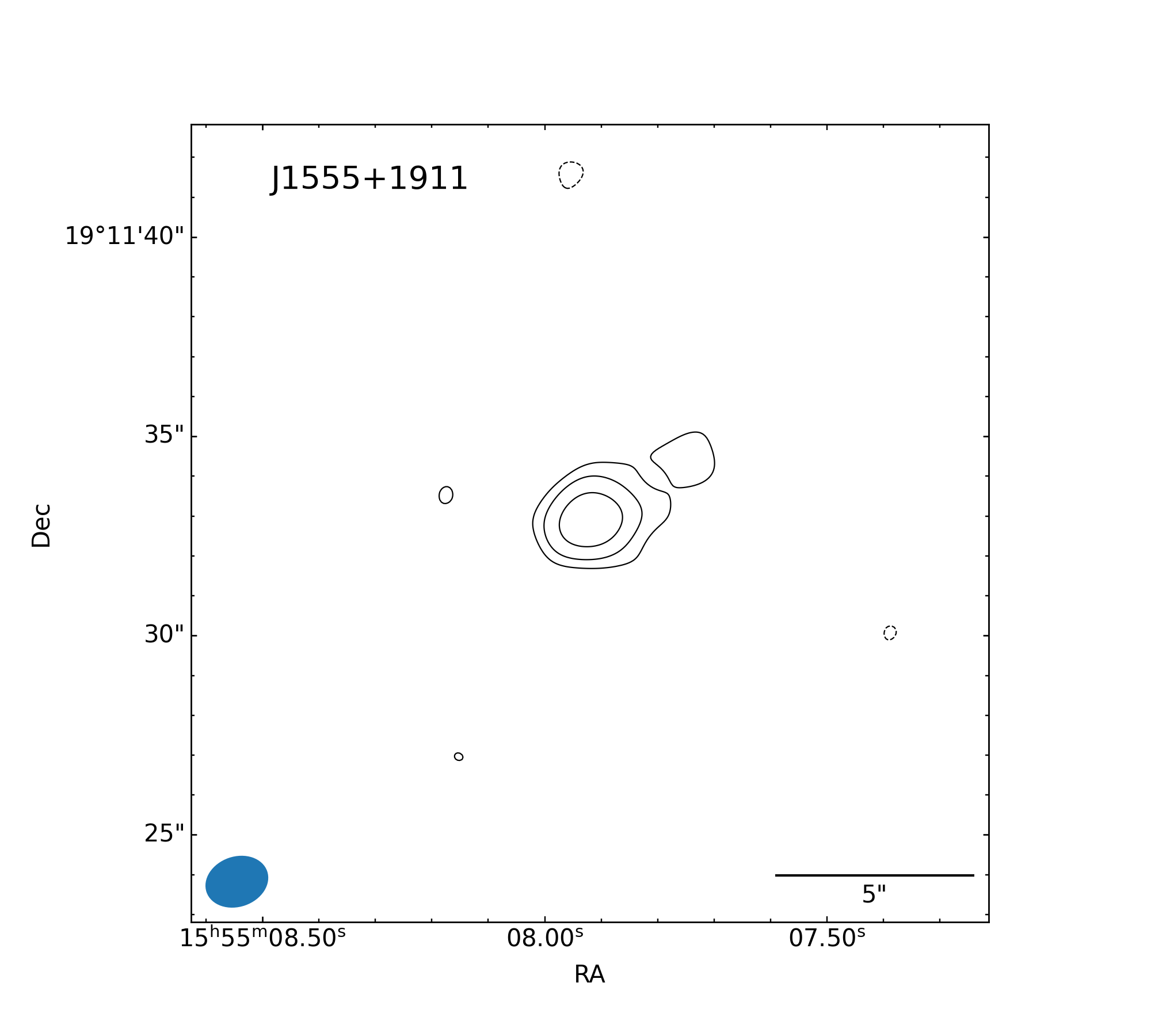}
         \caption{Tapered map with \texttt{uvtaper} = 90k$\lambda$, rms = 15$\mu$Jy beam$^{-1}$, contour levels at -3, 3 $\times$ 2$^n$, $n \in$ [0, 2], beam size 1.12 $\times$ 0.88~kpc.} \label{fig:J1555-90k}
     \end{subfigure}
          \hfill
     \begin{subfigure}[b]{0.47\textwidth}
         \centering
         \includegraphics[width=\textwidth]{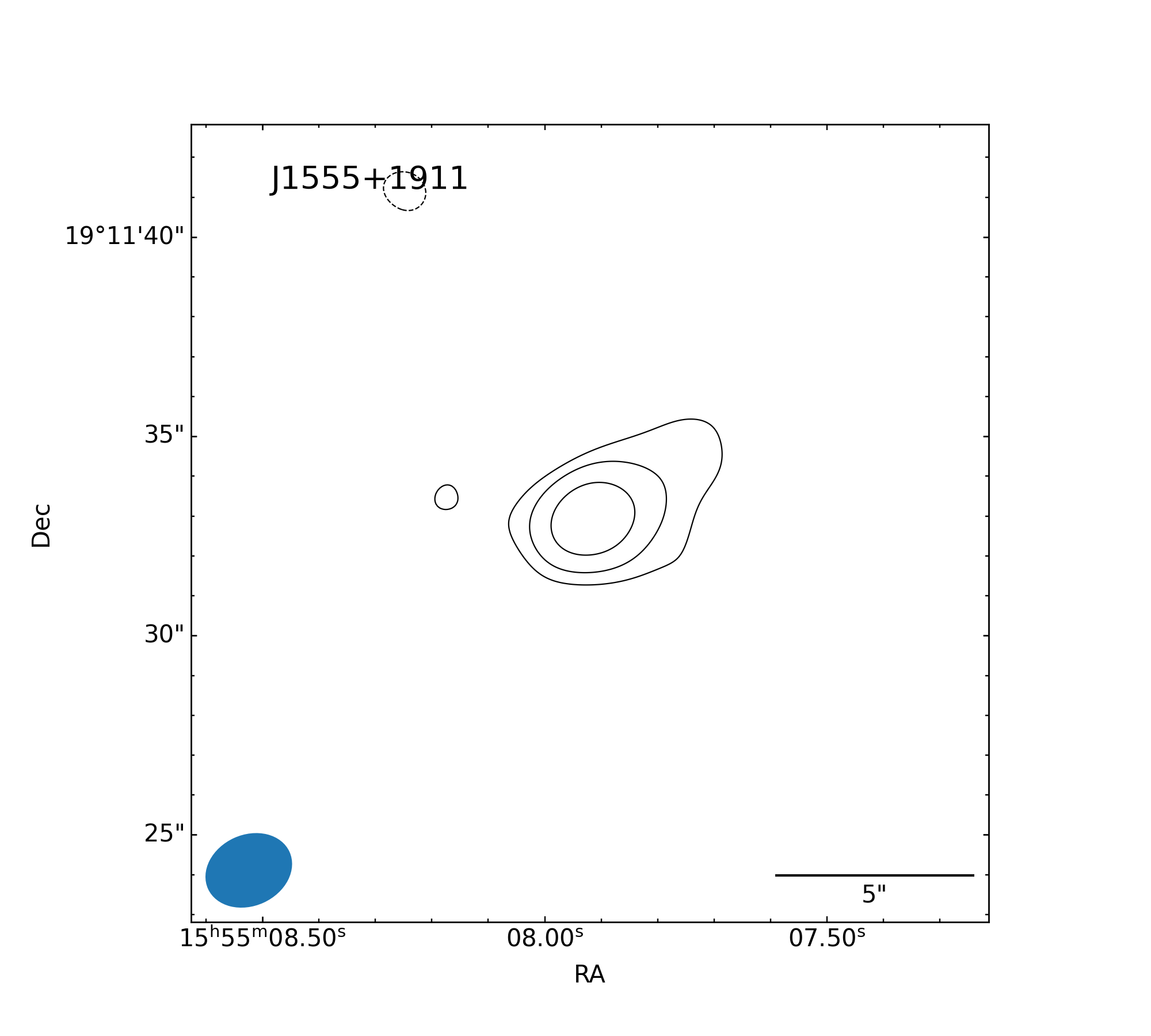}
         \caption{Tapered map with \texttt{uvtaper} = 60k$\lambda$, rms = 17$\mu$Jy beam$^{-1}$, contour levels at -3, 3 $\times$ 2$^n$, $n \in$ [0, 2], beam size 1.57 $\times$ 1.24~kpc.} \label{fig:J1555-60k}
     \end{subfigure}
          \hfill
     \\
     \begin{subfigure}[b]{0.47\textwidth}
         \centering
         \includegraphics[width=\textwidth]{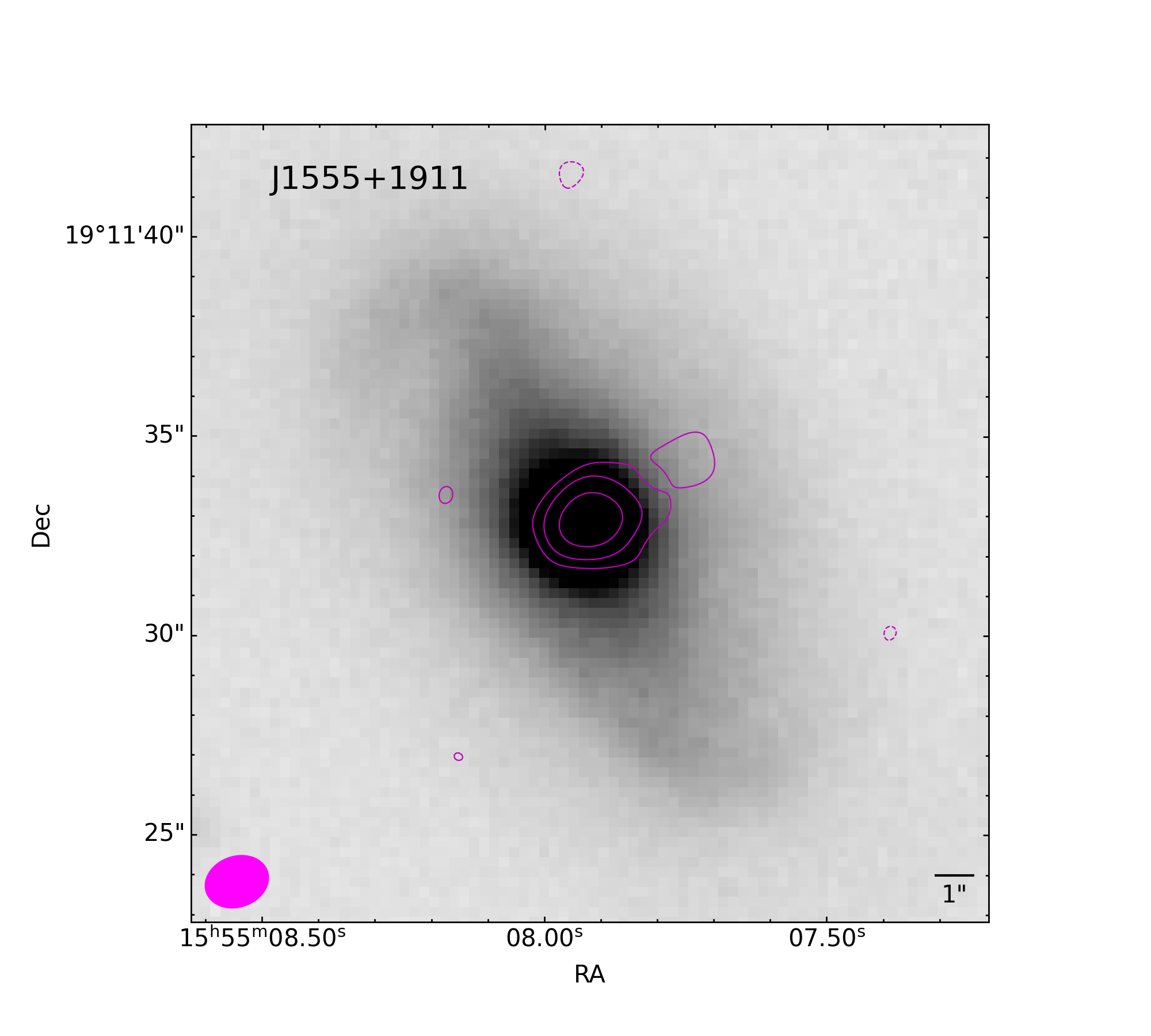}
         \caption{PanSTARRS $i$ band image of the host galaxy overlaid with the 90k$\lambda$ tapered map. Radio map properties as in Fig.~\ref{fig:J1555-90k}}. \label{fig:J1555-host}
     \end{subfigure}
        \caption{}
        \label{fig:J1555}
\end{figure*}


\begin{figure*}
     \centering
     \begin{subfigure}[b]{0.47\textwidth}
         \centering
         \includegraphics[width=\textwidth]{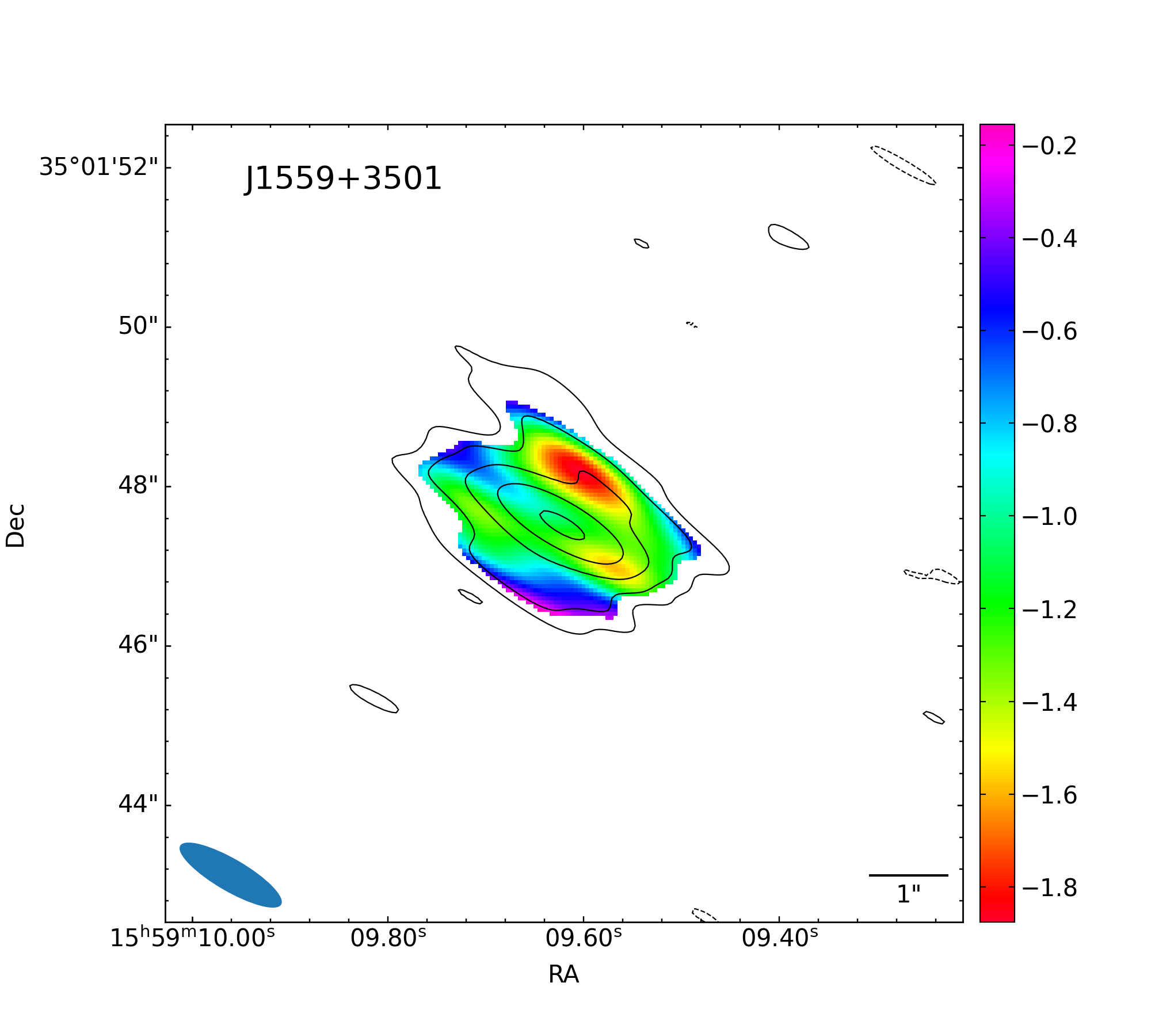}
         \caption{Spectral index map, rms = 11$\mu$Jy beam$^{-1}$, contour levels at -3, 3 $\times$ 2$^n$, $n \in$ [0, 4], beam size 0.91 $\times$ 0.25~kpc. } \label{fig:J1559spind}
     \end{subfigure}
     \hfill
     \begin{subfigure}[b]{0.47\textwidth}
         \centering
         \includegraphics[width=\textwidth]{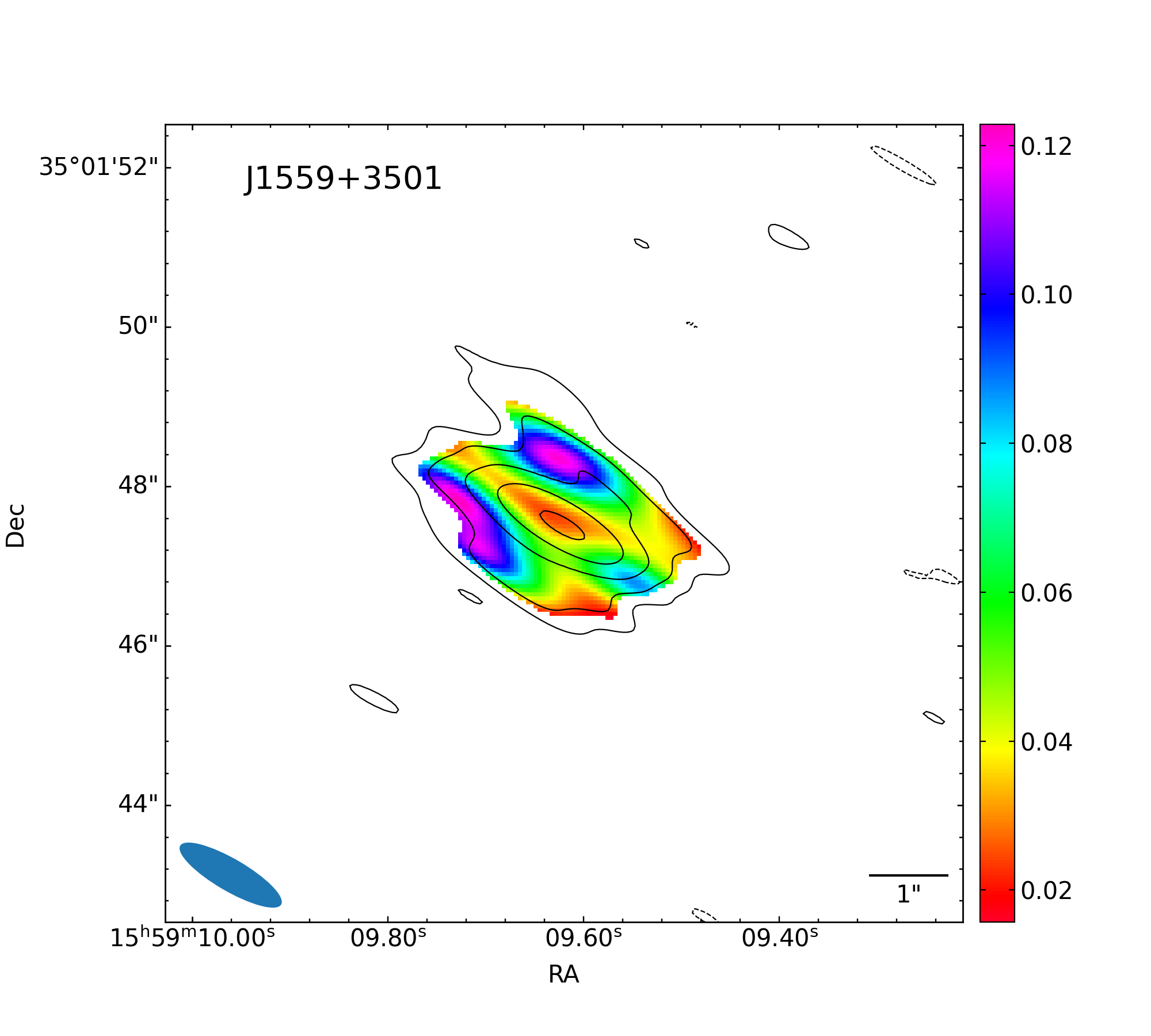}
         \caption{Spectral index error map, rms, contour levels, and beam size as in Fig.~\ref{fig:J1559spind}.} \label{fig:J1559spinderr}
     \end{subfigure}
     \hfill
     \\
     \begin{subfigure}[b]{0.47\textwidth}
         \centering
         \includegraphics[width=\textwidth]{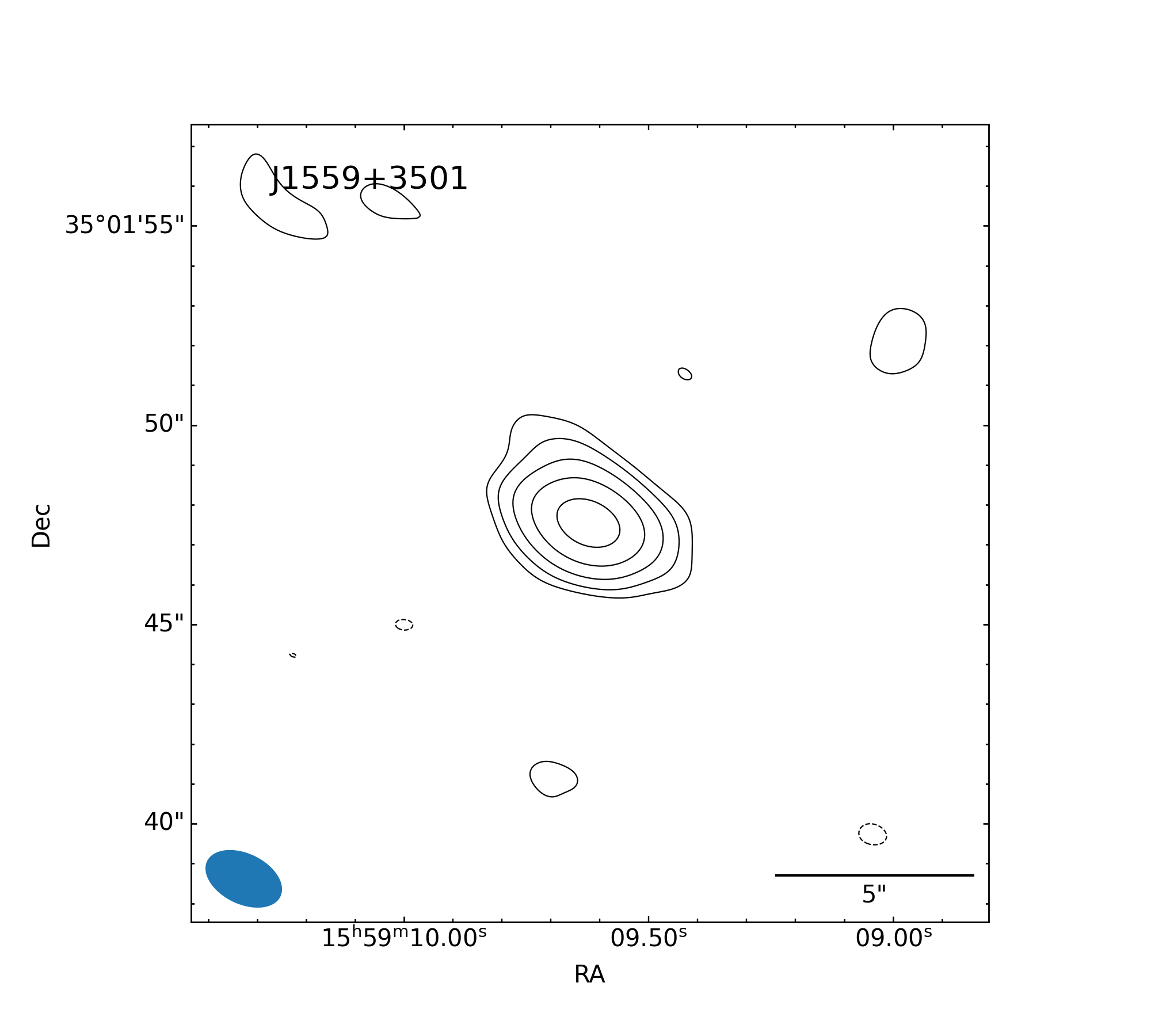}
         \caption{Tapered map with \texttt{uvtaper} = 90k$\lambda$, rms = 14$\mu$Jy beam$^{-1}$, contour levels at -3, 3 $\times$ 2$^n$, $n \in$ [0, 4], beam size 1.26 $\times$ 0.79~kpc.} \label{fig:J1559-90k}
     \end{subfigure}
          \hfill
     \begin{subfigure}[b]{0.47\textwidth}
         \centering
         \includegraphics[width=\textwidth]{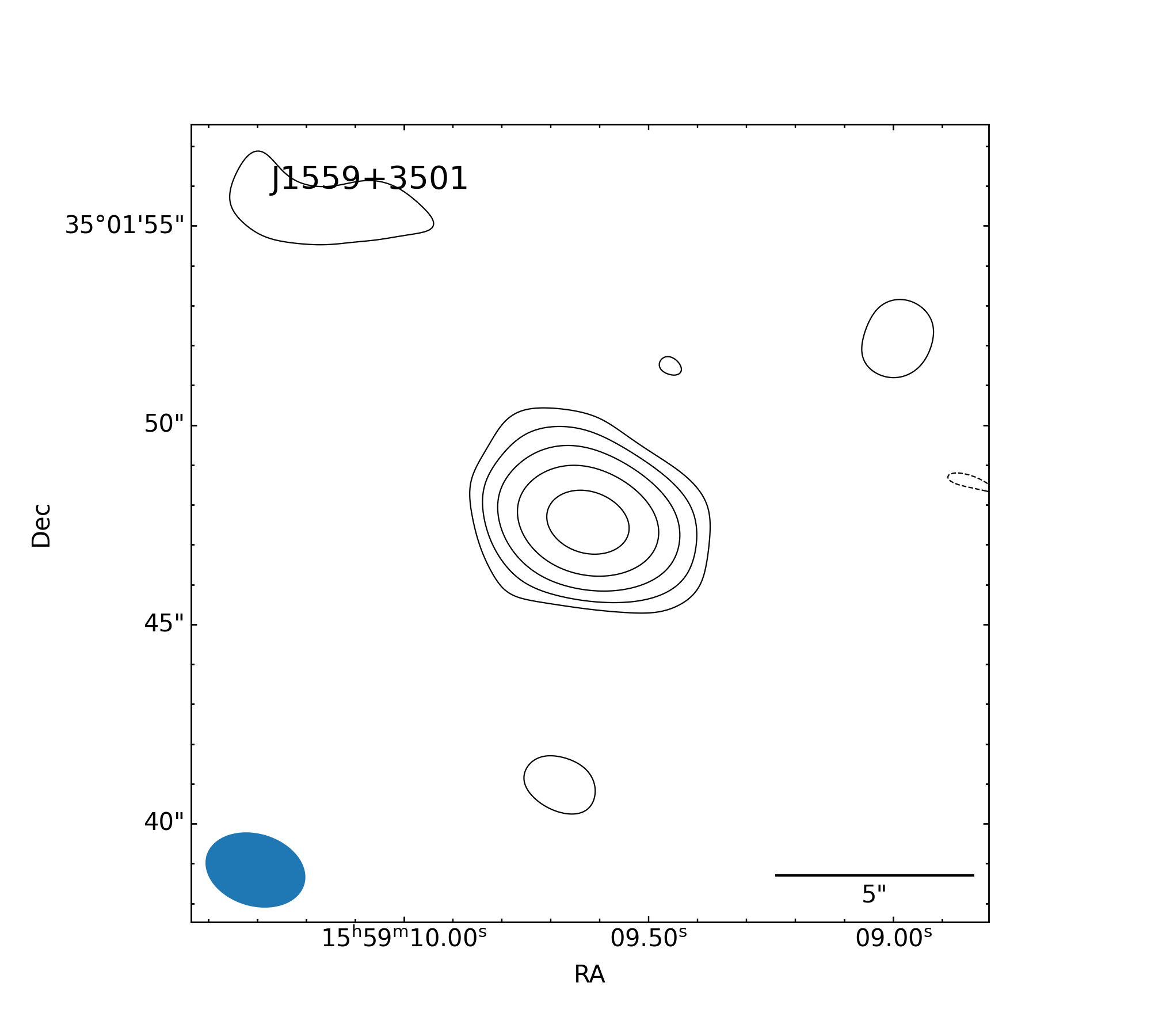}
         \caption{Tapered map with \texttt{uvtaper} = 60k$\lambda$, rms = 16$\mu$Jy beam$^{-1}$, contour levels at -3, 3 $\times$ 2$^n$, $n \in$ [0, 4], beam size 1.59 $\times$ 1.13~kpc.} \label{fig:J1559-60k}
     \end{subfigure}
          \hfill
     \\
     \begin{subfigure}[b]{0.47\textwidth}
         \centering
         \includegraphics[width=\textwidth]{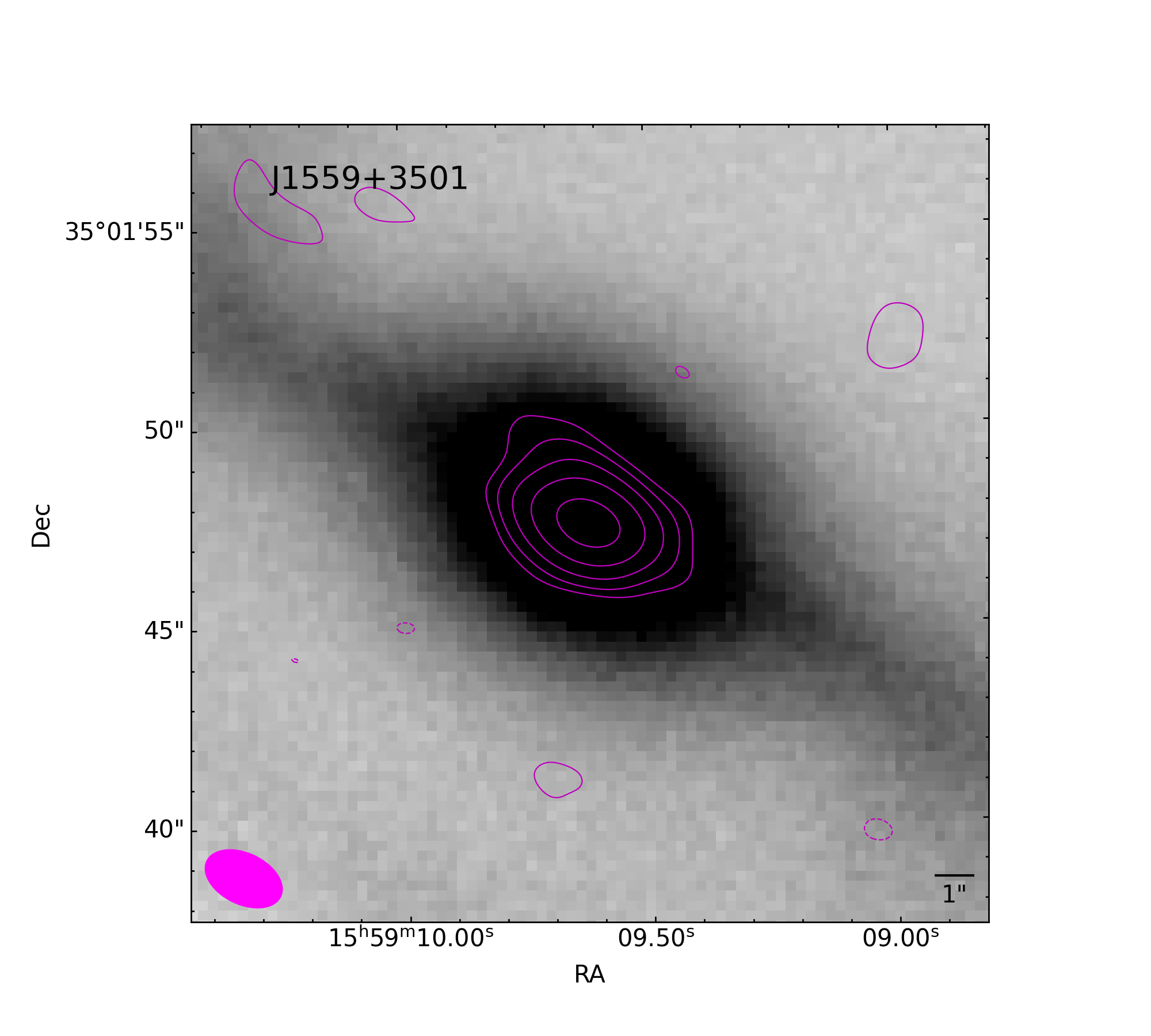}
         \caption{PanSTARRS $i$ band image of the host galaxy overlaid with the 90k$\lambda$ tapered map. Radio map properties as in Fig.~\ref{fig:J1559-90k}}. \label{fig:J1559-host}
     \end{subfigure}
        \caption{}
        \label{fig:J1559}
\end{figure*}


\begin{figure*}
     \centering
     \begin{subfigure}[b]{0.47\textwidth}
         \centering
         \includegraphics[width=\textwidth]{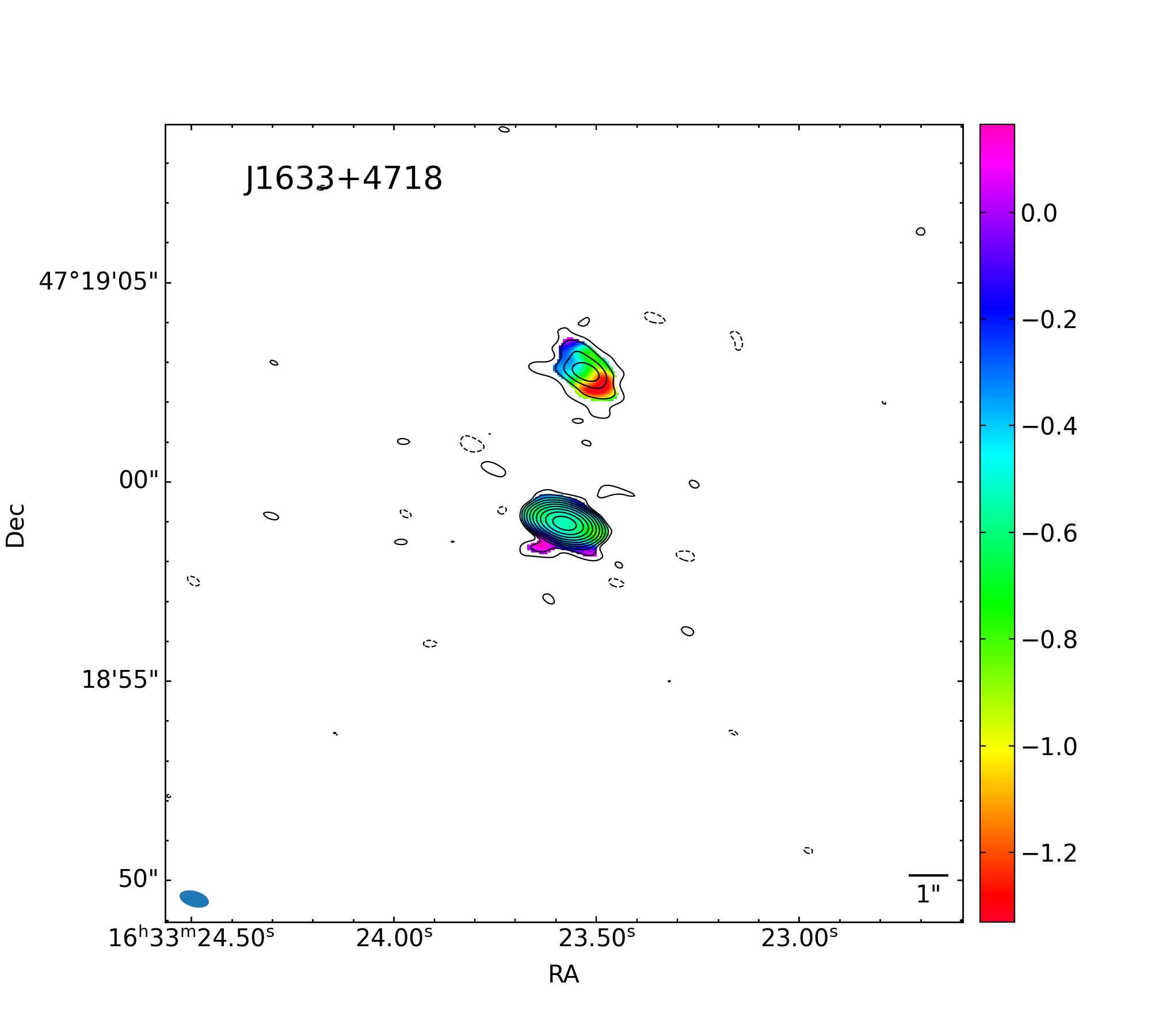}
         \caption{Spectral index map, rms = 10$\mu$Jy beam$^{-1}$, contour levels at -3, 3 $\times$ 2$^n$, $n \in$ [0, 9], beam size 1.60 $\times$ 0.84~kpc. } \label{fig:J1633spind}
     \end{subfigure}
     \hfill
     \begin{subfigure}[b]{0.47\textwidth}
         \centering
         \includegraphics[width=\textwidth]{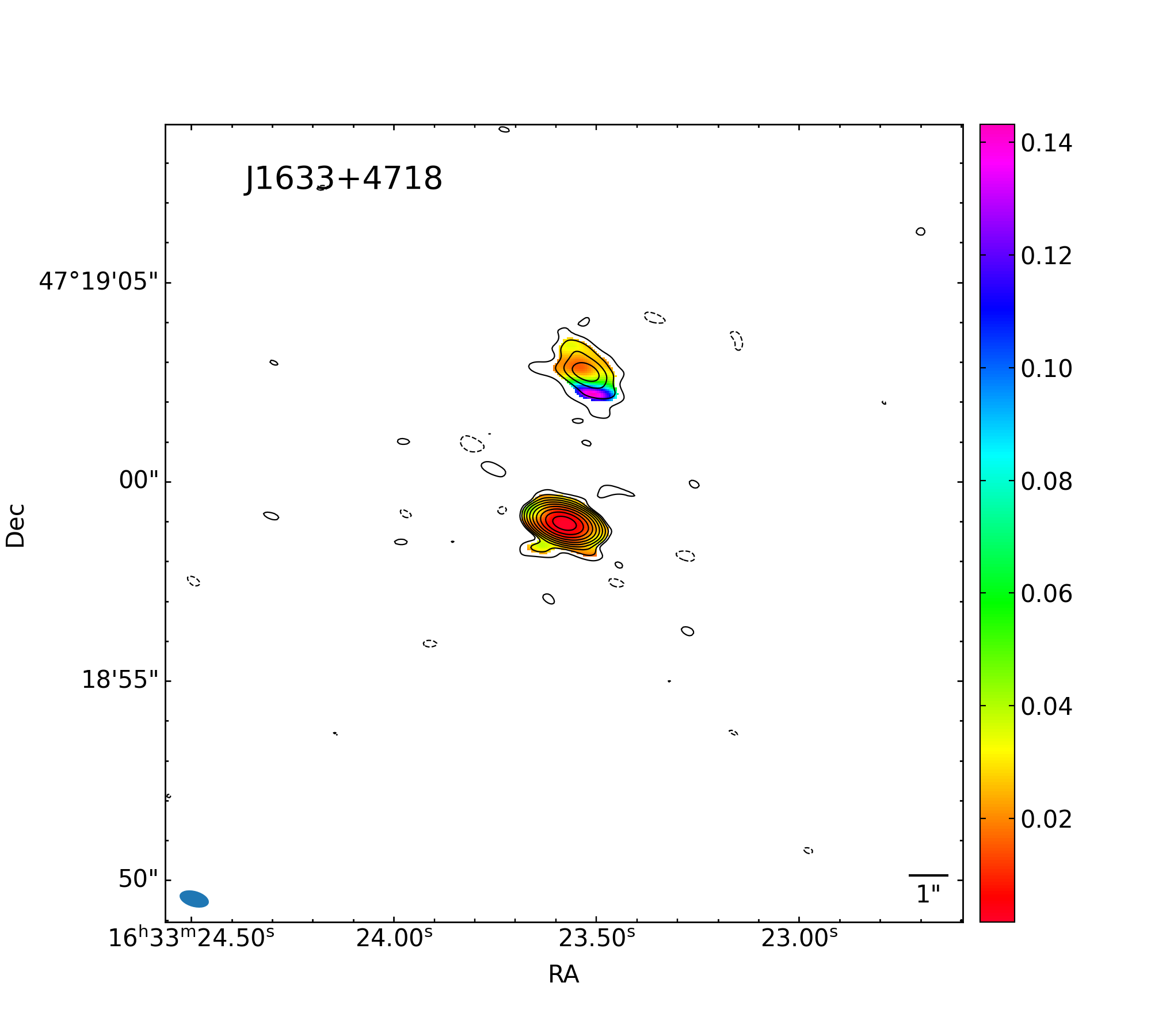}
         \caption{Spectral index error map, rms, contour levels, and beam size as in Fig.~\ref{fig:J1633spind}.} \label{fig:J1633spinderr}
     \end{subfigure}
     \hfill
     \\
     \begin{subfigure}[b]{0.47\textwidth}
         \centering
         \includegraphics[width=\textwidth]{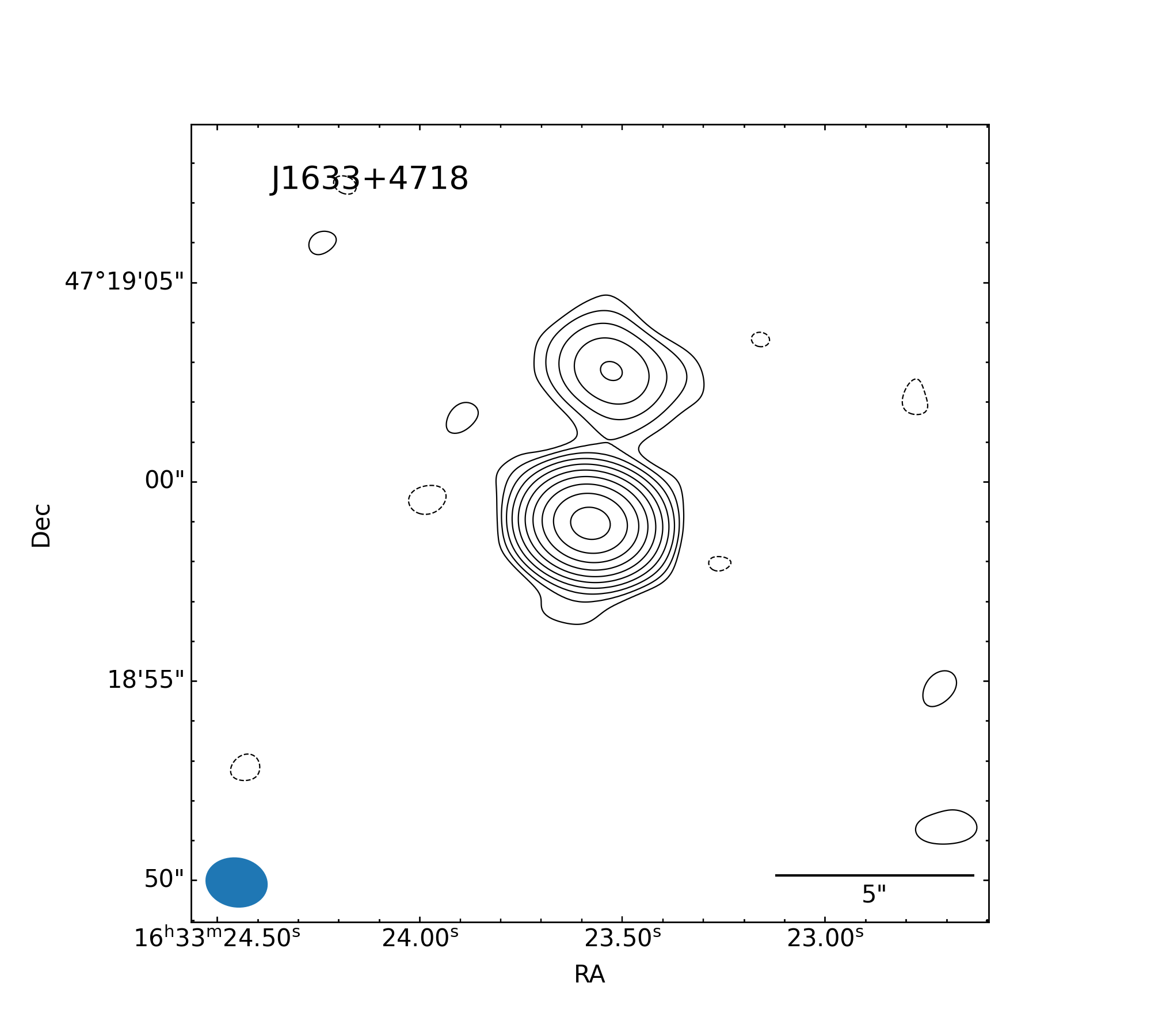}
         \caption{Tapered map with \texttt{uvtaper} = 90k$\lambda$, rms = 12$\mu$Jy beam$^{-1}$, contour levels at -3, 3 $\times$ 2$^n$, $n \in$ [0, 9], beam size 3.30 $\times$ 2.63~kpc.} \label{fig:J1633-90k}
     \end{subfigure}
          \hfill
     \begin{subfigure}[b]{0.47\textwidth}
         \centering
         \includegraphics[width=\textwidth]{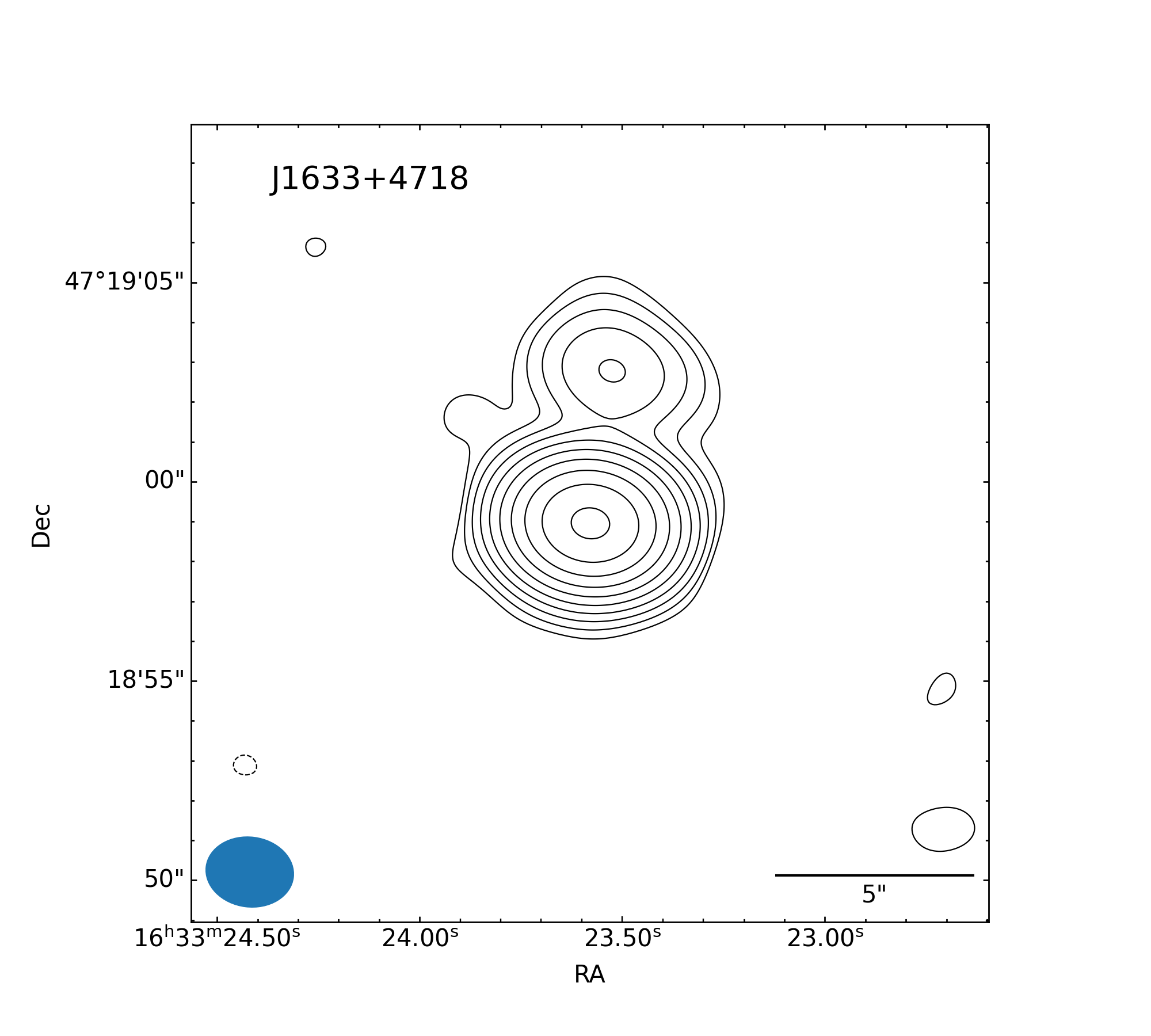}
         \caption{Tapered map with \texttt{uvtaper} = 60k$\lambda$, rms = 14$\mu$Jy beam$^{-1}$, contour levels at -3, 3 $\times$ 2$^n$, $n \in$ [0, 9], beam size 4.69 $\times$ 3.74~kpc.} \label{fig:J1633-60k}
     \end{subfigure}
          \hfill
     \\
     \begin{subfigure}[b]{0.47\textwidth}
         \centering
         \includegraphics[width=\textwidth]{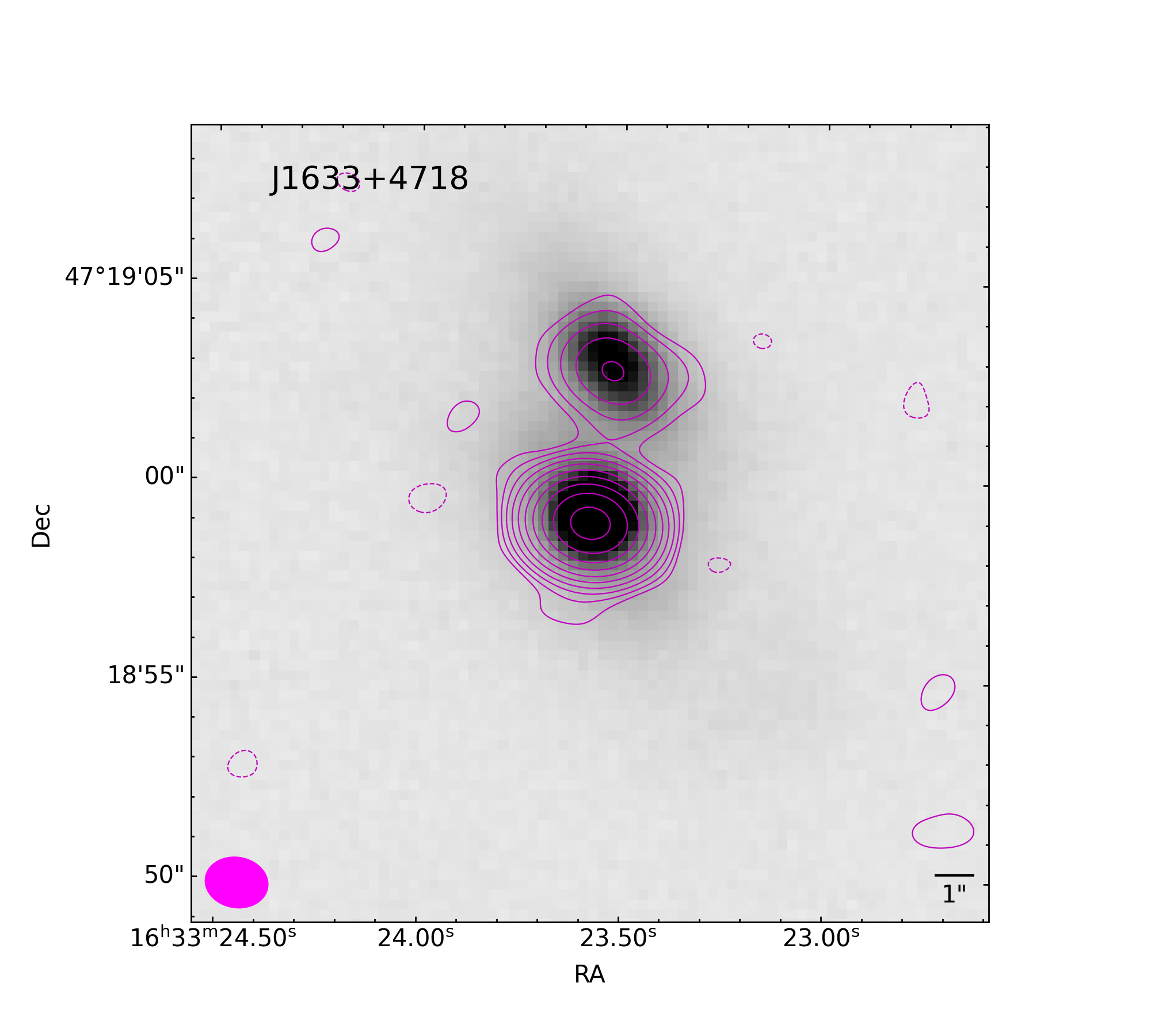}
         \caption{PanSTARRS $i$ band image of the host galaxy overlaid with the 90k$\lambda$ tapered map. Radio map properties as in Fig.~\ref{fig:J1633-90k}}. \label{fig:J1633-host}
     \end{subfigure}
        \caption{}
        \label{fig:J1633}
\end{figure*}


\begin{figure*}
     \centering
     \begin{subfigure}[b]{0.47\textwidth}
         \centering
         \includegraphics[width=\textwidth]{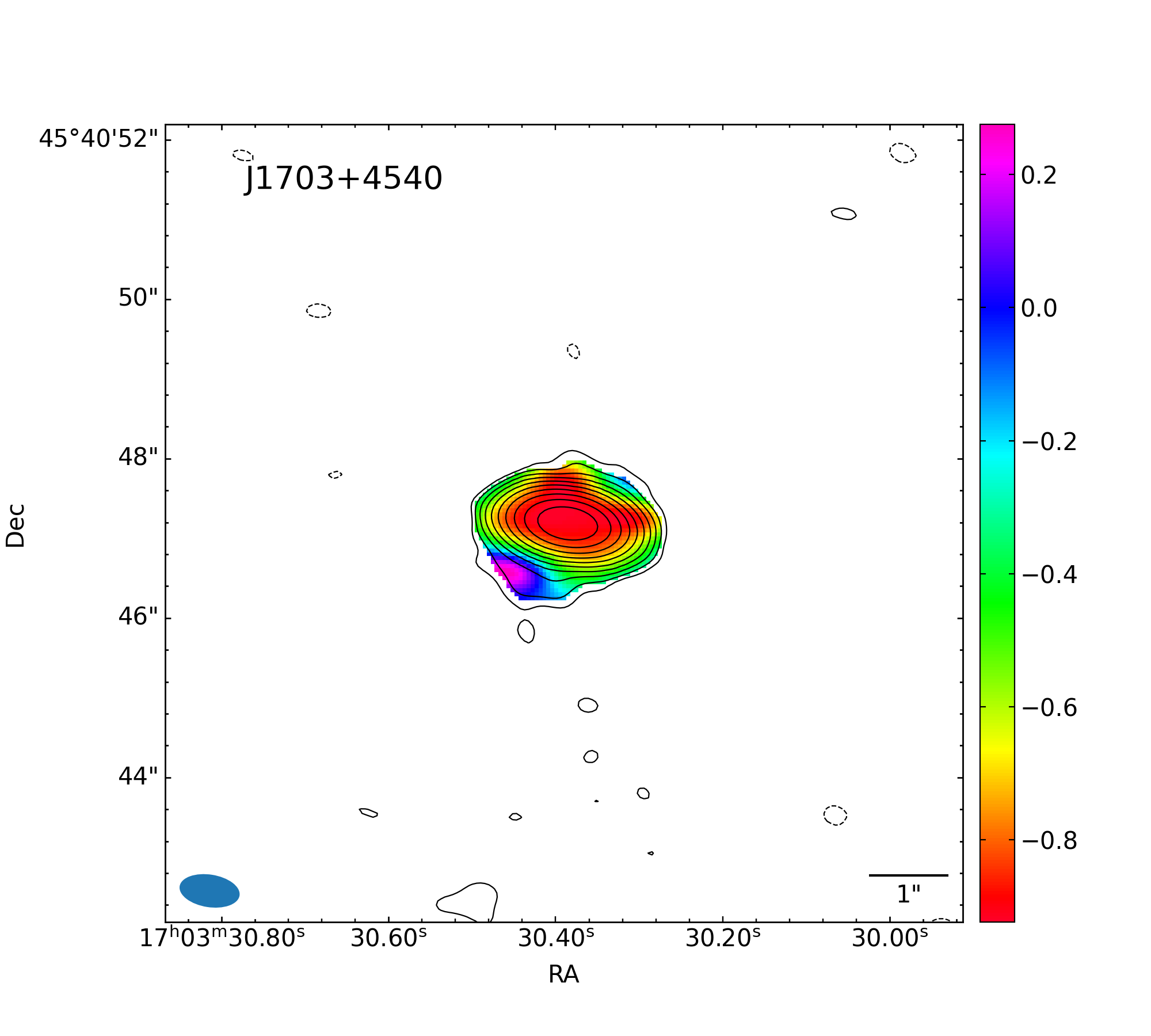}
         \caption{Spectral index map, rms = 11$\mu$Jy beam$^{-1}$, contour levels at -3, 3 $\times$ 2$^n$, $n \in$ [0, 9], beam size 0.88 $\times$ 0.48~kpc. } \label{fig:J1703spind}
     \end{subfigure}
     \hfill
     \begin{subfigure}[b]{0.47\textwidth}
         \centering
         \includegraphics[width=\textwidth]{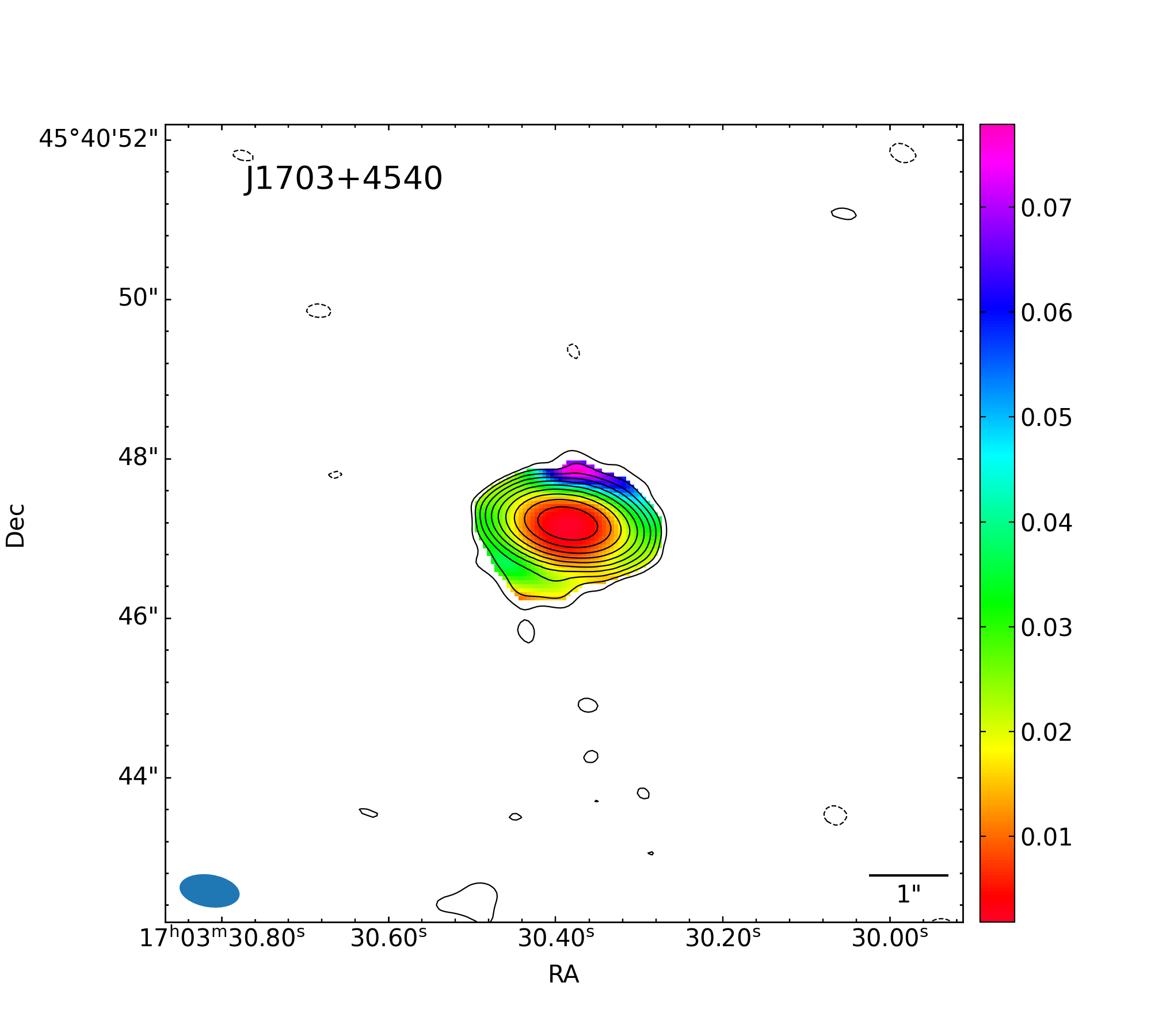}
         \caption{Spectral index error map, rms, contour levels, and beam size as in Fig.~\ref{fig:J1703spind}.} \label{fig:J1703spinderr}
     \end{subfigure}
     \hfill
     \\
     \begin{subfigure}[b]{0.47\textwidth}
         \centering
         \includegraphics[width=\textwidth]{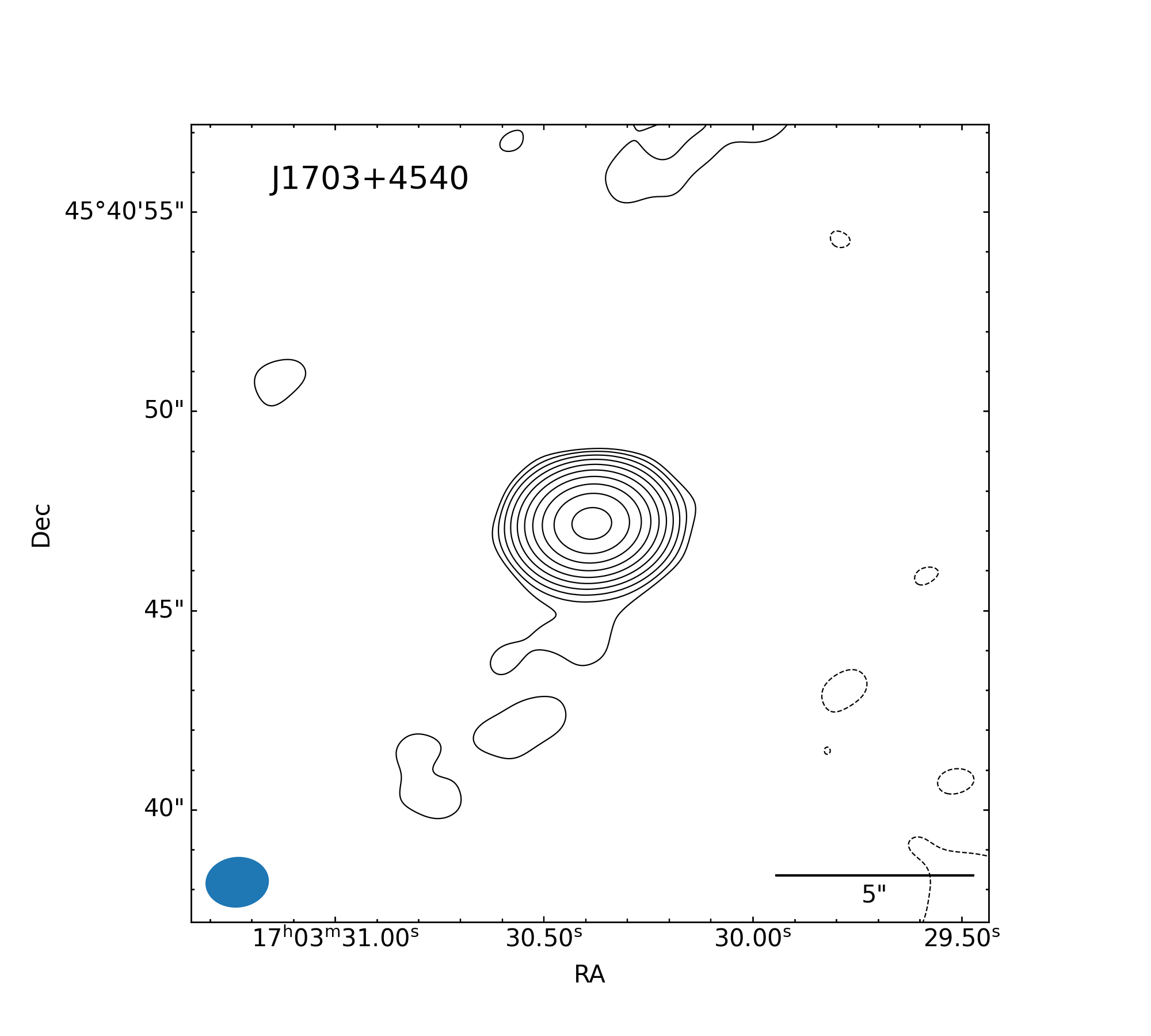}
         \caption{Tapered map with \texttt{uvtaper} = 90k$\lambda$, rms = 17$\mu$Jy beam$^{-1}$, contour levels at -3, 3 $\times$ 2$^n$, $n \in$ [0, 9], beam size 1.85 $\times$ 1.47~kpc.} \label{fig:J1703-90k}
     \end{subfigure}
          \hfill
     \begin{subfigure}[b]{0.47\textwidth}
         \centering
         \includegraphics[width=\textwidth]{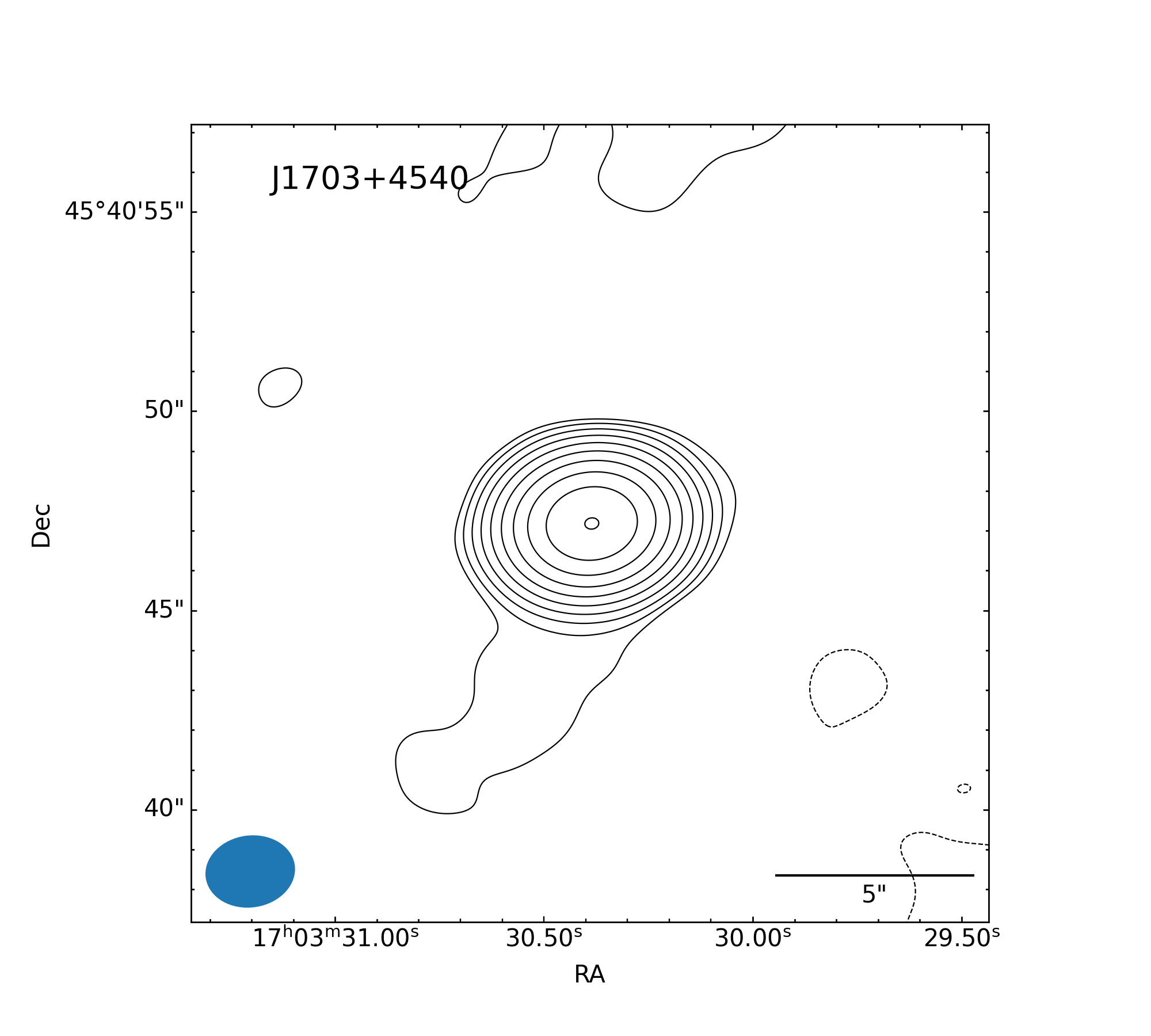}
         \caption{Tapered map with \texttt{uvtaper} = 60k$\lambda$, rms = 22$\mu$Jy beam$^{-1}$, contour levels at -3, 3 $\times$ 2$^n$, $n \in$ [0, 9], beam size 2.62 $\times$ 2.09~kpc.} \label{fig:J1703-60k}
     \end{subfigure}
          \hfill
     \\
     \begin{subfigure}[b]{0.47\textwidth}
         \centering
         \includegraphics[width=\textwidth]{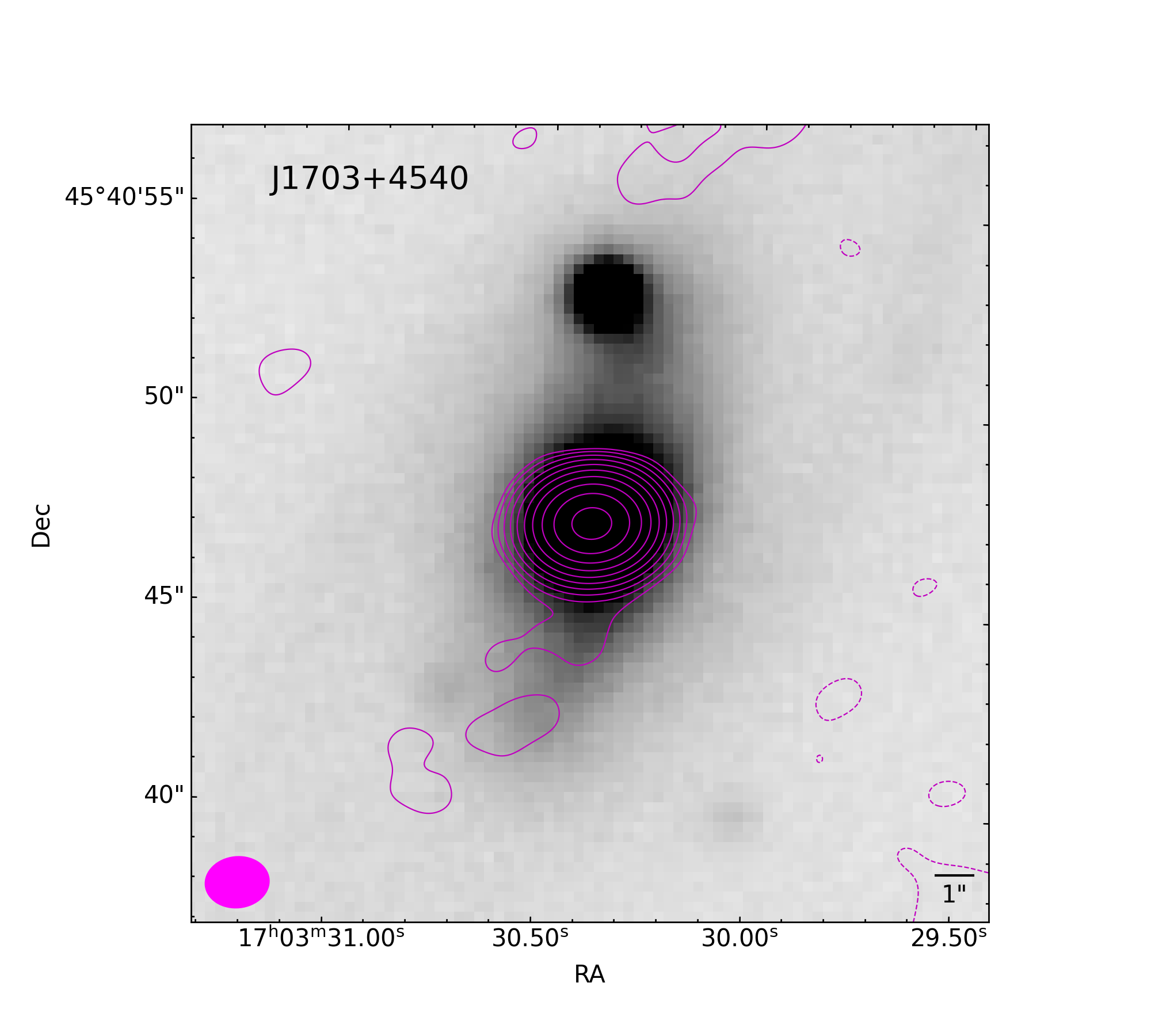}
         \caption{PanSTARRS $i$ band image of the host galaxy overlaid with the 90k$\lambda$ tapered map. Radio map properties as in Fig.~\ref{fig:J1703-90k}}. \label{fig:J1703-host}
     \end{subfigure}
        \caption{}
        \label{fig:J1703}
\end{figure*}


\begin{figure*}
     \centering
     \begin{subfigure}[b]{0.47\textwidth}
         \centering
         \includegraphics[width=\textwidth]{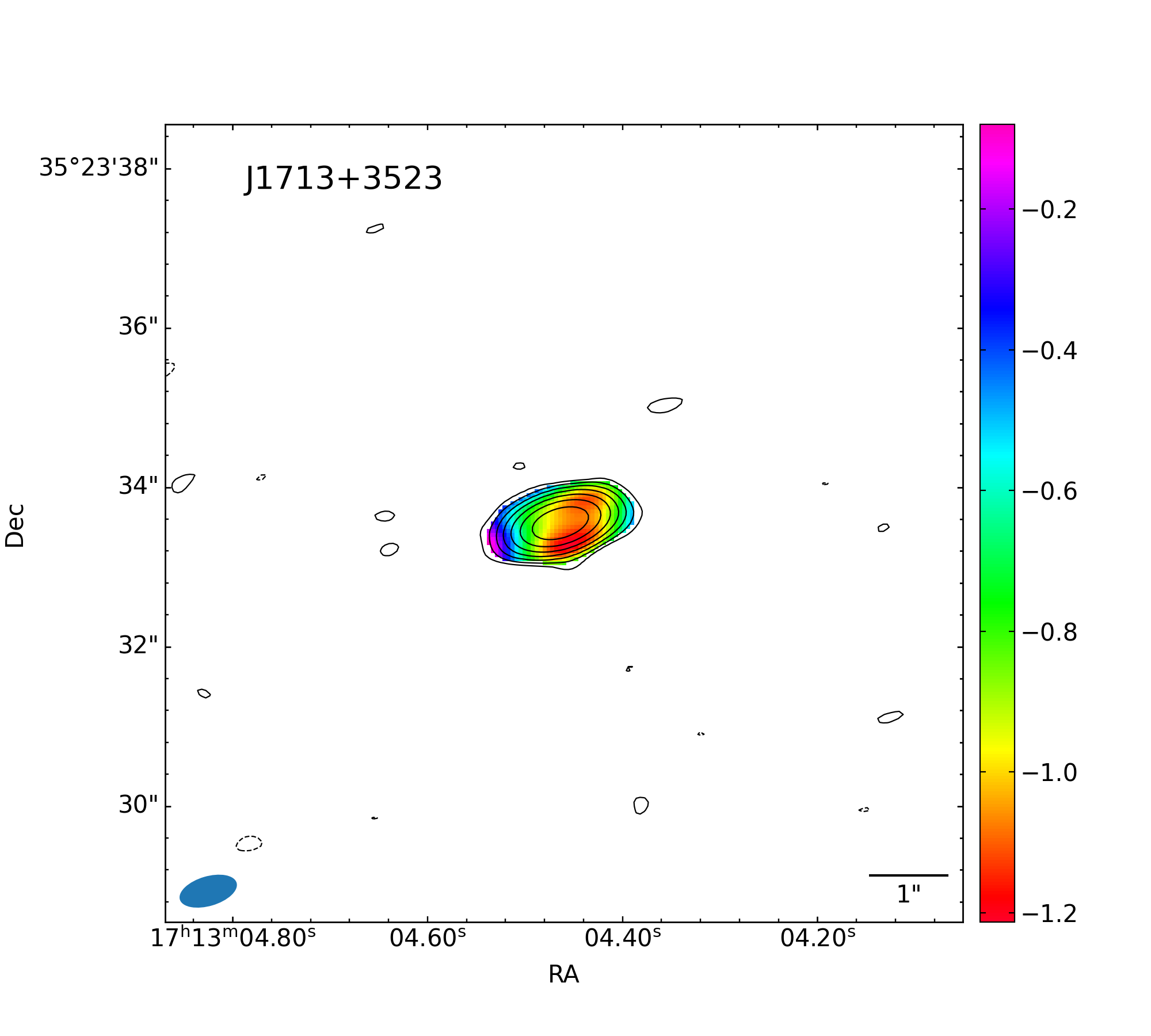}
         \caption{Spectral index map, rms = 10$\mu$Jy beam$^{-1}$, contour levels at -3, 3 $\times$ 2$^n$, $n \in$ [0, 6], beam size 1.16 $\times$ 0.58~kpc. } \label{fig:J1713spind}
     \end{subfigure}
     \hfill
     \begin{subfigure}[b]{0.47\textwidth}
         \centering
         \includegraphics[width=\textwidth]{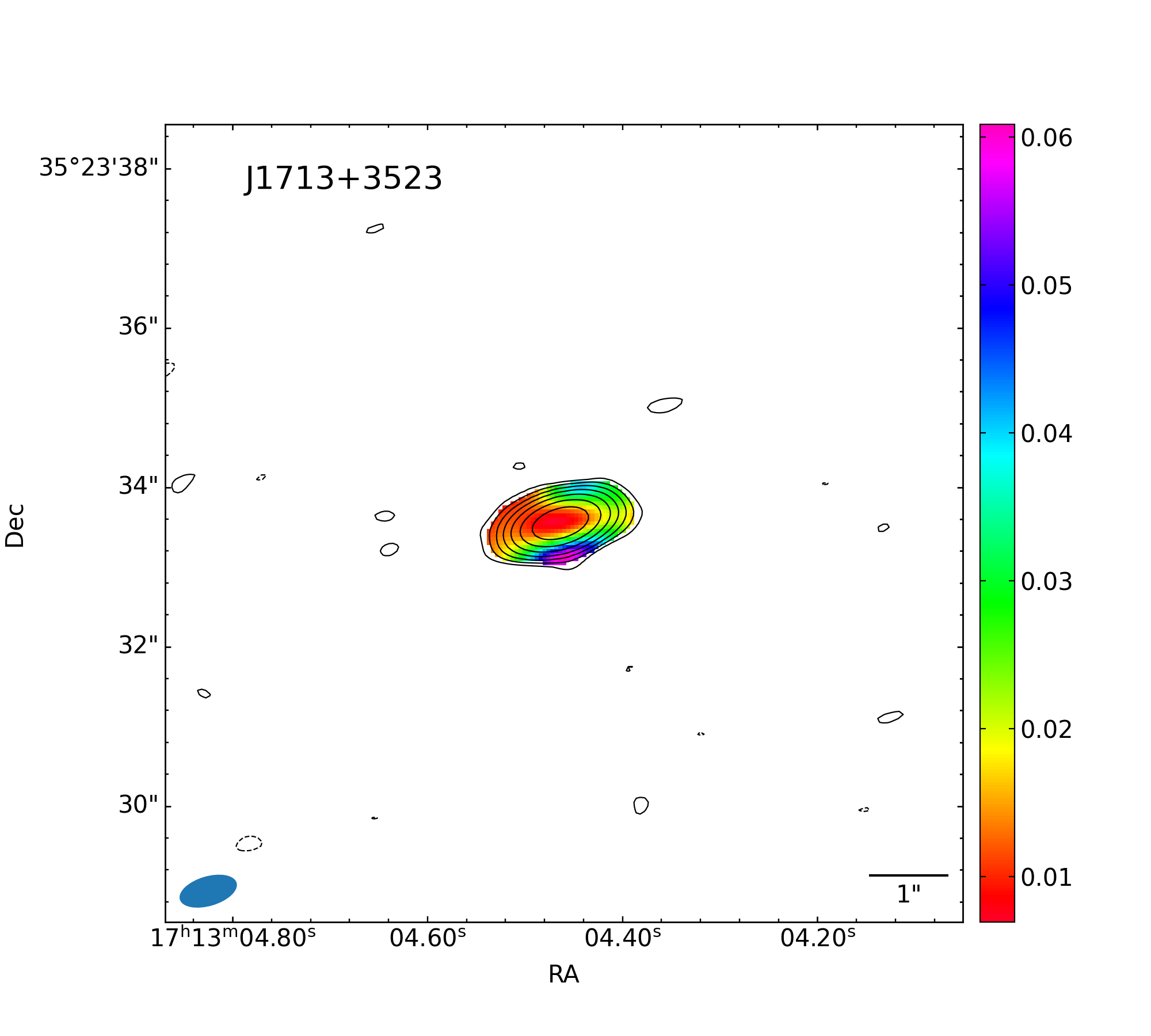}
         \caption{Spectral index error map, rms, contour levels, and beam size as in Fig.~\ref{fig:J1713spind}.} \label{fig:J1713spinderr}
     \end{subfigure}
     \hfill
     \\
     \begin{subfigure}[b]{0.47\textwidth}
         \centering
         \includegraphics[width=\textwidth]{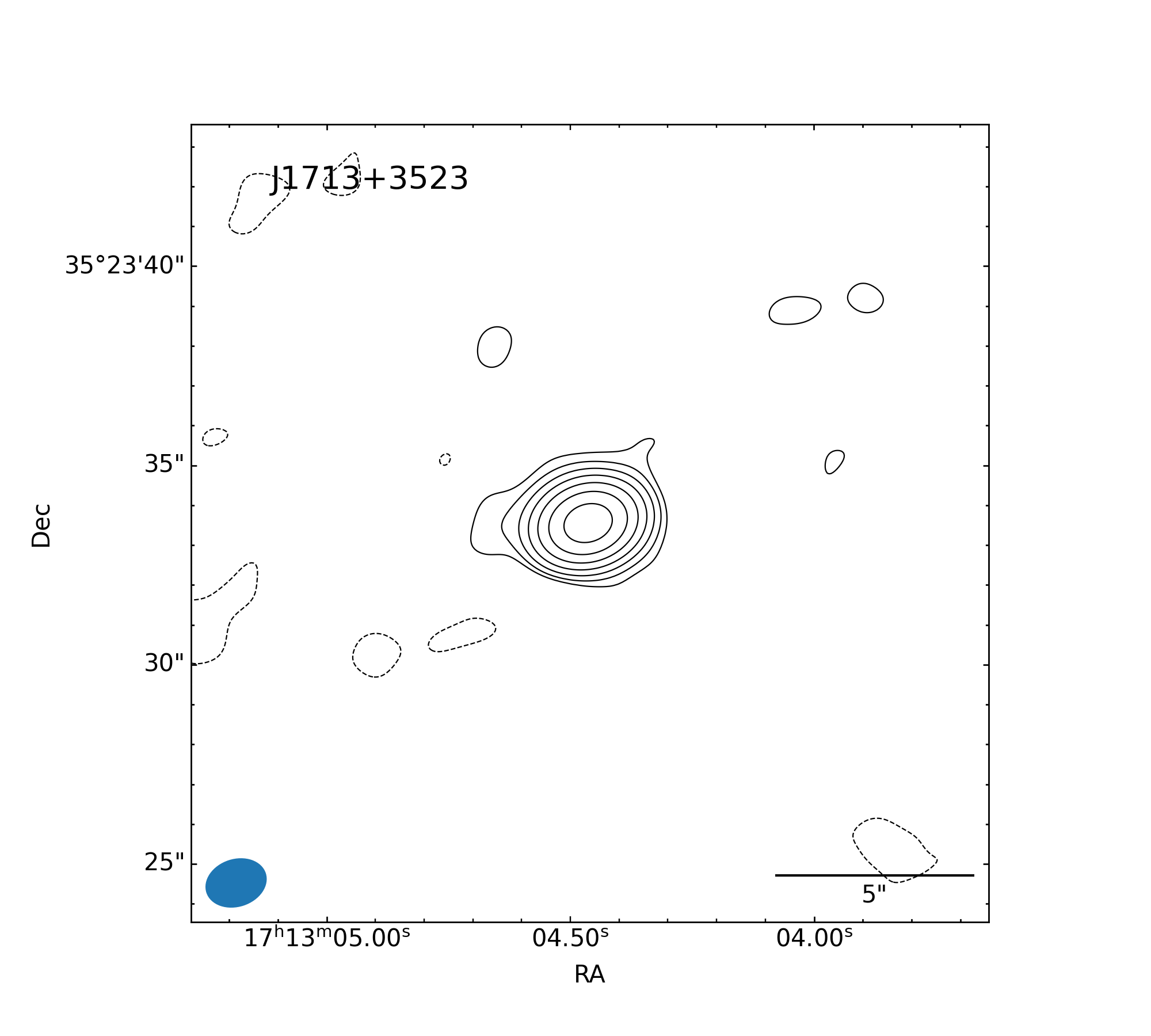}
         \caption{Tapered map with \texttt{uvtaper} = 90k$\lambda$, rms = 13$\mu$Jy beam$^{-1}$, contour levels at -3, 3 $\times$ 2$^n$, $n \in$ [0, 6], beam size 2.45 $\times$ 1.87~kpc.} \label{fig:J1713-90k}
     \end{subfigure}
          \hfill
     \begin{subfigure}[b]{0.47\textwidth}
         \centering
         \includegraphics[width=\textwidth]{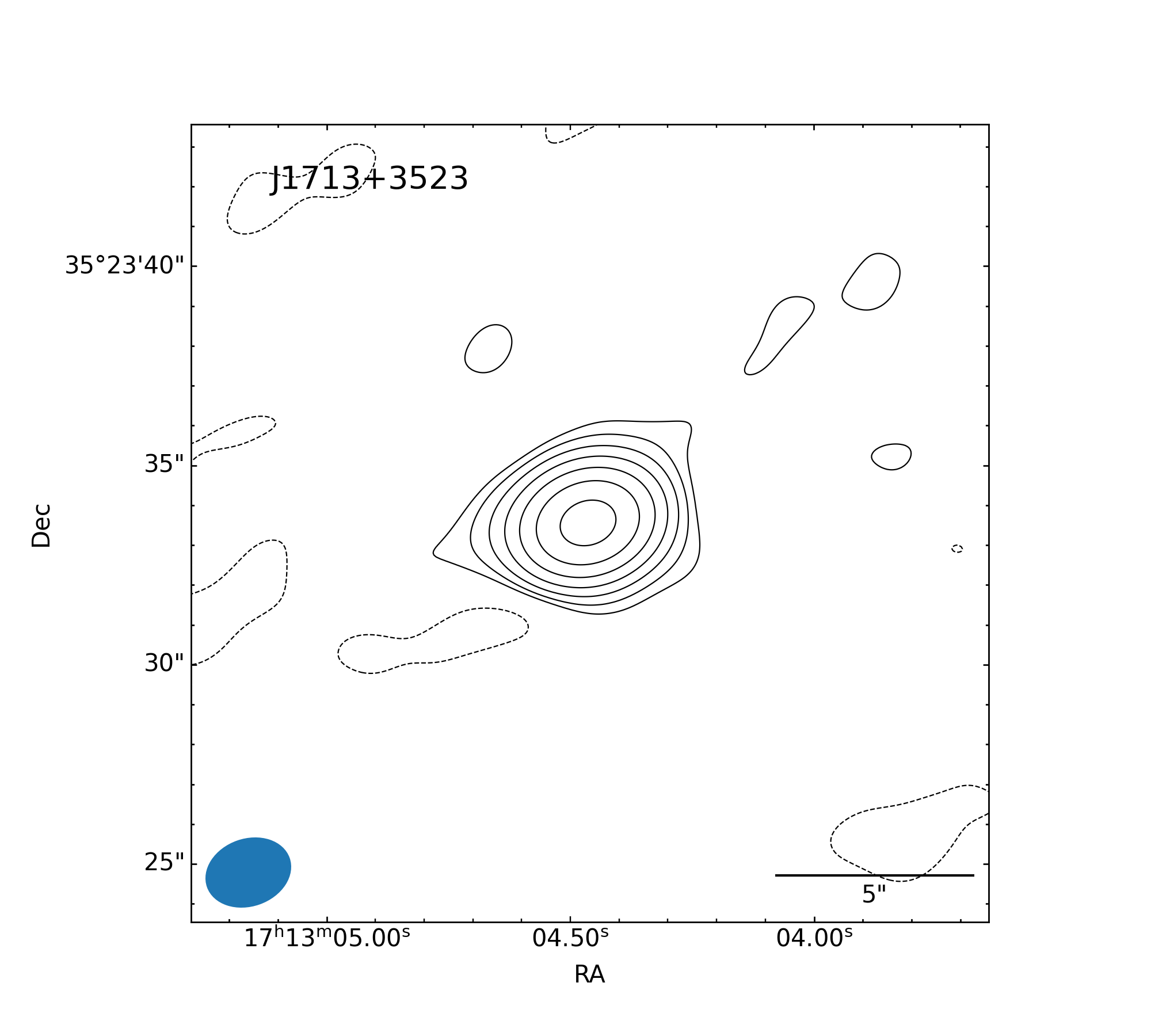}
         \caption{Tapered map with \texttt{uvtaper} = 60k$\lambda$, rms = 15$\mu$Jy beam$^{-1}$, contour levels at -3, 3 $\times$ 2$^n$, $n \in$ [0, 6], beam size 3.43 $\times$ 2.65~kpc.} \label{fig:J1713-60k}
     \end{subfigure}
          \hfill
     \\
     \begin{subfigure}[b]{0.47\textwidth}
         \centering
         \includegraphics[width=\textwidth]{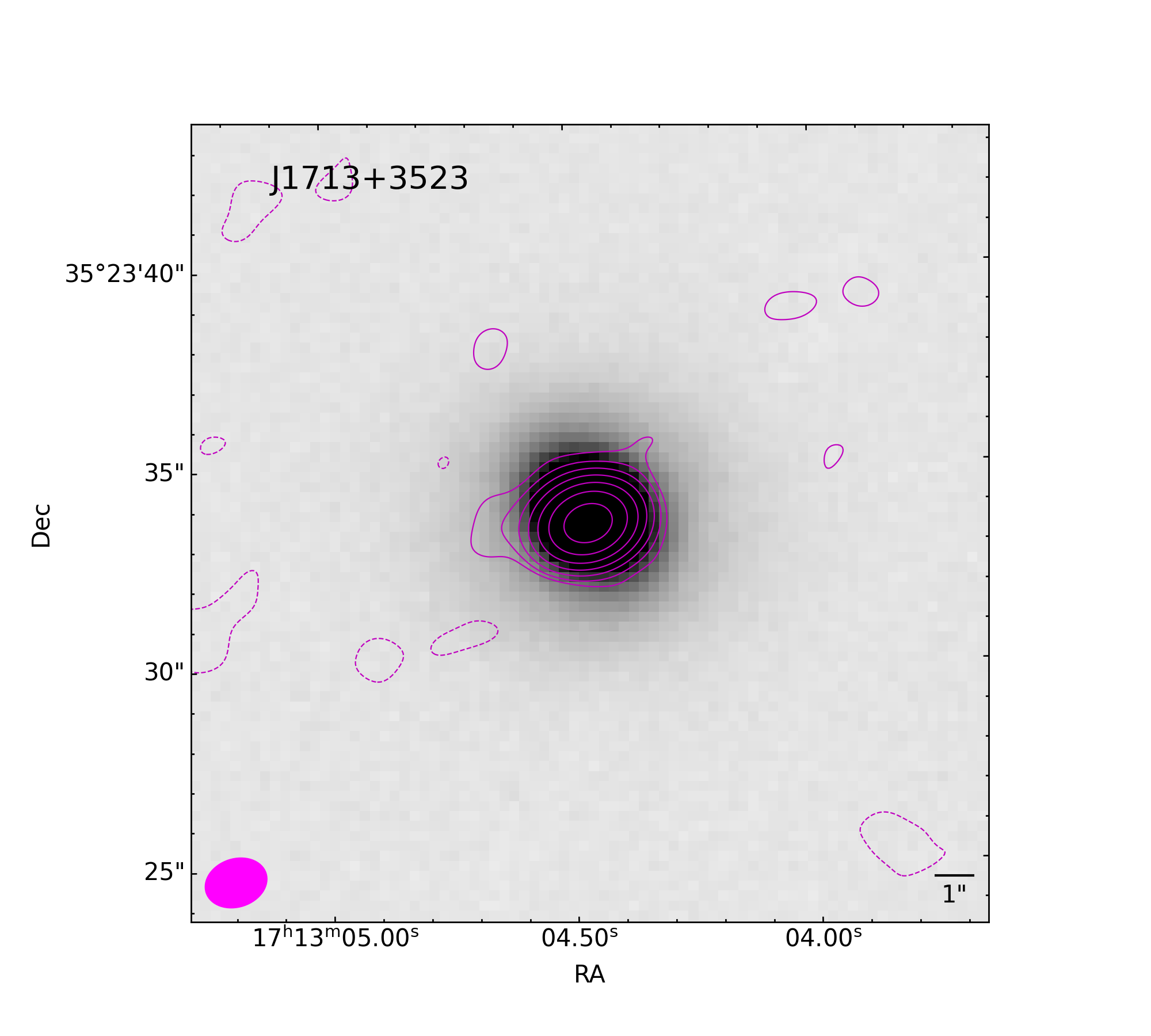}
         \caption{PanSTARRS $i$ band image of the host galaxy overlaid with the 90k$\lambda$ tapered map. Radio map properties as in Fig.~\ref{fig:J1713-90k}}. \label{fig:J1713-host}
     \end{subfigure}
        \caption{}
        \label{fig:J1713}
\end{figure*}


\begin{figure*}
     \centering
     \begin{subfigure}[b]{0.47\textwidth}
         \centering
         \includegraphics[width=\textwidth]{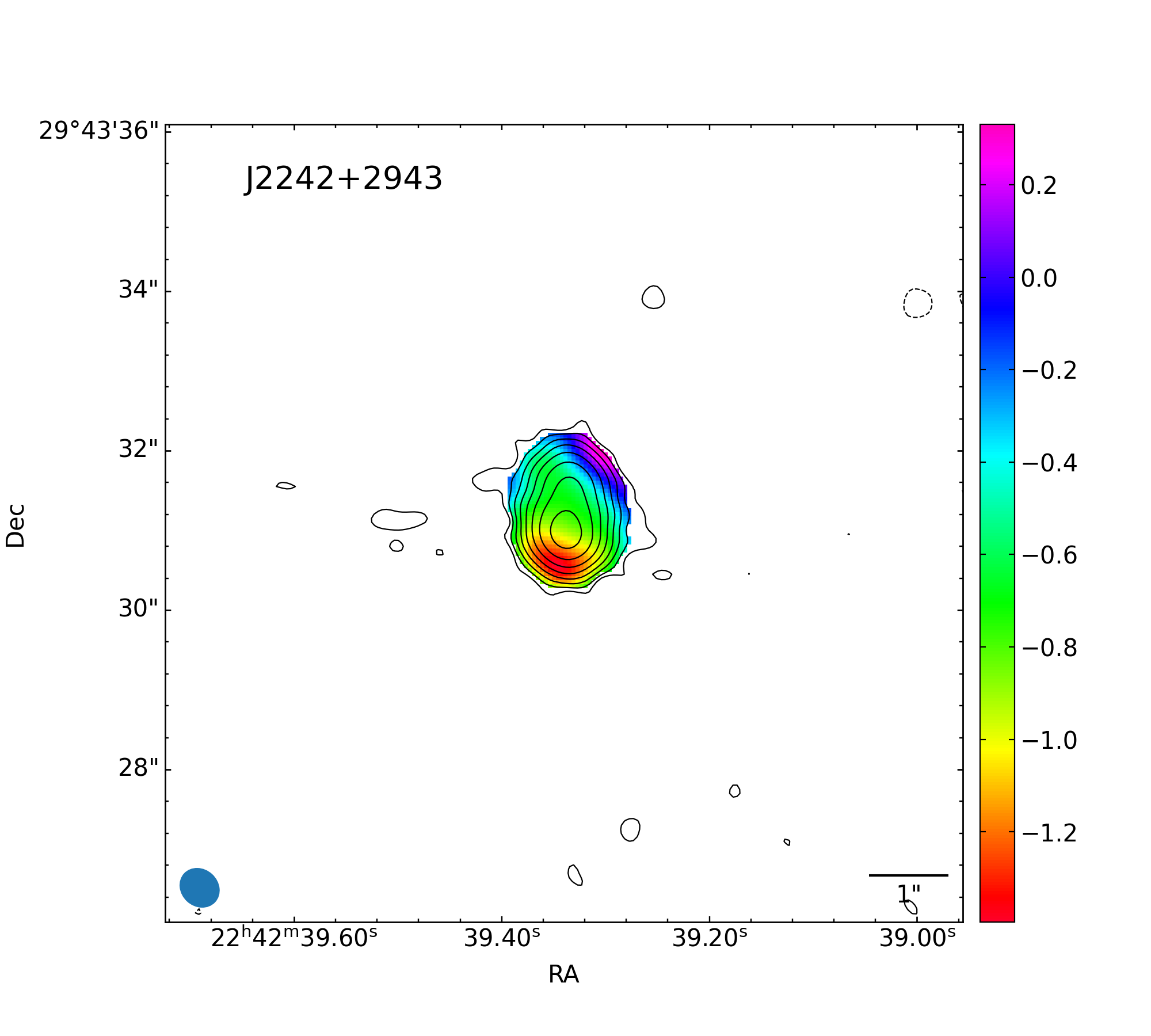}
         \caption{Spectral index map, rms = 11$\mu$Jy beam$^{-1}$, contour levels at -3, 3 $\times$ 2$^n$, $n \in$ [0, 7], beam size 0.27 $\times$ 0.24~kpc. } \label{fig:J2242spind}
     \end{subfigure}
     \hfill
     \begin{subfigure}[b]{0.47\textwidth}
         \centering
         \includegraphics[width=\textwidth]{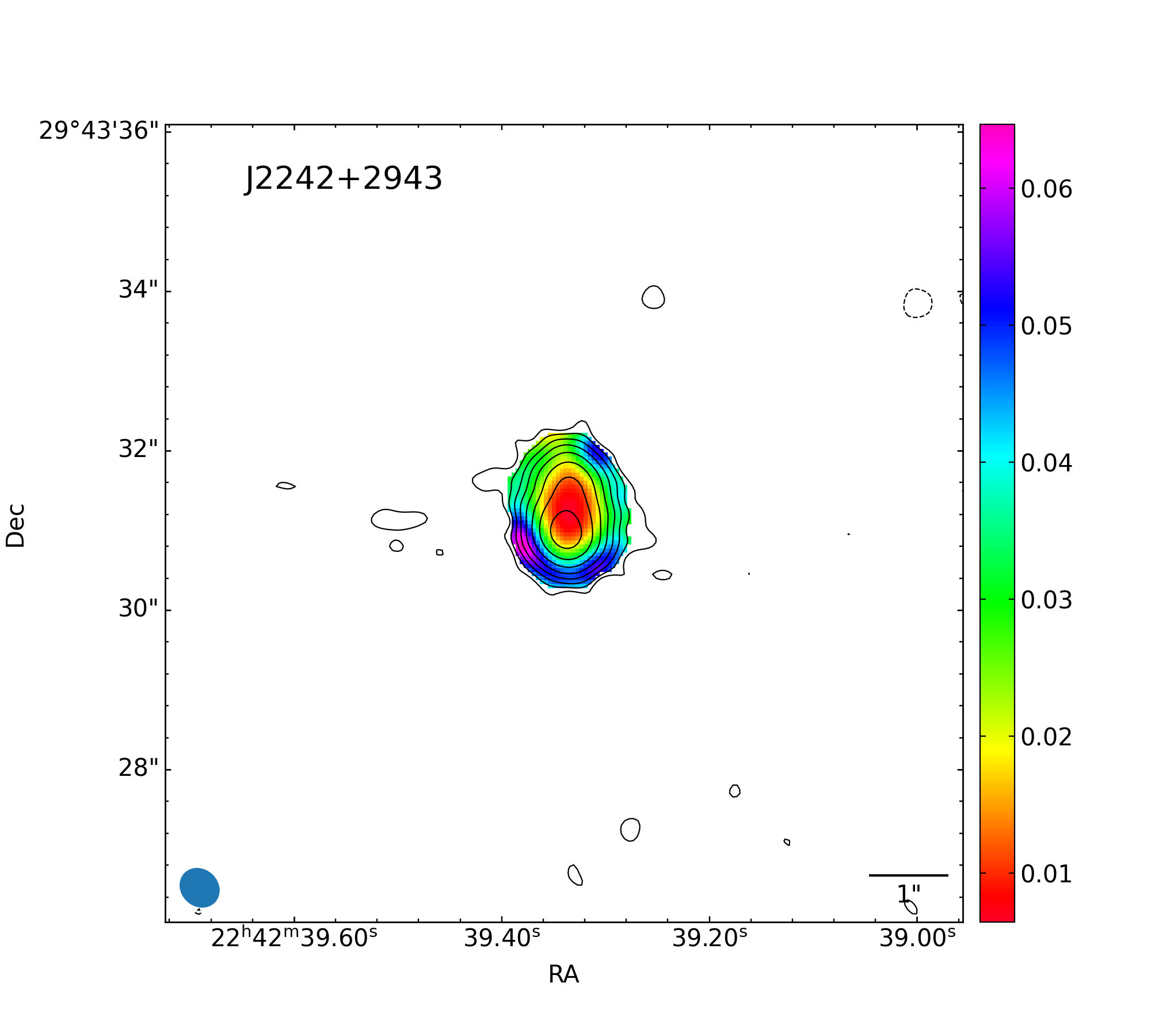}
         \caption{Spectral index error map, rms, contour levels, and beam size as in Fig.~\ref{fig:J2242spind}.} \label{fig:J2242spinderr}
     \end{subfigure}
     \hfill
     \\
     \begin{subfigure}[b]{0.47\textwidth}
         \centering
         \includegraphics[width=\textwidth]{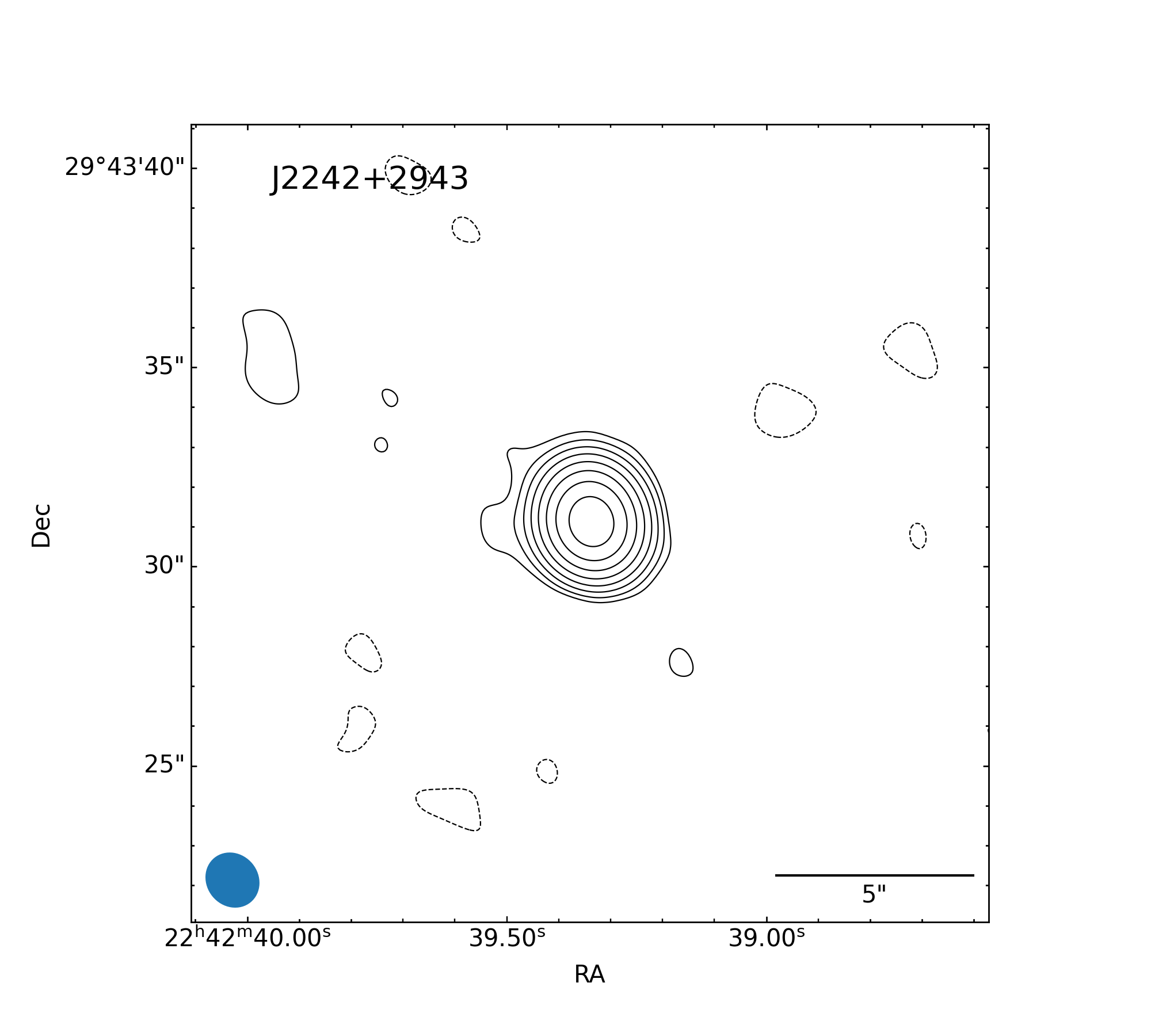}
         \caption{Tapered map with \texttt{uvtaper} = 90k$\lambda$, rms = 15$\mu$Jy beam$^{-1}$, contour levels at -3, 3 $\times$ 2$^n$, $n \in$ [0, 7], beam size 0.73 $\times$ 0.66~kpc.} \label{fig:J2242-90k}
     \end{subfigure}
          \hfill
     \begin{subfigure}[b]{0.47\textwidth}
         \centering
         \includegraphics[width=\textwidth]{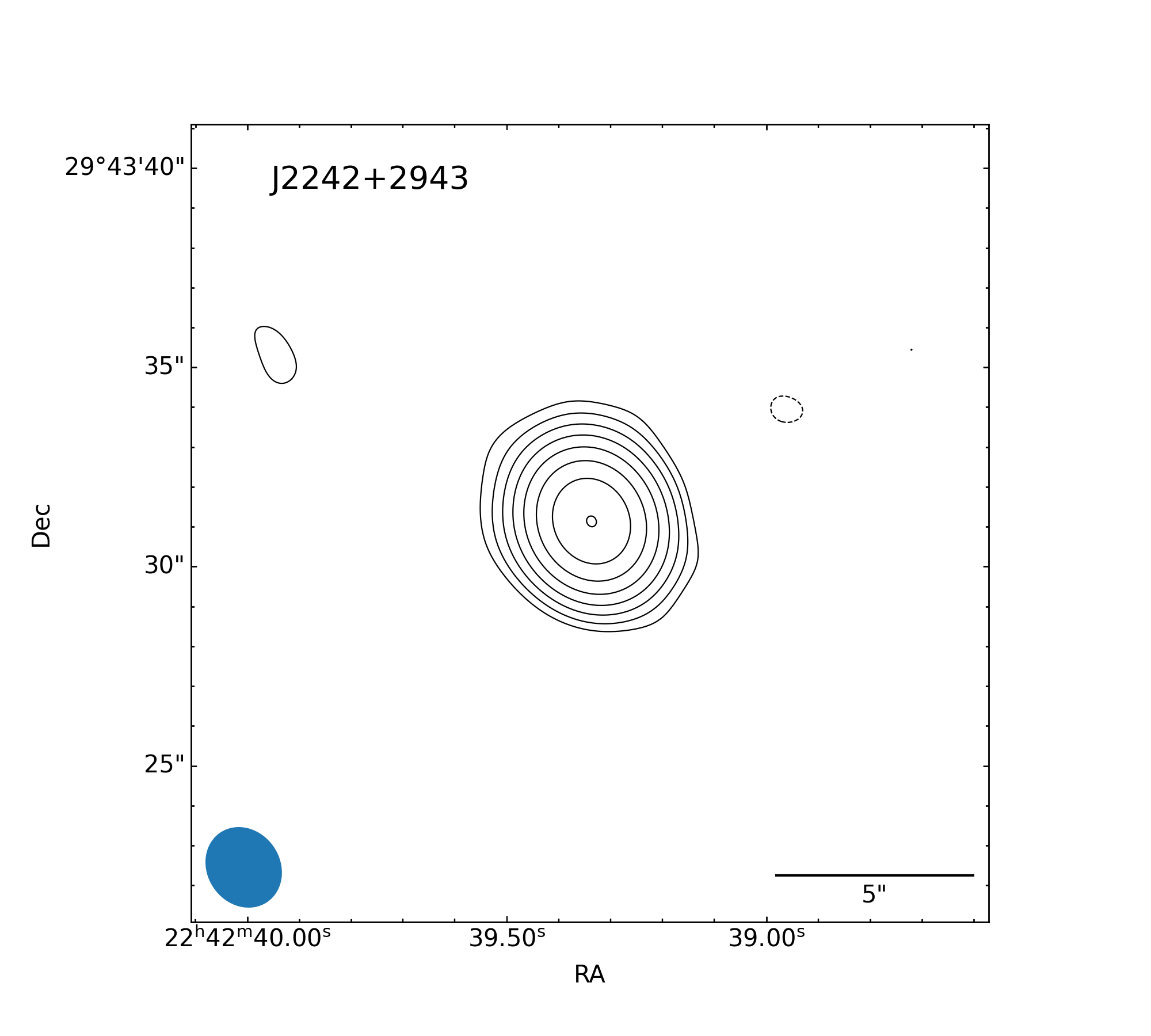}
         \caption{Tapered map with \texttt{uvtaper} = 60k$\lambda$, rms = 25$\mu$Jy beam$^{-1}$, contour levels at -3, 3 $\times$ 2$^n$, $n \in$ [0, 7], beam size 1.06 $\times$ 0.92~kpc.} \label{fig:J2242-60k}
     \end{subfigure}
          \hfill
     \\
     \begin{subfigure}[b]{0.47\textwidth}
         \centering
         \includegraphics[width=\textwidth]{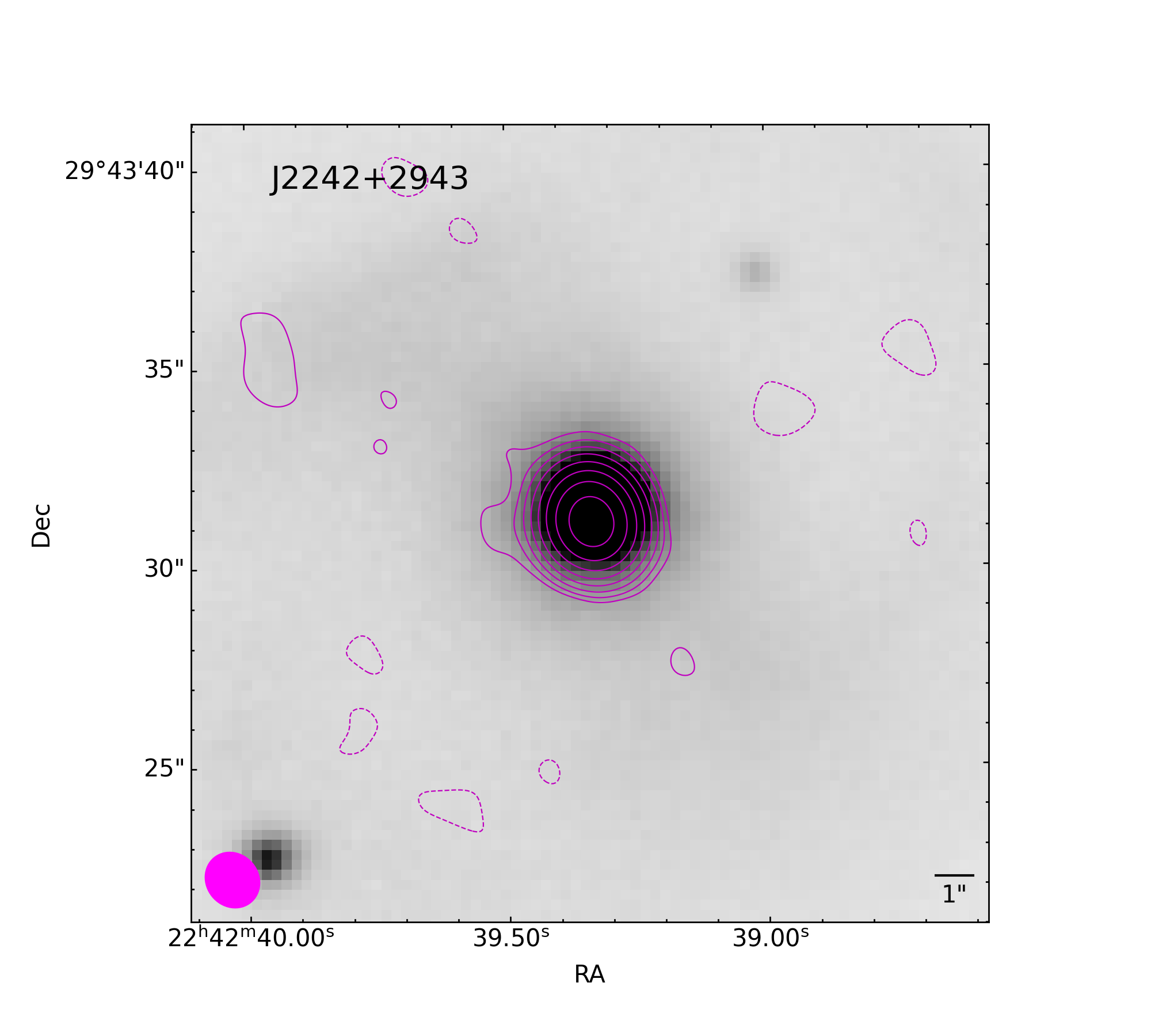}
         \caption{PanSTARRS $i$ band image of the host galaxy overlaid with the 90k$\lambda$ tapered map. Radio map properties as in Fig.~\ref{fig:J2242-90k}}. \label{fig:J2242-host}
     \end{subfigure}
        \caption{}
        \label{fig:J2242}
\end{figure*}

\clearpage
\begin{figure*}
     \centering
     \begin{subfigure}[b]{0.47\textwidth}
         \centering
         \includegraphics[width=\textwidth]{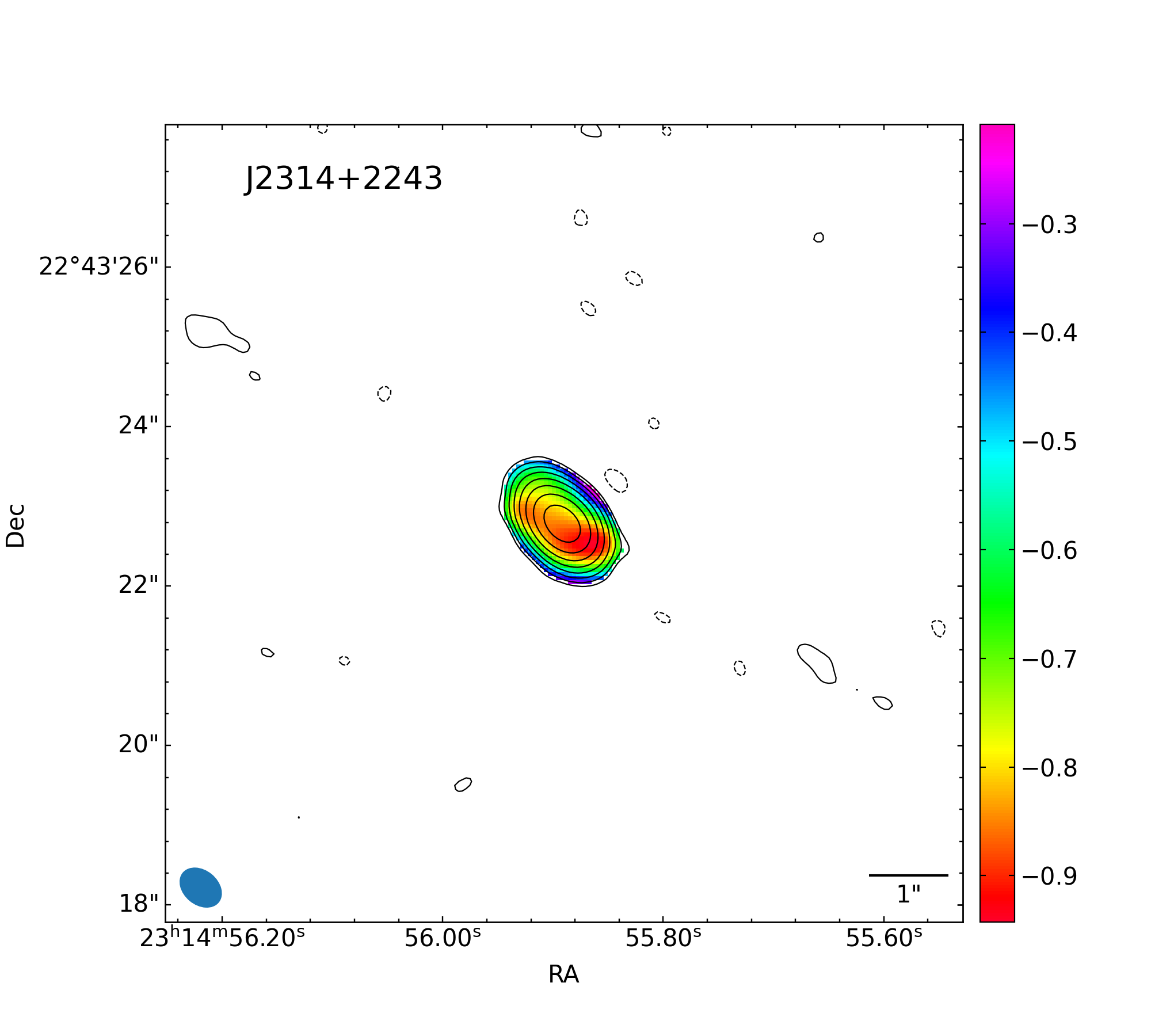}
         \caption{Spectral index map, rms = 10$\mu$Jy beam$^{-1}$, contour levels at -3, 3 $\times$ 2$^n$, $n \in$ [0, 7], beam size 1.70 $\times$ 1.27~kpc. } \label{fig:J2314spind}
     \end{subfigure}
     \hfill
     \begin{subfigure}[b]{0.47\textwidth}
         \centering
         \includegraphics[width=\textwidth]{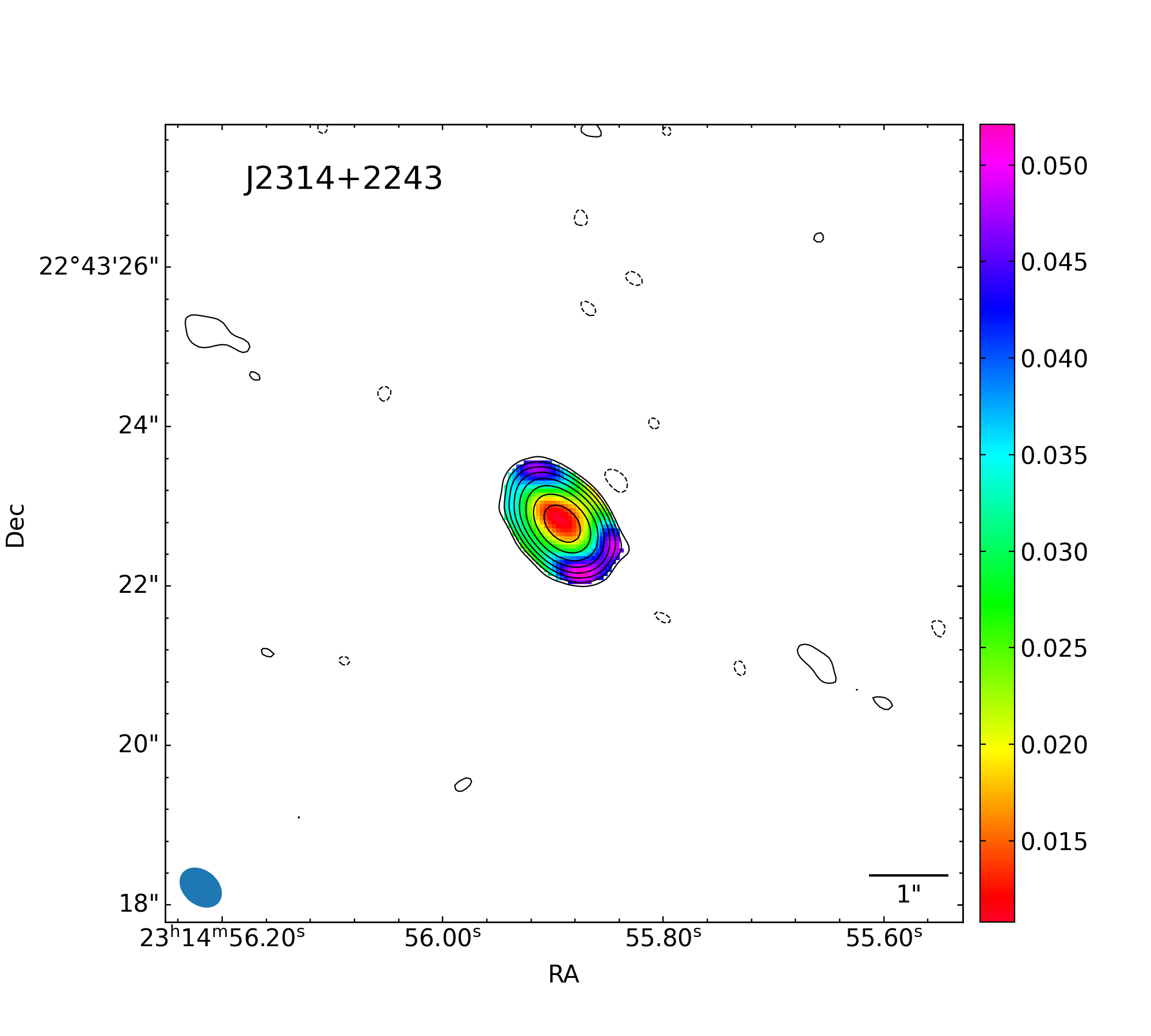}
         \caption{Spectral index error map, rms, contour levels, and beam size as in Fig.~\ref{fig:J2314spind}.} \label{fig:J2314spinderr}
     \end{subfigure}
     \hfill
     \\
     \begin{subfigure}[b]{0.47\textwidth}
         \centering
         \includegraphics[width=\textwidth]{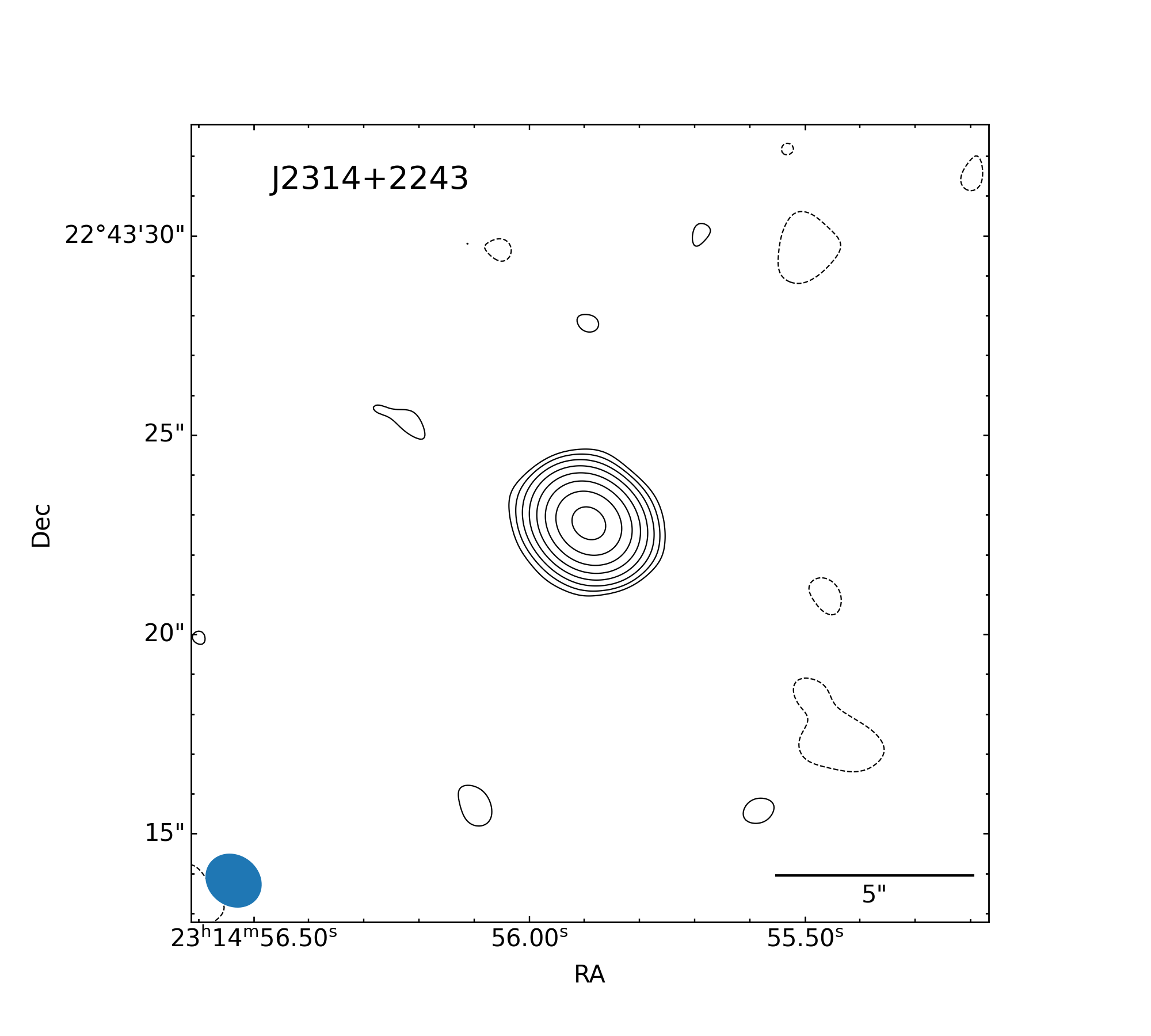}
         \caption{Tapered map with \texttt{uvtaper} = 90k$\lambda$, rms = 14$\mu$Jy beam$^{-1}$, contour levels at -3, 3 $\times$ 2$^n$, $n \in$ [0, 7], beam size 4.27 $\times$ 3.69~kpc.} \label{fig:J2314-90k}
     \end{subfigure}
          \hfill
     \begin{subfigure}[b]{0.47\textwidth}
         \centering
         \includegraphics[width=\textwidth]{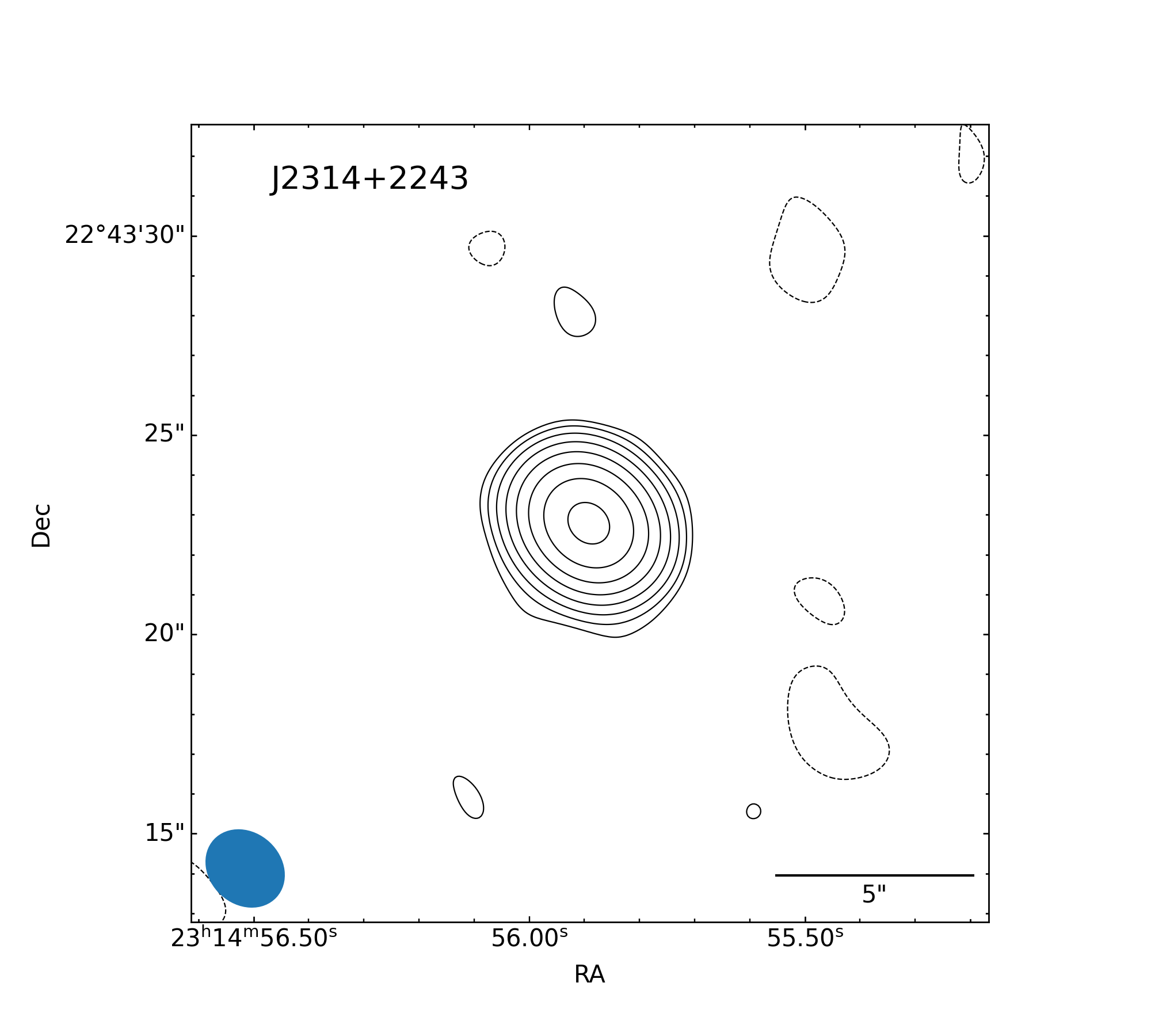}
         \caption{Tapered map with \texttt{uvtaper} = 60k$\lambda$, rms = 15$\mu$Jy beam$^{-1}$, contour levels at -3, 3 $\times$ 2$^n$, $n \in$ [0, 7], beam size 6.17 $\times$ 5.19~kpc.} \label{fig:J2314-60k}
     \end{subfigure}
          \hfill
     \\
     \begin{subfigure}[b]{0.47\textwidth}
         \centering
         \includegraphics[width=\textwidth]{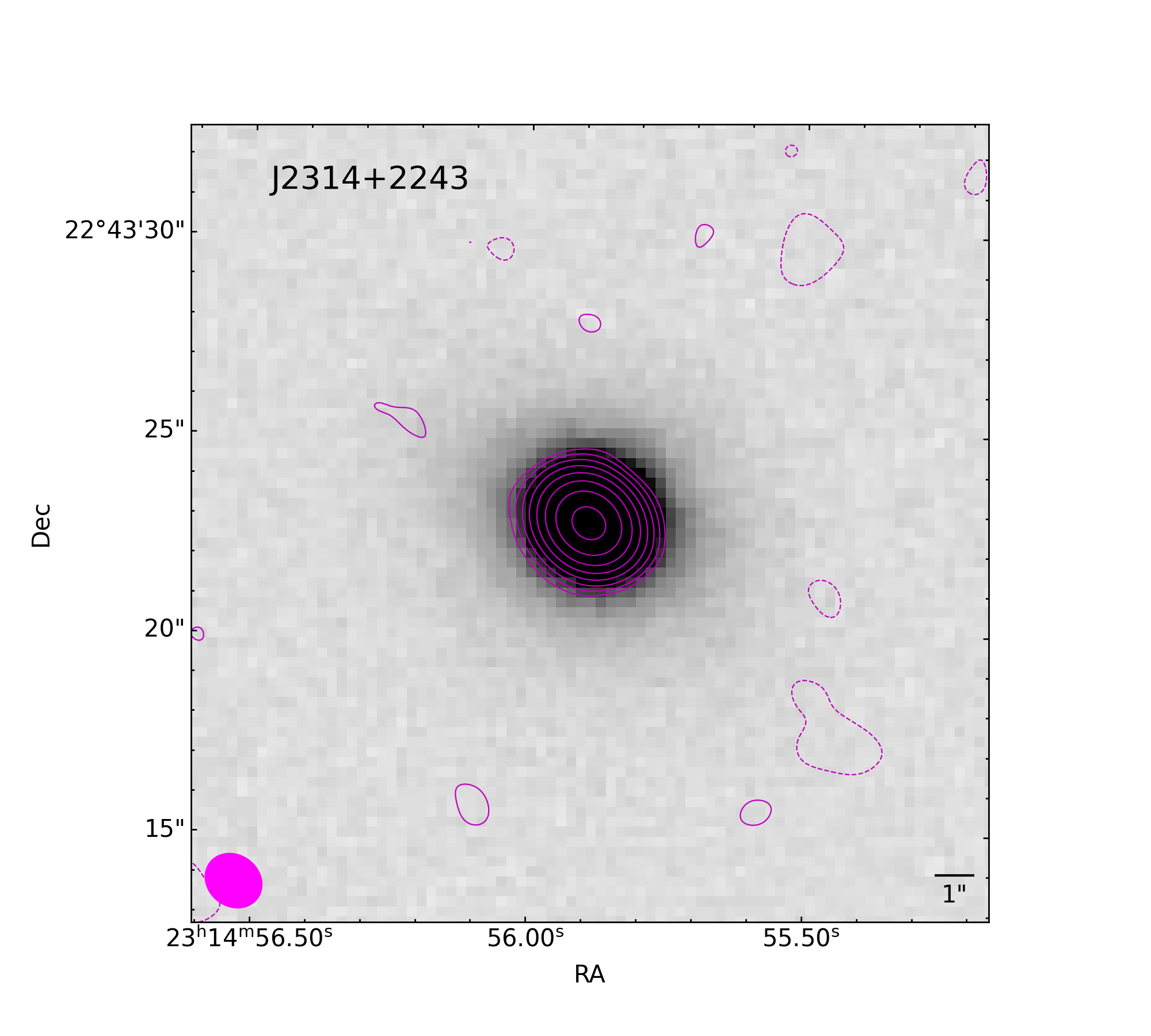}
         \caption{PanSTARRS $i$ band image of the host galaxy overlaid with the 90k$\lambda$ tapered map. Radio map properties as in Fig.~\ref{fig:J2314-90k}}. \label{fig:J2314-host}
     \end{subfigure}
        \caption{}
        \label{fig:J2314}
\end{figure*}

\end{document}